\def\ThesisType{mgr}

\def\StudyLanguage{en}

\def\ThesisTitle{Page image classification for content-specific data processing}

\def\ThesisAuthor{Bc. Kateryna Lutsai}

\def\YearSubmitted{2025}

\def\Department{Institute of Formal and Applied Linguistics}

\def\DeptType{Institute}

\def\Supervisor{Mgr. Pavel Straňák, Ph.D.}

\def\SupervisorsDepartment{Institute of Formal and Applied Linguistics}

\def\StudyProgramme{Computer Science – Language Technologies and Computational Linguistics}

\def\Dedication{%
I would like to thank Mgr. David Novák, Ph.D., and Ing. Dana Křivánková from \ac{arup} for providing data and practical feedback.

At \ac{ufal}, Mgr. Pavel Straňák, Ph.D., supervised the work, and doc. RNDr. Pavel Pecina, Ph.D., provided advice on thesis writing.

\ac{ufal} provided essential computational infrastructure.

Although this thesis itself was not funded by it, I would like to acknowledge the ATRIUM project as the overarching framework for collaboration between \ac{arup} and \ac{ufal}, which received funding from the European Commission’s HORIZON Research and Innovation Actions under the grant agreement GAP-101132163 --- \href{https://atrium-research.eu/}{ATRIUM} --- HORIZON-INFRA-2023-SERV-01-02 --- \textbf{A}dvancing Fron\textbf{T}ier \textbf{R}esearch \textbf{I}n the Arts and h\textbf{UM}anities.
}

\def\Abstract{%
Digitization projects in the humanities, specifically within the archaeological domain, often generate vast quantities of page images from historical documents, presenting significant challenges for manual curation and analysis. The primary goal of this project is to address this need by developing and evaluating an automated image classification system designed to categorize historical document pages based on their content, thereby enabling tailored downstream analysis pipelines. These archives contain diverse content, including various text types (handwritten, typed, printed), graphical elements (drawings, maps, photos), and layouts (plain text, tables, forms). By leveraging advancements in neural network architectures, this system facilitates content-specific workflows, such as separating pages requiring Optical Character Recognition (OCR) from those needing graphical analysis. The final models, datasets, and software are released under open-source licenses to support the broader digital humanities community.
}

\def\ThesisKeywords{
Image-based Document Processing
\sep Archival Digitization
\sep Page classification
\sep Model finetuning
\sep Layout elements detection
}

\def\ThesisAuthorXMP{\ThesisAuthor}
\def\ThesisTitleXMP{\ThesisTitle}
\def\ThesisKeywordsXMP{\ThesisKeywords}
\def\AbstractXMP{\Abstract}

\def\InfoPageFont{}
{
\catcode`\%=12
\global\edef\percenthack{%}
}

{
\def\sep{\string\sep\space}
\let~=\space

\newwrite\xmp
\immediate\openout\xmp=\jobname.xmpdata
\immediate\write\xmp{\percenthack\space Generated automatically from metadata.tex, please don't edit here.}
\def\xmpitem#1#2{\immediate\write\xmp{\string#1{#2}}}
\xmpitem\Author\ThesisAuthorXMP
\xmpitem\Title\ThesisTitleXMP
\xmpitem\Keywords\ThesisKeywordsXMP
\xmpitem\Subject\AbstractXMP
\xmpitem\Publisher{Charles University}
\immediate\closeout\xmp
}

\documentclass[12pt,a4paper]{report}
\setlength\textwidth{145mm}
\setlength\textheight{247mm}
\setlength\oddsidemargin{10mm}
\setlength\evensidemargin{10mm}
\setlength\topmargin{0mm}
\setlength\headsep{0mm}
\setlength\headheight{0mm}
\let\openright=\clearpage

\usepackage[a-2u]{pdfx}

\usepackage{lmodern}
\usepackage{iftex}
\ifpdftex
\usepackage[utf8]{inputenc}
\usepackage[T1]{fontenc}
\usepackage{textcomp}
\fi

\usepackage{amsmath}        % extensions for typesetting of math
\usepackage{amsfonts}       % math fonts
\usepackage{amsthm}         % theorems, definitions, etc.
\usepackage{bm}             % boldface symbols (\bm)
\usepackage{booktabs}       % improved horizontal lines in tables
\usepackage{caption}        % custom captions of floating objects
\usepackage{dcolumn}        % improved alignment of table columns
\usepackage{floatrow}       % custom float environments
\usepackage{graphicx}       % embedding of pictures
\usepackage{indentfirst}    % indent the first paragraph of a chapter
\usepackage[nopatch=item]{microtype}   % micro-typographic refinement
\usepackage{paralist}       % improved enumerate and itemize
\usepackage[nottoc]{tocbibind} % makes sure that bibliography and the lists
\usepackage{xcolor}         % typesetting in color

\usepackage{todonotes}
\usepackage{subcaption} % For subfigures
\usepackage{booktabs}
\usepackage{array}
\usepackage{longtable}
\usepackage{geometry}
\usepackage[style=toc]{acro}

\hypersetup{unicode}
\hypersetup{breaklinks=true}

\usepackage{algpseudocode}  % part of algorithmicx package
\usepackage{algorithm}
\usepackage{fancyvrb}       % improved verbatim environment
\usepackage{listings}       % pretty-printer of source code

\usepackage{xcolor}

\usepackage{cleveref}
\usepackage{multirow}

\usepackage[natbib,style=iso-alphabetic]{biblatex}
\addbibresource{bibliography.bib}

\def\TypeBc{bc}
\def\TypeMgr{mgr}
\def\TypePhD{phd}
\def\TypeRig{rig}

\ifx\ThesisType\TypeBc
\def\ThesisTypeName{bachelor}
\def\ThesisTypeTitle{BACHELOR THESIS}

\fi

\ifx\ThesisType\TypeMgr
\def\ThesisTypeName{master}
\def\ThesisTypeTitle{MASTER THESIS}

\fi

\ifx\ThesisType\TypePhD
\def\ThesisTypeName{doctoral}
\def\ThesisTypeTitle{DOCTORAL THESIS}

\fi

\ifx\ThesisType\TypeRig
\def\ThesisTypeName{rigorosum}
\def\ThesisTypeTitle{RIGOROSUM THESIS}

\fi

\ifx\ThesisTypeName\undefined
\PackageError{thesis}{Unknown thesis type.}{Please check the definition of ThesisType in metadata.tex.}
\fi

\def\LangCS{cs}
\def\LangEN{en}

\ifx\StudyLanguage\LangCS
\else\ifx\StudyLanguage\LangEN
\else\PackageError{thesis}{Unknown study language.}{Please check the definition of StudyLanguage in metadata.tex.}
\fi\fi

\makeatletter
\def\@makechapterhead#1{
  {\parindent \z@ \raggedright \normalfont
   \Huge\bfseries \thechapter\quad #1
   \par\nobreak
   \vskip 20\p@
}}
\def\@makeschapterhead#1{
  {\parindent \z@ \raggedright \normalfont
   \Huge\bfseries #1
   \par\nobreak
   \vskip 20\p@
}}
\makeatother

\overfullrule=1mm

\theoremstyle{plain}

\theoremstyle{remark}

\ifcsname DeclareCaptionStyle\endcsname
\DeclareCaptionStyle{thesis}{style=base,font=small,labelfont=bf,labelsep=quad}
\captionsetup{style=thesis}
\captionsetup[algorithm]{style=thesis,singlelinecheck=off}
\captionsetup[listing]{style=thesis,singlelinecheck=off}
\fi

\DefineVerbatimEnvironment{code}{Verbatim}{fontsize=\small, frame=single}

\ifcsname lstset\endcsname
\lstset{
  language=C++,
  tabsize=2,
  showstringspaces=false,
  basicstyle=\footnotesize\tt\color{black!75},
  identifierstyle=\bfseries\color{black},
  commentstyle=\color{green!50!black},
  stringstyle=\color{red!50!black},
  keywordstyle=\color{blue!75!black}}
\fi

\definecolor{darkbg}{gray}{0.95}

\lstdefinestyle{promptstyle}{
  language={}, % Disable syntax highlighting
  basicstyle=\ttfamily\small\color{black},
  backgroundcolor=\color{darkbg},
  breaklines=true, % Automatically break long lines
  frame=single,    % A single-line frame
  framerule=0pt,     % Frame line width
  framesep=5pt,      % Padding
  rulesepcolor=\color{darkbg}, % Color for padding area
  showstringspaces=false,
  commentstyle={},
  stringstyle={},
  keywordstyle={},
  identifierstyle={}
} 

\ifcsname DeclareNewFloatType\endcsname
\DeclareNewFloatType{listing}{}
\floatsetup[listing]{style=ruled}
\floatname{listing}{Program}
\fi

\ifx\citet\undefined\else

\ExecuteBibliographyOptions{maxcitenames=2}
\ExecuteBibliographyOptions{giveninits}
\DeclareDelimFormat[textcite]{multinamedelim}{\addcomma\space}
\DeclareDelimFormat[textcite]{finalnamedelim}{\space and~}

\fi

\acsetup{make-links = true, link-only-first = true}

\DeclareAcronym{eg}{
    short = {e.g.,},
    long = for example ,
    foreign = exempli gratia ,
    foreign-babel = latin,
    first-style=short
}

\DeclareAcronym{cpu}{
  short=CPU,
  long=Central Processing Unit,
  first-style=short-long,
  subsequent-style=short,
  long-plural=s,
  short-plural=s,
}
\DeclareAcronym{gpu}{
  short=GPU,
  long=Graphics Processing Unit,
  first-style=short-long,
  subsequent-style=short,
  long-plural=s,
  short-plural=s,
}
\DeclareAcronym{png}{
  short=PNG,
  long=Portable Network Graphics,
  first-style=short-long,
  subsequent-style=short
}
\DeclareAcronym{ai}{
  short=AI,
  long=Artificial Intelligence,
  first-style=long-short,
  subsequent-style=short,
  long-plural=s,
  short-plural=s,
}
\DeclareAcronym{cli}{
  short=CLI,
  long=Command-line Interface,
  first-style=long-short,
  subsequent-style=short,
  long-plural=s,
  short-plural=s,
}
\DeclareAcronym{api}{
  short=API,
  long=Application Programming Interface,
  first-style=long-short,
  subsequent-style=short,
  long-plural=s,
  short-plural=s,
}
\DeclareAcronym{pdf}{
  short=PDF,
  long=Portable Document File,
  first-style=short-long,
  subsequent-style=short
}
\DeclareAcronym{jpeg}{
  short=JPEG,
  long=Joint Photographic Experts Group,
  first-style=short-long,
  subsequent-style=short
}
\DeclareAcronym{tiff}{
  short=TIFF,
  long=Tagged Image File Format,
  first-style=short-long,
  subsequent-style=short
}
\DeclareAcronym{csv}{
  short=CSV,
  long=Comma-separated Values,
  first-style=short-long,
  subsequent-style=short
}
\DeclareAcronym{lda}{
  short=LDA,
  long=Latent Dirichlet Analysis,
  first-style=long-short,
  subsequent-style=short,
  extra=model type
}
\DeclareAcronym{knn}{
  short=k-NN,
  long=k-nearest neighbors,
  first-style=long-short,
  subsequent-style=short,
  extra=model type
}
\DeclareAcronym{svm}{
  short=SVM,
  long=Support Vector Machine,
  first-style=long-short,
  subsequent-style=short,
  long-plural=s,
  short-plural=s,
  extra=model type
}
\DeclareAcronym{cuda}{
  short=CUDA,
  long=Compute Unified Device Architecture,
  first-style=long-short,
  subsequent-style=short,
  extra={CUDA — is a parallel computing platform and programming model developed by NVIDIA. It allows software developers to use NVIDIA GPUs for general-purpose processing, rather than just graphics rendering.}
}
\DeclareAcronym{dla}{
  short=DLA,
  long=Document Layout Analysis,
  first-style=long-short,
  subsequent-style=short,
  extra={DLA — is a field focused on automatically identifying and categorizing different elements within a document's visual structure using computer vision and machine learning.}
}
\DeclareAcronym{ufal}{
  short=IFAL,
  long=Institute of Formal and Applied Linguistics,
  foreign=Ústav formální a aplikované lingvistiky (ÚFAL),
  foreign-babel=czech,
  first-style=long-short,
  subsequent-style=short,
  extra={IFAL — is a research department at Charles University in Prague, Czech Republic. It focuses on language technologies, computational linguistics, and machine learning.}
}
\DeclareAcronym{arup}{
  short=IAP,
  long=Institute of Archaeology of the Czech Academy of Sciences in Prague,
  foreign=Archeologický ústav AV ČR Praha v. v. i. (ARÚP),
  foreign-babel=czech,
  first-style=long-short,
  subsequent-style=short,
  extra={IAP — is a leading research institution dedicated to the study of the human past in its entirety, focusing on the Czech Republic and abroad. It is part of the Czech Academy of Sciences and is known for its comprehensive approach to archaeological research, encompassing fieldwork, analysis of material remains, and the development of general models of past societies}
}
\DeclareAcronym{arub}{
  short=IAB,
  long=Institute of Archaeology of the Czech Academy of Sciences in Brno,
  foreign=Archeologický ústav AV ČR Brno v. v. i. (ARÚB),
  foreign-babel=czech,
  first-style=long-short,
  subsequent-style=short,
  extra={IAB — is comprised of Czech and foreign researchers, technicians and service employees who are connected by a professional interest in investigating the history of humankind from the Paleolithic to the Middle Ages.}
}
\DeclareAcronym{dd}{
  short=DD,
  long=DeepDoctection,
  first-style=long-short,
  subsequent-style=short,
  extra={DeepDoctection — is a Python library that orchestrates Scan and PDF document layout analysis and extraction for RAG (Retrieval-Augmented Generation}
}
\DeclareAcronym{vit}{
  short=ViT,
  long=Vision Transformer,
  first-style=long-short,
  subsequent-style=short,
  long-plural=s,
  short-plural=s,
  extra={ViT — is a type of neural network architecture that applies the Transformer model, originally designed for natural language processing (NLP), to computer vision tasks, particularly image classification. The models break down an image into patches, treat them as a sequence of tokens, and process them using a Transformer encoder.}
}
\DeclareAcronym{dit}{
  short=DiT,
  long=Document Image Transformer,
  first-style=long-short,
  subsequent-style=short,
  long-plural=s,
  short-plural=s,
  extra={DiT — is an image transformer pretrained on large-scale unlabeled document images. It learns to predict the missing visual tokens from a corrupted input image. The pretrained DiT model can be used as a backbone in other models for visual document tasks like document image classification and table detection.}
}
\DeclareAcronym{cnn}{
  short=CNN,
  long=Convolutional Neural Network,
  first-style=long-short,
  subsequent-style=short,
  long-plural=s,
  short-plural=s,
  extra={CNN — is a type of deep learning model particularly well-suited for analyzing visual data like images and videos. It excels at identifying patterns and features within these images, making it a cornerstone for computer vision tasks such as image recognition and object detection.}
}
\DeclareAcronym{clip}{
  short=CLIP,
  long=Contrastive Language-Image Pretraining,
  first-style=long-short,
  subsequent-style=short,
  extra={CLIP — is designed to understand the connection between images and text. The models achieve this by mapping image and text embeddings into the same vector space, allowing for comparison and similarity assessment between the two modalities.}
}
\DeclareAcronym{ocr}{
  short=OCR,
  long=Optical Character Recognition,
  first-style=long-short,
  subsequent-style=short,
  extra={OCR — is a technology that converts images of text into machine-readable text.}
}
\DeclareAcronym{rfc}{
  short=RFC,
  long=Random Forest Classifier,
  first-style=long-short,
  subsequent-style=short,
  extra={RFC — is a powerful ensemble learning method that combines multiple decision trees to make predictions. It's a versatile tool used for both classification and regression tasks, known for its accuracy, robustness, and ability to handle large datasets with many features.}
}
\DeclareAcronym{mit}{
  short=MIT,
  long={Licensing policy created in Massachusetts Institute of Technology},
  first-style=short-long,
  subsequent-style=short,
  extra={MIT license — is a permissive free software license, widely used in the open-source community. It allows for the use, modification, and distribution of software, including commercial use, with minimal restrictions.}
}
\DeclareAcronym{llm}{
  short=LLM,
  long=Large Language Model,
  long-plural=s,
  short-plural=s,
  first-style=long-short,
  subsequent-style=short,
  extra={LLM — is a type of artificial intelligence (AI) program, deep learning model trained on massive amounts of text data. LLMs excel at understanding, generating, and manipulating human language, enabling them to perform various tasks like text generation, translation, question answering, etc.}
}
\DeclareAcronym{html}{
  short=HTML,
  long=HyperText Markup Language,
  first-style=long-short,
  subsequent-style=short,
  extra={HTML — is the standard markup language for structuring content on the Web.}
}
\DeclareAcronym{json}{
  short=JSON,
  long=JavaScript Object Notation,
  first-style=long-short,
  subsequent-style=short,
  extra={JSON — is a lightweight data-interchange format, commonly used for APIs and configuration.}
}
\DeclareAcronym{rest}{
  short=REST,
  long=Representational State Transfer,
  first-style=long-short,
  subsequent-style=short,
  extra={REST — is an architectural style for designing networked applications; a \ac{api} that follows these constraints is often called a REST API.}
}
\DeclareAcronym{http}{
  short=HTTP,
  long=Hypertext Transfer Protocol,
  first-style=long-short,
  subsequent-style=short,
  extra={HTTP — is the application-layer protocol used for communication on the Web.}
}

\definecolor{darkbg}{gray}{0.95} % Color for the prompt background

\lstdefinestyle{promptstyle}{
  basicstyle=\ttfamily\small\color{black}, % Use small tt font, solid black
  backgroundcolor=\color{darkbg}, % Set the background color
  breaklines=true, % Automatically break long lines
  frame=single, % A single-line frame
  framerule=0pt,  % Frame line width
  framesep=5pt,   % Padding
  rulesepcolor=\color{darkbg}, % Color for padding area
  showstringspaces=false, % Don't show spaces
  language={}, % Disable syntax highlighting
  commentstyle={},
  stringstyle={},
  keywordstyle={},
  identifierstyle={}
}

\begin{document}
%%% Title page of the thesis and other mandatory pages

%%% Inscriptions at the opening page of the hard cover

% We usually do not typeset the hard cover, but if you want to do it, change \iffalse to \iftrue
\iffalse

\pagestyle{empty}
\hypersetup{pageanchor=false}
\begin{center}

\large
Charles University

\medskip

Faculty of Mathematics and Physics

\vfill

{\huge\bf\ThesisTypeTitle}

\vfill

{\huge\bf\ThesisTitle\par}

\vfill
\vfill

\hbox to \hsize{\YearSubmitted\hfil \ThesisAuthor}

\end{center}

\newpage\openright
\setcounter{page}{1}

\fi

\pagestyle{empty}
\hypersetup{pageanchor=false}
\begin{center}

\centerline{\mbox{\includegraphics[width=166mm]{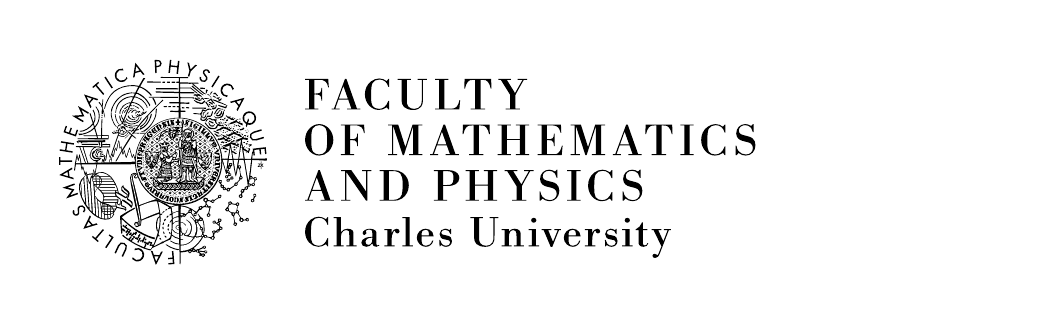}}}

\vspace{-8mm}
\vfill

{\bf\Large\ThesisTypeTitle}

\vfill

{\LARGE\ThesisAuthor}

\vspace{15mm}

{\LARGE\bfseries\ThesisTitle\par}

\vfill

\Department

\vfill

{
\centerline{\vbox{\halign{\hbox to 0.45\hsize{\hfil #}&\hskip 0.5em\parbox[t]{0.45\hsize}{\raggedright #}\cr
Supervisor of the \ThesisTypeName{} thesis:&\Supervisor \cr
\ifx\ThesisType\TypeRig\else
\noalign{\vspace{2mm}}
Study programme:&\StudyProgramme \cr
\fi
}}}}

\vfill

Prague \YearSubmitted

\end{center}

\newpage

\openright
\hypersetup{pageanchor=true}
\vglue 0pt plus 1fill

\noindent
I declare that I carried out this \ThesisTypeName{} thesis on my own, and only with the cited
sources, literature and other professional sources.
I understand that my work relates to the rights and obligations under the Act No.~121/2000 Sb.,
the Copyright Act, as amended, in particular the fact that the Charles
University has the right to conclude a license agreement on the use of this
work as a school work pursuant to Section 60 subsection 1 of the Copyright~Act.

\vspace{10mm}

\hbox{\hbox to 0.5\hsize{%
In \hbox to 6em{\dotfill} date \hbox to 6em{\dotfill}
\hss}\hbox to 0.5\hsize{\dotfill\quad}}
\smallskip
\hbox{\hbox to 0.5\hsize{}\hbox to 0.5\hsize{\hfil Author's signature\hfil}}

\vspace{20mm}
\newpage

\openright

\noindent
\Dedication

\newpage

\openright
{\InfoPageFont

\vtop to 0.5\vsize{
\setlength\parindent{0mm}
\setlength\parskip{5mm}

Title:
\ThesisTitle

Author:
\ThesisAuthor

\DeptType:
\Department

Supervisor:
\Supervisor, \SupervisorsDepartment

Abstract:
\Abstract

Keywords:
{\def\sep{\unskip, }\ThesisKeywords}

\vfil
}

}

\newpage

\pagestyle{plain}

\tableofcontents

\chapter*{Introduction} \label{chap-intro}
\addcontentsline{toc}{chapter}{Introduction}

\noindent This thesis develops an automated system for classifying page images from historical archives based on visual content and layout. Such a tool is particularly relevant for institutions like \acf{arup}, which manage large digital collections and need scalable ways to organize and process them.

\begin{figure}[h]
    \centering
    \caption{One of the oldest and one of the newest pages in our collection. Both contain graphical objects of interest, but the modern page is annotated with a structured data format (table).}
    \label{fig:intro-time}
    \begin{subfigure}{0.45\textwidth}
        \centering
        \includegraphics[width=\textwidth]{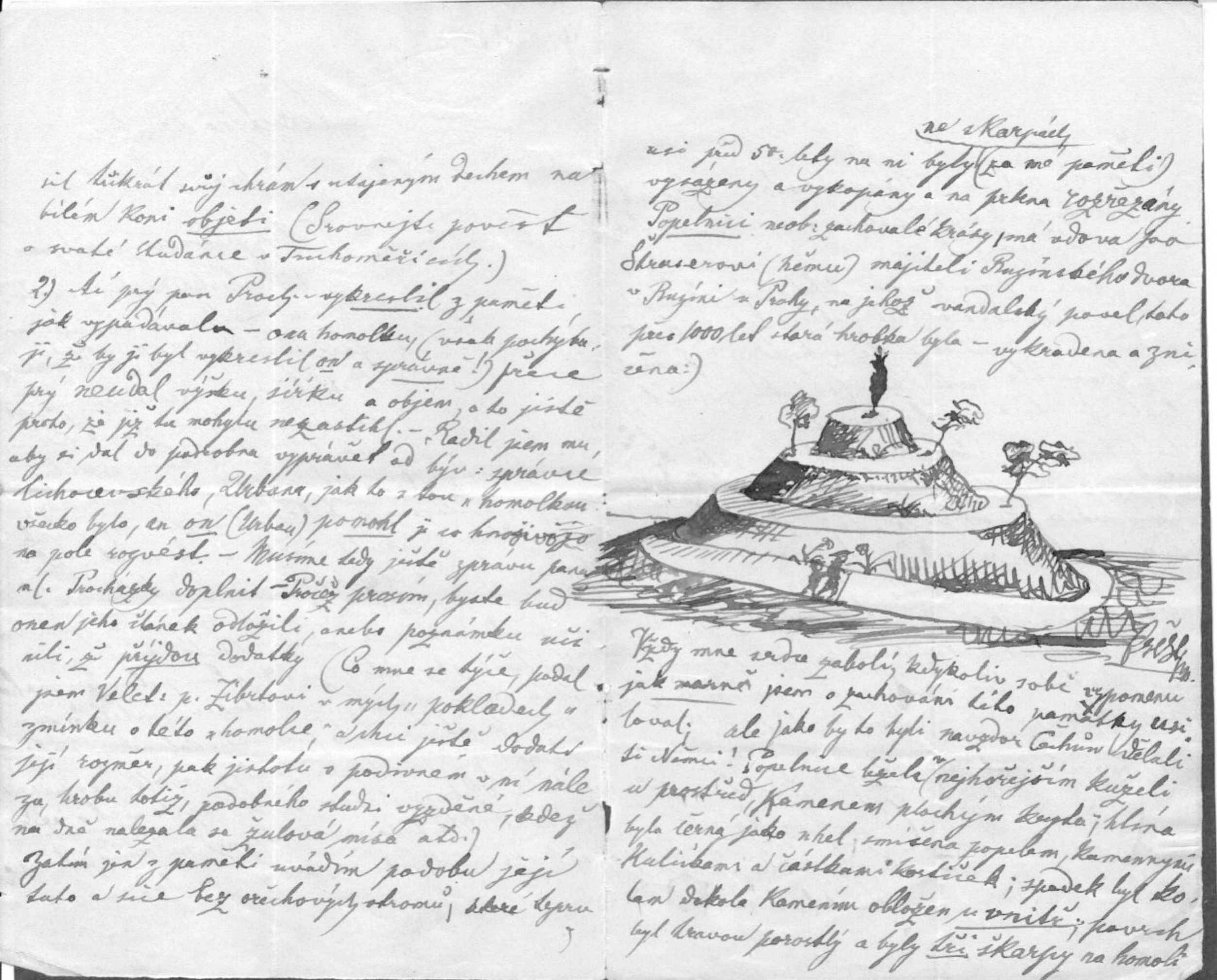}
        \caption{Notebook with a freehand sketch \& skewed lines \& grayish background}
        \label{fig:time1}
    \end{subfigure}
    % \hfill
    \begin{subfigure}{0.45\textwidth}
        \centering
        \includegraphics[width=\textwidth]{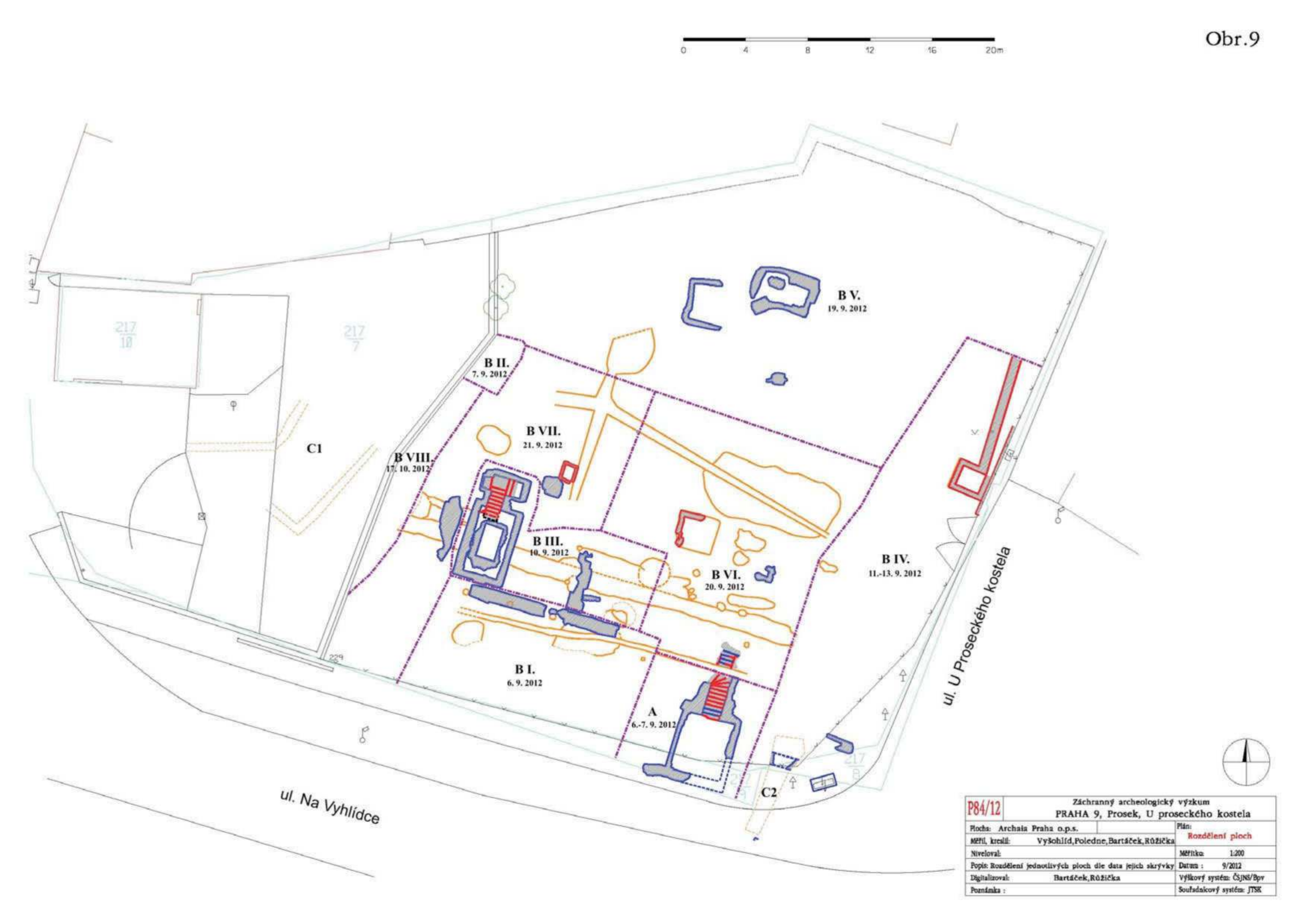}
        \caption{Modern digital-born (printed and then scanned for some reason) map \& tabular legend in the corner}
        \label{fig:time2}
    \end{subfigure}
\end{figure}

The goal is to enable targeted downstream processing on both historical and modern page scans (Figure \ref{fig:intro-time}), such as routing text-heavy pages to \ac{ocr} while sending pages with tables or graphics to specialized extraction pipelines. Different page types require different tools; for example, \acf{ocr} is appropriate for text, while layout analysis is needed for tables \cite{zhong2019publaynet,xu2020layoutlm} and image analysis for photographs. Without an initial classification step, downstream pipelines (\ac{eg} table parsing, image segmentation, text recognition) cannot be applied efficiently.

This leads to two central research questions. First, how can modern deep learning methods (\ac{eg} \acs{cnn}, \acs{clip}, \acs{vit}) be adapted to a massive, heterogeneous archive of scanned archaeological reports? Second, how should this heterogeneity be represented through classification labels (\ac{eg} which visual cues distinguish page classes)? 

A practical objective of this thesis is to evaluate whether existing open-source tools are sufficient for this task. When they are not, we define a domain-specific label scheme and train models that better match the needs of archaeologists. The result is intended to be published as an open-source toolset consisting of annotated data, model weights, and source code for content-aware page classification of archival materials.

\section*{Nature of digitized archival collection} \label{sec-data-nature}

\noindent The dataset for this project consists of scanned pages from \ac{arup} (primary source) and \acf{arub} (secondary source), initially supplied as multi-page \acp{pdf}. These archives are marked by profound heterogeneity. A single collection can contain everything from handwritten manuscripts and typewritten correspondence to printed articles, technical drawings, maps, and annotated photographs.

\begin{figure}[!h]
\centering
   \caption{Prague and Brno data from \acs{arup} and \acs{arub} (unlabeled): Page-scan counts over time in the archives of scanned documents}
    \label{fig:timeline}
     \includegraphics[width=140mm]{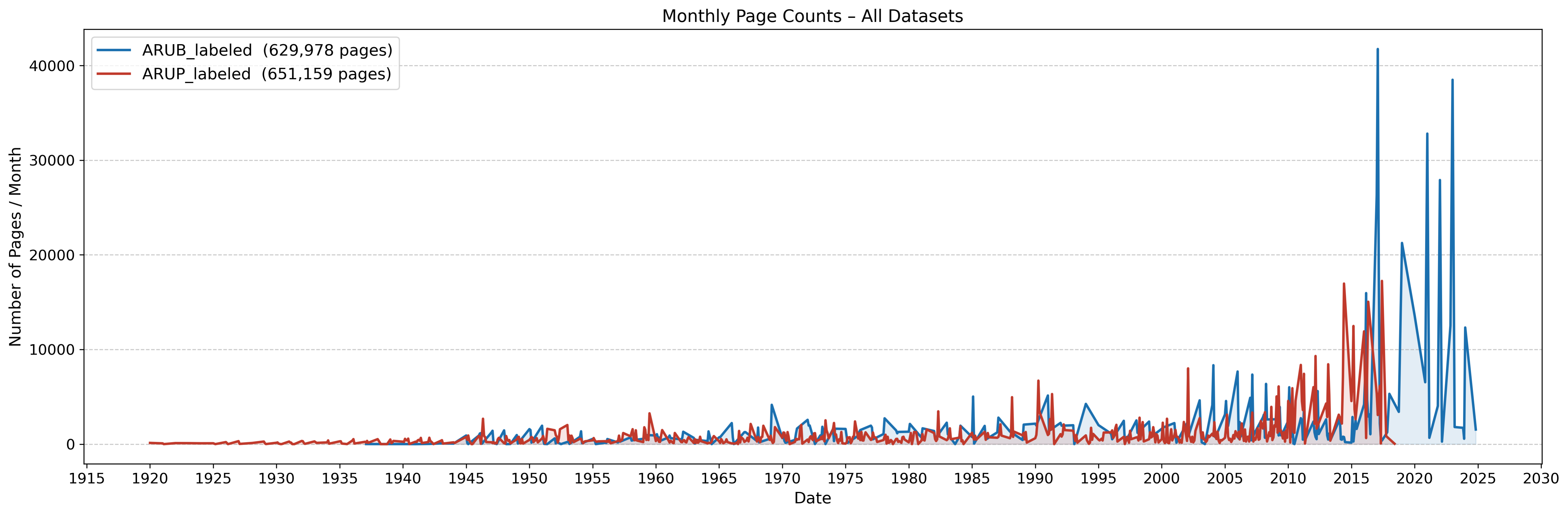}
\end{figure}

Our collection spans more than a century, with document creation ranging from the early 20th century to the present day. The incoming data volume has increased substantially over time (Figure \ref{fig:timeline}). In practice, the only consistently available metadata are the year and month recorded in filenames; scanning campaigns therefore appear as spikes in page counts along the timeline.

Because these collections were originally paper-based and digitized via scanning, they often lack descriptive metadata. Critical fields such as author, title, or document type are frequently missing, which complicates automated processing and retrieval.

\begin{figure}[h]
    \centering
    \caption{Page examples derived from the same collection that differ substantially in size, content, and paper condition.}
    \label{fig:intro-page-2}
    \begin{subfigure}{0.3\textwidth}
        \centering
        \includegraphics[width=\textwidth]{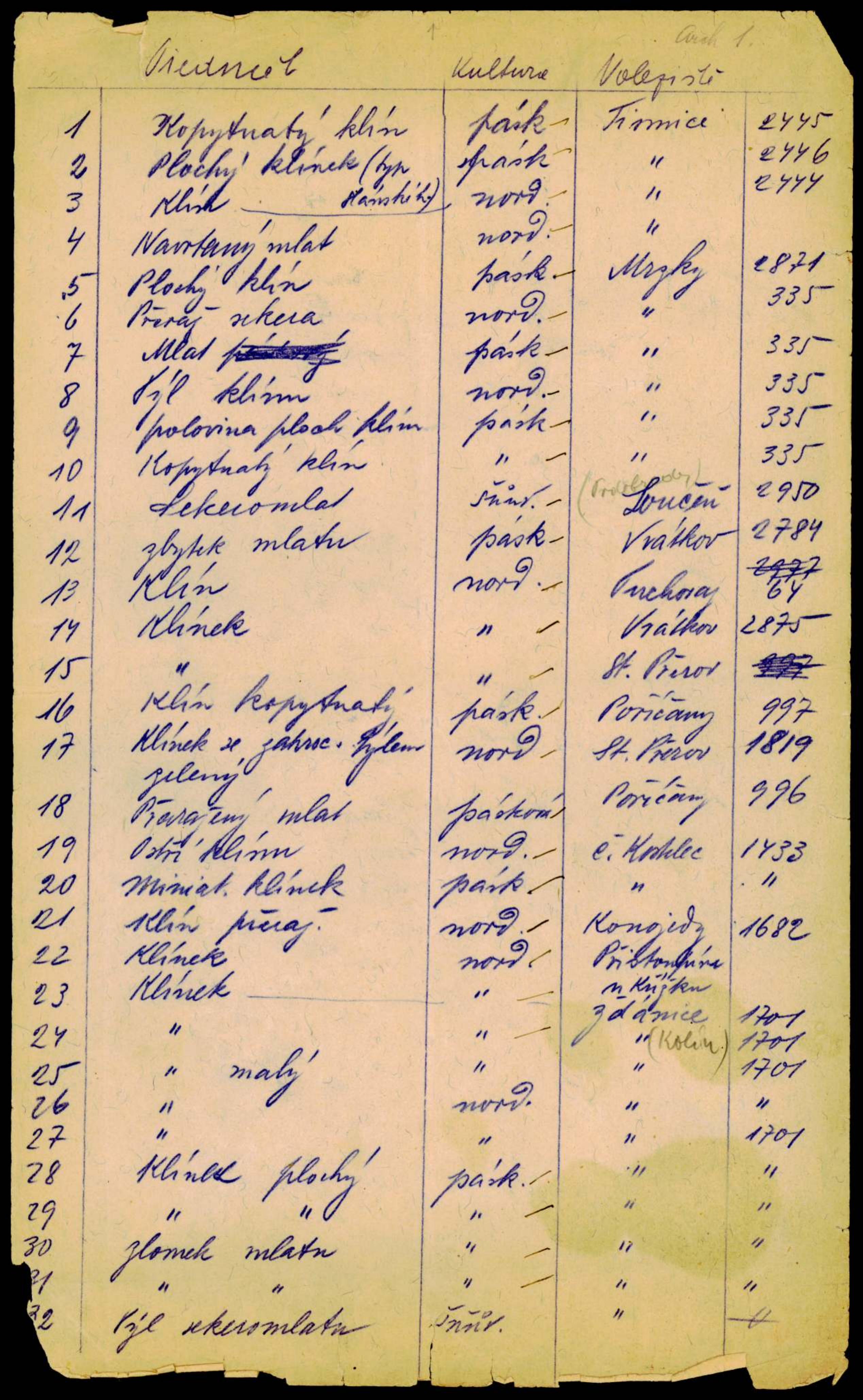}
        \caption{Handwritten table on damaged paper}
        \label{fig:intro3}
    \end{subfigure}
    % \hfill
    \begin{subfigure}{0.3\textwidth}
        \centering
        \includegraphics[width=\textwidth]{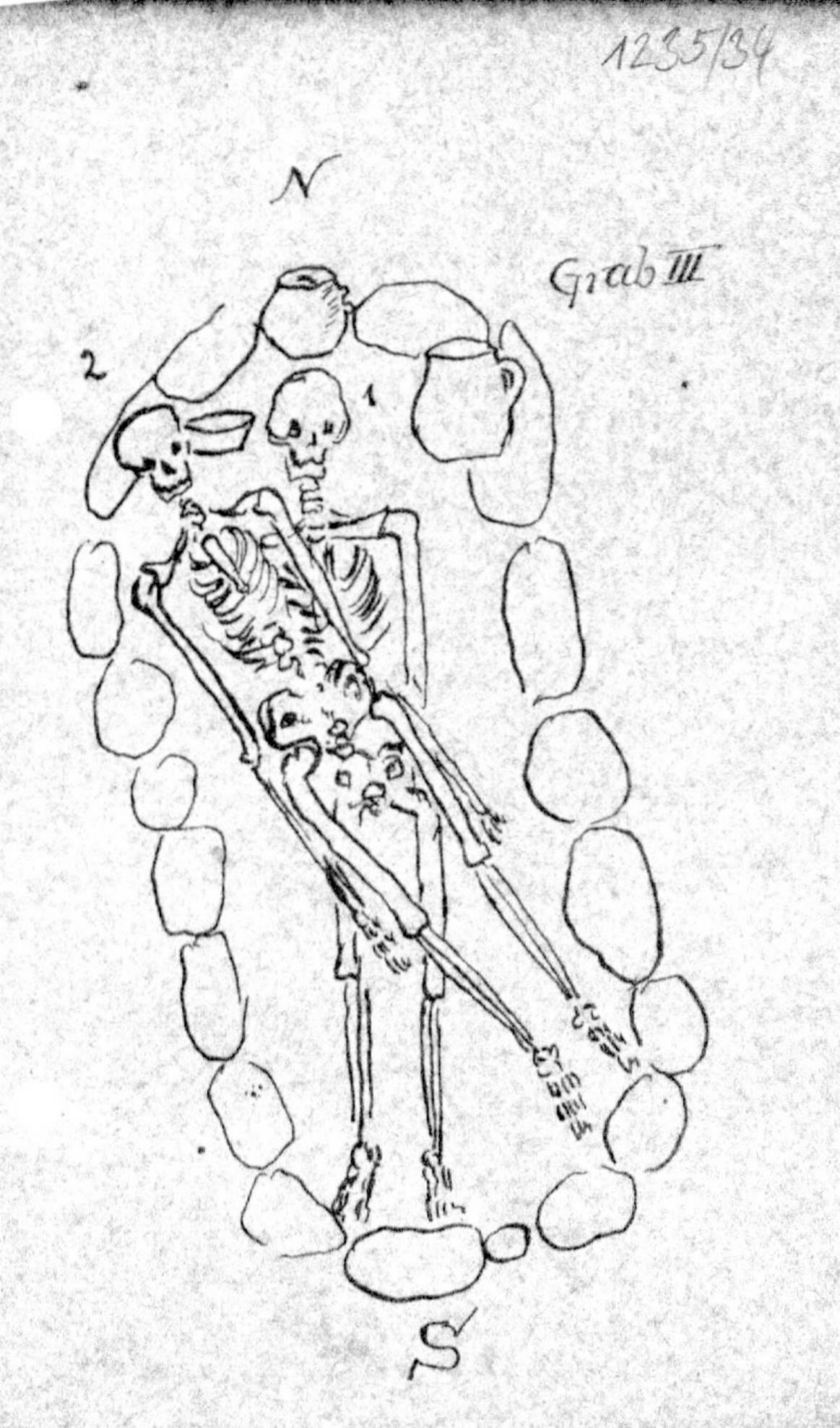}
        \caption{Tiny-scale drawing}
        \label{fig:intro4}
    \end{subfigure}
    % \hfill
    \begin{subfigure}{0.3\textwidth}
        \centering
        \includegraphics[width=\textwidth]{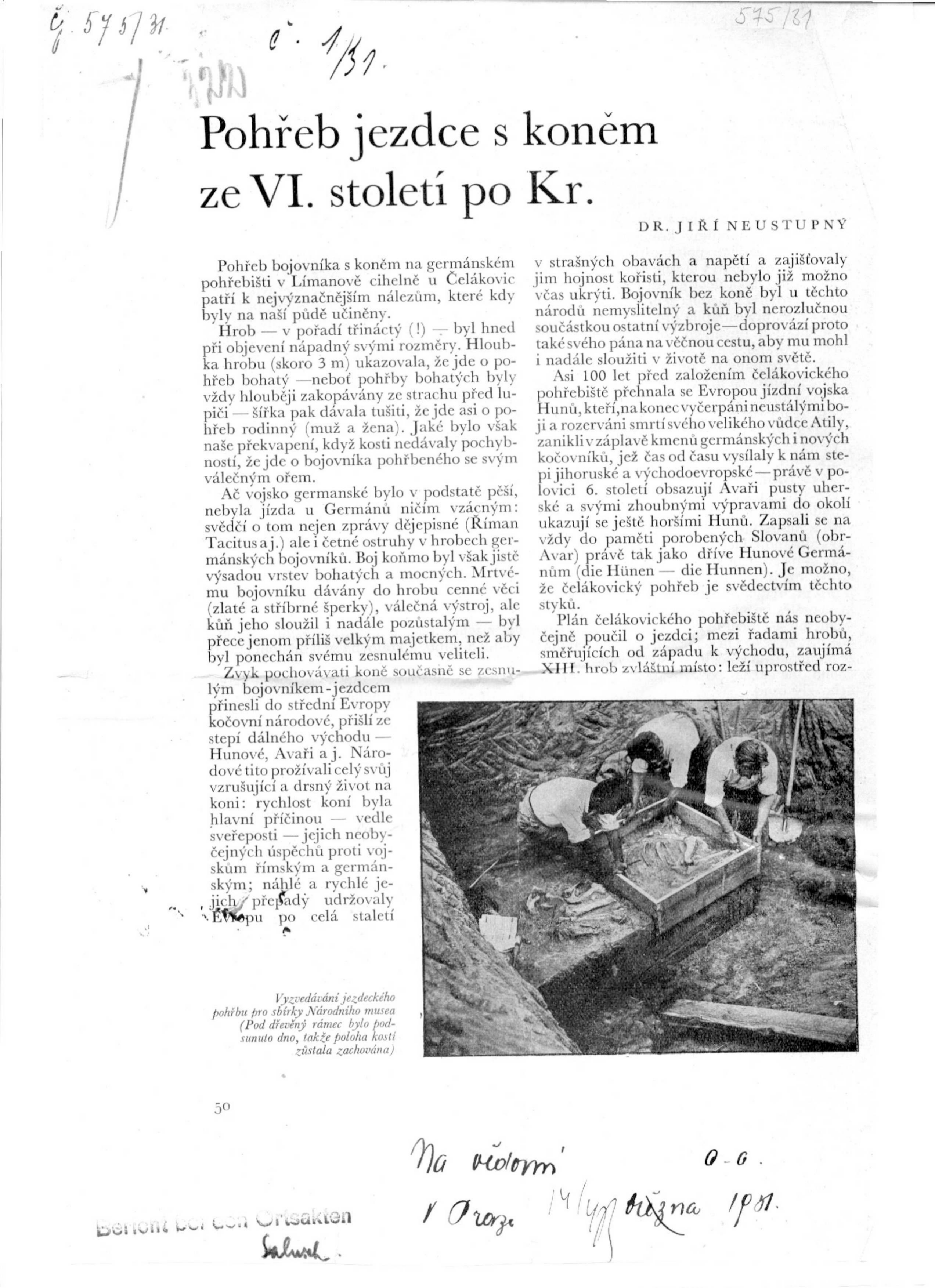}
        \caption{Article scan with a photo \& handwritten notes}
        \label{fig:intro5}
    \end{subfigure}
\end{figure}

\begin{figure}[h]
    \centering
    \caption{Examples of scans with different physical sizes from our annotated subset}
    \label{fig:intro-page-1}
    \begin{subfigure}{0.45\textwidth}
        \centering
        \includegraphics[width=\textwidth]{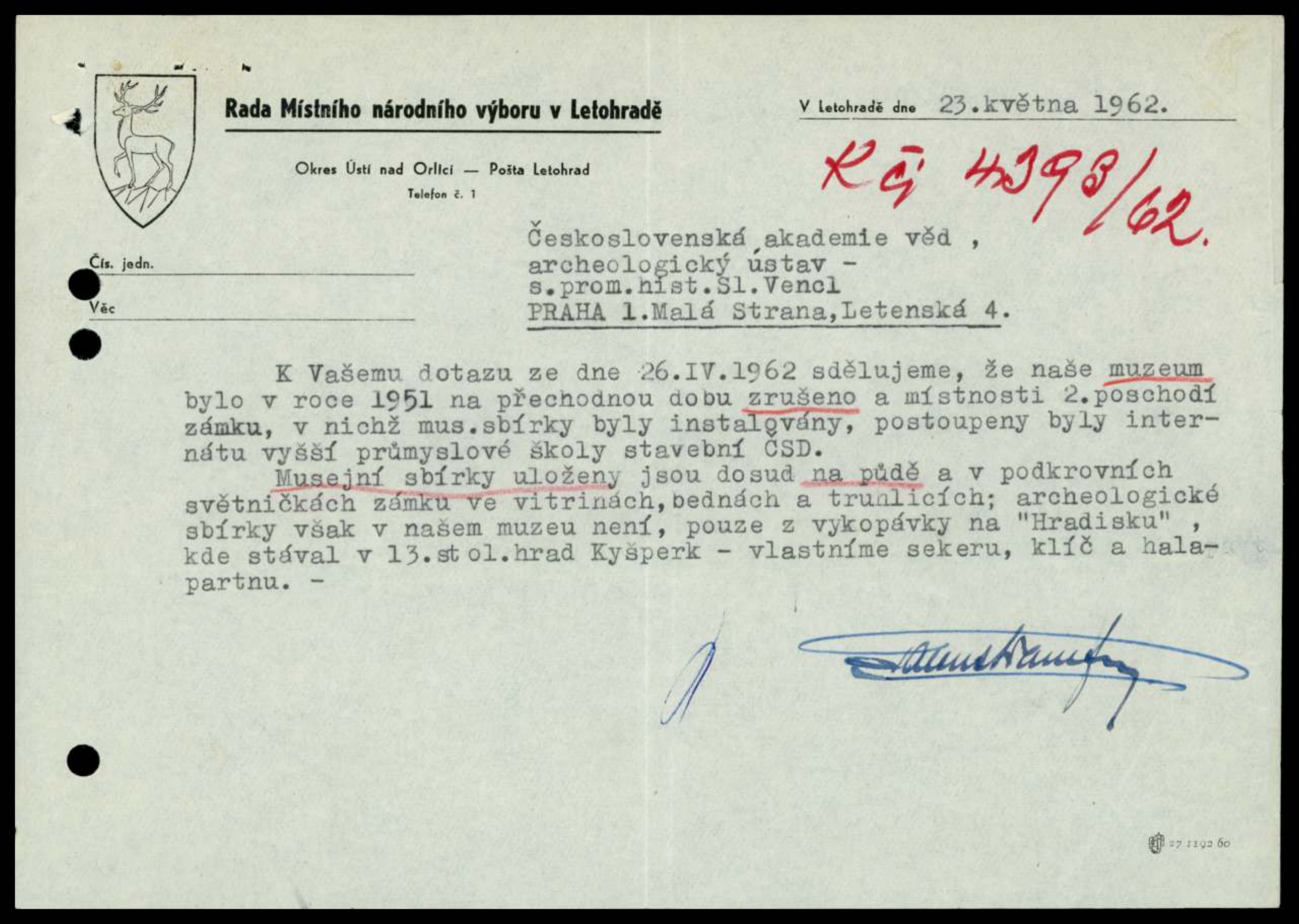}
        \caption{Manually commented typewritten report with a small logo}
        \label{fig:intro1}
    \end{subfigure}
    % \hfill
    \begin{subfigure}{0.45\textwidth}
        \centering
        \includegraphics[width=\textwidth]{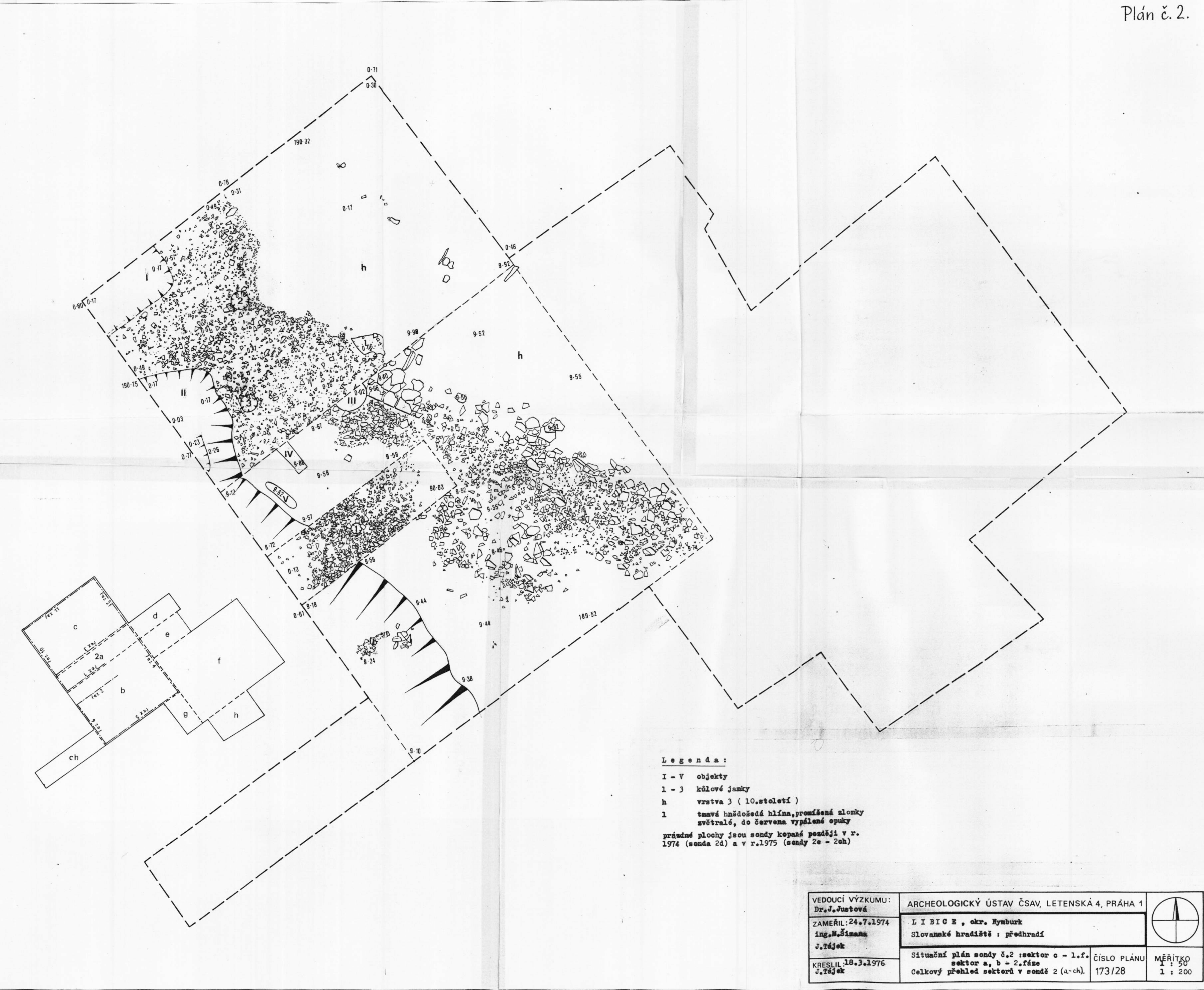}
        \caption{Large-scale canvas with a map and a legend table}
        \label{fig:intro2}
    \end{subfigure}
\end{figure}

At the same time, the archive contains many content types---often mixed within a single document---as demonstrated in Figures \labelcref{fig:intro-time,fig:intro-ocr,fig:intro-page-1,fig:intro-page-2}. High-resolution scans produce large files, and physical degradation (\acs{eg} stains, skew, torn edges) introduces visual artifacts that complicate analysis and motivate robustness to common visual corruptions \cite{hendrycks2019corruptions}. These properties make automated, content-aware page classification a practical prerequisite for scalable downstream processing.

One basic system requirement addresses the challenge of processing large volumes of data: the system must accommodate different use cases by accepting multiple input types:

\begin{itemize}
    \item Single image files for individual page classification (any standard image format, such as \ac{png}, \ac{jpeg}, or \ac{tiff})
    \item Directories of image files for batch processing
    \item Additionally:
    \begin{itemize}
        \item Directories with nested subdirectories of image files
        \item Handle various page numbering conventions for both Unix (pdftoppm) and Windows (ImageMagick) open-source \ac{pdf}-to-\ac{png} conversion tools
    \end{itemize}
\end{itemize}

% The primary contribution of this thesis is the implementation and rigorous evaluation of multiple classification approaches — from classical feature-based methods to state-of-the-art deep learning networks — to build a robust system capable of navigating the unique challenges posed by heterogeneous archival data. 

\section*{Challenges in management of scanned documents} \label{sec-scan-docs}

\noindent The characteristics outlined below create practical hurdles for archive management:
\begin{description}
    \item \textbf{Manual collection organization:} The volume and content diversity make manual curation impractical. Scanning campaigns typically produce large batches, which increases the risk of human error and makes consistent organization difficult. Manually reviewing each page to determine its content category is prohibitively time-consuming.

    \item \textbf{File classification without metadata:} The lack of descriptive metadata---often a result of scanning with default equipment settings---prevents straightforward automated grouping and complicates meaningful database organization.

\end{description}

These challenges motivate automated methods that distinguish page types before specialized processing.

\section*{Related work}

\begin{figure}[h]
    \centering
    \caption{Pages with background artifacts that degrade \acs{ocr} performance}
    \label{fig:intro-ocr}
    \begin{subfigure}{0.32\textwidth}
        \centering
        \includegraphics[width=\textwidth]{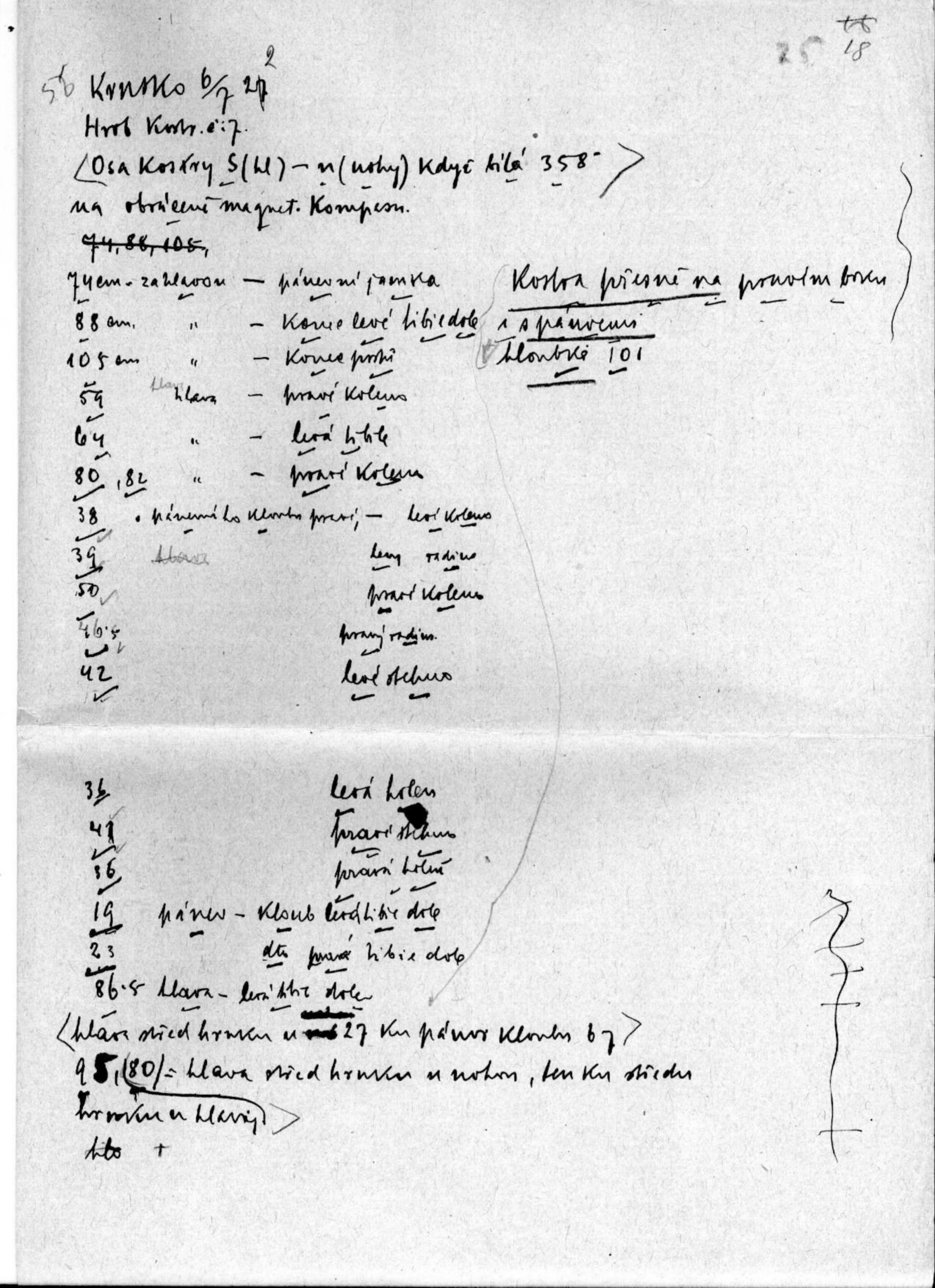}
        \caption{Handwritten text on a gray paper}
        \label{fig:ocr1}
    \end{subfigure}
    % \hfill
    \begin{subfigure}{0.32\textwidth}
        \centering
        \includegraphics[width=\textwidth]{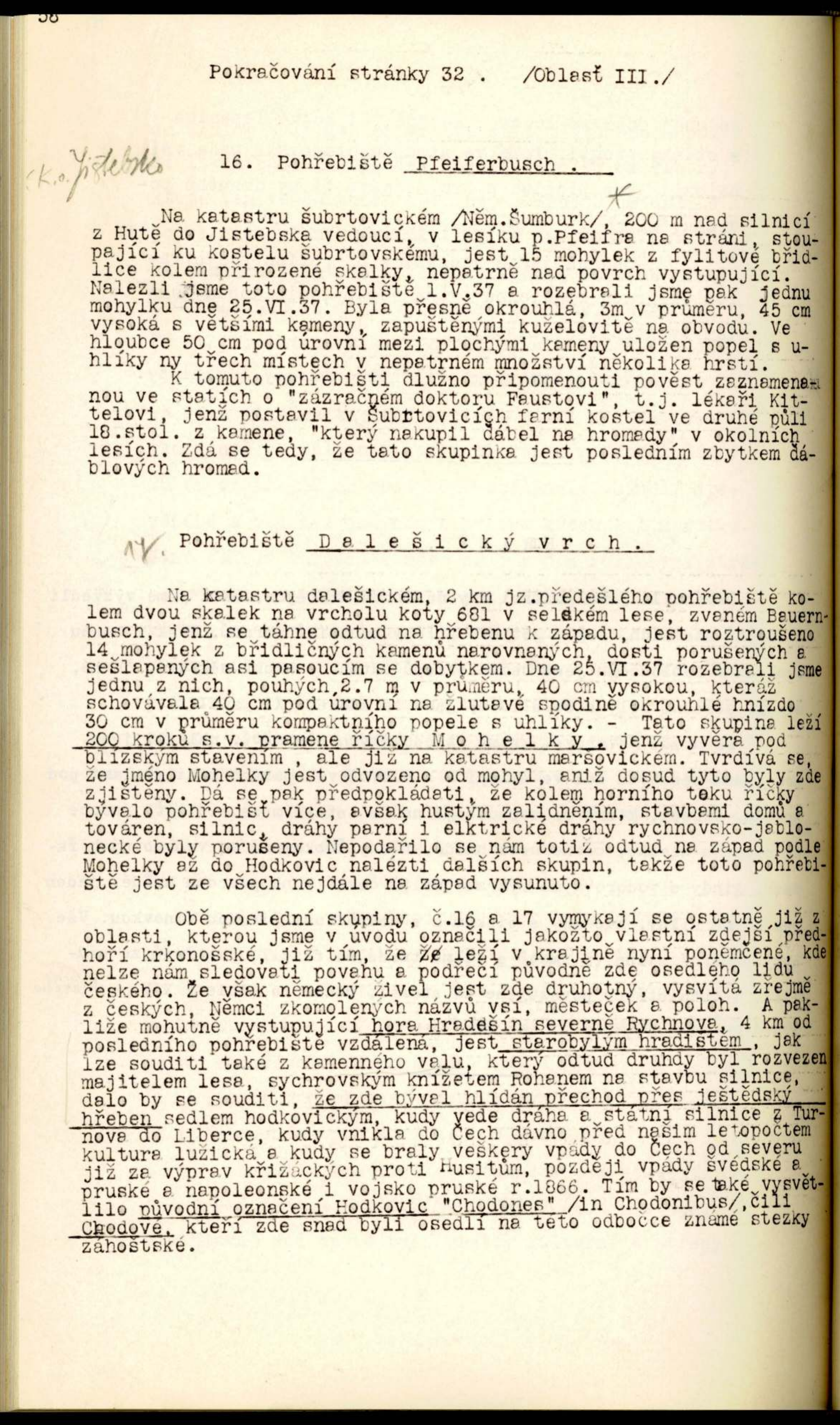}
        \caption{Page from a large volume journal}
        \label{fig:ocr2}
    \end{subfigure}
    % \hfill
    \begin{subfigure}{0.32\textwidth}
        \centering
        \includegraphics[width=\textwidth]{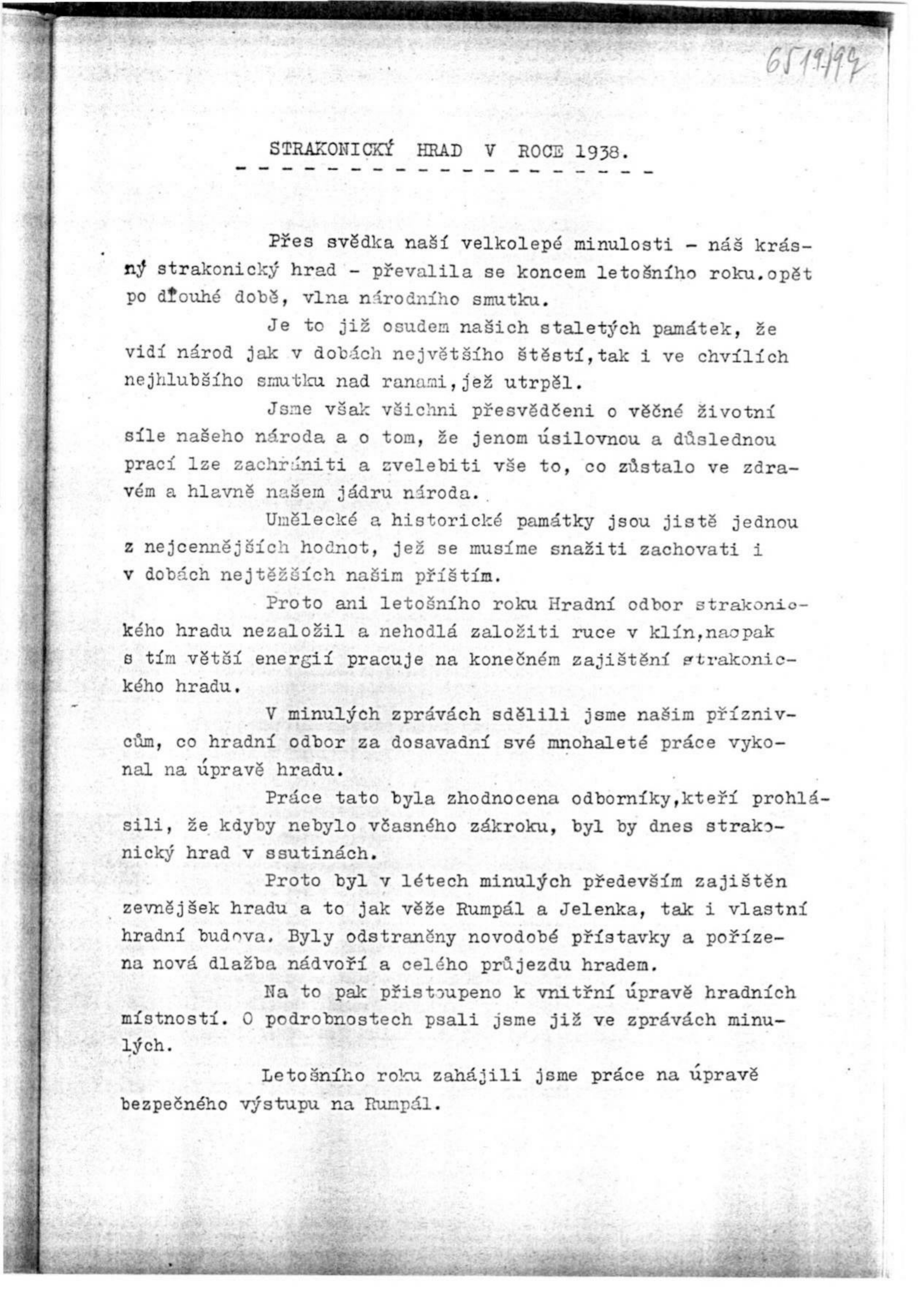}
        \caption{Scanned copy with printing defects}
        \label{fig:ocr3}
    \end{subfigure}
\end{figure}

Prior research in document analysis has typically addressed cleaner, more uniform printed documents; for instance, open-source \ac{ocr} engines like Tesseract \cite{smith2007overview} perform well on standard printed text but struggle with the noisy, handwritten, or structurally complex pages common in historical archives (see Figure \ref{fig:intro-ocr}). The limitations of traditional feature-based classifiers on such heterogeneous data have prompted a shift toward more robust methods.

The combination of diversity and poor documentation is a well-known characteristic of large-scale digitization efforts \cite{nikolaidou2022survey}. Initial consultations with our data providers from \acs{arup} and \acs{arub} confirmed that our collection was perceived as highly disorganized, reflecting a common reality in digitization projects where the scale of data acquisition outstrips the resources available for curation.

Furthermore, the burden of manual curation is a widely recognized challenge in digital humanities and archival sciences \cite{nikolaidou2022survey}. For a large-scale project, this manual process becomes a repetitive pattern-recognition task that is inefficient at scale, directly impeding progress towards a fully analyzed collection.

Recent surveys highlight the success of deep learning models, particularly Transformer-based architectures, in document image analysis \cite{liu2021document,dosovitskiy2020image,touvron2021deit}. Adopting these advanced models, this work develops a page image classification system specifically tailored to the complexities of historical archives. 

\section*{Methodology}

The development process consisted of several stages of design and implementation:

\begin{enumerate}
    \item First, we designed experiments with freely available \ac{dla} tools to justify annotating a new dataset for supervised image classification based on statistics of images' visual features, rather than recognized text and graphical elements pre-extracted from pages. This phase included studying raw data samples and defining the visual elements considered in the analysis.
    \item Next, we designed a set of classification labels to be recognized by the system. We conducted experiments with classic image classification algorithms (\ac{eg} \acs{rfc}, \acs{knn}, \acs{svm}) on a preliminary label set to determine a division of pages into categories that matched end-user needs and the technical capabilities of the applied models. Visual elements that fit into low-resolution patches used by the models (handwriting, table layout, drawings) were structured into distinguishing groups. The resulting annotation scheme was then used to compose a dataset of manually classified pages for fine-tuning more advanced models.
    \item We then fine-tuned state-of-the-art image classification models on the annotated data (several cross-validation folds) and evaluated them on a test set for comparison. This stage included defining model selection criteria to compare fine-tuned models across architectures and select the best models based on accuracy and size--efficiency trade-offs across \acp{cnn} \cite{tan2021efficientnetv2,radosavovic2020designing}, Transformers \cite{dosovitskiy2020image,li2022dit}, and multimodal \ac{clip} models \cite{radford2021learning}.
    \item Selecting a representative data subset for accuracy evaluation was a separate task. Because filenames refer mainly to the year \& month of creation (see Chapter \ref{chap-data-form}), and more than a third of pages in our archive originate from the 21st century, the subset selection algorithm must account for chronological order while allowing randomization for cross-validation model fine-tuning.
\end{enumerate}

Overall, end users' experience of model usage were considered a major factor in the final model selection. Beyond accuracy scores, model size parameters, and common model-specific mistakes across categories, there were no other indicators we could compute automatically.

One possible mitigation for fine-tuned model errors was to run several models per input image and return multiple labels. End users expected models to make consistent mistakes in some categories; comparing outputs from multiple models could help balance these errors.

The intended end users required that the open-source system run efficiently on local infrastructure (\ac{eg} office desktop computers) and offer a user-friendly interface. Because archaeologists are often tied to Windows-specific software in their work, a relatively lightweight and fast tool (models with fewer parameters) that remains accurate and is supported on both Windows and Unix systems was a key objective.

Finally, the project should be reusable from raw data through to a working image classification model. The user-friendly interface should support dataset development and model management and be reusable as the label scheme evolves.

% The remainder of this thesis is organized as follows. 

% Chapter \ref{chap-data-explore} explores the raw data and its properties; Chapter \ref{chap-data-form} details dataset annotation and formation; Chapter \ref{chap-img-class} presents the classification methodology. In Chapter \ref{chap-architect}, the proposed system architecture is detailed. Chapter \ref{chap-results} discusses the practical integration of the system into archival workflows. Finally, the Conclusion with a summary of the findings and suggestions of directions for future research, including a note explaining how \ac{ai} tools were used in the process of writing this thesis.

This thesis will show 4 outcomes:
\begin{enumerate}
    \item in Chapter \ref{chap-data-form} a development of an annotated dataset of almost 50{,}000 pages and its set of categories is shown
    \item in Section \ref{sec-split} data-aware subset selection algorithm is described and justified
    \item in Chapter \ref{chap-results} model architecture comparison based on test-set accuracy and parameters size is tabulated and visualized
    \item in Chapter \ref{chap-architect} practical deployment guidelines released along with the program are described 
\end{enumerate}

\chapter{Exploration of the raw data} \label{chap-data-explore}

\noindent The \ac{arup} archive available for this thesis totals approximately 400--420~GB (more than 60{,}000 \acp{pdf} and almost 650{,}000 pages). All \acp{pdf} were converted into individual \ac{png} image files and organized into directories corresponding to the source documents. This structure formed the basis for exploratory analysis and for manually annotating a representative subset of the collection.

This chapter summarizes the key characteristics of the collection identified during initial exploration and explains how they informed the annotation scheme for manual page classification.

\section{Characteristics of the source data} \label{sec-source-data-char}

The total collection contains 29{,}590 documents holding 649{,}723 image files, with a mean of $\approx 23$ files per folder (median: 4). Additionally, 35{,}978 documents are single-page samples. Most folders are small (75\% contain 15 or fewer pages; 90\% contain 50 or fewer), but a long tail of large documents exists (99th percentile: 240 pages per document).

In practice, this means that most inputs are short (one to several pages), while a small number of very large documents dominate storage and variance. We therefore treat the dataset as many small documents with rare large outliers. Downstream pipelines were designed to process data folder-by-folder and, for large outlier directories, in chunks of 1{,}000 pages.

The 100 largest images range from 61 to 169~MB, which is typical for maps digitized with specialized canvas-size scanning equipment. To handle these large files, we increased the cache memory limit of the Python image-reading library based on observed failures.

Finally, we implemented safeguards for truncated files so that damaged images can be filtered out before creating data loaders (which operate on scaled images represented as numerical vectors).

\subsection{Visual defects of the scanned pages} \label{sec-vis-defects}

\noindent The scanned pages frequently exhibit visual defects caused by both the physical condition of the source documents and the scanning process. These defects make naive document processing unreliable and motivate classifiers that are robust to visual noise.

Artifacts range from minor blemishes (\acs{eg} stains) to severe degradation that complicates automated content extraction and helps explain why general-purpose document analysis pipelines often perform poorly on historical scans.

The primary defect types are summarized below.

\begin{itemize}
    \item \textbf{Background Artifacts and Low Contrast:} A common issue is the presence of aged, yellowed, or gray paper backgrounds, which diminishes the contrast between text and page (Figures \labelcref{fig:defect_cornerhole,fig:defect_gray1,fig:defect_tablehw,fig:e_draw_3,fig:e_draw_5,fig:e_draw_7,fig:e_draw_l_5,fig:e_draw_l_6,fig:e_line_hw_7,fig:e_line_p_5,fig:e_line_t_2,fig:e_line_t_7,fig:e_photo_3,fig:e_photo_l_6,fig:e_text_3,fig:e_text_hw_3,fig:e_text_t_9,fig:e_draw_2,fig:e_line_t_3,fig:e_line_t_6,fig:e_text_7,fig:e_text_hw_2,fig:e_text_t_2}). This degradation is directly caused by paper aging and the quality of the original materials.
    
    \begin{figure}[!h]
        \centering
        \caption{Examples of content defects that reduce readability.}
        \label{fig:defects-1}
        \begin{subfigure}{0.3\textwidth}
            \centering
            \includegraphics[width=\textwidth]{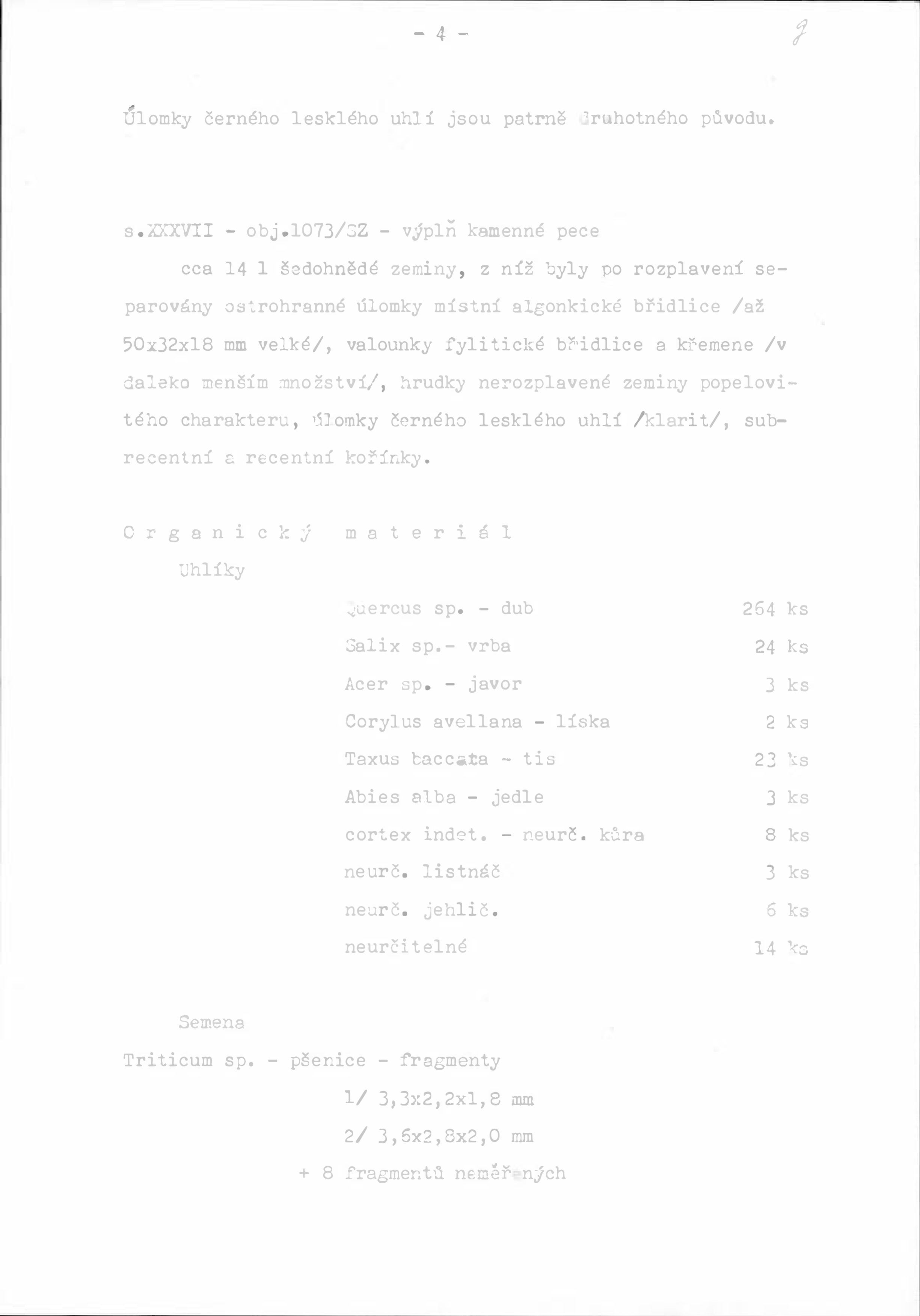}
            \caption{Low-contrast text}
            \label{fig:def1}
        \end{subfigure}
        % \hfill
        \begin{subfigure}{0.3\textwidth}
            \centering
            \includegraphics[width=\textwidth]{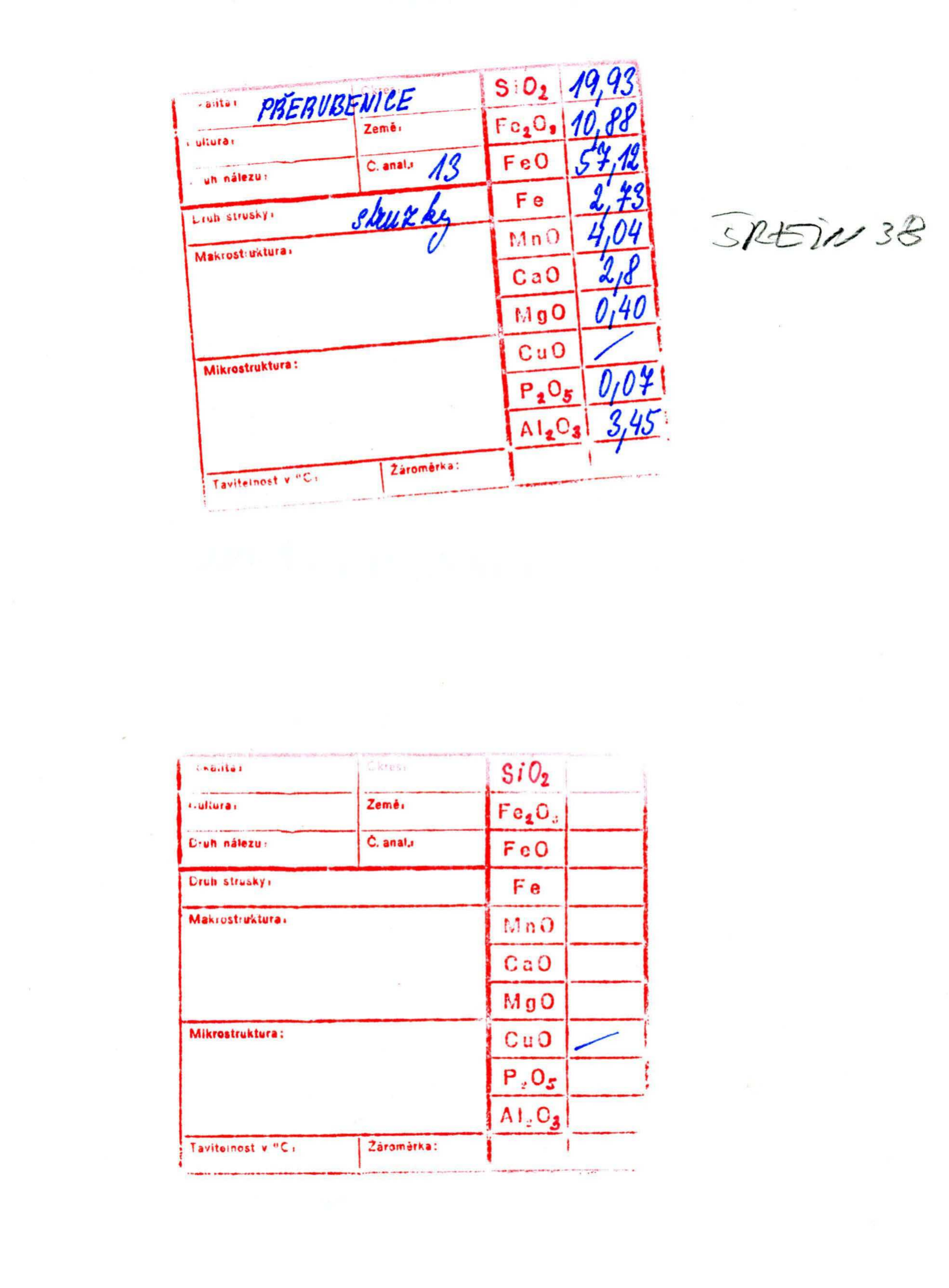}
            \caption{Skewed table}
            \label{fig:def2}
        \end{subfigure}
        % \hfill
        \begin{subfigure}{0.3\textwidth}
            \centering
            \includegraphics[width=\textwidth]{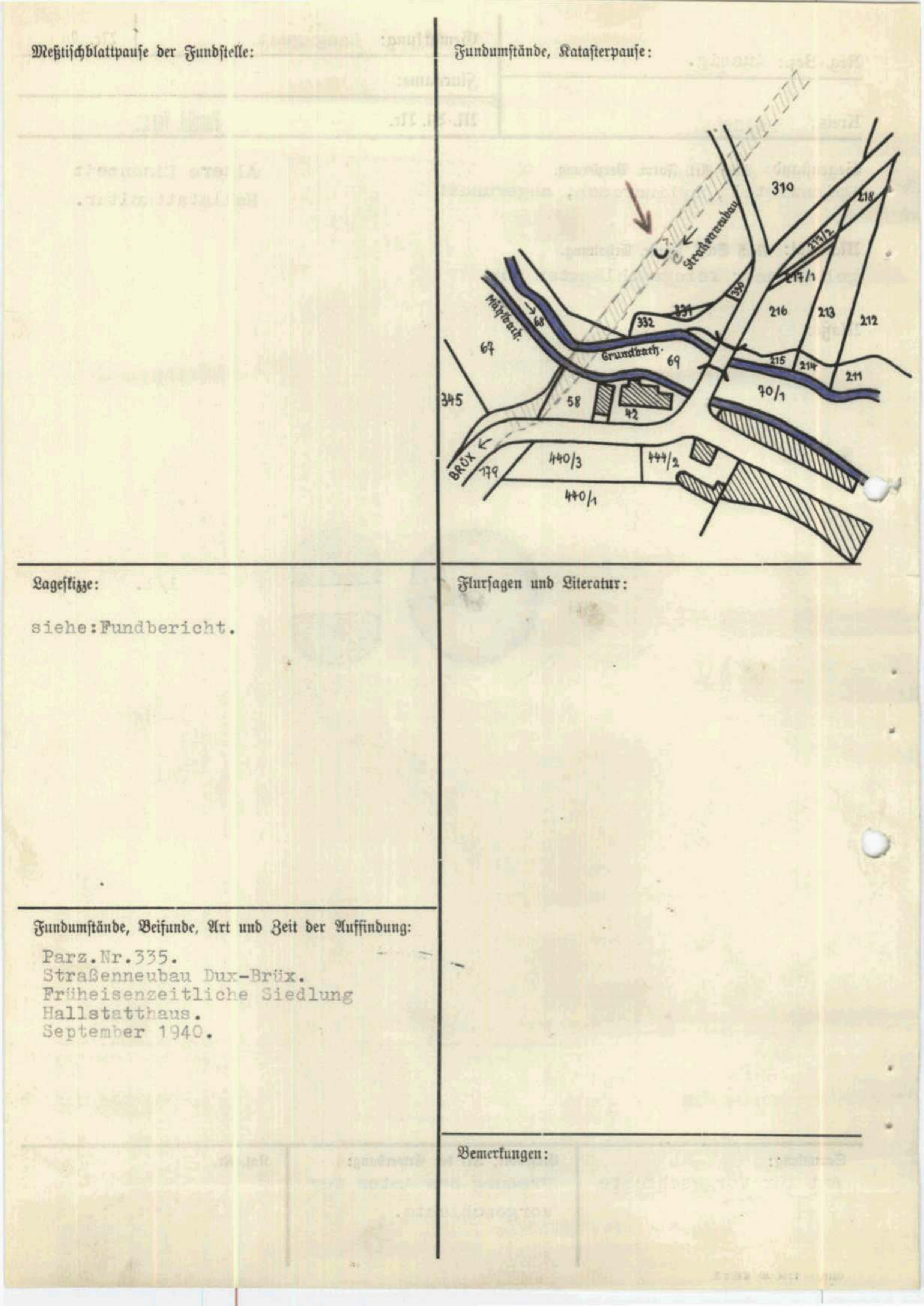}
            \caption{Bleed-through tables}
            \label{fig:def3}
        \end{subfigure}
    \end{figure}
    
    \item \textbf{Page Skew and Alignment Issues:} Many pages suffer from skew, where content is not aligned horizontally (Figures \labelcref{fig:defect_skew,fig:e_draw_7,fig:e_line_hw_3,fig:e_line_p_3,fig:e_line_p_7,fig:e_line_t_4,fig:e_line_t_6,fig:e_photo_1,fig:e_text_4,fig:e_text_5,fig:e_text_t_3,fig:e_text_t_7,fig:e_text_t_9}). This is a well-documented \cite{biswas2023document} problem in \ac{ocr} literature that can arise from improper paper feeding during scanning or the document's original state and often requires specialized preprocessing to correct.
    
    \item \textbf{Text Bleed-Through:} On documents printed on thin paper, ink on the reverse side is often visible, creating superimposed text that interferes with primary content (Figures \labelcref{fig:defect_bleed,fig:e_text_hw_2}). This phenomenon, known as bleed-through, is a significant challenge for \ac{ocr} systems, as it introduces noise that can be difficult to segment from the foreground text.
    
    \item \textbf{Water Damage:} Some documents show clear signs of water damage, resulting in blurred ink, stains, and overlapping text (Figures \labelcref{fig:defect_water,fig:e_text_hw_7}). This type of degradation is particularly severe in documents that have been exposed to events such as floods.
    
    \item \textbf{Physical Damage:} Prevalent physical damage includes tears, holes, and worn edges (Figures \labelcref{fig:defect_cornerhole,fig:e_line_t_6,fig:e_photo_1,fig:e_text_7,fig:e_text_t_7,fig:defect_edgeholes,fig:e_text_t_4,fig:e_text_t_6}). This ranges from simple corner tears to more significant edge damage and punch holes from binding.

    \begin{figure}[!h]
        \centering
        \caption{Defects and physical page features transferred to digitized scans as transparent or black fragments.}
        \label{fig:defects-3}
        \begin{subfigure}{0.3\textwidth}
            \centering
            \includegraphics[width=\textwidth]{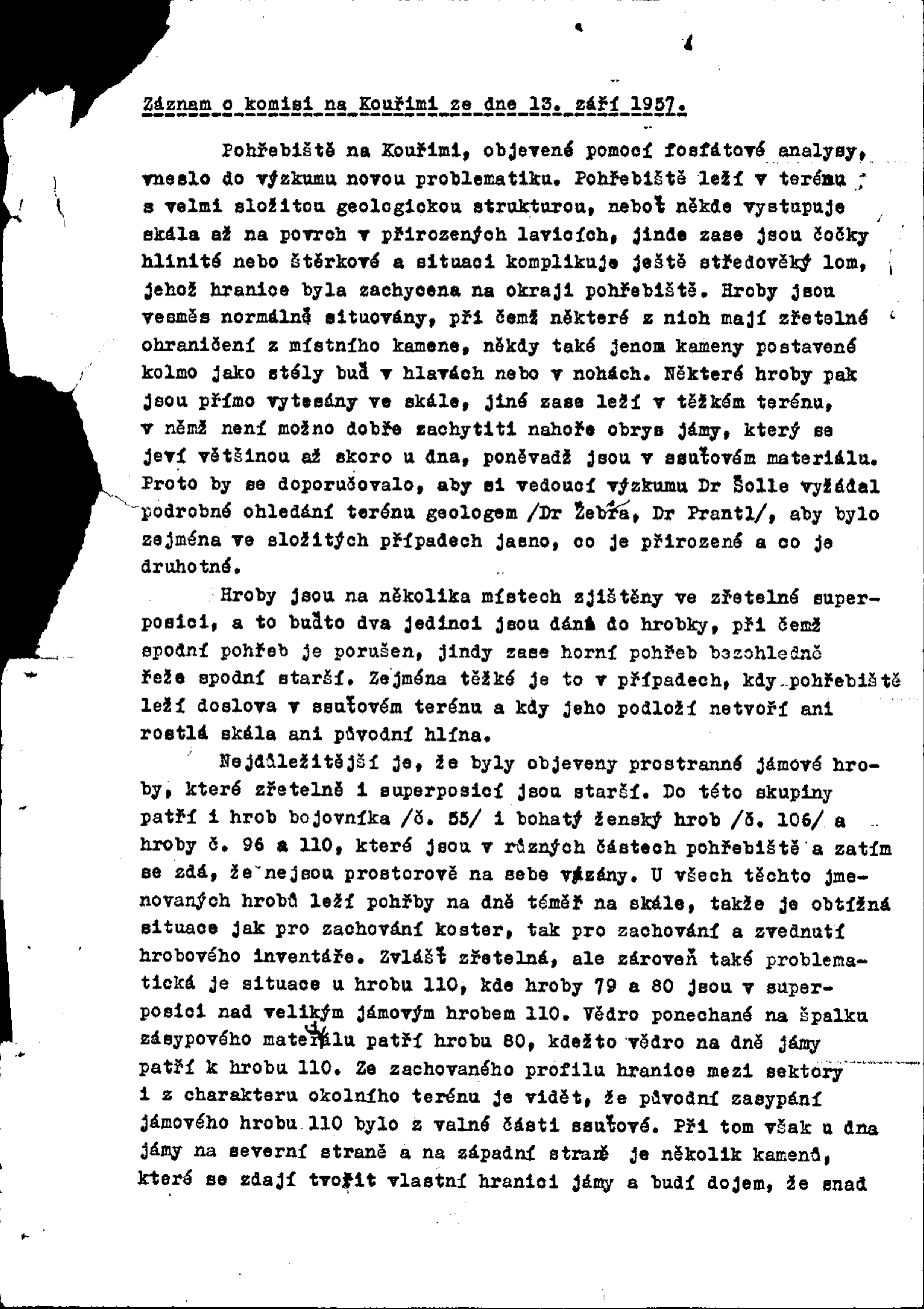}
            \caption{Torn page corner}
            \label{fig:def6}
        \end{subfigure}
        % \hfill
        \begin{subfigure}{0.3\textwidth}
            \centering
            \includegraphics[width=\textwidth]{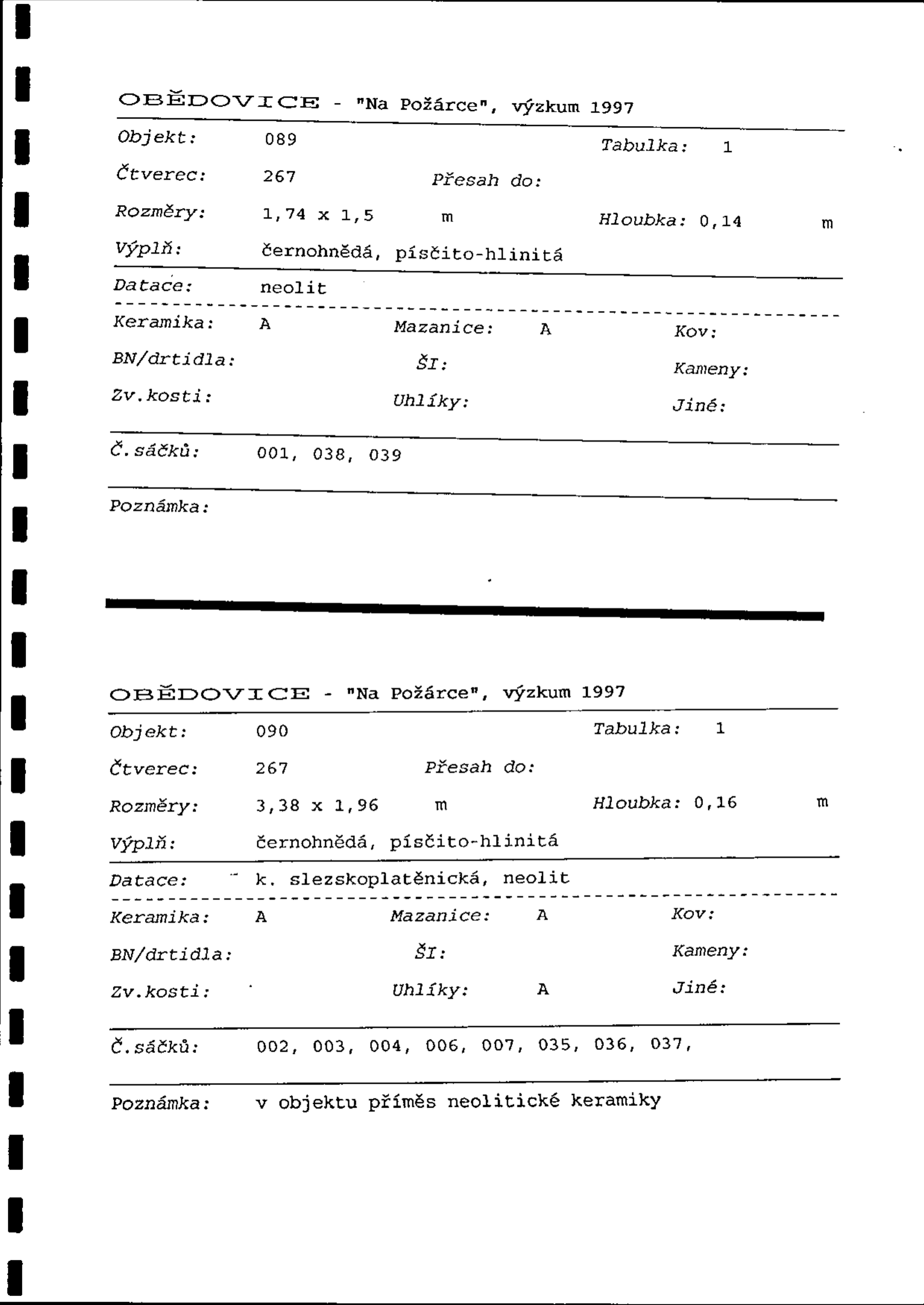}
            \caption{Large volume bound \& skewed table}
            \label{fig:def7}
        \end{subfigure}
        % \hfill
        \begin{subfigure}{0.3\textwidth}
            \centering
            \includegraphics[width=\textwidth]{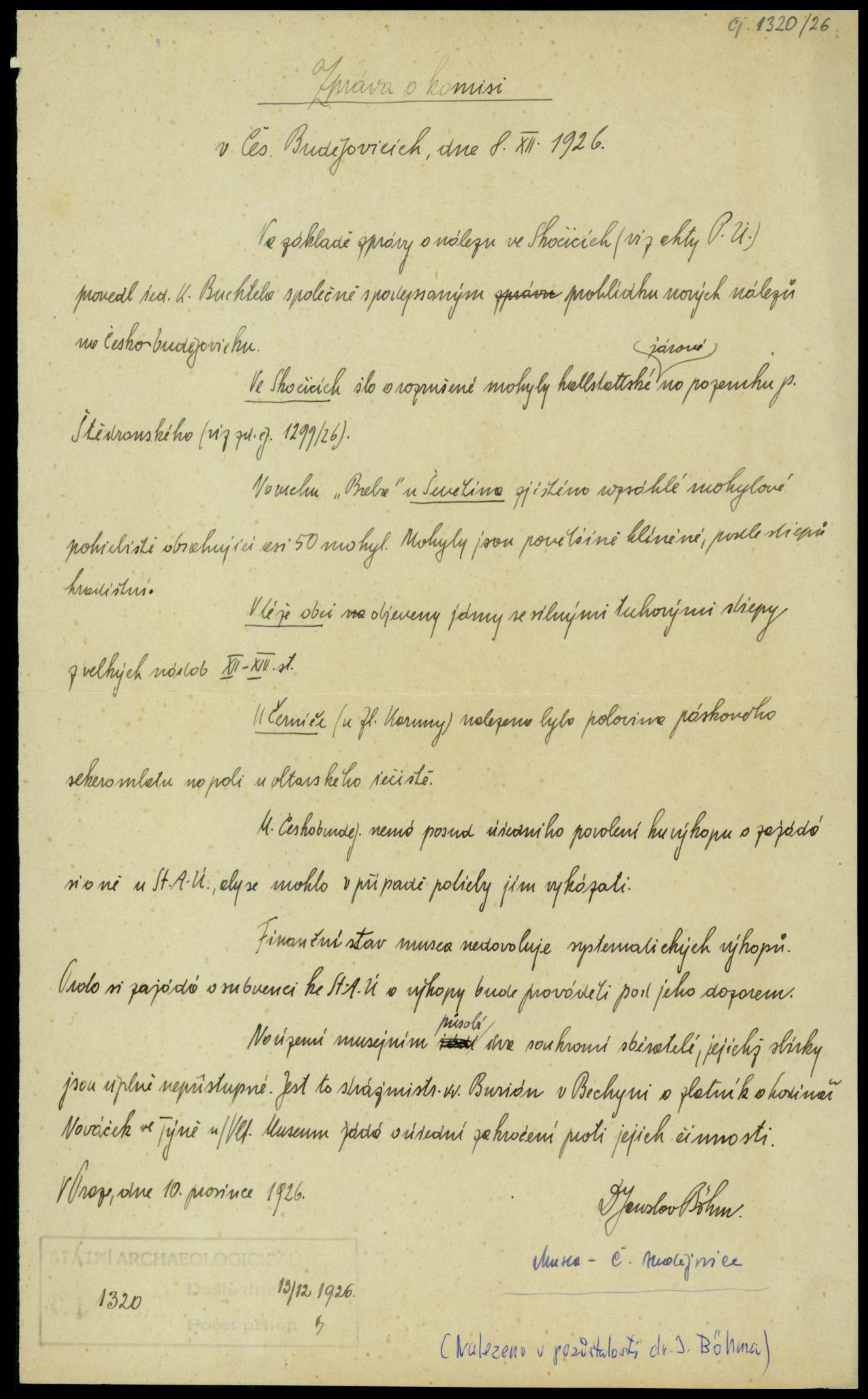}
            \caption{Corrections \& filled-in stamp}
            \label{fig:def8}
        \end{subfigure}
    \end{figure}
    
    \item \textbf{Scanning artifacts from Bound Volumes:} Scanning pages from thick, bound journals that cannot lie flat often introduces page curl and a dark gradient near the inner margin (Figures \labelcref{fig:defect_fatjournal,fig:e_text_t_2}).

\end{itemize}

\subsection{Textual Variations and Annotations} \label{sec-data-variants}

\noindent The documents also show substantial variation in textual presentation and annotation. Our objective is therefore to capture visual features that computer vision models can detect reliably and use as signals for page classification.

The diversity of page layouts---complicated by stamps (often small, fillable forms) placed at arbitrary locations---motivated an approach that is largely independent of heavy preprocessing. The model must generalize across visual features and capture differences in textual elements, including mixtures of printed, typewritten, and handwritten text within a single image.

In addition, the model must recognize sketches and handwritten content on grainy paper and distinguish graphical elements of interest from the visual noise described in Subsection~\ref{sec-vis-defects}.

\begin{itemize}
    \item \textbf{Stamps:} Official stamps and other preformatted ink annotations are frequently found on the documents, sometimes appearing as faint graphical elements. Figures \labelcref{fig:defect_stamp,fig:e_line_t_1,fig:e_line_t_2,fig:e_photo_3,fig:e_text_1,fig:e_text_5,fig:e_text_6,fig:e_text_hw_6,fig:e_text_hw_7} illustrate various pages with stamp impressions. 

    \begin{figure}[!h]
        \centering
        \caption{Variability of handwritten font sizes based on the format of physical pages}
        \label{fig:defects-2}
        \begin{subfigure}{0.45\textwidth}
            \centering
            \includegraphics[width=\textwidth]{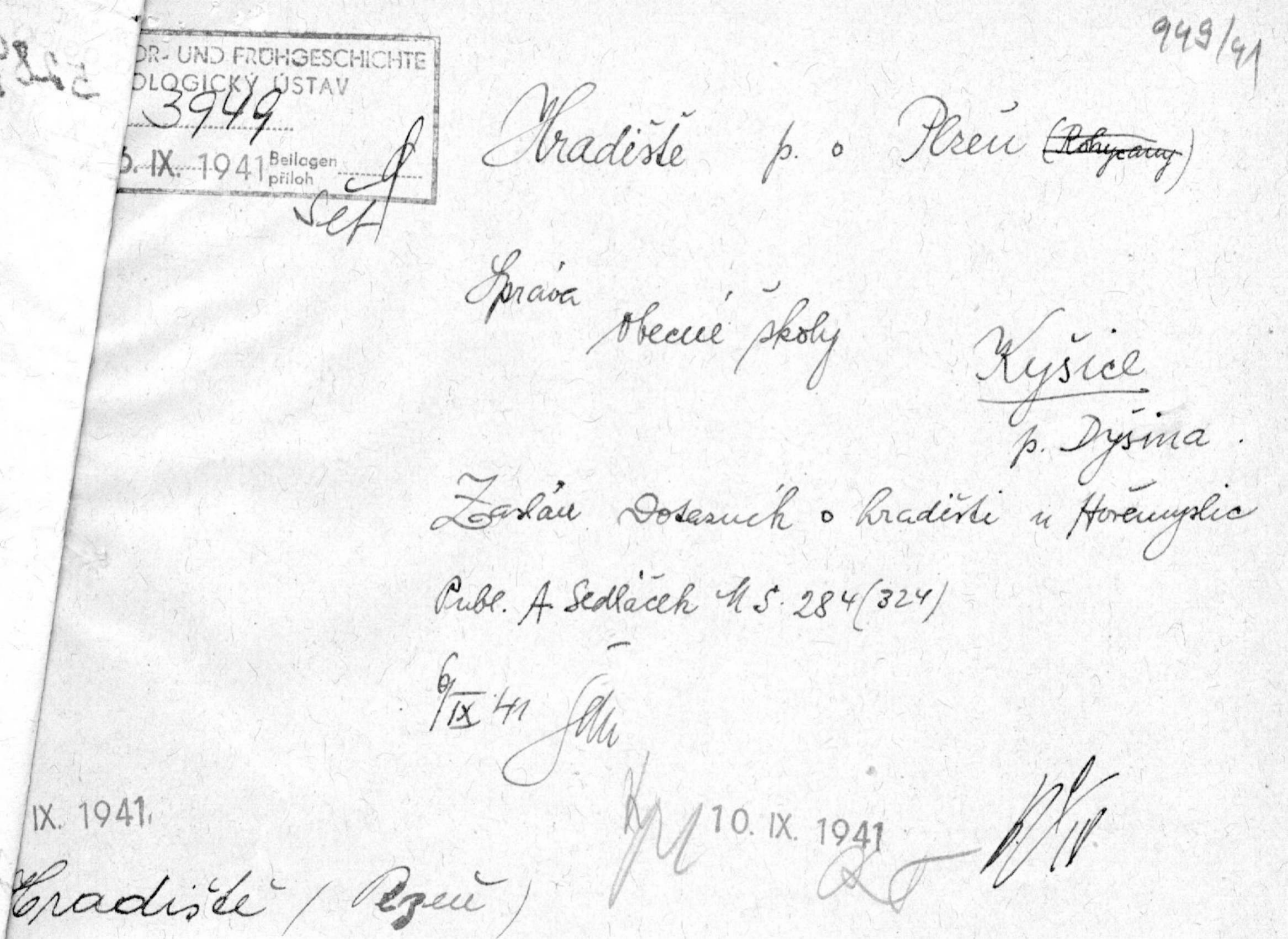}  
            \caption{Wet paper, Czech writing \& a German stamp}
            \label{fig:def4}
        \end{subfigure}
        % \hfill
        \begin{subfigure}{0.45\textwidth}
            \centering
            \includegraphics[width=\textwidth]{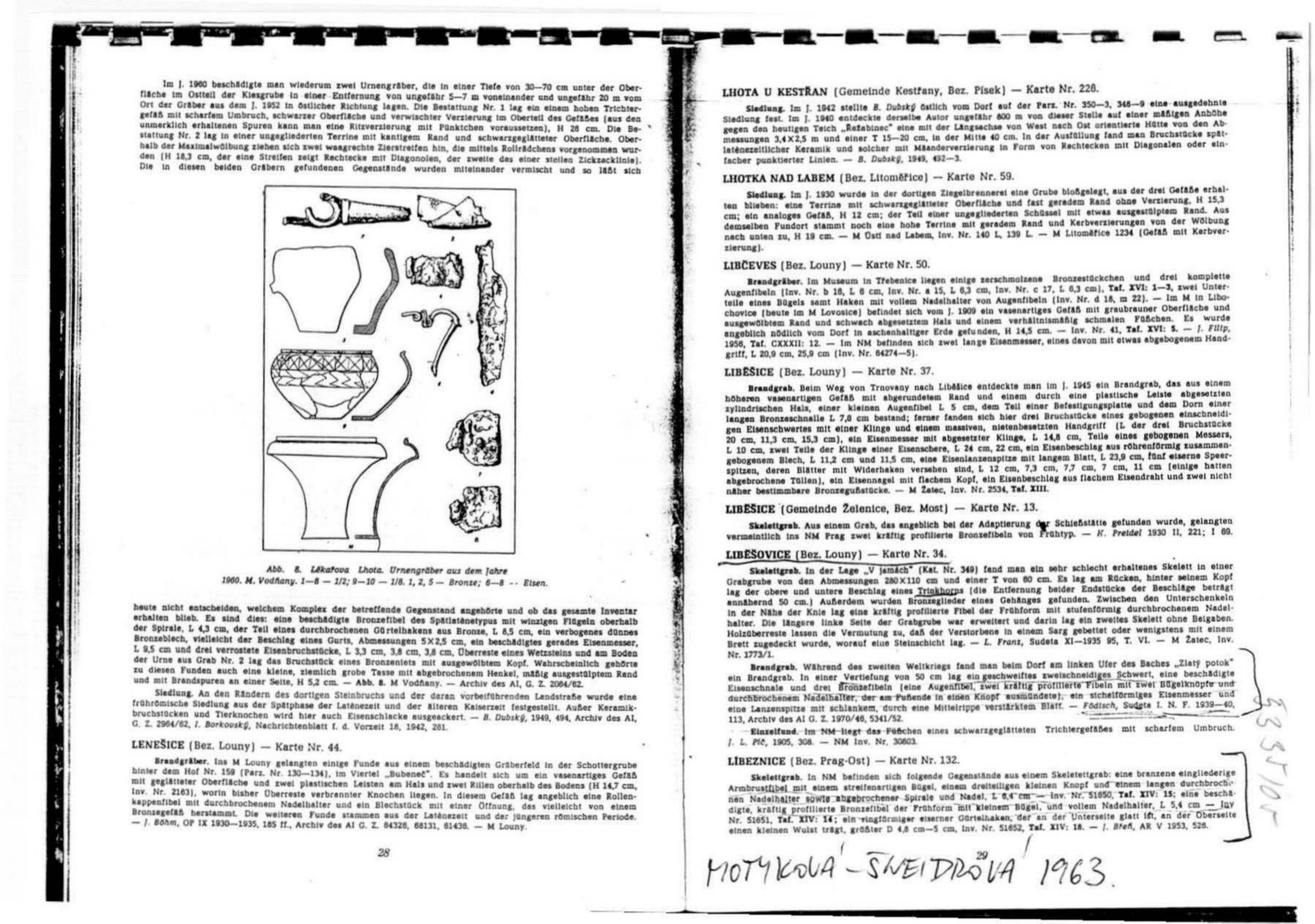}
            \caption{Commented book scan with a drawing}
            \label{fig:def5}
        \end{subfigure}
    \end{figure}

    \item \textbf{Manual Corrections:} Handwritten modifications were common, ranging from simple strikeouts of characters and words to the removal of entire paragraphs (see Figures \labelcref{fig:defect_textt,fig:e_line_t_2,fig:e_photo_l_1,fig:e_line_hw_3,fig:e_text_3,fig:e_text_hw_3,fig:e_text_hw_7}) to interlinear notes on typewritten documents (Figures \labelcref{fig:textt1,fig:e_line_hw_5,fig:e_line_hw_6,fig:e_line_t_2,fig:e_text_3}). 
    
    \item \textbf{Scribbles and annotations:} Beyond formal corrections, pages frequently contained scribbles, underlines, and other margin notes, as shown in Figures \labelcref{fig:tablehw1,fig:e_draw_6,fig:e_line_t_1,fig:e_photo_1}. These marginal annotations are often ambiguous, making it difficult to determine their relevance or relationship to the primary content.

    \item \textbf{Mixed Text Styles:} Pages often combined multiple text formats. For instance, typewritten documents frequently had handwritten page numbers or comments (Figures \labelcref{fig:skew1a,fig:e_text_1,fig:e_text_2,fig:e_text_3}). Front pages might mix printed letterheads, stamps (Figures \labelcref{fig:stamp2,fig:e_line_t_1,fig:e_line_t_2,fig:e_photo_3,fig:e_text_1,fig:e_text_5,fig:e_text_6,fig:e_text_hw_6,fig:e_text_hw_7}), and handwritten notations, followed by typewritten text, drawings, or forms filled in by hand. Figures \labelcref{fig:defect_gray1,fig:example_text} illustrate such mixed-content pages.

    \item \textbf{Text within Graphics:} Finally, textual elements were commonly embedded within graphical content. Maps and technical drawings included labels and captions (Figures \labelcref{fig:large1,fig:large2,fig:example_draw,fig:example_draw_l}), while photographs were often accompanied by typewritten or handwritten descriptions (Figures \labelcref{fig:mixed2,fig:mixed3,fig:e_photo_4,fig:e_photo_5}). Consequently, purely graphical pages devoid of any text were relatively rare.
\end{itemize}

The range of text styles that can be distinguished is limited by the models' input resolution (i.e., the degree of downsampling), leaving mainly handwritten, typewritten, and printed genres of textual content. In contrast, layout (whether mixed in style or not) is expected to be captured more easily at typical model input sizes of 200--400 pixels in width.

These varied annotations and mixtures of typewritten and handwritten text mean that any classification system must be robust to diverse, combined content types on a single page.

All samples in Figures \labelcref{fig:defects-1,fig:defects-2,fig:defects-3} and additional examples are provided in Appendices \labelcref{app-pages,app-labels}.

In summary, the visual characteristics of the source pages vary too much for a one-size-fits-all preprocessing strategy (\ac{eg} globally increasing contrast or sharpening edges). Instead, an effective system must be robust to page color, paper grain, and low contrast between text and background, and it should tolerate pencil drawings and informal annotations.

\section{Application of the available \acs{dla} framework} \label{sec-dla}

\noindent \acl{dla} aims to identify and categorize components within a document image, such as text blocks, images, and tables. We evaluated \ac{dla} tools as a potential ``off-the-shelf'' solution and as a way to understand which visual signals are realistically detectable in our data.

The primary purpose of these experiments was to justify the need for a problem-specific labeled dataset and to inform the label scheme used later for supervised model training. To establish a baseline, we applied the \href{https://github.com/deepdoctection/deepdoctection/blob/master/README.md}{DeepDoctection} framework to detect structural elements (examples in Figures \labelcref{fig:dd_newspaper,fig:dd_table,fig:dd_text,fig:dd_mis_draw}). In parallel, we used the Tesseract \ac{ocr} engine \cite{smith2007overview} to assess text extraction quality.

These experiments did not rely on manual ground-truth labels. Instead, they used automatically detected elements (tables, figures, text blocks) to infer page layout and to estimate whether rule-based categories could support reliable classification.

\begin{table}[h!]
\centering
\begin{tabular}{p{2cm}p{12cm}}
\toprule
\textbf{Category} & \textbf{Description} \\
\midrule
Form & Pages characterized by horizontal lines (\texttt{H\_line}) but lacking the high vector line counts associated with technical drawings. Also includes pages where detected table content outweighs text content (\texttt{TXT < TAB}), implying a structured layout without a strict grid. \\
Form-figure & A hybrid category where pages contain both horizontal lines and moderate vector line complexity (\texttt{maybe\_picture}), or where both Table and Image detectors return positive values. \\
Table & Pages strictly defined by the presence of vertical or horizontal separators (\texttt{V\_line}, \texttt{H\_line}) where text content is negligible or non-existent, representing clean grids. \\
Text-body & Pages dominated by recognized text blocks (\texttt{TXT > 0.9}) or explicitly flagged as \texttt{Manuscript} (handwritten content). This category also applies when headers exist but make up a smaller portion of the page than the body text. \\
Gallery & Pages identified primarily by high counts of ``long'' or ``short'' vector lines (exceeding 1000--3000 lines), indicating technical drawings, maps, or blueprints. Also includes pages where image content occupies the majority of the page area (\texttt{IMG > 0.9}). \\
Figure-text & Pages containing a mixture of text and visual elements where the image content is present but does not dominate the page (\texttt{TXT > IMG}), or where text is present but significantly less than the image area (\texttt{TXT < IMG}). \\
Table-text & Mixed-content pages where a table is detected but is smaller in proportion to the accompanying text block (\texttt{TAB < TXT}). \\
Headers-text & Pages where text is detected, but the header regions are calculated to be larger or more significant than the remaining text body (\texttt{TXT < HDR}). \\
Mixed & A fallback category for pages containing a combination of Images, Headers, Text, and Tables, where no single element meets the dominance threshold to trigger a specific category. \\
Neither & A default state for pages that do not trigger any specific detection rules. \\
Empty-text & Pages with very low content scores and vector line counts below the threshold (\ac{eg} $<$ 100 lines), representing (nearly) blank pages. \\
\bottomrule
\end{tabular}
\caption{Classification categories based on detected content features and line complexity. No ground-truth labels existed for this scheme.}
\label{tab:categories_draft}
\end{table}

Table~\ref{tab:categories_draft} shows our initial, rule-based label set, derived from detected layout elements and counts of long/short horizontal and vertical lines.

End users reviewed the proposed labels and the per-page element counts produced by these handcrafted rules (\ac{eg} text lines, headers, tables, images).

\begin{figure}[h]
    \centering
    \caption{\ac{dla} application samples}
    \label{fig:dla-sample}
    \begin{subfigure}{0.3\textwidth}
        \centering
        \includegraphics[width=\textwidth]{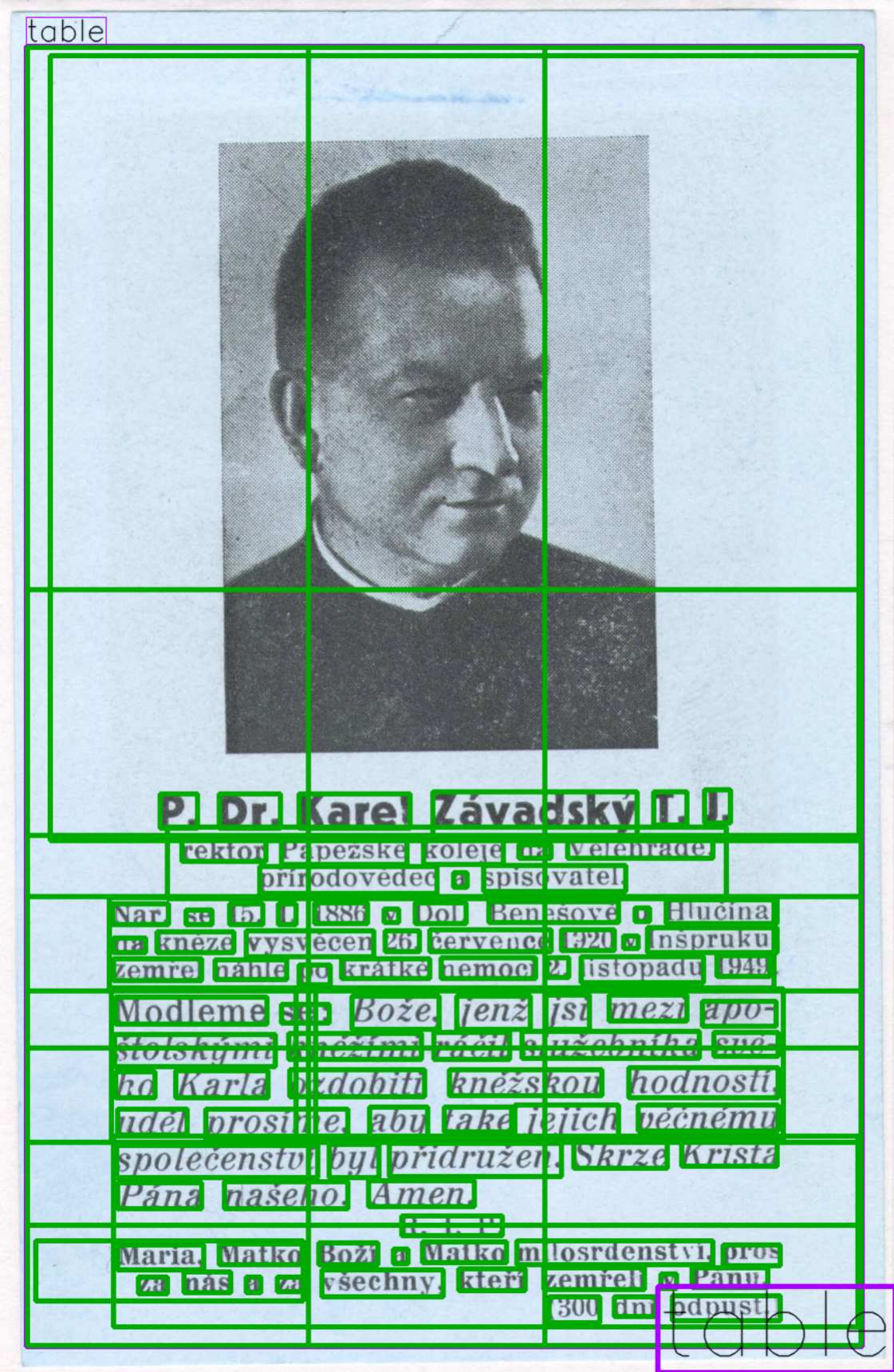}
        \caption{Imaginary tables \& ignored figure}
        \label{fig:dla1}
    \end{subfigure}
    % \hfill
    \begin{subfigure}{0.3\textwidth}
        \centering
        \includegraphics[width=\textwidth]{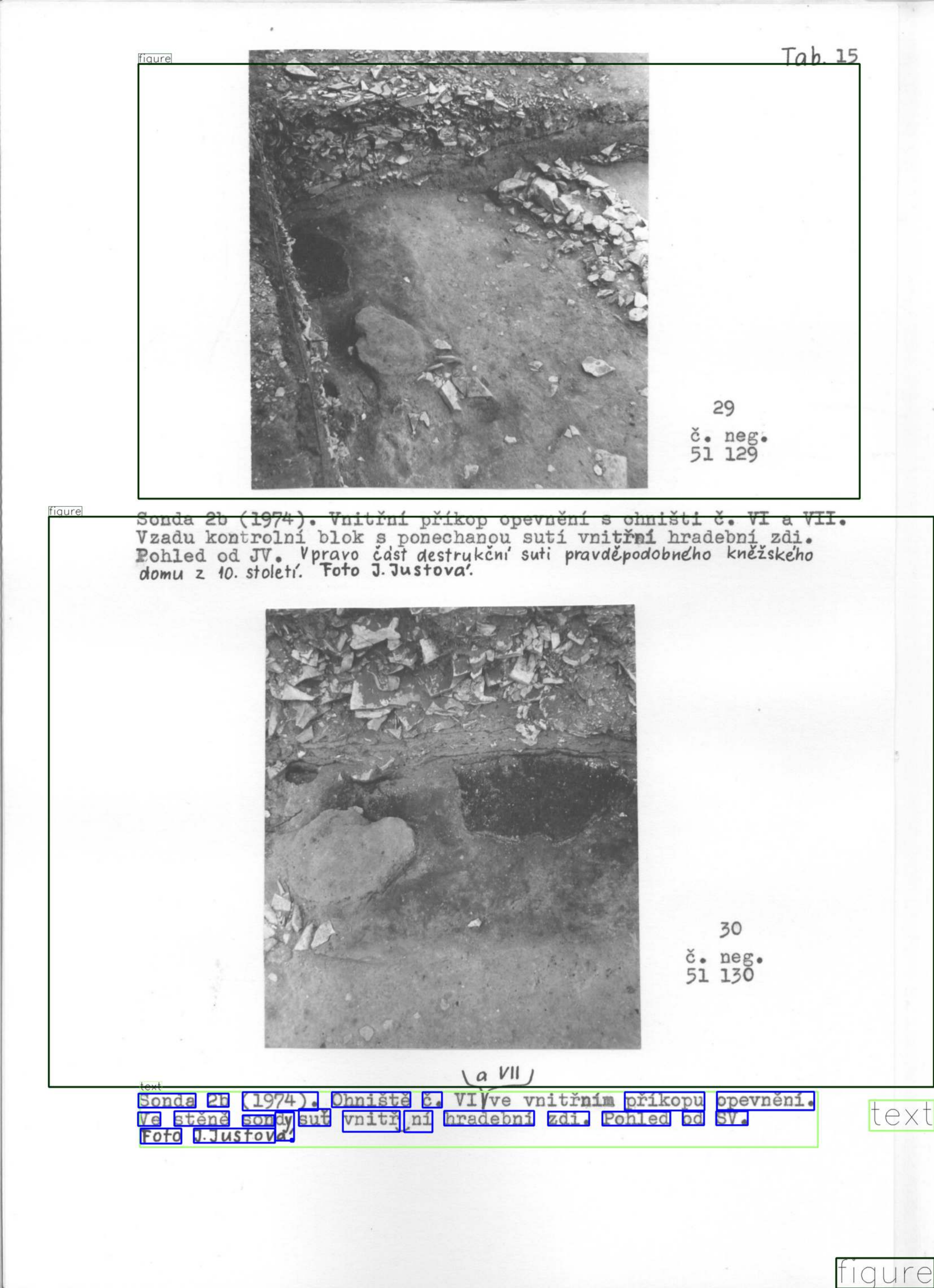}
        \caption{Ignored text \& imaginary figures}
        \label{fig:dla2}
    \end{subfigure}
    % \hfill
    \begin{subfigure}{0.3\textwidth}
        \centering
        \includegraphics[width=\textwidth]{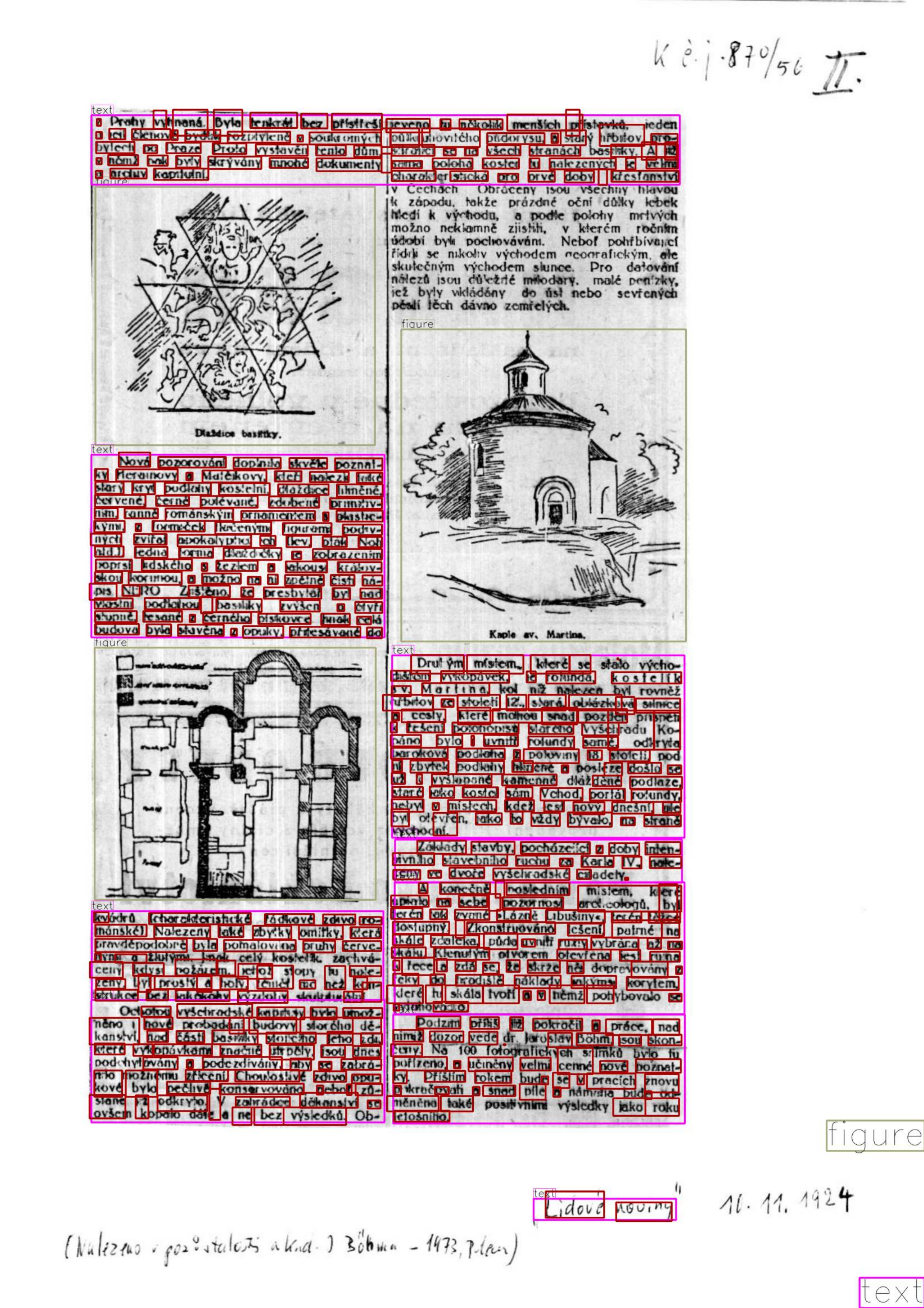}
        \caption{Ignored text paragraph}
        \label{fig:dla3}
    \end{subfigure}
\end{figure}

These trials indicated that off-the-shelf tools could not meet the project's content-specific classification goals (Figure~\ref{fig:dla-sample}), motivating a tailored solution.

\subsection{\ac{ocr} performance} \label{sec-dla-ocr}
\noindent \acf{ocr} converts images of text into machine-readable text, enabling search and downstream analysis.

We applied the Tesseract \cite{TessOverview,smith2007overview} \ac{ocr} engine (via the \ac{dla} framework) to a sample of pages. Recognition was accurate on clean, high-contrast scans (Figure~\ref{fig:dd_text}) but degraded on dark or noisy backgrounds, often producing incomplete or garbled output (Figure~\ref{fig:dd_mis_texts}).

\subsection{Structured data detection and extraction} \label{sec-dla-struct}

\begin{figure}[h]
    \centering
    \caption{\ac{dd} mistakes on pages with tables and figures}
    \label{fig:dd_mistakes}
    \begin{subfigure}{0.3\textwidth}
        \centering
        \includegraphics[width=\textwidth]{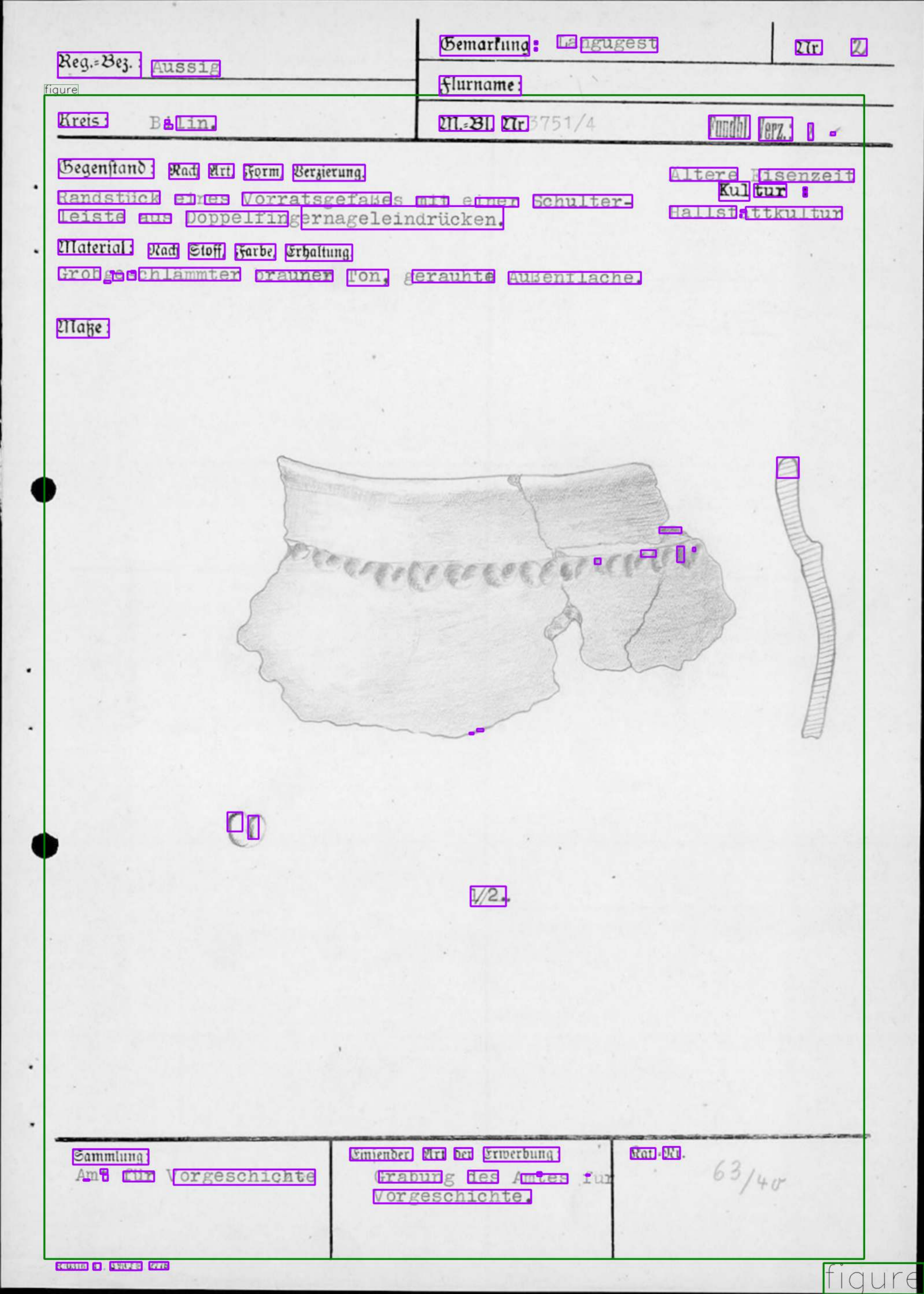}
        \caption{Table as a figure}
        \label{fig:dd_mis1}
    \end{subfigure}
    % \hfill
    \begin{subfigure}{0.3\textwidth}
        \centering
        \includegraphics[width=\textwidth]{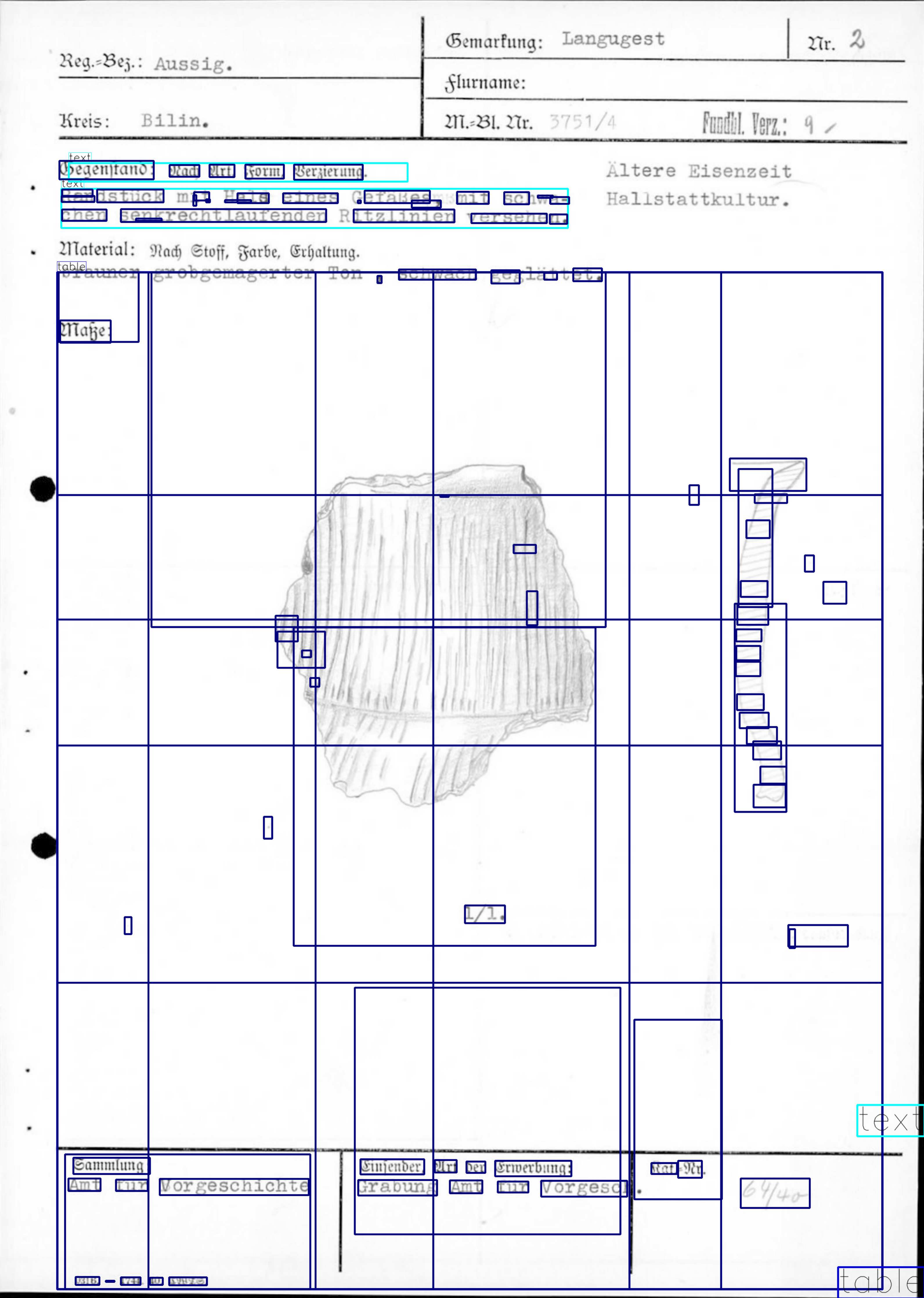}
        \caption{Drawing as a table}
        \label{fig:dd_mis2}
    \end{subfigure}
    % \hfill
    \begin{subfigure}{0.3\textwidth}
        \centering
        \includegraphics[width=\textwidth]{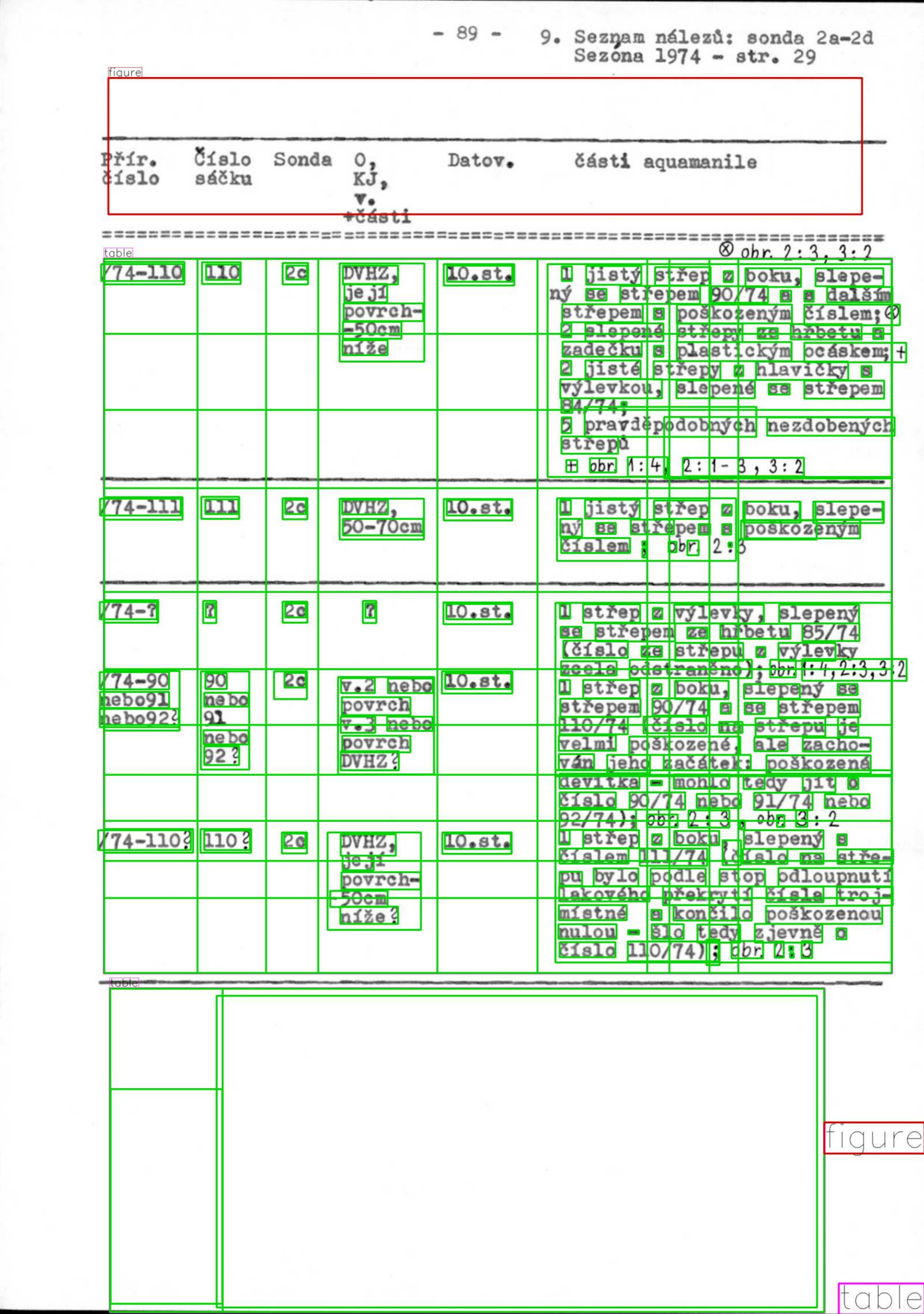}
        \caption{Header as a figure}
        \label{fig:dd_mis3}
    \end{subfigure}
\end{figure}

\noindent For general \ac{dla} and table recognition, we used \href{https://github.com/facebookresearch/detectron2/blob/18f69583391e5040043ca4f4bebd2c60f0ebfde0/README.md}{Detectron2} (Facebook \acs{ai} Research), which also serves as the table-recognition module in \acl{dd}.

Table detection and structure extraction were unreliable: \ac{dd} merged rows or missed cells when borders were faint, incomplete, or skewed (Figures~\labelcref{fig:dd_table,fig:dd_mis_tables}), and it sometimes confused tables with other page elements. A recurring failure case involved tables placed near page corners, which were often ignored.

\subsection{Graphic elements detection} \label{sec-dla-img}

\noindent Detection of graphical elements (maps, drawings, photographs) was often inaccurate: items were missed or misclassified (\ac{eg} maps labeled as tables; Figures~\labelcref{fig:dd_mis_draw,fig:dd_mix,fig:dd_newspaper}), and handwritten annotations further confused the detector. Overall, the tested tools did not reliably capture the archive's graphical diversity.

\begin{figure}[h]
    \centering
    \caption{\ac{dd} mistakes on pages with maps and drawings}
    \label{fig:dd_mistakes_draw}
    \begin{subfigure}{0.3\textwidth}
        \centering
        \includegraphics[width=\textwidth]{img/pages/dd_mis_map.pdf}
        \caption{Map as a table}
        \label{fig:dd_mis4}
    \end{subfigure}
    % \hfill
    \begin{subfigure}{0.3\textwidth}
        \centering
        \includegraphics[width=\textwidth]{img/pages/dd_mis_map_legend.pdf}
        \caption{Map as a table \& ignored legend}
        \label{fig:dd_mis5}
    \end{subfigure}
    % \hfill
    \begin{subfigure}{0.3\textwidth}
        \centering
        \includegraphics[width=\textwidth]{img/pages/dd_mis_rocks.pdf}
        \caption{Figures as a table}
        \label{fig:dd_mis6}
    \end{subfigure}
\end{figure}

\subsection{Human expert feedback} \label{sec-dla-feedback}

\noindent A domain expert from \ac{arup} reviewed the \ac{dla} outputs as an intended end user, focusing on whether the results were inspectable and trustworthy.

Given the heterogeneity described in Sections~\labelcref{sec-vis-defects,sec-data-variants}, the expert expected layout predictions from open-source models to be an unreliable basis for content classification.

The handcrafted label set (Table~\ref{tab:categories_draft}) showed inconsistent assignments for visually similar pages, motivating clearer category definitions in the next scheme (Tables~\labelcref{tab:categories_initial,tab:categories_proposal}).

Feedback emphasized misclassification of high-priority content, particularly full-page tables (\ac{eg} Figure~\ref{fig:dd_table}) and large drawings (\ac{eg} Figure~\ref{fig:dd_mix}); \ac{dla} sometimes swapped these classes (Figure~\ref{fig:dd_mistakes}).

While preprocessing (\acs{eg} binarization, thresholding) may improve some cases, manual validation at archive scale is impractical; meaningful evaluation therefore required a manually annotated dataset, which motivated provider involvement in annotation review.

\section{Critical human expert feedback} \label{sec-dla-critical}

Further collaboration with the domain expert clarified the classification criteria. Evaluation on a held-out set highlighted systematic limitations of the automated approach and helped refine the project goals and final categories.

The \ac{dla} test set was relatively small (fewer than 2{,}000 pages) and intentionally contained difficult samples (\acs{eg} water damage, copy scans), allowing worst-case scenarios to be identified early.

The data providers summarized their conclusions about the applied \ac{dla} tool as follows:
\begin{itemize}
    \item \textbf{Classification consistency:} Pages with similar content must be assigned to the same category. Consistency was prioritized over isolated instances of correctness.

    \item \textbf{Primacy of structured data:} Pages containing tabular or form-like data must be classified as such, even when substantial plain text is present (\ac{eg} Figure \ref{fig:dd_mistakes}). Only fillable stamps are of interest (as structured data), although they may be ignored if values are barely legible or the page contains clearer mixed content.

    \item \textbf{Priority of graphical content:} A significant graphical element (\ac{eg} photograph, map, drawing) should take precedence over text. The definition of ``significant'' was refined from an initial one-third-of-page threshold to a smaller, stamp-sized element, reflecting the actual archive content.

    \item \textbf{Handling of handwritten annotations:} A consistent policy is needed: minor peripheral notes (\acs{eg} page numbers) can be ignored, while prominent handwritten elements should influence classification.

    \item \textbf{Robustness to defects:} The system must tolerate background noise and degradation, which caused the \ac{dla} framework to hallucinate tables and figures (\ac{eg} Figures \labelcref{fig:dd_mistakes_draw,fig:dd_mis_tables,fig:dd_mis_photo}).
\end{itemize}

This expert input clarified the annotation guidelines and established an initial six-label scheme proposed by the data providers (Table~\ref{tab:categories_initial}). This scheme was a pragmatic starting point aligned with downstream tools for table/graphics extraction and text recognition, rather than a final taxonomy.

\begin{table}[h!]
\centering
\begin{tabular}{p{2.8cm}p{11cm}}
\toprule
\textbf{Category} & \textbf{Description} \\ 
\midrule
REST & Mixture of printed, handwritten, and/or typewritten text, potentially with minor graphical elements (contained all ambiguous cases, including drawings considered insignificant at that time), as illustrated in Figure \ref{fig:e_text_2}. \\ 
TEXT\_LINE & Pages primarily consisting of typewritten, printed, or handwritten text organized in a tabular or form-like structure, illustrated in Figures \labelcref{fig:example_line_hw,fig:example_line_p,fig:example_line_t,fig:e_draw_l_3}. \\ 
PHOTO & Pages dominated by photographs or photographic cutouts (maybe maps, paintings, schematics), with few text captions. Illustrated in Figures \labelcref{fig:e_draw_1,fig:e_draw_2,fig:e_photo_l_2}. \\ 
PHOTO\_TEXT & Similar to \texttt{PHOTO}, but the visual content is presented along with a text block(s) of any style (Figures \labelcref{fig:example_photo,fig:example_draw_l}). \\ 
TEXT & Pages containing plain corpora of almost pure printed, or handwritten, or typewritten text, as illustrated in Figures \labelcref{fig:example_text_p,fig:example_text_hw,fig:example_text_t,fig:e_line_t_2}. \\ 
TEXT\_OTHER & Pages containing mixtures of printed, handwritten, and/or typewritten text, potentially with minor graphical elements. Demonstrated in Figure \ref{fig:e_text_1}.\\
\bottomrule
\end{tabular}
\caption{Category definitions initially designed by the data provider, inspired by the previously observed \ac{dla} attempts}
\label{tab:categories_initial}
\end{table}

The six categories were the minimum granularity the providers considered feasible to annotate, since adding more categories would mean a definition of new classes' visual distinctions and a justification their separation from the existing classes. For supervised learning, we then split visually heterogeneous classes and redefined labels to better match what computer vision models can distinguish.

\begin{figure}[h!]
    \centering
    \caption{\acs{rfc} (Section \ref{sec-rfc}) confusion matrices of early annotation schemes (fewer than 3{,}000 samples in total)}
    \label{fig:rfc}
    
    \begin{subfigure}[b]{0.48\textwidth}
        \centering
        \includegraphics[width=\textwidth]{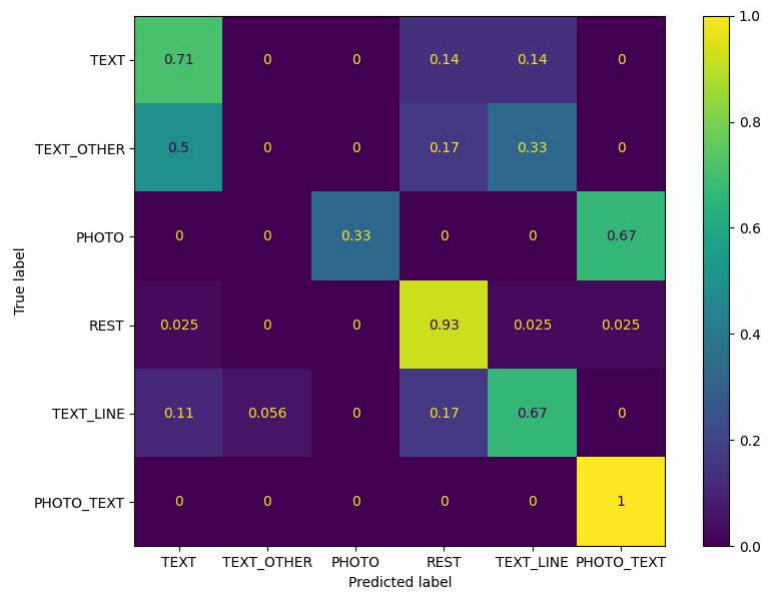}
        \caption{Annotation proposed by our data provider}
        \label{fig:rfc-provider}
    \end{subfigure}
    % \hfill
    \begin{subfigure}[b]{0.48\textwidth}
        \centering
        \includegraphics[width=\textwidth]{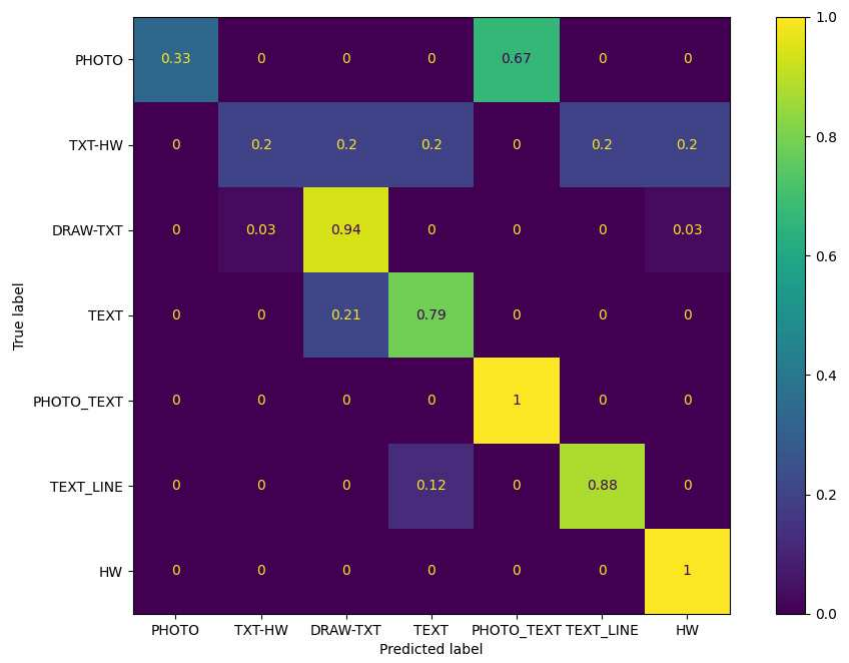}
        \caption{Annotation variant proposed by us}
        \label{fig:rfc-our}
    \end{subfigure}

\end{figure}

The revised definitions (Table~\ref{tab:categories_proposal}) supported an initial training set and a comparison between a seven-label variant and the original six labels. The confusion matrices in Figure~\ref{fig:rfc} suggest that separating handwritten from typewritten text improves performance (Figure~\ref{fig:rfc-our}, \texttt{HW}) and that contour-heavy drawings are distinct from photographs (Figure~\ref{fig:rfc-our}, \texttt{DRAW-TXT}), motivating the refined categories.

\begin{table}[h!]
\centering
\begin{tabular}{p{2.8cm}p{11.1cm}}
\toprule
\textbf{Category} & \textbf{Description} \\ \midrule
DRAW\_TEXT & Pages dominated by drawings, maps, paintings, schematics, or graphics that include text labels or captions (Figures \labelcref{fig:example_draw,fig:example_draw_l}). \\ 
TEXT\_LINE & Pages consisting primarily of typewritten, printed, or handwritten text organized in a tabular or form-like structure (Figures \labelcref{fig:example_line_hw,fig:example_line_p,fig:example_line_t}). \\
PHOTO & Pages dominated by photographs or photographic cutouts (and related graphics), with few text captions (Figures \labelcref{fig:e_photo_3,fig:e_photo_l_2}). \\
PHOTO\_TEXT & Similar to \texttt{PHOTO}, but the visual content is accompanied by a substantial text block (Figures \labelcref{fig:e_photo_1,fig:e_photo_2,fig:e_photo_l_3,fig:e_photo_l_1}). \\
TEXT & Pages containing plain corpora of almost pure printed or typewritten text (Figures \labelcref{fig:example_text_p,fig:example_text_t}). \\ 
HW & Pages consisting purely of handwritten text in paragraph or block form (non-tabular) (Figure \ref{fig:example_text_hw}). \\ 
TEXT\_HW & Pages containing mixtures of handwritten and typewritten text, potentially with minor graphical elements (Figures \labelcref{fig:e_line_t_2,fig:e_text_3}). \\ 
\bottomrule
\end{tabular}%
\caption{Overview of the revised intermediate category definitions derived from the initial provider's proposal}
\label{tab:categories_proposal}
\end{table}

Early low-compute experiments and expert feedback motivated finer label granularity. We refined the scheme to reflect (i) dominant text style, (ii) structured layouts, and (iii) graphical content (photos versus drawings); for example, \texttt{PHOTO\_TEXT} was split into \texttt{PHOTO\_L} and \texttt{DRAW\_L}, and \texttt{TEXT}/\texttt{TEXT\_LINES} were subdivided into handwritten (\texttt{\_HW}), typewritten (\texttt{\_T}), and printed (\texttt{\_P}) variants. Further details are given in Section~\ref{sec-img-classes}. We then finalized the eleven-label training scheme (Table~\ref{tab:categories_used}); Tables~\labelcref{tab:categories_draft,tab:categories_initial,tab:categories_proposal} reflect intermediate steps toward this taxonomy.

\chapter{Dataset formation}
\label{chap-data-form}

\noindent To train (fine-tune) classifiers for page types, we needed to create a representative dataset of page contents discussed in Chapter \ref{chap-data-explore}. The collection required a tailored approach to annotation and dataset splitting into training, development, and performance test subsets. Pages are alphabetically ordered by filename and therefore only approximately chronological, because filenames encode scan dates rather than document creation dates, and sometimes contain only month and internal identifier which come after the main portion of the pages (around 90\%) also sorted in ascending order.

This approximation reflects the archivists' scanning workflow: they record a document's year and month only when the paper comes from a box explicitly labeled with that date. In all other cases (around 20\% of the pages), the filename captures the scan date rather than the document's creation date, as is typical for the unsystematized portion of the archive where materials from different decades are stored together by location or authorship origin.

Because the collection is heterogeneous (Section~\ref{sec-source-data-char}) and pages cluster by scanning campaign or document type, we avoided random shuffling and instead used deterministic periodic sampling with a small randomized offset (select every \(S\)-th page) to preserve category proportions while maintaining coverage across the archive's 100-year span.

This chapter describes the split algorithm (Section~\ref{sec-subset}), the annotation scheme development (Section~\ref{sec-img-classes}), and dataset modifications based on development-set mistakes (Section~\ref{sec-categ-modif}).

The following abbreviations used in category names refer to page content genres:
\begin{itemize}
    \item \texttt{*\_L} --- filled-in form lines; linear outer frame of a table (tabular legend)
    \item \texttt{*\_T} --- typewritten; typed on a machine with a monospaced font
    \item \texttt{*\_P} --- printed; printed using a laser or ink printer
    \item \texttt{*\_HW} --- handwritten; manual writing
\end{itemize}
These subcategories capture visual features that we expect the fine-tuned models to distinguish.

\section{Image classification categories} \label{sec-img-classes}

\noindent The initial category scheme followed the structure of the tested \ac{dla} framework, distinguishing figures, tables, and text (Table~\ref{tab:categories_initial}). This \ac{dla}-based division was expanded with an ambiguous category, and the mixed content category was considered sufficient from the data provider's perspective. The scheme was subsequently adjusted to better align with the capabilities of statistical models. Based on the observation in Figure~\ref{fig:rfc} that distinguishing handwritten from typewritten text was beneficial (Figure~\ref{fig:rfc-our}, category \texttt{HW}), and that drawings with contour lines appeared visually distinct from photographs (Figure~\ref{fig:rfc-our}, category \texttt{DRAW-TXT}), we proposed a refined set of categories (Table~\ref{tab:categories_proposal}).

\begin{table}[h!]
\centering
\begin{tabular}{p{2cm}p{11.8cm}}
\toprule
\textbf{Category} & \textbf{Description} \\ 
\midrule
DRAW & Pages dominated by drawings, maps, paintings, schematics, or graphics, potentially containing text labels or captions, as illustrated in Figure~\ref{fig:example_draw}. \\ 
DRAW\_L & Similar to \texttt{DRAW}, but presented within a table-like layout or including a legend formatted as a table (Figure~\ref{fig:example_draw_l}). \\ 
LINE\_HW & Handwritten text organized in a tabular or form-like structure (Figure~\ref{fig:example_line_hw}). \\ 
LINE\_P & Printed text organized in a tabular or form-like structure (Figure~\ref{fig:example_line_p}). \\ 
LINE\_T & Typewritten text organized in a tabular or form-like structure (Figure~\ref{fig:example_line_t}). \\ 
PHOTO & Pages dominated by photographs or photographic cutouts, potentially with text captions (Figure~\ref{fig:example_photo}). \\ 
PHOTO\_L & Similar to \texttt{PHOTO}, but presented within a table-like layout or accompanied by tabular annotations (Figure~\ref{fig:example_photo_l}). \\ 
TEXT & Mixtures of printed, handwritten, and/or typewritten text, potentially with minor graphical elements (Figure~\ref{fig:example_text}). \\ 
TEXT\_HW & Handwritten text in paragraph or block form (non-tabular), as demonstrated in Figure~\ref{fig:example_text_hw}. \\ 
TEXT\_P & Printed text in paragraph or block form (non-tabular), as demonstrated in Figure~\ref{fig:example_text_p}. \\ 
TEXT\_T & Typewritten text in paragraph or block form (non-tabular), as demonstrated in Figure~\ref{fig:example_text_t}. \\
\bottomrule
\end{tabular}
\caption{Overview of categories used in the fine-tuned models; unless otherwise specified, each category includes pages primarily dominated by the described content type.}
\label{tab:categories_used}
\end{table}

After demonstrating initial results with these categories on low-compute models, the data providers agreed to expand the set of target categories. Subsequent expert feedback led to finer granularity.

This collaborative process ultimately produced a final set of eleven distinct categories designed to capture relevant variations for downstream processing pipelines. These eleven target classes are defined in Table~\ref{tab:categories_used}, which covers almost half of all content type combinations shown in Table~\ref{tab:11-label}.

A priority order was established to handle pages that could fit multiple categories, prioritizing visually distinct or structured content requiring specific processing:
\begin{enumerate}
    \item PHOTOs (\texttt{PHOTO}, \texttt{PHOTO\_L}): highest priority during graphic extraction.
    \item DRAWs (\texttt{DRAW}, \texttt{DRAW\_L}): second priority for graphic extraction.
    \item LINEs (\texttt{LINE\_HW}, \texttt{LINE\_P}, \texttt{LINE\_T}): third priority for structured data extraction (\ac{eg} fields as keys and contents as values).
    \item TEXTs (\texttt{TEXT}, \texttt{TEXT\_HW}, \texttt{TEXT\_P}, \texttt{TEXT\_T}): lowest priority, targeting font-specific \ac{ocr}.
\end{enumerate}

This hierarchy ensures that visually dominant or structured content is prioritized during subsequent pipeline processing.

Importantly, the priority order above was established as common ground for data annotators. Data annotation was carried out by me and Dana from \ac{arup}, a domain expert and end-user representative.

Multiple labels per page were disallowed because categories were defined as mutually exclusive. The data provider decided that a single label would be sufficient for further page aggregation.

\begin{table}[h]
\begin{tabular}{rcccc}
\toprule
 & Printed & Typewritten & HandWritten & Mixed \\ 
 \midrule
Photos & \multicolumn{4}{c}{\textbf{PHOTO}} \\ 
Drawings etc. & \multicolumn{4}{c}{\textbf{DRAW}} \\ 
Photo in table & \multicolumn{4}{c}{\textbf{PHOTO\_L}} \\ 
Drawing in table & \multicolumn{4}{c}{\textbf{DRAW\_L}} \\ 
Tables \& Forms & \multicolumn{1}{c}{\textbf{LINE\_P}} & \multicolumn{1}{c}{\textbf{LINE\_T}} & \multicolumn{2}{c}{\textbf{LINE\_HW}} \\ 
Plain texts & \multicolumn{1}{c}{\textbf{TEXT\_P}} & \multicolumn{1}{c}{\textbf{TEXT\_T}} & \multicolumn{1}{c}{\textbf{TEXT\_HW}} & \textbf{TEXT} \\
\bottomrule
\end{tabular}
\caption{Coverage of data feature variability, summarizing the mapping between content type and writing mode.}
\label{tab:11-label}
\end{table}

To capture the full variability of the data, we created an expanded label scheme of 24 distinct types by separating each core category into printed, typewritten, and handwritten variants (Table~\ref{tab:24-label}). This comprehensive set was used for analysis, while the 11-category set was used for fine-tuning the final models.

\begin{table}[h]
\begin{tabular}{rcccc}
\toprule
 & Printed & Typewritten & Handwritten & Mixed \\ 
\midrule
Photos & \textbf{P\_P} & \textbf{P\_T} & \textbf{P\_HW} & \textbf{PHOTO} \\ 
Drawings & \textbf{D\_P} & \textbf{D\_T} & \textbf{D\_HW} & \textbf{DRAW} \\ 
Photos in table & \textbf{P\_L\_P} & \textbf{P\_L\_T} & \textbf{P\_L\_HW} & \textbf{PHOTO\_L} \\ 
Drawings in table & \textbf{D\_L\_P} & \textbf{D\_L\_T} & \textbf{D\_L\_HW} & \textbf{DRAW\_L} \\ 
Tables \& Forms & \textbf{LINE\_P} & \textbf{LINE\_T} & \textbf{LINE\_HW} & \textbf{LINE} \\ 
Plain texts & \textbf{TEXT\_P} & \textbf{TEXT\_T} & \textbf{TEXT\_HW} & \textbf{TEXT} \\
\bottomrule
\end{tabular}
\caption{Expanded label scheme illustrating coverage of data variability (analytical only).}
\label{tab:24-label}
\end{table}

\section{Representative subset selection}
\label{sec-subset}

\noindent The primary goal of subset selection was to preserve the proportional size of each category across the training, validation, and test sets. We therefore selected samples independently within each category. However, matching category proportions alone does not guarantee representativeness: each category contains substantial internal variability, and many page types are clustered by scanning campaign.

To reduce bias toward specific templates or time periods, we designed a time-aware selection procedure based on alphabetical filename order (which is approximately chronological because filenames may encode scan dates). The key motivations were:

\begin{description}
    \item[Clustering in the source data] Pages are often clustered by scanning campaign or document type. A naive random shuffle can place too many near-duplicate pages from the same source into the development or test sets, inflating or destabilizing evaluation.

    \item[Long-term variability] The collection spans 100 years and includes systematic changes in scan appearance (\acs{eg} black-and-white vs. color scans, yellowed paper, and common defects like damage from floods). These factors are not uniformly distributed over time.

    \item[Evolving document features] Fonts, tabular templates, and annotation practices change over time (from early typewritten pages to modern printed layouts). Preserving chronological coverage helps the model generalize across these shifts.

    \item[Deterministic periodic sampling] Selecting every \(S\)-th element (with a bounded random offset) provides controlled randomness while ensuring that each subset receives samples across the full timeline.
\end{description}

Concretely, the selection procedure must balance several varying factors, including color scheme, font, content type clustering, form templates, graphical objects, defects, and the prevalence of annotations.

These interacting sources of variability---combined with clustering---make a simple random shuffle inadequate. A shuffle can easily create a development or test set dominated by a single scanning campaign, template, or era, skewing the estimated performance. We therefore use the structured splitting procedure below.

\subsection{Split procedure}
\label{sec-split}

\noindent Instead of a simple random shuffle, we employ deterministic periodic sampling with a randomized offset. To keep the training set as large as possible, the development and test subsets are selected first, and the training subset contains the remaining pages. For each category, we proceed as follows:

\begin{enumerate}
  \item Compute the desired subset size \(k\) as a fixed proportion of the category \(N\).
  \item Compute the selection step \(S\) as \(S \approx \frac{N}{k}\).
  \item Select every \(S\)-th element from the alphabetically ordered sequence, but perturb each selected index with a small bounded random offset (\ac{eg} within \(\pm \frac{S}{4}\)) to avoid strict periodicity.
  \item Apply boundary checks to handle out-of-range indices.
\end{enumerate}

This procedure (a) respects the original ordering and local clustering, (b) preserves category proportions, and (c) adds controlled randomness so that selected samples are not perfectly periodic. Crucially, it maintains coverage of the full chronological and structural variability of the collection.

We used random seeds 420--424. Document-level grouping was not preserved. For each fold, the final subset counts were 38{,}625 / 4{,}823 / 4{,}823 for train / development / test. Overall, 43{,}050 images were used across the five training subsets; the remaining 5{,}449 images (Figure~\ref{fig:label_time_test}) formed the final performance test subset used for results visualization.

\begin{figure}[!h]
    \centering
    \caption{Proportions of categories in train, development, and test subsets (columns) across five folds (rows) on a timescale of document creation dates.}
    \includegraphics[width=140mm]{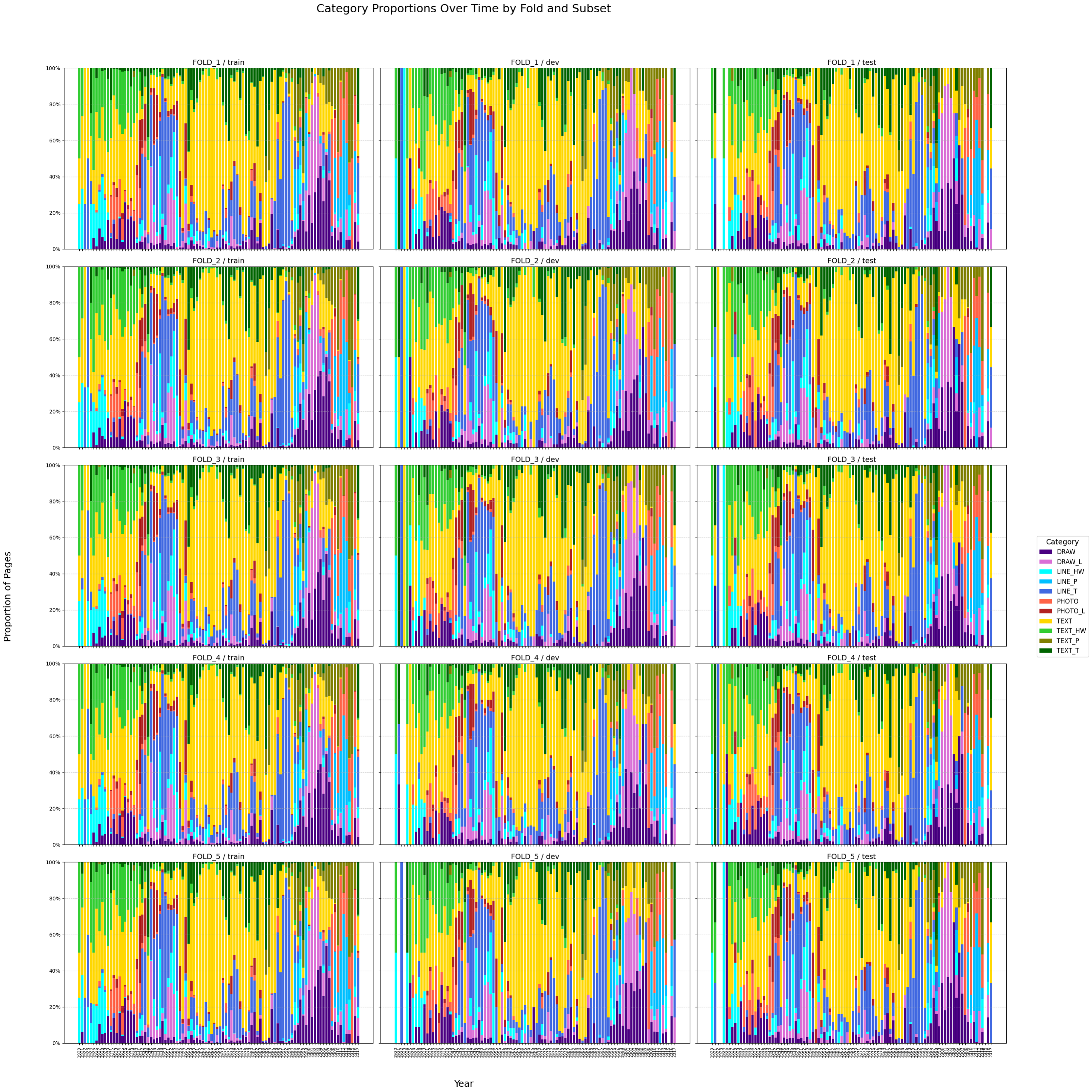}
    \label{fig:folds-categ-distribs}
\end{figure}

To preserve historical and structural variability, we leverage chronological ordering despite its weak correlation with exact document creation dates. Deterministic periodic sampling (with a bounded random offset) maintains local clustering while ensuring that all subsets cover the full chronological spectrum (Figure \ref{fig:folds-categ-distribs}). This reduces bias toward any single time period and enables evaluation against long-term shifts in document appearance.

\section{Datasets and annotation summary}
\label{sec:annotation-distrib}

\noindent This section details the dataset composition and annotation metadata used for training (fine-tuning the pretrained models) and evaluating the image classification models.

\paragraph{Training sets}
The iterative development of the models relied on four progressively larger training sets:
\begin{itemize}
    \item \textbf{8950} images for initial \texttt{vX.0}.
    \item \textbf{10745} images for poor selection \texttt{vX.1}.
    \item \textbf{14565} images for refined \texttt{vX.2}.
    \item \textbf{38625} images for the final (largest) \texttt{vX.3}.
\end{itemize}

\paragraph{Evaluation (test) sets}
Model performance was assessed using distinct evaluation sets corresponding to development stages:
\begin{itemize}
    \item \textbf{1290} images (taken from \texttt{vX.2} annotations and used for all older models).
    \item \textbf{4823} images for every \texttt{vX.3} model.
    \item \textbf{5449} images for the final cross-validation of \texttt{vX.3} models, including the 227 excluded \texttt{TEXT} samples (see Table \ref{tab:category-distribution-datasets}).
\end{itemize}

The manual annotation process was time-intensive, drawing from a diverse collection of archival documents spanning 1920--2020. While the specific textual content and language of the source data are irrelevant given the model's visual-only resolution, the data source itself introduces potential bias. All samples were sourced from \emph{archaeological reports}, which frequently contain freehand sketches and digitized illustrations of common object forms (\ac{eg} ceramic pieces, arrowheads, rocks). This provenance may bias the model toward drawing detection.

\begin{table}[h!]
\centering
\small
\begin{tabular}{p{3cm}p{2.1cm}p{2.2cm}p{2.3cm}p{2.4cm}}
\toprule
\textbf{Category} & \textbf{Dataset 0} & \textbf{Dataset 1} & \textbf{Dataset 2} & \textbf{Dataset 3} \\ 
\midrule
\texttt{DRAW}    & 1090 (9.1\%)  & 1368 (8.8\%)   & 1472 (9.3\%)   & 2709  (5.6\%)  \\
\texttt{DRAW\_L} & 1091 (9.1\%)  & 1383 (8.9\%)   & 1402 (8.8\%)   & 2921  (6.0\%)  \\
\texttt{LINE\_HW}& 1055 (8.8\%)  & 1113 (7.2\%)   & 1115 (7.0\%)   & 2514  (5.2\%)  \\
\texttt{LINE\_P} & 1092 (9.1\%)  & 1540 (9.9\%)   & 1580 (10.0\%)  & 2439  (5.0\%)  \\
\texttt{LINE\_T} & 1098 (9.2\%)  & 1664 (10.7\%)  & 1668 (10.5\%)  & 9883 (20.4\%)  \\
\texttt{PHOTO}   & 1081 (9.1\%)  & 1632 (10.5\%)  & 1730 (10.9\%)  & 2691  (5.5\%)  \\
\texttt{PHOTO\_L}& 1087 (9.1\%)  & 1087 (7.0\%)   & 1088 (6.9\%)   & 2830  (5.8\%)  \\
\texttt{TEXT}    & 1091 (9.1\%)  & 1587 (10.3\%)  & 1592 (10.0\%)  & 14227 (29.3\%) \\
\texttt{TEXT\_HW}& 1091 (9.1\%)  & 1092 (7.1\%)   & 1092 (6.9\%)   & 2008  (4.1\%)  \\
\texttt{TEXT\_P} & 1083 (9.1\%)  & 1540 (9.9\%)   & 1633 (10.3\%)  & 2312  (4.8\%)  \\
\texttt{TEXT\_T} & 1081 (9.1\%)  & 1476 (9.5\%)   & 1482 (9.3\%)   & 3965  (8.2\%)  \\ 
\midrule
\textbf{Unique \acp{pdf}} & 5001    & 5694     & 5729     & 37328    \\
\textbf{Total Pages}  & 11,940  & 15,482   & 15,854   & 48,499   \\ 
\bottomrule
\end{tabular}
\caption{Category distribution across the models' dataset partitions. The set of eleven categories tabulated here refers to the final labels scheme described in Section \ref{sec-img-classes}}
\label{tab:category-distribution-datasets}
\end{table}

Table~\ref{tab:category-distribution-datasets} summarizes category distributions for successive versions of the annotated dataset (columns are ordered chronologically). The final column, \texttt{Dataset 3}, represents the full partition used for the \texttt{vX.3} models. The table reports the total number of available samples per category in each version; the main exception is \texttt{TEXT}, which was capped at 14{,}000 pages in practice, leaving 227 \texttt{TEXT} samples without a subset marker (\ac{eg} train, development, or test).

For the latest models (\texttt{Dataset 3}), we used an \textbf{80\%}/\textbf{10\%}/\textbf{10\%} train/development/test split. Earlier dataset versions (\texttt{Dataset 0}--\texttt{Dataset 2}) represent only a fraction of the final annotated collection (roughly a quarter to a third of the labeled pages) and are therefore best understood as development snapshots. We mainly reference these versions as concrete examples of insufficient per-category sample sizes, used to assess whether the number of manually annotated pages was adequate and whether the selected pages provided sufficient variability within and across categories.

The category imbalance in \texttt{Dataset 3} is not a design choice; it reflects the natural composition of the archive. The increase in short documents in the final dataset is largely due to a deliberate effort to process and manually correct all single-page \acp{pdf}.

\begin{figure}[!h]
    \centering
    \caption{Distribution of categories in the final annotated dataset (\texttt{Dataset 3}) based on the document creation year}
    \label{fig:label_time_train}
        \includegraphics[width=\textwidth]{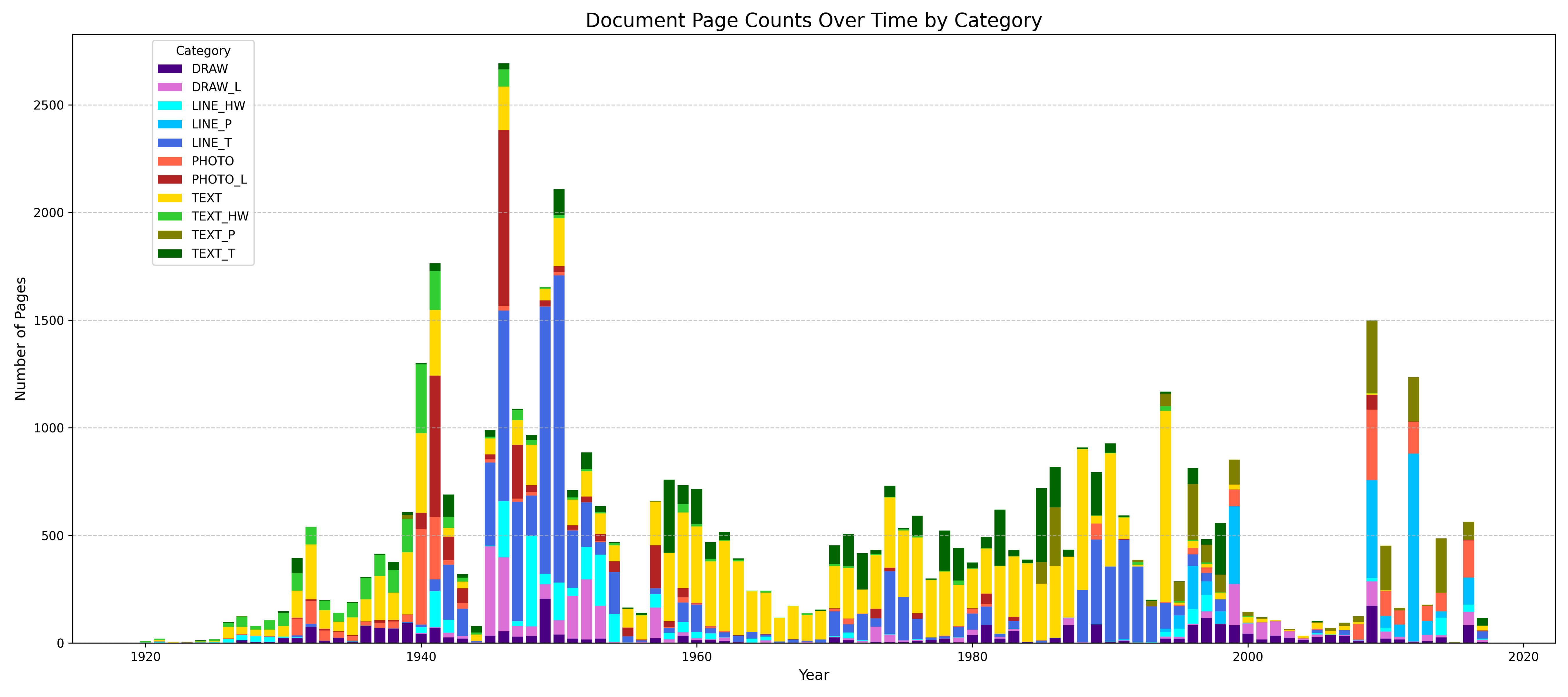}
\end{figure}

\begin{figure}[!h]
    \centering
    \caption{Distribution of categories in the performance test subset (samples not included in training subsets of the five cross-validation folds), based on the document creation year.}
    \label{fig:label_time_test}
        \includegraphics[width=\textwidth]{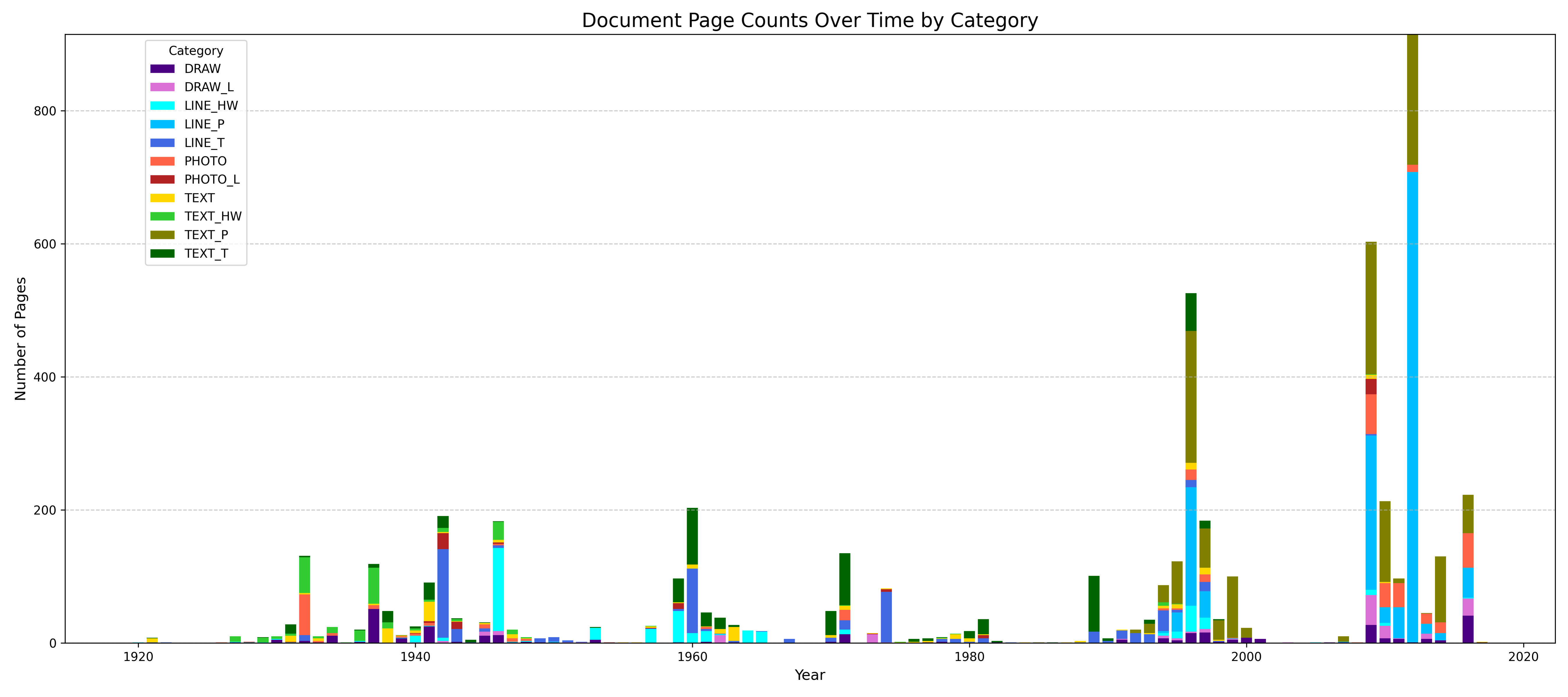}
\end{figure}

\begin{figure}[!h]
    \centering
    \caption{Temporal distributions of categories across annotation versions (Datasets 0, 2, and 3), based on the document creation year}
    \label{fig:label_timeline_triple}
        \includegraphics[width=140mm]{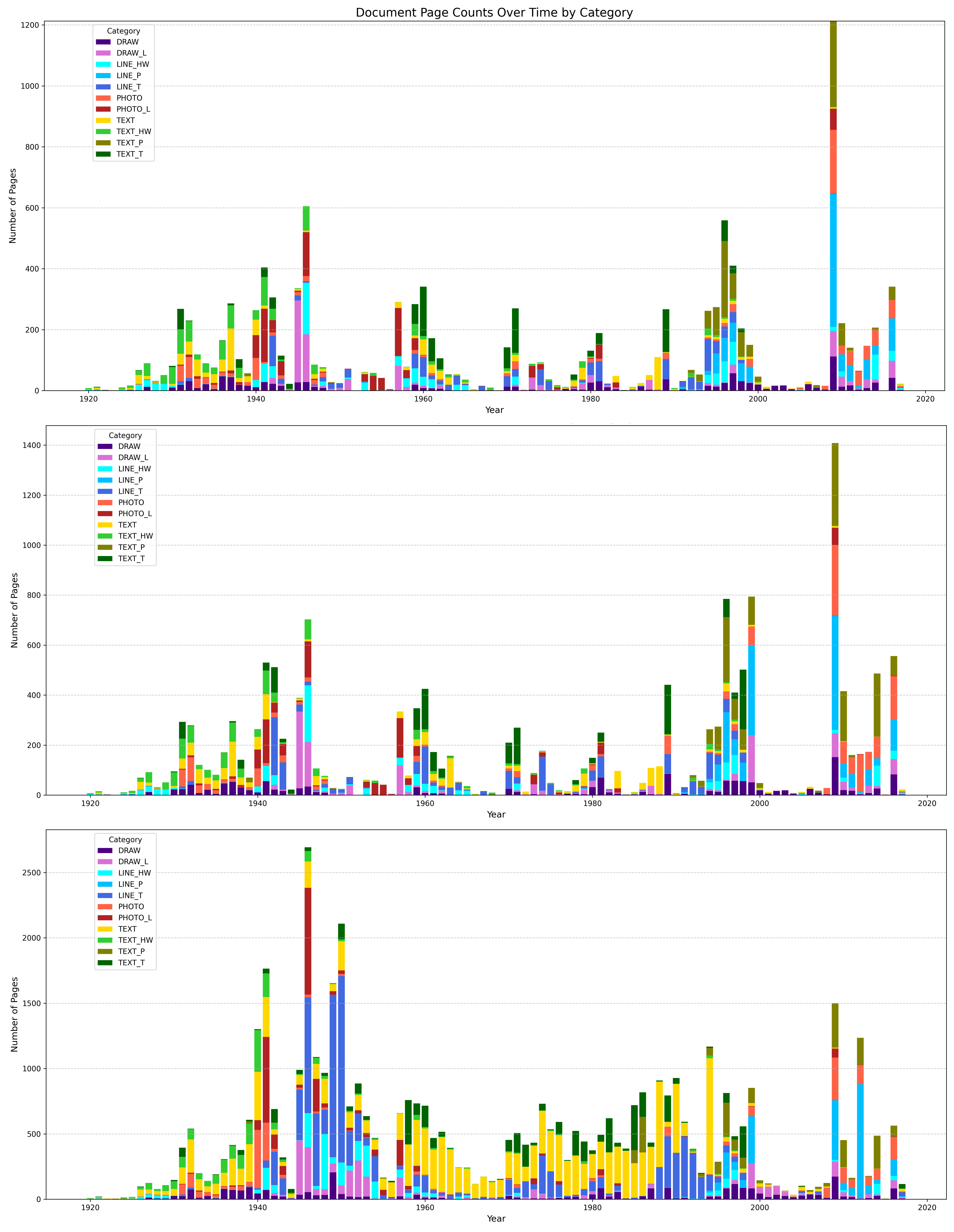}
\end{figure}

Figures \labelcref{fig:label_time_train,fig:label_time_test,fig:label_timeline_triple} visualize how the annotations vary over time. In these plots, labeled pages are sorted by document creation year and colored by category.

Figure~\ref{fig:label_time_train} illustrates the temporal distribution of the final annotated data, while Figure~\ref{fig:label_time_test} shows the same for the final performance test set (a subset of the full annotations). This test subset includes all annotated samples not used in the training subset of any cross-validation fold.

Figure~\ref{fig:label_timeline_triple} compares the temporal category distributions across annotation stages: the initial (\texttt{Dataset 0}), refined (\texttt{Dataset 2}), and final (\texttt{Dataset 3}) versions. The intermediate \textit{poor selection} version (\texttt{Dataset 1}) is omitted.

Samples were selected manually, largely influenced by the page count of the source \acp{pdf}. As a result, many samples were drawn from the same multi-page documents, and pages from a single document could be assigned to either the same or different categories.

All three temporal plots reveal a gap in the 1990s. Documents from this period were difficult to classify because printed text often used monospaced fonts that resemble typewritten text. We intentionally excluded these ambiguous pages from the so-called ground-truth collection, forcing the models to learn and infer the distinction.

Finally, the prominent spikes in early-year samples (Figure \ref{fig:label_time_train} and the bottom graph of Figure \ref{fig:label_timeline_triple}) are a direct result of the single-page document enrichment step. This initiative substantially increased the proportion of short documents in the final annotated collection compared to earlier versions (\texttt{Dataset 0}--\texttt{Dataset 2}).

\section{Data modifications in categories} \label{sec-categ-modif}

\noindent Achieving high accuracy was not only a matter of model selection but also of careful dataset curation. The training data evolved through successive stages of annotation and refinement. We measured the impact of these changes by evaluating a baseline \ac{vit} model (ViT-B/16-224) fine-tuned for each dataset version using the evaluation subset of the refined (\texttt{Dataset 2} in Table~\ref{tab:category-distribution-datasets}) annotation version.

Initially, our \textit{initial} annotation set had limited document variety. We expanded samples within each category, corrected misclassifications, and reclassified ambiguous items such as pages containing stamps. One iteration, referred to as \textit{poor selection}, involved removing samples that seemed too noisy or redundant from the data provider's perspective; however, evaluation revealed that removing these semi-repetitive but distinct examples was detrimental to performance.

Consequently, the \textit{refined} annotation phase involved restoring many removed pages to ensure sufficient sample diversity, particularly for the \texttt{PHOTO}, \texttt{TEXT\_P}, and \texttt{TEXT\_T} categories. This iterative cycle of training, evaluation, expert review, and dataset refinement was critical to achieving the final high-performing models.

\begin{figure}[!h]
    \centering
    \caption{Confusion matrices of Top-1 predictions for \ac{vit}-Base at $224\times224$ resolution, fine-tuned on successive dataset versions and evaluated on the refined test set (before the final test set was created). Accuracy scores are not included. Dataset numbers refer to the versions in Table~\ref{tab:category-distribution-datasets}.}
    \label{fig:vits}
    \begin{subfigure}[b]{0.32\textwidth}
        \centering
        \includegraphics[width=\textwidth]{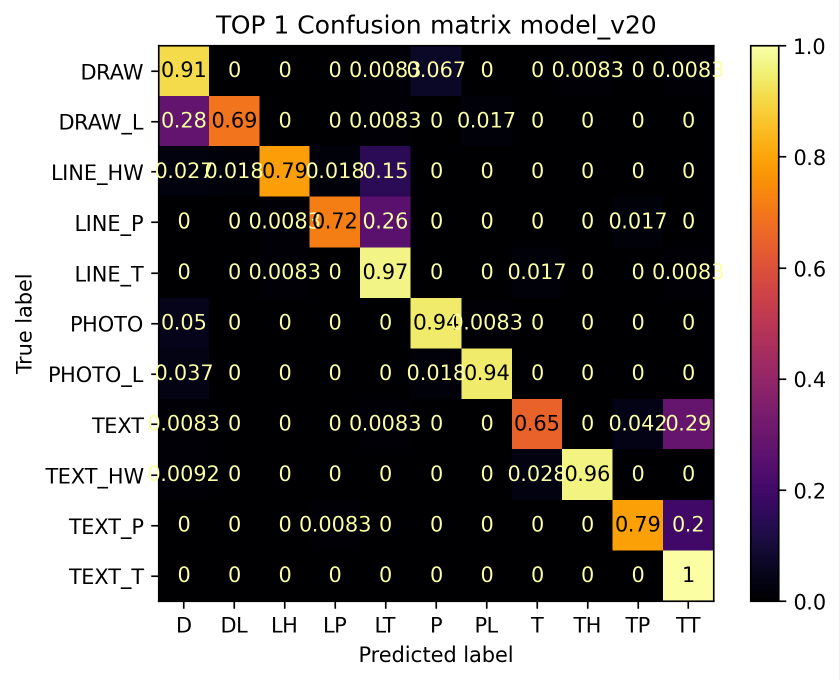}
        \caption{Initial Dataset 0}
        \label{fig:init}
    \end{subfigure}
    \begin{subfigure}[b]{0.32\textwidth}
        \centering
        \includegraphics[width=\textwidth]{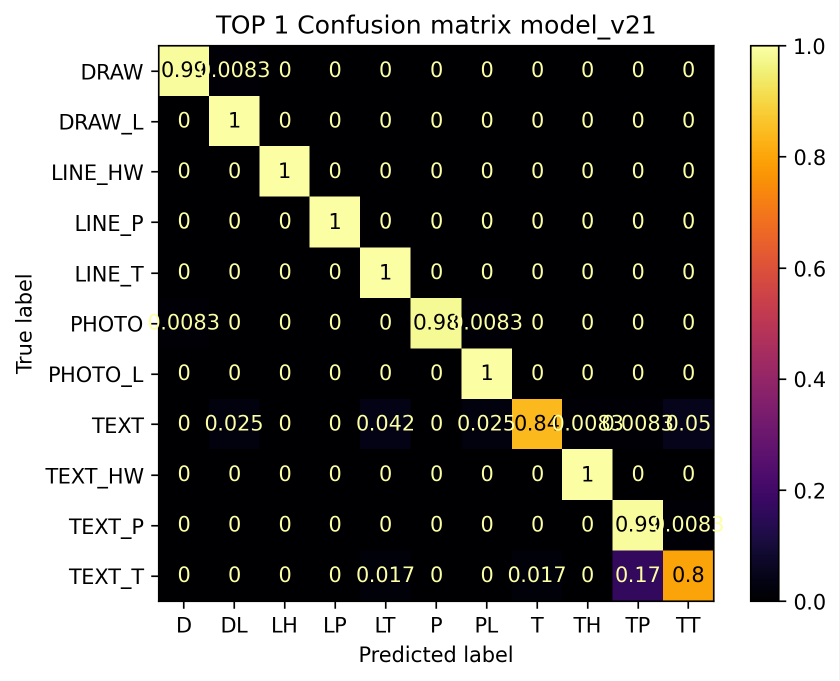}
        \caption{Poor selection Dataset 1}
        \label{fig:poor}
    \end{subfigure}
    \begin{subfigure}[b]{0.34\textwidth}
        \centering
        \includegraphics[width=0.95\textwidth]{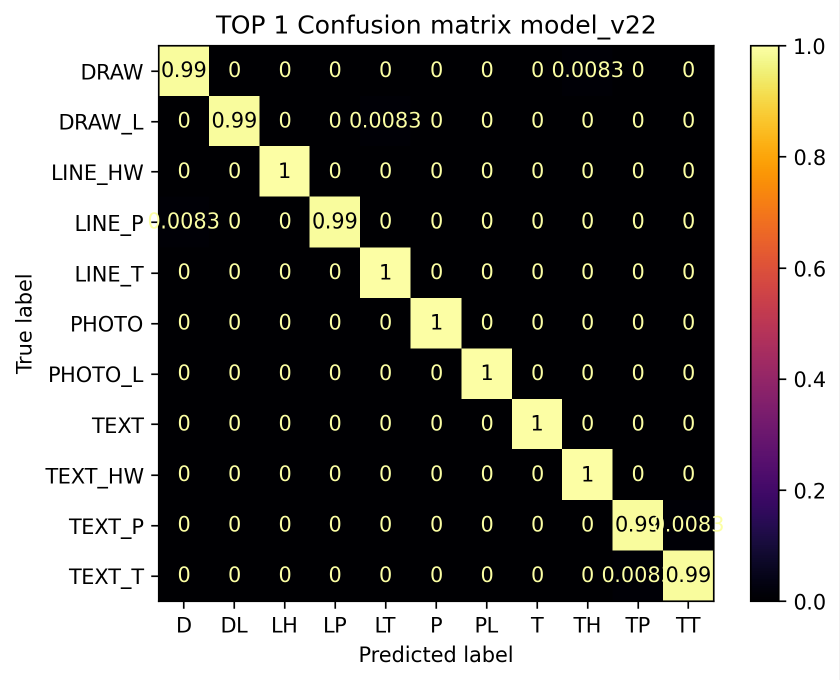}
        \caption{Refined (not Final) Dataset 2}
        \label{fig:final}
    \end{subfigure}
\end{figure}

Figure \ref{fig:vits} shows the Top-1 prediction confusion matrices for the \ac{vit}-Base (224px) model fine-tuned on the \textit{initial}, so-called \textit{poor selection} (\texttt{Dataset 1}), and \textit{refined} (but not \textit{final}) versions of the dataset, illustrating the category-specific performance improvements gained through careful data curation.

Key takeaways from developing the annotated dataset of digitized pages:

\begin{itemize}
    \item Transformer models such as \acs{vit}, \acs{dit}, and \acs{clip} tokenize images into $16\times16$ or $32\times32$ patches. Because we did not use rotation or flip augmentations, it was important to include diverse page layouts so that the model learned to attend to the whole page (not only the most likely location of a photograph or the borders of a form).
    \item Distinguishing typewritten from printed text remains challenging at common input resolutions ($224\times224$ to $384\times384$). As a result, pairs such as \texttt{TEXT\_T}--\texttt{TEXT\_P} and \texttt{LINE\_T}--\texttt{LINE\_P} are likely separated using subtle cues (\acs{eg} whitespace between lines and page margins).
    \item Freehand sketches are highly variable; it is unlikely that a finite dataset captures all relevant drawing styles.
    \item Mixed-content pages (\texttt{TEXT}) often contain stamps or small graphics (\ac{eg} newspaper logos), which can resemble \texttt{LINE\_T} or \texttt{DRAW}.
\end{itemize}

A direct mitigation is to expand the dataset with more examples per category. Given time constraints and the number of pages already labeled by hand, we stopped expansion before reaching 50{,}000 records. The smallest category size reached slightly above 2{,}000 pages, while the largest exceeded 14{,}000 pages.

The category-size imbalance reflects the source data and the addition of single-page documents during refined annotation. To reduce the impact of imbalance during training, we used a batch sampler that fixes the batch size to the number of categories (i.e., one sample per category per batch; Section~\ref{sec-finetune}). During each epoch, the sampler revisits already-seen samples in a randomized order, similar to the \ac{clip} few-shot fine-tuning example.

\chapter{Image classification} \label{chap-img-class}

\noindent As the previous Chapter \ref{chap-data-form} describes procedures implemented to form a representative training dataset, we were ready to automatically evaluate models on the selected test sets. 

Given the limitations of unsupervised \acf{dla} methods on heterogeneous historical document data (Section~\ref{sec-dla-critical}), we shifted to supervised image classification. This chapter presents (i) a low-compute baseline, (ii) a comparison of state-of-the-art deep learning architectures, and (iii) the iterative data refinement process that enabled high accuracy.

The latter Chapter \ref{chap-results} is closely connected to this part of the thesis, since it contains visualized and tabulated results of the models' evaluation.

\section{Low-compute approach} \label{sec-low-comp}

\noindent As a computationally efficient baseline, we first implemented a traditional computer vision pipeline using handcrafted features and a \acf{rfc}. This initial system was designed for resource efficiency, requiring no specialized \acp{gpu} and making it suitable for standard \ac{cpu}-based hardware.

\begin{figure}[h]
\centering
\includegraphics[width=140mm]{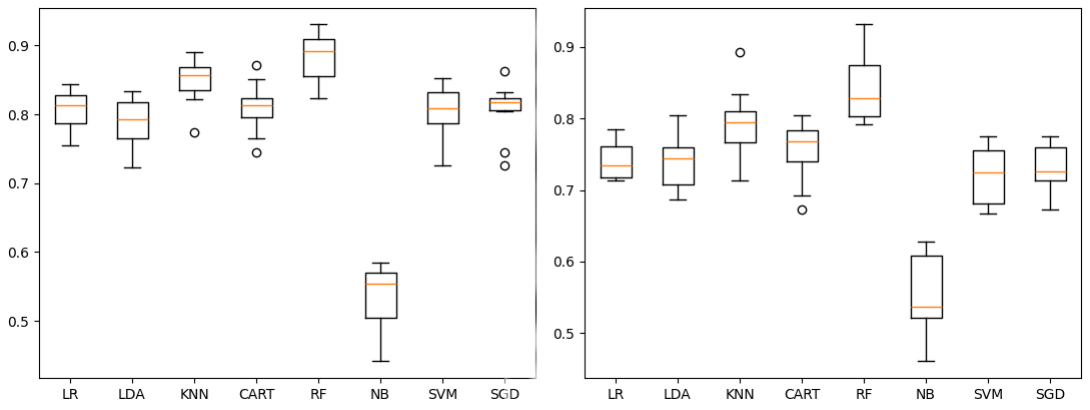}
    \caption{Low-compute models compared on the same data in cross-fold validation. Left: data-provider annotation. Right: our proposed annotation scheme.}
\label{fig:low-compared}
\end{figure}

\subsection{Image feature extraction}

\noindent The feature extraction step transforms each raw page image into a numerical feature vector. We engineered this vector by combining descriptors computed from both the grayscale image and a binarized version produced by Otsu thresholding:

\begin{itemize}
    \item \textbf{Preprocessing:} Grayscale conversion, Otsu thresholding \cite{yousefi2011image}, and basic image properties (dimensions, pixel ratios).
    \item \textbf{Hu moments \cite{hu1962visual}:} Seven invariant moments capturing shape information robust to translation, scale, and rotation.
    \item \textbf{Haralick texture features \cite{haralick1973textural}:} Statistics derived from gray-level co-occurrence matrices (GLCM), including contrast, correlation, and homogeneity.
    \item \textbf{Histogram features:} Pixel-intensity distributions (256 bins for grayscale; two bins for binary).
\end{itemize}

The resulting representation is low-dimensional (298 floats), enabling efficient processing on modest hardware. However, it is less expressive than deep features, especially for visually subtle distinctions beyond coarse categories such as photographs or handwriting.

\subsection{\ac{rfc}}
\label{sec-rfc}

\noindent A \acf{rfc} \cite{breiman2001random} is an ensemble learning method that aggregates the predictions of multiple decision trees. Each tree is trained on a random subset of the training data and features, and the final prediction is obtained by combining the trees' outputs.

We chose \ac{rfc} due to its effectiveness, interpretability, and efficiency on modest hardware. While we also evaluated other low-compute models (\ac{lda}, \ac{knn}, Naive Bayes, \ac{svm}, logistic regression), \ac{rfc} achieved the best preliminary accuracy (Figure~\ref{fig:low-compared}).

Our implementation included class weighting to mitigate imbalance, optional hierarchical classification, and Top-$N$ confidence outputs.

To evaluate the impact of the evolving label scheme, we trained \ac{rfc} on two early annotation variants: the data-provider proposal (Table~\ref{tab:categories_initial}) and our refined proposal (Table~\ref{tab:categories_proposal}). Development-set results are shown in Figure~\ref{fig:rfc}.

Overall accuracy was about 75\%, which is insufficient for reliable automated processing in production. Nevertheless, the baseline performed well on some categories, confirming that image-based classification is viable and motivating the transition to deep learning models trained on larger datasets using the final annotation scheme (Table~\ref{tab:categories_used}).

\section{Typical Models for Image Classification Fine-Tuning} \label{sec-typical-img-class}

\noindent Given the performance limitations of the low-compute approach, we turned to deep neural network architectures, which are highly effective for complex image recognition tasks. We evaluated several architectures by fine-tuning them on our dataset, initializing each with ImageNet-pretrained weights. We experimented with \acp{cnn} (EfficientNetV2 \cite{tan2021efficientnetv2,tan2019efficientnet}, RegNetY \cite{radosavovic2020designing}) and Transformer-based models (\ac{dit} \cite{li2022dit}, \ac{vit} \cite{dosovitskiy2020image}, \ac{clip} \cite{radford2021learning}).

All models used a consistent set of data augmentations (Section~\ref{sec-preproc}). Hyperparameters such as the learning rate and scheduler were kept at default values (chosen by the training code) across models. All image-based models were fine-tuned for three epochs, while hybrid (\ac{clip}) models were fine-tuned for seven epochs.

For image-based models, the cross-validation training procedure involved five folds of randomized data selection (Subsection~\ref{sec-split}). For models with the same base architecture, the resulting weights were averaged, as described in Section~\ref{sec-img-classes}. Hybrid models were fine-tuned using the split generated from the first cross-validation seed. Finally, we composed a performance test subset consisting of images not included in any fold's training subset and used it to evaluate each final model.

\subsection{EfficientNetV2 and RegNetY approaches} \label{sec-effnet} \label{sec-regnet}

\noindent EfficientNetV2 is a convolutional neural network family optimized for faster training and improved parameter efficiency \cite{tan2021efficientnetv2} and is related to the original EfficientNet compound-scaling formulation \cite{tan2019efficientnet}, pretrained on ImageNet-21k \cite{ridnik2021imagenet}. RegNetY is a family of ResNet-like architectures that defines a parameterized design space optimizing depth and width \cite{radosavovic2020designing}.

Model specifications are provided in Table~\ref{tab:efficientnetv2_regnety}, and evaluation results are summarized in Table~\ref{tab:model_ranking}.

EfficientNetV2 and RegNetY are \ac{cnn} architectures that achieve strong performance on image classification benchmarks. We fine-tuned several variants of each family, ranging in size and input resolution. For all \ac{cnn} models, we replaced the final classification layer to match the eleven classes and trained using cross-entropy loss. These models achieved high accuracies: the large variants exceeded 98\%, and the best (RegNetY-16GF) surpassed 99\%.

At that point, we discovered a single \acs{cnn} model that reaches 99\% accuracy, unlike other \acp{cnn}, which led us to further exploration of the available base model architectures. Firstly, to support predictions of RegNetY-16GF, and secondly, to compare equally accurate models of different resolutions and sizes in practice.

\begin{table}[!h]
\centering
\begin{tabular}{@{}lcccccc@{}}
\toprule
\textbf{Model} & \textbf{Resolution} & \textbf{Pretraining dataset} & \textbf{Params (M)} \\ 
\midrule
EfficientNet-v2-S & $300\times300$ & ImageNet-21k & 48.2 \\
EfficientNet-v2-M & $384\times384$ & ImageNet-21k \& ImageNet-1k & 54.1 \\
EfficientNet-v2-L & $384\times384$ & ImageNet-21k \& ImageNet-1k & 118.5 \\ \midrule
RegNetY-12GF & $224\times224$ & ImageNet-12k \& ImageNet-1k & 51.8 \\
RegNetY-16GF & $224\times224$ & ImageNet-12k & 83.6 \\
RegNetY-64GF & $384\times384$ & SEER \& ImageNet-1k & 281.4 \\ 
\bottomrule
\end{tabular}
\caption{Specifications of EfficientNetV2 and RegNetY Models}
\label{tab:efficientnetv2_regnety}
\end{table}

\subsection{\ac{dit} and \ac{vit} approaches} \label{sec-dit} \label{sec-vit}

\noindent \acf{dit} builds upon BEiT-style masked visual token prediction but was specifically pretrained on large-scale document images. The \ac{dit}-base and \ac{dit}-large models \cite{li2022dit} were self-supervised on 42M pages from IIT-CDIP (Lewis et al.~\cite{lewis2006building}) and optionally fine-tuned on RVL-CDIP \cite{harley2015icdar} (400K grayscale images across 16 document types). Input images were resized to $224\times224$, patchified into $16\times16$ tokens, and augmented via the standard image transform.

Similarly, \ac{vit} applies Transformers directly to image patches \cite{dosovitskiy2020image} and was pretrained on ImageNet-21k \cite{ridnik2021imagenet} and ImageNet-1k \cite{beyer2022better}.

\begin{table}[h]
\centering
\begin{tabular}{@{}lcccr@{}}
\toprule
\textbf{Model} & \textbf{Resolution} & \textbf{Pretraining dataset} & \textbf{Params (M)} \\ 
\midrule
dit‑base-rvlcdip & $224\times224$ & IIT‑CDIP (42M) \& RVL‑CDIP & 86 \\
dit‑large & $224\times224$ & IIT‑CDIP (42M) & 304 \\
dit‑large-rvlcdip & $224\times224$ & IIT‑CDIP (42M) \& RVL‑CDIP & 304 \\
\midrule
vit‑base‑patch16 & $224\times224$ & ImageNet‑21k \& ImageNet‑1k & 86.6 \\
vit‑base‑patch16 & $384\times384$ & ImageNet‑21k \& ImageNet‑1k & 86.9 \\
vit‑large‑patch16 & $384\times384$ & ImageNet‑21k \& ImageNet‑1k & 304.7 \\ 
\bottomrule
\end{tabular}
\caption{Specifications of \ac{dit} and \ac{vit} models}
\label{tab:dit_vit_models}
\end{table}
Model specifications are provided in Table~\ref{tab:dit_vit_models}, and evaluation results are shown in Figures~\ref{fig:vit} and~\ref{fig:dit}.

\ac{dit} and \ac{vit} apply the Transformer mechanism directly to image patches, in contrast to traditional \acp{cnn}. Because \ac{dit} was pretrained on large-scale document images, it is a strong candidate for our task. We fine-tuned several variants of both \ac{dit} and \ac{vit}.

All variants consistently achieved accuracies above 98\%, performing on par with \acp{cnn}, but typically at the cost of 3--4 times more parameters for transformers.

\ac{vit} yielded one of the highest accuracies among image-based models. We therefore used a \ac{vit}-based model as the default classifier for annotation refinements (Section~\ref{sec-categ-modif}). However, the best-performing Transformer model was much larger than the best \ac{cnn} model; thus, we evaluated a hybrid approach next.

\subsection{Comparative Analysis and Error Patterns} \label{sec-model-comparison}

\noindent While all deep learning architectures achieved high performance, their trade-offs differ in accuracy, parameter efficiency, and error patterns. Table \ref{tab:model_ranking} summarizes the top-performing models.

\begin{table}[h]
\centering
\caption{Consolidated model performance ranking (Top-1 accuracy)}
\label{tab:model_ranking}
\begin{tabular}{@{}llcccr@{}}
\toprule
\textbf{Rank} & \textbf{Model} & \textbf{Type} & \textbf{Resolution} & \textbf{Params (M)} & \textbf{Accuracy} \\
\midrule
\textbf{1} & \textbf{RegNetY-16GF} & \textbf{CNN} & \textbf{$224\times224$} & \textbf{83.6} & \textbf{99.21\%} \\
2 & ViT-Large-384 & Transformer & $384\times384$ & 304.7 & 99.12\% \\
3 & ViT-Base-384 & Transformer & $384\times384$ & 86.9 & 98.92\% \\
4 & EfficientNet-V2-M & CNN & $384\times384$ & 54.1 & 98.90\% \\
5 & RegNetY-64GF & CNN & $384\times384$ & 281.4 & 98.79\% \\
6 & EfficientNet-V2-L & CNN & $384\times384$ & 118.5 & 98.77\% \\
7 & DiT-Base-RVL & Transformer & $224\times224$ & 86.0 & 98.72\% \\
\bottomrule
\end{tabular}
\end{table}

\noindent RegNetY-16GF is the most efficient performer: despite having far fewer parameters (83.6M) than large Transformer models (304M), it achieves the highest overall accuracy. To understand this gap, we inspected the main confusion patterns (see Appendix \labelcref{confusion-matrices}).

\begin{description}
    \item[The \texttt{TEXT} ambiguity] The largest error source across models is distinguishing generic \texttt{TEXT} from typeset/transcribed text (\texttt{TEXT\_T}).
        \begin{itemize}
            \item \textbf{CNN scaling:} EfficientNet-V2-S struggles most, misclassifying 13\% of \texttt{TEXT}. Increasing resolution (EfficientNet-V2-L) reduces this error to 5.6\%.
            \item \textbf{Architecture efficiency:} RegNetY-16GF handles this ambiguity best, reaching 94\% accuracy on \texttt{TEXT} with 5.1\% confusion.
            \item \textbf{Transformer limitations:} Even domain-specific Transformers such as DiT-Large achieve slightly lower accuracy on this distinction (91\%) than the best CNNs.
        \end{itemize}

    \item[Structural detection (drawings)] A secondary pattern appears when separating \texttt{DRAW} (freehand) from \texttt{LINE} (technical/line drawings). CNNs tend to be sharper on this edge-driven distinction: RegNetY-16GF reaches 99\% accuracy on \texttt{DRAW}. Transformers show slightly fuzzier boundaries; for example, DiT-Base-RVL reaches 95\% accuracy and confuses a small fraction with other line-based categories.
\end{description}

Based on these results, RegNetY-16GF is the preferred deployment model: it combines the highest accuracy (99.21\%) with lower computational cost than the ViT-Large alternatives, and it performs best on the dataset's hardest distinction (\texttt{TEXT} vs. \texttt{TEXT\_T}).

\subsection{\acs{clip}-based approach} \label{sec-clip}

\noindent Since transformer image-only models showed near-perfect accuracy, we decided to also investigate the integration of text features into the classification. Would it help the models to understand the data or would it only confuse them more?

We adapted a multimodal approach using \ac{clip} \cite{radford2021learning}, pretrained on WebImageText \cite{xu2023demystifying}. \ac{clip} supports zero-shot classification by comparing image features to the features of textual category descriptions. In our domain, however, zero-shot performance was limited (below $\approx50\%$ accuracy; Figure \ref{fig:clip-zero}), highlighting the difficulty of applying general models to specialized archival data without adaptation.

\begin{table}[h]
\centering
\begin{tabular}{@{}lcccccc@{}}
\toprule
\textbf{Model} & \textbf{Resolution} & \textbf{Pretraining dataset} & \textbf{Params (M)} \\ 
\midrule
CLIP-ViT-B/32 & $224\times224$ & WebImageText (400M) & 151 \\
CLIP-ViT-B/16 & $224\times224$ & WebImageText (400M) & 150 \\
CLIP-ViT-L/14 & $224\times224$ & WebImageText (400M) & 428 \\
CLIP-ViT-L/14@336px & $336\times336$ & WebImageText (400M) & 428 \\ 
\bottomrule
\end{tabular}
\caption{Specifications of OpenAI's \ac{clip}-\ac{vit} models}
\label{tab:clip_vit}
\end{table}

After fine-tuning for seven epochs on our annotated dataset, \ac{clip} achieved results comparable to the image-only \ac{cnn} and Transformer baselines. Accuracy was typically above 98.0\%, but only ViT-B/16 among the \ac{clip}-based models exceeded 99.0\%.

The smaller of the two \ac{vit}-B variants reached 99.0\% accuracy (Figure \ref{fig:clip-best5}), while the larger \ac{vit}-B and both \ac{vit}-L models scored in the 98--99\% range (Figure \ref{fig:clip-base4}). Overall, these results show that targeted few-shot-style fine-tuning can make \ac{clip} competitive for archival page classification.

\subsubsection{Description of Categories} \label{sec-clip-categ}

\begin{table}[h!]
\centering
\small
\caption{Summary of \ac{clip} category description sets. ``Rev.'' denotes the revision (model version) fine-tuned to a specific label set of text features.}
\label{tab:summary_classification}
\begin{tabular}{p{0.6cm}p{1.7cm}p{11.1cm}}
\toprule
\textbf{Rev.} & \textbf{Label set} & \textbf{Characteristics of category descriptions} \\
\midrule
\texttt{vX.1} & init is in Table \ref{tab:init_classification} &
Provides the full, initial set of classification categories with detailed distinctions between drawings, photos, and text, and further separates content by handwritten, printed, and typewritten forms both inside and outside tables or forms. \\[1ex]

\texttt{vX.2} & detailed in Table \ref{tab:enhanced_classification} &
Delivers the most detailed annotation‑driven version, augmenting each label with notes on layout (legends, tables), annotation styles, and use‑case examples. \\[1ex]

\texttt{vX.3} & extra in Table \ref{tab:comprehensive_classification} &
Presents a balanced yet thorough taxonomy enriched with illustrative examples, combining clarity and completeness with real‑world document‑type scenarios. \\[1ex]

\texttt{vX.4} & gemini in Table \ref{tab:long_classification} &
Expands upon the page‑based description with verbose, researcher‑oriented vocabulary, adding examples and elaborations per category. \\[1ex]

\texttt{vX.5} & gpt's from Table \ref{tab:page_classification} &
Reframes the classification around page‑level context, explicitly noting that each label applies to an entire page containing the specified content, thereby shifting focus from isolated elements to page composition. \\[1ex]

\texttt{vX.6} & large from Table \ref{tab:extended_classification} &
Like \textit{gpt}'s, but page mentions are moved to the end, and it also provides an extension that includes specific composite labels for graphics combined with tables or legends. \\[1ex]

\texttt{vX.7} & mid from Table \ref{tab:short_classification} &
Offers a more concise restatement of the \textit{init} taxonomy, trimming phrasing while preserving the same category distinctions, emphasizing brevity  - shortness \& minimalism - in the labels’ descriptions. \\[1ex]

\texttt{vX.8} & min from Table \ref{tab:minimal_classification} &
Distills the taxonomy to its bare essentials, reducing each description to the minimal wording needed to convey whether an element is a drawing, table, photo, or text and its modality, completely eliminating any redundancies in phrasings. \\[1ex]

\texttt{vX.9} & short from Table \ref{tab:crop_classification} &
Adapts the minimal set by omitting the word “page” and refining descriptions to emphasize cropped or cell‑level occurrences of drawings, tables, photos, or text without referencing full pages. \\
\bottomrule
\end{tabular}
\end{table}

\begin{figure}[!h]
\centering
\includegraphics[width=140mm]{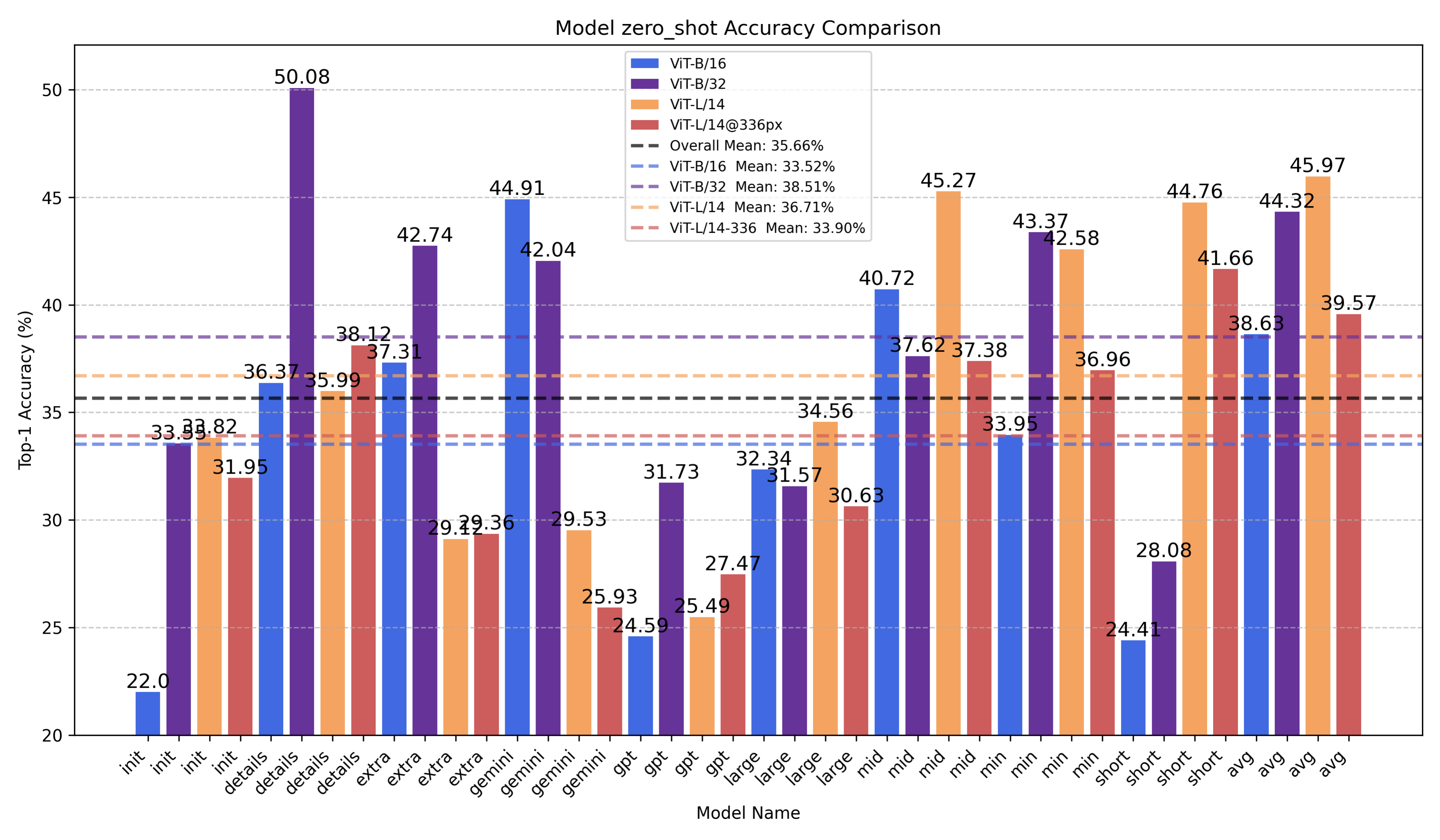}
\caption{Zero-shot \ac{clip} models comparison of classification accuracy per category descriptions set plus averaged text features variants of all four base models}
\label{fig1:clip-acc-categ-zero}
\end{figure}

\noindent We evaluated \ac{clip} using multiple sets of category descriptions, where each set defines the text features used at inference time. The goal was to assess how sensitive \ac{clip} is to linguistic variations in the category prompts. Table \ref{tab:summary_classification} summarizes the tested description sets (details in Tables \labelcref{tab:init_classification,tab:enhanced_classification}).

\begin{figure}[!h]
\centering
\includegraphics[width=140mm]{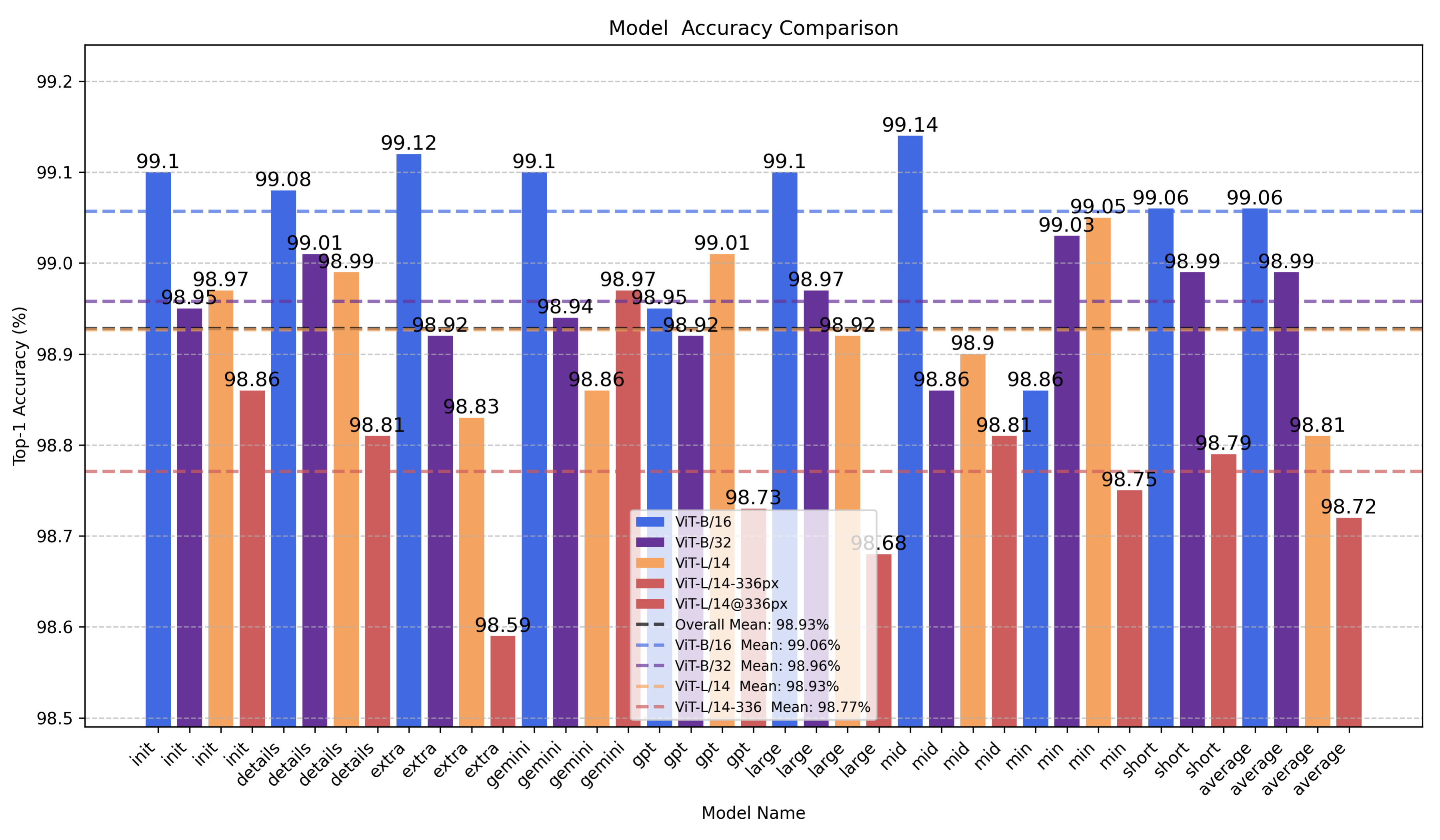}
\caption{Comparison of fine-tuned \ac{clip} models by classification accuracy across category description sets (label set names in alphabetical order). ``Average'' denotes averaged text features.}
\label{fig1:clip-acc-categ}
\end{figure}

Most description sets were written manually and ranged from brief to highly detailed. To broaden the linguistic variation, we also tested two sets proposed by GPT-4 Deep Research and Gemini 2.5 Deep Research (Tables \labelcref{tab:page_classification,tab:long_classification}). These prompts were based on the manual descriptions (Tables \labelcref{tab:init_classification,tab:minimal_classification,tab:short_classification}) and included a request to follow best practices for \ac{clip} prompting (Appendix \ref{app-llm-prompt}).

Early experiments showed that isolated prompt variants can yield inconsistent error rates. We therefore also evaluated an averaging strategy (``Average'' in Figures \labelcref{fig1:clip-acc-categ-zero,fig1:clip-acc-categ} and Figure \ref{fig1:clip-acc-matrix}), where the text feature for each category is computed as the mean of features from multiple description sets. This was intended to reduce prompt sensitivity and stabilize predictions.

\subsubsection{Zero-shot model inference} \label{sec-zero-clip}

\begin{figure}[!h]
    \centering
    \caption{\ac{clip} zero-shot confusion matrices (averaged text features)}
    \label{fig:clip-zero}
    
    \begin{subfigure}[b]{0.48\textwidth}
        \centering
        \includegraphics[width=\textwidth]{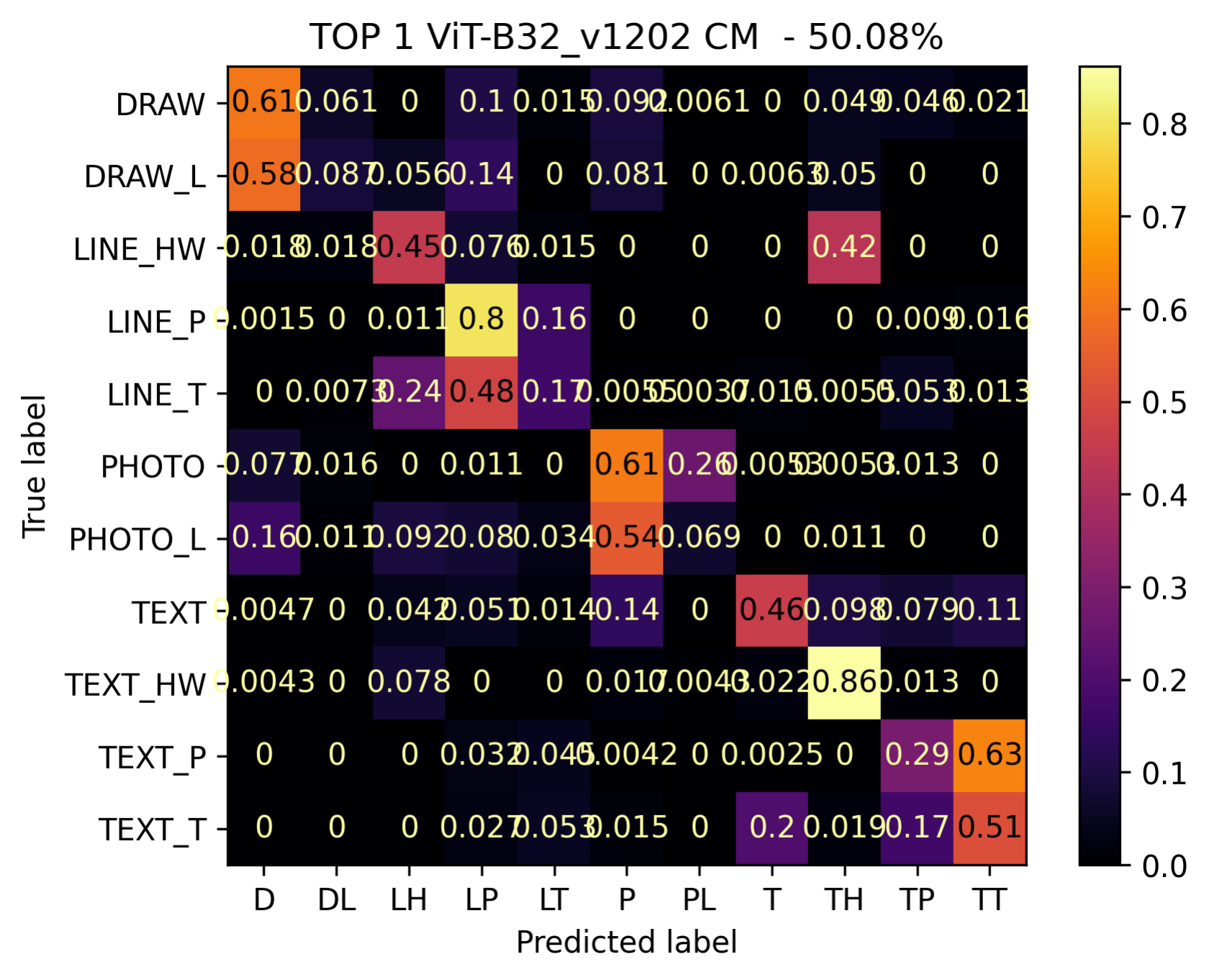}
        \caption{ViT-B/32 (detailed; Table \ref{tab:enhanced_classification}): 50.08\%}
        \label{fig:zero-b32}
    \end{subfigure}
    % \hfill
    \begin{subfigure}[b]{0.48\textwidth}
        \centering
        \includegraphics[width=\textwidth]{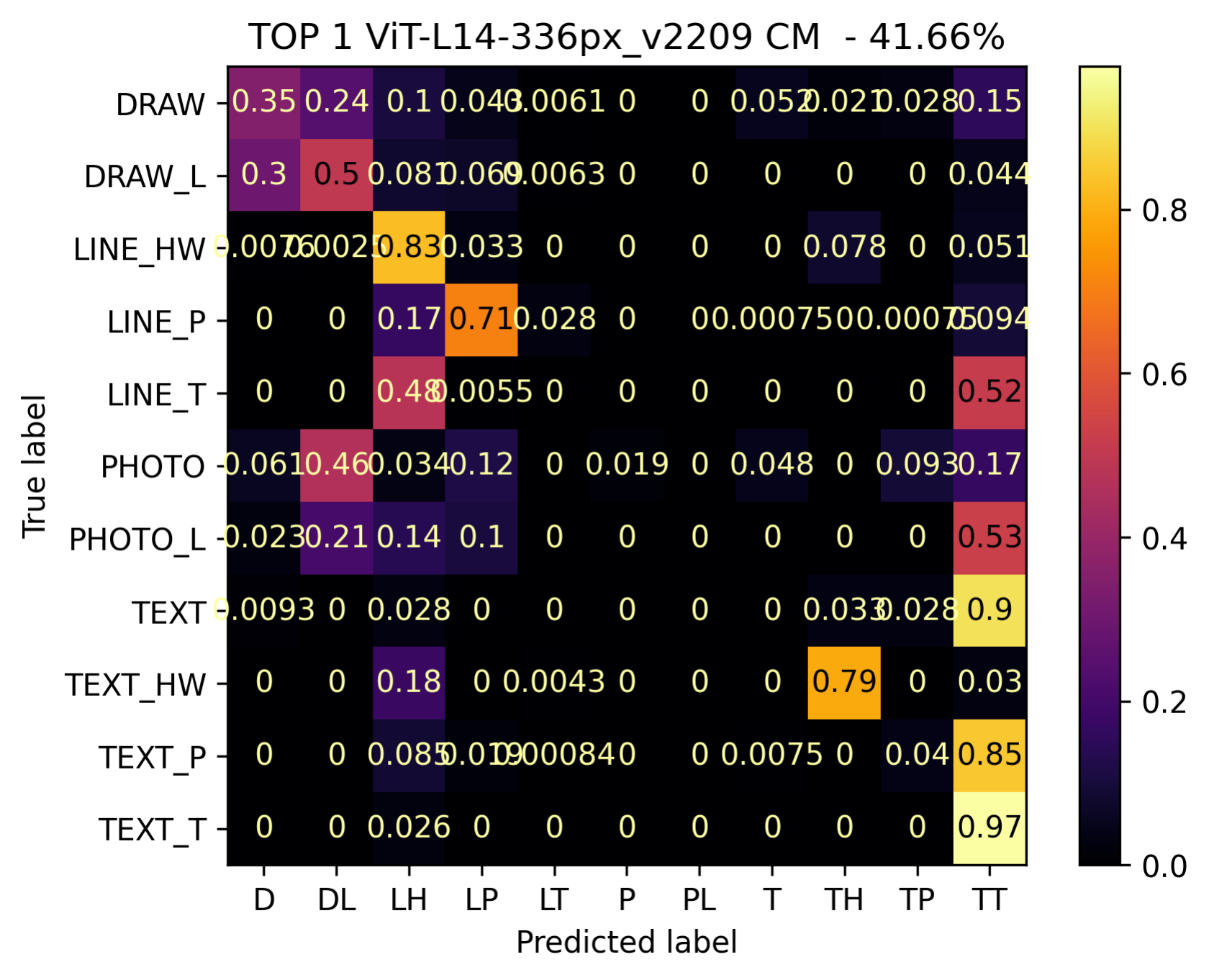}
        \caption{ViT-L/14-336 (short; Table \ref{tab:crop_classification}): 41.66\%}
        \label{fig:zero-l336}
    \end{subfigure}

    \hspace{0.5em}

    \begin{subfigure}[b]{0.48\textwidth}
        \centering
        \includegraphics[width=\textwidth]{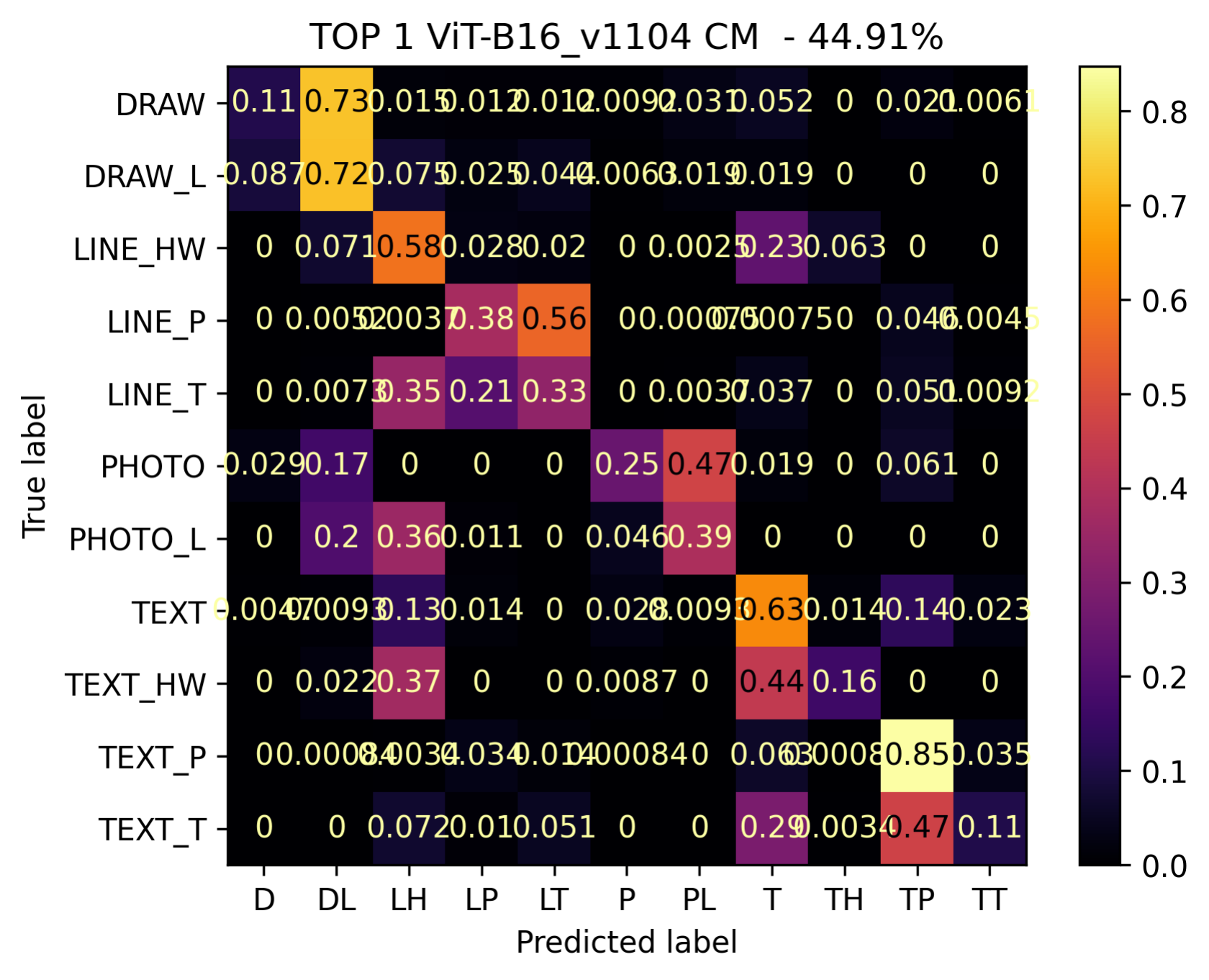}
        \caption{ViT-B/16 (Gemini; Table \ref{tab:long_classification}): 44.91\%}
        \label{fig:zero-b16}
    \end{subfigure}
    % \hfill
    \begin{subfigure}[b]{0.48\textwidth}
        \centering
        \includegraphics[width=\textwidth]{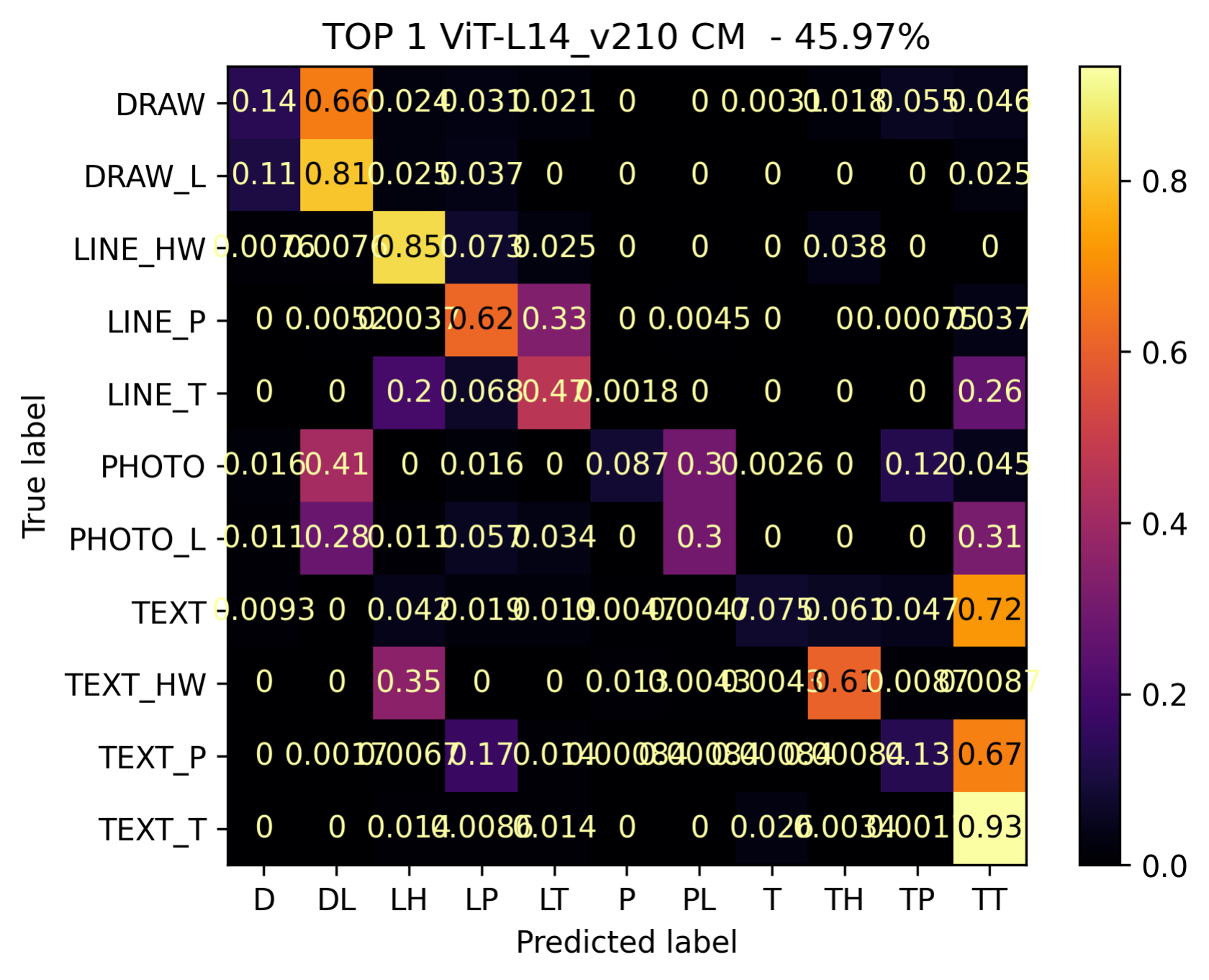}
        \caption{ViT-L/14 (averaged): 45.97\%}
        \label{fig:zero-l}
    \end{subfigure}
  
\end{figure}

\noindent The zero-shot inference process involves using an off-the-shelf pretrained \ac{clip} model to classify images based on their similarity to the provided textual descriptions of the image categories.

While \ac{clip} can separate broad visual concepts, its domain-specific zero-shot performance was limited: the mean accuracy across all label sets in Table \ref{tab:summary_classification} (including averaged-feature variants) was 35.69\%. The ambiguity of natural-language prompts and the presence of domain-specific page types led to many errors across model setups (Figure \ref{fig1:clip-acc-categ-zero}). The best variant of each base model family is illustrated with confusion matrices in Figure \ref{fig:clip-zero}.

In zero-shot settings, accuracy variations do not correlate clearly with either the base model type or the description set. Moving to few-shot fine-tuning, we found that zero-shot averaged models and models using the \textit{mid} description set (Table \ref{tab:short_classification}) consistently outperform the overall mean score. We therefore used these text feature sets as reliable defaults in subsequent experiments.

\subsubsection{Fine-tuned model inference} \label{sec-few-clip}

\begin{figure}[!h]
    \centering
    \caption{\ac{clip} best models (ViT-B/16 variations) confusion matrices, seven epochs}
    \label{fig:clip-best5}
        \includegraphics[width=\textwidth]{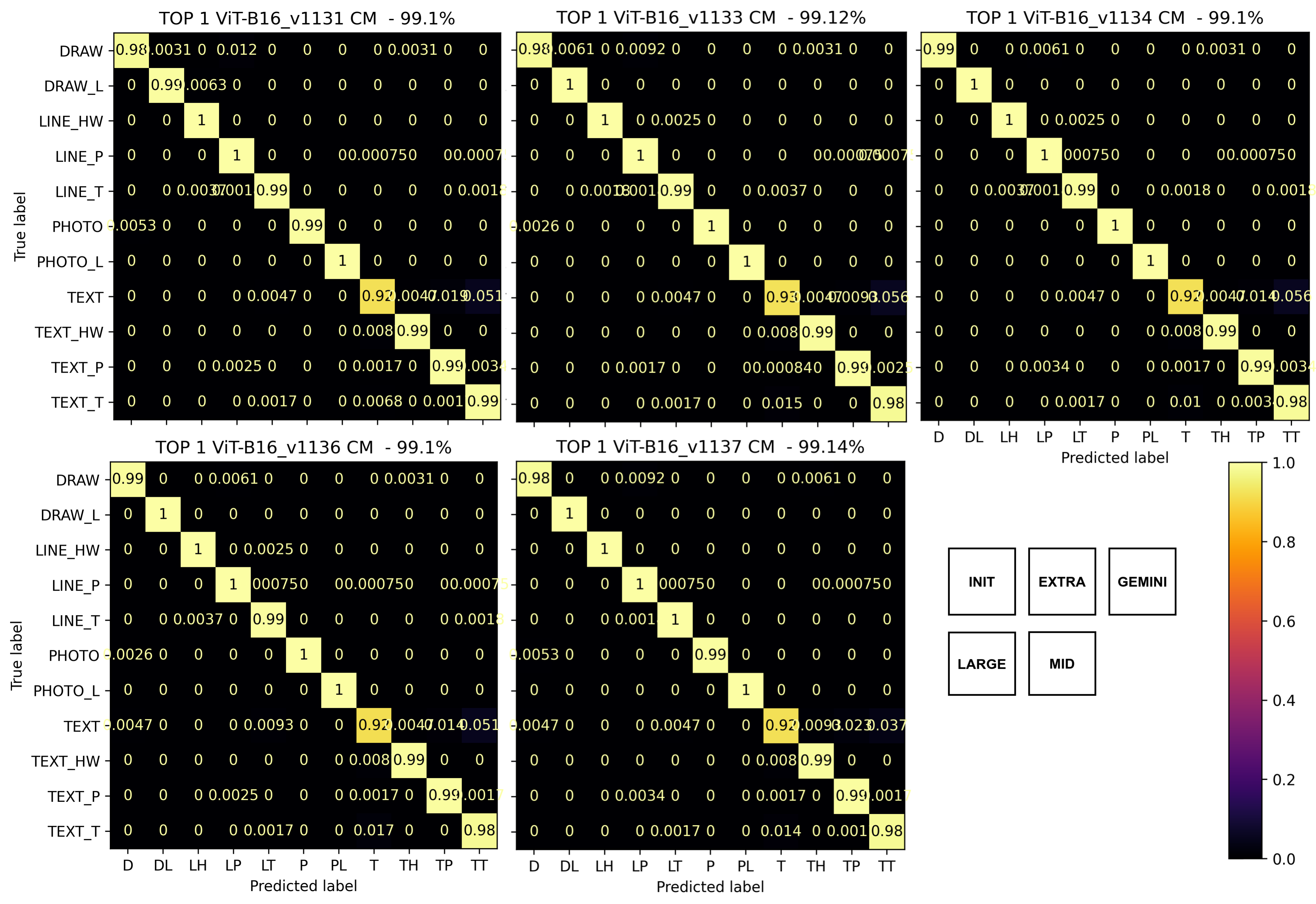}
\end{figure}

\noindent We explored versions of \ac{clip} through few-shot fine-tuning for seven epochs on the first fold (of five cross-validation runs) of the data split, in addition to testing its inherent zero-shot capabilities. Differences in predictions among the same base models are attributed to the category-specific text features used during fine-tuning and inference.

Figures \labelcref{fig1:clip-acc-categ,fig1:clip-acc-matrix} present a comprehensive analysis of classification accuracy among models tailored to different category description sets, highlighting the modest benefit of the averaging strategy observed in the zero-shot setting. In few-shot fine-tuning experiments, only the ViT-B/32 averaged model consistently outperformed its non-averaged counterparts. The \textit{mid} set of category descriptions (Table \ref{tab:short_classification}), used for ViT-B/16 fine-tuning and also scoring above the global average in all four zero-shot models (Figure \ref{fig1:clip-acc-categ-zero}), achieved the highest score among all fine-tuned \ac{clip} models (99.14\%), as shown in Figure \ref{fig:clip-best5}.

Unexpectedly, the largest, highest-resolution model (ViT-L/14 at 336px) struggled with the mixed-content category \texttt{TEXT}, which it often misclassified as pure typewritten pages (Figure \ref{fig:clip-base4}). The other base models showed similar behavior, which is expected given the visual similarity between \texttt{TEXT} and \texttt{TEXT\_*}, \texttt{LINE\_T}, or \texttt{DRAW}.

\begin{figure}[!h]
\centering
\includegraphics[width=140mm]{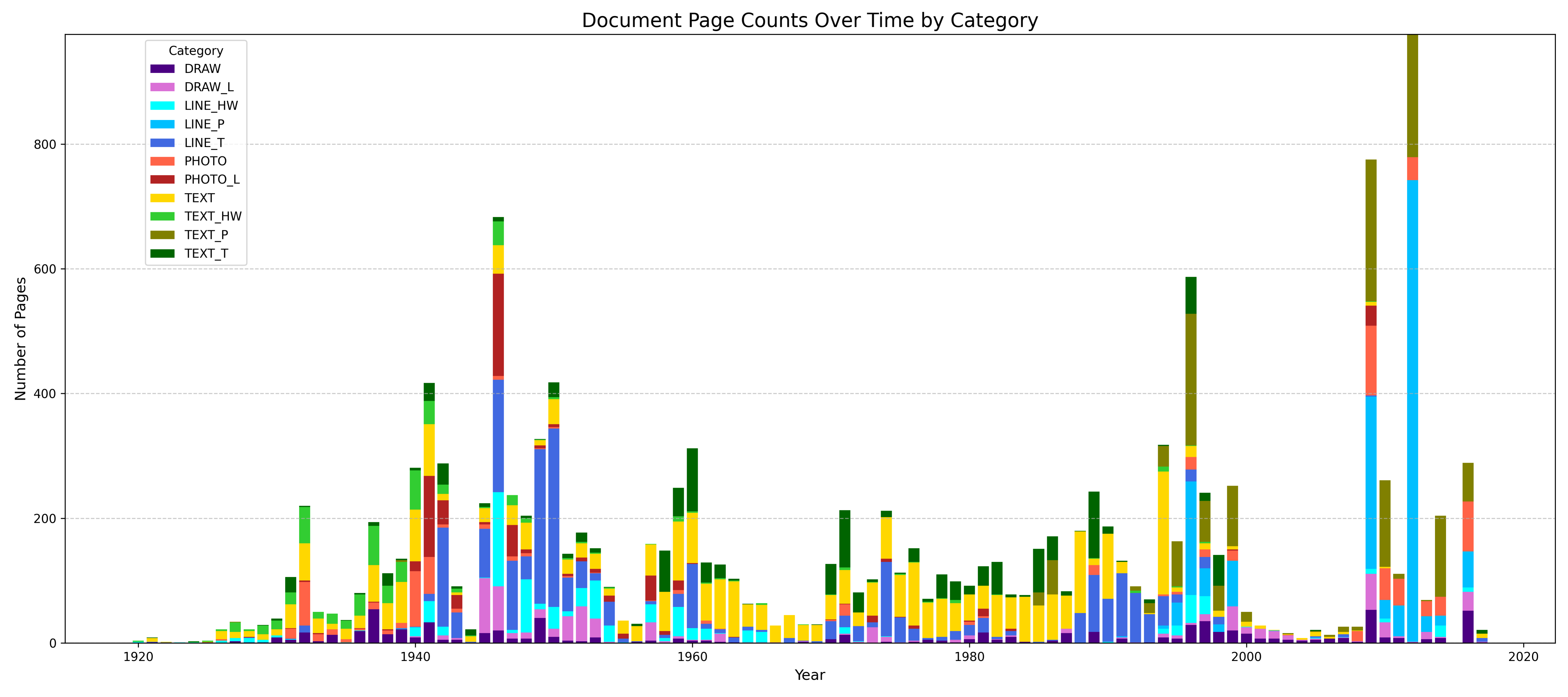}
\caption{Category distribution in the expanded test dataset (all non-training samples from the first-fold split, random seed 420; 14{,}162 pages).}
\label{fig1:time-f1}
\end{figure}

We also observed a dataset effect (Figures \labelcref{fig1:clip-acc-matrix,fig1:clip-acc-matrix-big}). Average accuracy per base model dropped by 3--4\%, and some label sets changed rank, likely because the expanded test dataset contains a wider variety of samples in the smaller categories. Among the fine-tuned \ac{clip} models, B/16 and B/32 scored 99.06\% and 98.96\% on average on the standard performance test set, but on the expanded test dataset (Figure \ref{fig1:time-f1}), only B/16 kept pace with the larger L/14 and L/14-336px variants.

Accuracy varied across both label sets and base models. For example, L/14-336px improved, whereas B/32 decreased relative to other models. A likely reason is that the expanded test dataset includes a wider variety of samples in the smaller categories. As shown in Figures \labelcref{fig:label_time_test,fig1:time-f1}, the pool of available non-training annotations differs substantially across folds: the expanded test dataset includes development and test subsets from the first fold, plus additional pages labeled \texttt{TEXT} that were not filtered out from the training subsets of folds 2--5.

To sum up the analysis of differences and commonalities between all of the investigated base models, we provide extended results in Chapter \ref{chap-results}. There you will find a plot of models' parameter size and test accuracy (see Figure \ref{fig:acc-trend}), comparison of the inter-model agreement for the predictions of unlabeled pages (Figure \ref{fig:similarity}), and a tabulated ambiguity of averaged predictions in Table \ref{tab:arup-arub-labeled-pages-by-category}. In general, finalized justifications for the selection of the models released can be found there.

\begin{figure}[!h]
\centering
\includegraphics[width=120mm]{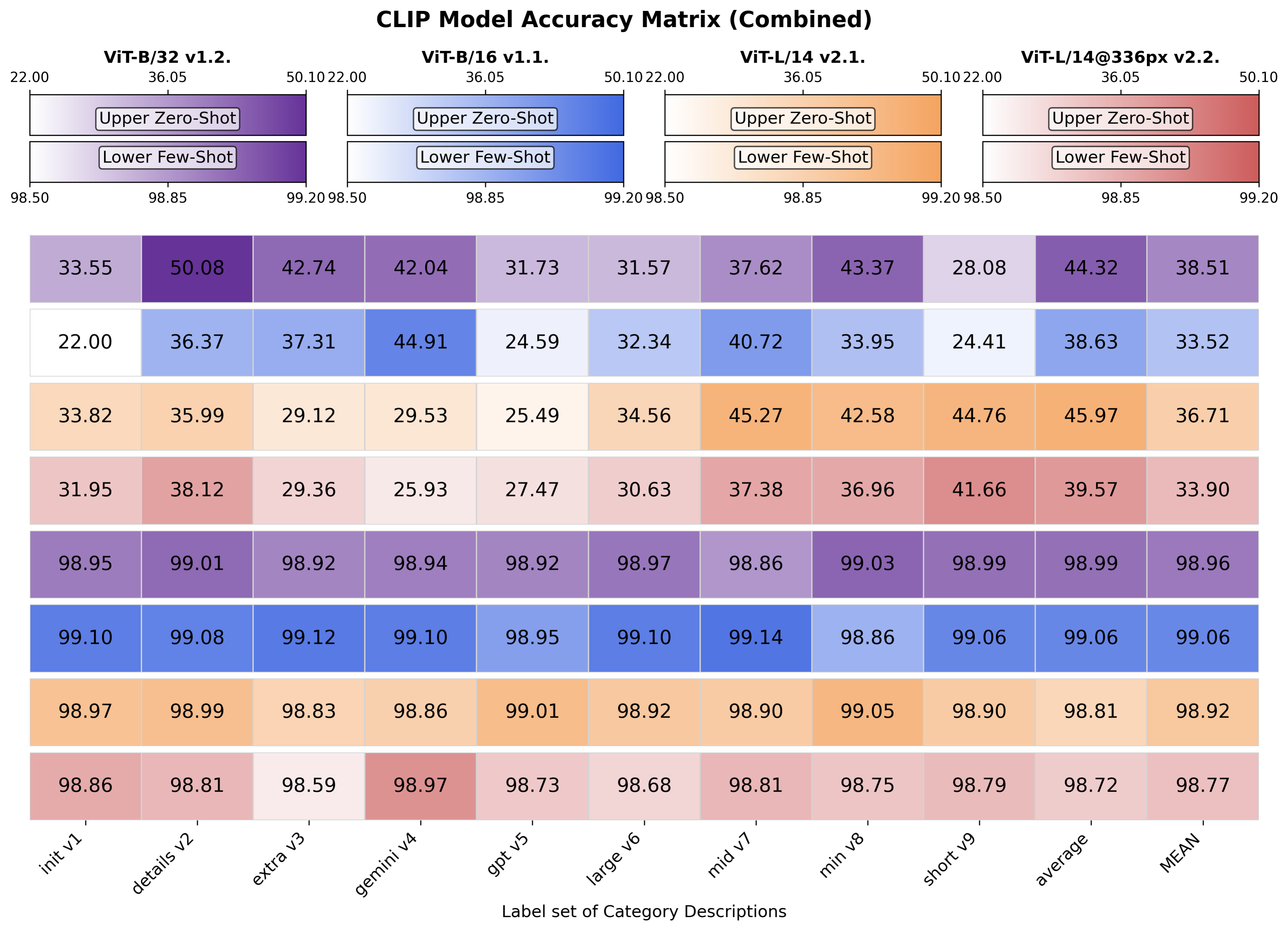}
\caption{Combined zero-shot and fine-tuned \ac{clip} models: comparison of classification accuracy across category description sets on the standard test dataset (5{,}449 pages).}
\label{fig1:clip-acc-matrix}
\end{figure}

\begin{figure}[!h]
\centering
\includegraphics[width=120mm]{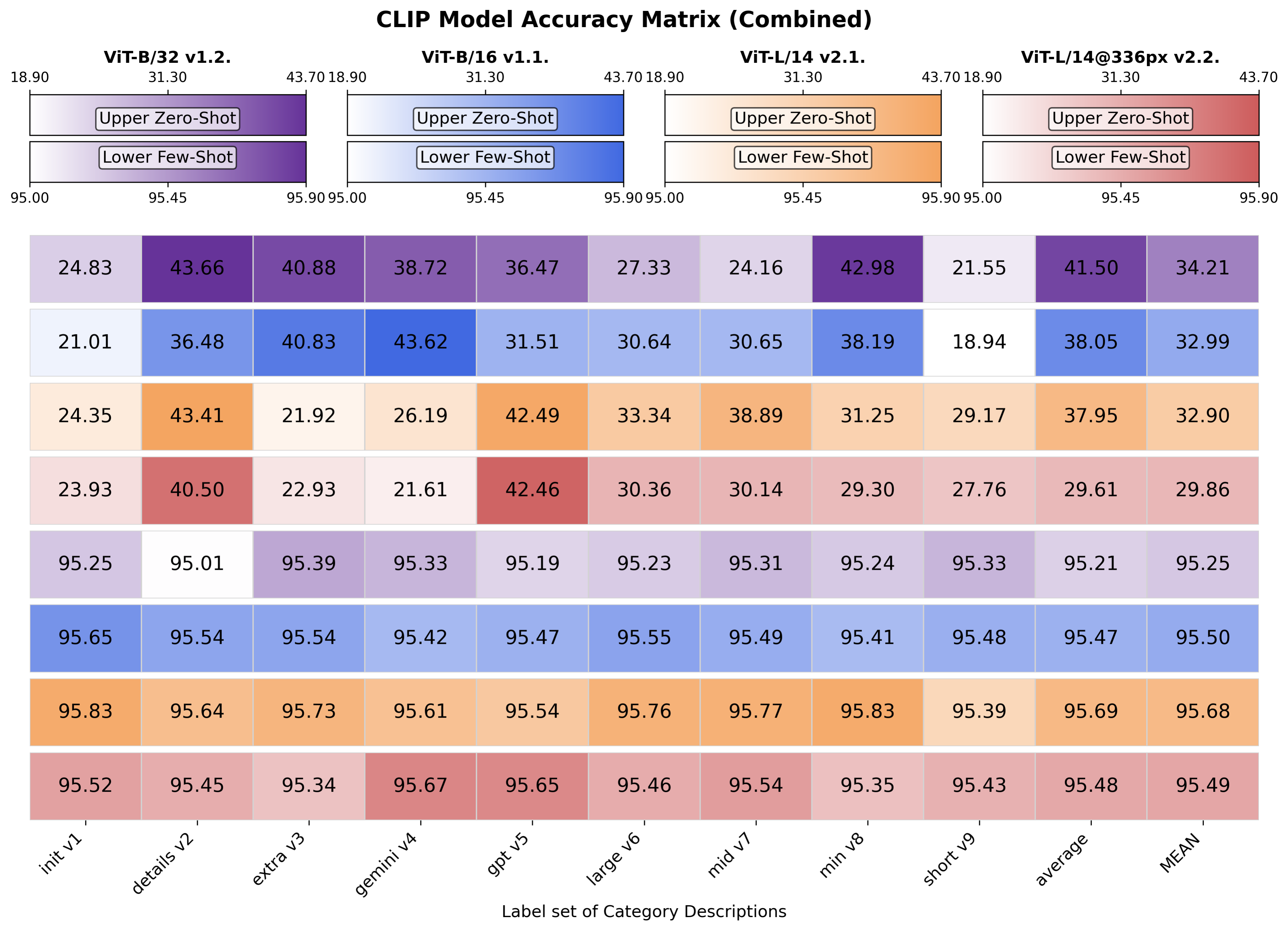}
\caption{Combined zero-shot and fine-tuned \ac{clip} models: comparison of classification accuracy across category description sets on an expanded test dataset (all non-training samples from the first-fold split, random seed 420; 14{,}162 pages).}
\label{fig1:clip-acc-matrix-big}
\end{figure}

\chapter{System architecture} \label{chap-architect}

\noindent The classification system is implemented as a modular pipeline that transforms raw, disorganized archival documents into a categorized, searchable repository. It is designed to be platform-agnostic, supporting both the Unix environments common in research and the Windows systems typically used in archival offices.

\begin{figure}[h]
\centering
\includegraphics[width=125mm]{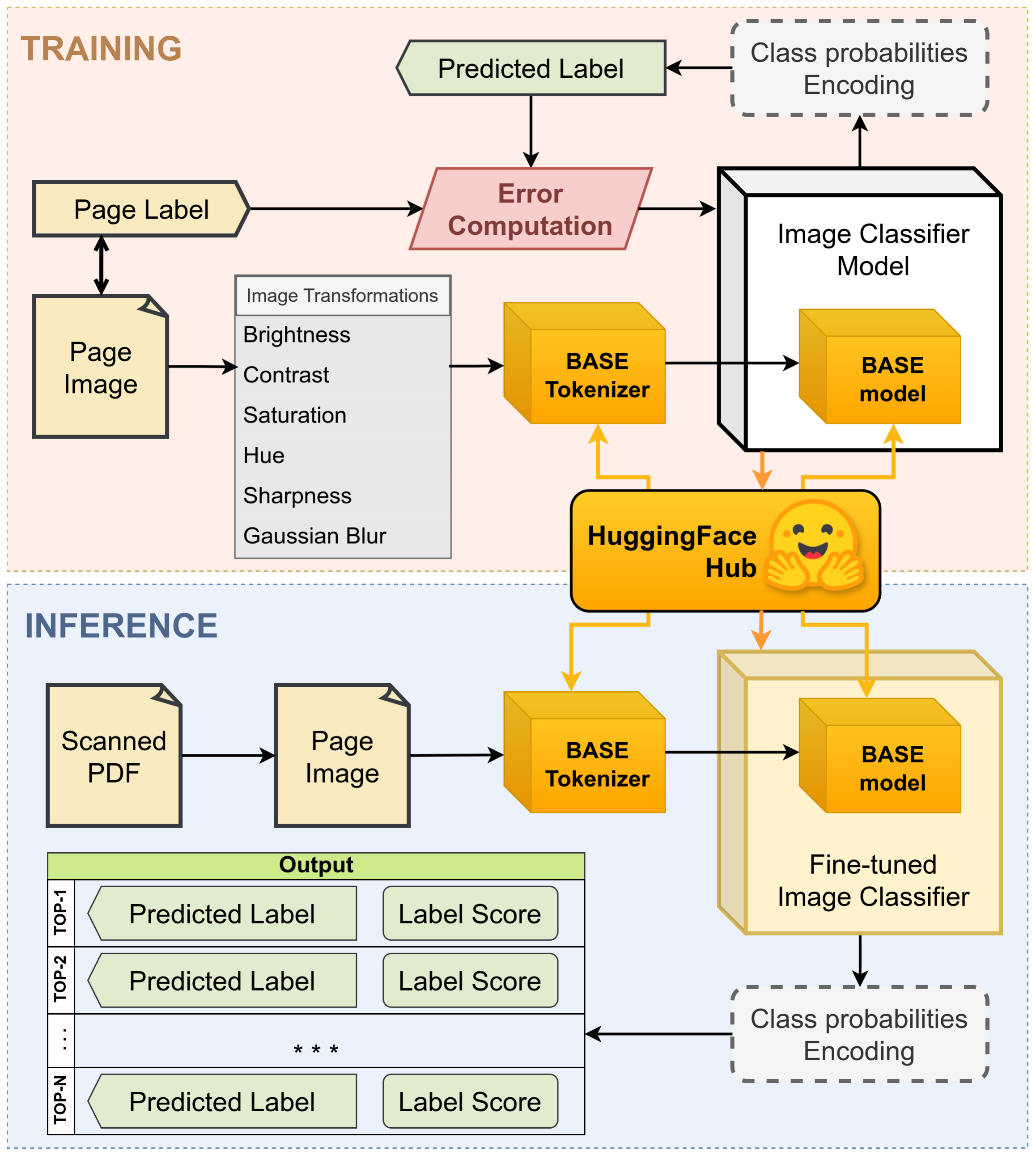}
\caption{System overview for the image-only (\acp{cnn} and Transformer) variants}
\label{fig:architect}
\end{figure}

\begin{figure}[!hp]
\centering
\includegraphics[width=125mm]{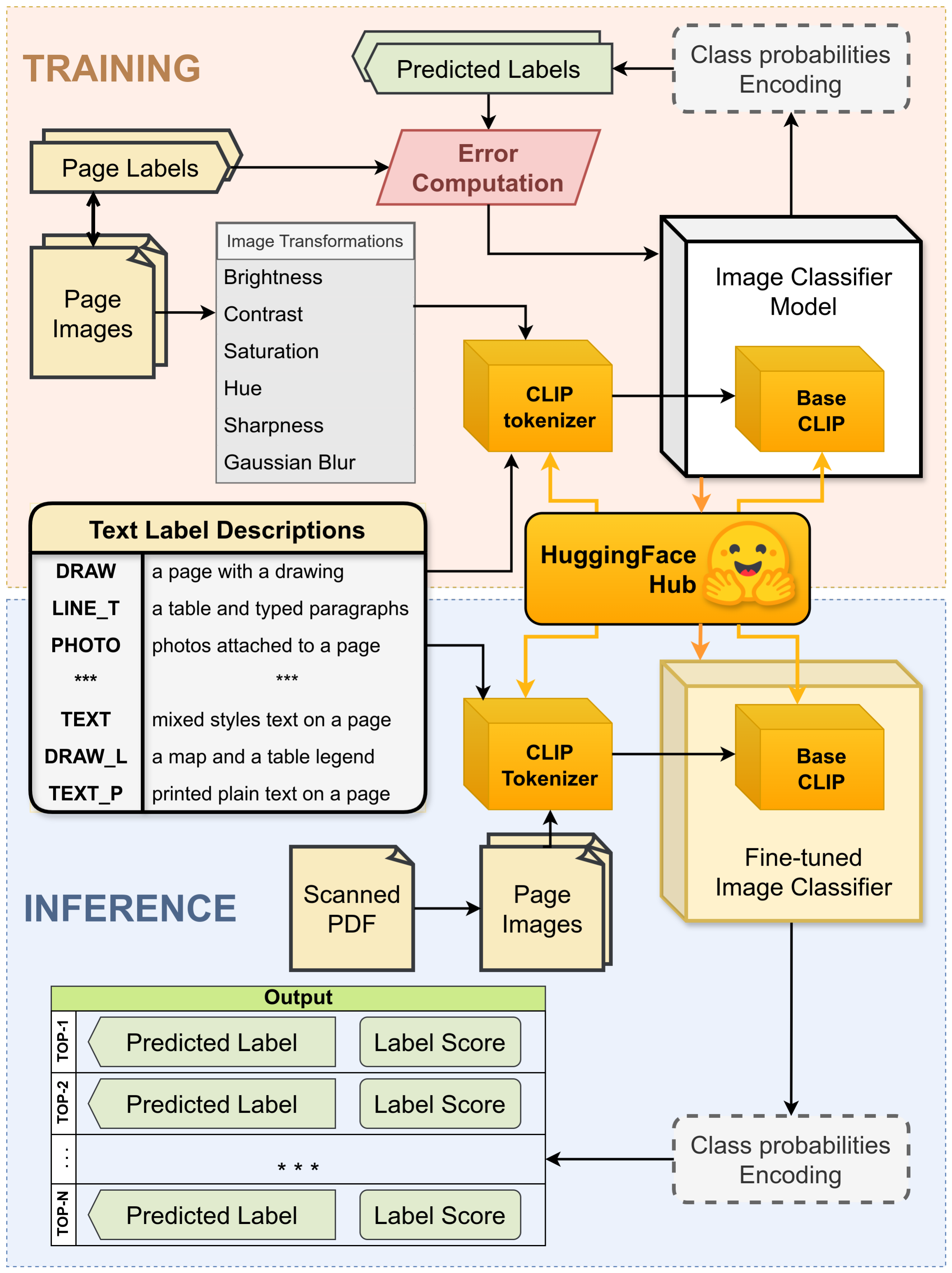}
\caption{Model overview for the hybrid \ac{clip} variant}
\label{fig:architect_clip}
\end{figure}

At its core, the pipeline applies fine-tuned image classification models---\acp{cnn} (\acs{eg} EfficientNetV2 \cite{tan2021efficientnetv2}, RegNetY \cite{radosavovic2020designing}), Transformers (\ac{dit} \cite{li2022dit}, \ac{vit} \cite{dosovitskiy2020image}), and hybrid multimodal models (\ac{clip} \cite{radford2021learning})---to sort historical page scans (Figures \labelcref{fig:architect,fig:architect_clip}).

The architecture consists of (i) configuration and interfaces (Section \ref{sec-interface}), (ii) input processing (Section \ref{sec-streamline}), (iii) model training (fine-tuning; Section \ref{sec-finetune}), (iv) evaluation and output generation (Section \ref{sec-output}), and (v) data preparation utilities (Section \ref{sec-data-prepare}). The system is implemented in Python, with helper scripts for input/output processing on both Windows and Unix platforms.

The system is built in Python using PyTorch \cite{paszke2019pytorch} and HuggingFace Transformers \cite{wolf2019huggingface} to manage model loading and input processing across model types. The complete codebase (including both \ac{vit} and \ac{clip} branches) is available in a public GitHub repository \cite{Lutsai_ATRIUM_s_page_classifier_2025}.

Recognizing the sensitivity of many archival collections, the architecture is strictly on-premises: processing occurs entirely on local hardware, ensuring that restricted or unpublished data never leaves the institution's secure network.

Additional implementation details are provided in Appendix \ref{architecture-system}, including hyperparameters, image preprocessing, dataset splitting, and output options.

\section{Interface} \label{sec-interface}

\noindent To accommodate both end users and developers, the system provides a configuration file for persistent settings, a \ac{cli} for direct operation, and utility scripts for data preparation. To reduce the required programming expertise, the two primary interaction modes are an editable text configuration file and a command-line interface for automation.

\subsection{Configuration file} \label{sec-config}

\noindent The \texttt{config.txt} file serves as the central repository for user-adjustable parameters, allowing users to modify the system’s behavior without changing the source code. The settings are organized into the sections summarized in Table \ref{tab:config-settings}.

\begin{table}[h!]
\centering
\caption{Summary of configuration settings in \texttt{config.txt} sections}
\label{tab:config-settings}
\begin{tabular}{m{1.5cm}p{12.4cm}}
\toprule
\textbf{Section} & \textbf{Description of variables in section} \\
\midrule
\texttt{[INPUT]} & Specifies the default input directory and chunk size; controls whether processing is chunked or results are recorded at the end. \\
\texttt{[OUTPUT]} & Sets paths for result \ac{csv} files, model checkpoints, and visualization outputs. \\
\texttt{[SETUP]} & Contains operational parameters: batch size, input file format, top-$N$ value, base model architecture, and random seed. \\
\texttt{[TRAIN]} & Stores training settings such as dataset path, number of epochs, learning rate, validation split, and logging frequency. \\
\texttt{[EVAL]} & Sets the evaluation directory layout (category subdirectories with files named \{document\_name\}-\{page\_number\}). \\
\texttt{[HF]} & Stores HuggingFace integration settings and defines a global model revision. \\
\bottomrule
\end{tabular}
\end{table}

All system behaviors (\acs{eg} input discovery, chunk size, and model choice) are managed through \texttt{config.txt}. Users can toggle settings (\ac{eg} ``safe mode'' for corrupted files) without editing the Python code.

\subsection{Command line entry point} \label{sec-cmd}

\noindent The system's primary entry point is \texttt{run.py}, which exposes a \ac{cli} for fine-grained control.

For large-scale processing, a single command can process an entire directory (\texttt{--dir}) and optionally include nested folders (\texttt{--inner}). The framework is also designed to support rapid adaptation: launching fine-tuning on a new annotated dataset does not require code changes and can be done via a single flag (\texttt{--train}).

\begin{table}[h!]
\centering
\caption{Selected \ac{cli} options in \texttt{run.py} (representative subset)}
\label{tab:cli-options}
\begin{tabular}{p{3.2cm}p{10.4cm}}
\toprule
\textbf{Option} & \textbf{Purpose} \\
\midrule
\texttt{--file} & Classify a single image file (one page). \\
\texttt{--dir} & Classify all supported images in a directory. \\
\texttt{--inner} & Recursively include nested subdirectories under \texttt{--dir}. \\
\texttt{--chunk} & Enable chunked processing for large folders (memory-efficient batching). \\
\texttt{-b} & Set batch size (overrides \texttt{config.txt}). \\
\texttt{--train} & Run fine-tuning on an annotated dataset specified in \texttt{config.txt}. \\
\bottomrule
\end{tabular}
\end{table}

\subsection{Streamlined Input Processing} \label{sec-streamline}

The system bridges the gap between raw scans and model-ready inputs through a robust input layer. For large archives, chunked processing improves memory efficiency by loading and predicting on small batches of images; this can be enabled via the \texttt{--chunk} flag.

While the system can run on a \ac{cpu} for small batches, it is optimized for NVIDIA \acp{gpu} to increase throughput on large collections. Batch size can be configured in \texttt{config.txt} or via the \texttt{-b} command-line flag.

\subsection{Web service interface} \label{subsec-web}

\noindent To support interactive testing and integration with downstream applications, the system additionally exposes a lightweight \ac{rest} \ac{api} built with FastAPI, served from the \texttt{service/} subdirectory of the main project repository. The service provides three \ac{http} endpoints summarized in Table \ref{tab:api-endpoints}.

\begin{table}[h!]
\centering
\caption{REST API endpoints exposed by the web service}
\label{tab:api-endpoints}
\begin{tabular}{p{1.2cm} p{2.0cm} p{10.1cm}}
\toprule
\textbf{Method} & \textbf{Path} & \textbf{Description} \\
\midrule
\texttt{GET}  & \texttt{/}        & Serves the static \ac{html} frontend for manual image upload and testing. \\
\texttt{GET}  & \texttt{/info}    & Returns metadata about available model versions and the active computation device (\texttt{cpu} or \texttt{cuda}). \\
\texttt{POST} & \texttt{/predict} & Performs inference on a single uploaded image and returns structured \ac{json} predictions. \\
\bottomrule
\end{tabular}
\end{table}

The primary \texttt{/predict} endpoint accepts a multipart form-data request with three parameters: \texttt{file} (a \ac{jpeg} or \ac{png} image), \texttt{version} (the model version string, \acs{eg} \texttt{v5.3}, or \texttt{all} to invoke a five-model ensemble), and the optional integer \texttt{topn} controlling the number of returned predictions (default: 3). The \ac{json} response includes the best predicted category, its confidence score, and the full ranked Top-$N$ list, as illustrated in Listing~\ref{lst:api-response}.

\begin{figure}[h!]
\begin{verbatim}
{
  "model_version": "google/vit-large-patch16-384 (v5.3)",
  "best_category": "TEXT",
  "score": 0.985,
  "requested_topn": 3,
  "predictions": [
    {"label": "TEXT",   "score": 0.985},
    {"label": "TEXT_P", "score": 0.010},
    {"label": "LINE_P", "score": 0.002}
  ]
}
\end{verbatim}
\caption{Example JSON response from the \texttt{/predict} endpoint (model \texttt{v5.3},
         Top-3 predictions)}
\label{lst:api-response}
\end{figure}

The service supports all model architectures listed in Section~\ref{sec-typical-img-class} and automatically detects available \ac{cuda} devices to accelerate inference. A static \ac{html}/JavaScript frontend (\texttt{service/frontend/}) is bundled for immediate browser-based testing without additional tooling. Full client integration should be maintained separately in the LINDAT Service repository.

\section{Output formats} \label{sec-result-present}

\noindent The system provides several output formats to support different analysis and workflow needs:

\begin{enumerate}
\item \textbf{Console output}: For single-file classification, the system prints the Top-$N$ predictions with confidence scores.

\item \textbf{Top-$N$ tables}: \ac{csv} files sorted by file and page number, with columns \texttt{CLASS-N} and \texttt{SCORE-N} for each of the Top-$N$ predictions.

\item \textbf{Averaged Top-$N$ tables}: \ac{csv} files sorted by file and page number, with columns for Top-$1$ model guesses from \texttt{v1.3} to \texttt{v5.3}, then, \texttt{CLASS-N} and \texttt{SCORE-N} for each of the Top-$N$ predictions computed as average of five models' scores per category.

\item \textbf{Raw probability tables}: \ac{csv} files containing the model's raw probabilities for all categories, enabling detailed analysis of normalized logits.

\item \textbf{Confusion matrix plots}: Visualizations of model performance on annotated data, showing the relationship between predicted and true categories.
\end{enumerate}

Results are saved to locations specified in \texttt{config.txt}, with filenames that include timestamps and model identifiers for traceability.

For archival management, the outputs are designed to be immediately actionable: the system can generate tables listing each file and its predicted category (\acs{eg} \texttt{LINE\_P}, \texttt{PHOTO\_L}, \texttt{TEXT\_HW}) with a confidence score. It can also copy or move images into category-specific subfolders based on predictions, allowing an archivist to open a folder such as \texttt{DRAW} or \texttt{PHOTO\_L} and quickly verify results or begin targeted processing.

\section{Data preparation functionality} \label{sec-data-prepare}

\noindent The system includes a suite of utilities to help users prepare document data for both fine-tuning and classification, starting from raw \ac{pdf} files. Figure \ref{fig2:sorting} illustrates the end-to-end workflow for a typical model inference use case.

\begin{figure}[h]
\centering
\includegraphics[width=120mm]{img/sorting.pdf}
\caption{Model inference use-case: input \ac{pdf} is parsed into a folder of pages, then sorted into category-specific subdirectories}
\label{fig2:sorting}
\end{figure}

\subsection{\ac{pdf} documents to page images} \label{sec-pdf-2-png}
The data preparation pipeline includes cross-platform scripts to convert \ac{pdf} documents into \ac{png} images:

\begin{itemize}
    \item \textbf{Unix script} (\texttt{pdf2png.sh}): uses pdftoppm for conversion with zero-padded page numbers
    \item \textbf{Windows script} (\texttt{pdf2png.bat}): uses ImageMagick and Ghostscript with sequential page numbers
\end{itemize}

These scripts create a directory structure where each \ac{pdf} is converted into a subdirectory of page-specific images. Additional utilities facilitate the organization of single-page documents (any subdirectories containing a single file) for more efficient annotation.

\subsection{Annotated data systematization} \label{sec-data-sort}

\noindent To facilitate model training (and fine-tuning of pretrained models), the system provides tools for organizing annotated data into a structured format:

\begin{enumerate}
\item \textbf{Annotation process}: Users create \ac{csv} files with three required columns: \texttt{FILE}, \texttt{PAGE}, and \texttt{CLASS}.
\item \textbf{Data systematization}: The provided scripts (\texttt{sort.sh}/\texttt{sort.bat}) parse the annotation files and copy the corresponding image files into category-specific subdirectories. These scripts also support quick visual verification of annotations and model outputs.
\item \textbf{Category balancing}: The \texttt{max\_categ} parameter in the configuration file can be used to limit the number of samples in overrepresented categories, helping to address class imbalance that could otherwise bias model performance during fine-tuning.
\end{enumerate}

This structured workflow ensures that training data is properly organized according to the model's expected input format.

\chapter{Results} \label{chap-results}
\noindent This chapter evaluates the best fine-tuned models and discusses how the resulting classifiers can be used in archival workflows.

We report Top-1 accuracy and relate it to model size (efficiency). We also analyze agreement between model predictions on unlabeled pages.

\begin{figure}[!h]
\centering
    \includegraphics[width=140mm]{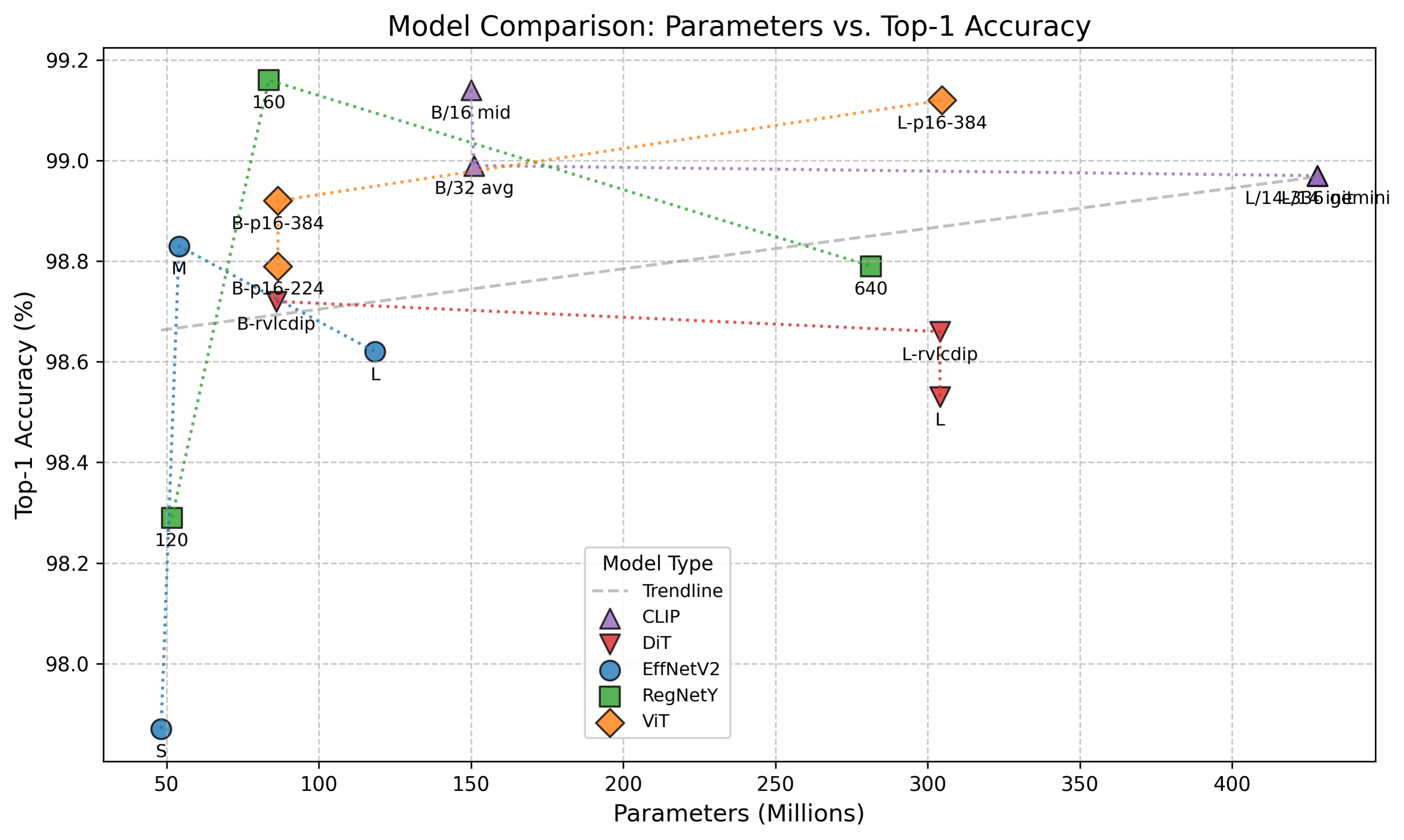}
    \caption{Accuracy vs. parameter count across evaluated models. Models above the trendline deliver superior efficiency.}
    \label{fig:acc-trend}
\end{figure}

Figure \ref{fig:acc-trend} shows that \acp{cnn}, \ac{vit} models, and fine-tuned hybrid (text+image) \ac{clip} models exceed $\approx98.5\%$ accuracy with parameter budgets below 200M. In comparison, similarly sized \ac{dit} variants perform acceptably but below the top-performing architectures.

\section{Accuracy of tested models} \label{sec-accuracy}
\noindent We compared \acp{cnn} (EfficientNetV2 \cite{tan2021efficientnetv2}, RegNetY \cite{radosavovic2020designing}), Transformers (\ac{vit} \cite{dosovitskiy2020image}, \ac{dit} \cite{li2022dit}), and hybrid \ac{clip} models \cite{radford2021learning} (Sections \ref{sec-effnet}, \ref{sec-dit}, \ref{sec-clip}). Models were fine-tuned and evaluated on the archival dataset using Top-1 accuracy.

The test set contains 5{,}449 images that were not used in any training subset, enabling fair evaluation of all cross-validation folds (and their averaged variants). Figure \ref{fig:label_time_test} illustrates the category distribution over time. Table \ref{tab:model_comparison} summarizes model size, input resolution, and performance.

\begin{table}[h]
\centering
\caption{Top-1 accuracy and model complexity for all evaluated architectures. \textit{Italic}: most efficient; \textbf{bold}: most accurate (and most efficient) per model type.}
\label{tab:model_comparison}
\begin{tabular}{@{}lccc@{}}
\toprule
\textbf{Base model} & \textbf{Parameters (M)} & \textbf{Resolution} & \textbf{Accuracy Top-1 (\%)} \\ \midrule
RFC baseline & 333 trees & $298\times1$ & 75.34 \\ \midrule
EfficientNetV2-S & 48.2 & $300\times300$ & 97.87 \\
\textbf{EfficientNetV2-M} & \textbf{54.1} & \textbf{$384\times384$} & \textbf{98.83} \\
EfficientNetV2-L & 118.5 & $384\times384$ & 98.62 \\
\midrule
RegNetY-12GF & 51.8 & $224\times224$ & 98.29 \\
\textbf{RegNetY-16GF} & \textbf{83.6} & \textbf{$224\times224$} & \textbf{99.16} \\
RegNetY-64GF & 281.4 & $384\times384$ & 98.79 \\
\midrule
\textbf{DiT-base-RVL} & \textbf{86} & \textbf{$224\times224$} & \textbf{98.72} \\
DiT-large & 304 & $224\times224$ & 98.53 \\
DiT-large-RVL & 304 & $224\times224$ & 98.66 \\
\midrule
\textit{ViT-base-patch16} & \textit{86.6} & \textit{$224\times224$} & \textit{98.79} \\
ViT-base-patch16 & 86.9 & $384\times384$ & 98.92 \\
\textbf{ViT-large-patch16} & \textbf{304.7} & \textbf{$384\times384$} & \textbf{99.12} \\
\midrule
\textbf{CLIP-ViT-B/16} & \textbf{150} & \textbf{$224\times224$} & \textbf{99.14}\\
CLIP-ViT-B/32 & 151 & $224\times224$ & 98.99 \\
CLIP-ViT-L/14 & 428 & $224\times224$ & 98.97 \\
CLIP-ViT-L/14 & 428 & $336\times336$ & 98.97 \\
\bottomrule
\end{tabular}
\end{table}

Table \ref{tab:model_comparison} and Figure \ref{fig:acc-trend} summarize the main results:

\begin{itemize}
  \item \textbf{RegNetY (\ac{cnn}):} RegNetY-16GF achieves $\approx99.16\%$ accuracy with a relatively small parameter count (83.6M) at $224\times224$ resolution \cite{radosavovic2020designing}. RegNetY-64GF (281.4M parameters) falls below the global efficiency trendline, but was published (\texttt{v6.3}) for comparison.
  \item \textbf{EfficientNetV2 (\ac{cnn}):} EfficientNetV2-M reaches $\approx98.83\%$ accuracy at 54.1M parameters with $384\times384$ inputs \cite{tan2021efficientnetv2}, making it a reasonable option for lower-resource machines (subject to throughput constraints).
  \item \textbf{Vision Transformers (\ac{vit}):} \ac{vit}-large reaches $\approx99.12\%$ Top-1 accuracy \cite{dosovitskiy2020image} but requires substantially more parameters (304.7M) than the best-performing \acp{cnn}.
  \item \textbf{Document image Transformers (\ac{dit}):} \ac{dit} variants achieve 98--99\% Top-1 accuracy \cite{li2022dit} but do not outperform the best \acp{cnn} and \ac{vit} models at comparable sizes.
  \item \textbf{\ac{clip}:} Zero-shot performance is inadequate for this domain ($\leq50\%$), highlighting the need for fine-tuning and carefully designed text features \cite{radford2021learning}. After fine-tuning (seven epochs on 80\% of annotated pages), all tested \ac{clip} variants exceed $\approx98.97\%$ accuracy; the best result is achieved by CLIP-ViT-B/16 ($\approx99.14\%$) with the \textit{mid} label descriptions (Table \ref{tab:short_classification}).
\end{itemize}

\section{Picking the best model} \label{sec-archive}
\noindent The primary objective of this research was to develop a classification system that combines accuracy with practical applicability within existing archival processes. This section outlines the deployment specifications and the expert-defined criteria the system is designed to meet.

\subsection{Deployment and Usability} \label{sec-deploy}

\noindent Comprehensive installation and usage instructions are provided in the project's public repository (\ac{eg} via the \href{https://github.com/ufal/atrium-page-classification/blob/vit/README.md}{GitHub README}). To facilitate adoption by archival institutions, the system is packaged for local deployment under the \ac{mit} license.

\begin{table}[h!]
\centering
\small
\caption{NVIDIA \acp{gpu} Specifications Comparison}
\label{tab:gpu_specs}
\begin{tabular}{p{2.6cm}p{2.5cm}p{2.5cm}p{2.5cm}p{2.5cm}}
\toprule
\textbf{Specification} & \textbf{\texttt{A30}} & \textbf{\texttt{A40}} & \textbf{\texttt{L40}} & \textbf{\texttt{RTX 3090}} \\ 
\midrule
\textbf{Architecture} & Ampere (GA100) & Ampere (GA102) & Ada Lovelace (AD102) & Ampere (GA102) \\ 
\textbf{Comp. Cap.} & 8.0 & 8.6 & 8.9 & 8.6 \\ 
\textbf{Mem. Size} & 24 GB & 48 GB & 48 GB & 24 GB \\ 
\textbf{Mem. Type} & HBM2 & GDDR6-ECC & GDDR6-ECC & GDDR6X \\ 
\textbf{Mem. Interf.} & 3072-bit & 384-bit & 384-bit & 384-bit \\ 
\textbf{Mem. BW} & 933 GB/s & 696 GB/s & 864 GB/s & 936 GB/s \\ 
\textbf{CUDA Cores} & 3,584 & 10,752 & 18,176 & 10,496 \\ 
\textbf{Tensor Cores} & 224 (3rd Gen) & 336 (3rd Gen) & 568 (4th Gen) & 328 (3rd Gen) \\ 
\textbf{RT Cores} & N/A (GA100) & 84 (2nd Gen) & 142 (3rd Gen) & 82 (2nd Gen) \\ 
\textbf{Max Power} & 165W & 300W & 300W & 350W \\ 
\textbf{MIG Support} & Yes $\leq4$ & No & No & No \\ 
\bottomrule
\end{tabular}%
\end{table}

Key operational considerations were addressed to ensure the system's practicality and ease of use. Designed for on-premises deployment, it is vital for archives with sensitive or restricted collections. While standard \ac{cpu} hardware suffices for inference, using a \ac{gpu} is advisable for fine-tuning with user-annotated datasets and for improving inference speed on large datasets.

For best performance with \ac{cuda} \ac{gpu} acceleration, an NVIDIA graphics card is required. The specifications of the \acp{gpu} we tested are listed in Table \ref{tab:gpu_specs}. Users must also allocate significant disk space (up to 30~GB) to accommodate Python dependencies, pretrained model weights, and the \ac{pdf}-to-page conversion outputs typical of large archival collections.

On a modern office desktop computer, the system can process hundreds of thousands of pages within a week, with throughput over 48 hours potentially exceeding 1,000,000 pages on higher-end hardware.

\begin{table}[h!]
\centering
\caption{Average training/evaluation throughput (batches of 11-16 samples per second) across NVIDIA \ac{gpu} types and model size (parameters) ranges. Values are reported as \textit{train / eval} pairs of the processing speed measurements.}
\label{tab:throughput_comparison}
\begin{tabular}{@{}lcccc@{}}
\toprule
\textbf{Model size} & \texttt{A30} & \texttt{A40} & \texttt{L40} & \texttt{RTX 3090} \\ 
\midrule
0--70M    & 1.58 / 5.21 & 1.15 / 3.24 &   & 1.30 / 4.53 \\
70--100M  & 1.50 / 4.68 & 1.20 / 3.42 & 1.58 / 5.19 &  \\
100--200M & 1.48 / 4.60 & 1.04 / 2.83 &  & 1.32 / 4.07 \\
200--350M & 1.27 / 3.63 & 1.07 / 2.85 &  &  \\
\bottomrule
\end{tabular}
\end{table}

\begin{table}[h!]
\centering
\caption{Average efficiency (accuracy increase per hour) across NVIDIA \ac{gpu} types and model size (parameters) ranges.}
\label{tab:efficiency_comparison}
\begin{tabular}{@{}lcccc@{}}
\toprule
\textbf{Model size} & \texttt{A30} & \texttt{A40} & \texttt{L40} & \texttt{RTX 3090} \\ 
\midrule
0--70M    & 4.81 & 3.42 &  & 3.98 \\
70--100M  & 4.58 & 3.67 & 4.87 &  \\
100--200M & 5.19 & 3.21 &  & 4.00 \\
200--350M & 3.82 & 3.29 &  &  \\
\bottomrule
\end{tabular}
\end{table}

Tables \labelcref{tab:throughput_comparison,tab:efficiency_comparison} report training-time measurements on the NVIDIA \acp{gpu} listed in Table \ref{tab:gpu_specs}. Inference (evaluation) is about 2--3 times faster. While throughput depends slightly on model size, it is primarily determined by \ac{gpu} characteristics.

Comprehensive installation and usage instructions are provided in the project's documentation, separately for image-only (\ac{vit}, \ac{dit}, \acp{cnn}) and hybrid (\ac{clip}) models. The core functionality is exposed via the \ac{cli} and is designed to work on both Unix and Windows systems.

In practice, the classifier can be integrated into a digitization pipeline: scanned images are processed automatically, and each page receives a category label. This label enables routing to downstream steps (\acs{eg} sending plain text pages to \ac{ocr}, sending tables to structured data extraction, and routing pages with illustrations to segmentation or human review). Automating this routing reduces manual curation effort and improves throughput and consistency.

\subsection{Agreement with field experts} \label{sec-experts}
\noindent We refined the classification logic based on criteria established in collaboration with archival experts (Section \ref{sec-dla-critical}). These requirements ensure that automated labels match practical archival needs for retrieval and downstream processing:

\begin{itemize}
    \item Photographs or drawings of significant size (at least postage-stamp sized) must be classified as \texttt{PHOTO} or \texttt{DRAW}, enabling targeted extraction of graphical content.
    \item Tables and forms must be detected even when borders are faint or missing; these are classified as \texttt{LINE\_T}, \texttt{LINE\_P}, or \texttt{LINE\_HW}.
    \item Only pages containing clean, uniform text should be classified as \texttt{TEXT\_P}, \texttt{TEXT\_T}, or \texttt{TEXT\_HW}. Such pages should be free of significant annotations or non-text elements; minor peripheral notes (\acs{eg} handwritten page numbers) may be ignored.
    \item Pages with mixed text styles (\texttt{TEXT}) may include minor graphical elements (\acs{eg} newspaper logos) that do not require separate extraction.
    \item Graphical elements embedded in tabular layouts or accompanied by table-like legends must be classified as \texttt{PHOTO\_L} or \texttt{DRAW\_L}, signaling the need to extract both graphics and structured data.
\end{itemize}

Given offline, batch-oriented workflows, the modest additional inference cost of transformer-based models is acceptable. To balance accuracy and model complexity, we selected \textbf{RegNetY 16GF (224)} for deployment. As an alternative for higher-memory \ac{gpu} systems, \textbf{vit-large-patch16-384} can be used.

\section{Similarity of predictions in different models} \label{sec-similarity}

In Figure \ref{fig1:clip-acc-categ}, both smaller and larger \ac{clip} models, after only seven epochs of fine-tuning with standard preprocessing, achieve accuracy comparable to the image-only models discussed in Sections \labelcref{sec-effnet,sec-dit,sec-vit,sec-regnet}. In other words, with targeted fine-tuning, \ac{clip} (especially ViT-B/16) can achieve high accuracy on annotated page images (Figure \ref{fig:clip-best5}).

\begin{figure}[h]
\centering
\includegraphics[width=145mm]{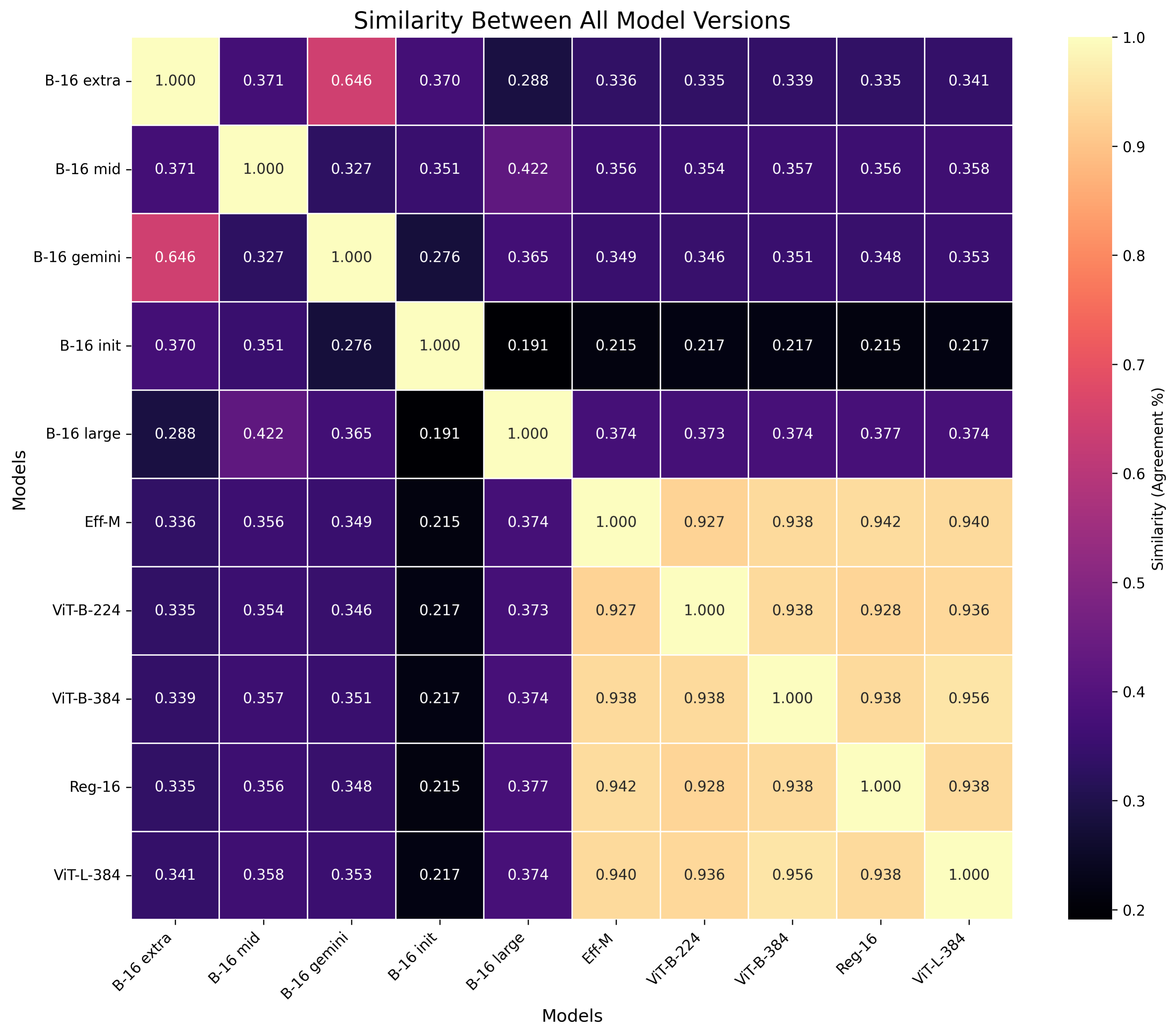}
\caption{Similarity of predictions for \ac{clip} (first five) and image-based (last five) models on the unlabeled collection (649{,}508 pages).}
\label{fig:similarity}
\end{figure}

However, when we applied the best image-only and hybrid models to unlabeled pages, we observed a clear difference in prediction agreement. As shown in Figure \ref{fig:similarity}, \ac{clip} models diverge markedly from image-only models and even from other \ac{clip} variants. Image-only models share over 90\% agreement in their predictions, whereas hybrid models show under 65\% similarity with any other model.

The data providers confirmed that this lack of agreement makes hybrid predictions harder to trust in practice. Consequently, image-only models were preferred for archival management because they assign consistent labels to most pages. In contrast, the hybrid models produced erratic predictions and failed to identify a common category for over 80\% of the unlabeled pages.

The low agreement on unlabeled pages contrasts with the high scores obtained by \ac{clip} on annotated samples. One hypothesis is that hybrid models need more fine-grained category descriptions; for example, \texttt{DRAW} currently merges maps, illustrations, and sketches and could benefit from more specific descriptions within the same label. Another possible explanation is overfitting to previously seen pages, which would inflate test-set estimates.

Overall, hybrid models require substantially more dataset preparation to compete with image-only approaches in practice. We therefore released fine-tuned \ac{clip} variants mainly for illustrative purposes; they were generally less useful for archival management than the image-only architectures.

\subsection{Analysis of Common Mistakes} \label{sec-anal-mistakes}

Across all models, errors are concentrated in a small number of confusions rather than being uniformly distributed. The dominant error source is the distinction between generic \texttt{TEXT} and typeset/transcribed text (\texttt{TEXT\_T}). A secondary confusion involves separating drawings (\texttt{DRAW}) from drawings embedded in tabular layouts (\texttt{DRAW\_L}) and other line-dominated categories.

The smallest \ac{cnn} baseline (EfficientNet-V2-S) struggles most with the \texttt{TEXT} ambiguity: it correctly classifies 85\% of \texttt{TEXT} samples and misclassifies 13\% as \texttt{TEXT\_T}. Scaling up reduces this confusion; EfficientNet-V2-L improves \texttt{TEXT} accuracy to 93\% and reduces the error to 5.6\%.

RegNetY-16GF performs best among the efficient models, reaching 94\% accuracy on \texttt{TEXT} with only 5.1\% confusion. It also reaches 99\% accuracy on \texttt{DRAW} (EfficientNet-V2-S reaches 97\%), suggesting that \acp{cnn} are sharper at edge- and structure-driven distinctions such as \texttt{DRAW} vs. \texttt{DRAW\_L}.

Among Transformers, DiT-Large (pretrained for documents) still struggles more with \texttt{TEXT} vs. \texttt{TEXT\_T} than the best \acp{cnn}, achieving 91\% accuracy on \texttt{TEXT}. More generally, Transformer variants appear slightly ``fuzzier'' on purely visual structural distinctions: DiT-Base-RVL reaches 95\% on \texttt{DRAW} and confuses 3.4\% with other line/drawing classes; DiT-Large-RVL similarly reaches 95\%.

\paragraph{CNNs (RegNetY / EfficientNet):}
\begin{itemize}
    \item \textbf{Strength and efficiency:} Strong edge/structure detection (high \texttt{DRAW} accuracy). RegNetY-16GF is a practical ``sweet spot,'' matching or beating much larger models.
    \item \textbf{Scaling behavior:} Increasing input resolution from 300 to 384 pixels per side (S$\rightarrow$M) substantially improves fine-grained text distinctions (85\%$\rightarrow$92\% accuracy on \texttt{TEXT}).
\end{itemize}

\paragraph{Transformers (DiT / ViT):}
\begin{itemize}
    \item \textbf{Strength:} Strong, consistent performance, but high parameter counts (300M+) are typically needed to compete with medium-sized \acp{cnn} (approximately 80M parameters).
    \item \textbf{Weakness:} Slightly weaker separation of visually similar structural categories (\acs{eg} \texttt{DRAW} vs. \texttt{DRAW\_L}) and less efficient resolution of the \texttt{TEXT} ambiguity.
\end{itemize}

Overall, if deployment efficiency is a priority, RegNetY-16GF is the strongest choice: it offers the highest accuracy (99.16\%) with a fraction of the parameters of large Transformer models and performs best on the hardest distinction in our dataset (\texttt{TEXT} vs. \texttt{TEXT\_T}).

\section{Labeled collections from Prague and Brno}

\begin{figure}[!h]
    \centering  
    \caption{Stacked annual counts for Prague and Brno data from \acs{arup} and \acs{arub}: Categorized page-scan counts over time in the archives of scanned documents (best guess of 5 fine-tuned models' averaged Top-3 predictions)}
    \label{fig:timeline_labeled}
    \includegraphics[width=135mm]{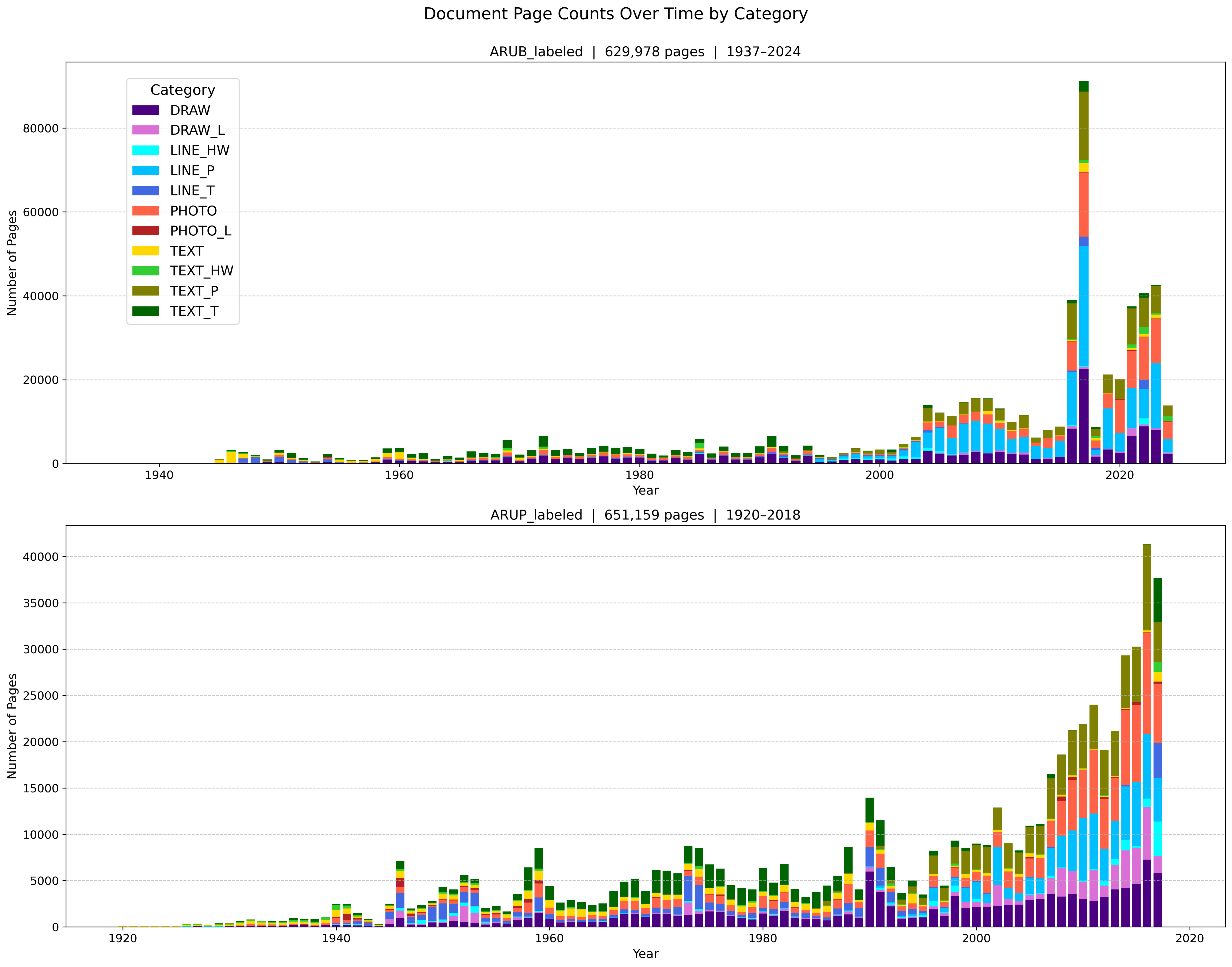}
\end{figure}

When we defined the five best image-only models, inference runs were performed for the whole Brno and Prague collections of page scans. After Top-3 predictions from all five models (\texttt{EffN2-M}, \texttt{ViT-B224}, \texttt{ViT-B384}, \texttt{RegNY-16GF}, \texttt{ViT-L384}) were obtained, an average of class scores from all models was computed for each category to derive Top-3 labels per input image file. Thus, Table \ref{tab:arup-arub-labeled-pages-by-category} and Figure \ref{fig:timeline_labeled} show statistics for the Top-1 guess of the averaged predictions.

\begin{table}[h!]
\centering
\small
\caption{Labeled page counts by category in the Prague and Brno archives (whole collection's best guess of 5 fine-tuned models' averaged Top-3 predictions). ``Ambiguous'' denotes pages with a Top-1 averaged class score $<0.9$.}
\label{tab:arup-arub-labeled-pages-by-category}
\begin{tabular}{@{}l|rr|rr@{}}
\toprule
\textbf{Category} & \textbf{Brno \acs{arub}} & \textbf{Ambiguous} & \textbf{Prague \acs{arup}} & \textbf{Ambiguous} \\
\midrule
DRAW     & 138{,}643 & 15{,}147 (10.9\%) & 128{,}364 & 15{,}504 (12.1\%) \\
DRAW\_L   &   7{,}782 &  3{,}978 (51.1\%) &  50{,}067 &  9{,}032 (18.0\%) \\
LINE\_HW  &   7{,}696 &  4{,}347 (56.5\%) &  15{,}285 &  5{,}375 (35.2\%) \\
LINE\_P   & 156{,}836 & 25{,}055 (16.0\%) &  73{,}457 &  5{,}828  (7.9\%) \\
LINE\_T   &  19{,}220 &  8{,}688 (45.2\%) &  44{,}167 &  8{,}987 (20.3\%) \\
PHOTO    & 112{,}704 &  8{,}913  (7.9\%) & 119{,}767 &  4{,}788  (4.0\%) \\
PHOTO\_L  &   1{,}163 &    500 (43.0\%) &   7{,}239 &  1{,}790 (24.7\%) \\
TEXT     &  24{,}249 &  8{,}091 (33.4\%) &  37{,}254 & 11{,}409 (30.6\%) \\
TEXT\_HW  &   9{,}287 &  4{,}024 (43.3\%) &   9{,}406 &  2{,}219 (23.6\%) \\
TEXT\_P   &  96{,}155 & 15{,}815 (16.4\%) &  87{,}938 &  9{,}455 (10.8\%) \\
TEXT\_T   &  56{,}243 & 13{,}841 (24.6\%) &  78{,}215 & 16{,}819 (21.5\%) \\
\midrule
\textbf{Year range} & 1937--2024 & --- & 1920--2018 & --- \\
\textbf{Total pages} & 629{,}978 & 108{,}399 (17.2\%) & 651{,}159 & 91{,}206 (14.0\%) \\
\bottomrule
\end{tabular}
\end{table}

Table \ref{tab:arup-arub-labeled-pages-by-category} summarizes the resulting labeled page counts in both collections, where the volume is concentrated in a small number of high-level document types: printed tables, drawings, photos, and text categories (printed and typewritten). \acs{arub} is dominated by printed tables, whereas \acs{arup} contains substantially more tabulated drawings (schemes) and typewritten texts. 

The highest ambiguity rates appear in visually ``borderline'' sub-classes with weak cues or frequent overlap with neighboring categories: handwritten tables and tabulated drawings (schemes) exceed 50\% ambiguity in \acs{arub}, and typewritten tables, mixed texts, handwritten texts, and tabulated photos remain consistently ambiguous in both archives. Practically, this suggests that some portion of the ``uncertainty'' is likely due to true multi-label structure rather than pure model error.

Both archives in Figure \ref{fig:timeline_labeled} show relatively low and stable activity, followed by a broad increase from the 1990s onward and a sharp expansion since the 2000s, where printed tables, photos, and printed texts drive most of the growth.

\chapter*{Conclusion} \label{chap-conclus}
\addcontentsline{toc}{chapter}{Conclusion}

\noindent This thesis addresses the need for automated page-image classification in heterogeneous historical archives. A reliable page classifier enables content-specific downstream processing---including \ac{ocr}, table extraction, and graphic analysis---without applying every processing pipeline to every page.

After characterizing the challenges of our dataset---visual defects, skew, and mixed content types (Introduction and Chapter~\ref{chap-data-explore})---we develop a subset-selection algorithm based on randomized periodic indexing (Chapter~\ref{chap-data-form}). We then demonstrate the limitations of existing document layout analysis tools (Chapter~\ref{chap-img-class}) and evaluate a lightweight \ac{rfc} baseline, which achieved only $\approx 75\%$ accuracy (Section~\ref{sec-rfc}). This result is consistent with surveys reporting the limitations of hand-crafted features for document understanding \cite{liu2021document}. Following these low-compute experiments, we define and annotate eleven categories (Section~\ref{sec-img-classes}), reaching nearly 50{,}000 labeled pages in the final dataset version.

Fine-tuning modern \acp{cnn} (RegNetY \cite{radosavovic2020designing} and EfficientNetV2 \cite{tan2021efficientnetv2} families) markedly improve performance to $\approx 97.9$--$99.1\%$ accuracy. Fine-tuned \acf{vit} \cite{dosovitskiy2020image} and \acf{dit} \cite{li2022dit} models consistently match the \acp{cnn}, achieving $\approx 98.5$--$99.1\%$ Top-1 accuracy (Table~\ref{tab:model_comparison}). We also show that \ac{clip} models \cite{radford2021learning} can reach $\approx 99\%$ classification accuracy after fine-tuning, although their inter-model agreement remained very low.

Based on this empirical evaluation, \textbf{RegNetY-16GF (224px)} is selected for deployment due to its balance of accuracy, generalization, and computational cost in an offline archival setting. In addition, \textbf{ViT-Large (384px)} is selected as the best-performing but more resource-demanding alternative among \ac{vit} models. Other smaller \ac{vit} variant \textbf{ViT-B/16 (224px)} and \acs{cnn} model \textbf{EfficientNetV2-M (384px)} are released as examples of models that perform above the accuracy--model-size trendline.

The final system is implemented in PyTorch and HuggingFace Transformers (Chapter~\ref{chap-architect}). It provides modular configuration, a \ac{cli}, and cross-platform scripts to support integration into existing archival pipelines (Chapter~\ref{chap-results}).

The released tool set directly addresses the previously identified gap: an open-source, content-based page-classification tool suitable for standard desktop hardware, cross-platform deployment, and practical use by archivists.

\subsection*{Contributions}
\begin{itemize}
    \item Analyzed archival page-image characteristics and documented failure modes of existing \ac{dla} tools (\ac{eg} \href{https://github.com/deepdoctection/deepdoctection/blob/c0a92be1b46c38c91557f02c78e133563bd5c1fe/docs/about.md}{DeepDoctection}) on the historical-archives data.

    \item Designed, annotated, and published the dataset of labeled pages \cite{Lutsai_ATRIUM_dataset_2025}; and developed and released the accompanying software solution \cite{Lutsai_ATRIUM_s_page_classifier_2025} in accordance with the \href{https://www.go-fair.org/fair-principles/}{FAIR} principles.

    \item Released eight fine-tuned \ac{clip} model-weight variants in the \href{https://huggingface.co/ufal/clip-historical-page}{HuggingFace repository}, accompanied by category-description sets in a separate \href{https://github.com/ufal/atrium-page-classification/tree/a66c906dfe929ac50c7f68d612a45d69fde4b5da/category_descriptions}{GitHub folder}.

    \item Published three medium-sized fine-tuned \ac{vit} models, two variants of EfficientNetV2-M, and one RegNetY-16GF model---all exceeding $\approx 98.79\%$ accuracy---in a separate \href{https://huggingface.co/ufal/vit-historical-page}{HuggingFace repository}.

    \item Identified RegNetY-16GF as the most effective model for this task by rigorously fine-tuning and comparing deep \acp{cnn} and transformer architectures. All image-only model versions above the accuracy--model-size trendline were released in the same \href{https://huggingface.co/ufal/vit-historical-page}{repository}.

    \item Provided deployment guidance for large-scale, on-premises archival processing, including a configurable pipeline and an extensible category taxonomy \cite{Lutsai_ATRIUM_s_page_classifier_2025}.

    \item Showed that a compact \ac{cnn} (EfficientNetV2-M) can outperform its larger counterpart (EfficientNetV2-L), and that RegNetY-16GF outperforms the 12GF and 64GF variants, making it a practical and efficient model for deployment.
\end{itemize}

\subsection*{Future work}
\begin{itemize}
\item Explore architectural optimizations to further improve efficiency and performance.
\item Integrate the classifier into broader digital-archive management frameworks; although developed for archaeological archives, our data characteristics overlap with many archival and library collections.
\item Expand the category taxonomy to cover additional document types (\ac{eg} Table~\ref{tab:24-label}).
\item Investigate advanced fine-tuning using predicted and user-refined annotations to improve utility and adaptability.
\end{itemize}

By delivering near-perfect classification accuracy and a practical deployment framework, this thesis reduces manual effort in historical-document workflows and supports digital humanities research and mass digitization initiatives \cite{nikolaidou2022survey, liu2021document}.

\chapter*{\acs{llm} assisted copy editing} \label{chap-ai}
\addcontentsline{toc}{chapter}{\acs{llm} assisted copy editing}

\noindent This thesis was written with assistance from \acp{llm}. Early drafts (factual bullet points and rough paragraphs) were provided to Gemini~2.5 and GPT-4 to generate a more fluent narrative.

These \ac{ai}-assisted edits primarily affected wording, structure, and formatting. I reviewed and revised the resulting text to ensure that it accurately reflects our work and that no incorrect assumptions remained.

In particular, Chapter~\ref{chap-architect} was initially generated and then manually post-edited based on the GitHub project materials (\href{https://github.com/ufal/atrium-page-classification/blob/8574f466a6fac45b940ad3dea5bfdc23cdd8a2f7/README.md}{README}, \href{https://github.com/ufal/atrium-page-classification/blob/8b9eac972607a18a9d94b14dd54f1ed64d23ad94/config.txt}{configuration file}, Python code, and OS-specific \href{https://github.com/ufal/atrium-page-classification/tree/8b9eac972607a18a9d94b14dd54f1ed64d23ad94/data_scripts}{data-processing scripts}). Because the code base and documentation already contained user-oriented system descriptions, the generated text served as a solid starting point for the system overview.

Chapter~\ref{chap-data-explore} was written manually from the beginning and only then stylized into a scientific-article format. Chapters~\labelcref{chap-img-class,chap-results} were generated from concise paragraph outlines and then manually post-edited to remove inaccuracies.

The Introduction was based on a Deep Research run using a section skeleton and short bullet points. When most chapters were complete and the technical details had been added to the \acp{llm}-generated drafts, the Conclusion was composed from narrative bullet points and the preceding thesis content. Some sources suggested by \acp{llm} were then manually verified and cited in the introduction or conclusion.

In later stages, the models were asked to produce a coherent overview of the thesis and the relevant materials; therefore, the requests focused on language and narrative composition rather than technical content.

Finally, during manual writing and post-editing, I used the free version of \href{https://www.grammarly.com/}{Grammarly} for grammar and spelling checks. In addition, \href{https://www.overleaf.com/learn/how-to/Writefull_integration}{Writefull} (integrated in Overleaf) was used to align the text with a scientific writing style. Similarly, Overleaf's \href{https://www.overleaf.com/about/ai-features}{AI Assistant} was used for light, medium, and heavy refinements of text (\acs{eg} removing repetitions, fixing sentence structures, aligning hyphen formats, shortening of paragraphs), and searching for several relevant works to include and cite in the thesis manuscript.

Although many edits were made during the initial drafting, additional refinements were applied throughout the thesis after incorporating the \ac{llm} outputs.

% We let bibliography items stick out of the right margin a little
\def\bibfont{\hfuzz=2pt}

\printbibliography[heading=bibintoc]

\listoffigures

\listoftables

\cleardoublepage
\addcontentsline{toc}{chapter}{List of Abbreviations}%
\printacronyms[name=List of Abbreviations]

\appendix
\chapter{Source data pages} \label{app-pages}

\begin{figure}[h]
    \centering
    \caption{Water damage resulting in blur and overlapping ink}
    \label{fig:defect_water}
    \begin{subfigure}{0.3\textwidth}
        \centering
        \includegraphics[width=\textwidth]{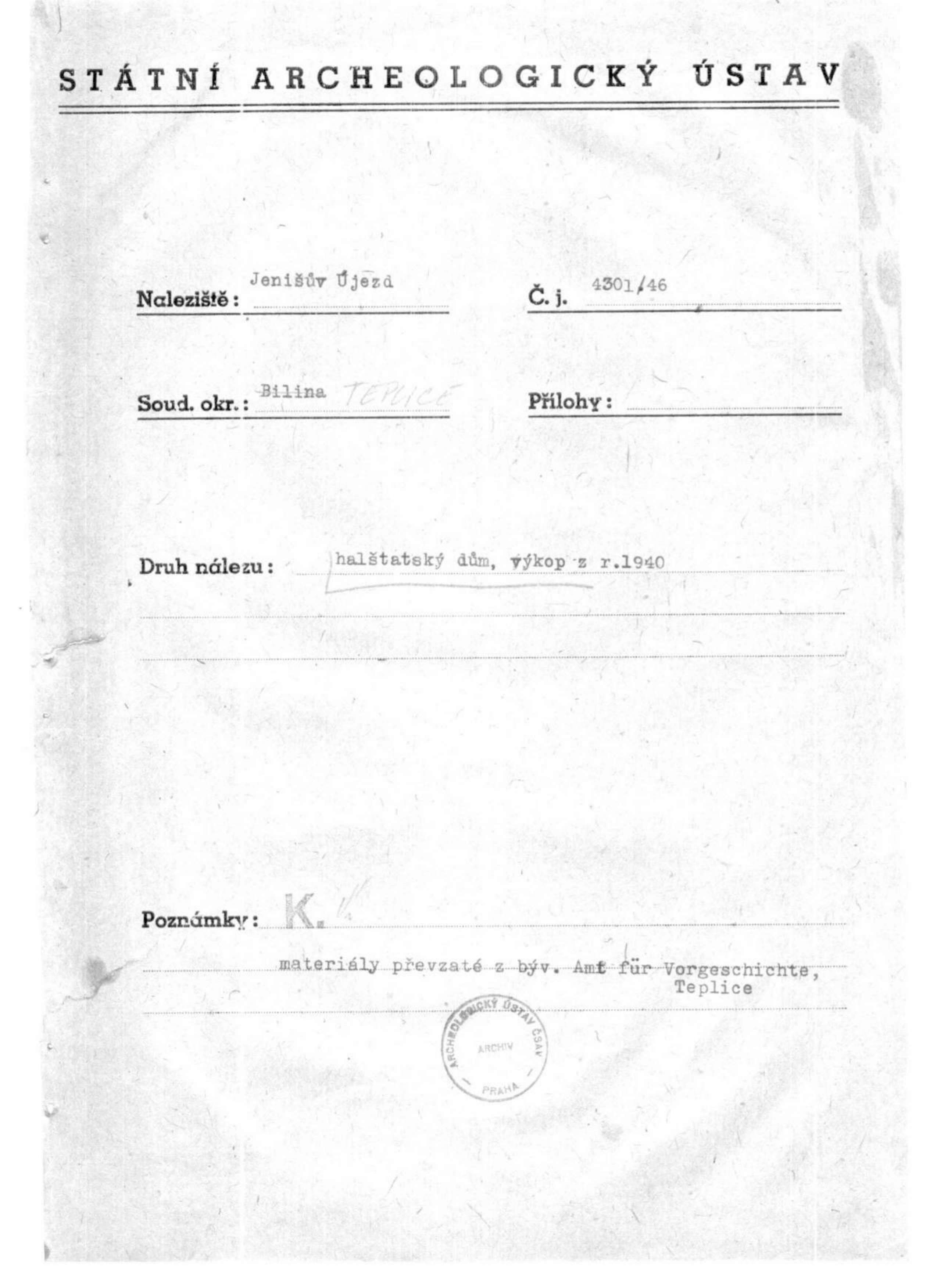}
        \caption{Mixed text with round stamp, water‐blurred}
        \label{fig:water1}
    \end{subfigure}
    % \hfill
    \begin{subfigure}{0.3\textwidth}
        \centering
        \includegraphics[width=\textwidth]{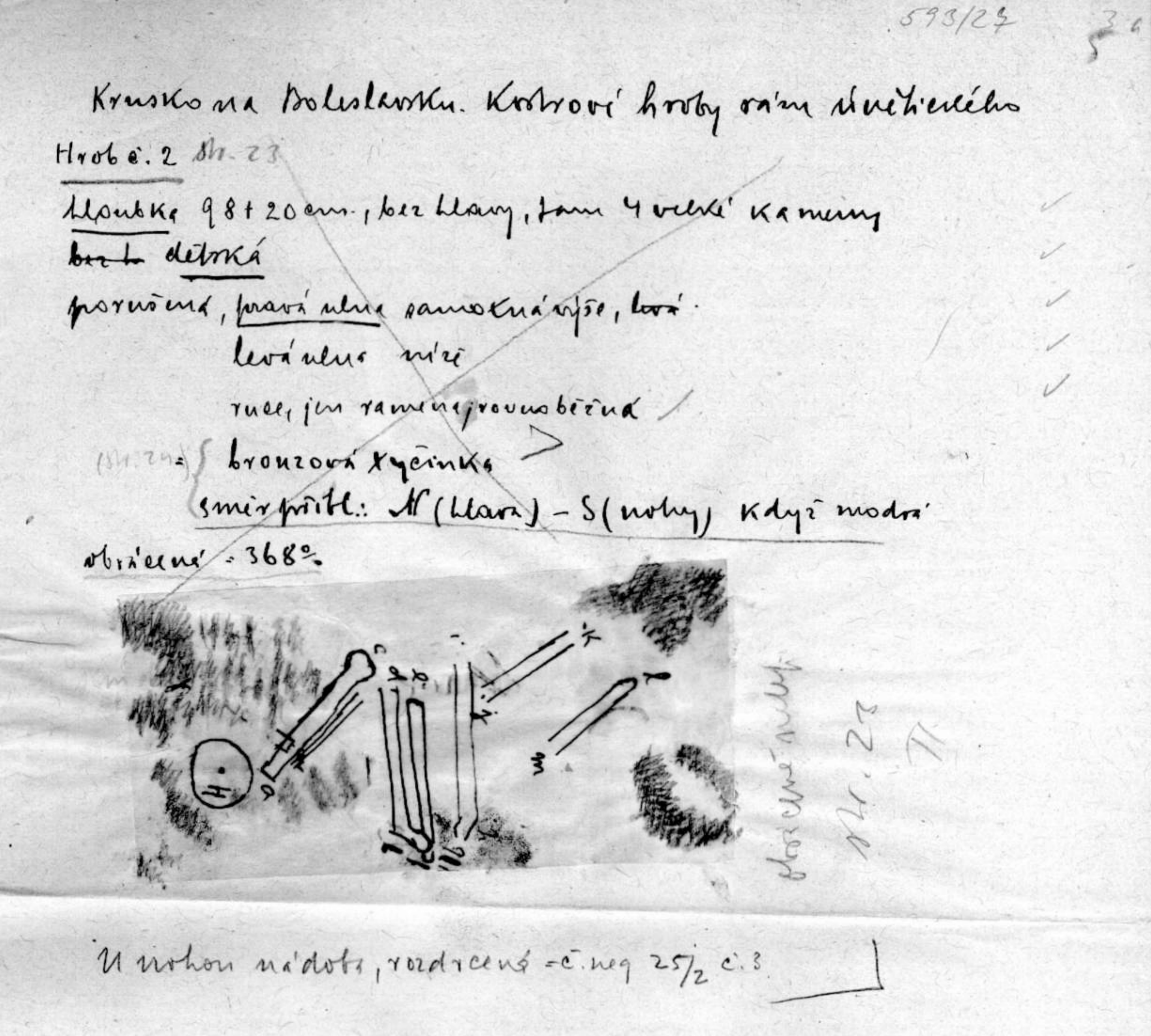}
        \caption{Handwritten text, water‐damaged}
        \label{fig:water2}
    \end{subfigure}
    % \hfill
    \begin{subfigure}{0.3\textwidth}
        \centering
        \includegraphics[width=\textwidth]{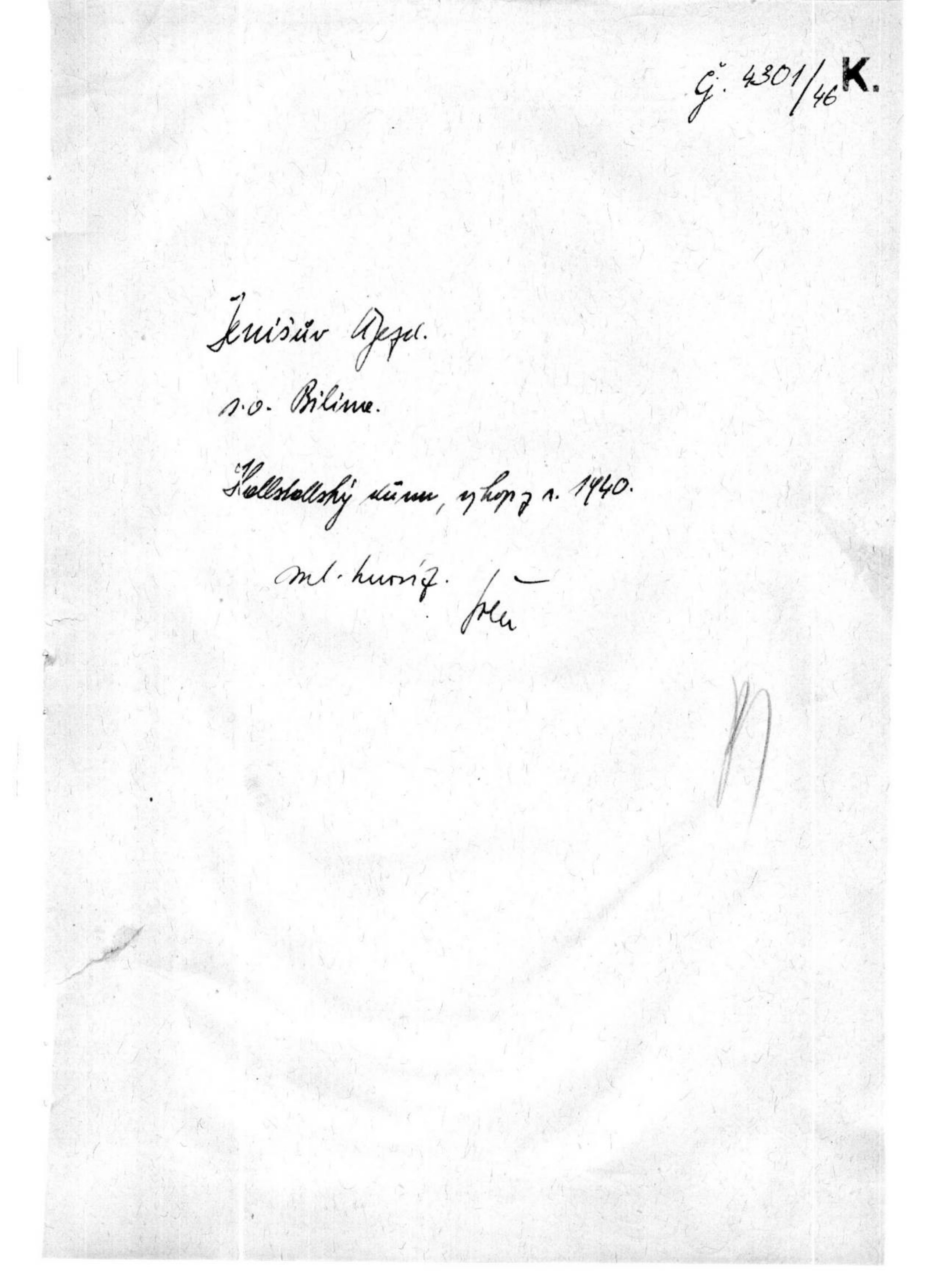}
        \caption{Typewritten text blurred by water stains}
        \label{fig:water3}
    \end{subfigure}
\end{figure}

\begin{figure}[h]
    \centering
    \caption{Corner-hole damage exposing paper fibers \& mixed content (yellowish paper texture)}
    \label{fig:defect_cornerhole}
    \begin{subfigure}{0.3\textwidth}
        \centering
        \includegraphics[width=\textwidth]{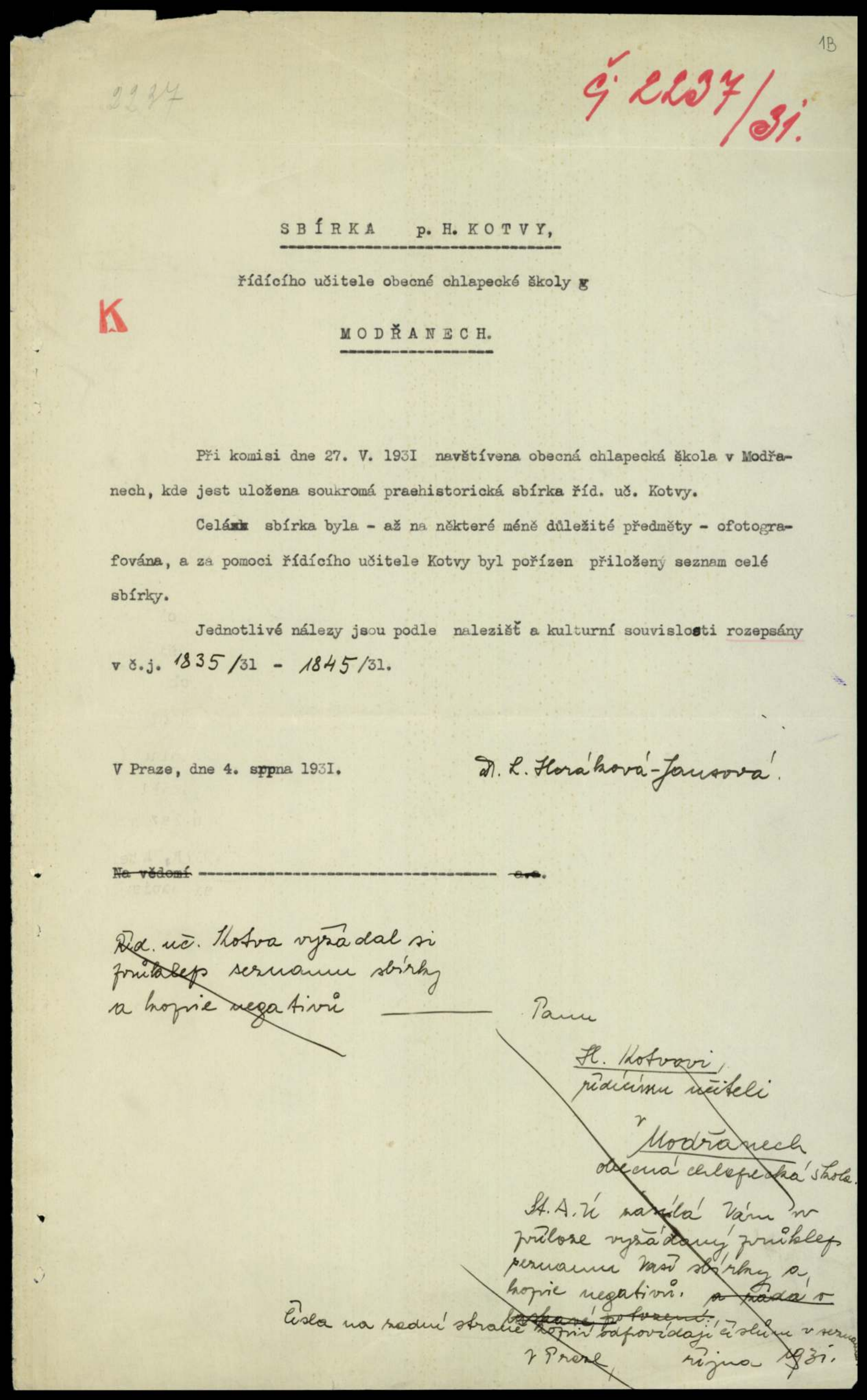}
        \caption{Major text correction with a corner-hole tear}
        \label{fig:corner1}
    \end{subfigure}
    % \hfill
    \begin{subfigure}{0.3\textwidth}
        \centering
        \includegraphics[width=\textwidth]{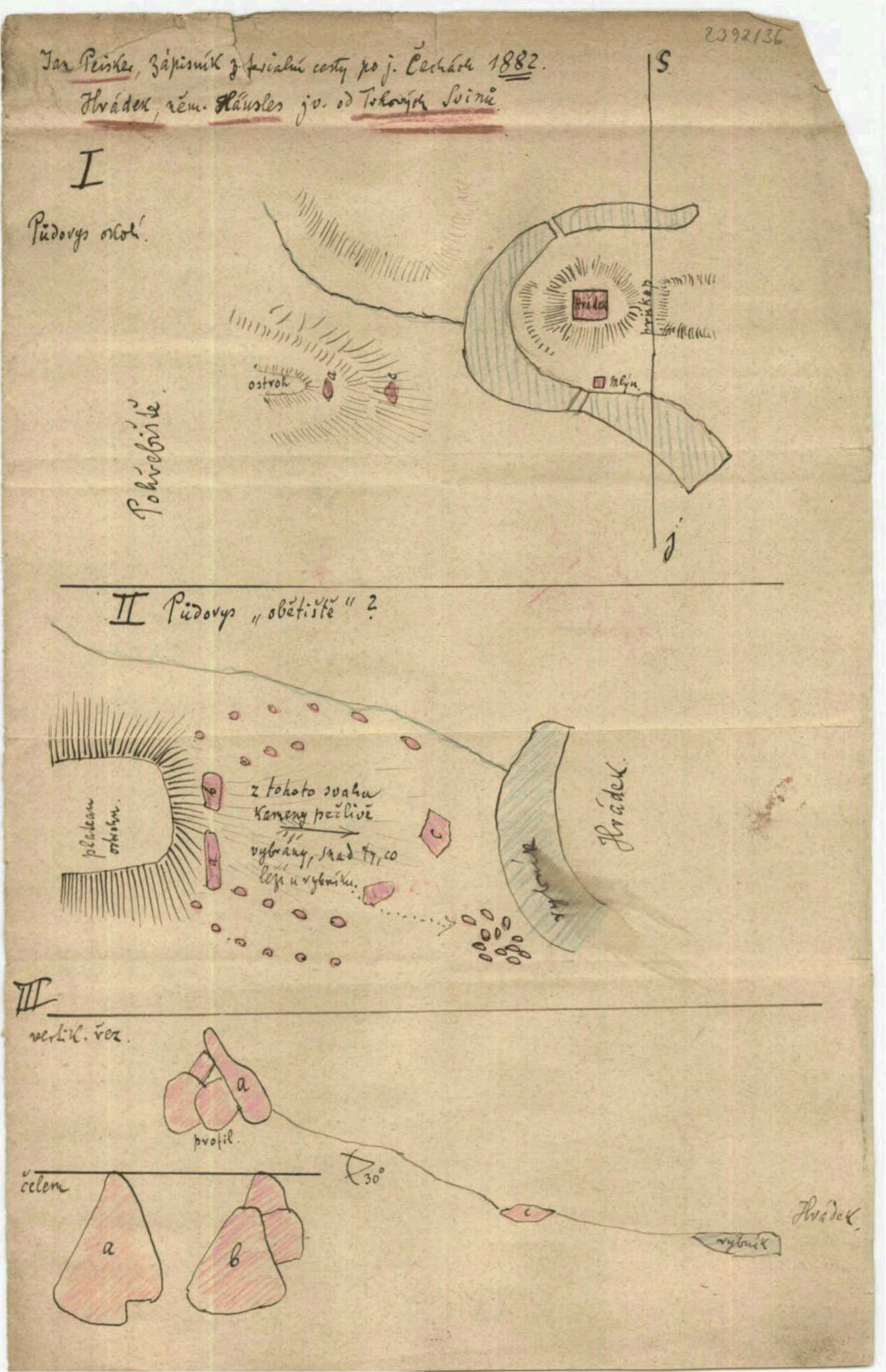}
        \caption{Handwritten drawing next to a corner-hole}
        \label{fig:corner2}
    \end{subfigure}
    % \hfill
    \begin{subfigure}{0.3\textwidth}
        \centering
        \includegraphics[width=\textwidth]{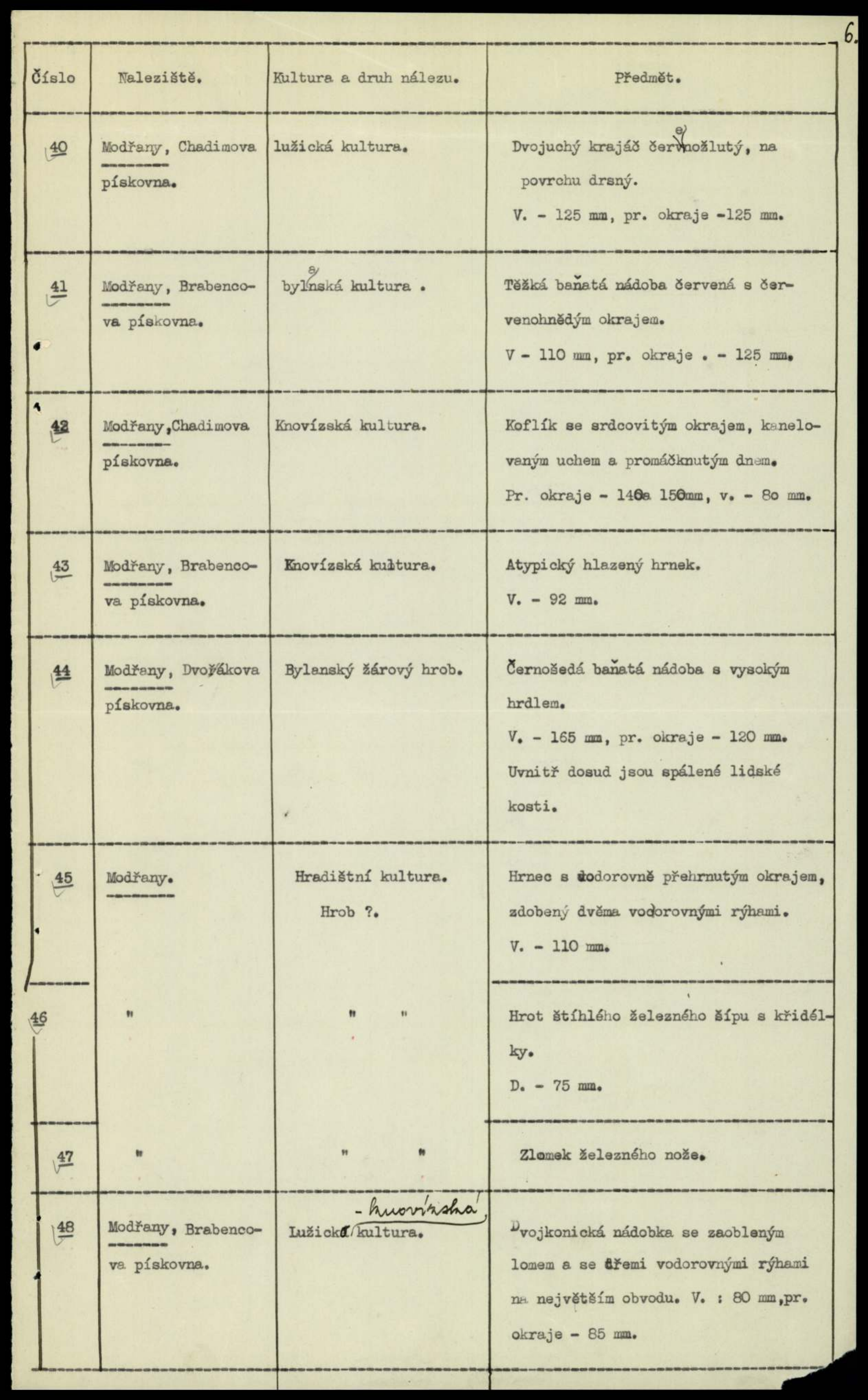}
        \caption{Typewritten table layout with a torn corner-hole}
        \label{fig:corner3}
    \end{subfigure}
\end{figure}

\begin{figure}[h]
    \centering
    \caption{Gray‐background artifacts and table/drawing elements}
    \label{fig:defect_gray1}
    \begin{subfigure}{0.45\textwidth}
        \centering
        \includegraphics[width=\textwidth]{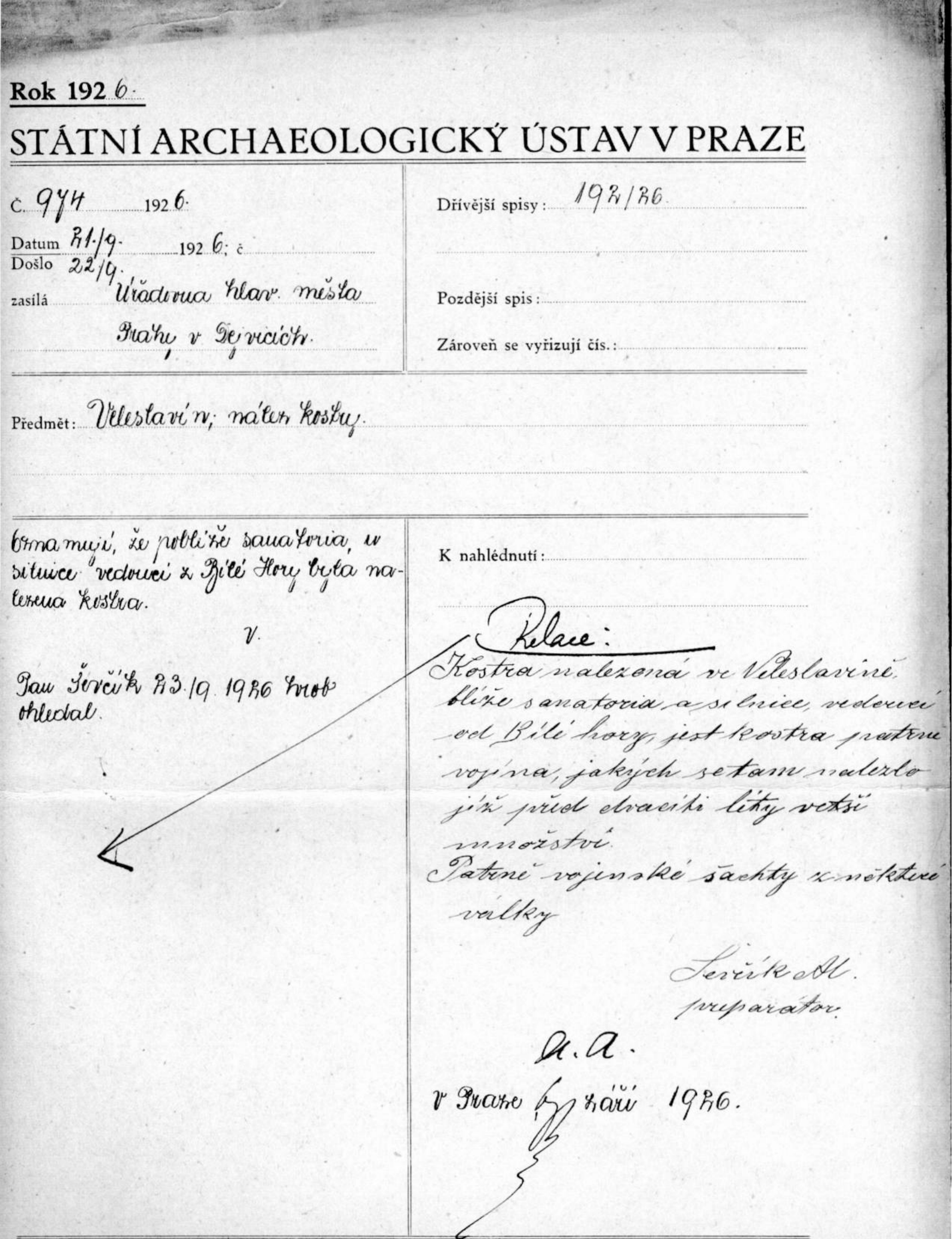}
        \caption{Skewed table \& arrow on gray paper}
        \label{fig:gray1a}
    \end{subfigure}
    \hfill
    \begin{subfigure}{0.45\textwidth}
        \centering
        \includegraphics[width=\textwidth]{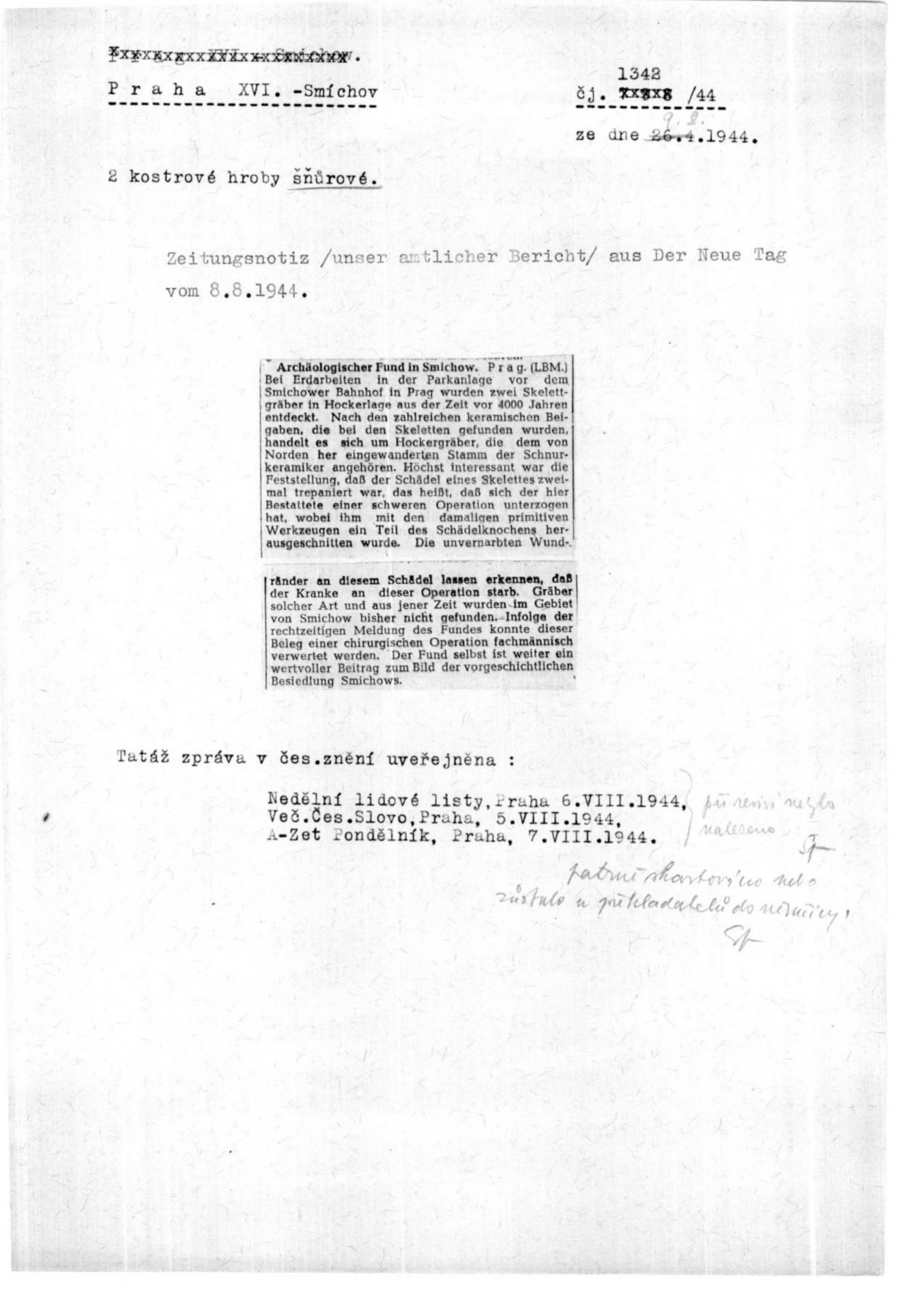}
        \caption{Mixed text on gray‐toned corners}
        \label{fig:gray1b}
    \end{subfigure}
    \vspace{1em}
    \begin{subfigure}{0.45\textwidth}
        \centering
        \includegraphics[width=\textwidth]{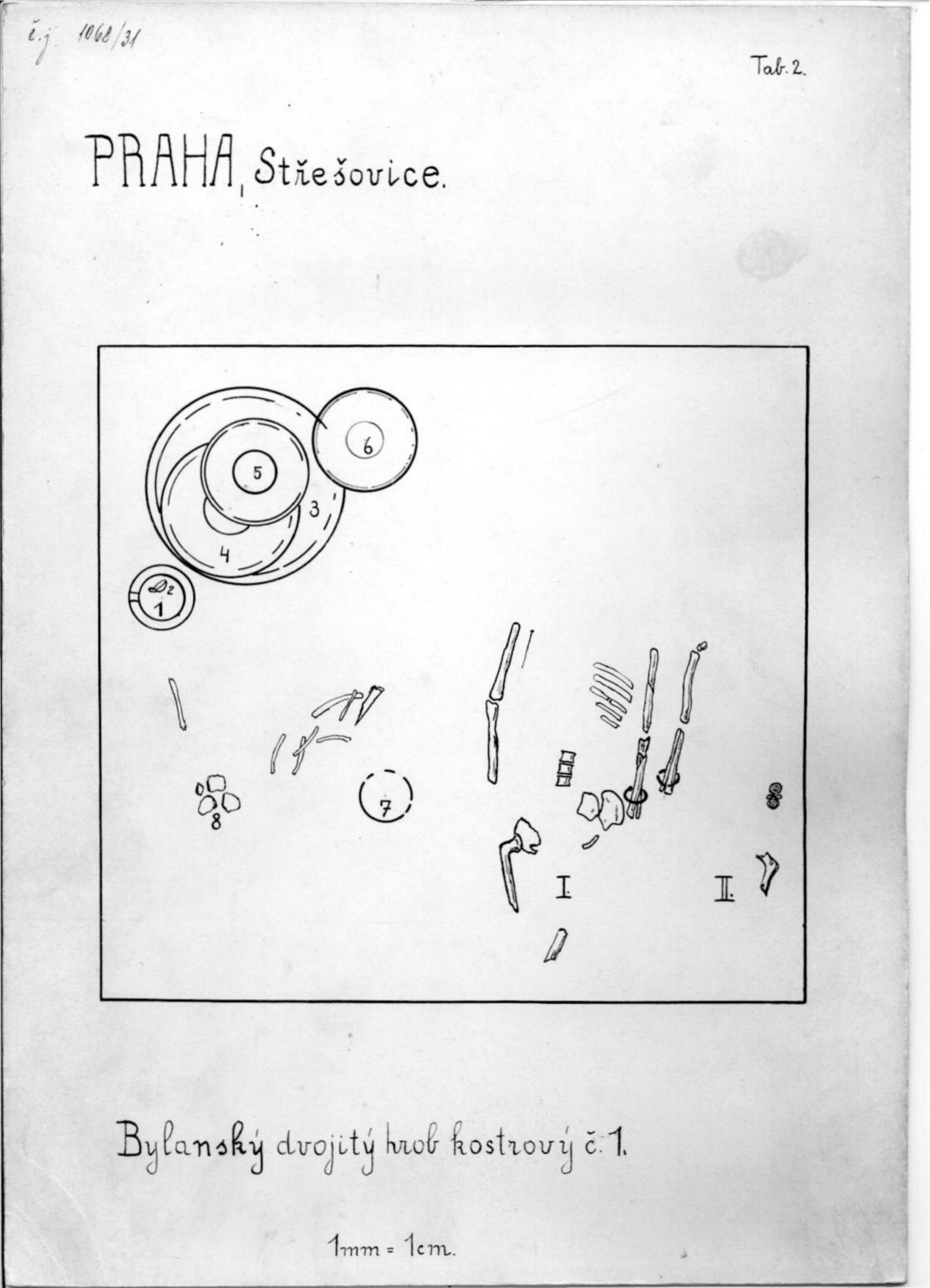}
        \caption{Grayish corner with handwritten drawing}
        \label{fig:gray2a}
    \end{subfigure}
    \hfill
    \begin{subfigure}{0.45\textwidth}
        \centering
        \includegraphics[width=\textwidth]{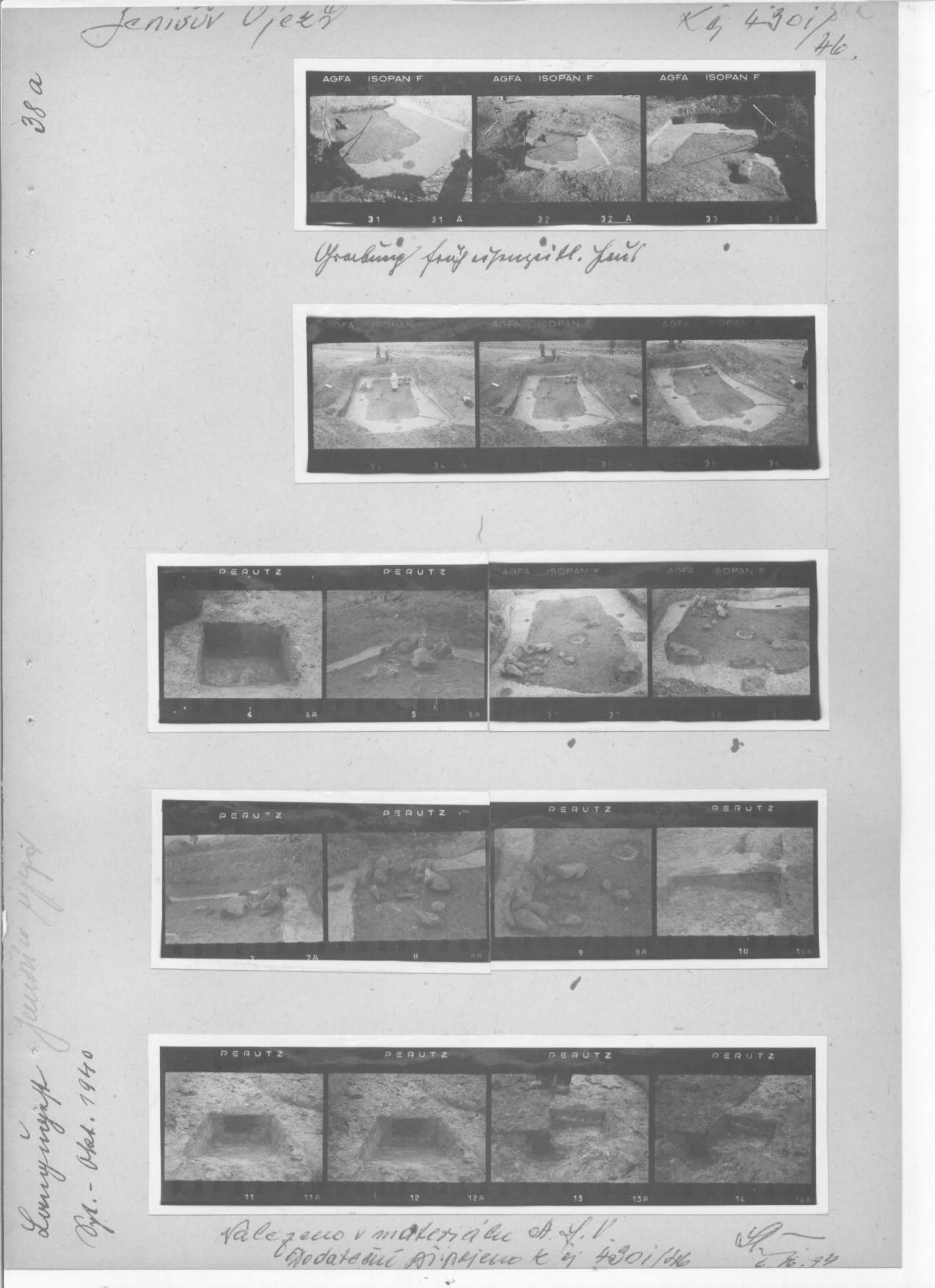}
        \caption{Gray paper with photograph edges}
        \label{fig:gray2b}
    \end{subfigure}
\end{figure}

\begin{figure}[h]
    \centering
    \caption{Stamp annotations and faded ink impressions}
    \label{fig:defect_stamp}
    \begin{subfigure}{0.3\textwidth}
        \centering
        \includegraphics[width=\textwidth]{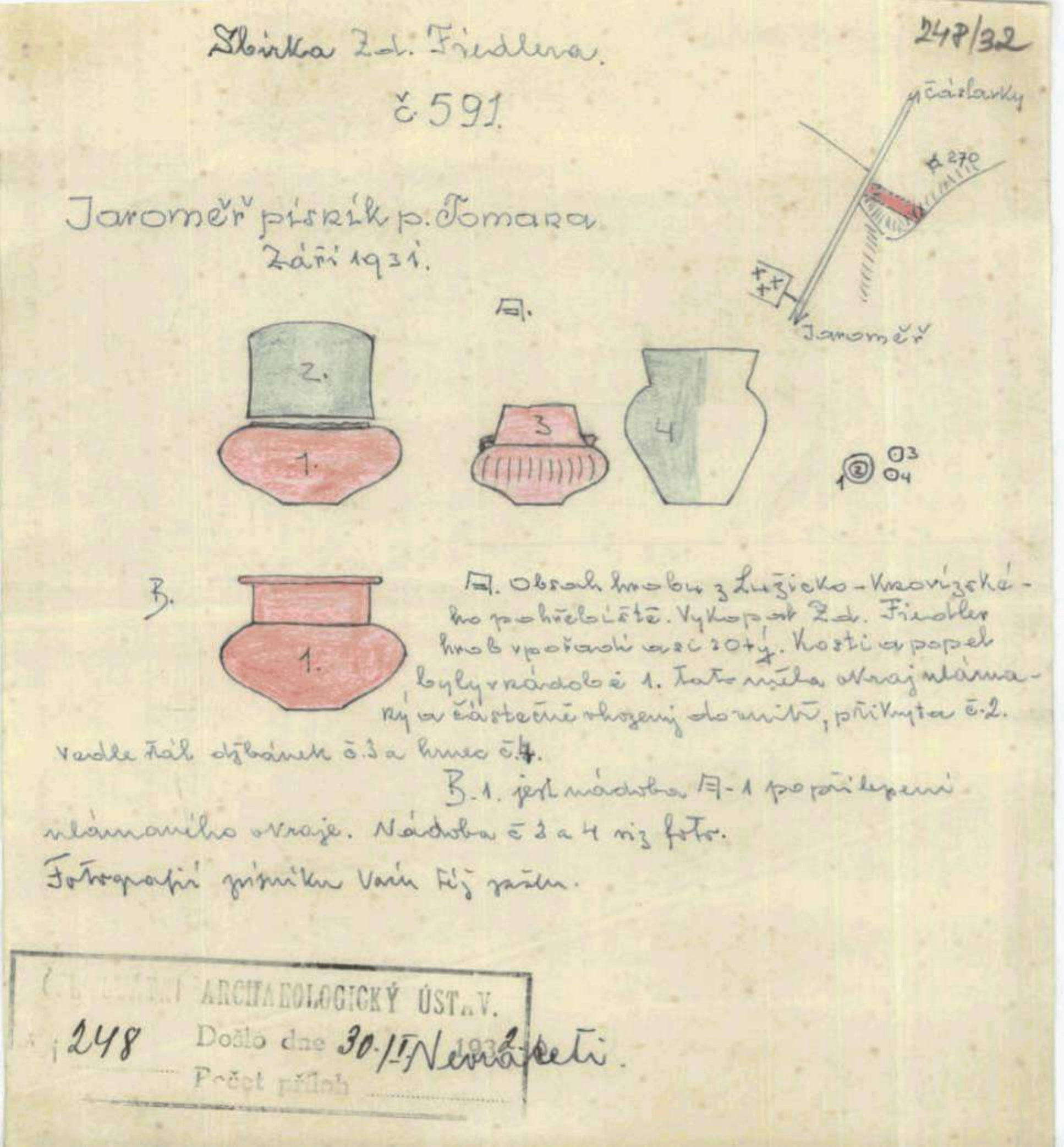}
        \caption{Hand‐written drawing with stamp overlap}
        \label{fig:stamp1}
    \end{subfigure}
    % \hfill
    \begin{subfigure}{0.3\textwidth}
        \centering
        \includegraphics[width=\textwidth]{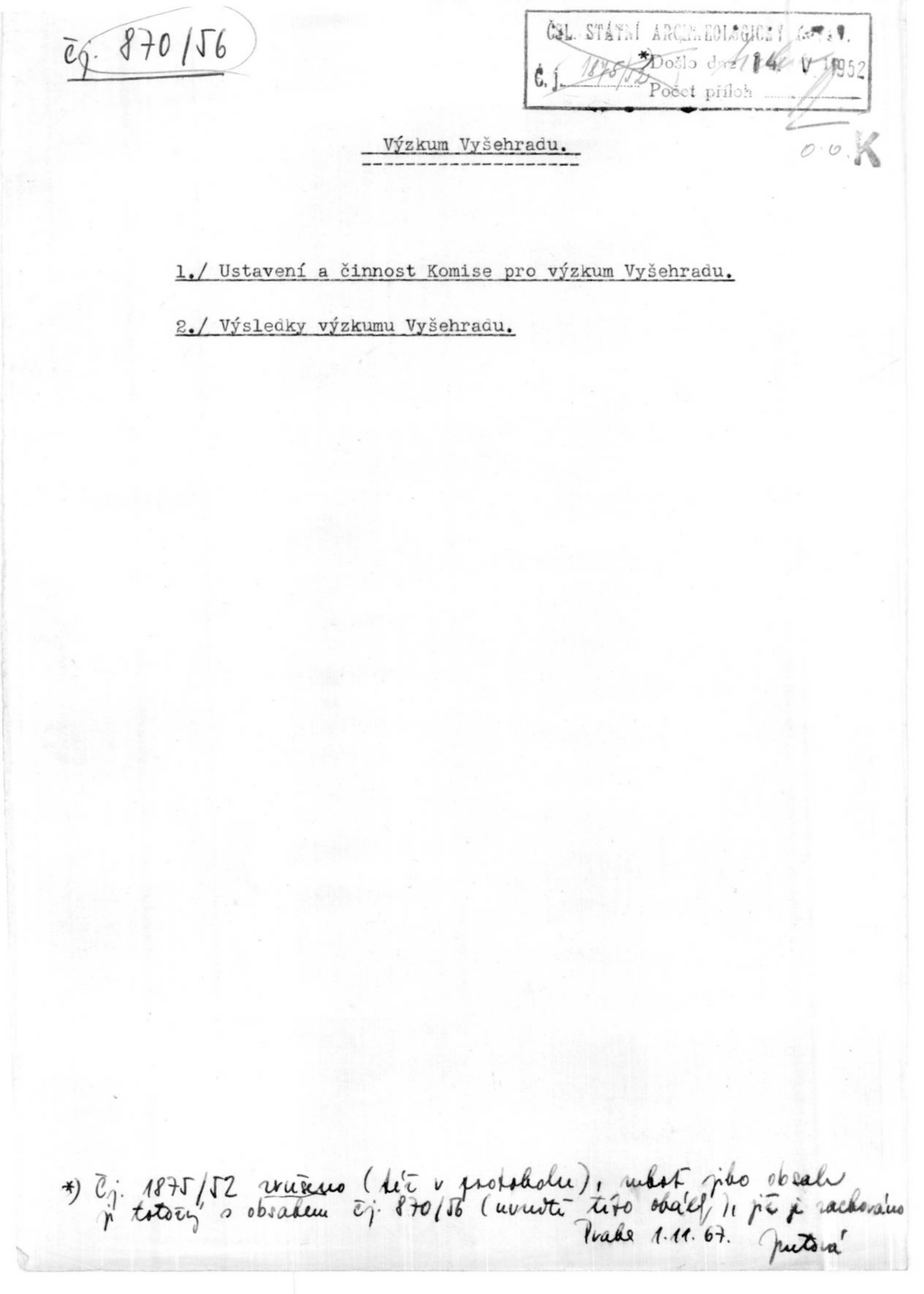}
        \caption{Mixed text, gray stamp at bottom}
        \label{fig:stamp2}
    \end{subfigure}
    % \hfill
    \begin{subfigure}{0.3\textwidth}
        \centering
        \includegraphics[width=\textwidth]{img/pages/stamp_text_hw_minor_mistake.pdf}
        \caption{Minor handwriting mistake with stamp}
        \label{fig:stamp3}
    \end{subfigure}
\end{figure}

\begin{figure}[h]
    \centering
    \caption{Bleed‐through artifacts on thin paper}
    \label{fig:defect_bleed}
    \begin{subfigure}{0.3\textwidth}
        \centering
        \includegraphics[width=\textwidth]{img/pages/bleed_draw_table.pdf}
        \caption{Drawn table showing text bleed}
        \label{fig:bleed1}
    \end{subfigure}
    % \hfill
    \begin{subfigure}{0.3\textwidth}
        \centering
        \includegraphics[width=\textwidth]{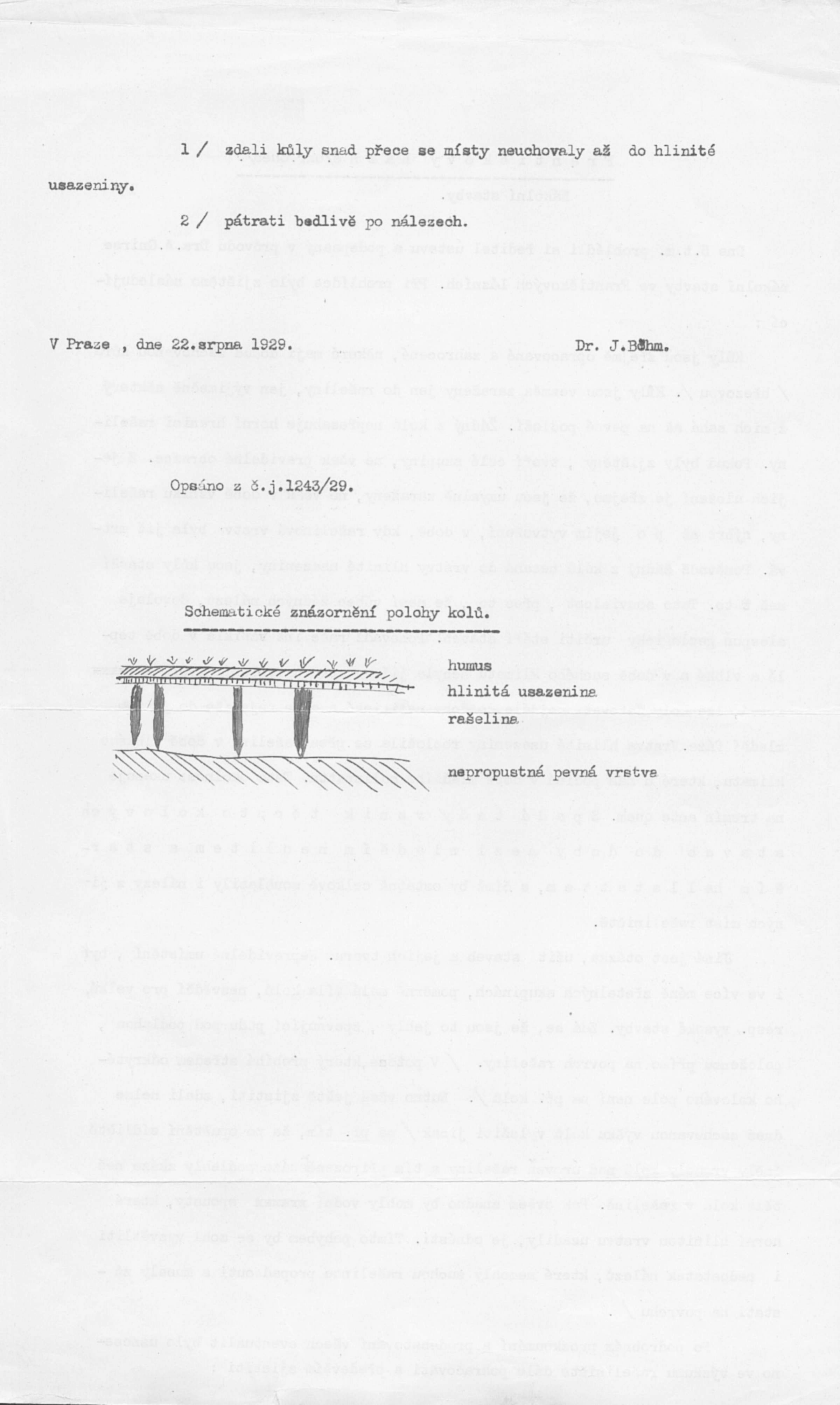}
        \caption{Handwritten text blurred by bleed}
        \label{fig:bleed2}
    \end{subfigure}
    % \hfill
    \begin{subfigure}{0.3\textwidth}
        \centering
        \includegraphics[width=\textwidth]{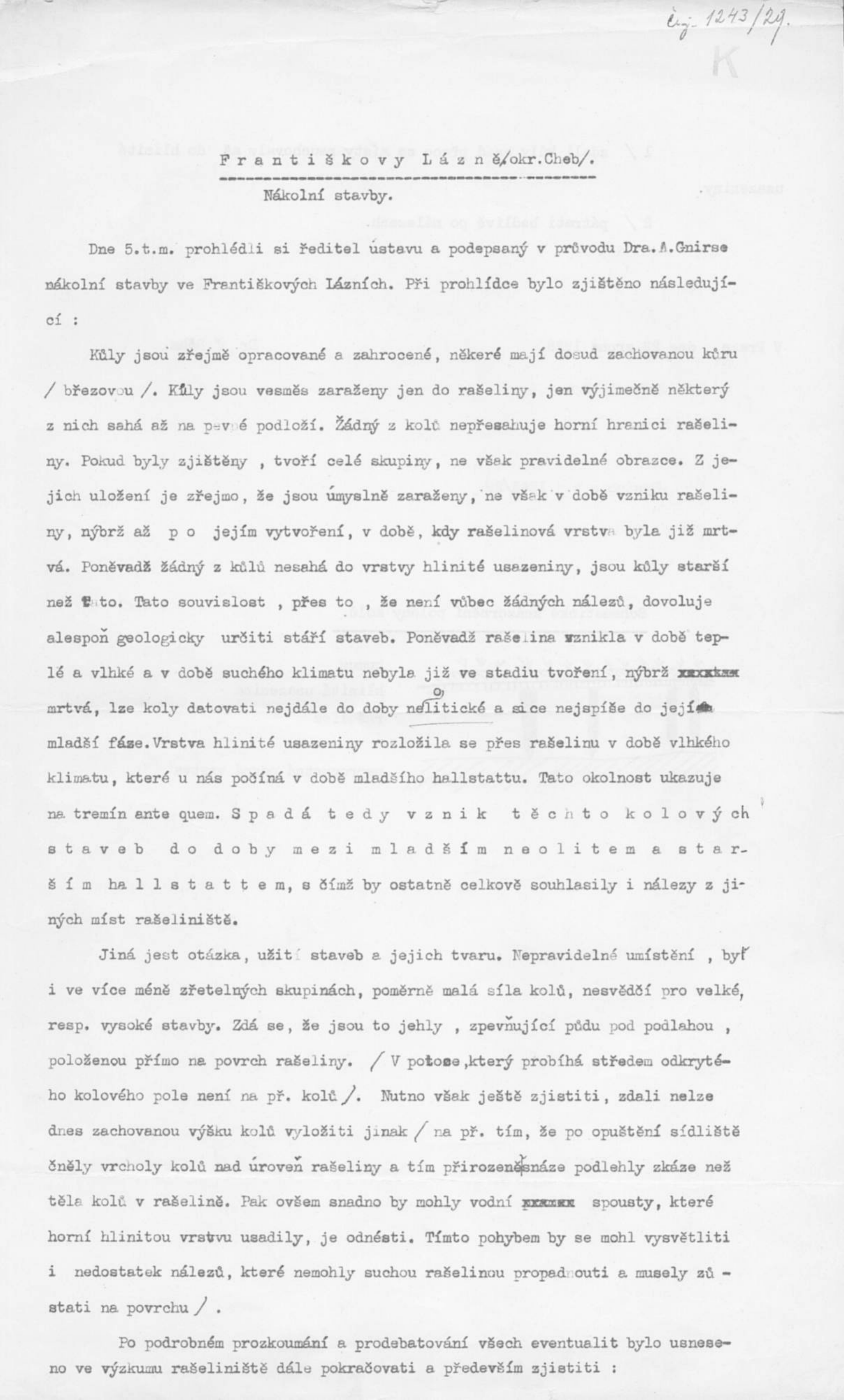}
        \caption{Typewritten text with minor bleed correction}
        \label{fig:bleed3}
    \end{subfigure}
\end{figure}

\begin{figure}[h]
    \centering
    \caption{Skewed table prints and alignment issues}
    \label{fig:defect_skew}
    \begin{subfigure}{0.45\textwidth}
        \centering
        \includegraphics[width=\textwidth]{img/pages/skew_table_hw_color_comment.pdf}
        \caption{Handwritten comment on skewed color table}
        \label{fig:skew1a}
    \end{subfigure}
    \hfill
    \begin{subfigure}{0.45\textwidth}
        \centering
        \includegraphics[width=\textwidth]{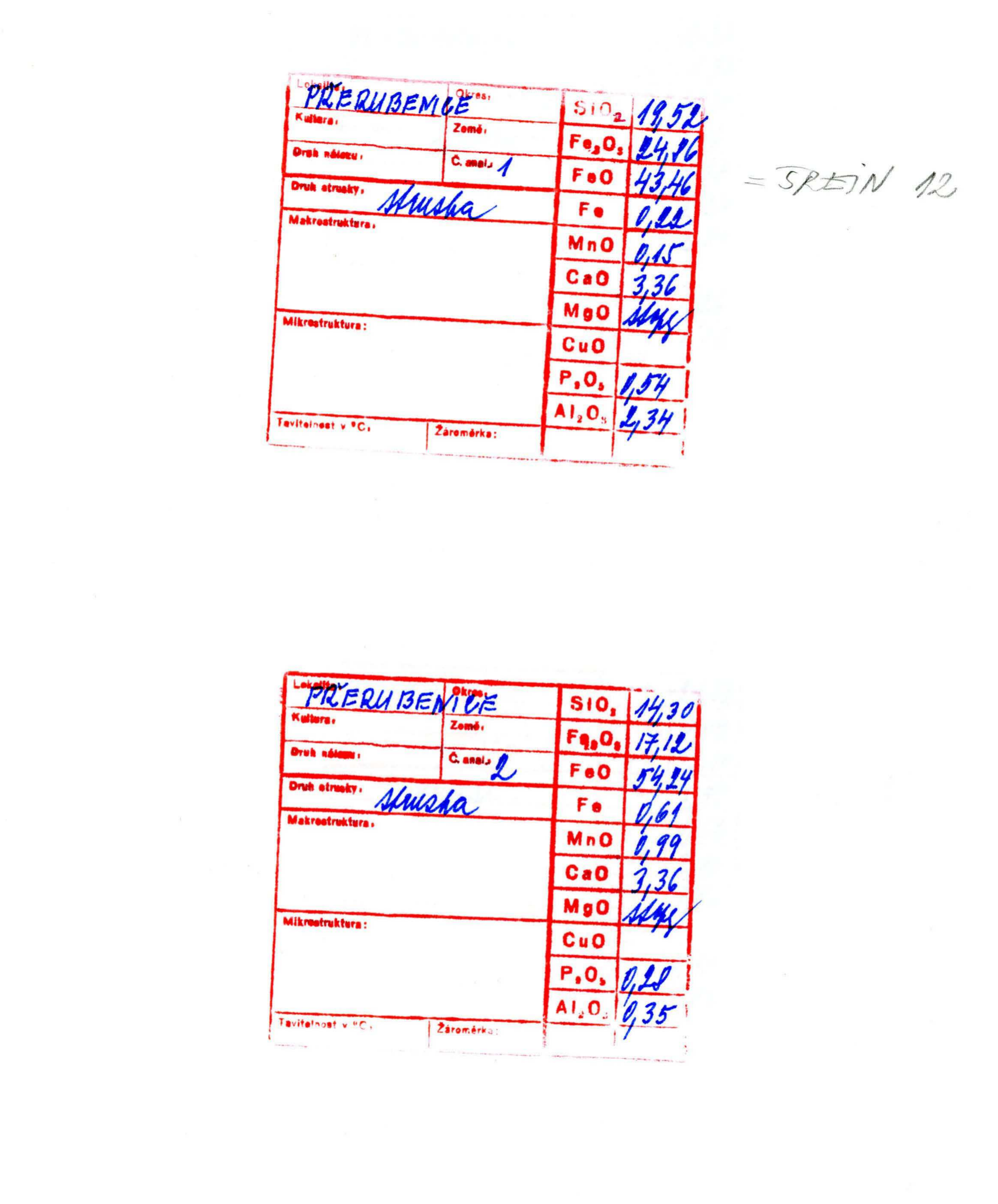}
        \caption{Color‐printed table, slight skew}
        \label{fig:skew1b}
    \end{subfigure}
    \vspace{1em}
    \begin{subfigure}{0.45\textwidth}
        \centering
        \includegraphics[width=\textwidth]{img/pages/table_t_skew.pdf}
        \caption{Typewritten table with visible skew}
        \label{fig:skew2a}
    \end{subfigure}
    \hfill
    \begin{subfigure}{0.45\textwidth}
        \centering
        \includegraphics[width=\textwidth]{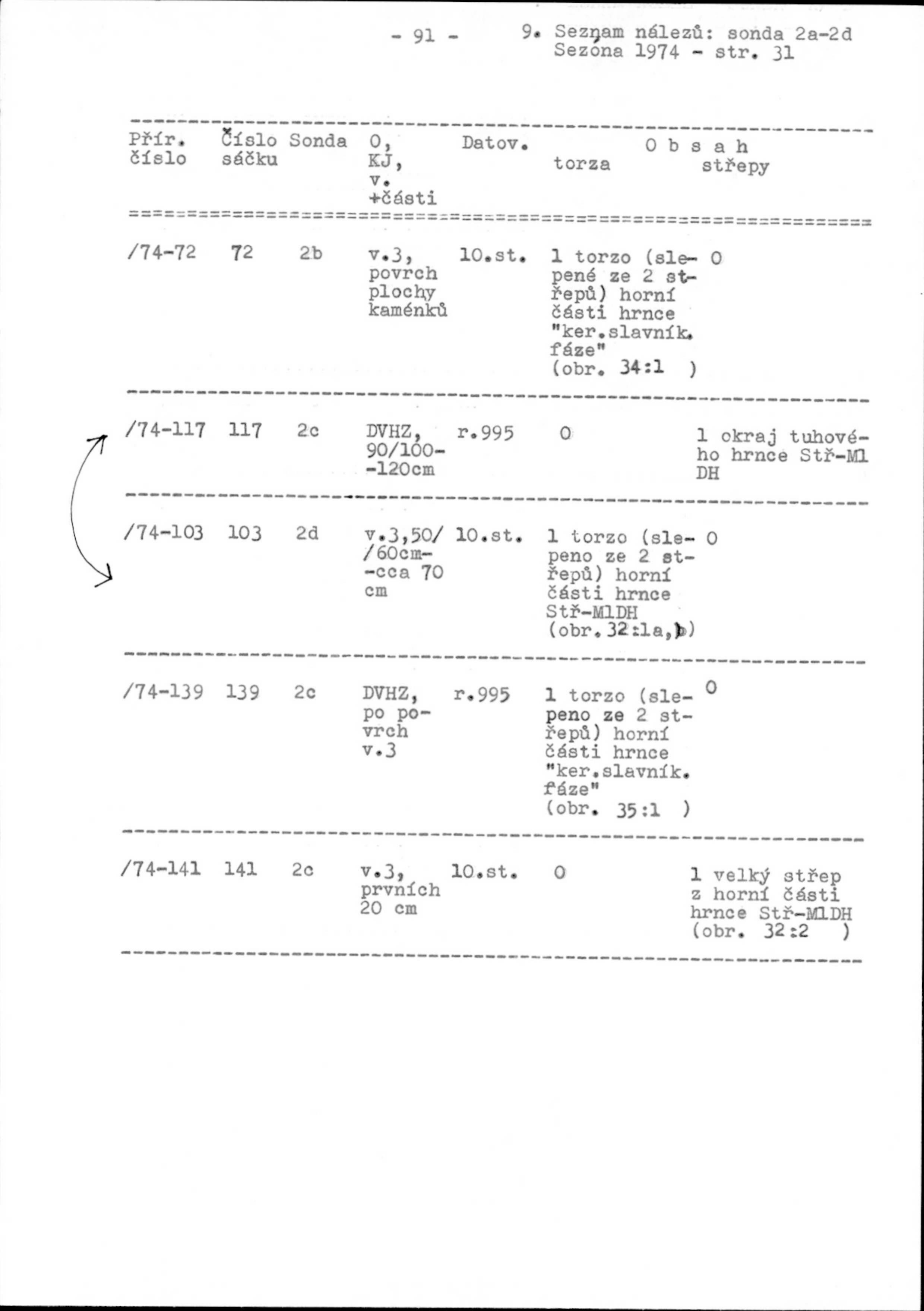}
        \caption{Arrow marking on a skewed table page}
        \label{fig:skew2b}
    \end{subfigure}
\end{figure}

\begin{figure}[h]
    \centering
    \caption{Edgehole damage and paper tearing}
    \label{fig:defect_edgeholes}
    \begin{subfigure}{0.45\textwidth}
        \centering
        \includegraphics[width=\textwidth]{img/pages/edgeholes_stains_table_hw.pdf}
        \caption{Table print with edgehole stains}
        \label{fig:edge1}
    \end{subfigure}
    \hfill
    \begin{subfigure}{0.45\textwidth}
        \centering
        \includegraphics[width=\textwidth]{img/pages/edgeholes_text_t.pdf}
        \caption{Typewritten text adjacent to edgeholes}
        \label{fig:edge2}
    \end{subfigure}
    \vspace{1em}
    \begin{subfigure}{0.45\textwidth}
        \centering
        \includegraphics[width=\textwidth]{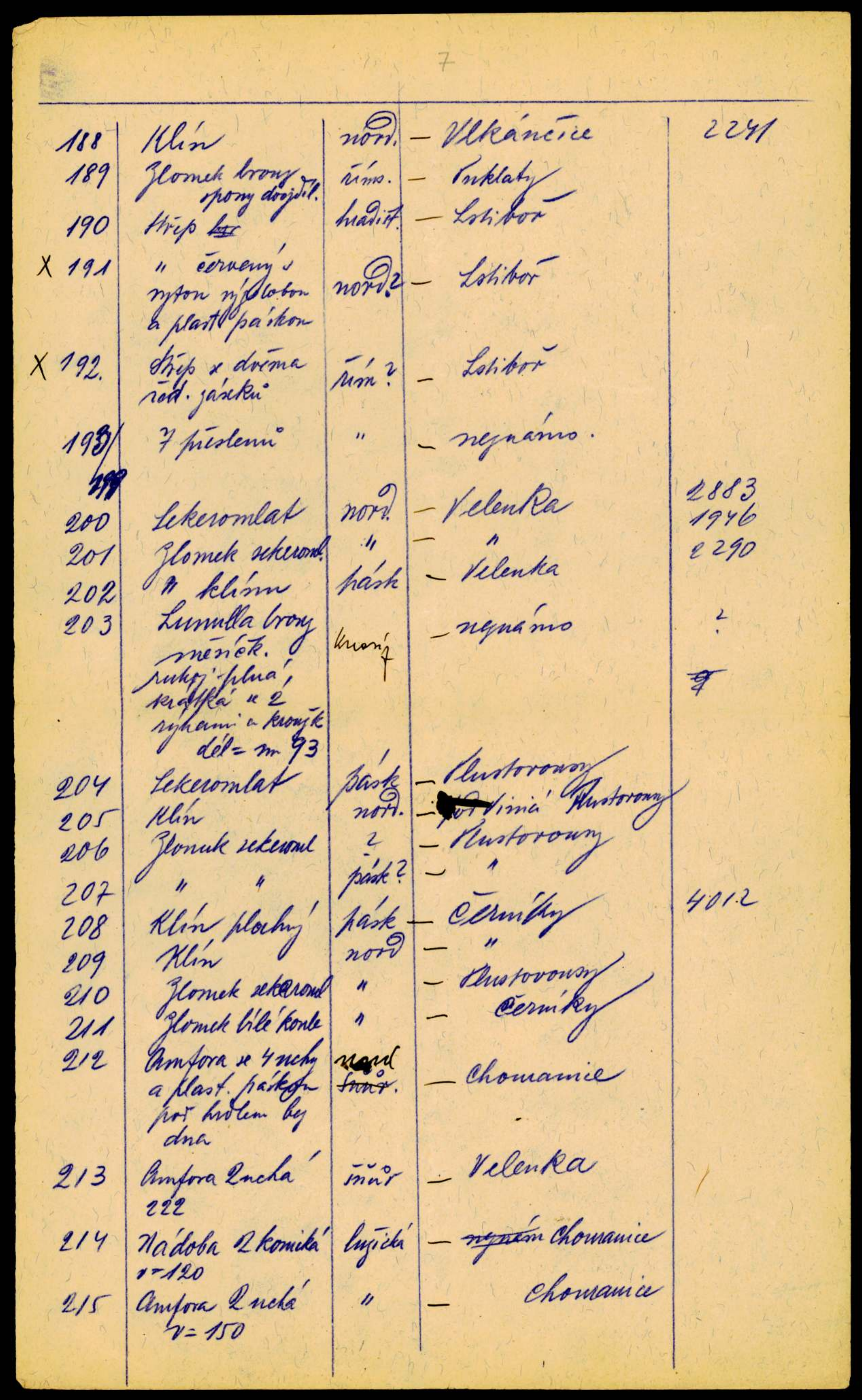}
        \caption{Hole through table on yellowed paper}
        \label{fig:edge3}
    \end{subfigure}
    \hfill
    \begin{subfigure}{0.45\textwidth}
        \centering
        \includegraphics[width=\textwidth]{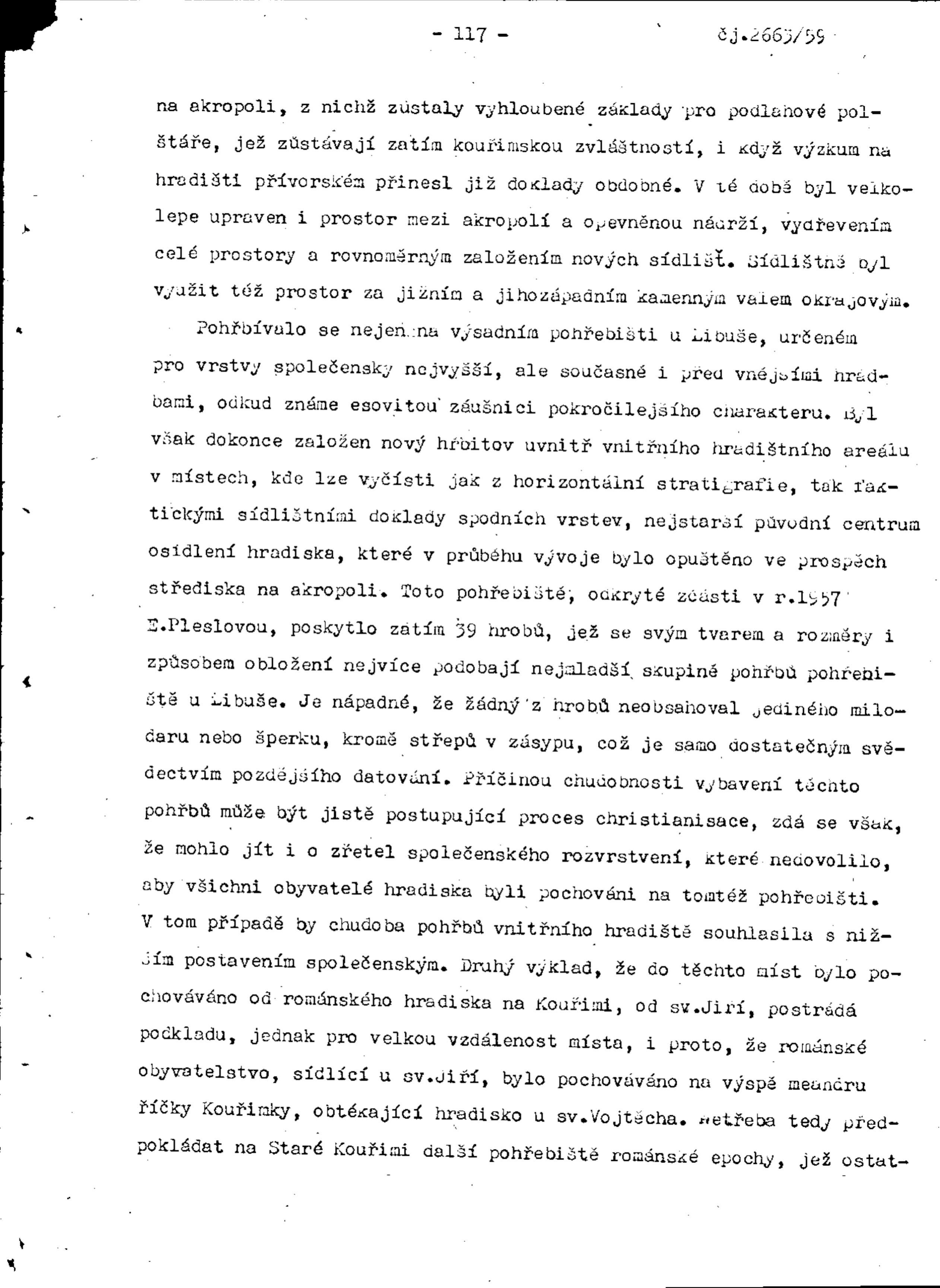}
        \caption{Cornerhole \& skewed text print}
        \label{fig:edge4}
    \end{subfigure}
\end{figure}

\begin{figure}[h]
    \centering
    \caption{Large drawn tables and grained‐paper scans}
    \label{fig:defect_largedraw}
    \begin{subfigure}{0.45\textwidth}
        \centering
        \includegraphics[width=\textwidth]{img/pages/large_draw_table_squared_paper.pdf}
        \caption{Squared‐table drawing on large page}
        \label{fig:large1}
    \end{subfigure}
    \hfill
    \begin{subfigure}{0.45\textwidth}
        \centering
        \includegraphics[width=\textwidth]{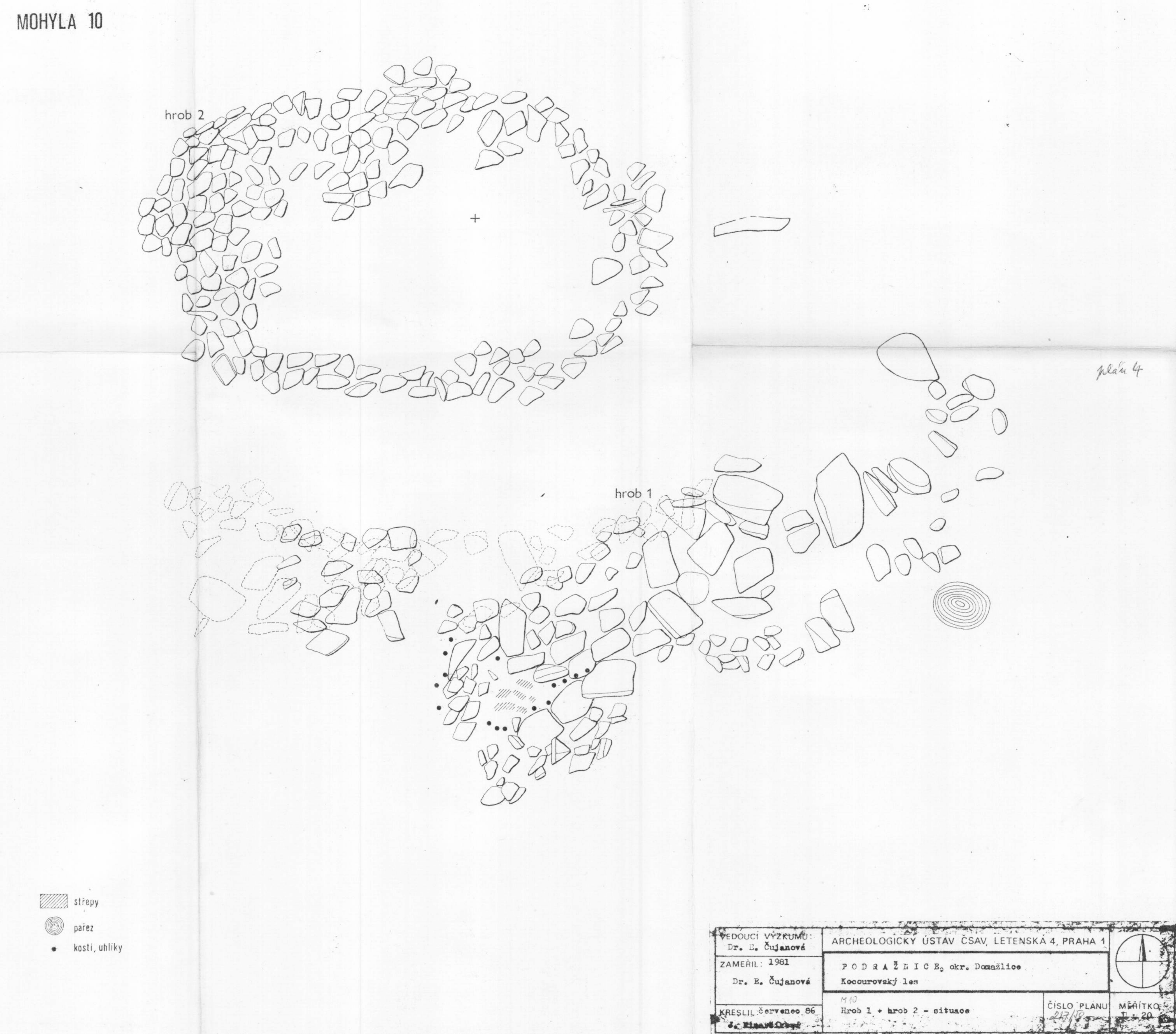}
        \caption{Packed‐table drawing on large paper}
        \label{fig:large2}
    \end{subfigure}

    \vspace{1em}

    \begin{subfigure}{0.45\textwidth}
        \centering
        \includegraphics[width=\textwidth]{img/pages/draw_paper_grained_scan.pdf}
        \caption{Hand-drawn image on grained paper scan}
        \label{fig:large3}
    \end{subfigure}
    \hfill
    \begin{subfigure}{0.45\textwidth}
        \centering
        \includegraphics[width=\textwidth]{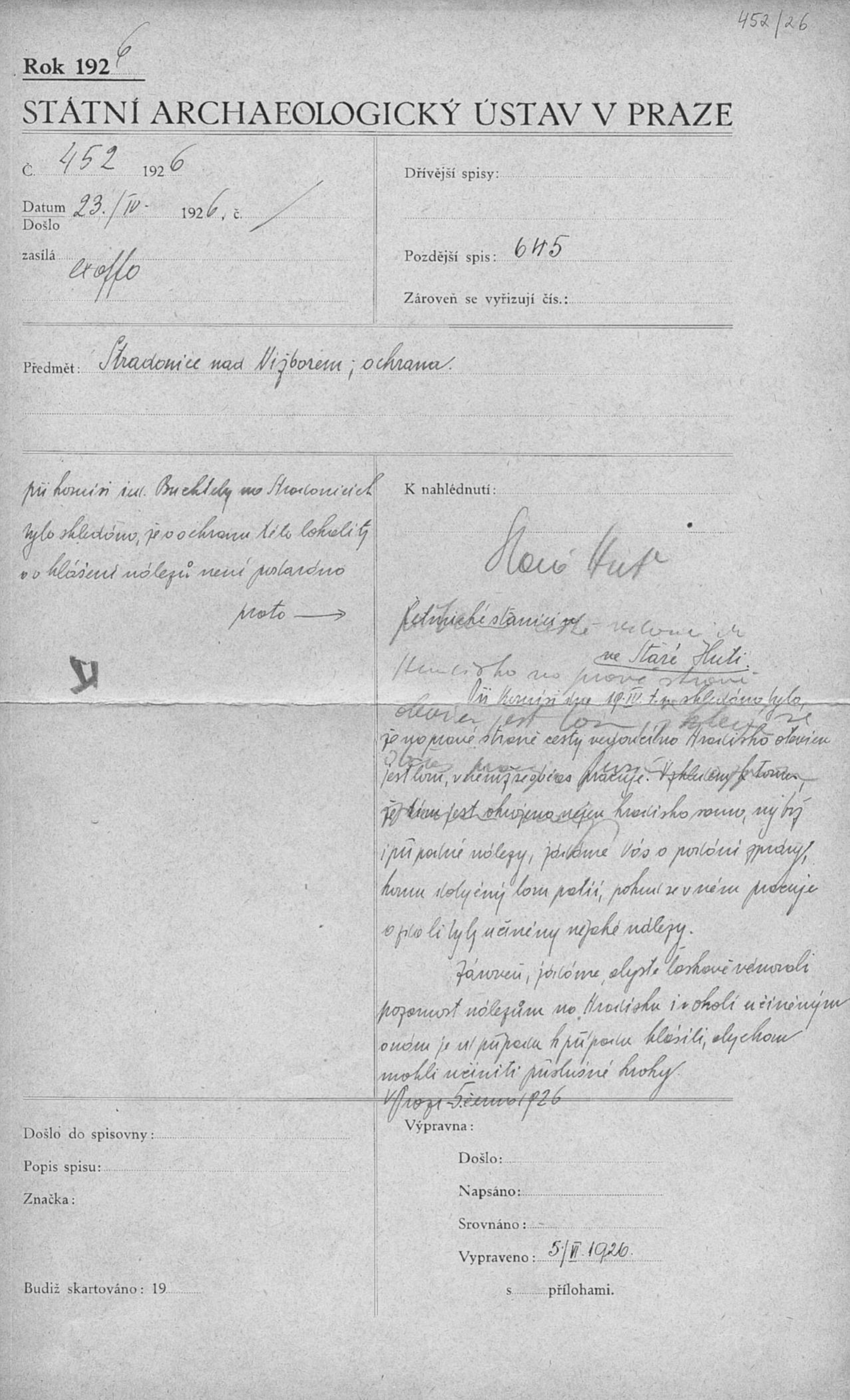}
        \caption{Extra‐edge table on paper (mixed content)}
        \label{fig:large4}
    \end{subfigure}
\end{figure}

\begin{figure}[h]
    \centering
    \caption{Scans from thick volumes and journals}
    \label{fig:defect_fatjournal}
    \begin{subfigure}{0.3\textwidth}
        \centering
        \includegraphics[width=\textwidth]{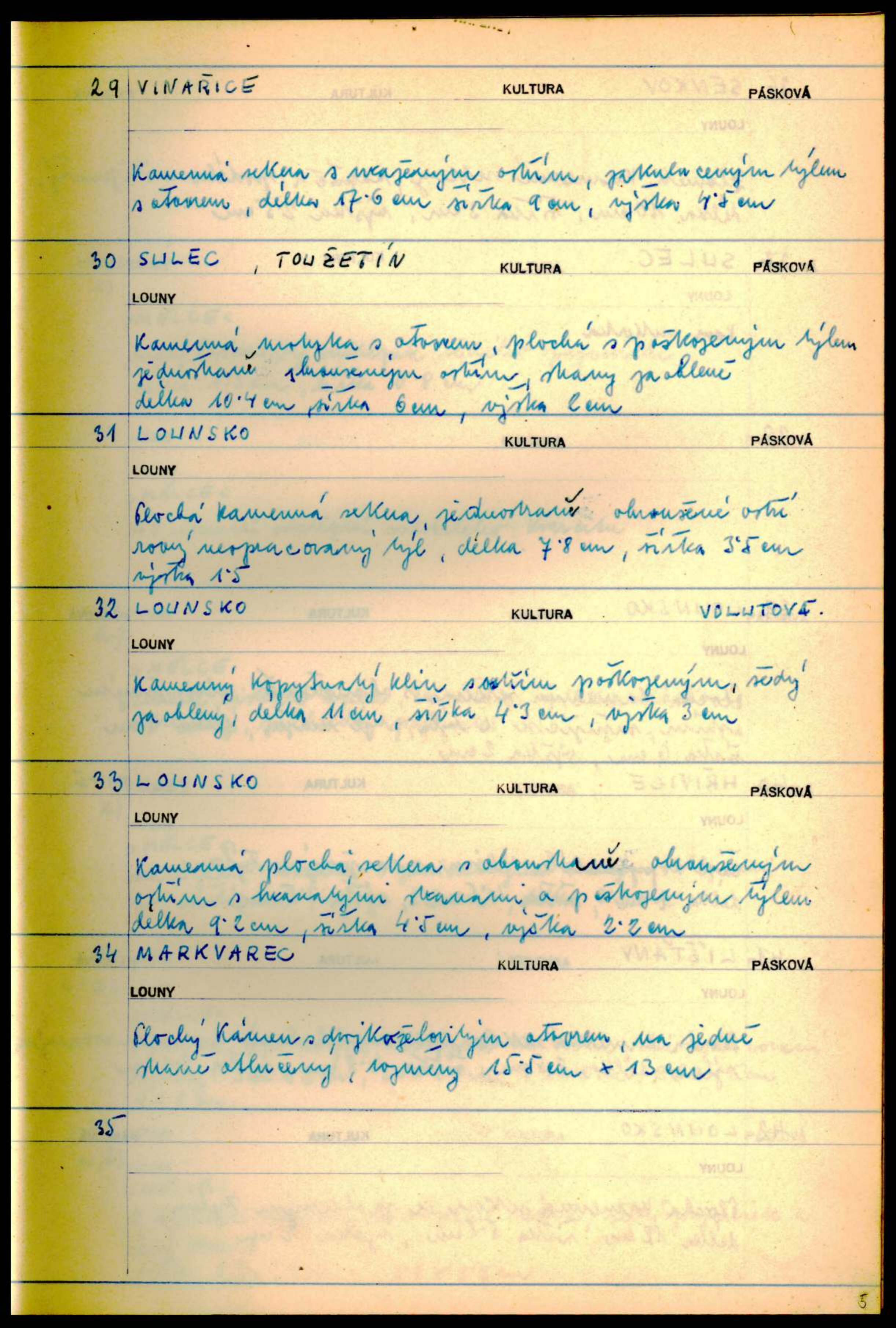}
        \caption{Color table in thick journal scan}
        \label{fig:fat1}
    \end{subfigure}
    % \hfill
    \begin{subfigure}{0.3\textwidth}
        \centering
        \includegraphics[width=\textwidth]{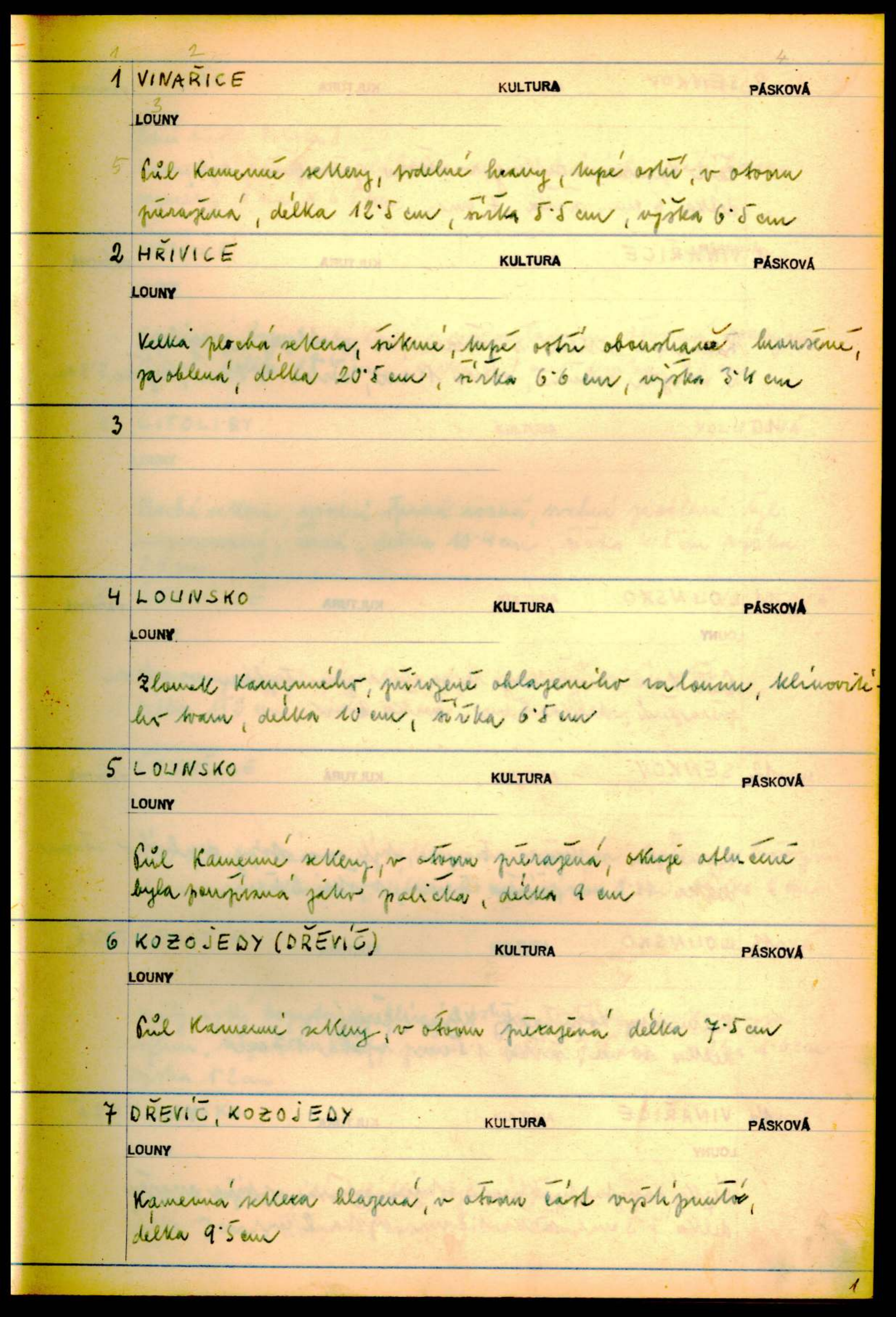}
        \caption{Black‐and‐white journal page with table}
        \label{fig:fat2}
    \end{subfigure}
    % \hfill
    \begin{subfigure}{0.3\textwidth}
        \centering
        \includegraphics[width=\textwidth]{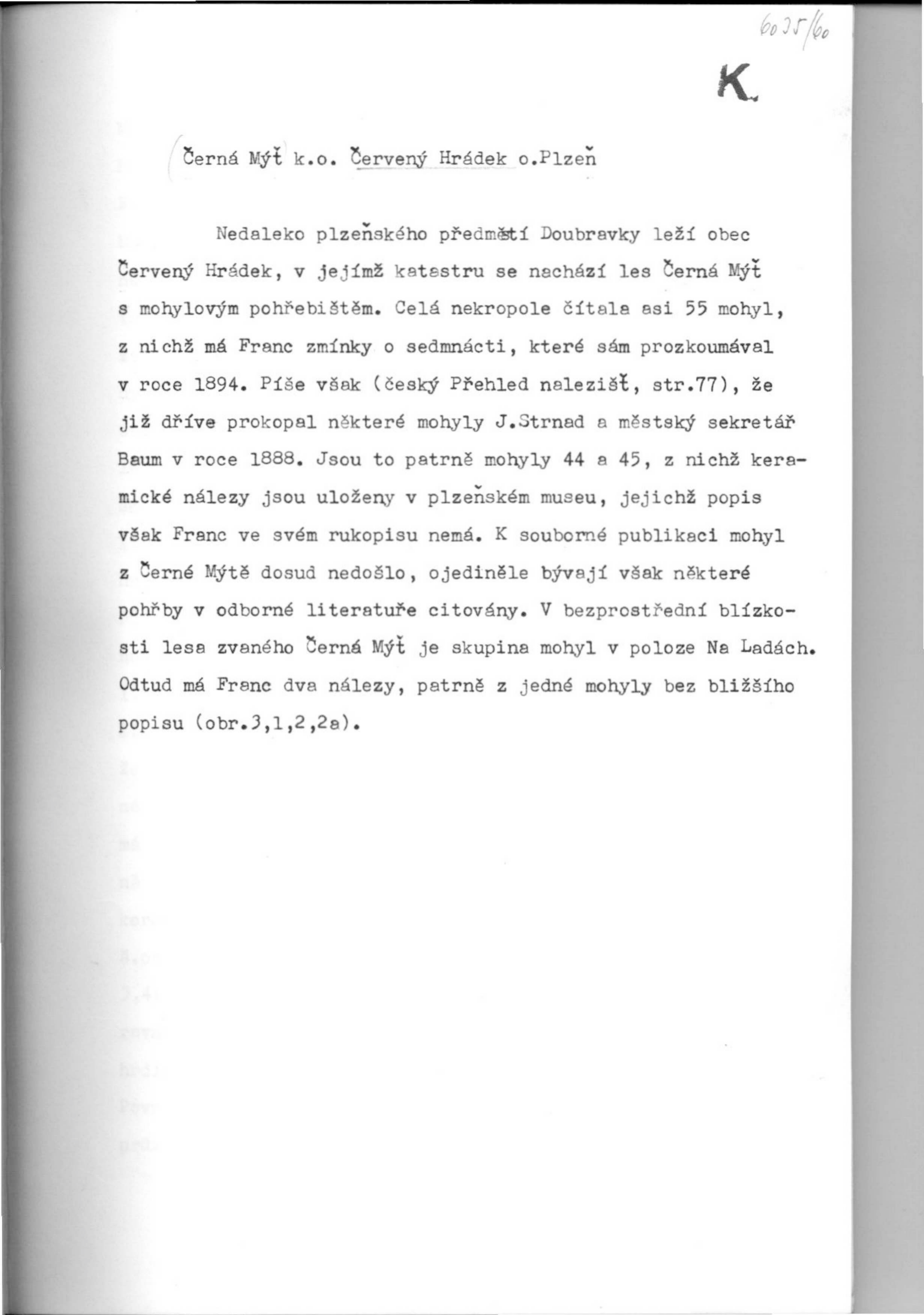}
        \caption{Text from a thick book scan}
        \label{fig:fat3}
    \end{subfigure}
\end{figure}

\begin{figure}[h]
    \centering
    \caption{Typewritten text corrections}
    \label{fig:defect_textt}
    \begin{subfigure}{0.3\textwidth}
        \centering
        \includegraphics[width=\textwidth]{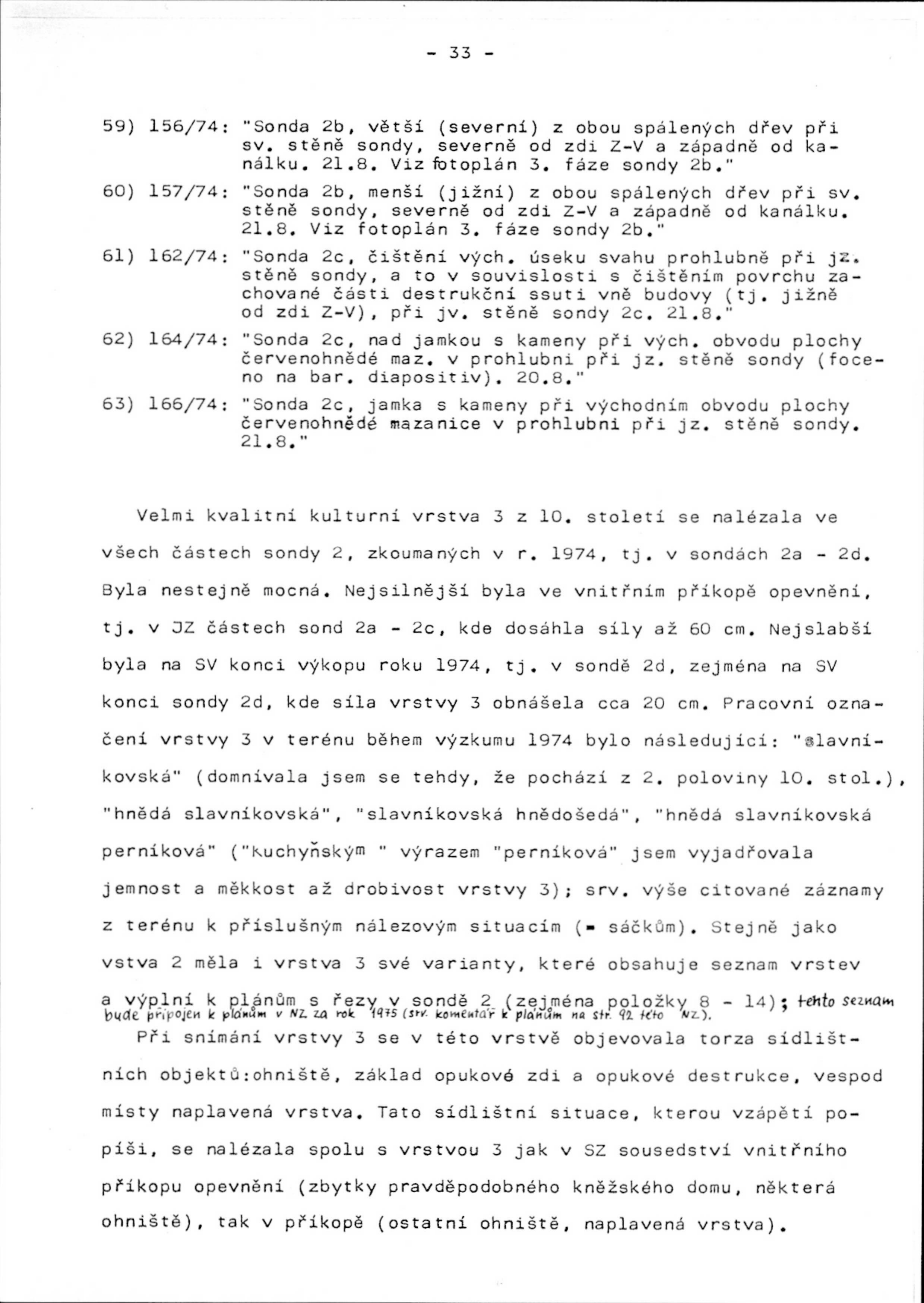}
        \caption{Corrected typewritten text}
        \label{fig:textt1}
    \end{subfigure}
    % \hfill
    \begin{subfigure}{0.3\textwidth}
        \centering
        \includegraphics[width=\textwidth]{img/pages/text_t_minor_correct_bleed.pdf}
        \caption{Minor typewritten correction}
        \label{fig:textt2}
    \end{subfigure}
    % \hfill
    \begin{subfigure}{0.3\textwidth}
        \centering
        \includegraphics[width=\textwidth]{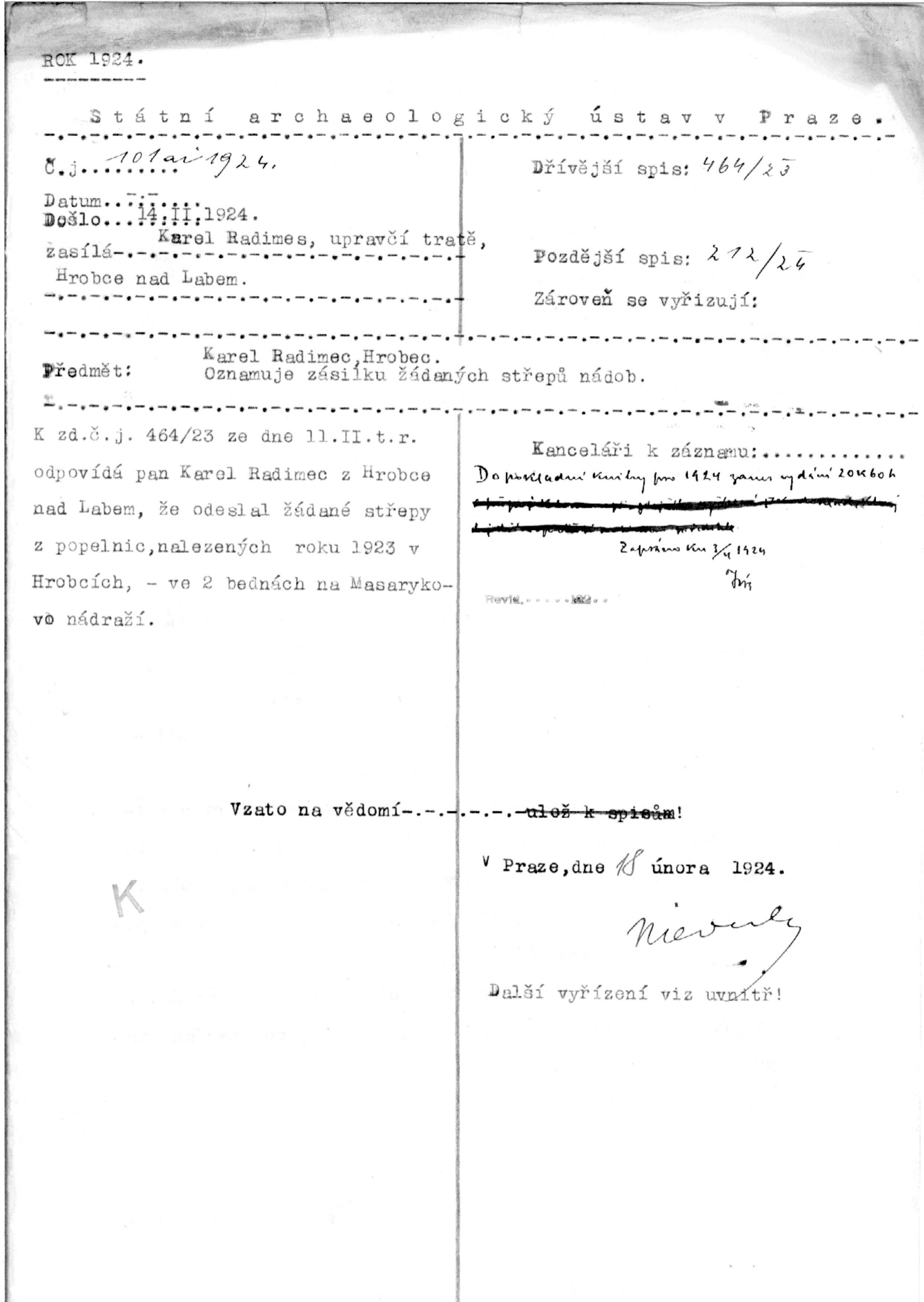}
        \caption{Crossed-out lines \& gray corners}
        \label{fig:textt3}
    \end{subfigure}
    
\end{figure}

\begin{figure}[h]
    \centering
    \caption{Tables with color artifacts (\ac{eg} scanning noise, pen stains, and paper discoloration)}
    \label{fig:defect_tablehw}
    \begin{subfigure}{0.3\textwidth}
        \centering
        \includegraphics[width=\textwidth]{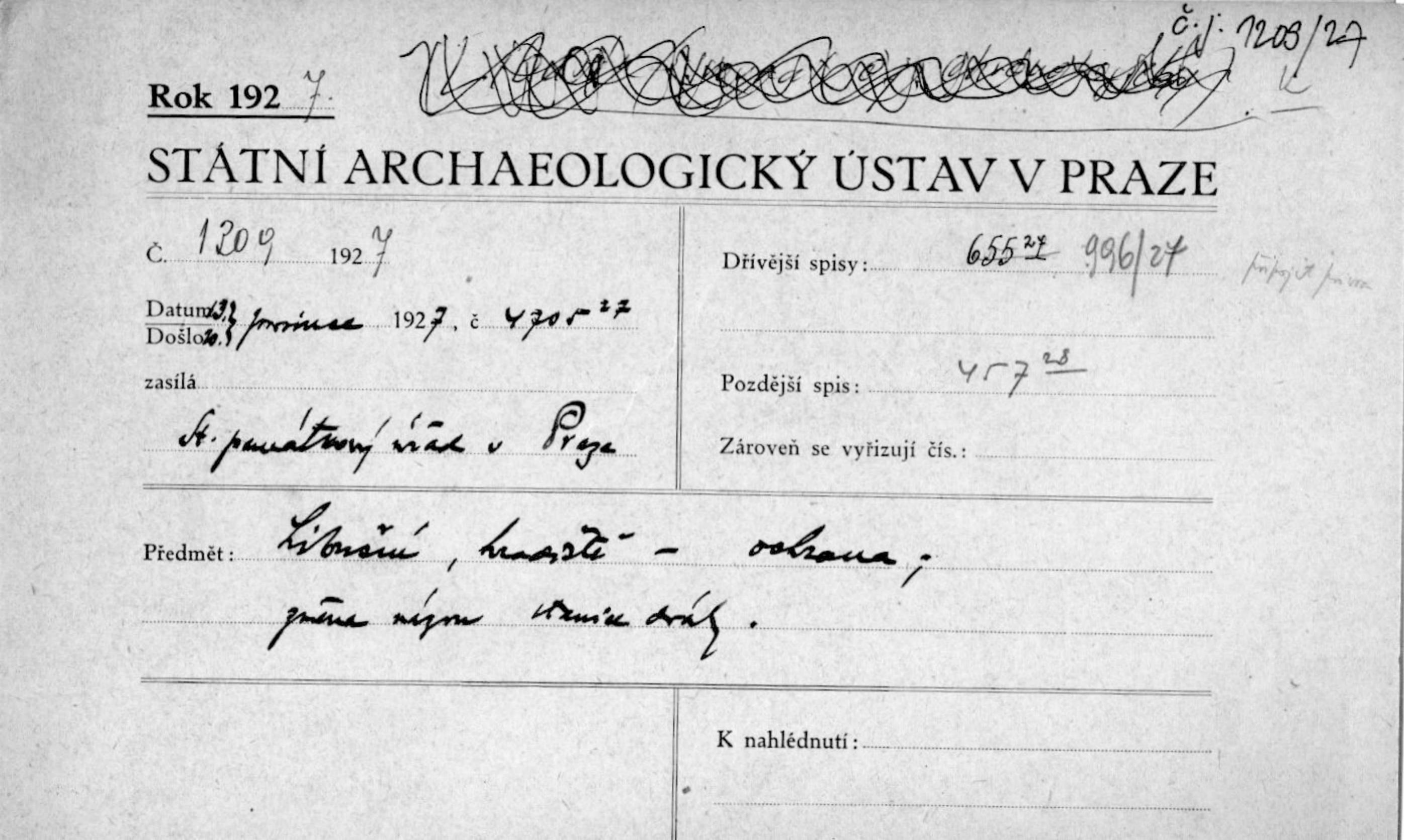}
        \caption{Table with heavy scribbles}
        \label{fig:tablehw1}
    \end{subfigure}
    % \hfill
    \begin{subfigure}{0.3\textwidth}
        \centering
        \includegraphics[width=\textwidth]{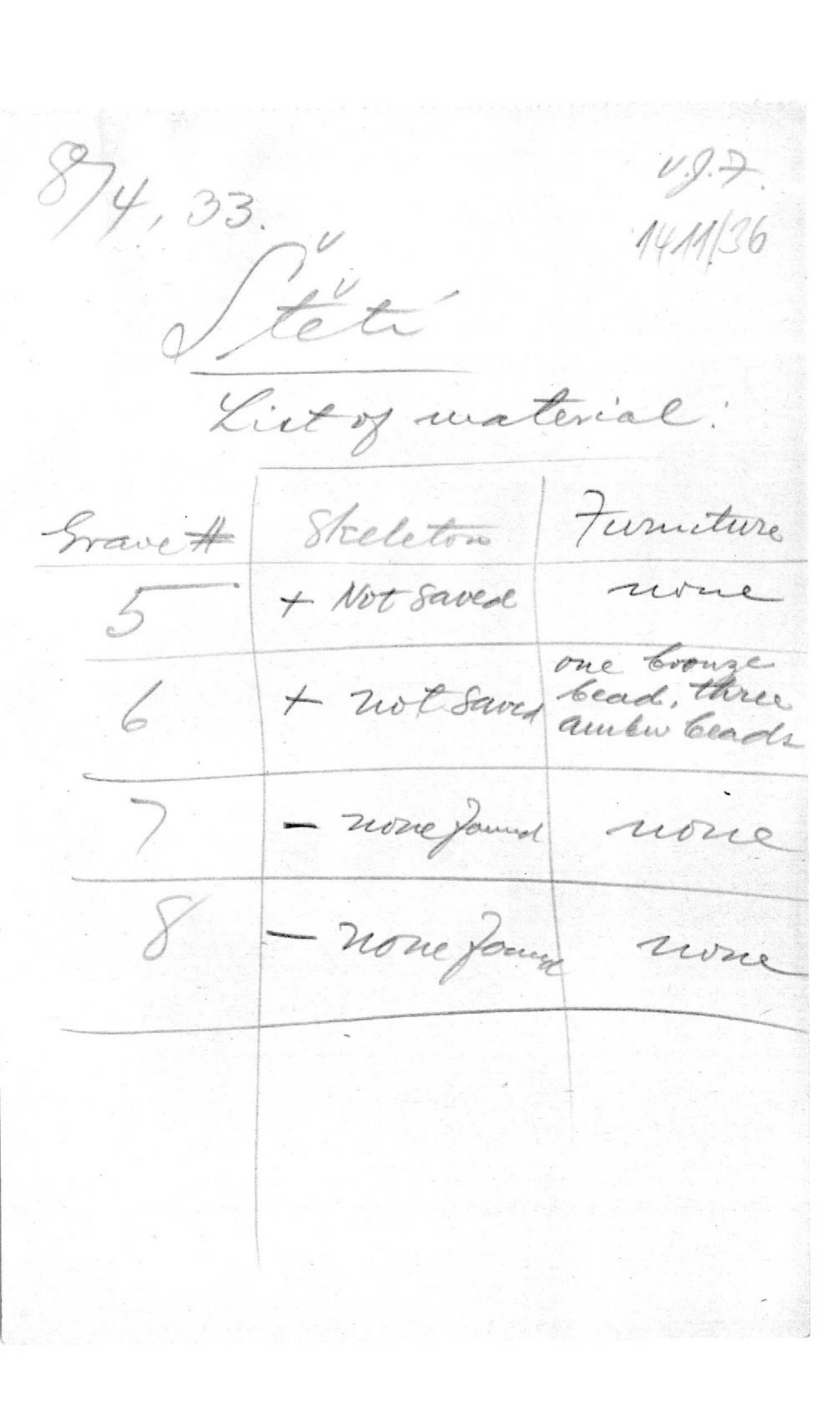}
        \caption{Small table on a page}
        \label{fig:tablehw2}
    \end{subfigure}
    % \hfill
    \begin{subfigure}{0.3\textwidth}
        \centering
        \includegraphics[width=\textwidth]{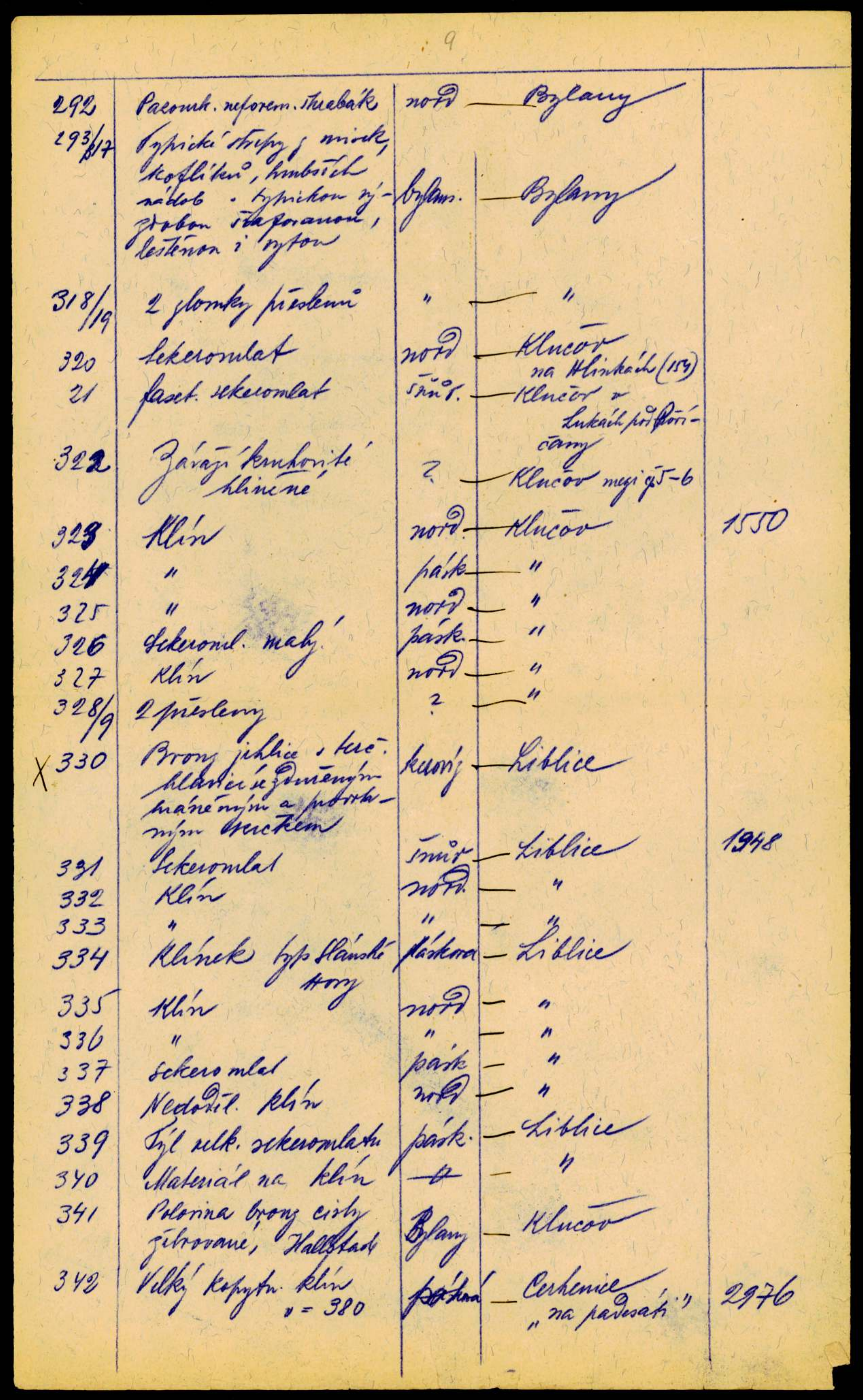}
        \caption{Pen stains and yellowed paper under the table}
        \label{fig:tablehw3}
    \end{subfigure}
\end{figure}

\begin{figure}[h]
    \centering
    \caption{Mixed‐text pages and scribbles}
    \label{fig:defect_mixedtext}
    \begin{subfigure}{0.3\textwidth}
        \centering
        \includegraphics[width=\textwidth]{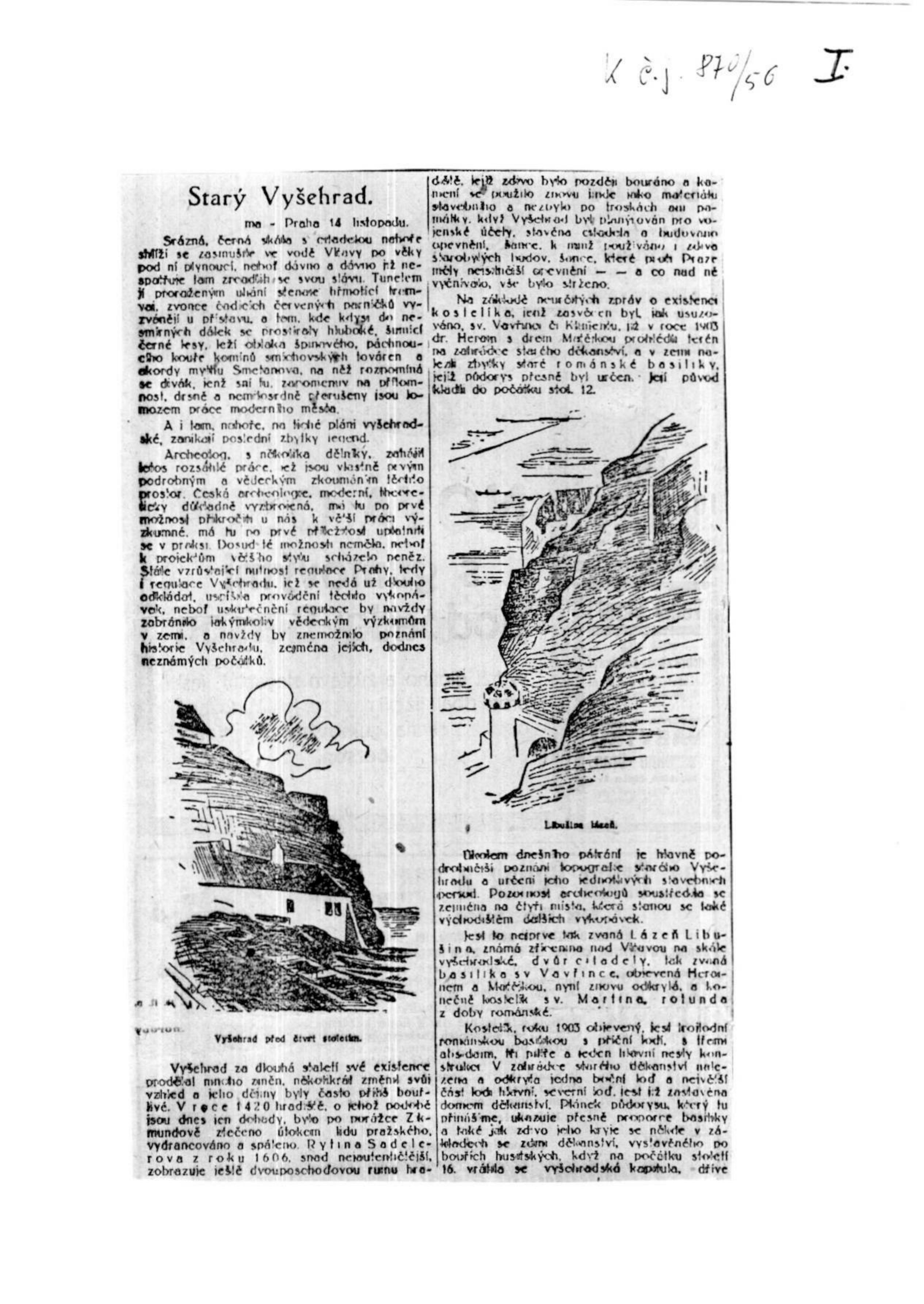}
        \caption{Hand-drawn mixed-text page}
        \label{fig:mixed1}
    \end{subfigure}
    % \hfill
    \begin{subfigure}{0.3\textwidth}
        \centering
        \includegraphics[width=\textwidth]{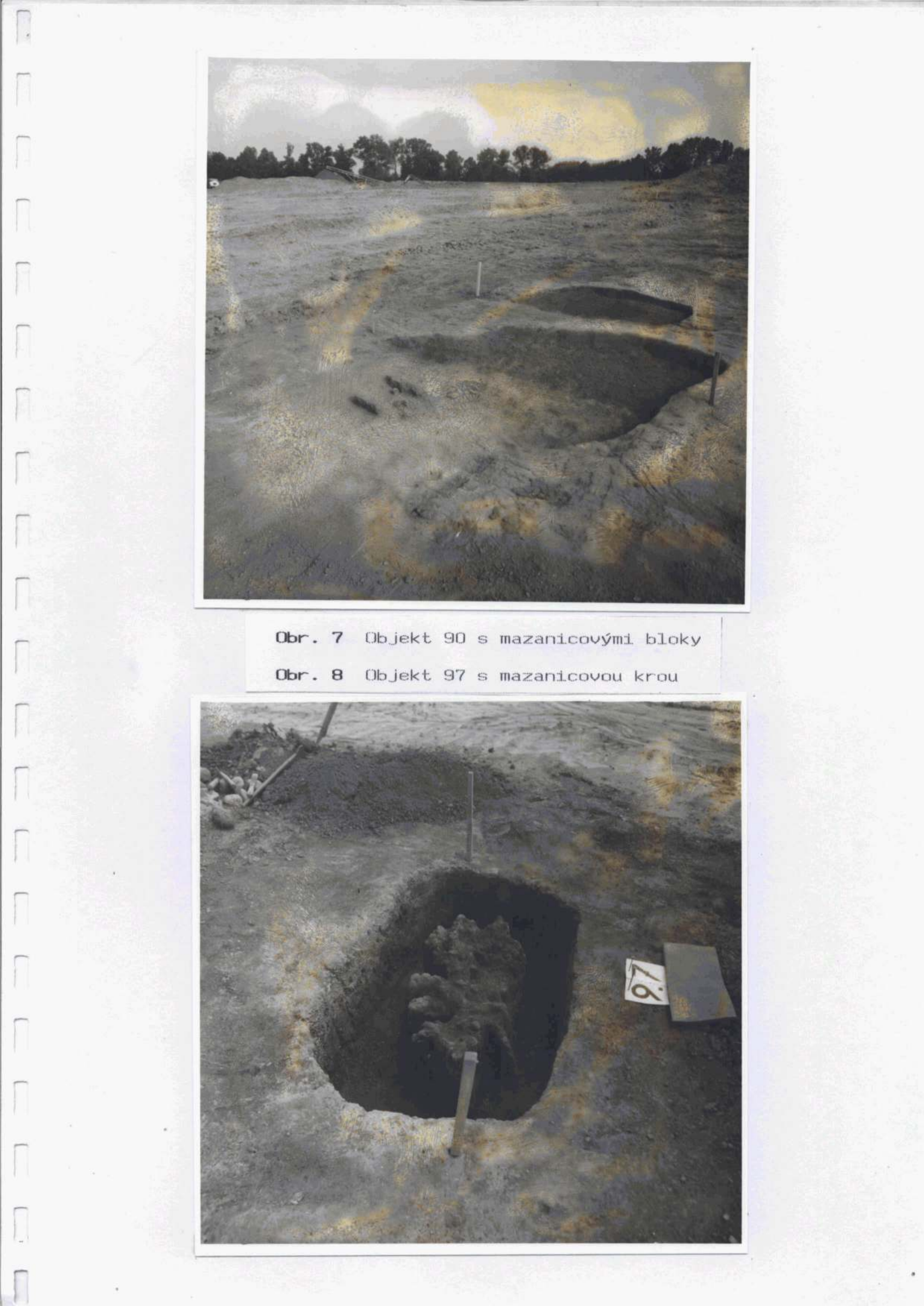}
        \caption{Yellowed photo with typewritten caption}
        \label{fig:mixed2}
    \end{subfigure}
    % \hfill
    \begin{subfigure}{0.3\textwidth}
        \centering
        \includegraphics[width=\textwidth]{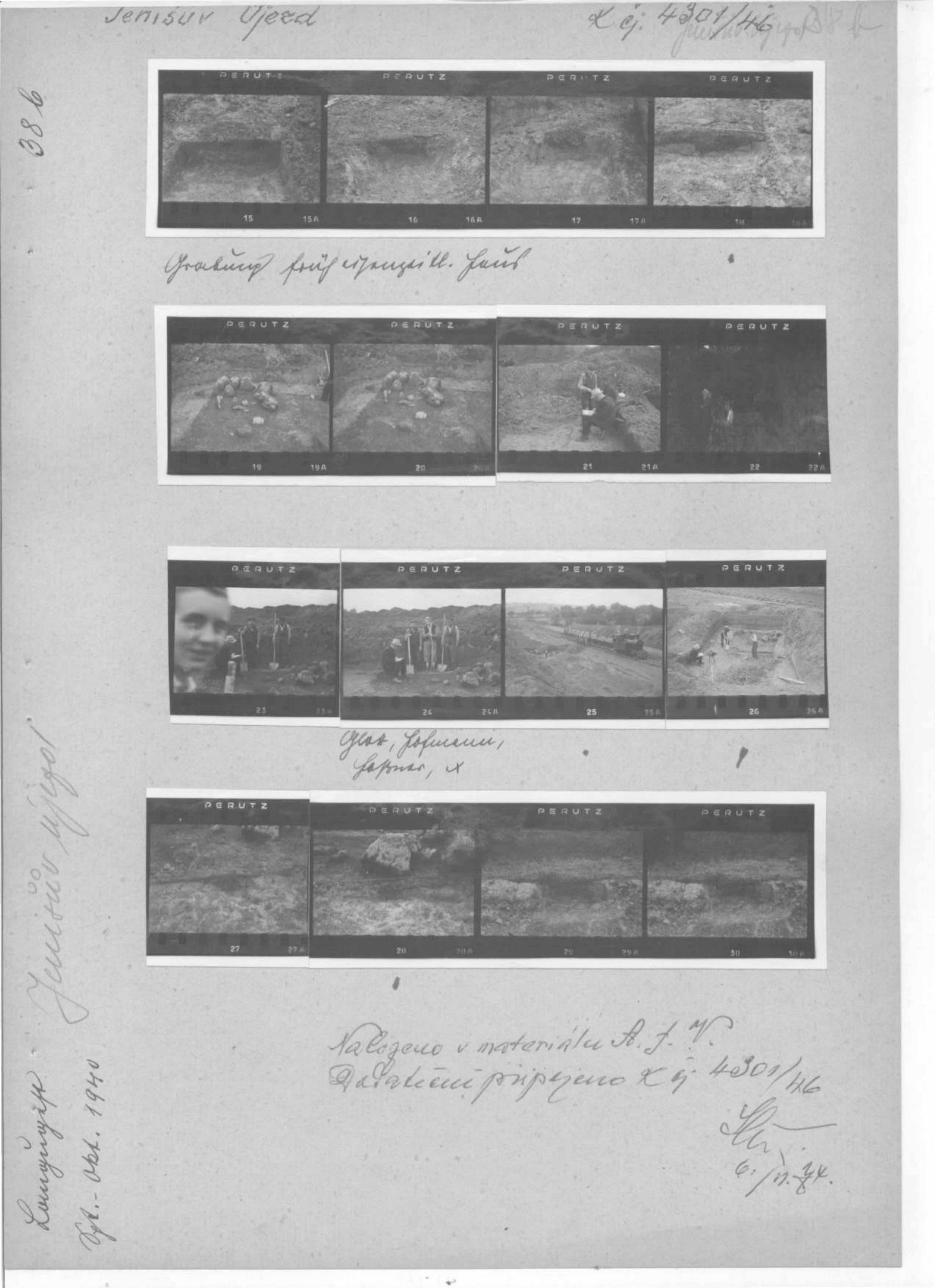}
        \caption{Gray paper with manual photo captions}
        \label{fig:mixed3}
    \end{subfigure}
\end{figure}

\chapter{Parsing attempts} \label{app-dd}

\begin{figure}[h]
    \centering
    \caption{\ac{dd} attempt on a newspaper page with drawings}
    \label{fig:dd_newspaper}
    \begin{subfigure}{0.3\textwidth}
        \centering
        \includegraphics[width=\textwidth]{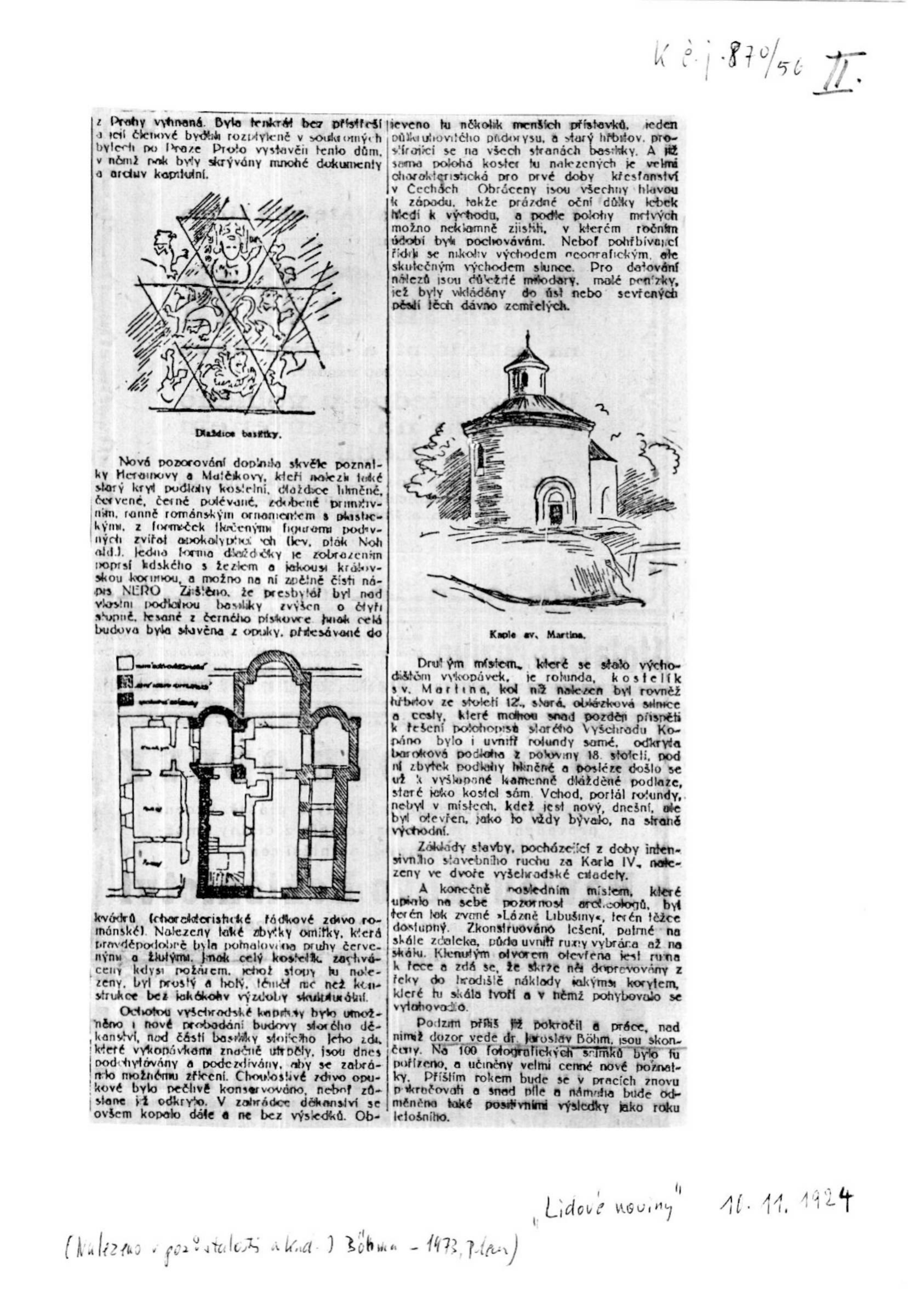}
        \caption{Source image}
        \label{fig:dd_news1}
    \end{subfigure}
    % \hfill
    \begin{subfigure}{0.3\textwidth}
        \centering
        \includegraphics[width=\textwidth]{img/pages/dd_newspaper.pdf}
        \caption{Recognized elements}
        \label{fig:dd_news2}
    \end{subfigure}
    % \hfill
    \begin{subfigure}{0.3\textwidth}
        \centering
        \includegraphics[width=\textwidth]{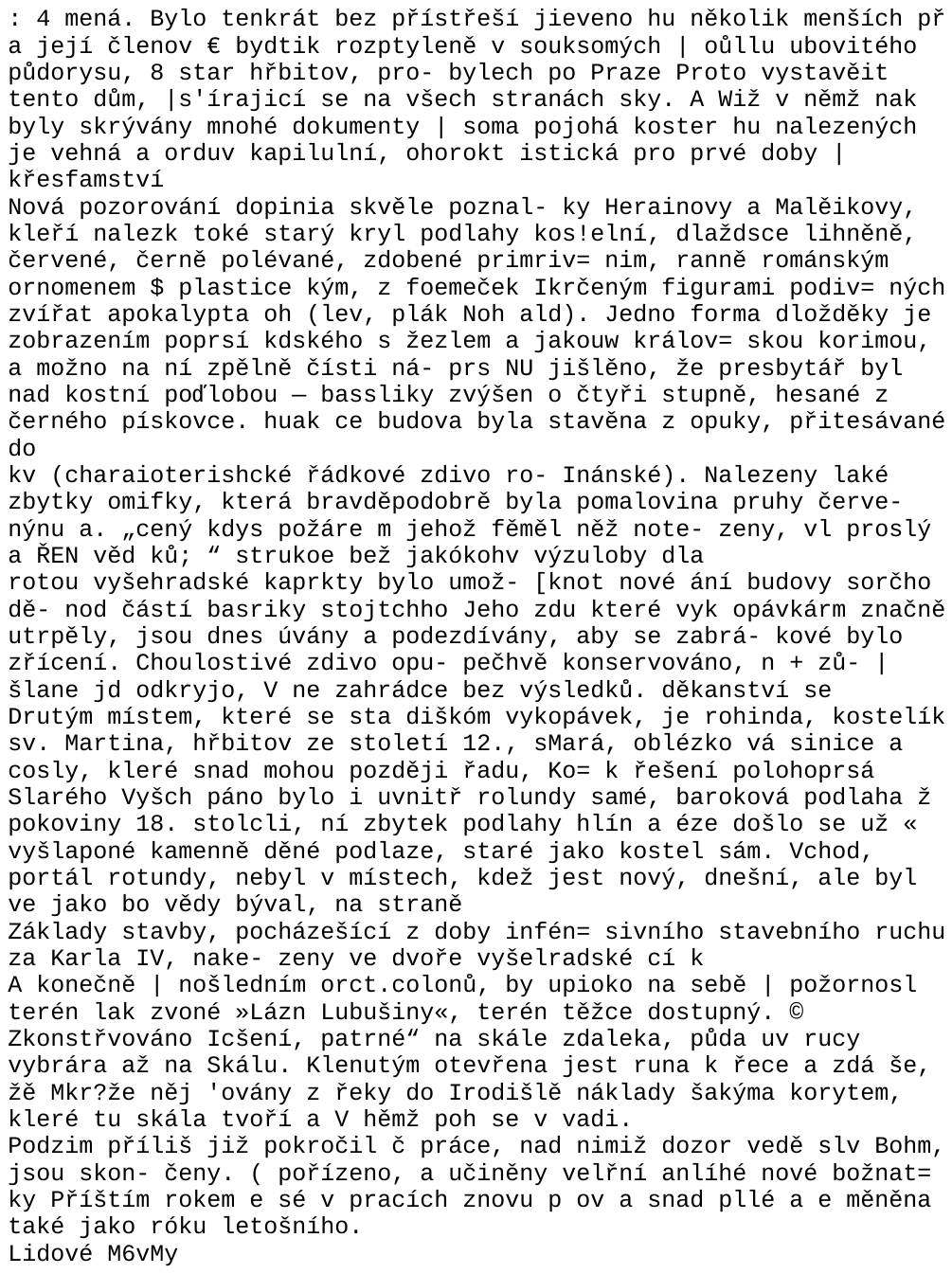}
        \caption{Extracted text}
        \label{fig:dd_news3}
    \end{subfigure}
\end{figure}

\begin{figure}[h]
    \centering
    \caption{\ac{dd} attempt on a table}
    \label{fig:dd_table}
    \begin{subfigure}{0.3\textwidth}
        \centering
        \includegraphics[width=\textwidth]{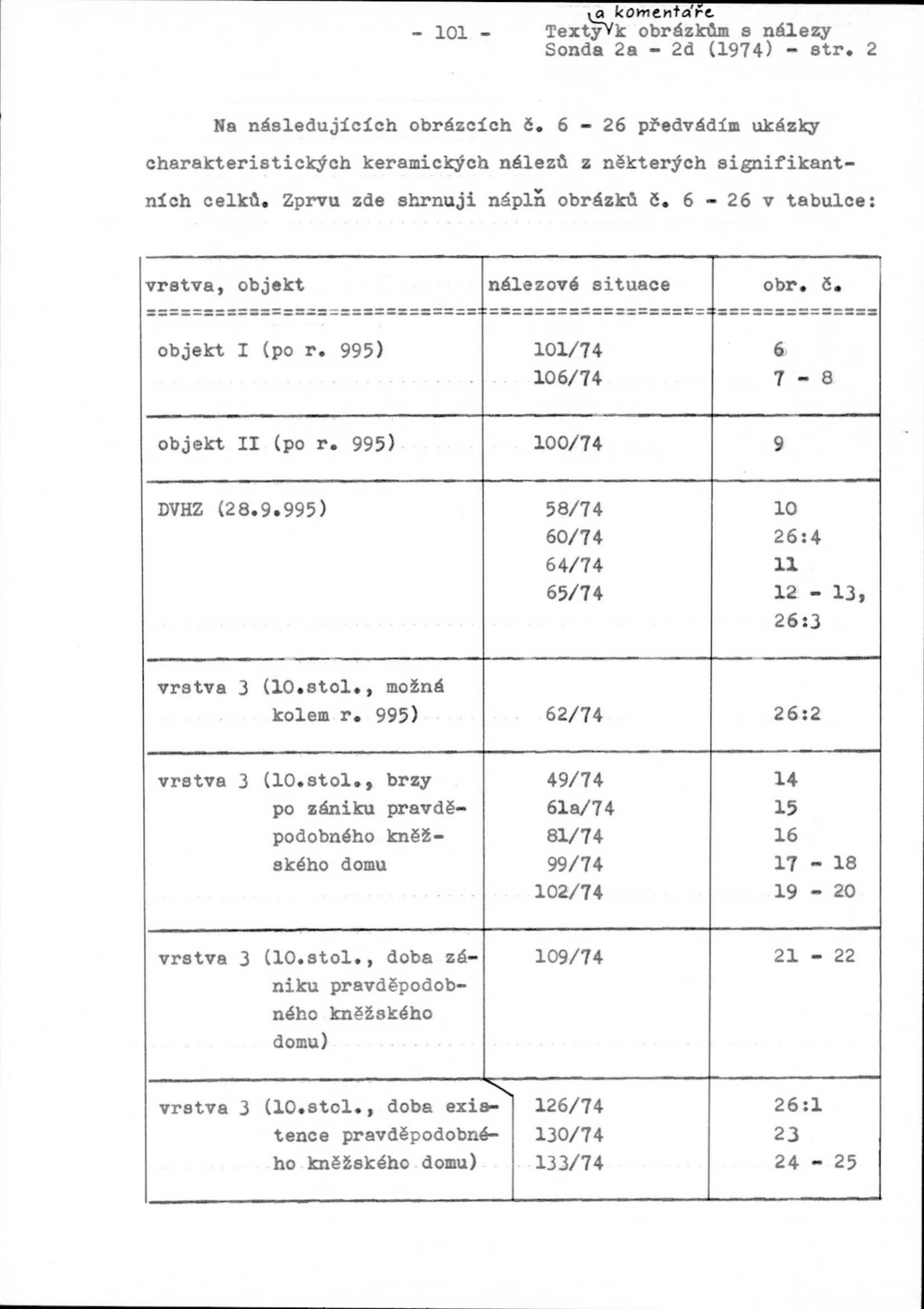}
        \caption{Source image}
        \label{fig:dd_tab1}
    \end{subfigure}
    % \hfill
    \begin{subfigure}{0.3\textwidth}
        \centering
        \includegraphics[width=\textwidth]{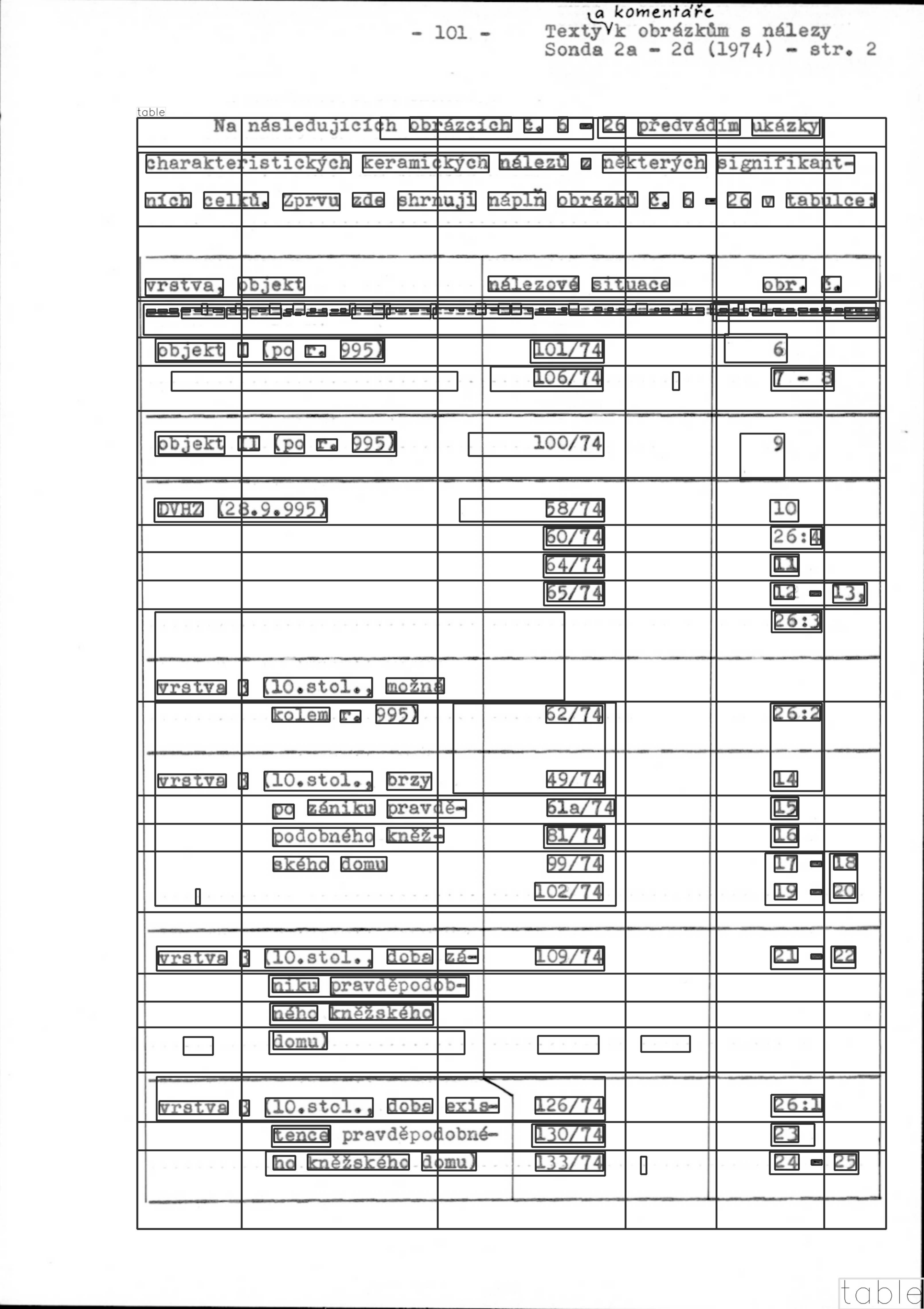}
        \caption{Recognized table cells}
        \label{fig:dd_tab2}
    \end{subfigure}
    % \hfill
    \begin{subfigure}{0.3\textwidth}
        \centering
        \includegraphics[width=\textwidth]{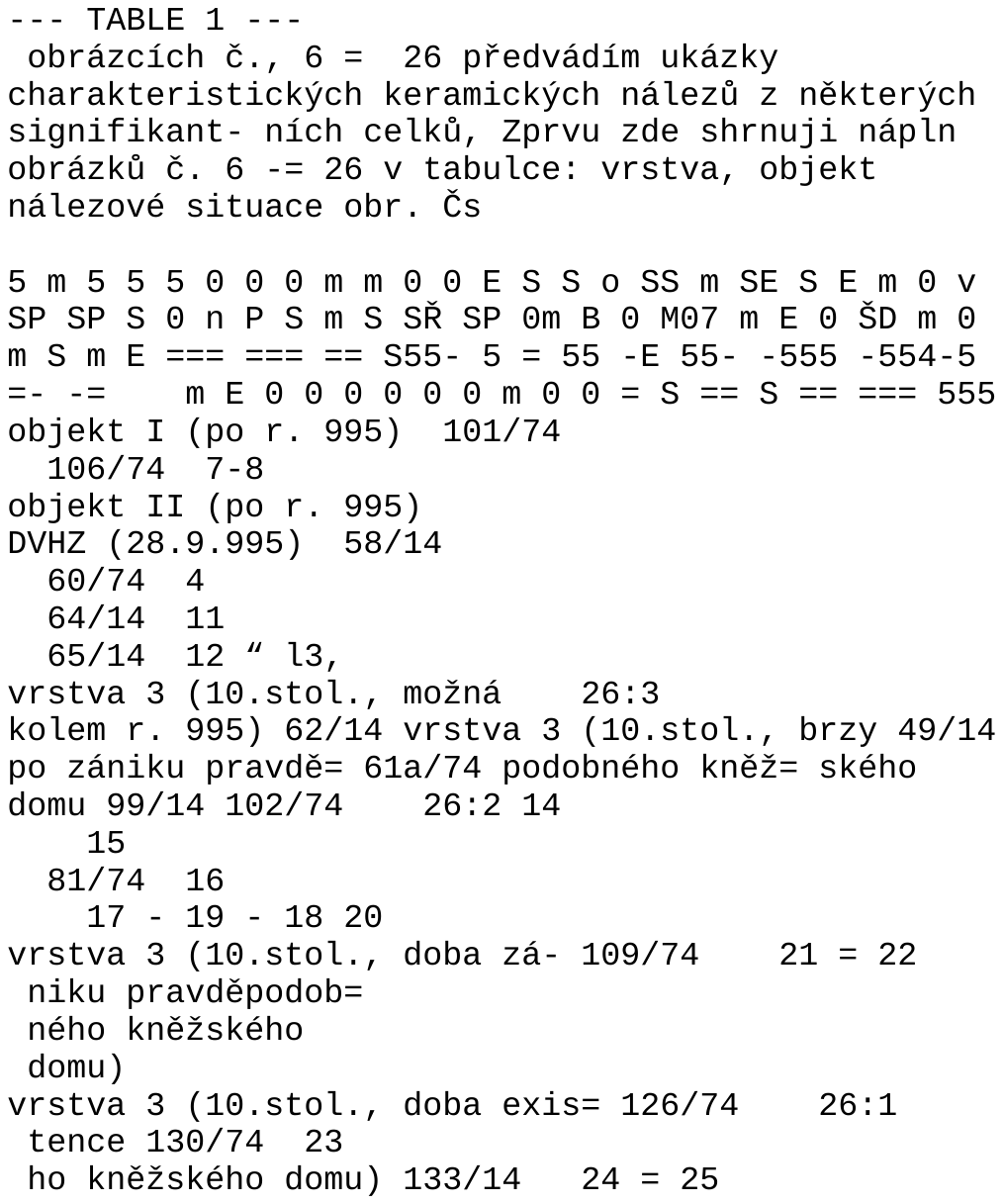}
        \caption{Extracted table}
        \label{fig:dd_tab3}
    \end{subfigure}
\end{figure}

\begin{figure}[h]
    \centering
    \caption{\ac{dd} attempt on a plain text}
    \label{fig:dd_text}
    \begin{subfigure}{0.3\textwidth}
        \centering
        \includegraphics[width=\textwidth]{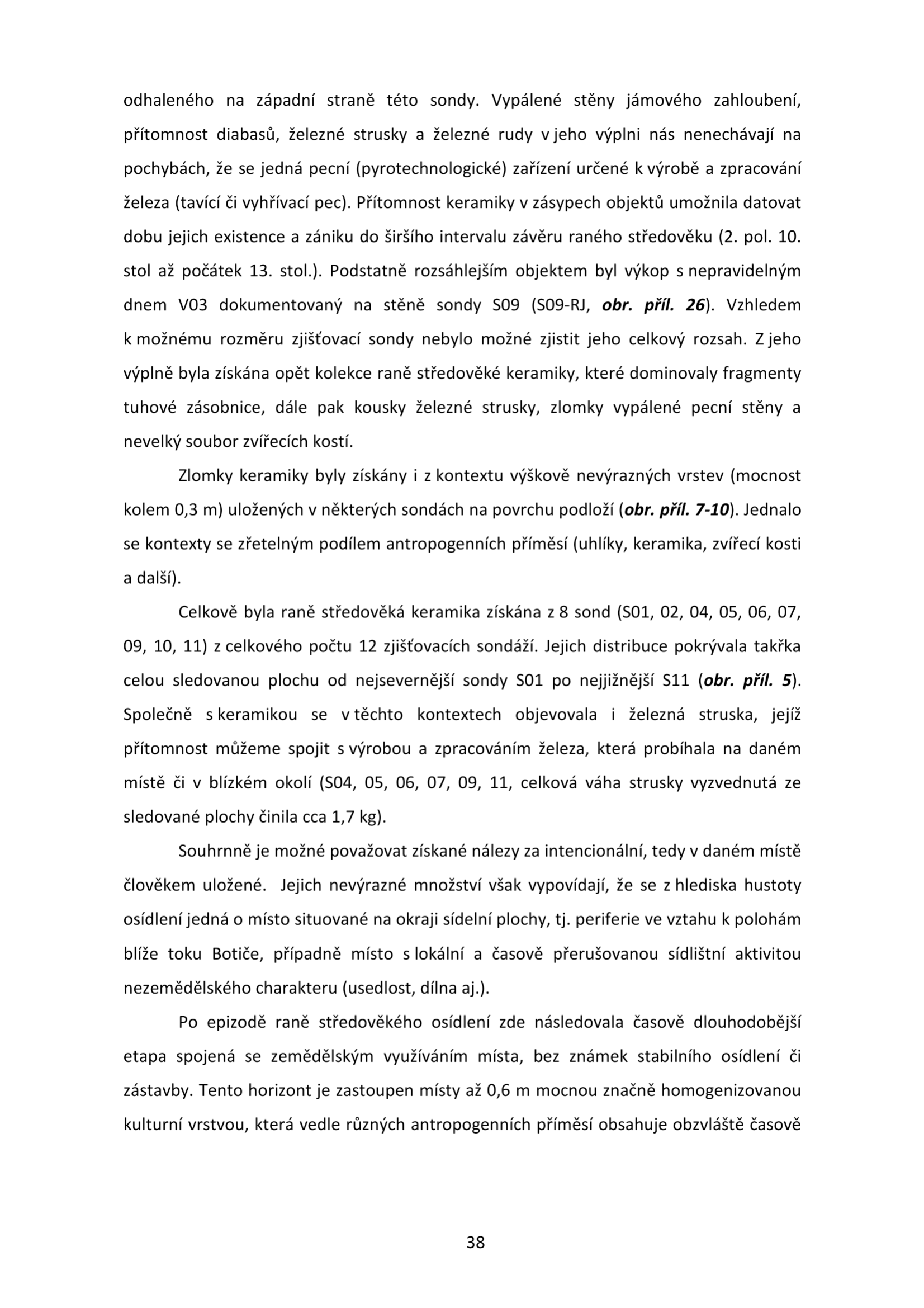}
        \caption{Source image}
        \label{fig:dd_txt1}
    \end{subfigure}
    % \hfill
    \begin{subfigure}{0.3\textwidth}
        \centering
        \includegraphics[width=\textwidth]{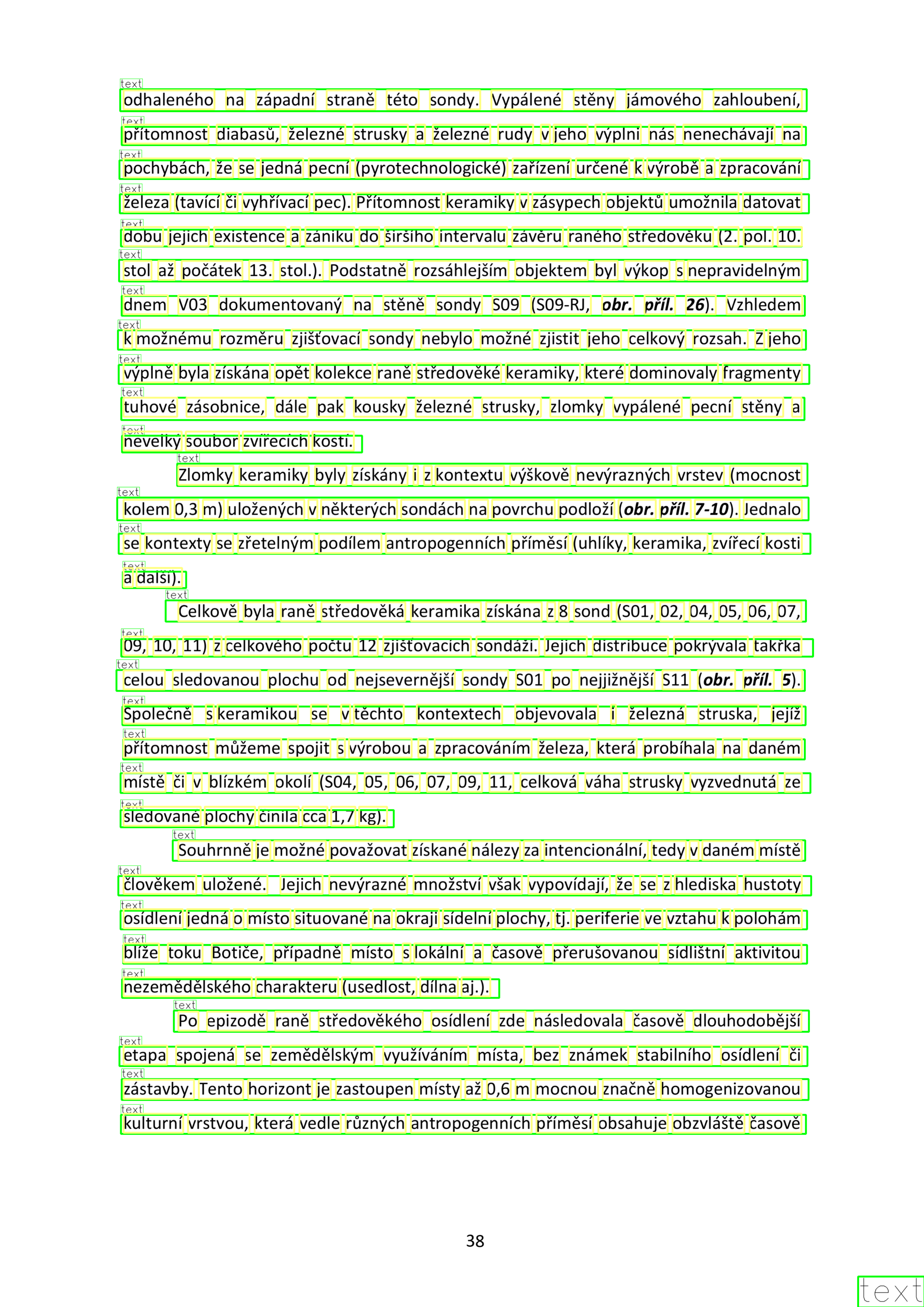}
        \caption{Recognized lines}
        \label{fig:dd_txt2}
    \end{subfigure}
    % \hfill
    \begin{subfigure}{0.3\textwidth}
        \centering
        \includegraphics[width=\textwidth]{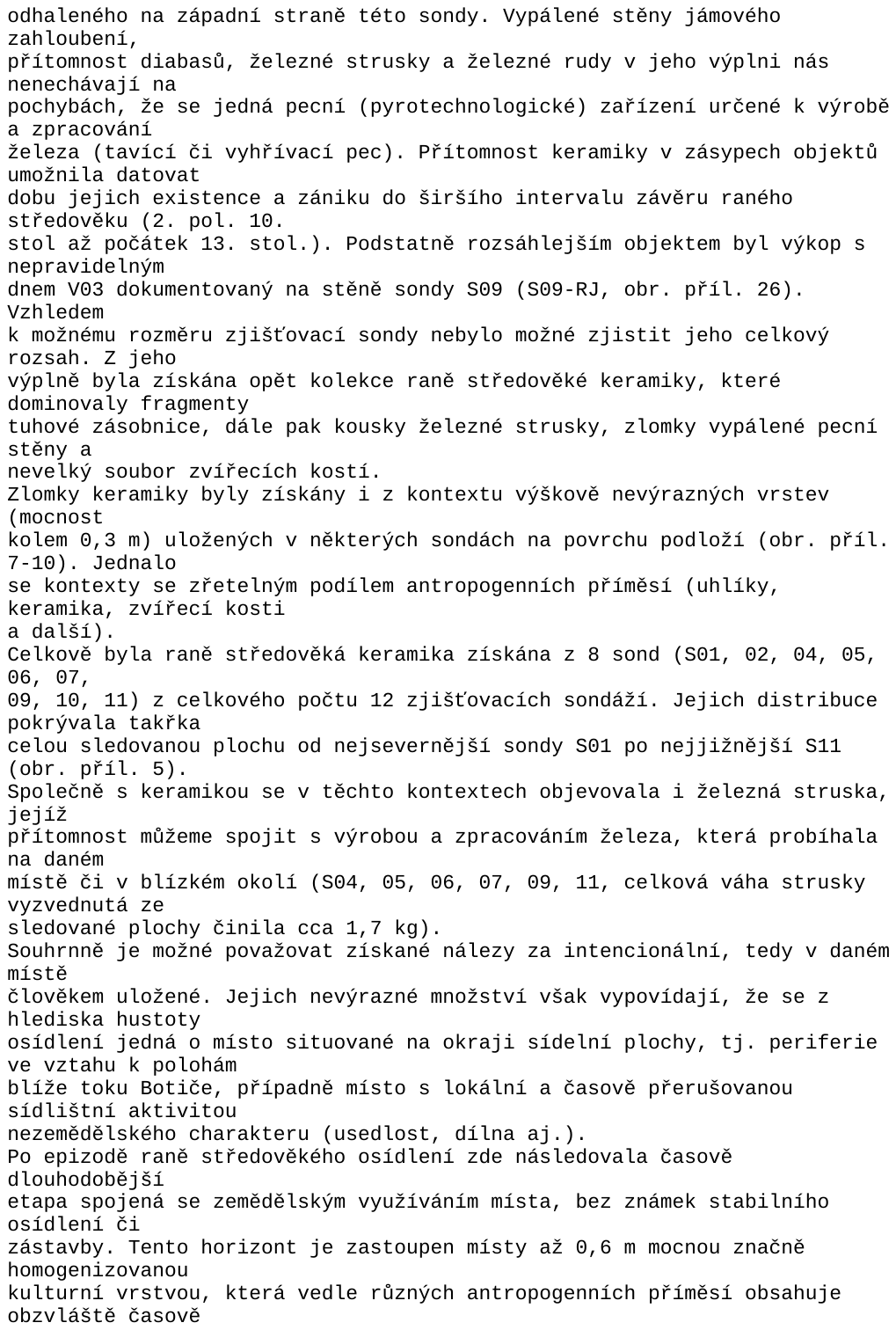}
        \caption{Extracted lines}
        \label{fig:dd_txt3}
    \end{subfigure}
\end{figure}

\begin{figure}[h]
    \centering
    \caption{\ac{dd} attempt on a photo with text}
    \label{fig:dd_mix}
    \begin{subfigure}{0.3\textwidth}
        \centering
        \includegraphics[width=\textwidth]{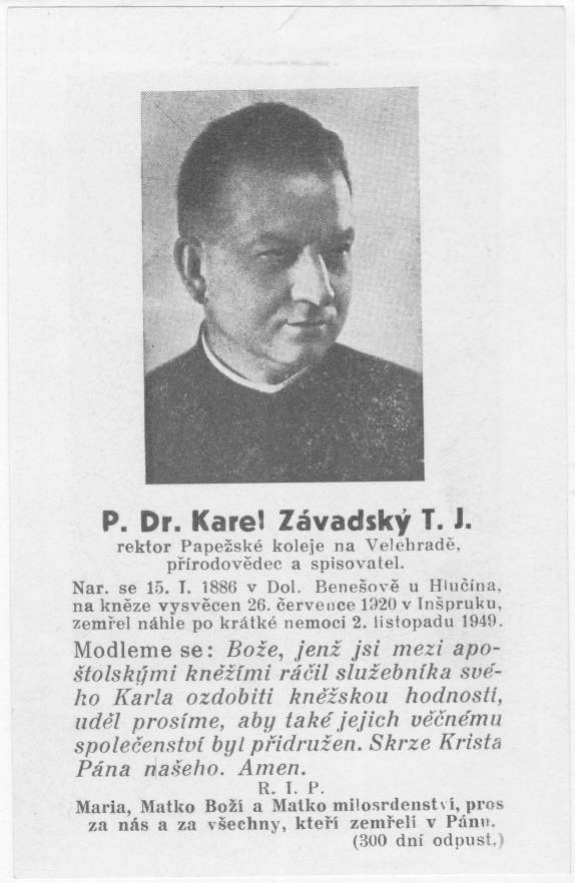}
        \caption{Source image}
        \label{fig:dd_mix1}
    \end{subfigure}
    % \hfill
    \begin{subfigure}{0.3\textwidth}
        \centering
        \includegraphics[width=\textwidth]{img/pages/dd_figure.pdf}
        \caption{Recognized elements}
        \label{fig:dd_mix2}
    \end{subfigure}
    % \hfill
    \begin{subfigure}{0.3\textwidth}
        \centering
        \includegraphics[width=\textwidth]{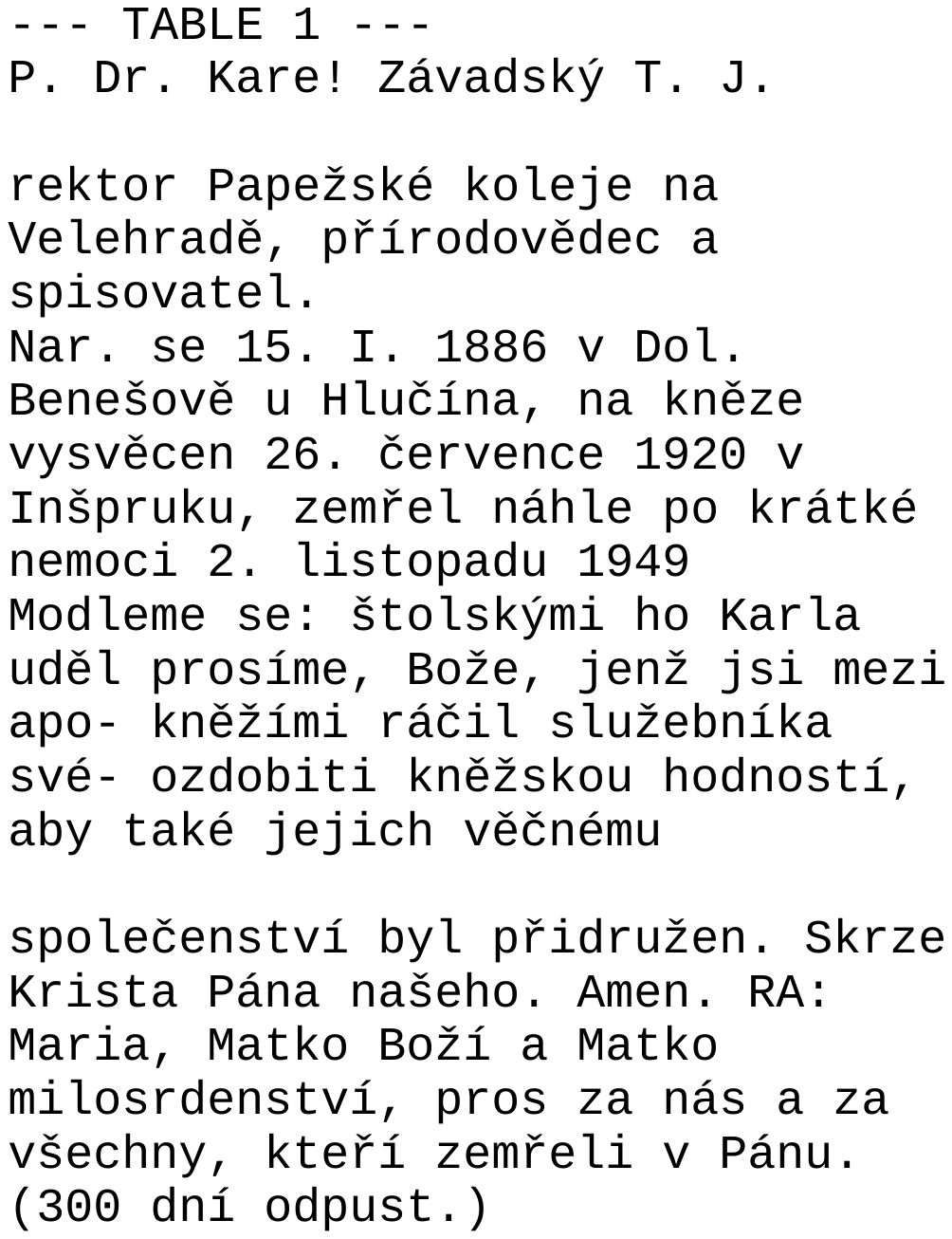}
        \caption{Extracted text}
        \label{fig:dd_mix3}
    \end{subfigure}
\end{figure}

\begin{figure}[h]
    \centering
    \caption{\ac{dd} mistakes on pages with photos}
    \label{fig:dd_mis_photo}
    \begin{subfigure}{0.3\textwidth}
        \centering
        \includegraphics[width=\textwidth]{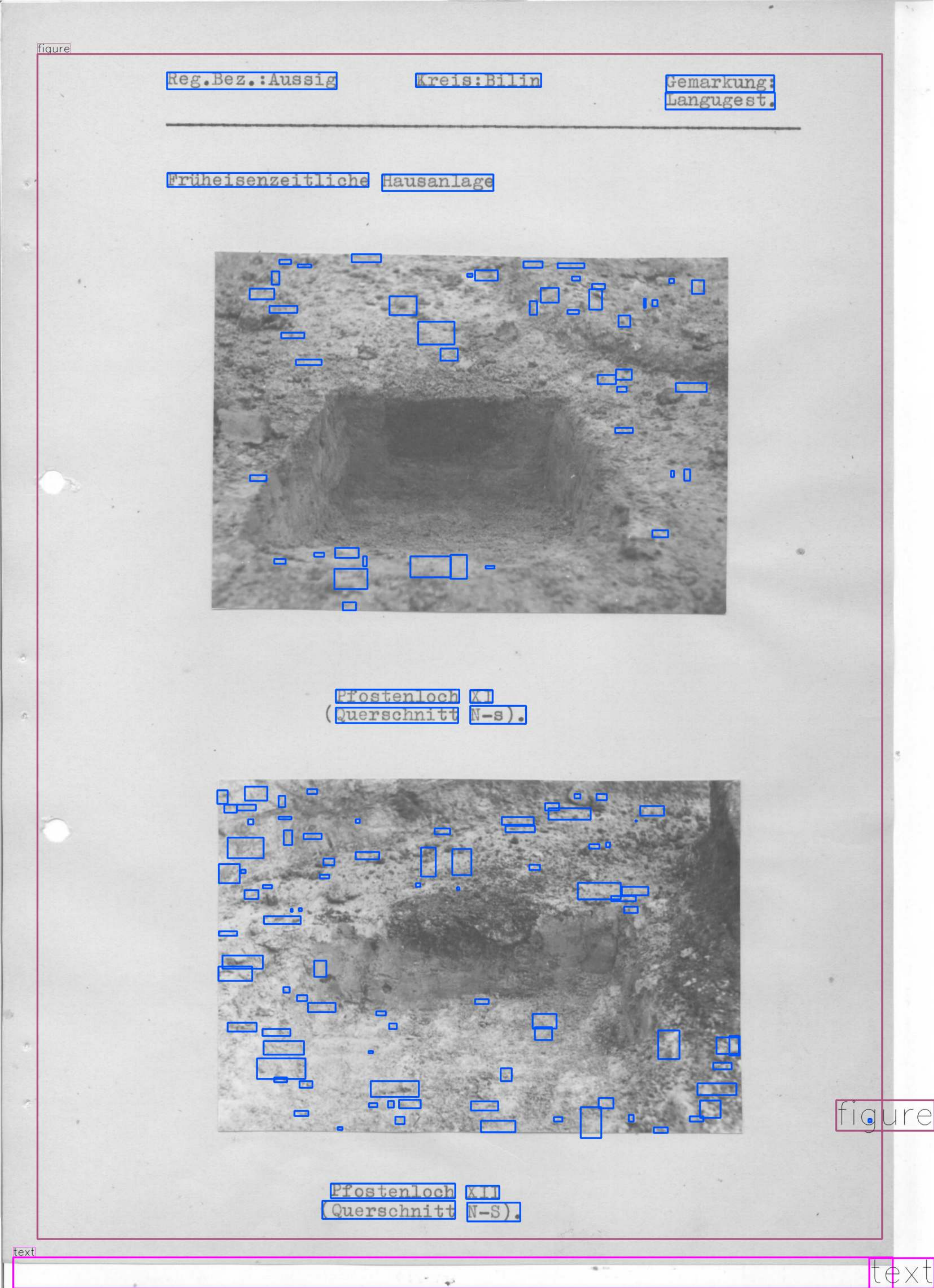}
        \caption{Imaginary characters}
        \label{fig:dd_mis_fig1}
    \end{subfigure}
    % \hfill
    \begin{subfigure}{0.3\textwidth}
        \centering
        \includegraphics[width=\textwidth]{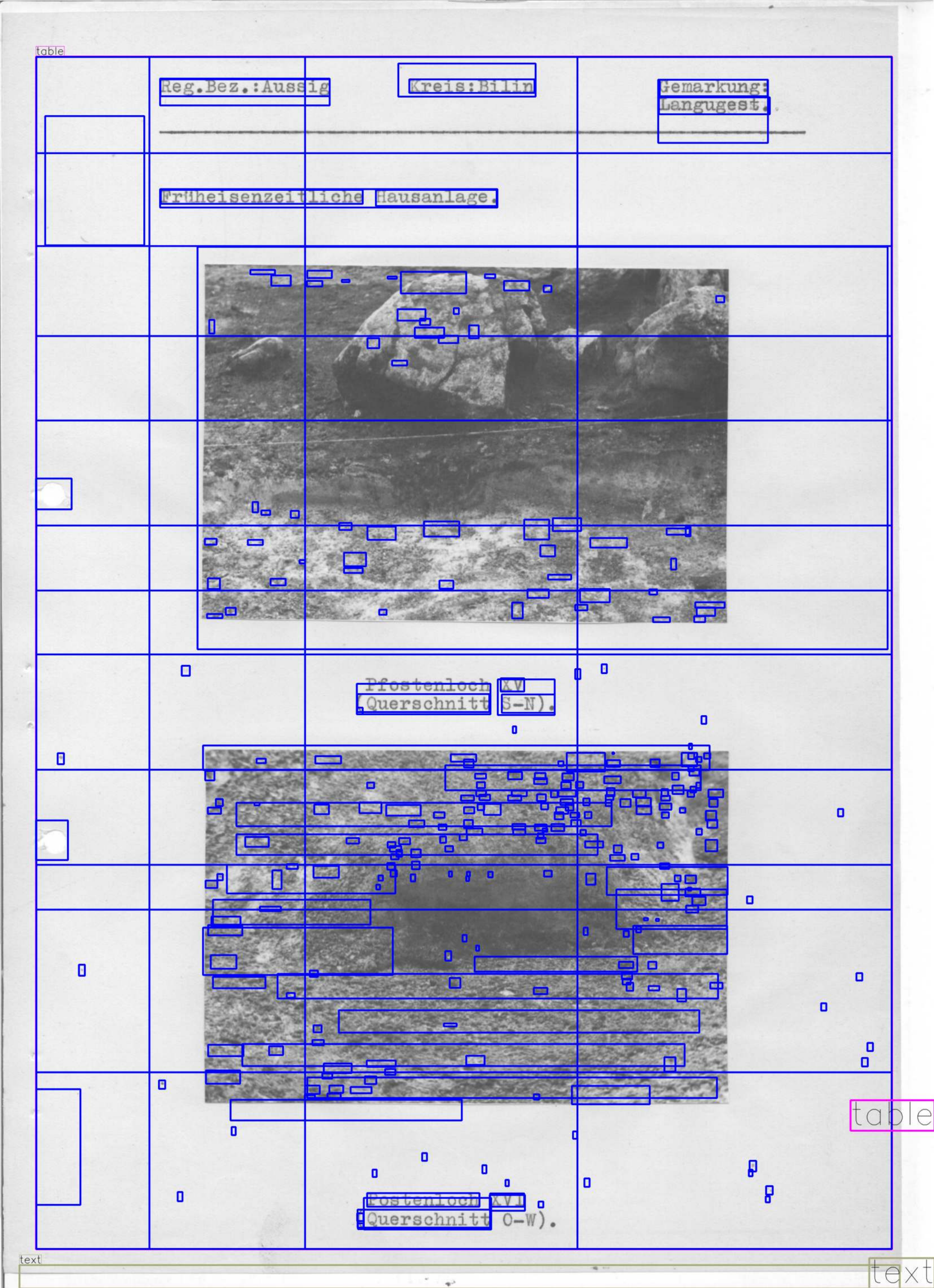}
        \caption{Imaginary table}
        \label{fig:dd_mis_fig2}
    \end{subfigure}
    % \hfill
    \begin{subfigure}{0.3\textwidth}
        \centering
        \includegraphics[width=\textwidth]{img/pages/dd_figure_mis_3.pdf}
        \caption{Missed paragraph}
        \label{fig:dd_mis_fig3}
    \end{subfigure}
\end{figure}

\begin{figure}[h]
    \centering
    \caption{\ac{dd} mistakes on pages with drawings}
    \label{fig:dd_mis_draw}
    \begin{subfigure}{0.3\textwidth}
        \centering
        \includegraphics[width=\textwidth]{img/pages/dd_draw_mis_1.pdf}
        \caption{Missed table \& drawing}
        \label{fig:dd_mis_drw1}
    \end{subfigure}
    % \hfill
    \begin{subfigure}{0.3\textwidth}
        \centering
        \includegraphics[width=\textwidth]{img/pages/dd_draw_mis_2.pdf}
        \caption{Partial table}
        \label{fig:dd_mis_drw2}
    \end{subfigure}
    % \hfill
    \begin{subfigure}{0.3\textwidth}
        \centering
        \includegraphics[width=\textwidth]{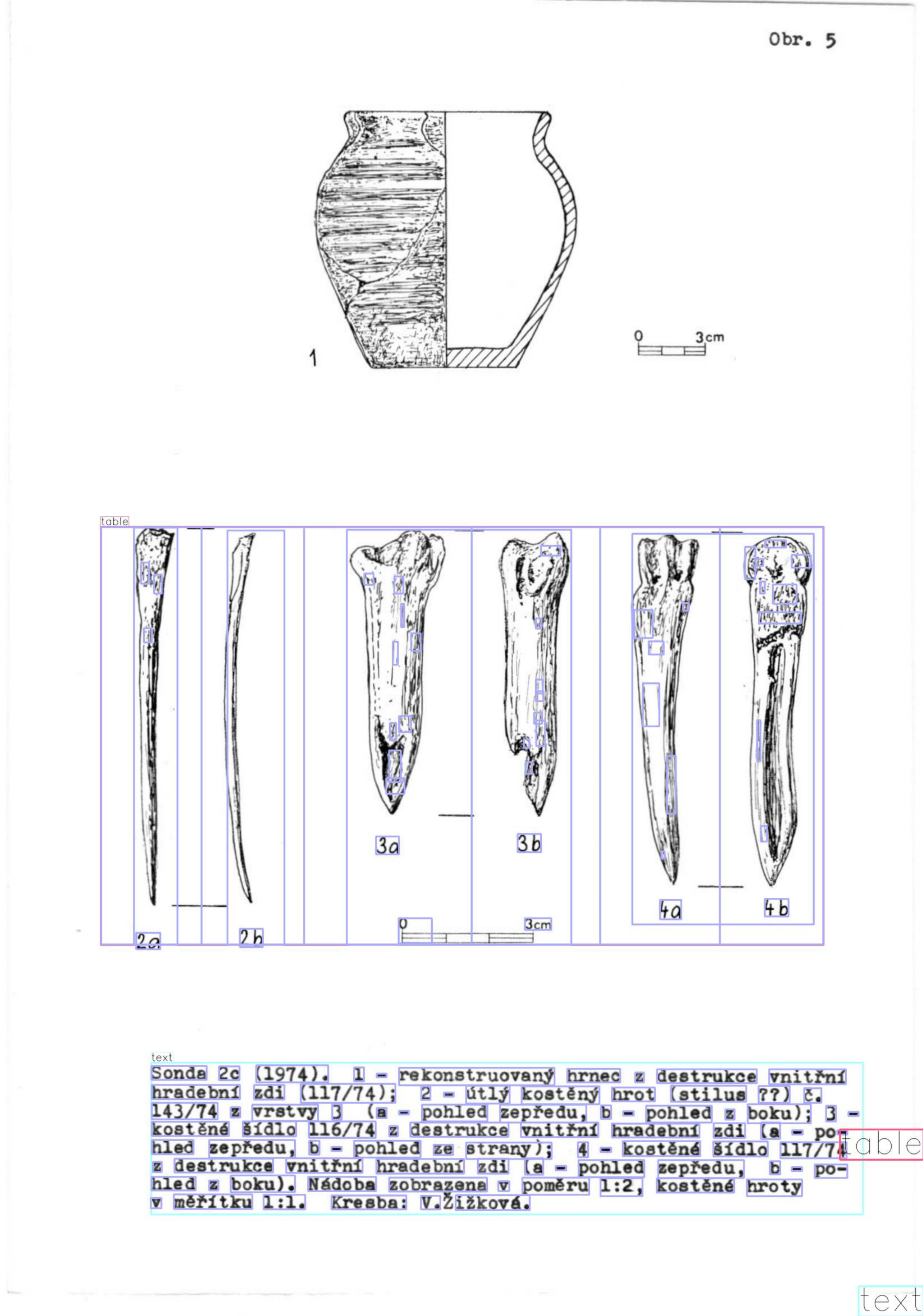}
        \caption{Missed drawing}
        \label{fig:dd_mis_drw3}
    \end{subfigure}
\end{figure}

\begin{figure}[h]
    \centering
    \caption{\ac{dd} mistakes on pages with tables}
    \label{fig:dd_mis_tables}
    \begin{subfigure}{0.3\textwidth}
        \centering
        \includegraphics[width=\textwidth]{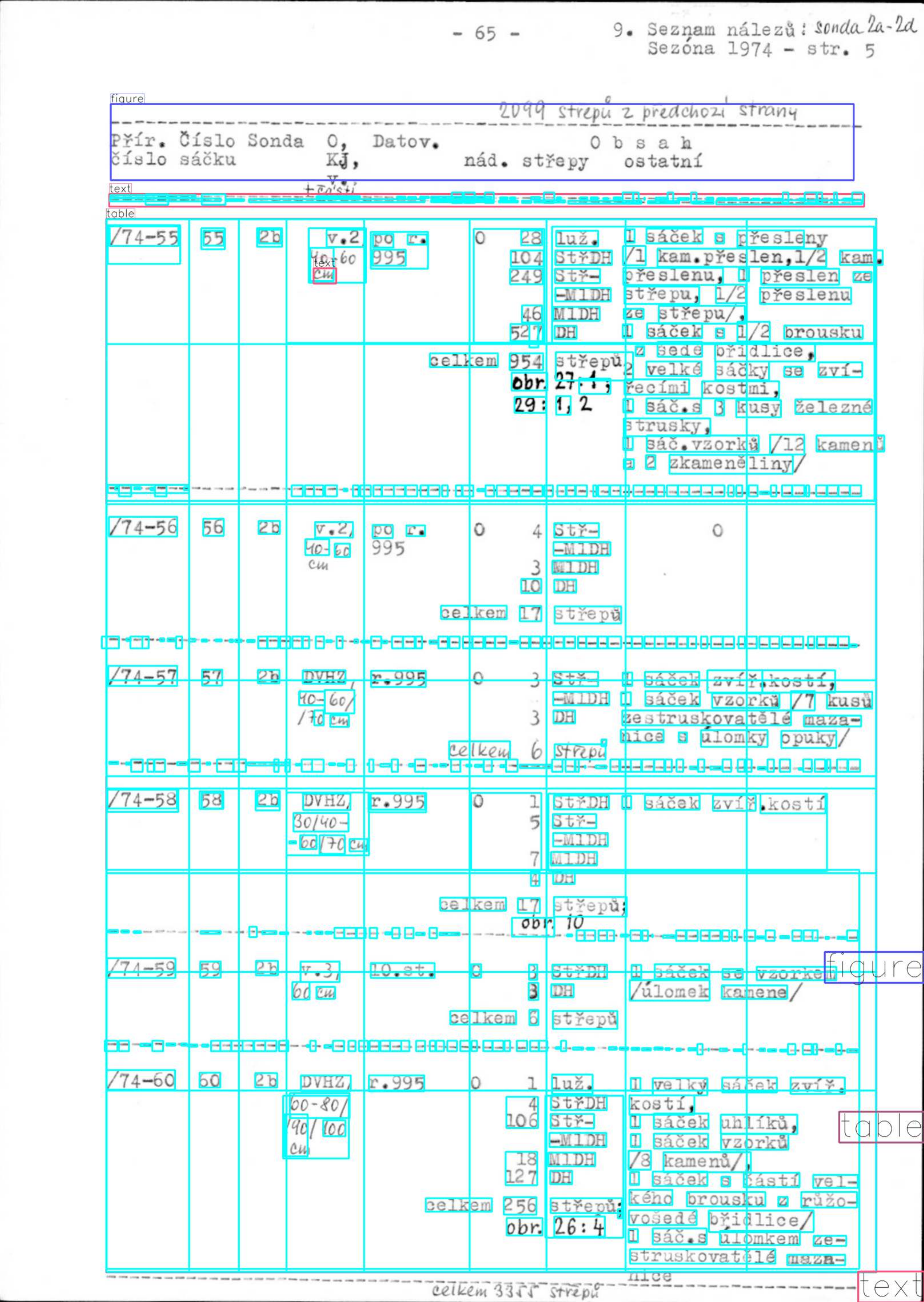}
        \caption{Text as a figure}
        \label{fig:dd_mis_tab1}
    \end{subfigure}
    % \hfill
    \begin{subfigure}{0.3\textwidth}
        \centering
        \includegraphics[width=\textwidth]{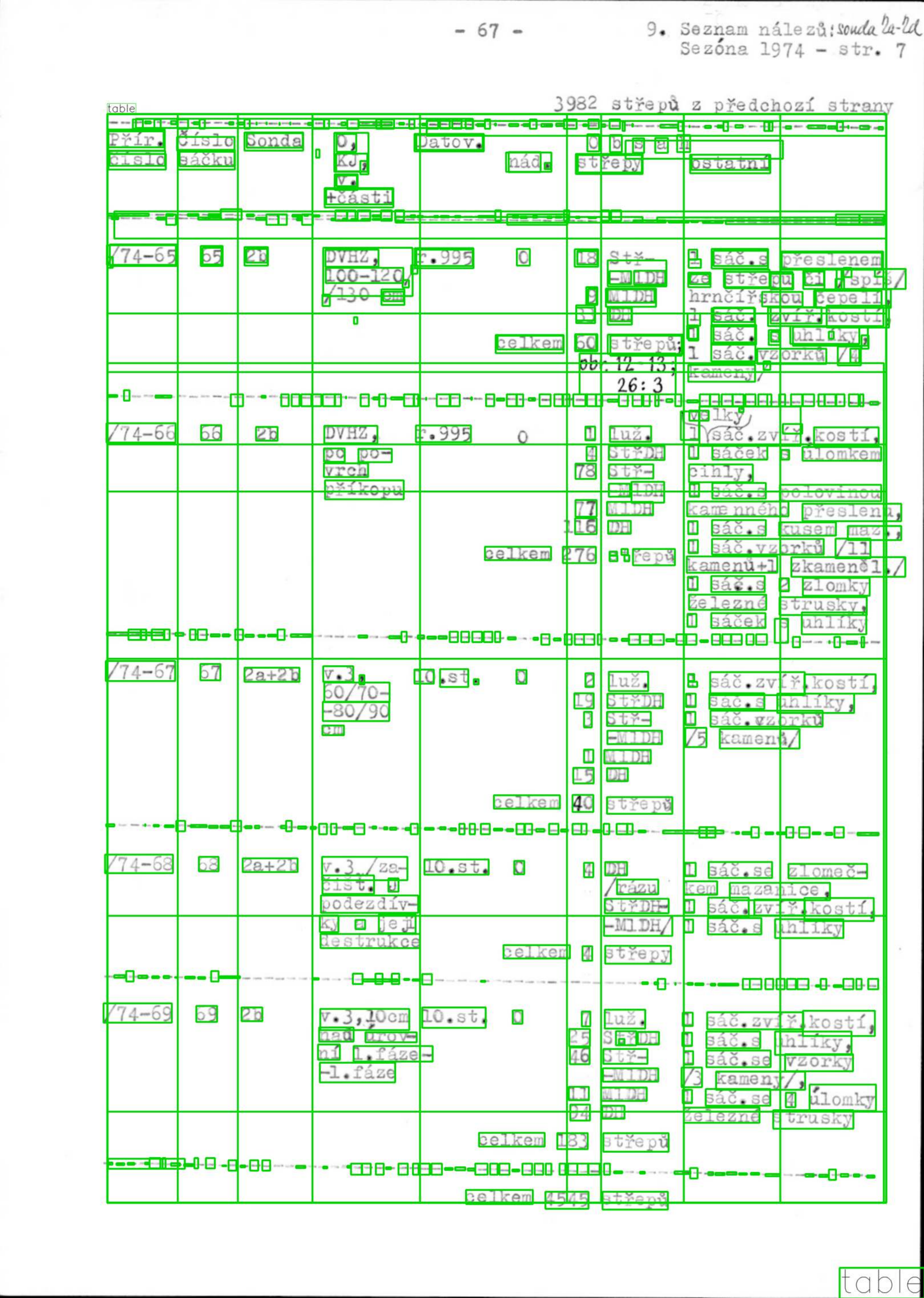}
        \caption{Text as a row}
        \label{fig:dd_mis_tab2}
    \end{subfigure}
    % \hfill
    \begin{subfigure}{0.3\textwidth}
        \centering
        \includegraphics[width=\textwidth]{img/pages/dd_table_mis_3.pdf}
        \caption{Imaginary cells}
        \label{fig:dd_mis_tab3}
    \end{subfigure}
\end{figure}

\begin{figure}[h]
    \centering
    \caption{\ac{dd} mistakes on pages with plain texts}
    \label{fig:dd_mis_texts}
    \begin{subfigure}{0.3\textwidth}
        \centering
        \includegraphics[width=\textwidth]{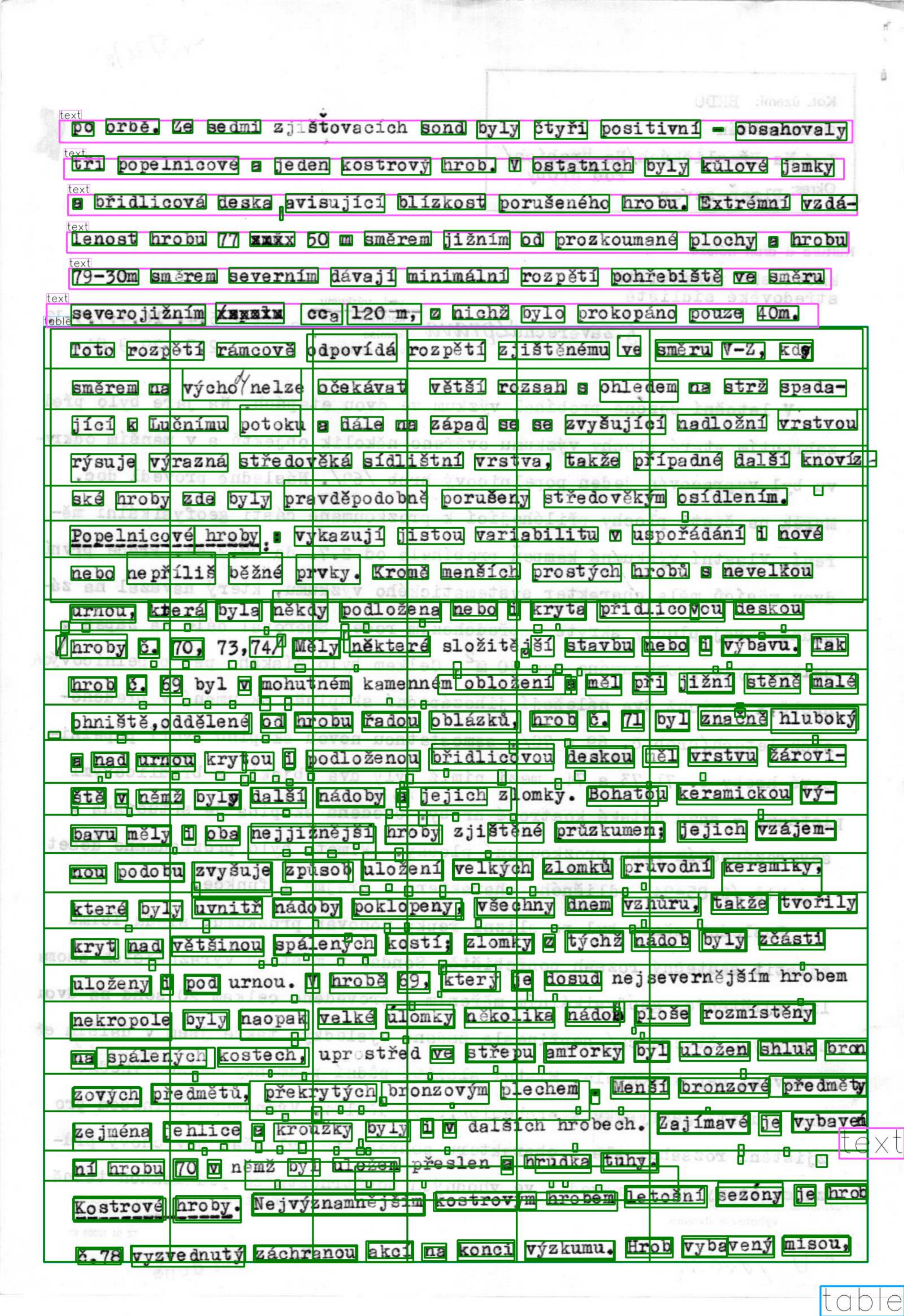}
        \caption{Text as table}
        \label{fig:dd_mis_txt1}
    \end{subfigure}
    % \hfill
    \begin{subfigure}{0.3\textwidth}
        \centering
        \includegraphics[width=\textwidth]{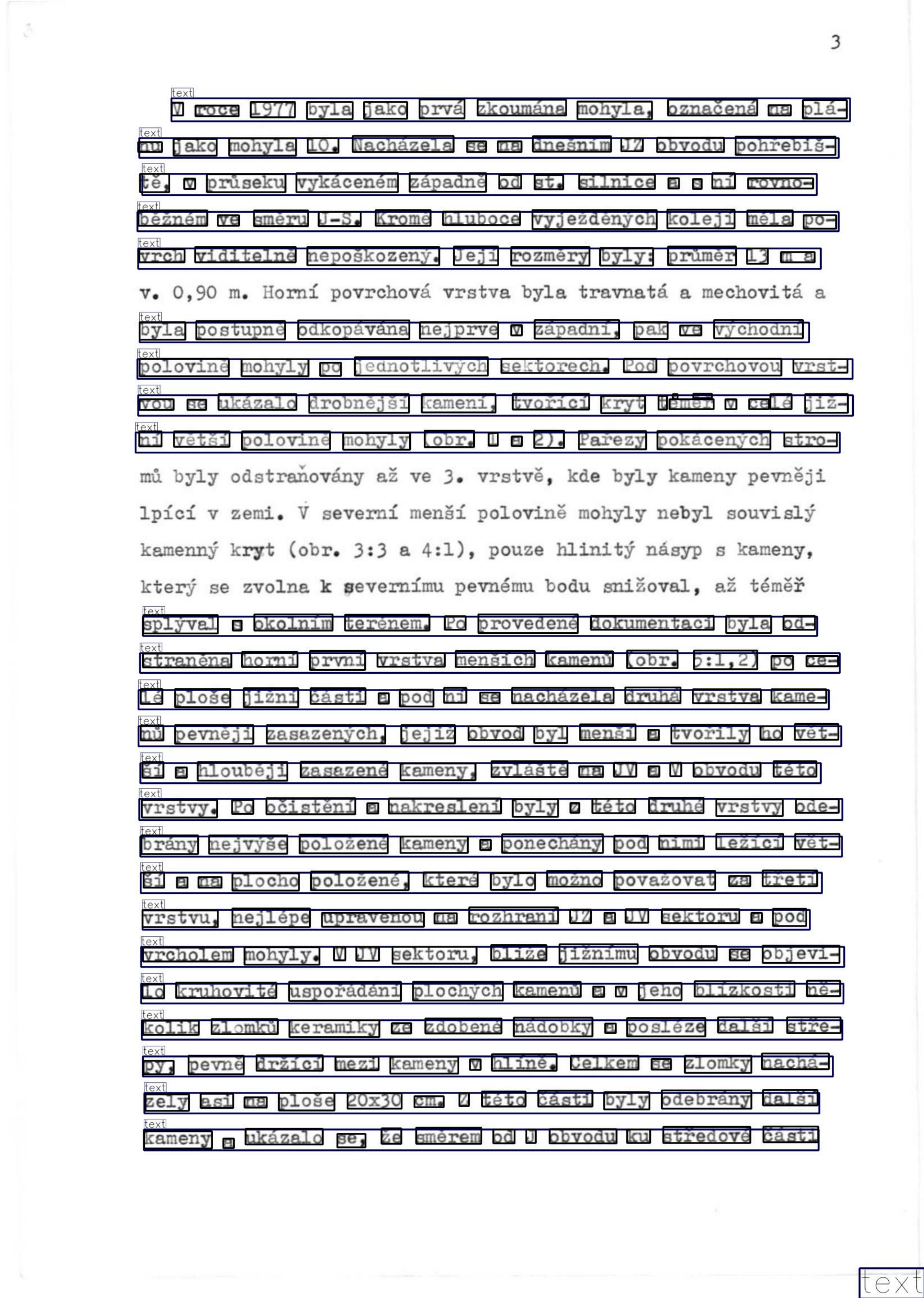}
        \caption{Missed font}
        \label{fig:dd_mis_txt2}
    \end{subfigure}
    % \hfill
    \begin{subfigure}{0.3\textwidth}
        \centering
        \includegraphics[width=\textwidth]{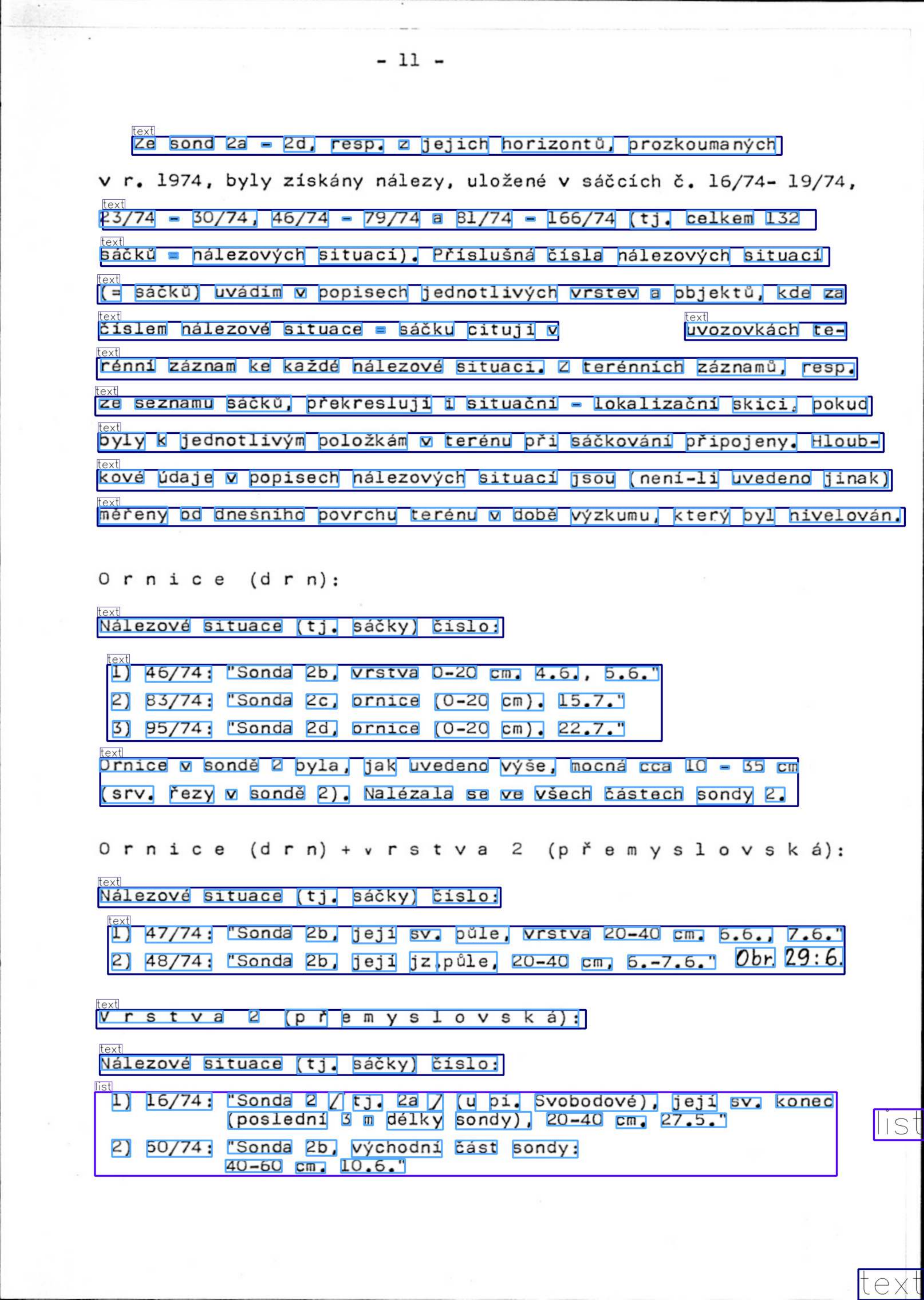}
        \caption{Missed lines}
        \label{fig:dd_mis_txt3}
    \end{subfigure}
\end{figure}

\chapter{Label examples} \label{app-labels}

% DRAW
\begin{figure}[h]
    \centering
    \caption{Label DRAW examples}
    \label{fig:example_draw}
    \begin{subfigure}{0.3\textwidth}
        \centering
        \includegraphics[width=\textwidth]{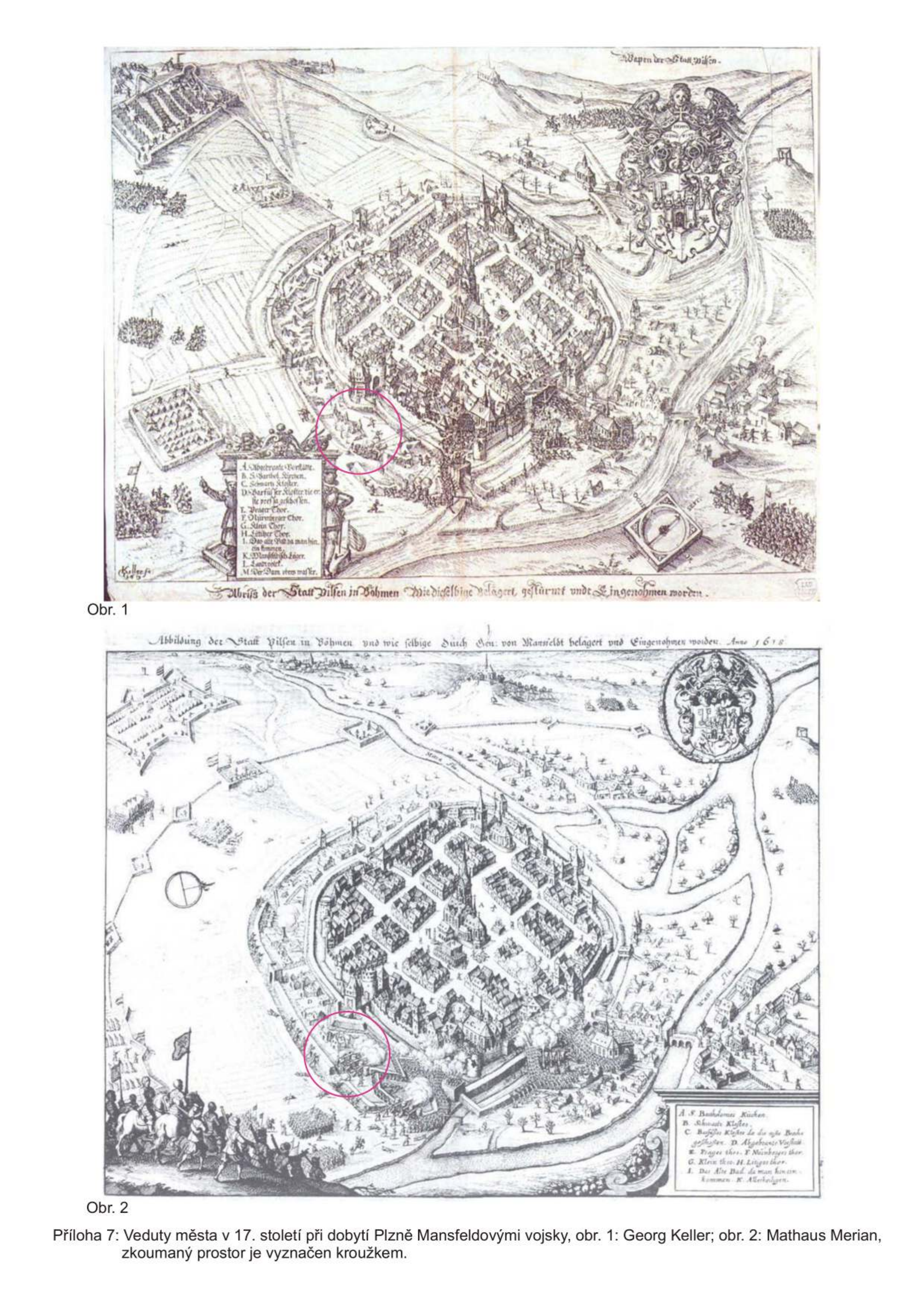}
        \caption{City drawings}
        \label{fig:e_draw_1}
    \end{subfigure}
    \begin{subfigure}{0.3\textwidth}
        \centering
        \includegraphics[width=\textwidth]{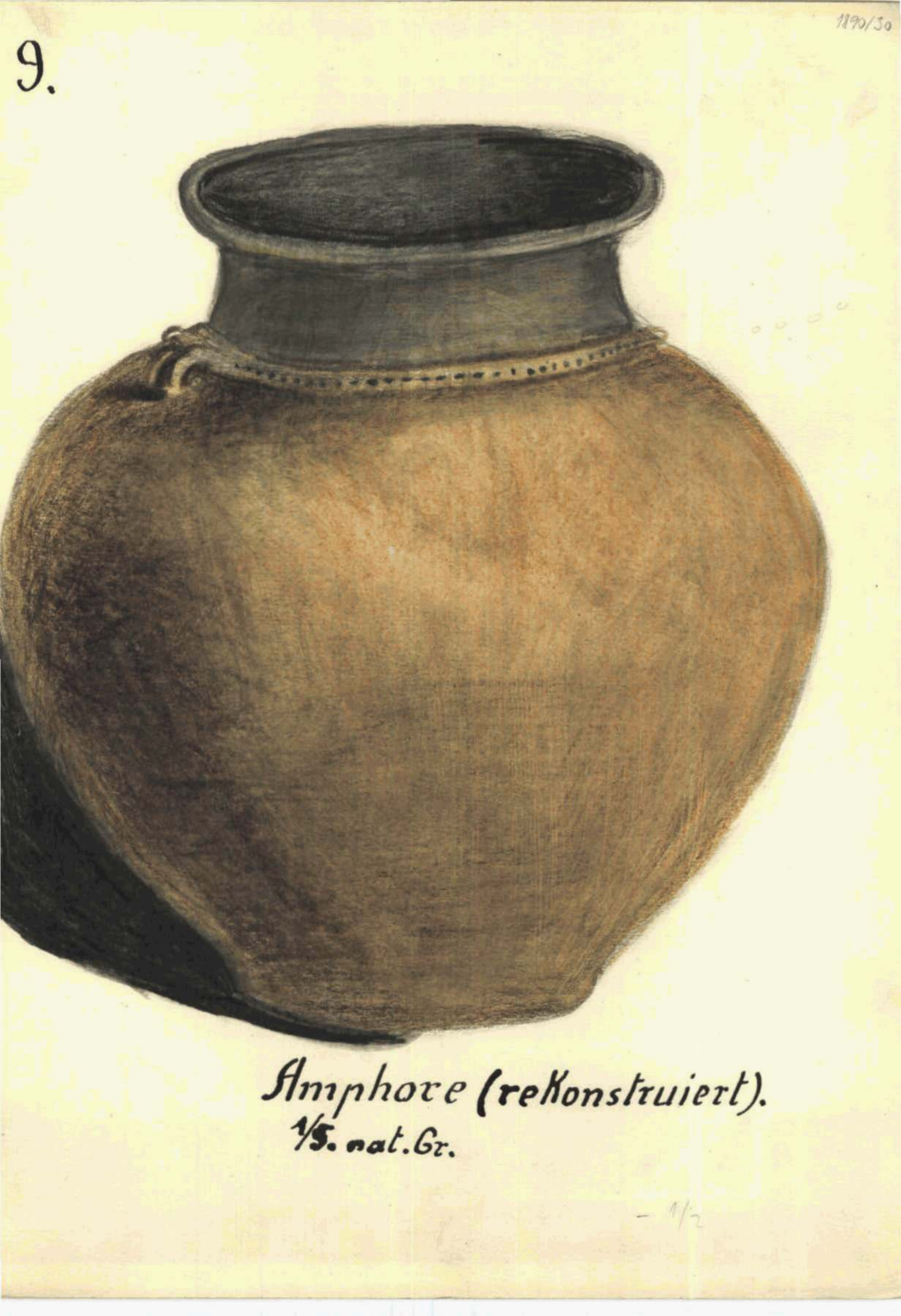}
        \caption{Realistic painting}
        \label{fig:e_draw_2}
    \end{subfigure}
    \begin{subfigure}{0.3\textwidth}
        \centering
        \includegraphics[width=\textwidth]{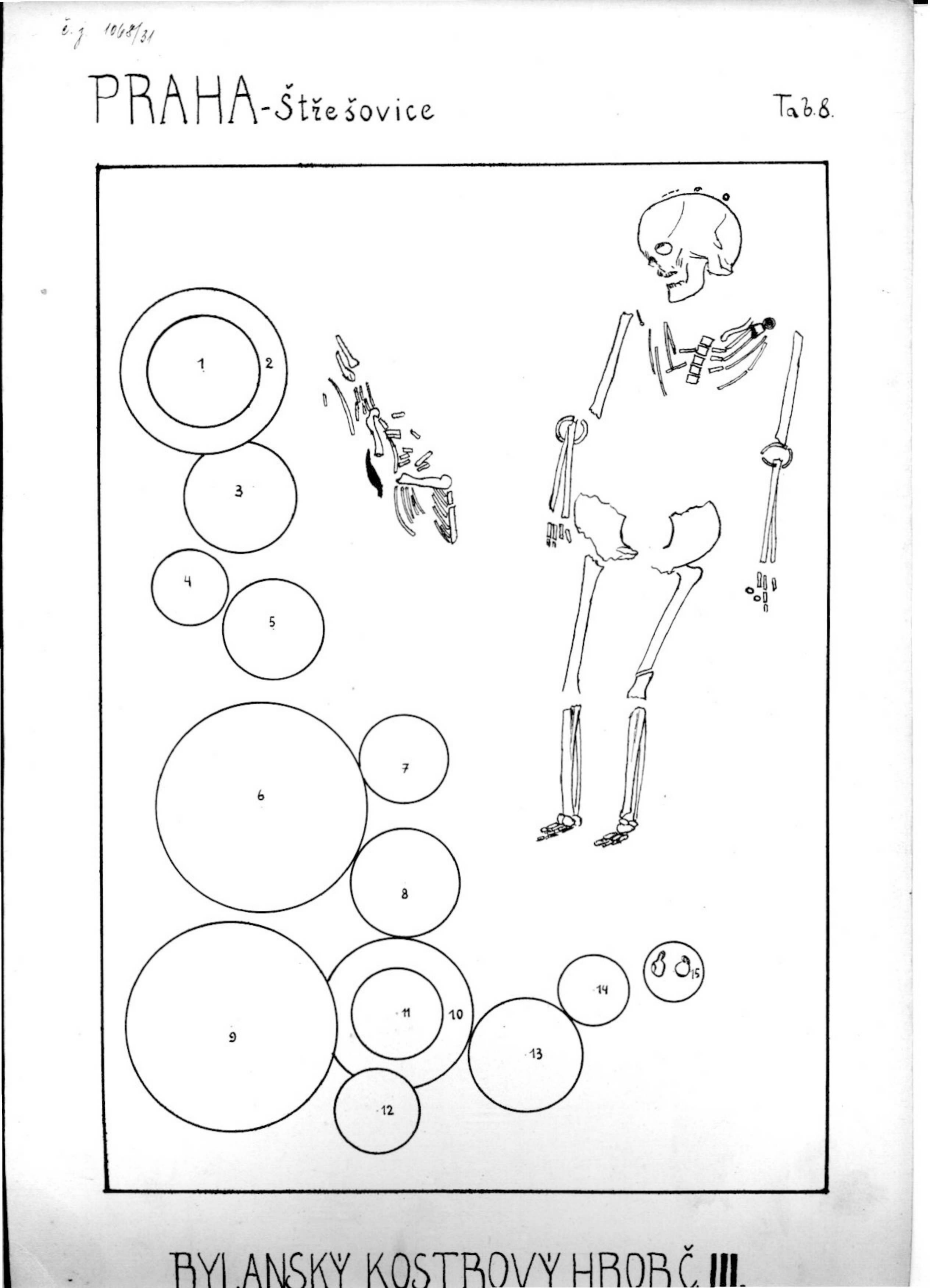}
        \caption{Ground schematic}
        \label{fig:e_draw_3}
    \end{subfigure}

    \vspace{0.5em}

    \begin{subfigure}{0.45\textwidth}
        \centering
        \includegraphics[width=\textwidth]{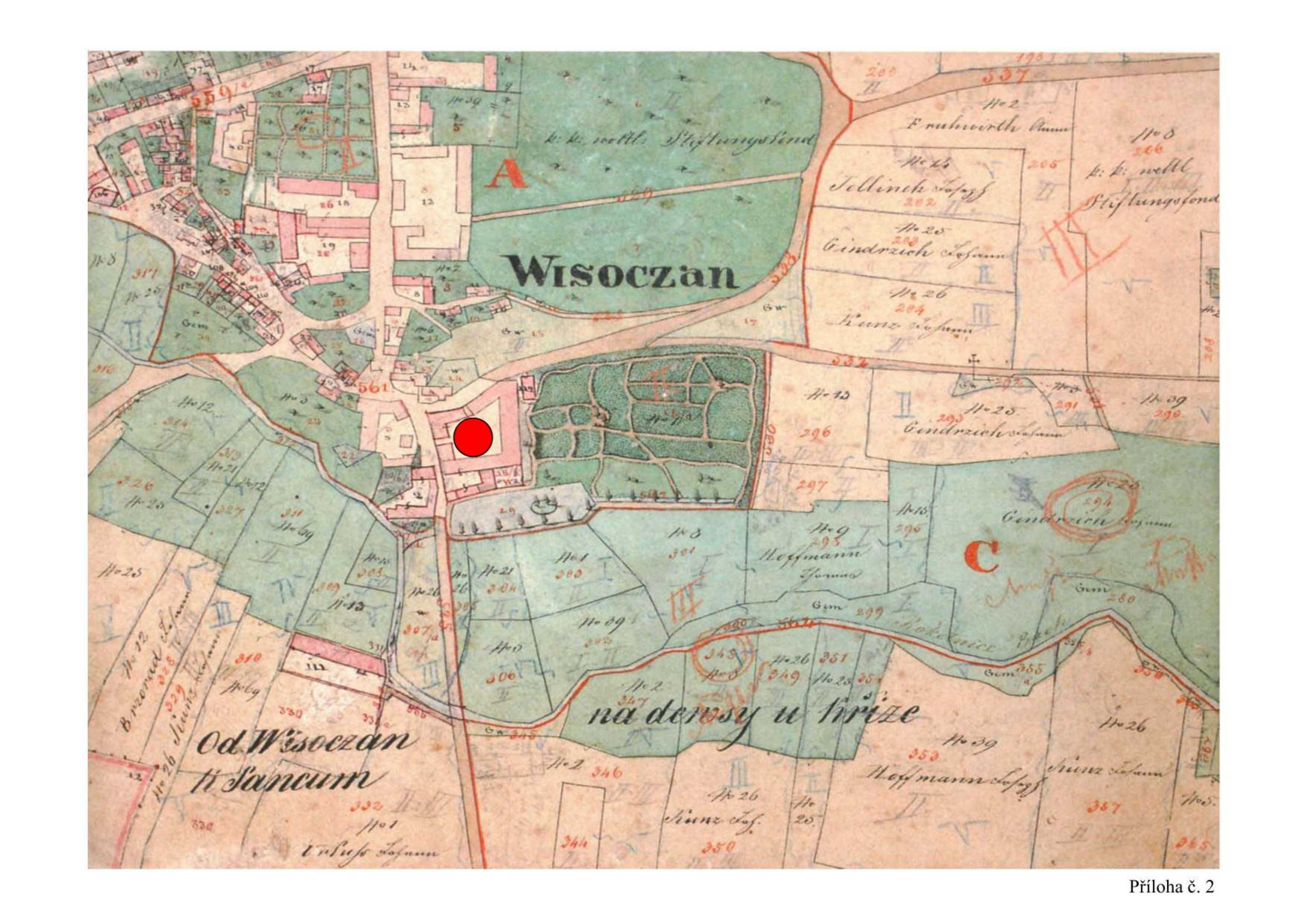}
        \caption{Territory map}
        \label{fig:e_draw_4}
    \end{subfigure}
    \begin{subfigure}{0.45\textwidth}
        \centering
        \includegraphics[width=\textwidth]{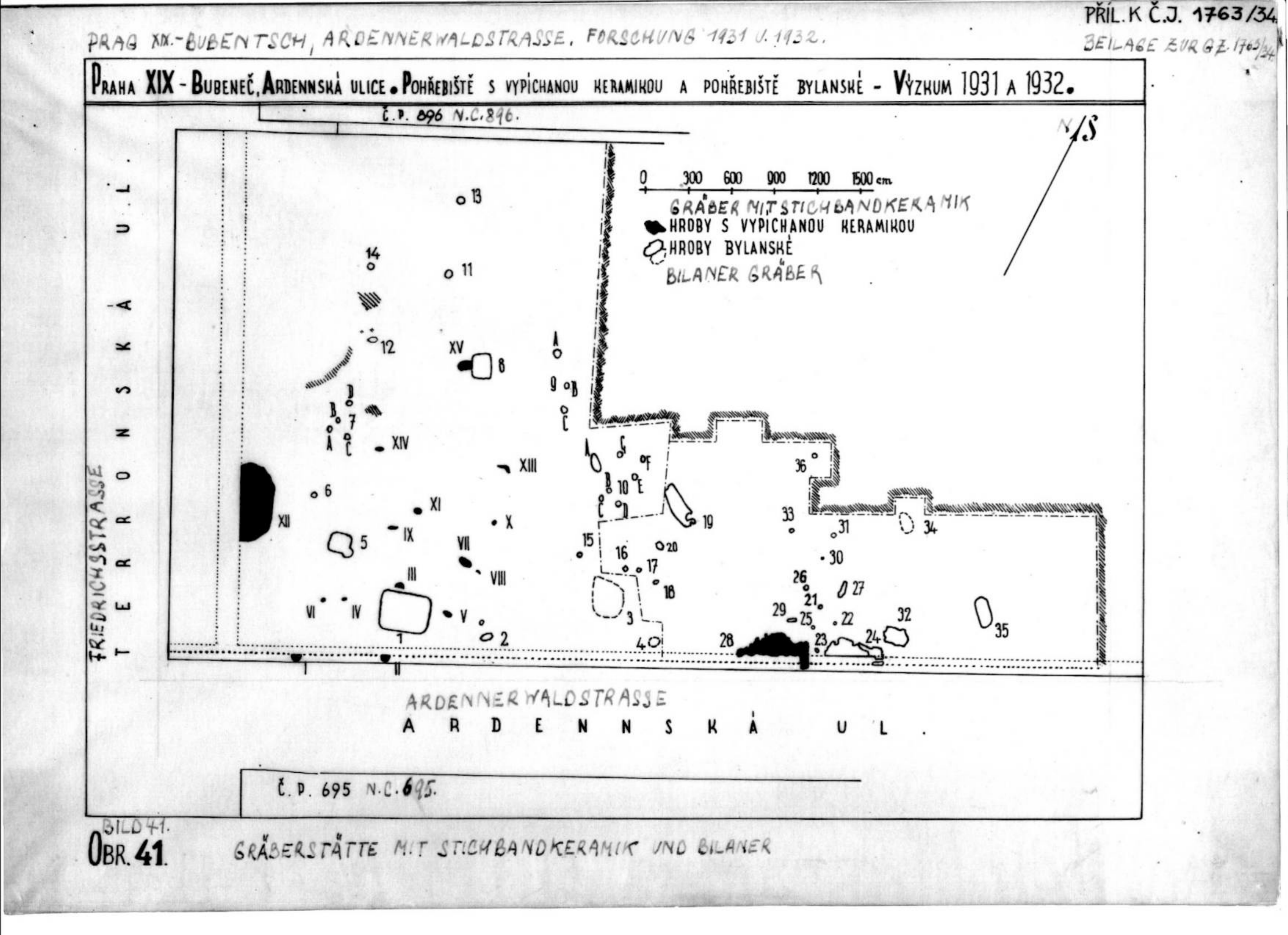}
        \caption{Building plan}
        \label{fig:e_draw_5}
    \end{subfigure}

    \vspace{0.5em}

    \begin{subfigure}{0.45\textwidth}
        \centering
        \includegraphics[width=\textwidth]{img/pages/example_DRAW_book.pdf}
        \caption{Within book scan}
        \label{fig:e_draw_6}
    \end{subfigure}
    \begin{subfigure}{0.45\textwidth}
        \centering
        \includegraphics[width=\textwidth]{img/pages/example_DRAW_hw.pdf}
        \caption{Within written notes}
        \label{fig:e_draw_7}
    \end{subfigure}
\end{figure}

% DRAW_L
\begin{figure}[h]
    \centering
    \caption{Label DRAW\_L examples}
    \label{fig:example_draw_l}
    \begin{subfigure}{0.3\textwidth}
        \centering
        \includegraphics[width=\textwidth]{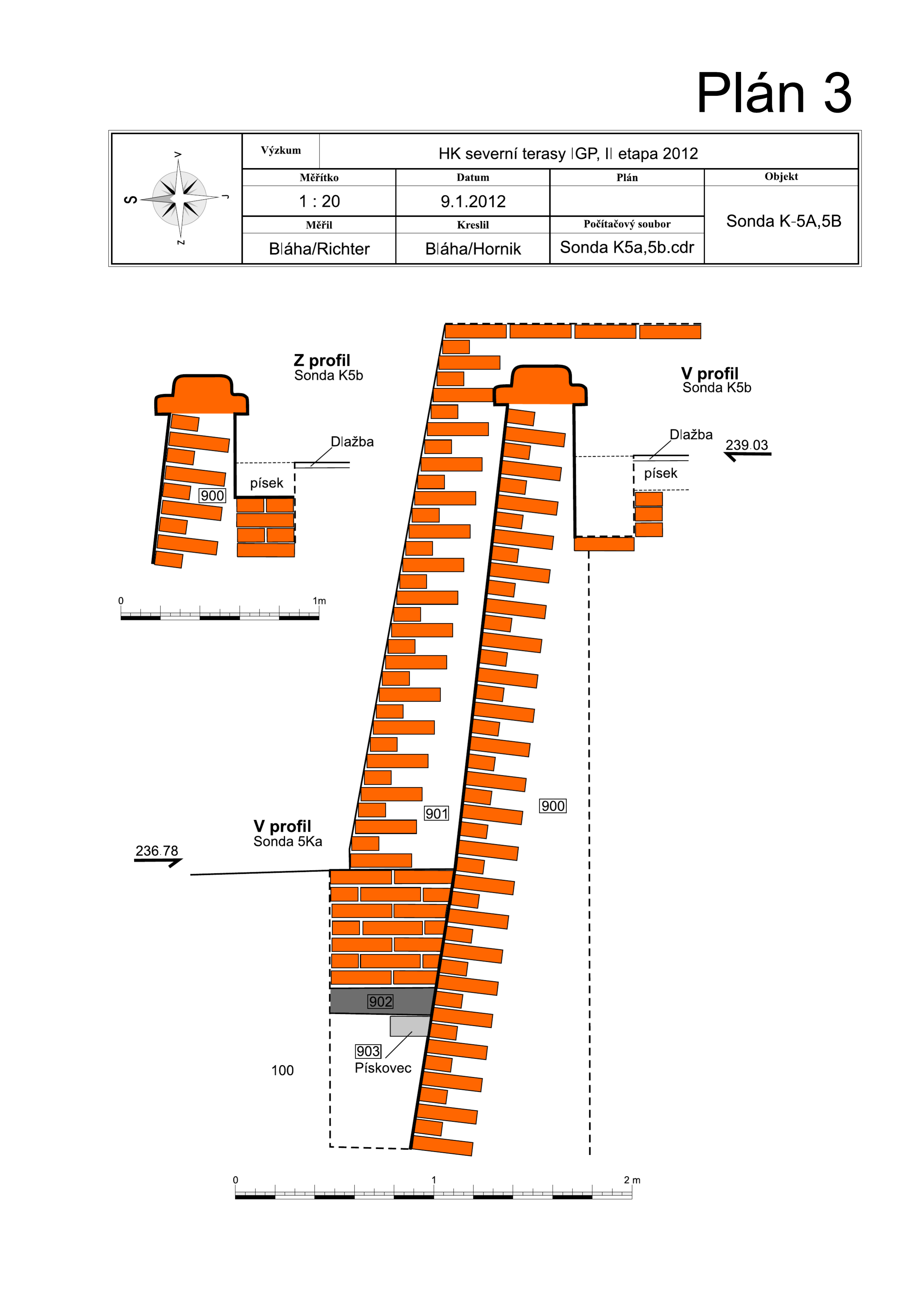}
        \caption{Scheme with a legend}
        \label{fig:e_draw_l_1}
    \end{subfigure}
    \begin{subfigure}{0.3\textwidth}
        \centering
        \includegraphics[width=\textwidth]{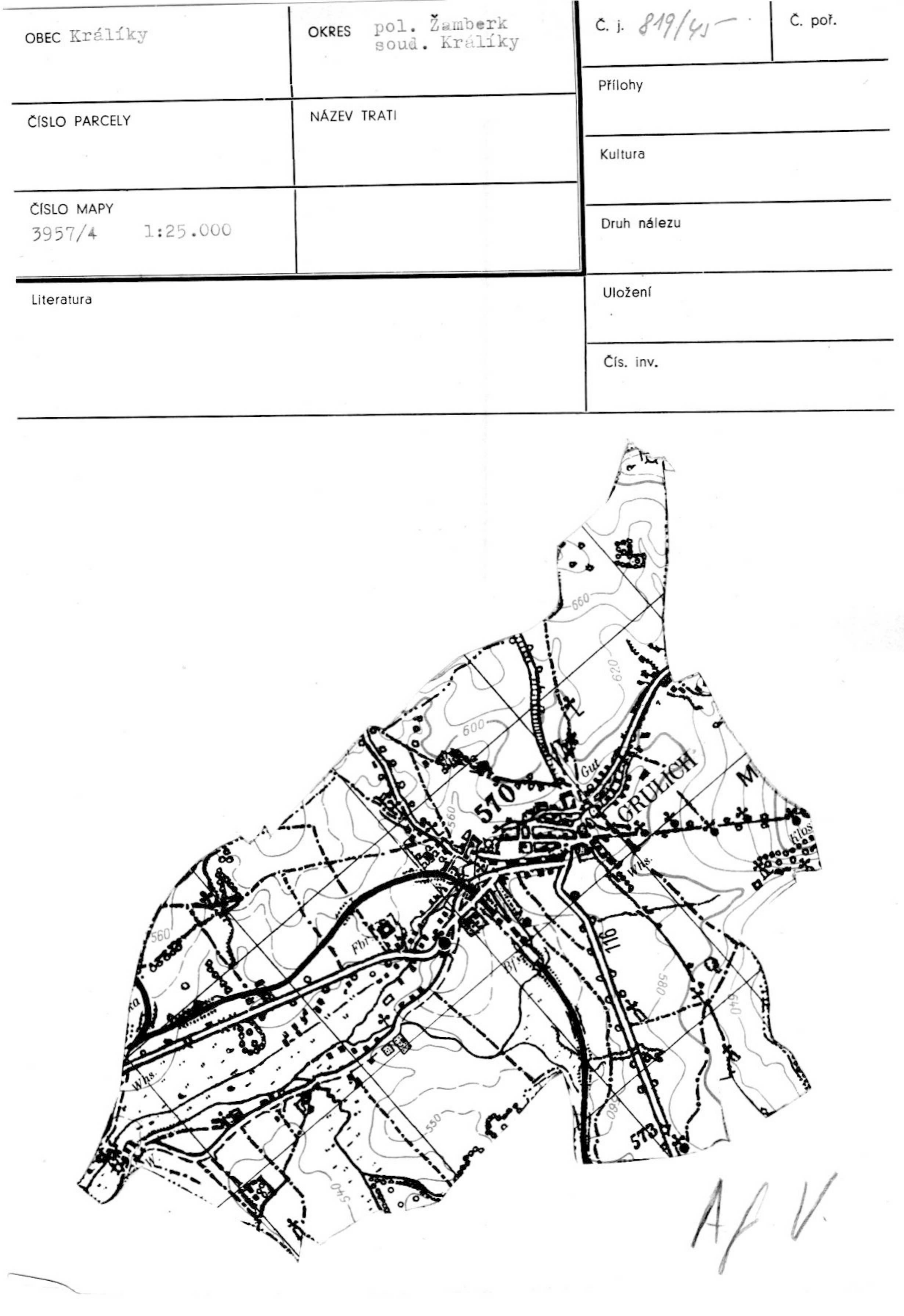}
        \caption{Map inside a form}
        \label{fig:e_draw_l_2}
    \end{subfigure}
    \begin{subfigure}{0.3\textwidth}
        \centering
        \includegraphics[width=\textwidth]{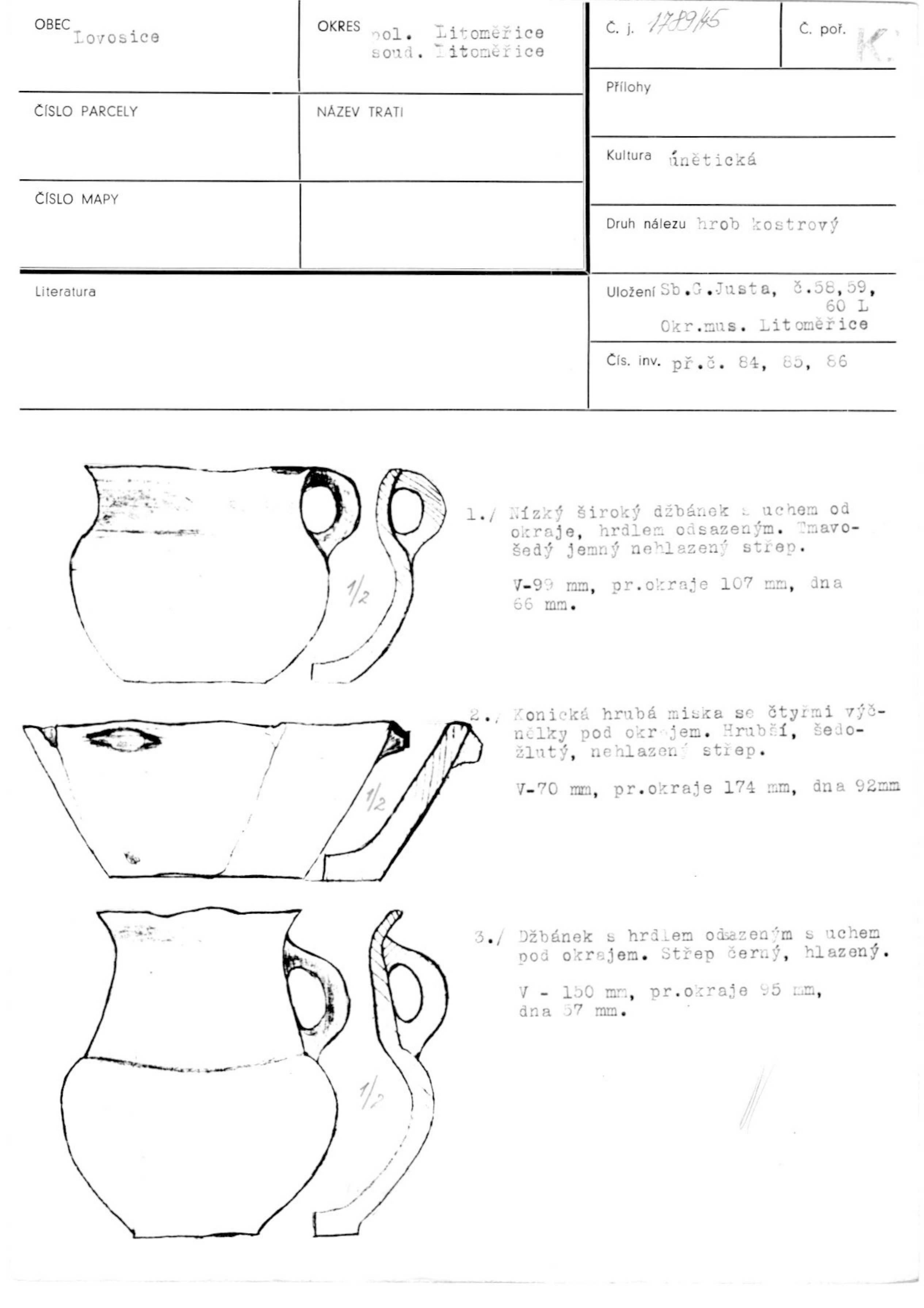}
        \caption{Drawing inside a form}
        \label{fig:e_draw_l_3}
    \end{subfigure}

    \vspace{0.5em}

    \begin{subfigure}{0.45\textwidth}
        \centering
        \includegraphics[width=\textwidth]{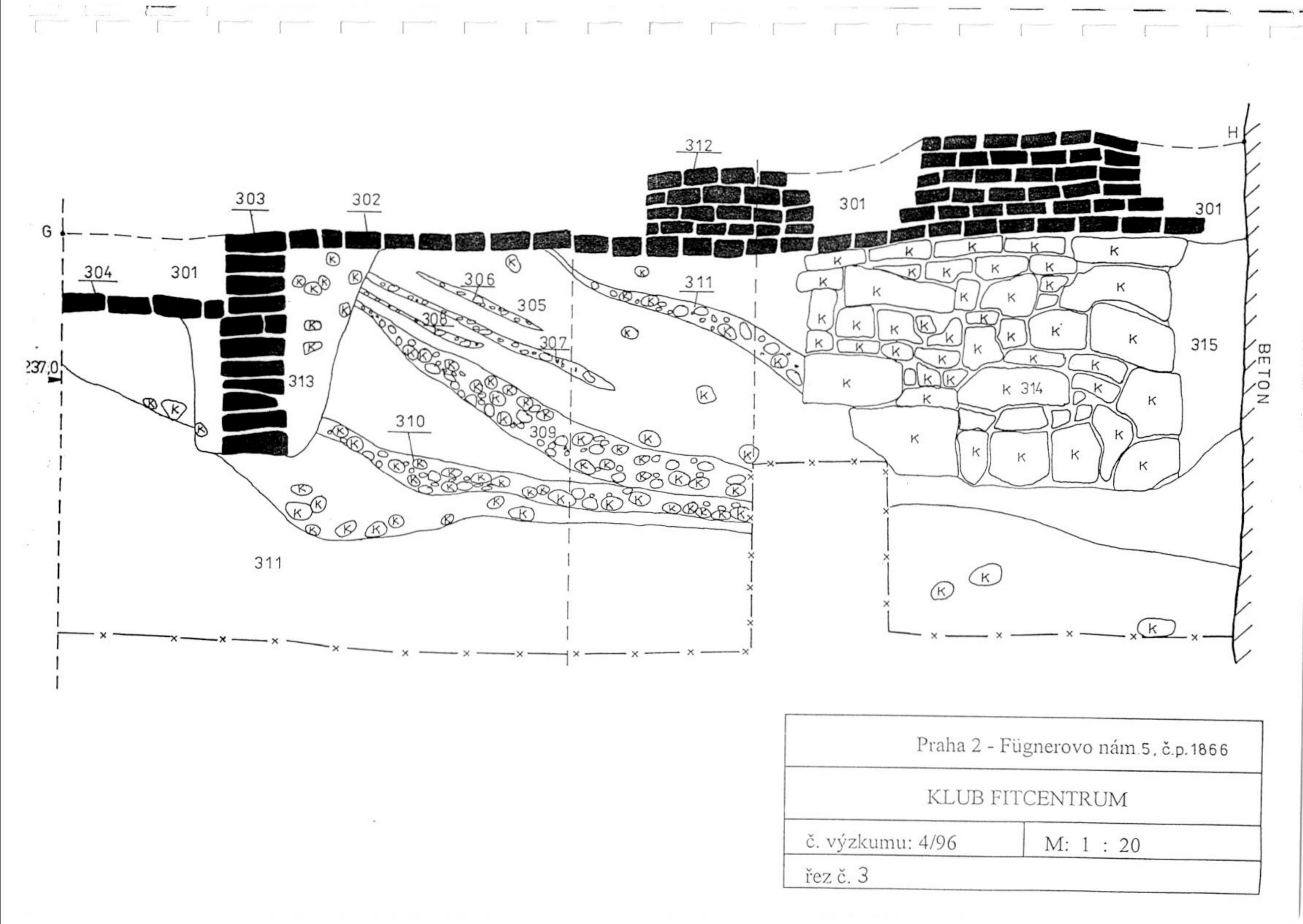}
        \caption{Wall drawing}
        \label{fig:e_draw_l_4}
    \end{subfigure}
    \begin{subfigure}{0.45\textwidth}
        \centering
        \includegraphics[width=\textwidth]{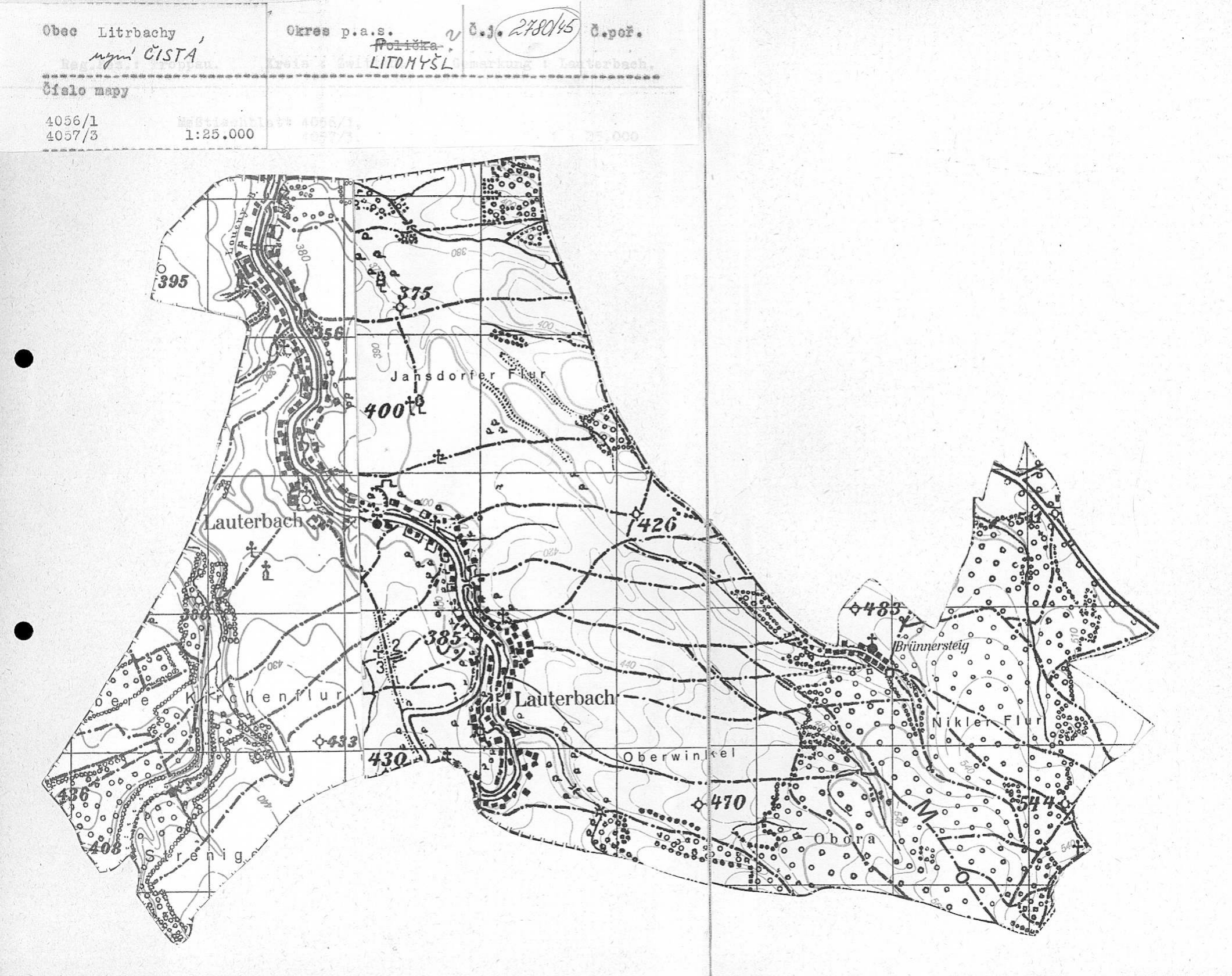}
        \caption{Territory map}
        \label{fig:e_draw_l_5}
    \end{subfigure}

    \vspace{0.5em}

    \begin{subfigure}{0.45\textwidth}
        \centering
        \includegraphics[width=\textwidth]{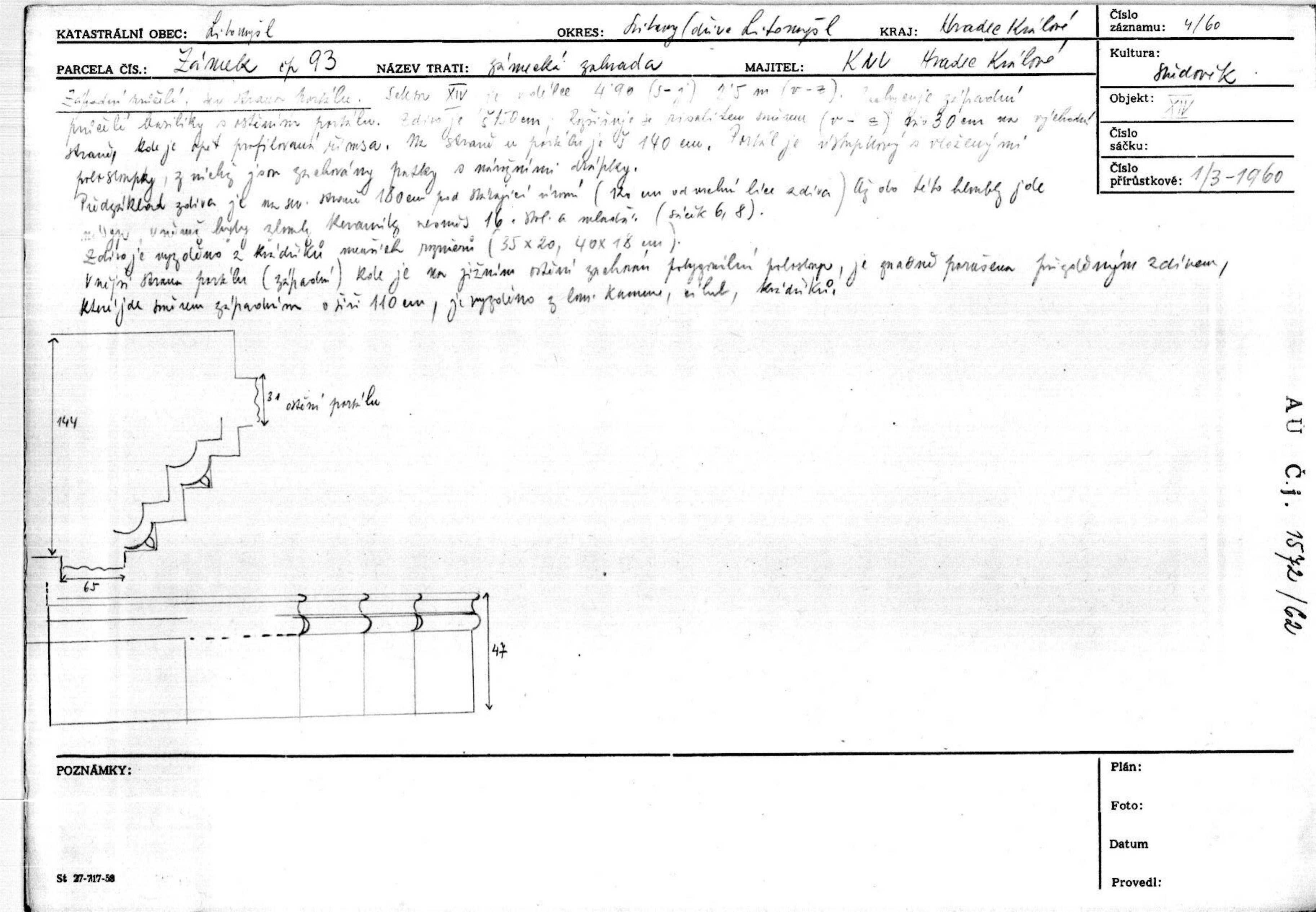}
        \caption{Schema in a form}
        \label{fig:e_draw_l_6}
    \end{subfigure}
    \begin{subfigure}{0.45\textwidth}
        \centering
        \includegraphics[width=\textwidth]{img/pages/example_DRAW_L_build.pdf}
        \caption{Buildings top view}
        \label{fig:e_draw_l_7}
    \end{subfigure}
\end{figure}

% LINE_HW
\begin{figure}[h]
    \centering
    \caption{Label LINE\_HW examples}
    \label{fig:example_line_hw}
    \begin{subfigure}{0.3\textwidth}
        \centering
        \includegraphics[width=\textwidth]{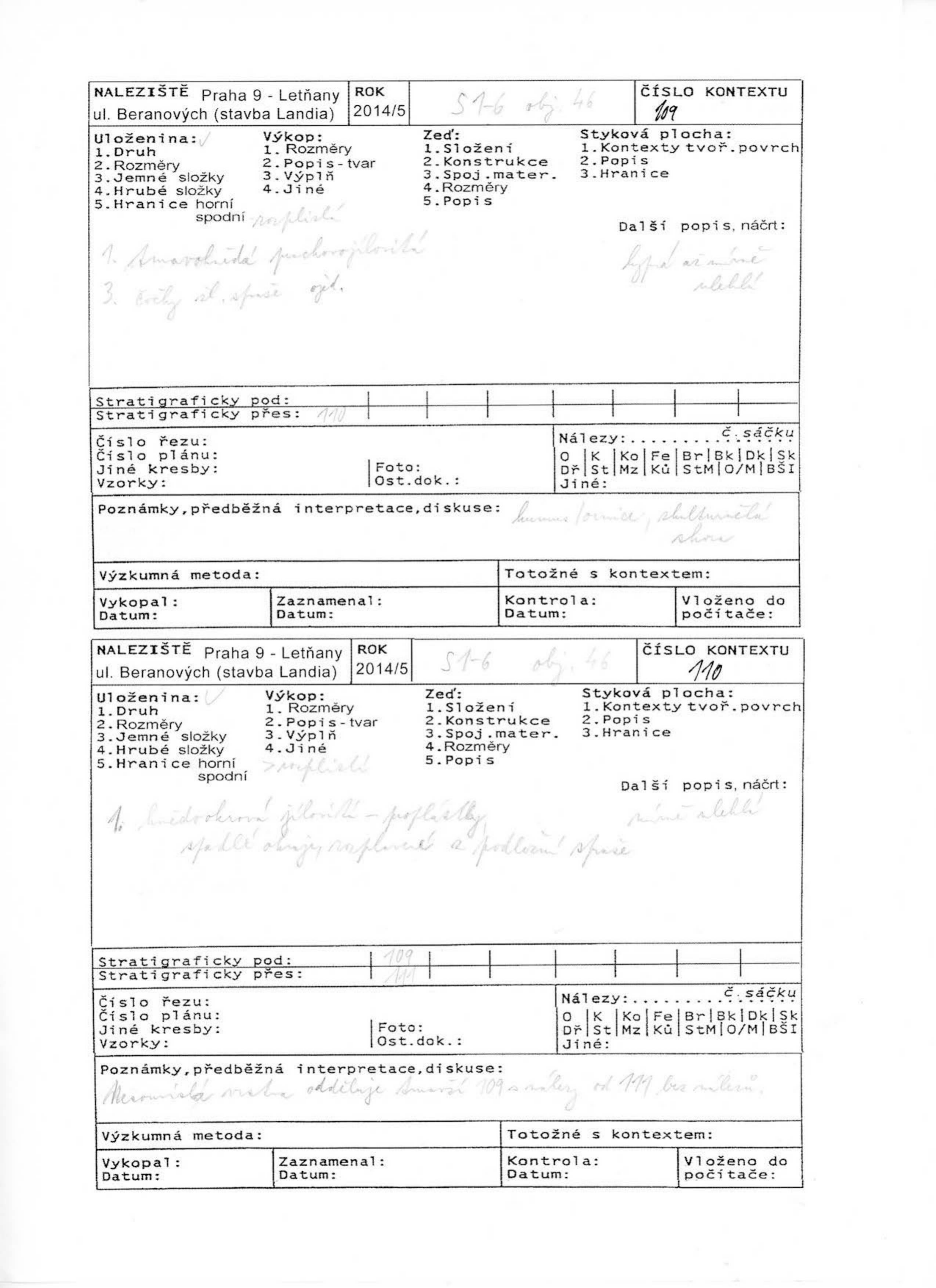}
        \caption{Manually filled-in form}
        \label{fig:e_line_hw_1}
    \end{subfigure}
    \begin{subfigure}{0.3\textwidth}
        \centering
        \includegraphics[width=\textwidth]{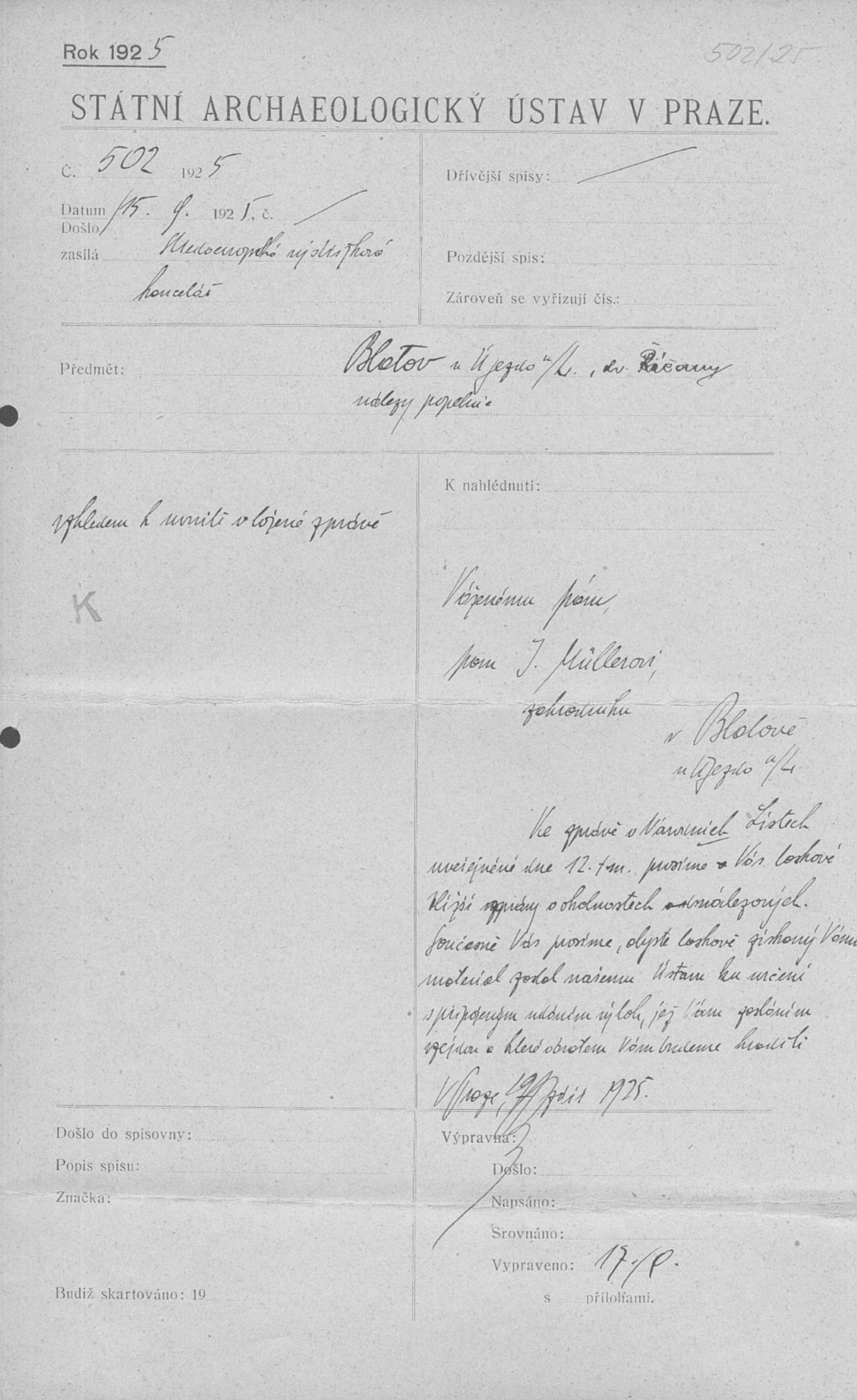}
        \caption{Gray paper form}
        \label{fig:e_line_hw_2}
    \end{subfigure}
    \begin{subfigure}{0.3\textwidth}
        \centering
        \includegraphics[width=\textwidth]{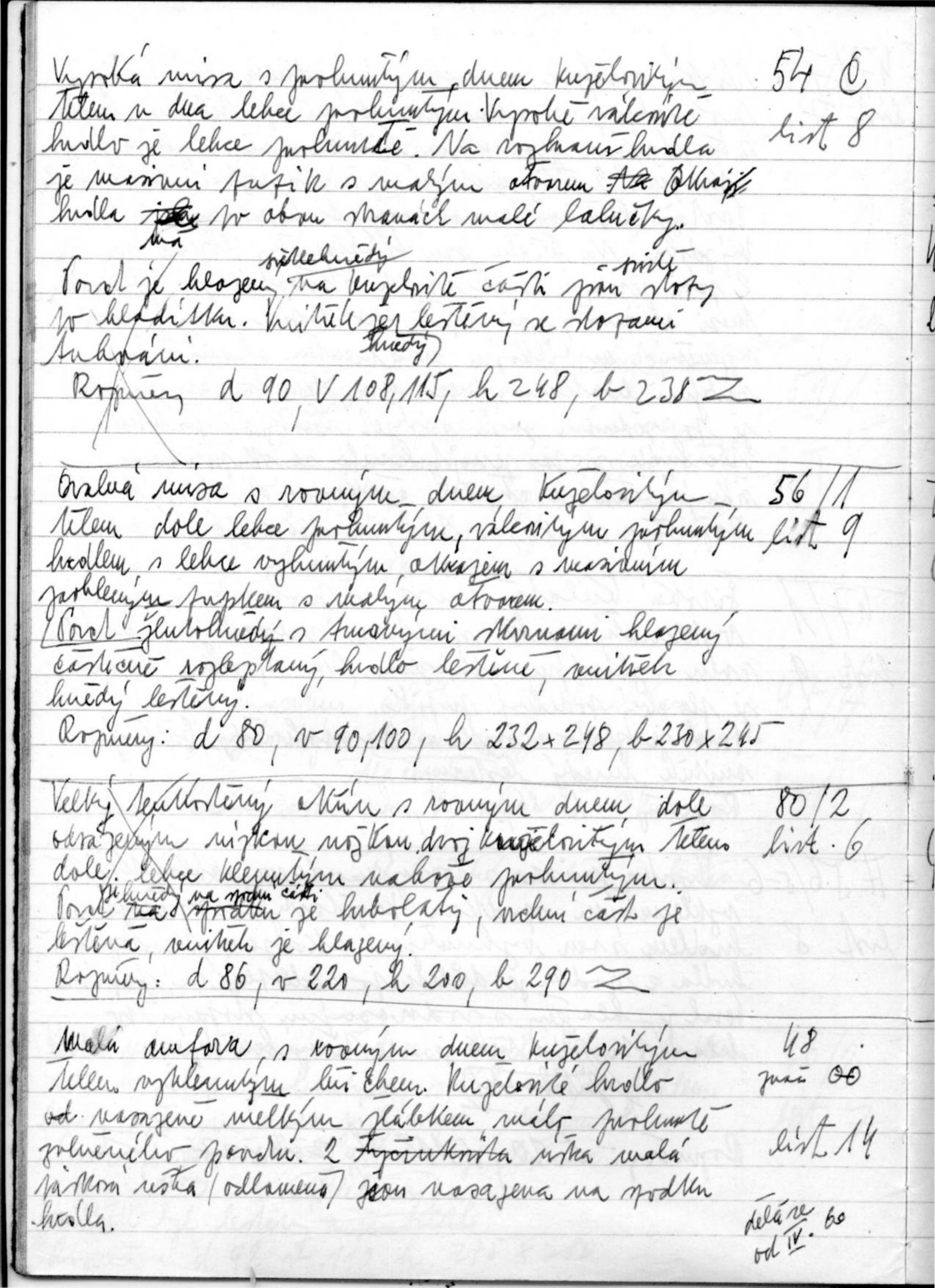}
        \caption{Handwritten journal}
        \label{fig:e_line_hw_3}
    \end{subfigure}

    \vspace{0.5em}

    \begin{subfigure}{0.45\textwidth}
        \centering
        \includegraphics[width=\textwidth]{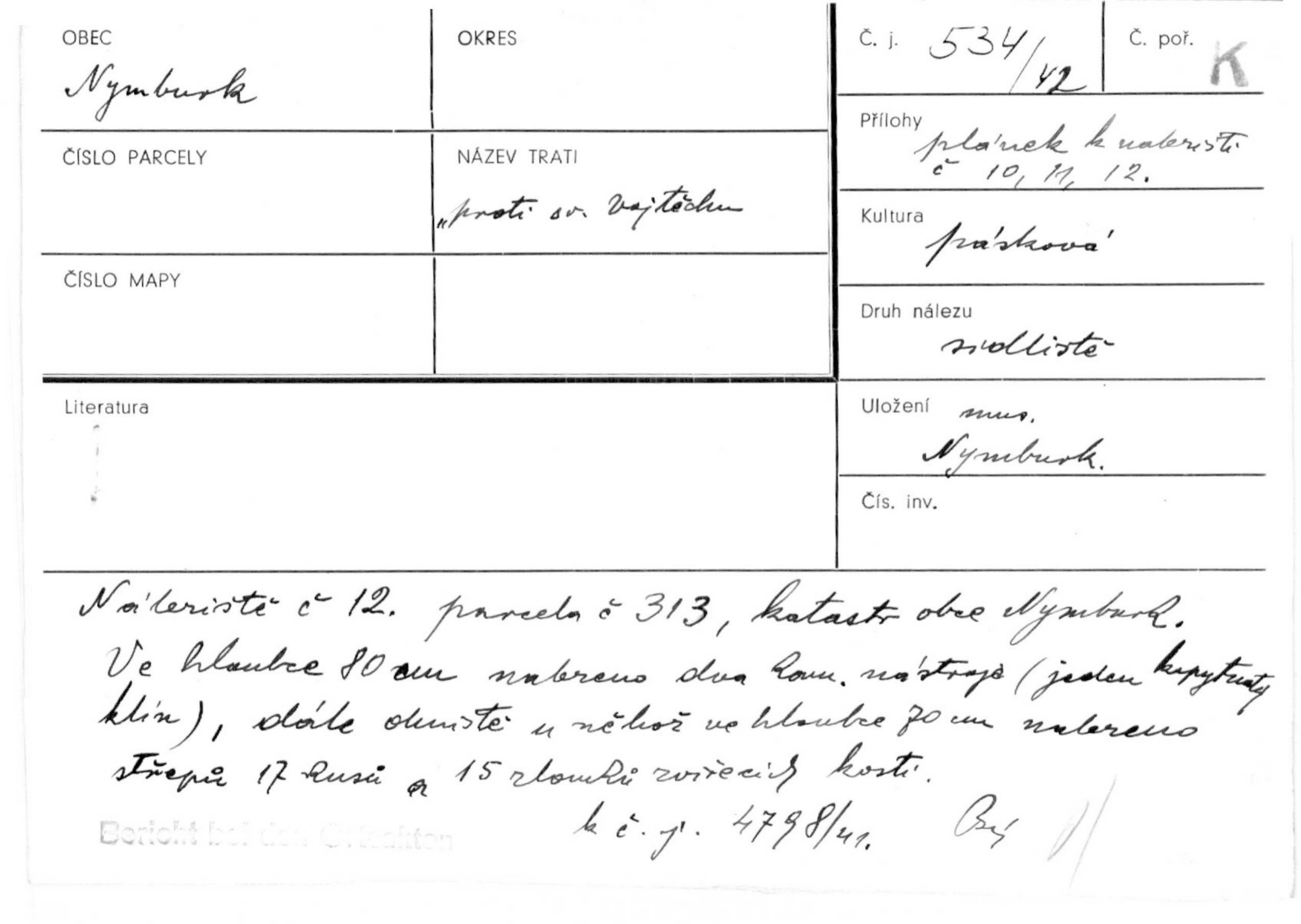}
        \caption{Filled form}
        \label{fig:e_line_hw_4}
    \end{subfigure}
    \begin{subfigure}{0.45\textwidth}
        \centering
        \includegraphics[width=\textwidth]{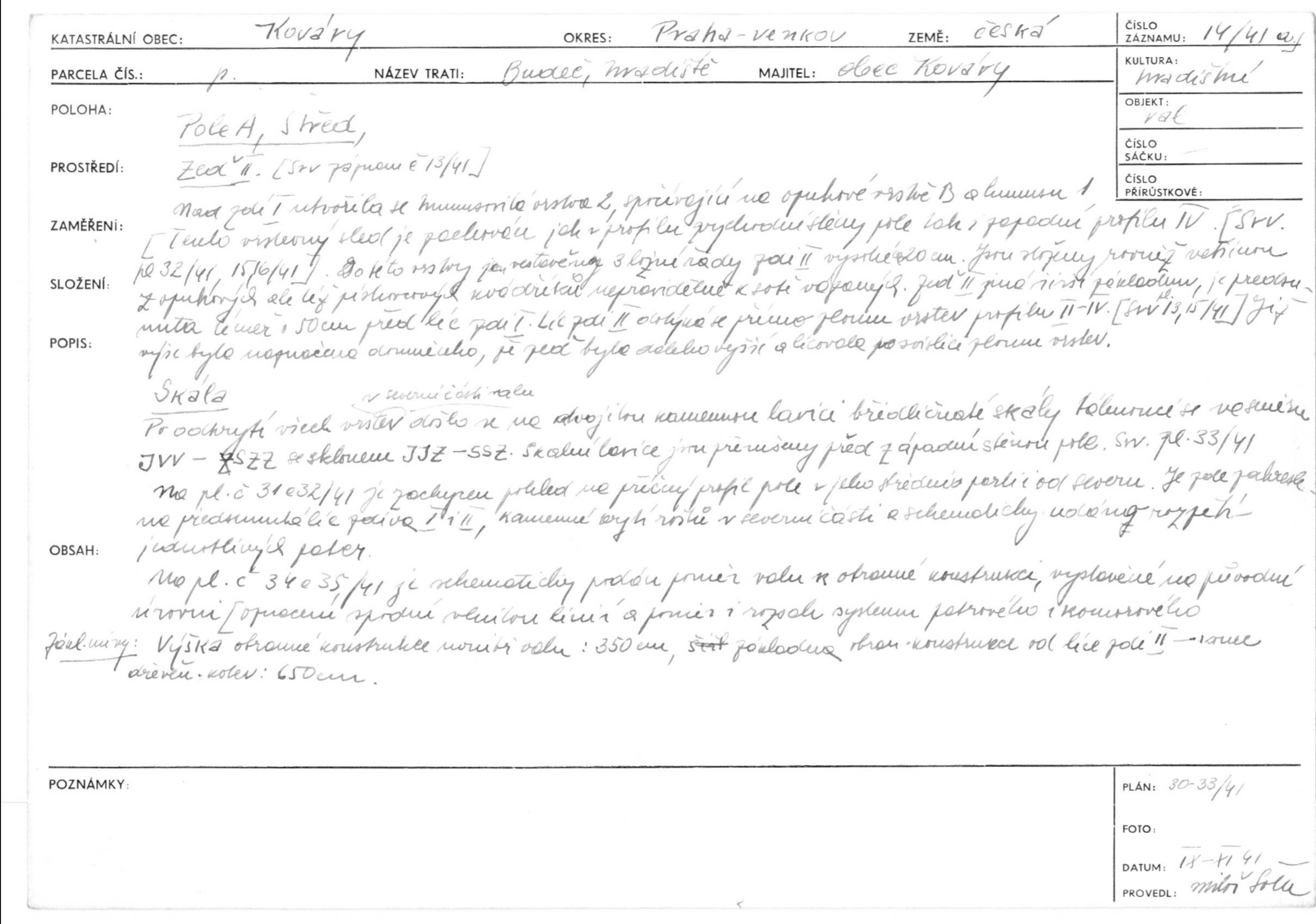}
        \caption{Filled object notes}
        \label{fig:e_line_hw_5}
    \end{subfigure}

    \vspace{0.5em}

    \begin{subfigure}{0.45\textwidth}
        \centering
        \includegraphics[width=\textwidth]{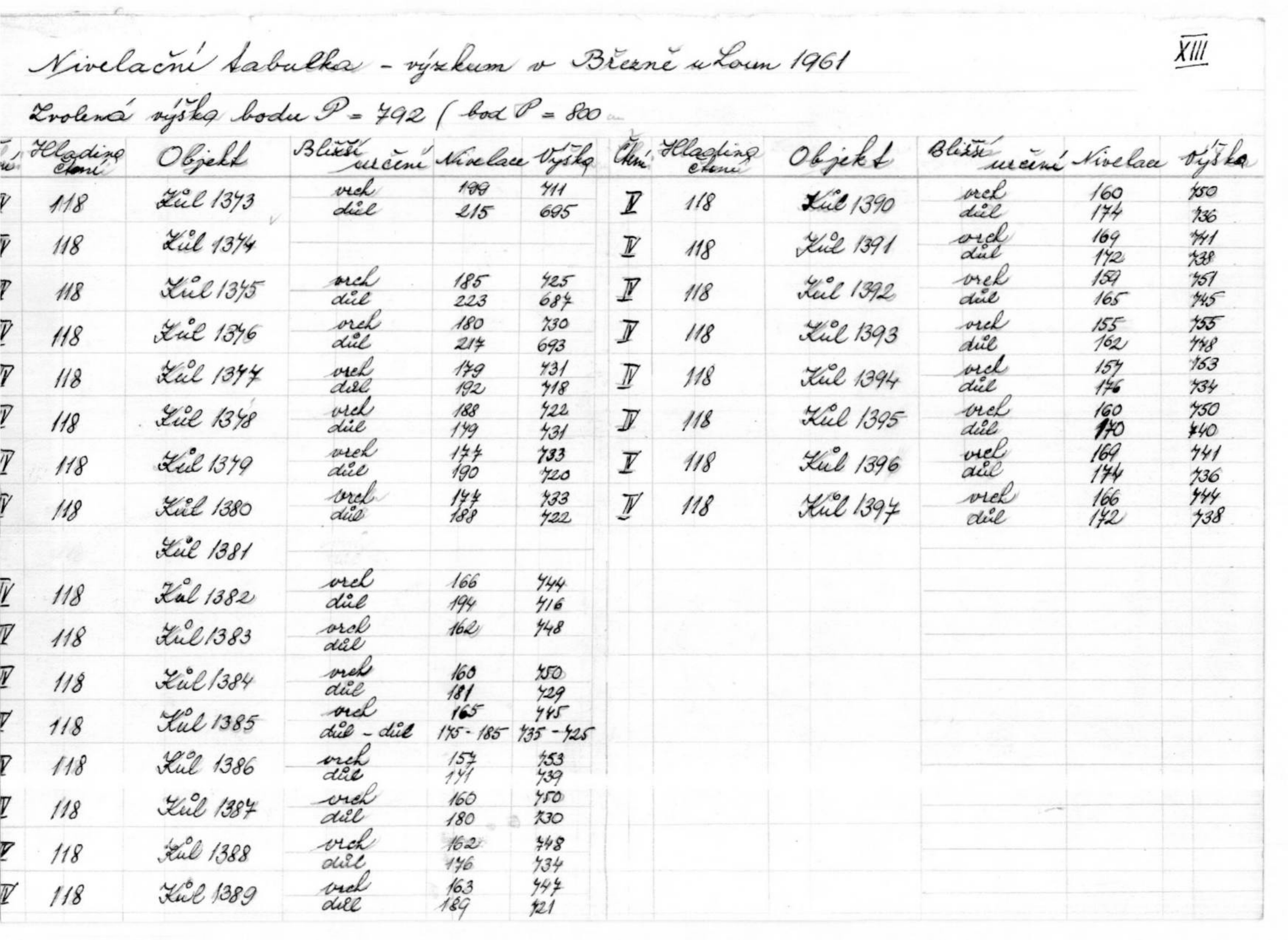}
        \caption{Handwritten table}
        \label{fig:e_line_hw_6}
    \end{subfigure}
    \begin{subfigure}{0.45\textwidth}
        \centering
        \includegraphics[width=\textwidth]{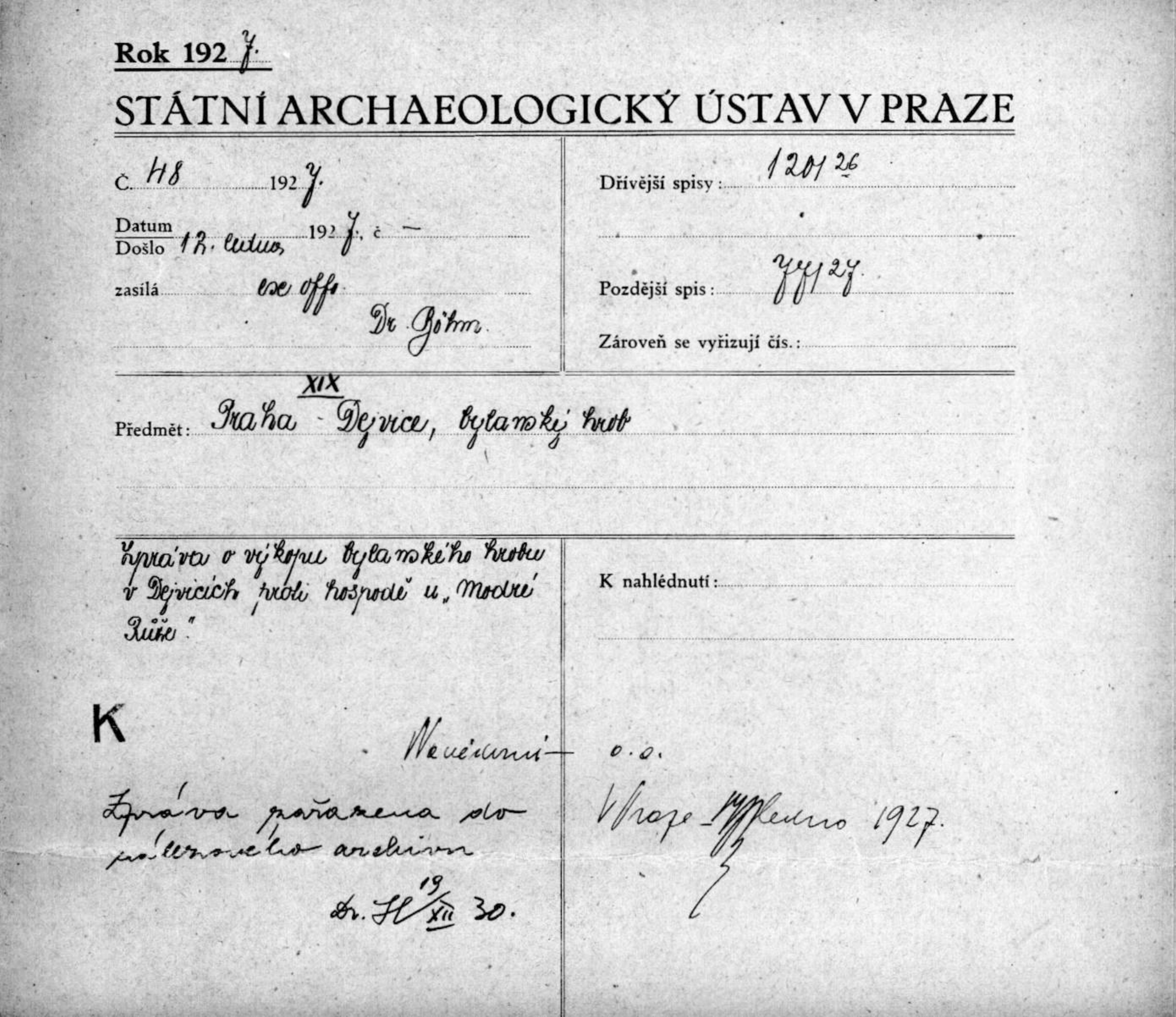}
        \caption{Front page header}
        \label{fig:e_line_hw_7}
    \end{subfigure}
\end{figure}

% LINE_P
\begin{figure}[h]
    \centering
    \caption{Label LINE\_P examples}
    \label{fig:example_line_p}
    \begin{subfigure}{0.3\textwidth}
        \centering
        \includegraphics[width=\textwidth]{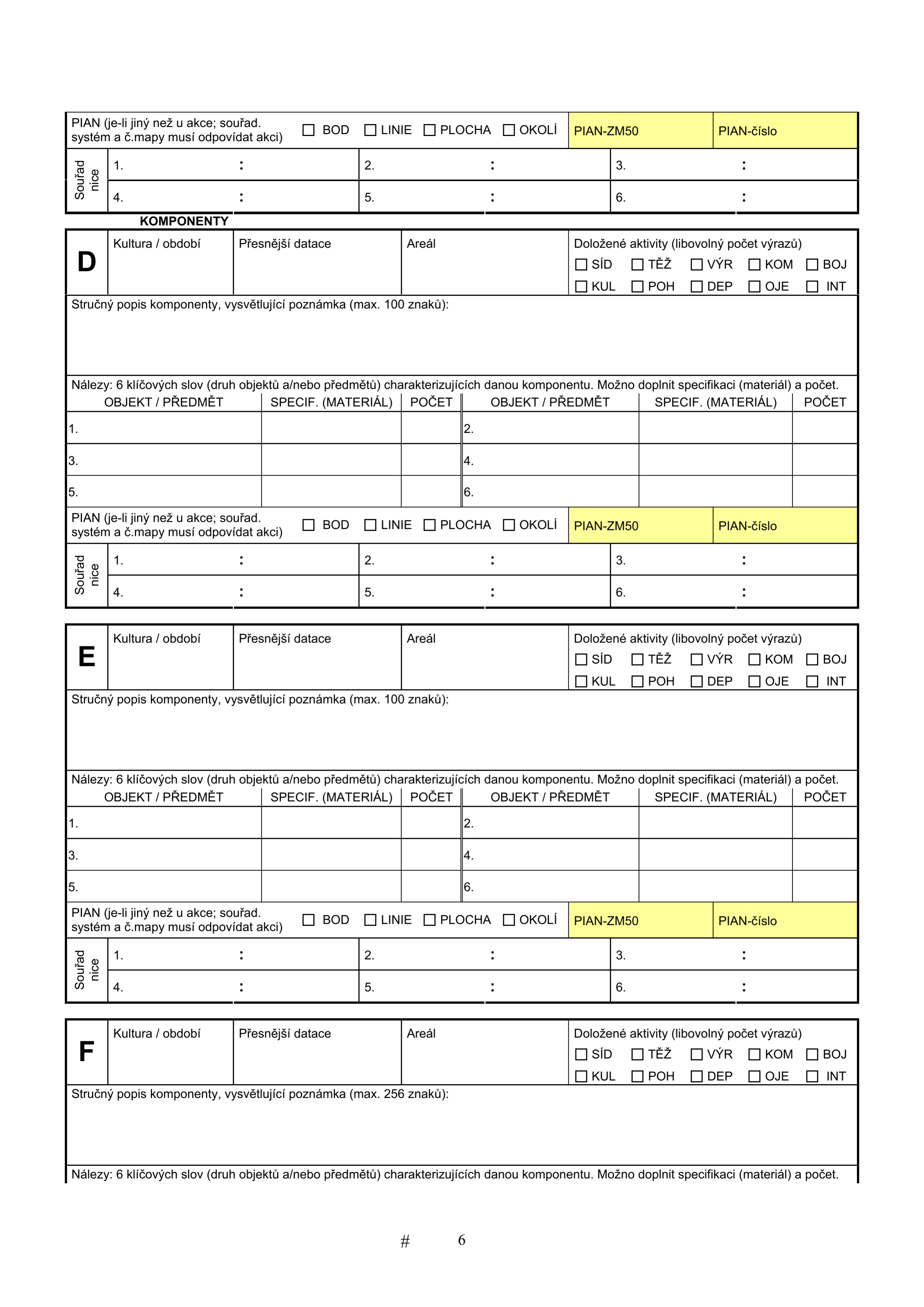}
        \caption{Colored cells}
        \label{fig:e_line_p_1}
    \end{subfigure}
    \begin{subfigure}{0.3\textwidth}
        \centering
        \includegraphics[width=\textwidth]{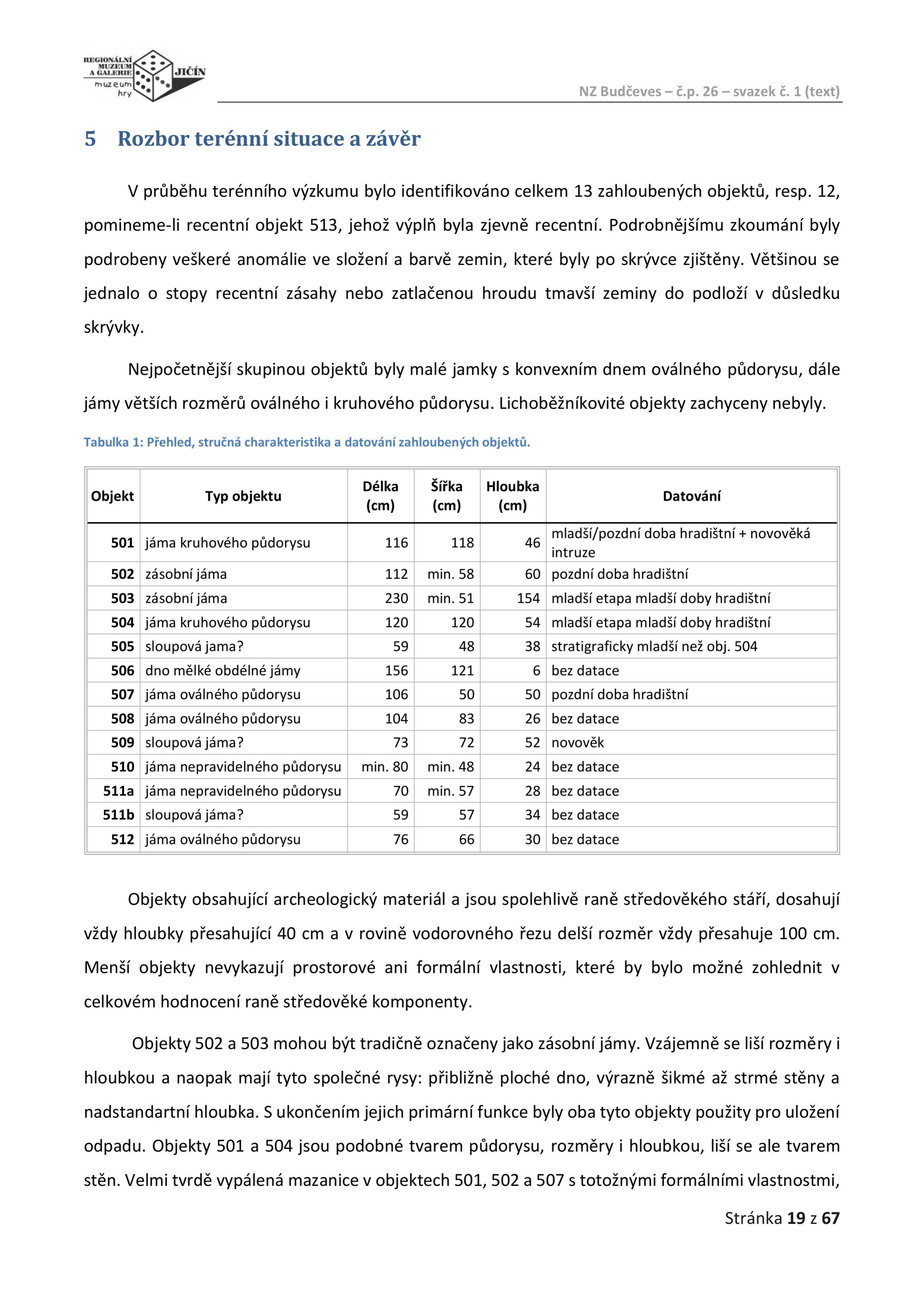}
        \caption{Table within text}
        \label{fig:e_line_p_2}
    \end{subfigure}
    \begin{subfigure}{0.3\textwidth}
        \centering
        \includegraphics[width=\textwidth]{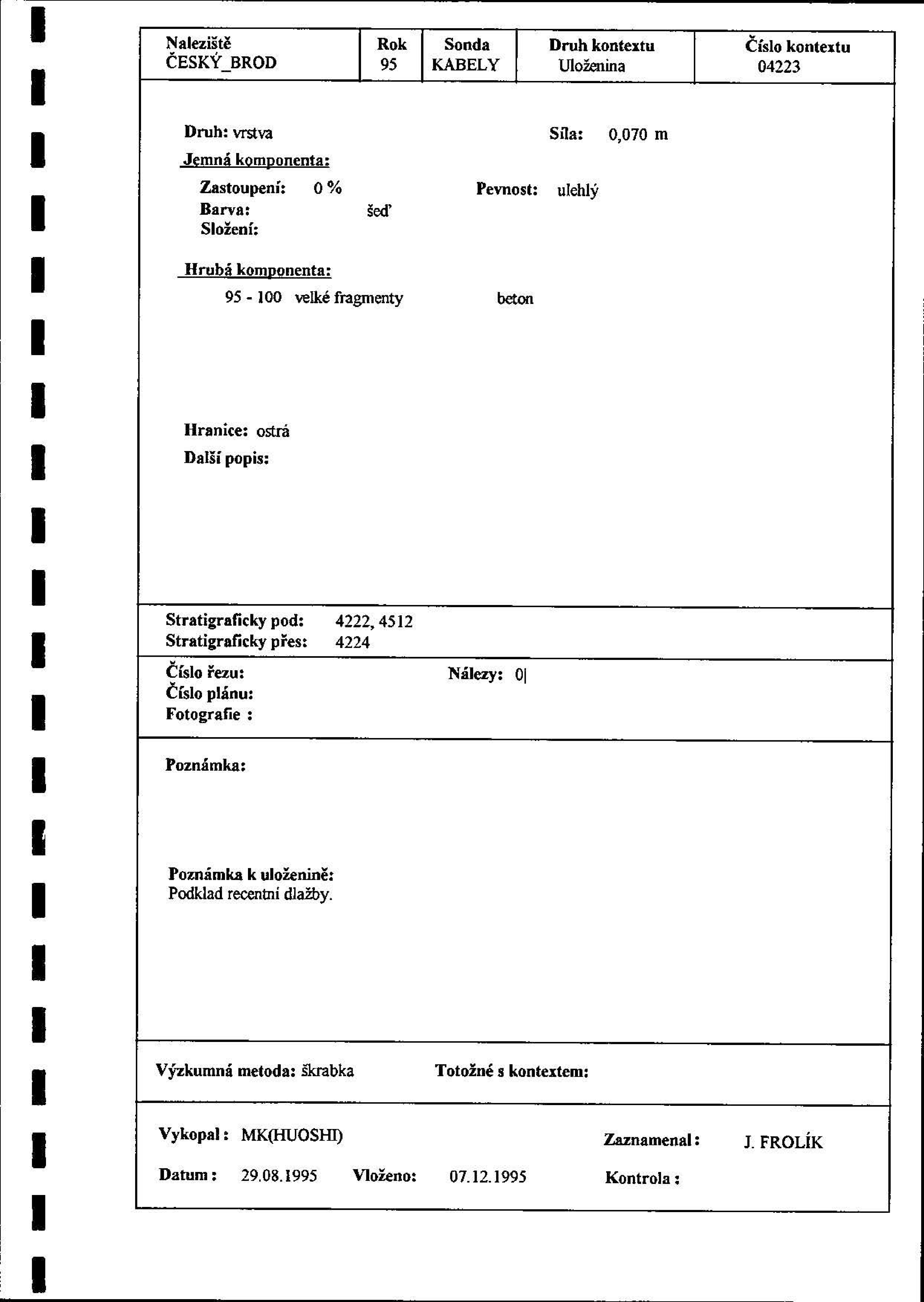}
        \caption{Journal page}
        \label{fig:e_line_p_3}
    \end{subfigure}

    \vspace{0.5em}

    \begin{subfigure}{0.45\textwidth}
        \centering
        \includegraphics[width=\textwidth]{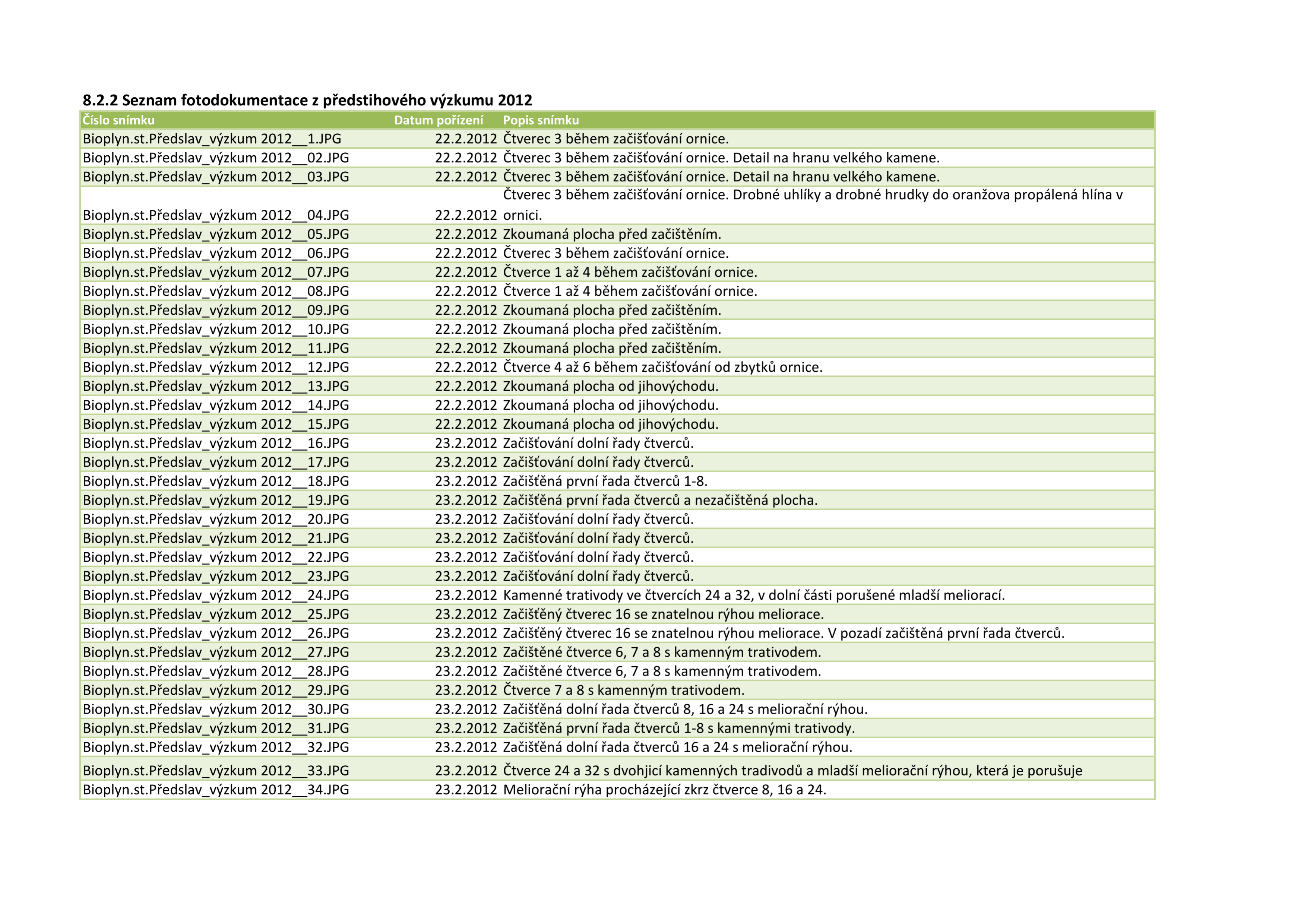}
        \caption{Colorful header}
        \label{fig:e_line_p_4}
    \end{subfigure}
    \begin{subfigure}{0.45\textwidth}
        \centering
        \includegraphics[width=\textwidth]{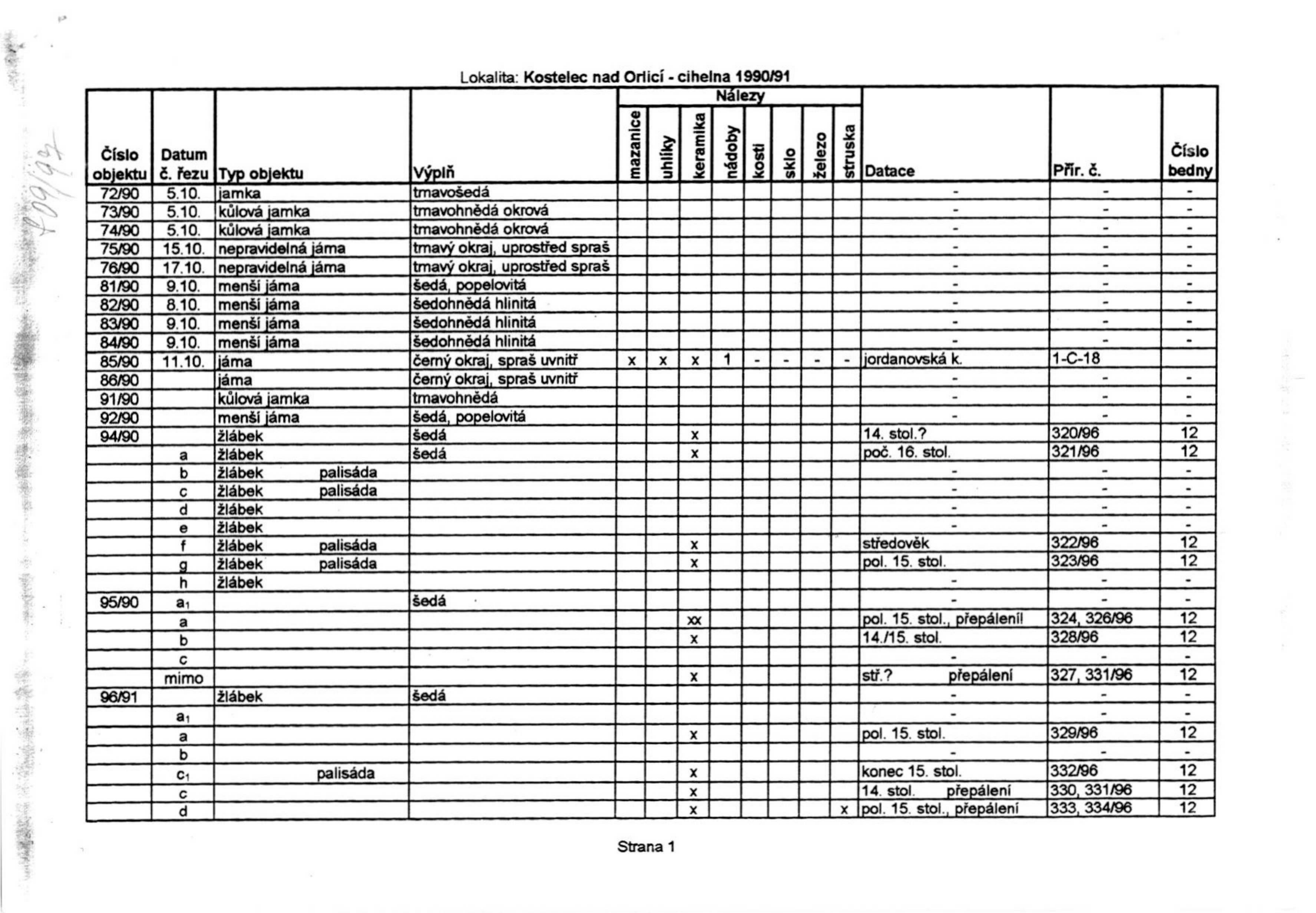}
        \caption{Old style print}
        \label{fig:e_line_p_5}
    \end{subfigure}

    \vspace{0.5em}

    \begin{subfigure}{0.45\textwidth}
        \centering
        \includegraphics[width=\textwidth]{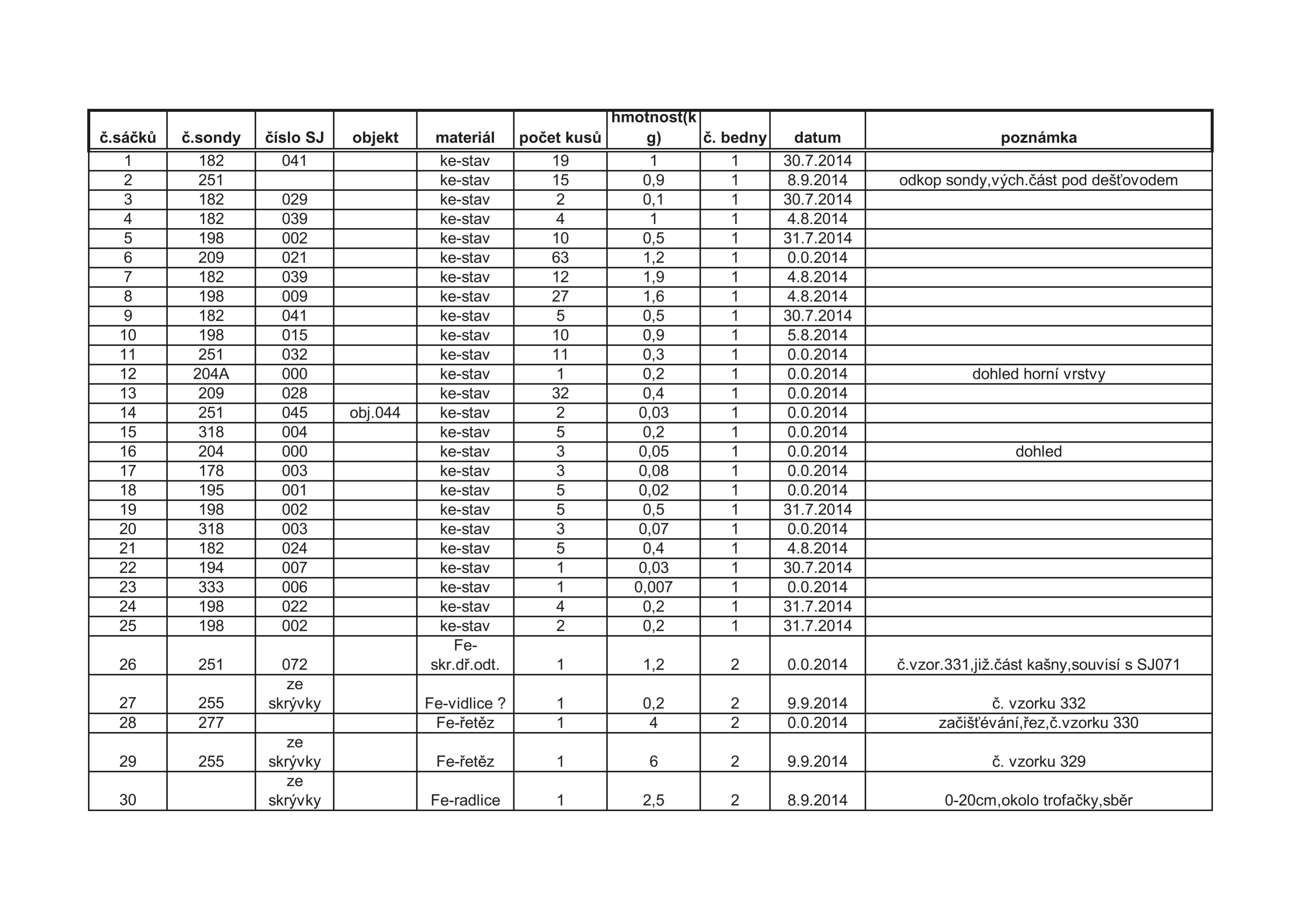}
        \caption{Full-page table}
        \label{fig:e_line_p_6}
    \end{subfigure}
    \begin{subfigure}{0.45\textwidth}
        \centering
        \includegraphics[width=\textwidth]{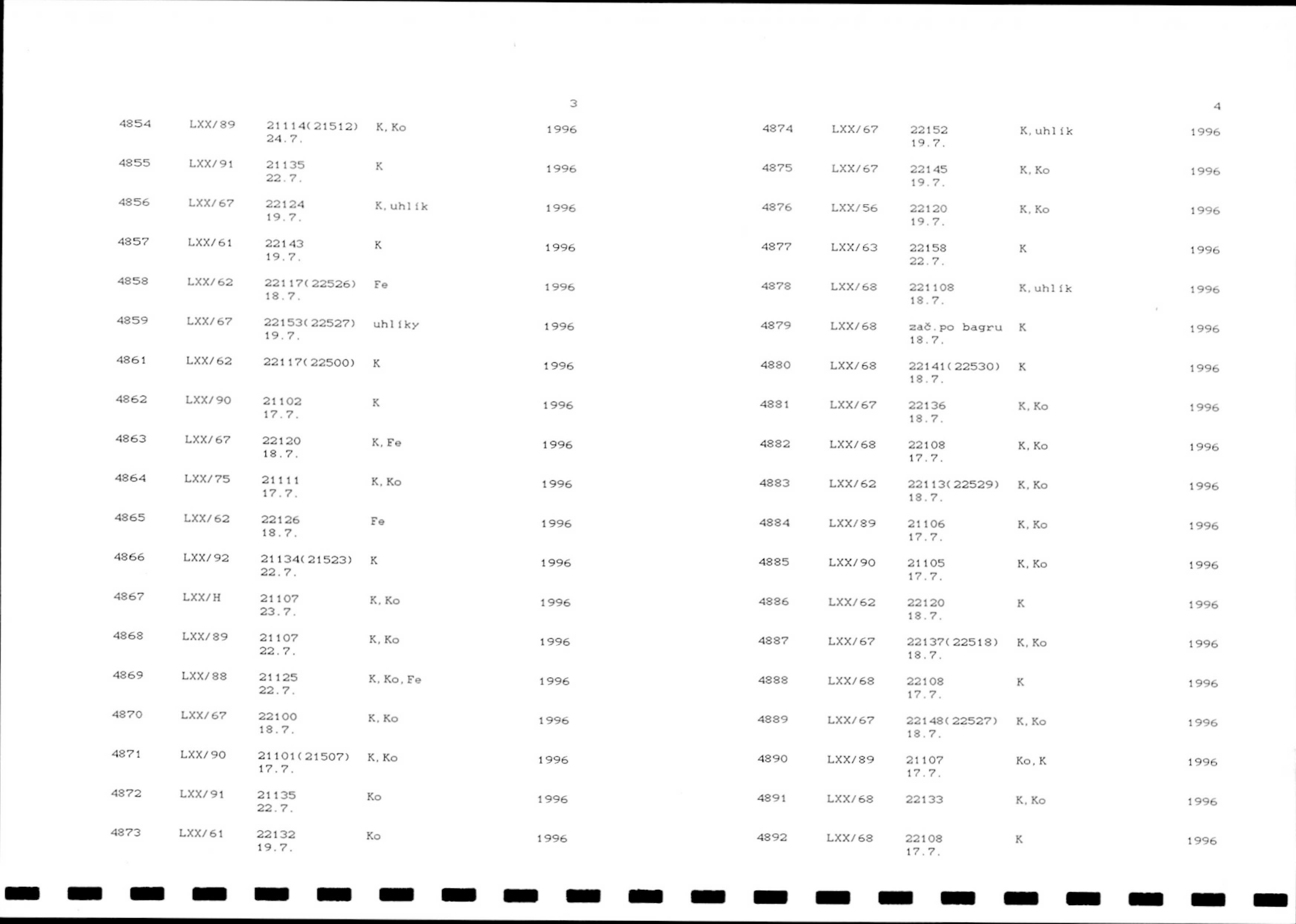}
        \caption{Widened print}
        \label{fig:e_line_p_7}
    \end{subfigure}
\end{figure}

% LINE_T
\begin{figure}[h]
    \centering
    \caption{Label LINE\_T examples}
    \label{fig:example_line_t}
    \begin{subfigure}{0.3\textwidth}
        \centering
        \includegraphics[width=\textwidth]{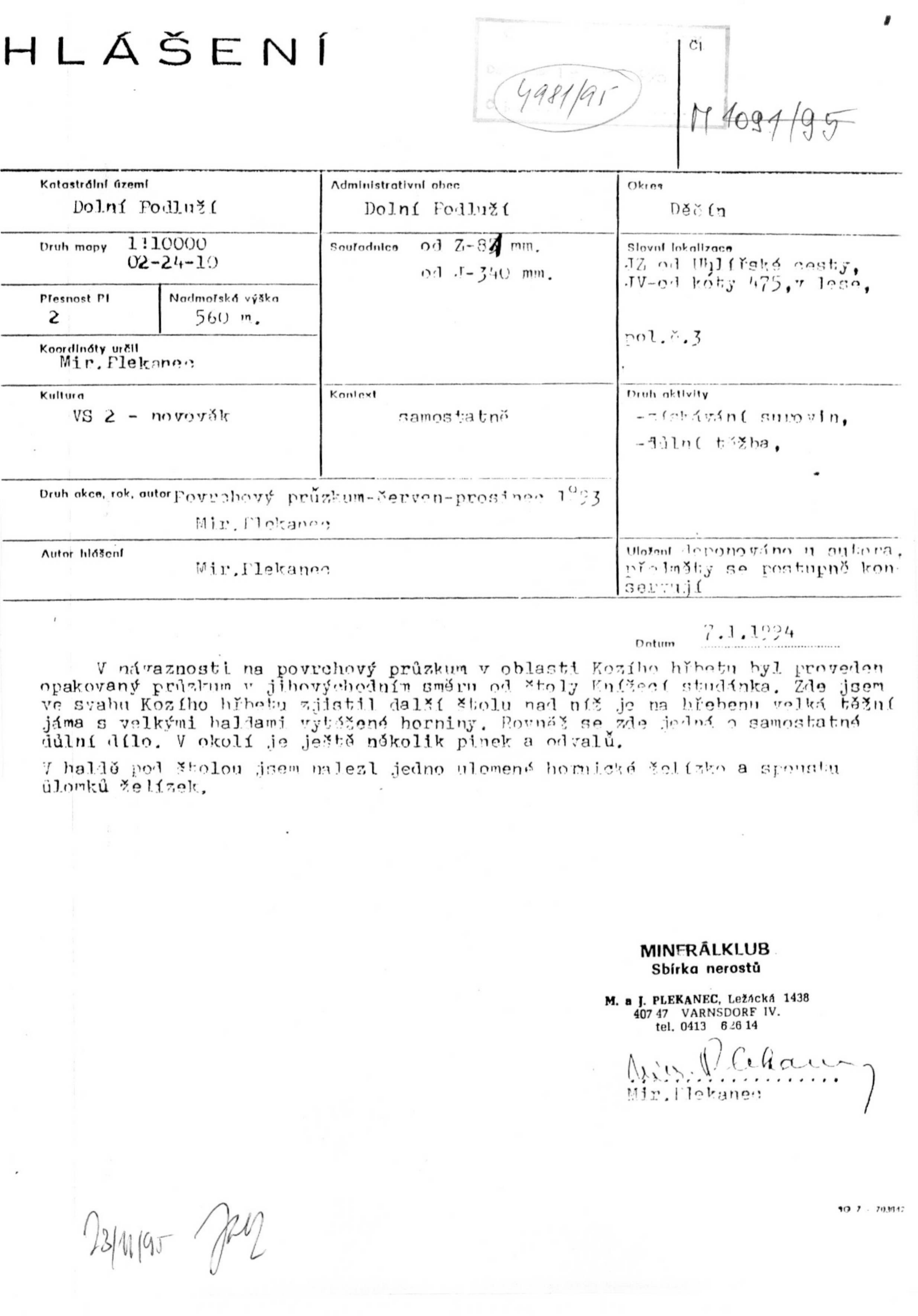}
        \caption{Filled-in form}
        \label{fig:e_line_t_1}
    \end{subfigure}
    \begin{subfigure}{0.3\textwidth}
        \centering
        \includegraphics[width=\textwidth]{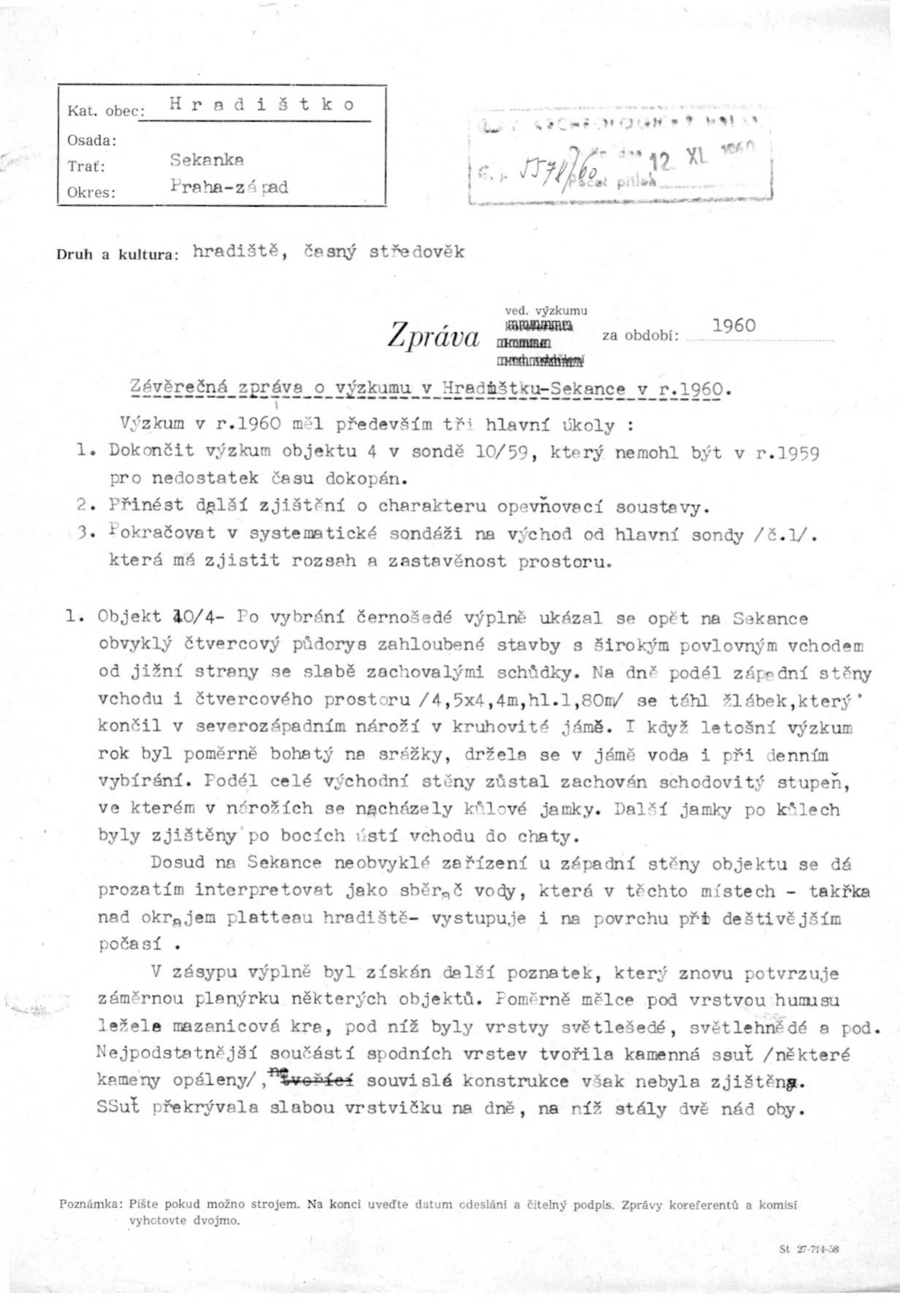}
        \caption{Stamped front page}
        \label{fig:e_line_t_2}
    \end{subfigure}
    \begin{subfigure}{0.3\textwidth}
        \centering
        \includegraphics[width=\textwidth]{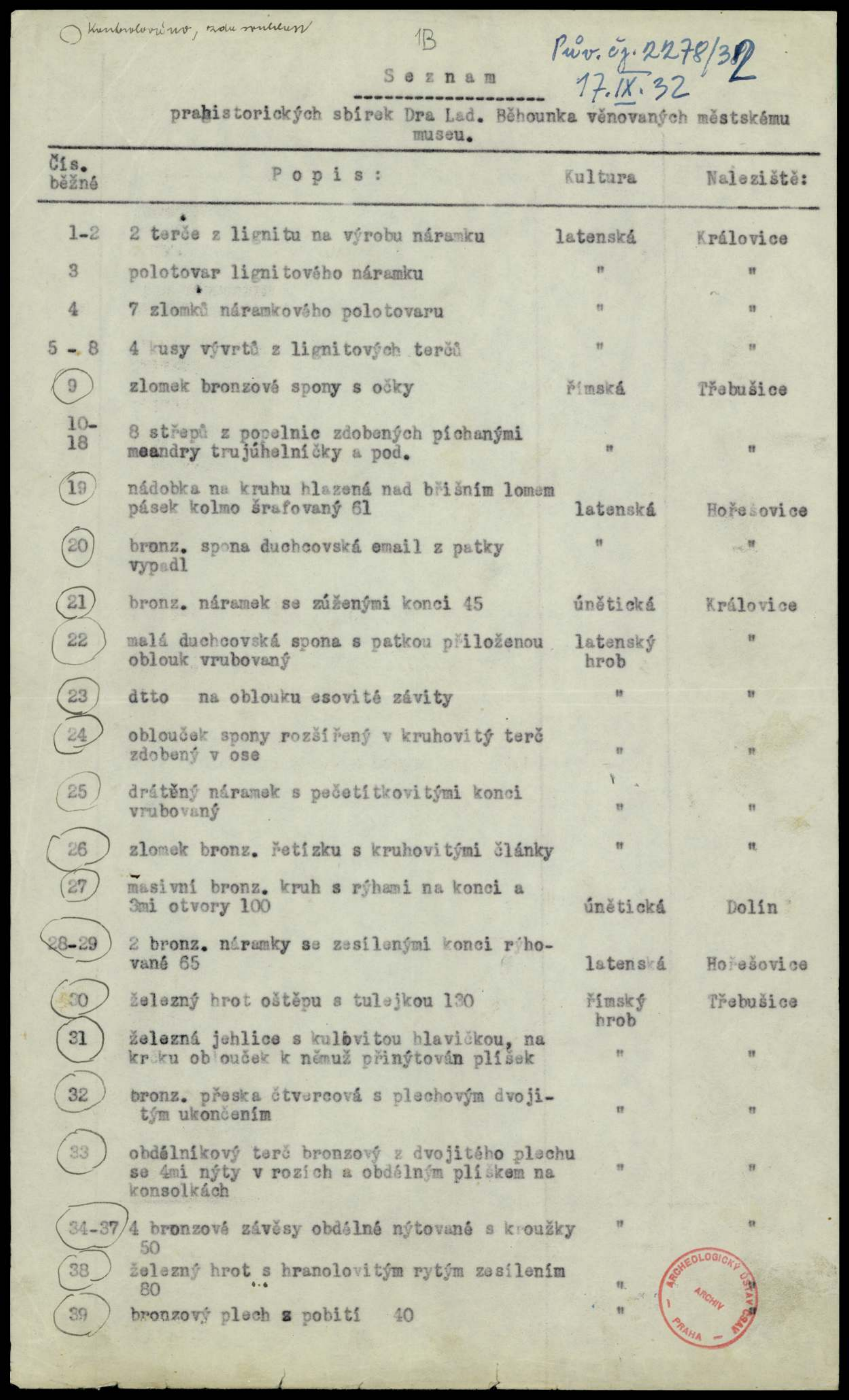}
        \caption{Grayish paper}
        \label{fig:e_line_t_3}
    \end{subfigure}

     \vspace{0.5em}

    \begin{subfigure}{0.45\textwidth}
        \centering
        \includegraphics[width=\textwidth]{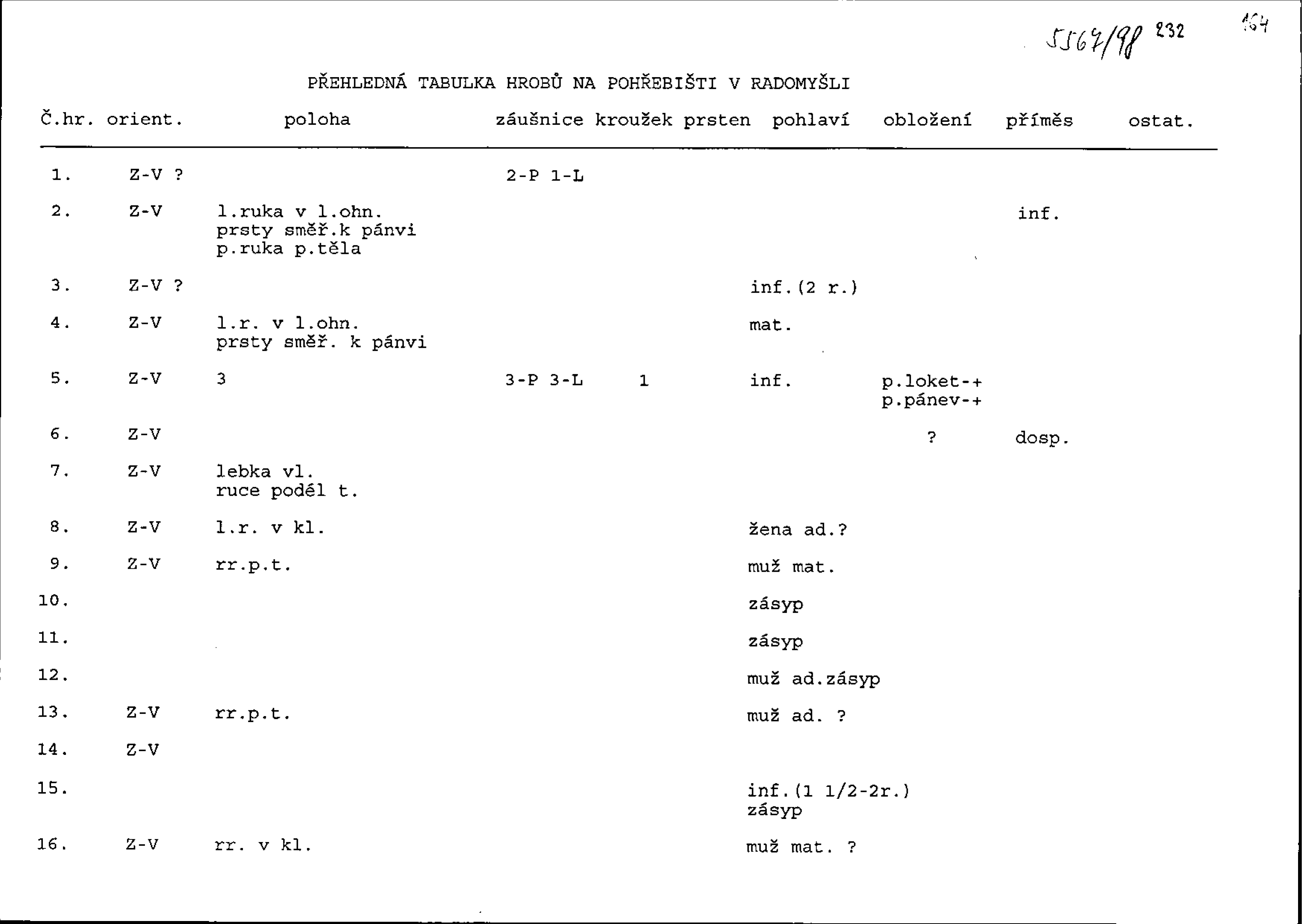}
        \caption{Edgeless table}
        \label{fig:e_line_t_4}
    \end{subfigure}
    \begin{subfigure}{0.45\textwidth}
        \centering
        \includegraphics[width=\textwidth]{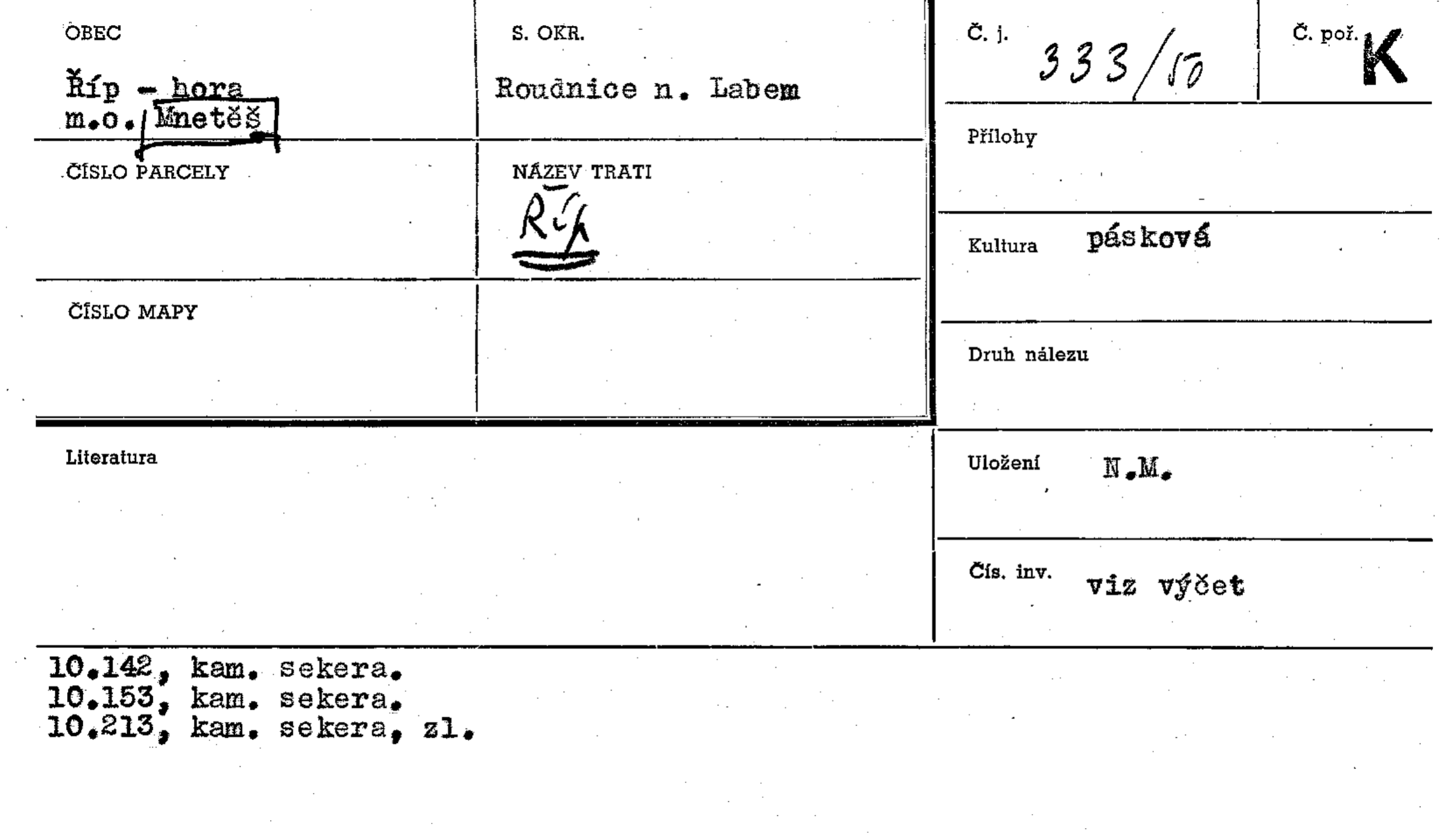}
        \caption{Filled-in form header}
        \label{fig:e_line_t_5}
    \end{subfigure}

    \vspace{0.5em}

    \begin{subfigure}{0.45\textwidth}
        \centering
        \includegraphics[width=\textwidth]{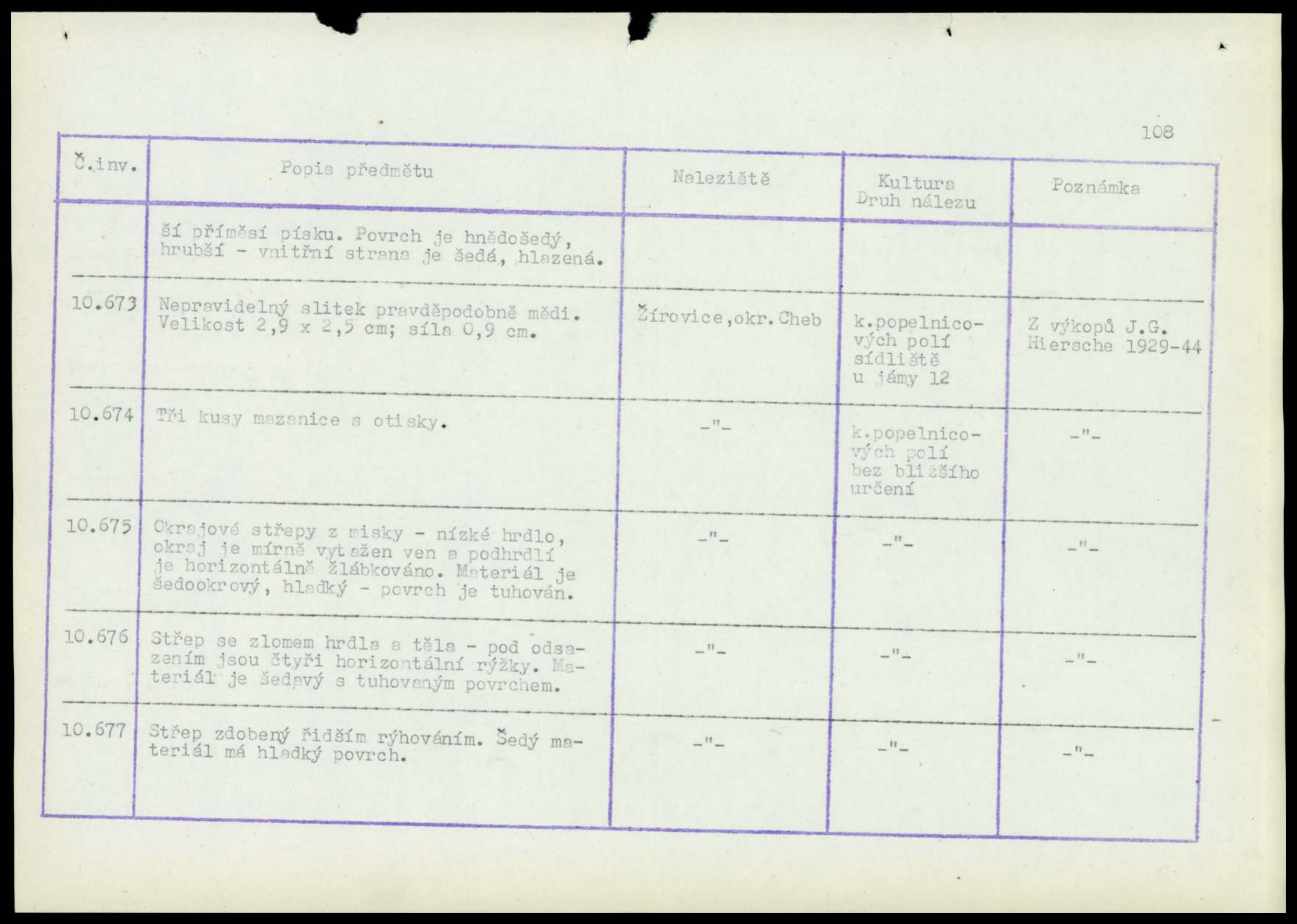}
        \caption{Page holes on the edge}
        \label{fig:e_line_t_6}
    \end{subfigure}
    \begin{subfigure}{0.45\textwidth}
        \centering
        \includegraphics[width=\textwidth]{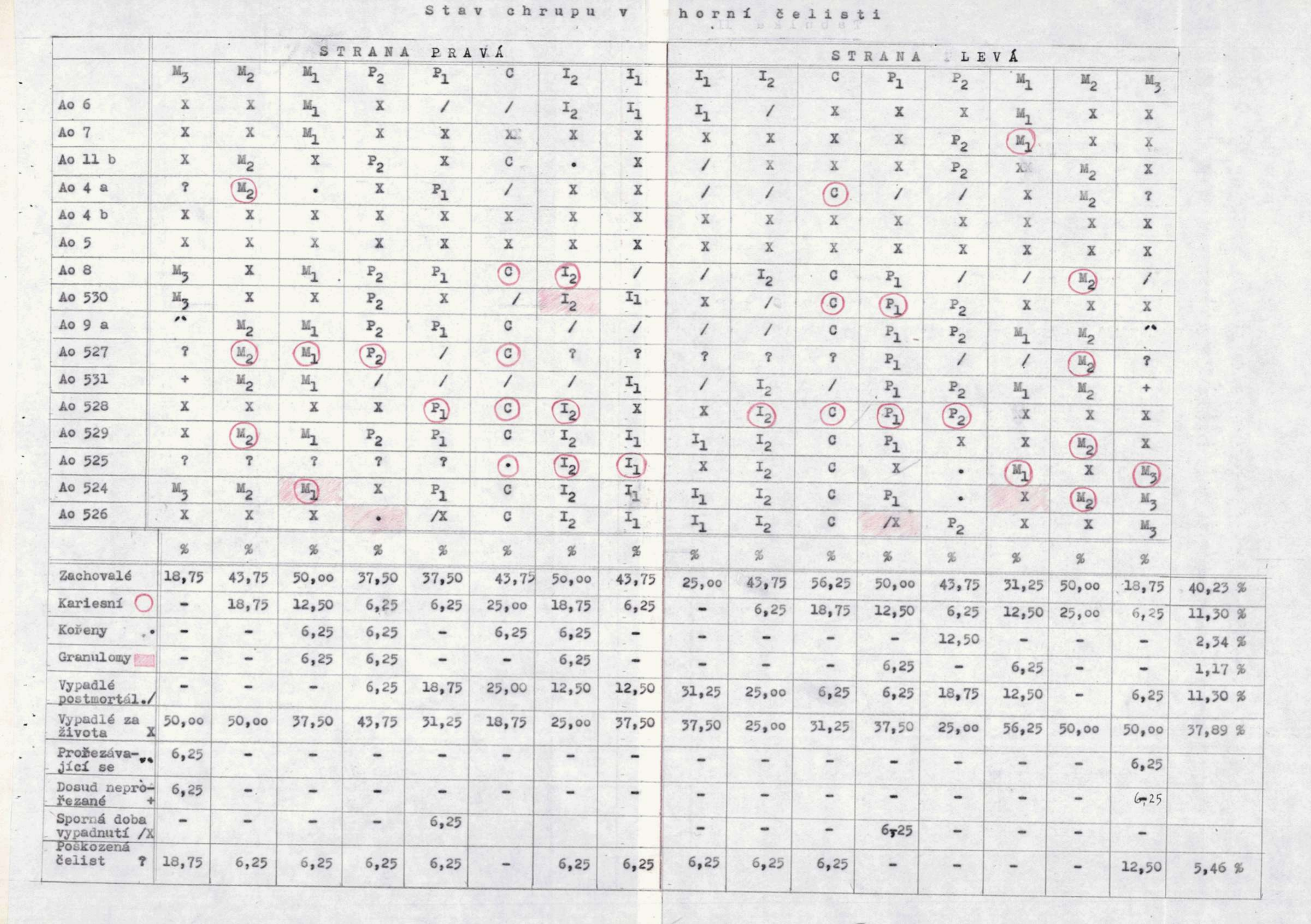}
        \caption{Full-page table}
        \label{fig:e_line_t_7}
    \end{subfigure}
\end{figure}

% PHOTO
\begin{figure}[h]
    \centering
    \caption{Label PHOTO examples}
    \label{fig:example_photo}
    \begin{subfigure}{0.3\textwidth}
        \centering
        \includegraphics[width=\textwidth]{img/pages/example_PHOTO_news.pdf}
        \caption{Newspaper page}
        \label{fig:e_photo_1}
    \end{subfigure}
    \begin{subfigure}{0.3\textwidth}
        \centering
        \includegraphics[width=\textwidth]{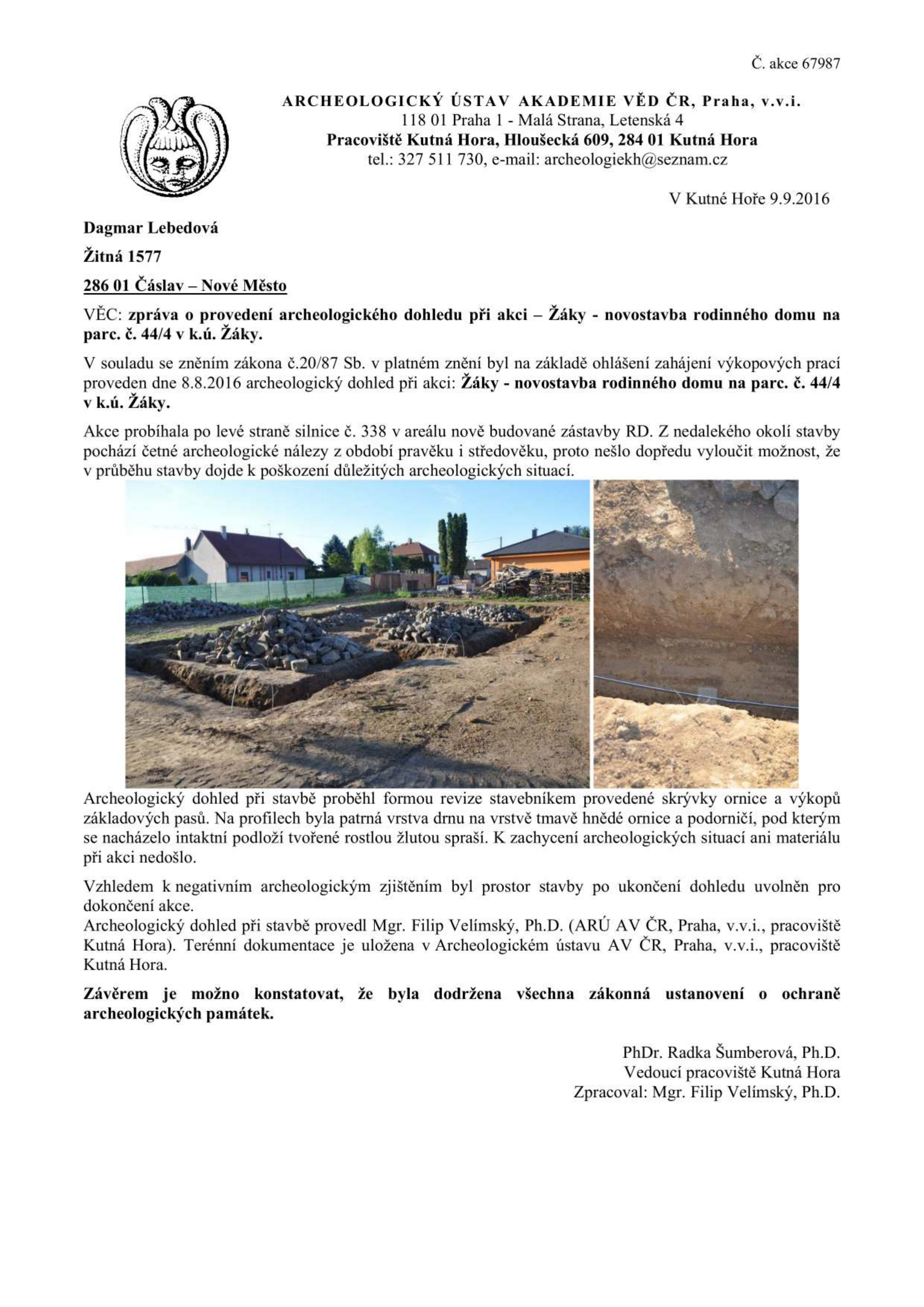}
        \caption{Front Page}
        \label{fig:e_photo_2}
    \end{subfigure}
    \begin{subfigure}{0.3\textwidth}
        \centering
        \includegraphics[width=\textwidth]{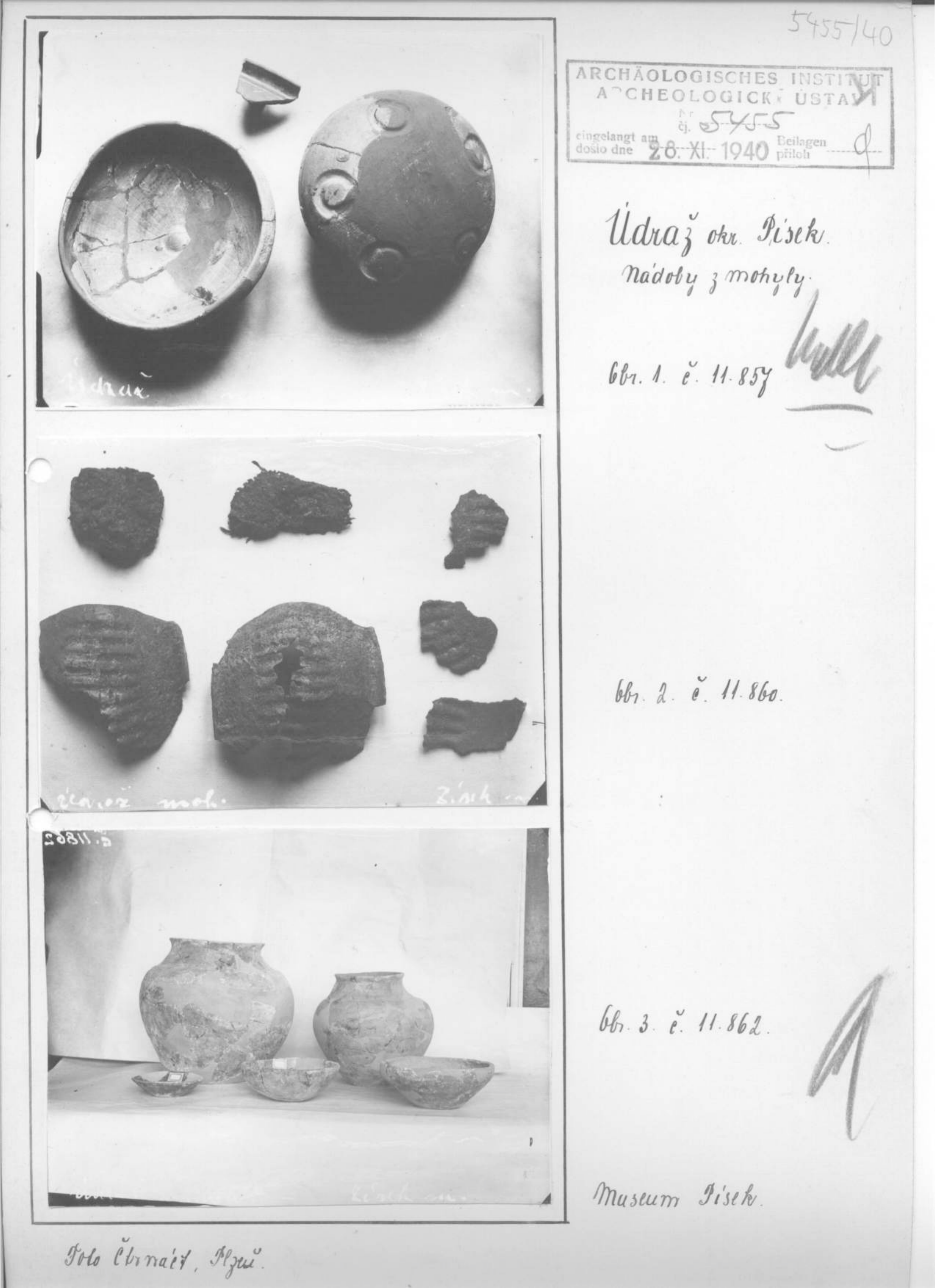}
        \caption{Artifact photos}
        \label{fig:e_photo_3}
    \end{subfigure}

    \vspace{0.5em}

    \begin{subfigure}{0.45\textwidth}
        \centering
        \includegraphics[width=\textwidth]{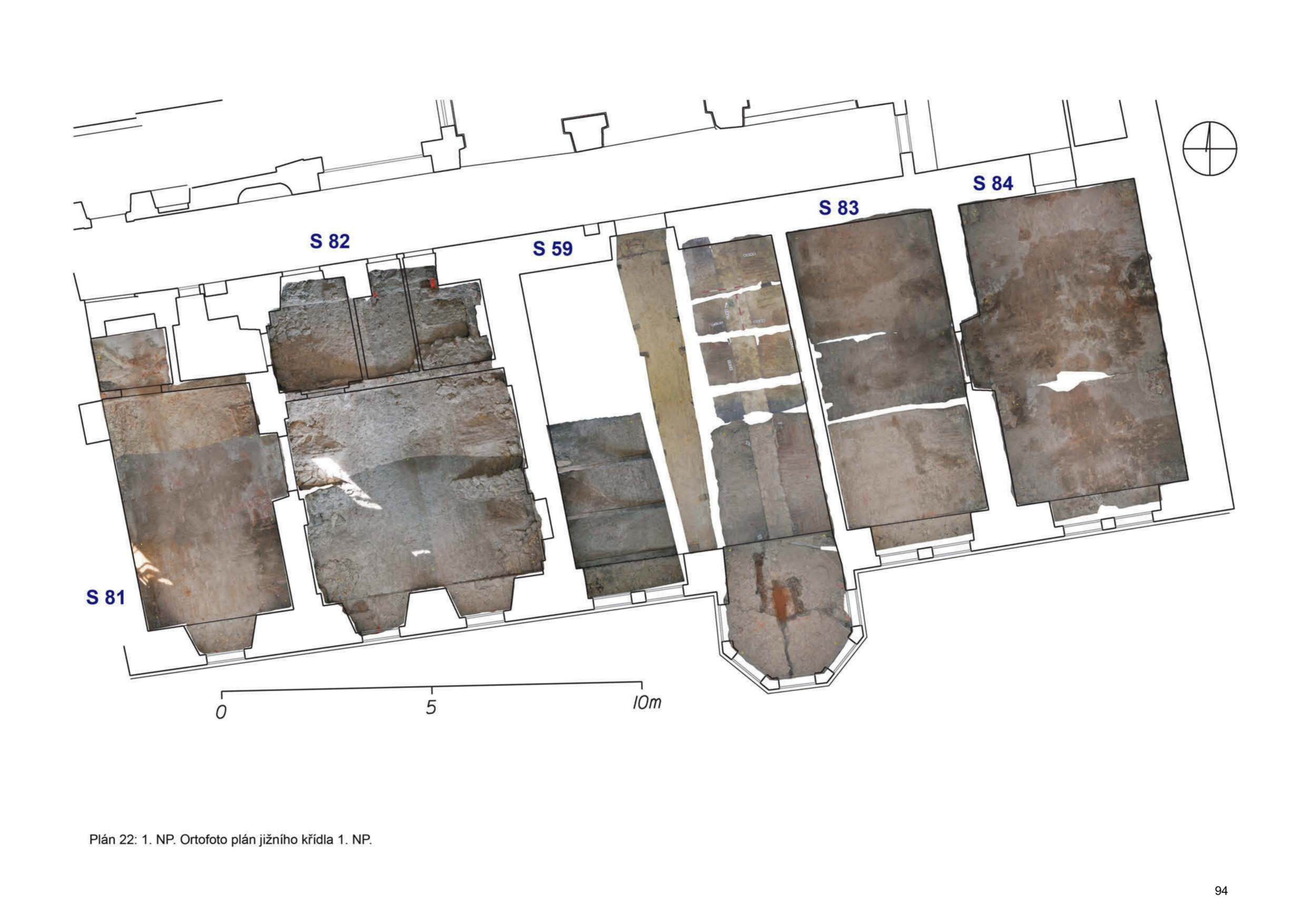}
        \caption{Texture cutouts}
        \label{fig:e_photo_4}
    \end{subfigure}
    \begin{subfigure}{0.45\textwidth}
        \centering
        \includegraphics[width=\textwidth]{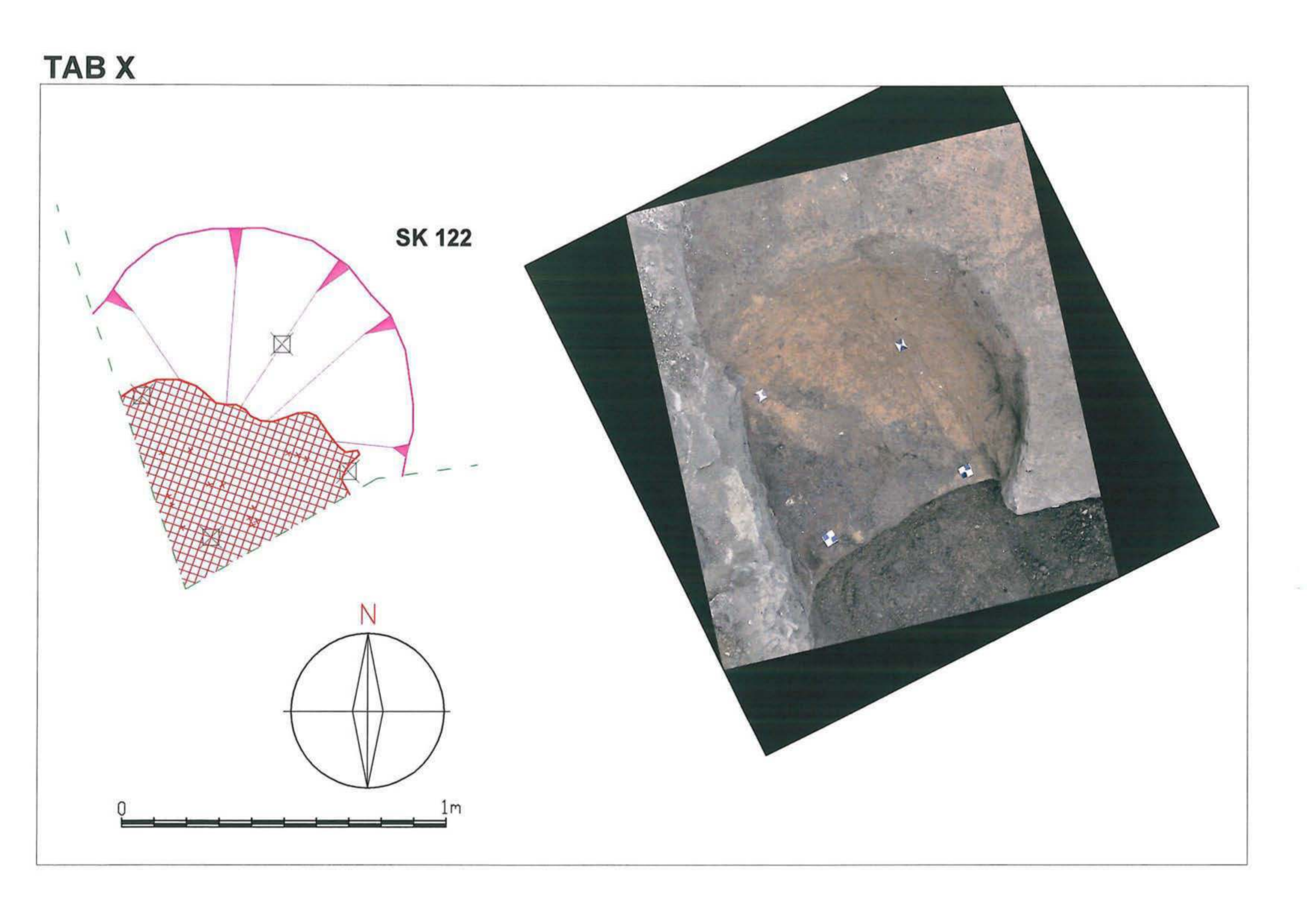}
        \caption{Schema and photo}
        \label{fig:e_photo_5}
    \end{subfigure}

    \vspace{0.5em}

    \begin{subfigure}{0.45\textwidth}
        \centering
        \includegraphics[width=\textwidth]{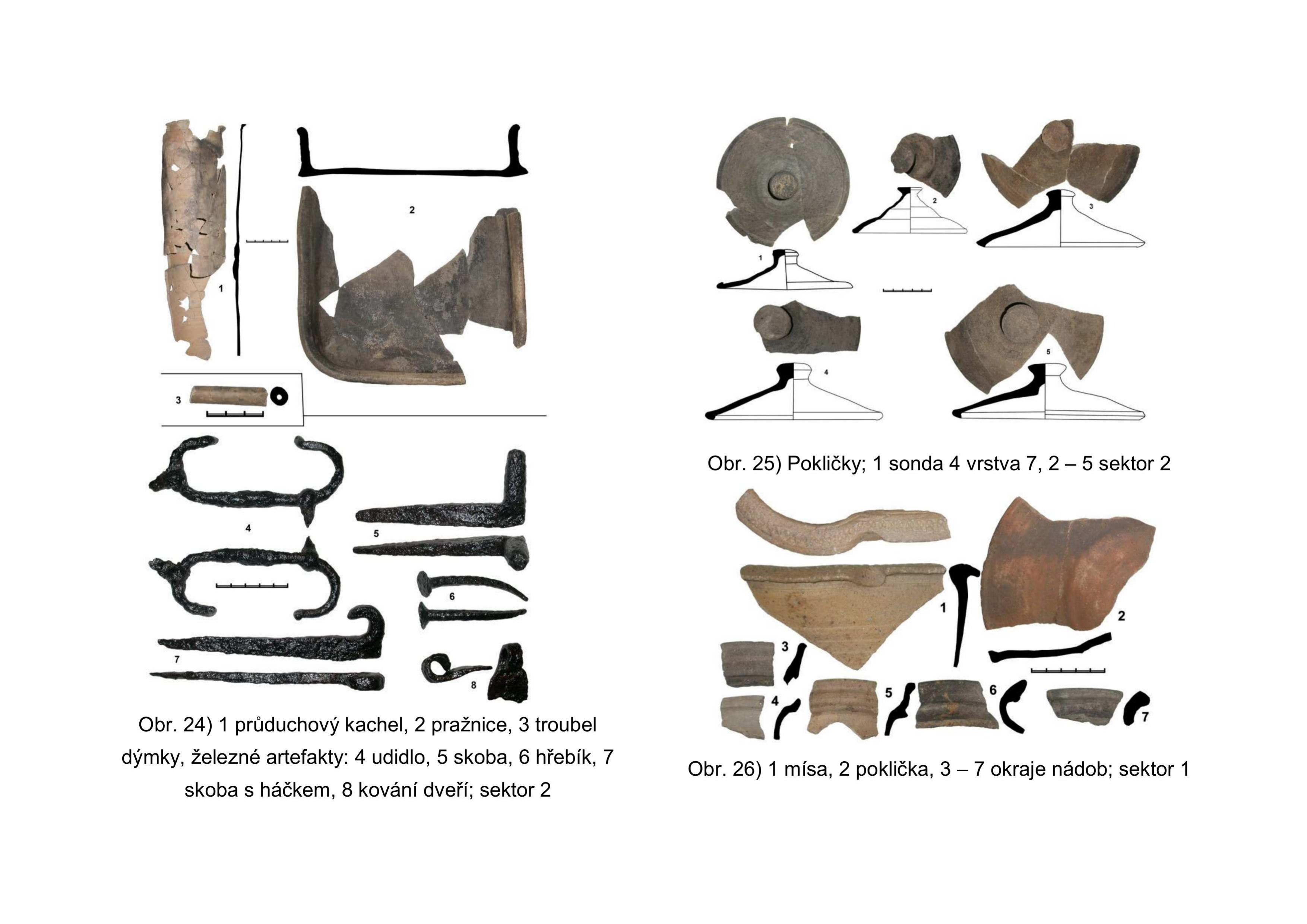}
        \caption{Multiple object cutouts}
        \label{fig:e_photo_6}
    \end{subfigure}
    \begin{subfigure}{0.45\textwidth}
        \centering
        \includegraphics[width=\textwidth]{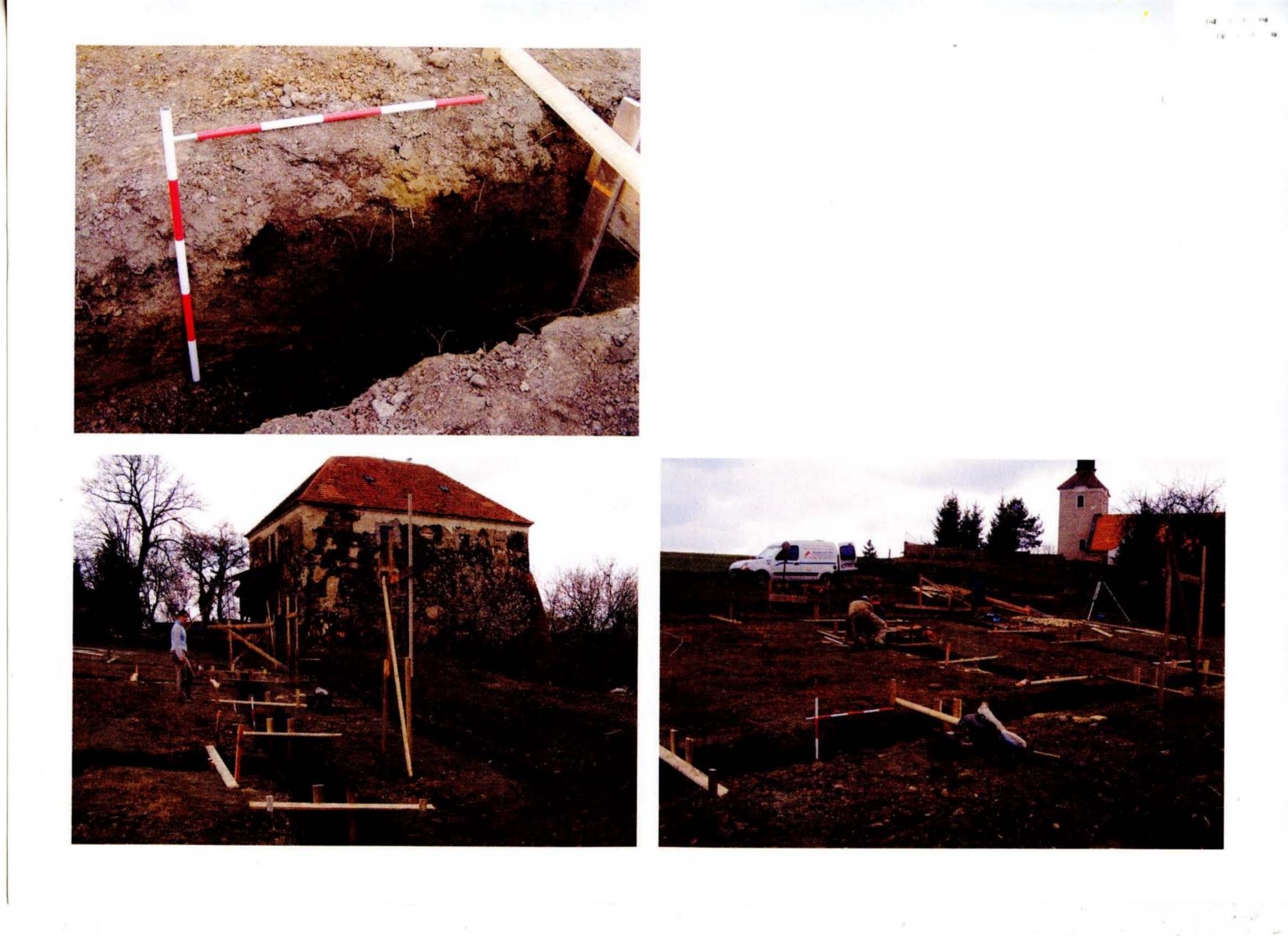}
        \caption{No captions}
        \label{fig:e_photo_7}
    \end{subfigure}
\end{figure}

% PHOTO_L
\begin{figure}[h]
    \centering
    \caption{Label PHOTO\_L examples}
    \label{fig:example_photo_l}
    \begin{subfigure}{0.3\textwidth}
        \centering
        \includegraphics[width=\textwidth]{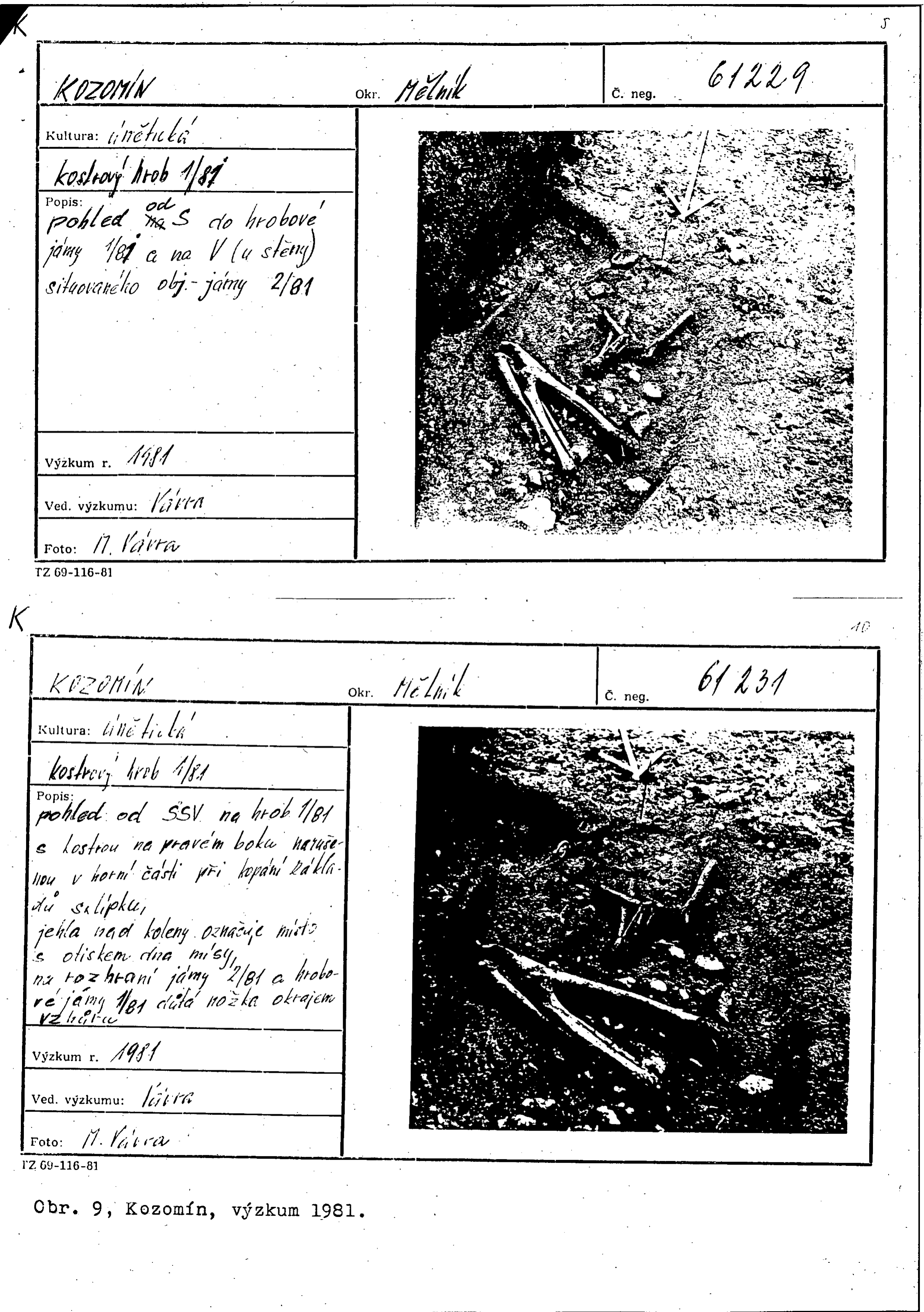}
        \caption{Manual captions}
        \label{fig:e_photo_l_1}
    \end{subfigure}
    \begin{subfigure}{0.3\textwidth}
        \centering
        \includegraphics[width=\textwidth]{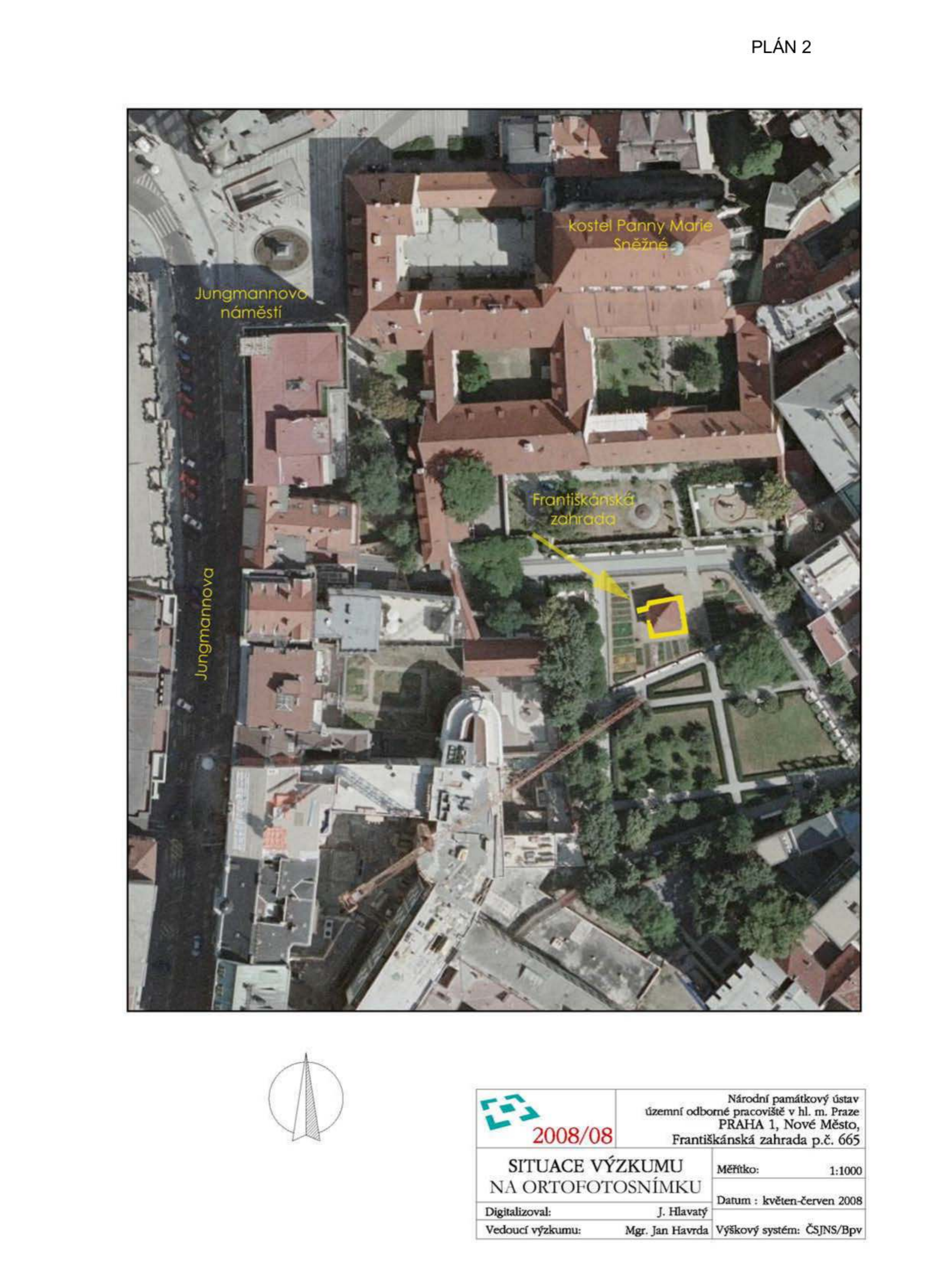}
        \caption{Map with legend}
        \label{fig:e_photo_l_2}
    \end{subfigure}
    \begin{subfigure}{0.3\textwidth}
        \centering
        \includegraphics[width=\textwidth]{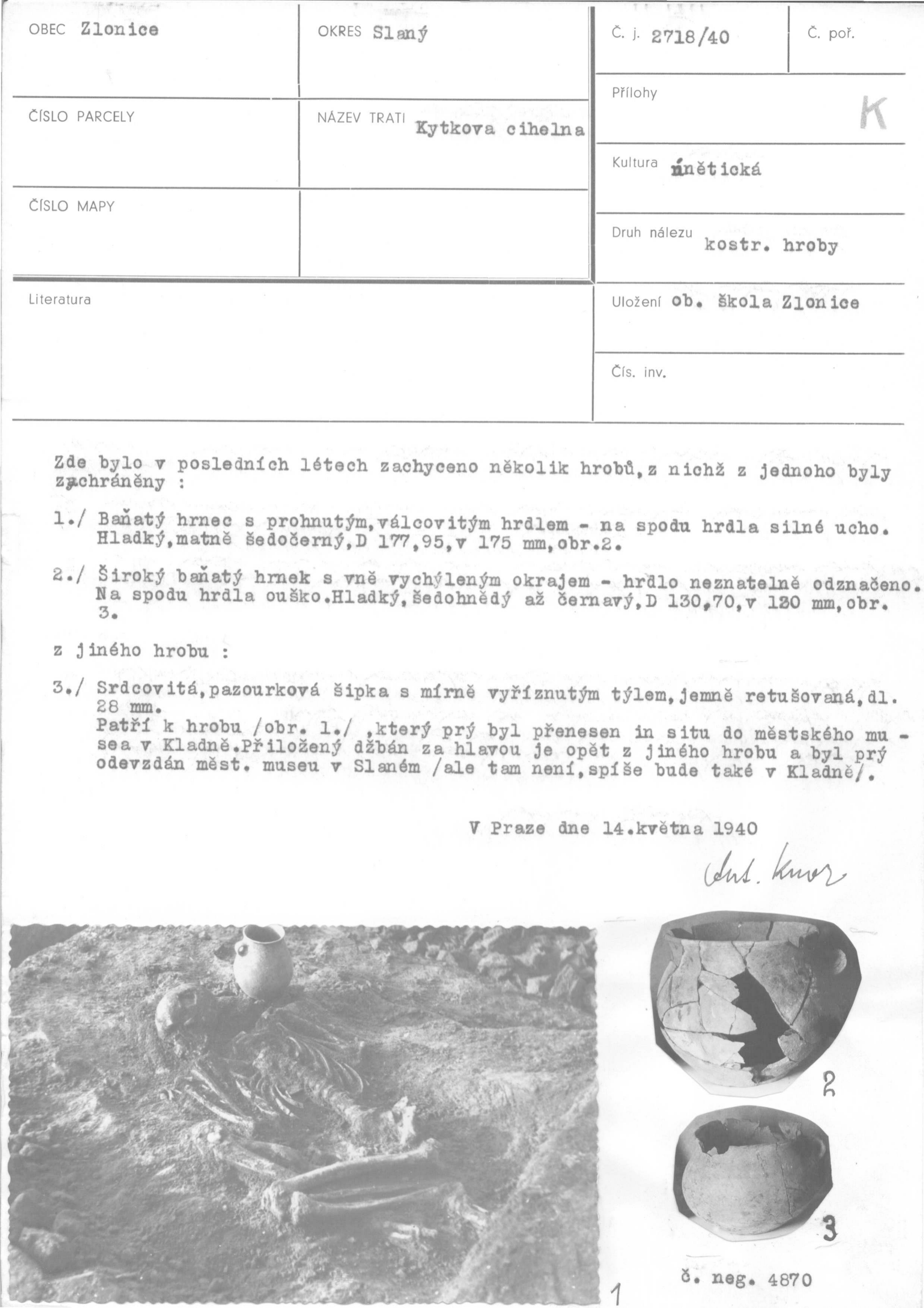}
        \caption{Photo cutouts}
        \label{fig:e_photo_l_3}
    \end{subfigure}

    \vspace{0.5em}

    \begin{subfigure}{0.45\textwidth}
        \centering
        \includegraphics[width=\textwidth]{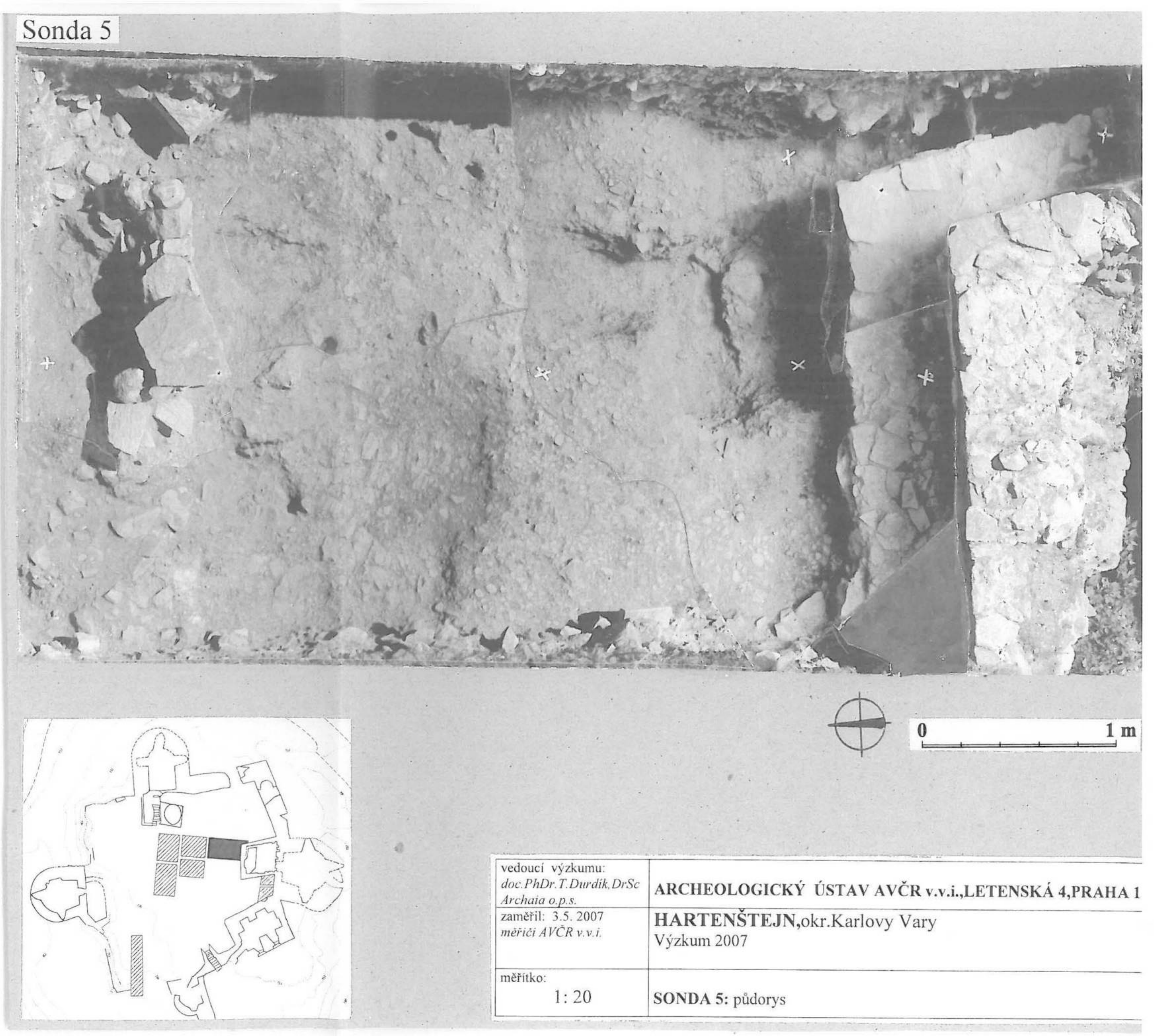}
        \caption{Photo and legend}
        \label{fig:e_photo_l_4}
    \end{subfigure}
    \begin{subfigure}{0.45\textwidth}
        \centering
        \includegraphics[width=\textwidth]{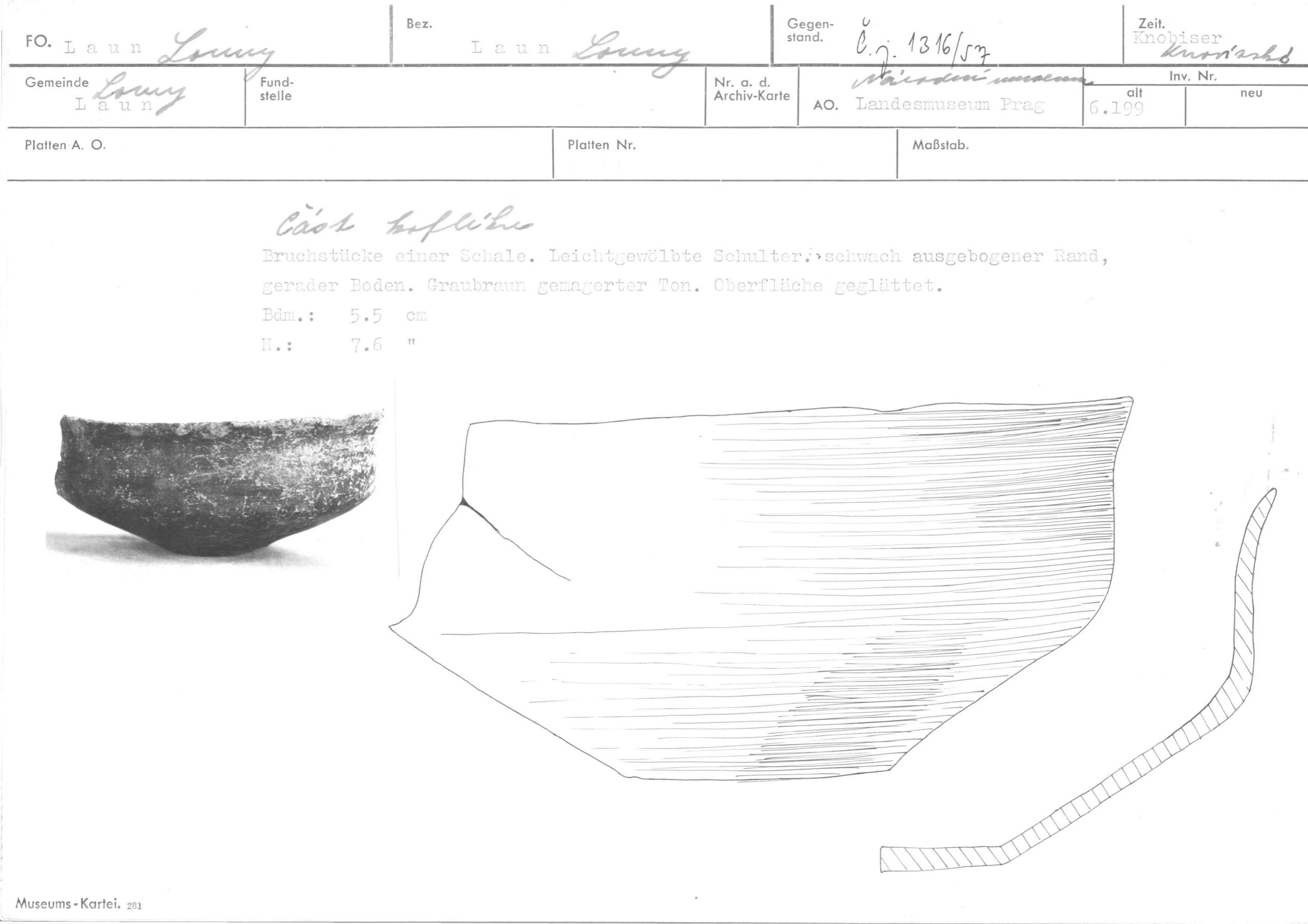}
        \caption{Near drawing}
        \label{fig:e_photo_l_5}
    \end{subfigure}

    \vspace{0.5em}

    \begin{subfigure}{0.45\textwidth}
        \centering
        \includegraphics[width=\textwidth]{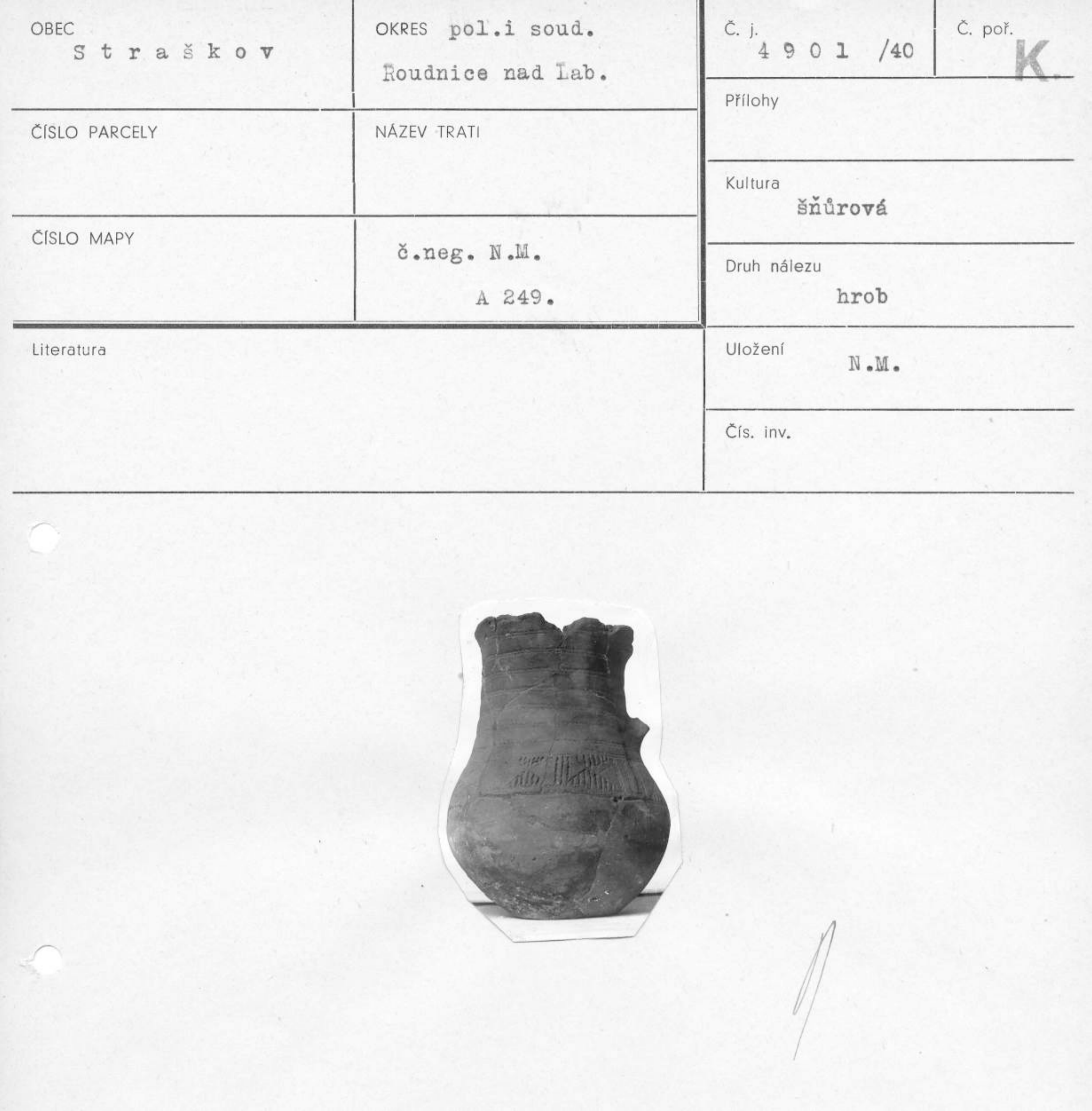}
        \caption{Within filled form}
        \label{fig:e_photo_l_6}
    \end{subfigure}
    \begin{subfigure}{0.45\textwidth}
        \centering
        \includegraphics[width=\textwidth]{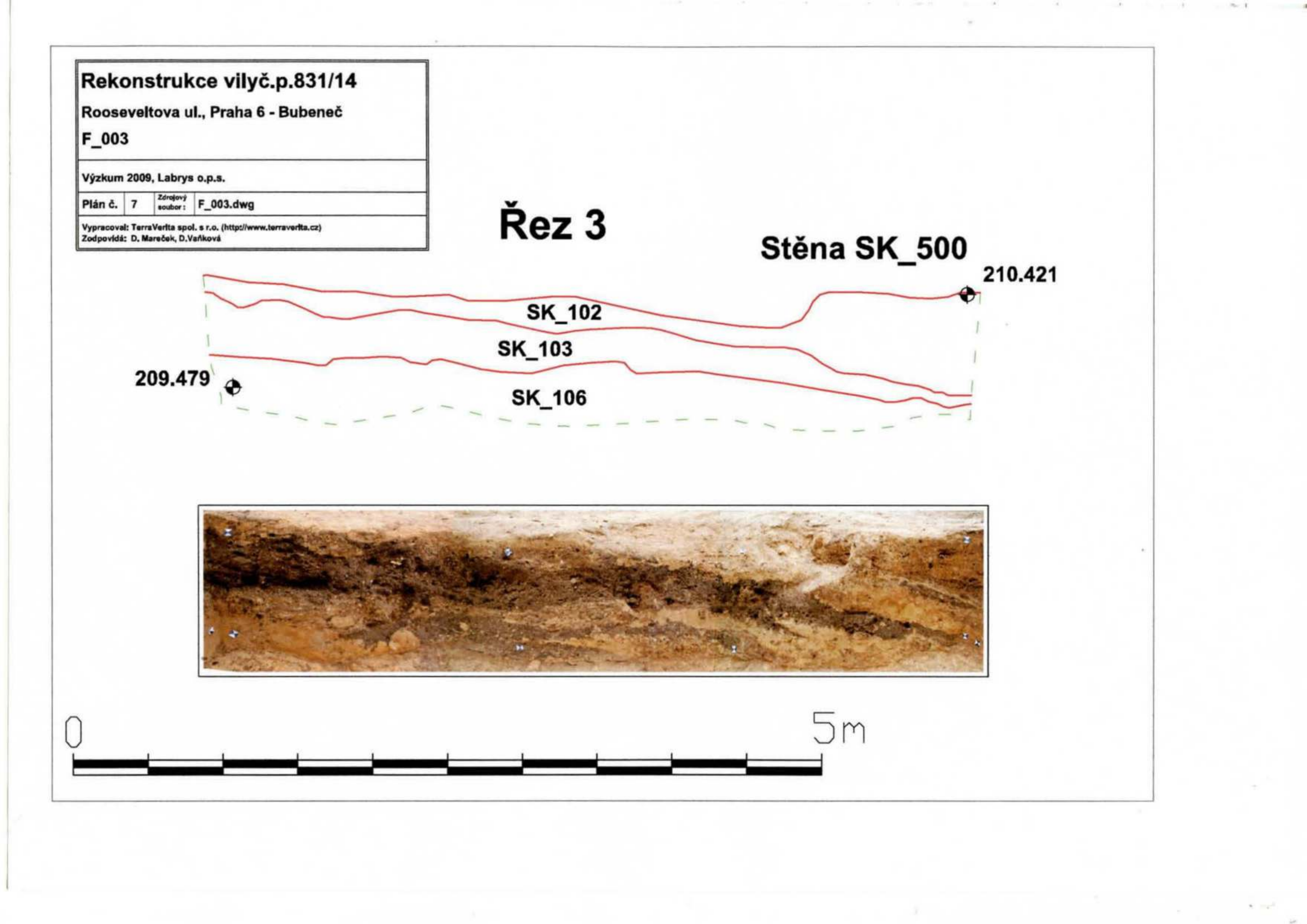}
        \caption{Object and legend}
        \label{fig:e_photo_l_7}
    \end{subfigure}
\end{figure}

% TEXT
\begin{figure}[h]
    \centering
    \caption{Label TEXT examples}
    \label{fig:example_text}
    \begin{subfigure}{0.3\textwidth}
        \centering
        \includegraphics[width=\textwidth]{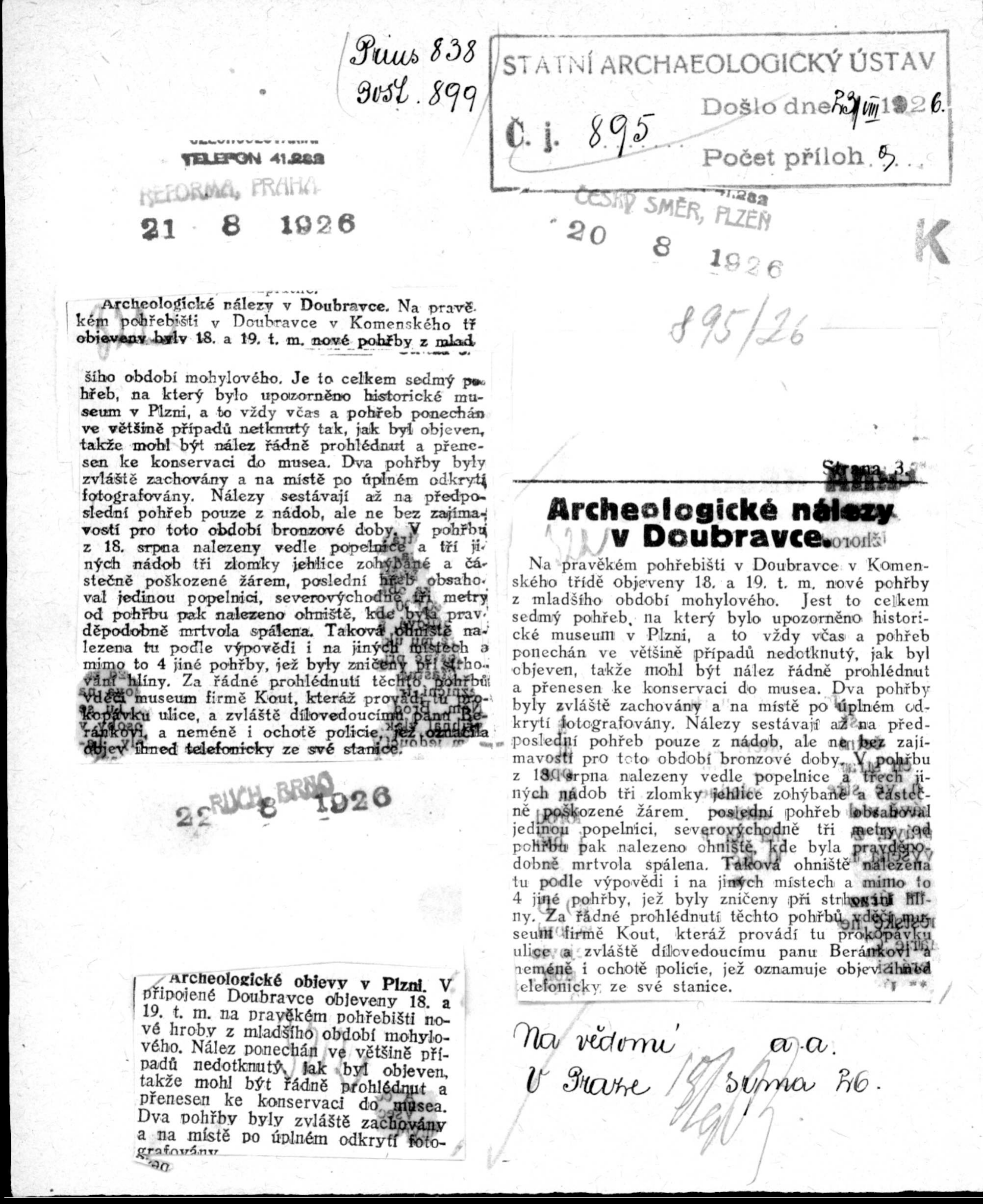}
        \caption{Newspaper cutouts}
        \label{fig:e_text_1}
    \end{subfigure}
    \begin{subfigure}{0.3\textwidth}
        \centering
        \includegraphics[width=\textwidth]{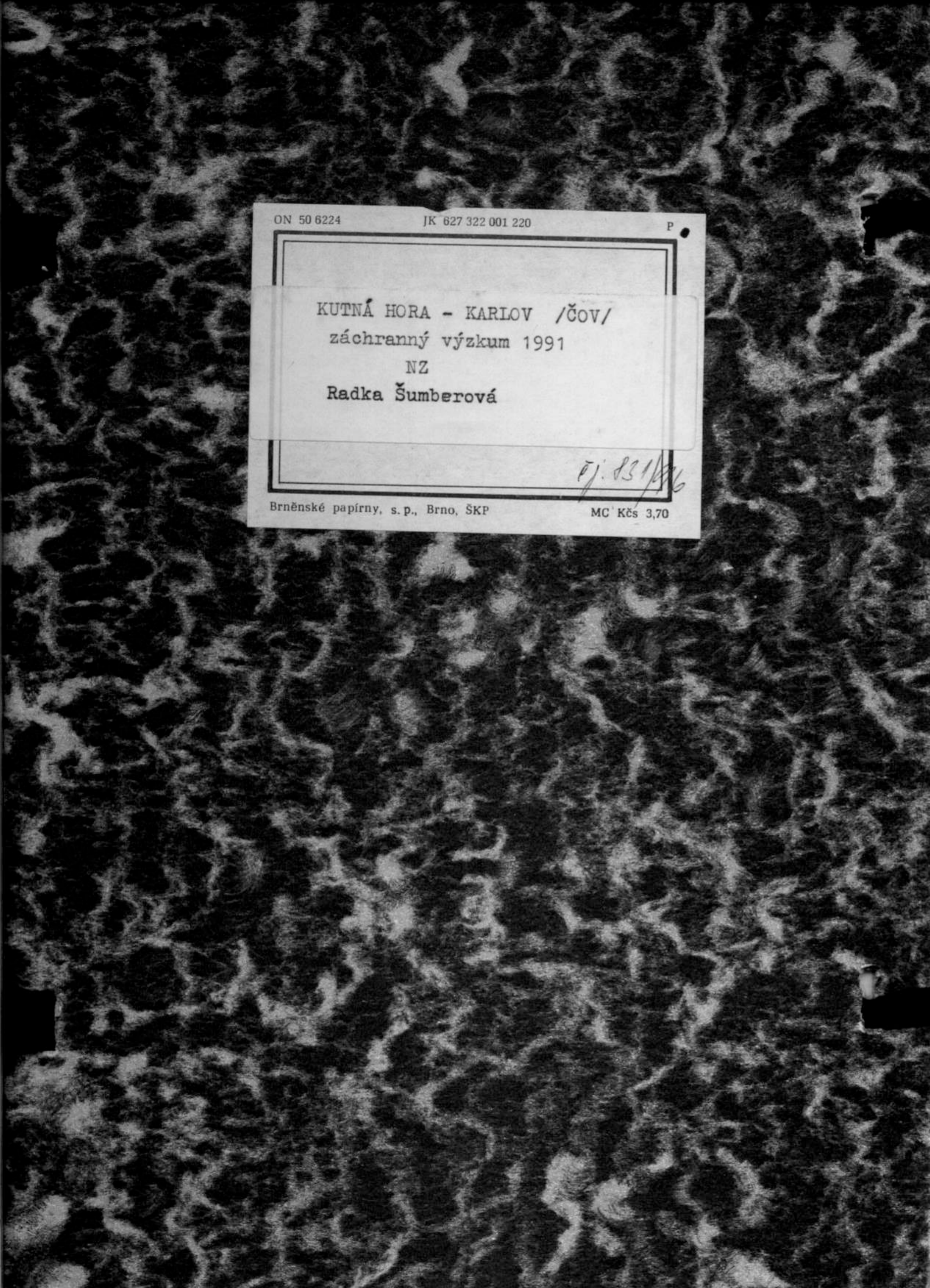}
        \caption{Journal cover}
        \label{fig:e_text_2}
    \end{subfigure}
    \begin{subfigure}{0.3\textwidth}
        \centering
        \includegraphics[width=\textwidth]{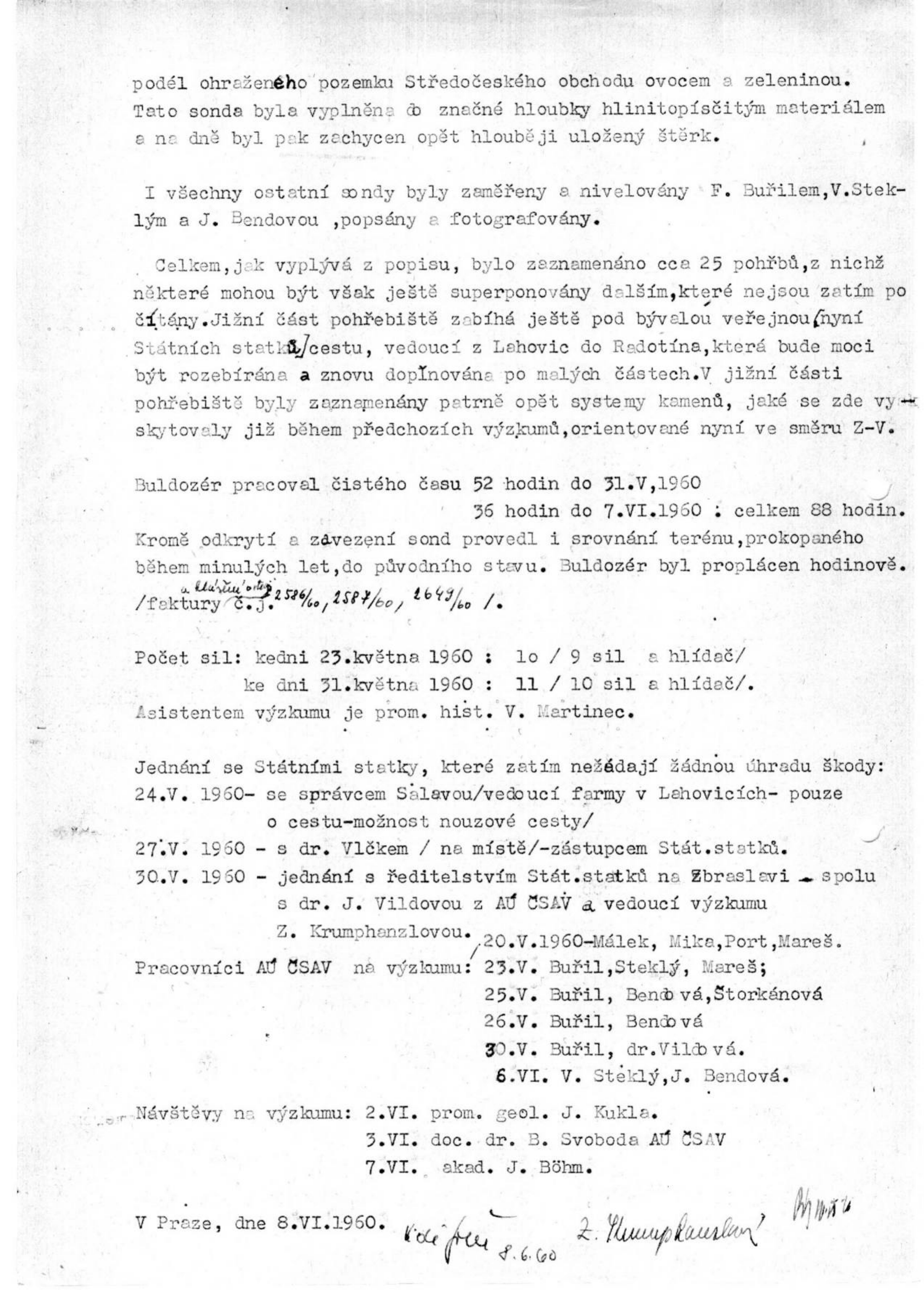}
        \caption{Manual correction}
        \label{fig:e_text_3}
    \end{subfigure}

    \vspace{0.5em}

    \begin{subfigure}{0.45\textwidth}
        \centering
        \includegraphics[width=\textwidth]{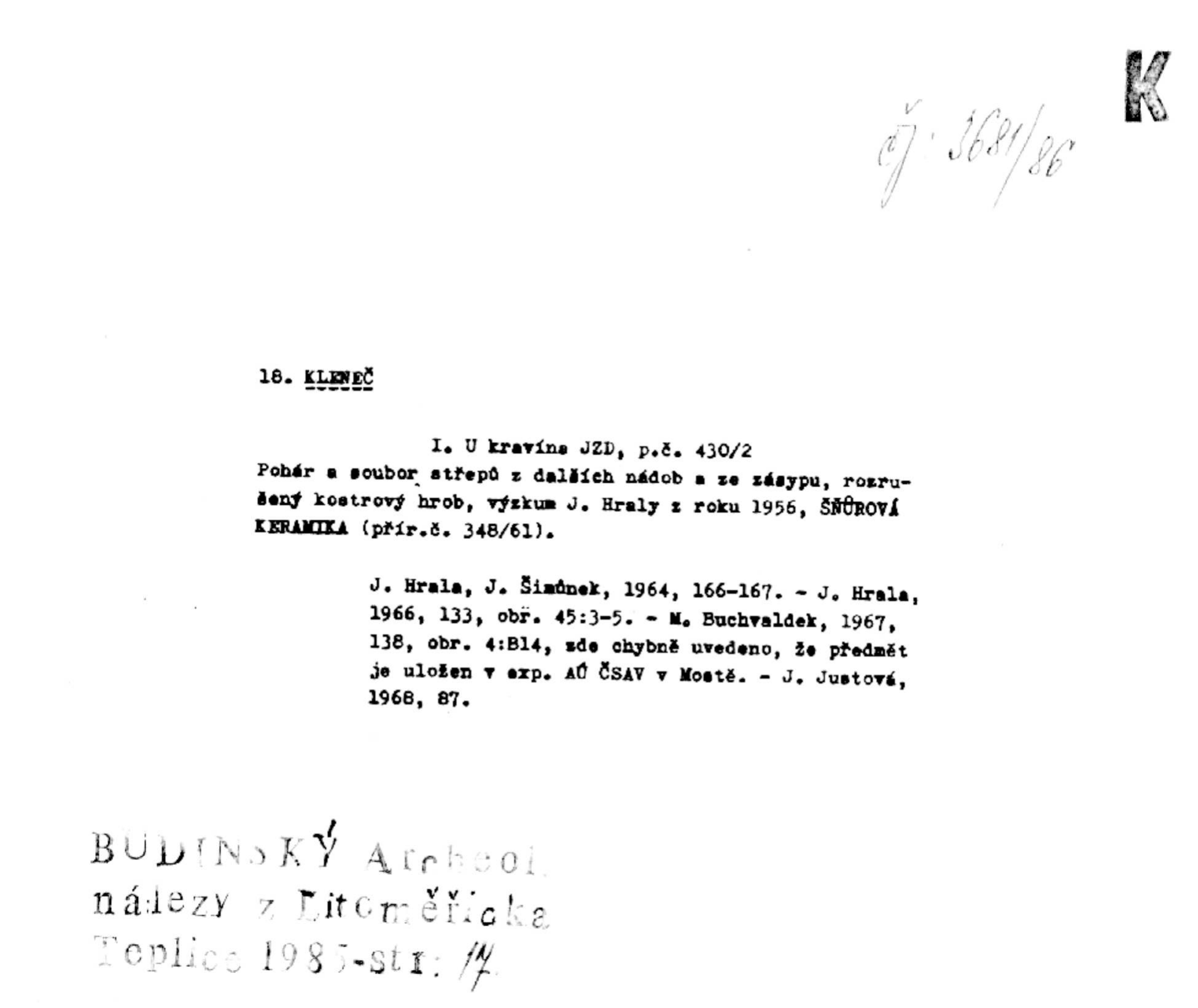}
        \caption{Text comment}
        \label{fig:e_text_4}
    \end{subfigure}
    \begin{subfigure}{0.45\textwidth}
        \centering
        \includegraphics[width=\textwidth]{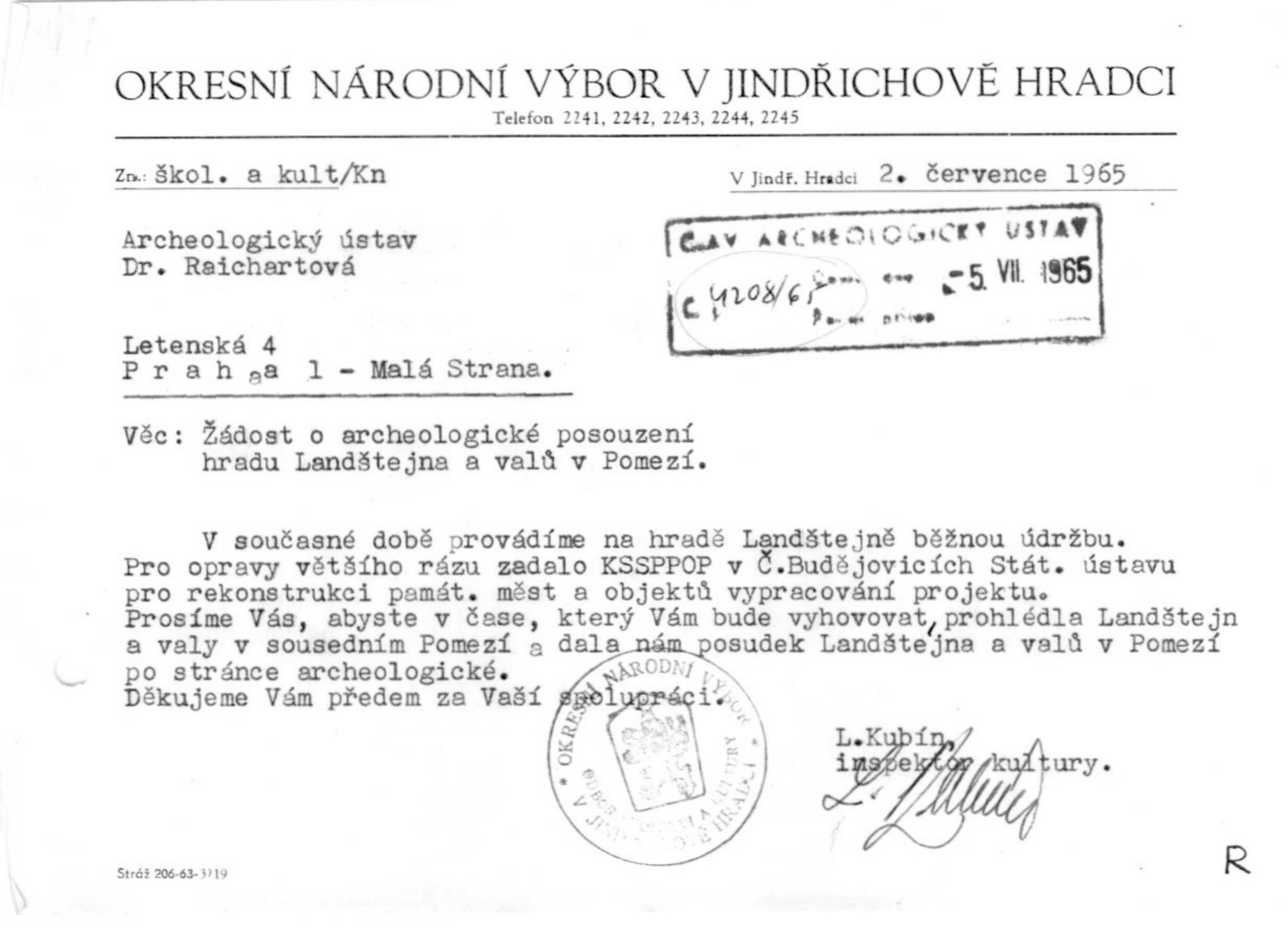}
        \caption{Front page}
        \label{fig:e_text_5}
    \end{subfigure}

    \vspace{0.5em}

    \begin{subfigure}{0.45\textwidth}
        \centering
        \includegraphics[width=\textwidth]{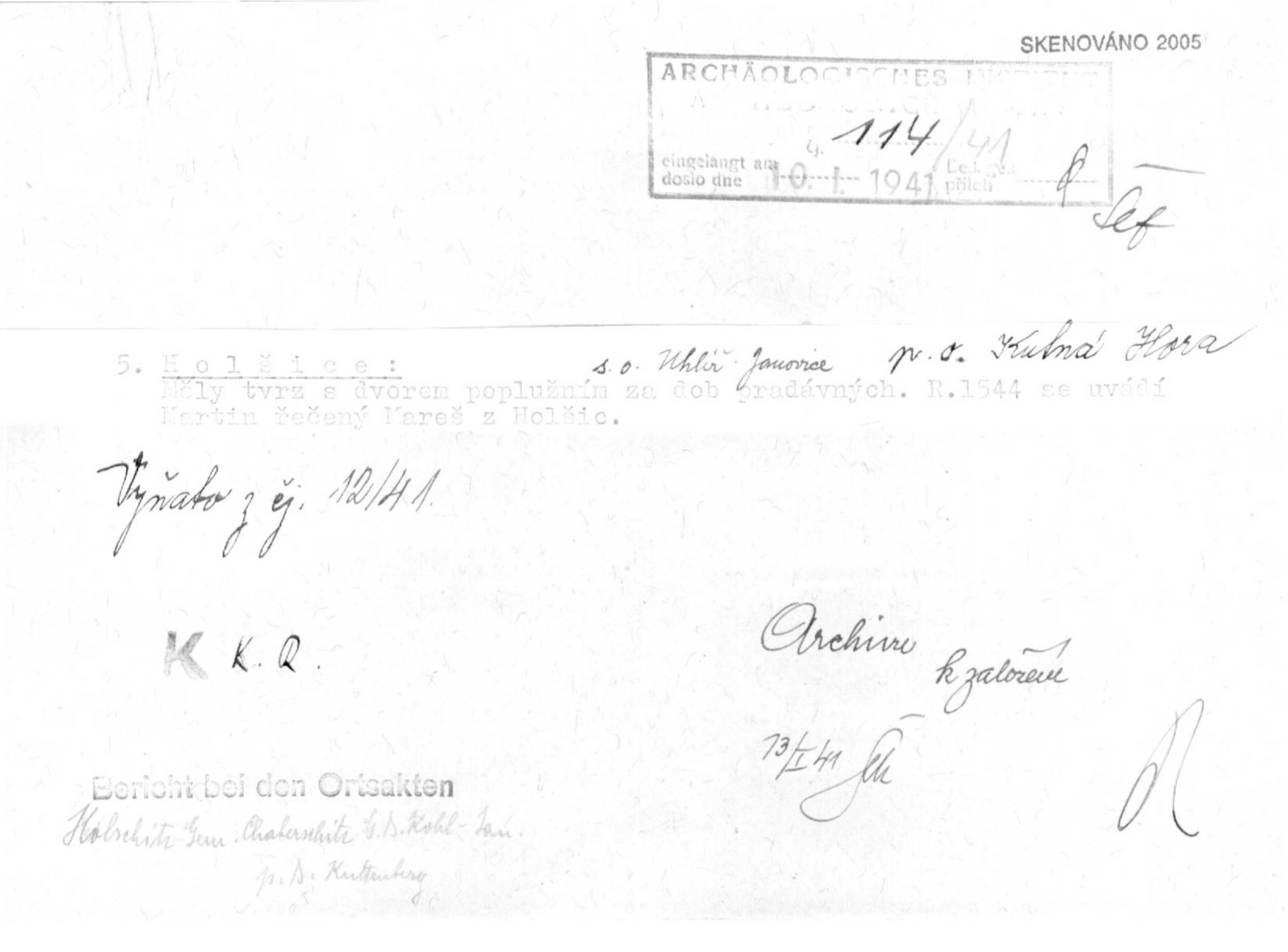}
        \caption{Whitespace}
        \label{fig:e_text_6}
    \end{subfigure}
    \begin{subfigure}{0.45\textwidth}
        \centering
        \includegraphics[width=\textwidth]{img/pages/example_TEXT_postcard.pdf}
        \caption{Postcard scan}
        \label{fig:e_text_7}
    \end{subfigure}
\end{figure}

% TEXT_HW
\begin{figure}[h]
    \centering
    \caption{Label TEXT\_HW examples}
    \label{fig:example_text_hw}
    \begin{subfigure}{0.3\textwidth}
        \centering
        \includegraphics[width=\textwidth]{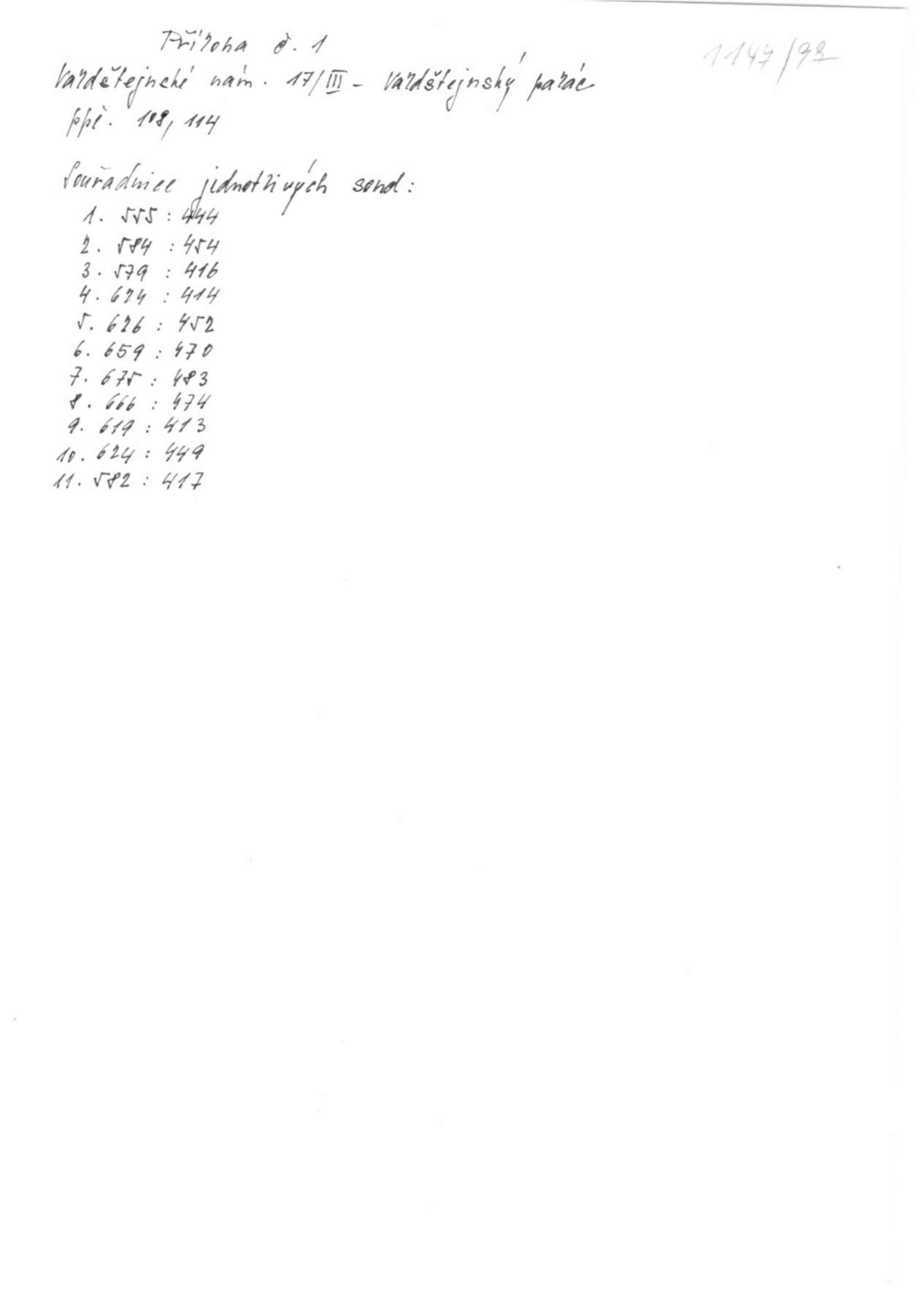}
        \caption{Clean list-like}
        \label{fig:e_text_hw_1}
    \end{subfigure}
    \begin{subfigure}{0.3\textwidth}
        \centering
        \includegraphics[width=\textwidth]{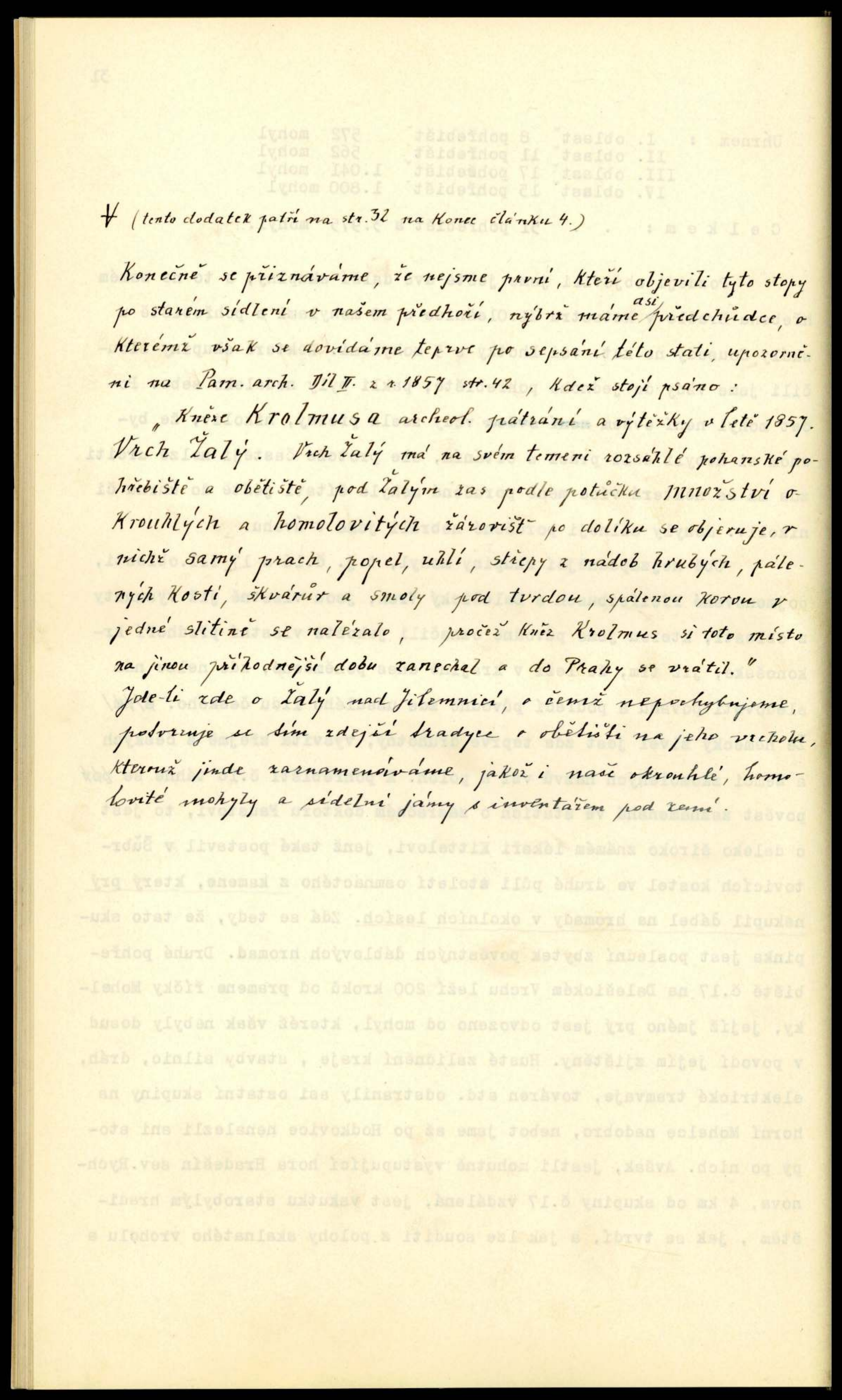}
        \caption{Yellow paper}
        \label{fig:e_text_hw_2}
    \end{subfigure}
    \begin{subfigure}{0.3\textwidth}
        \centering
        \includegraphics[width=\textwidth]{img/pages/example_TEXT_HW.pdf}
        \caption{Flipped paper}
        \label{fig:e_text_hw_3}
    \end{subfigure}

    \vspace{0.5em}

    \begin{subfigure}{0.45\textwidth}
        \centering
        \includegraphics[width=\textwidth]{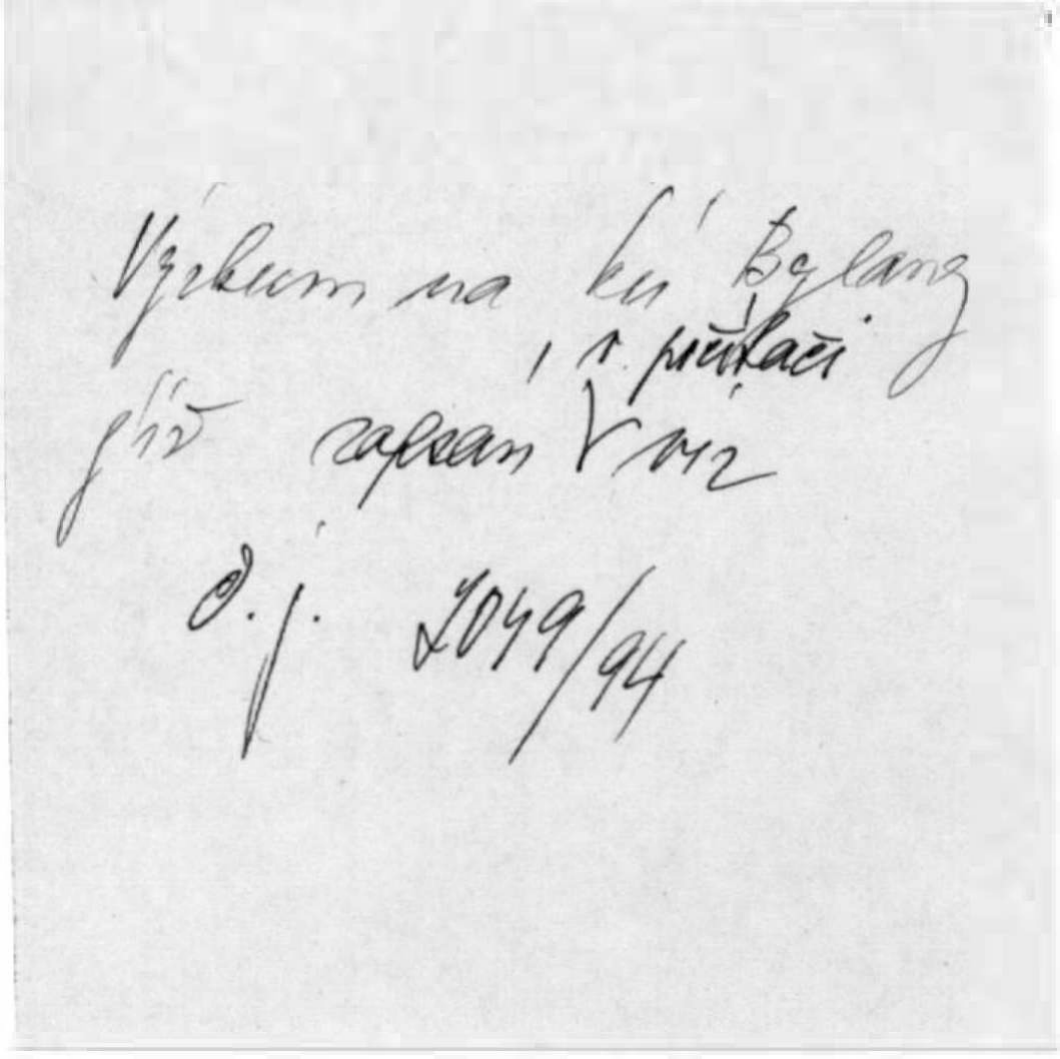}
        \caption{Tiny paper}
        \label{fig:e_text_hw_4}
    \end{subfigure}
    \begin{subfigure}{0.45\textwidth}
        \centering
        \includegraphics[width=\textwidth]{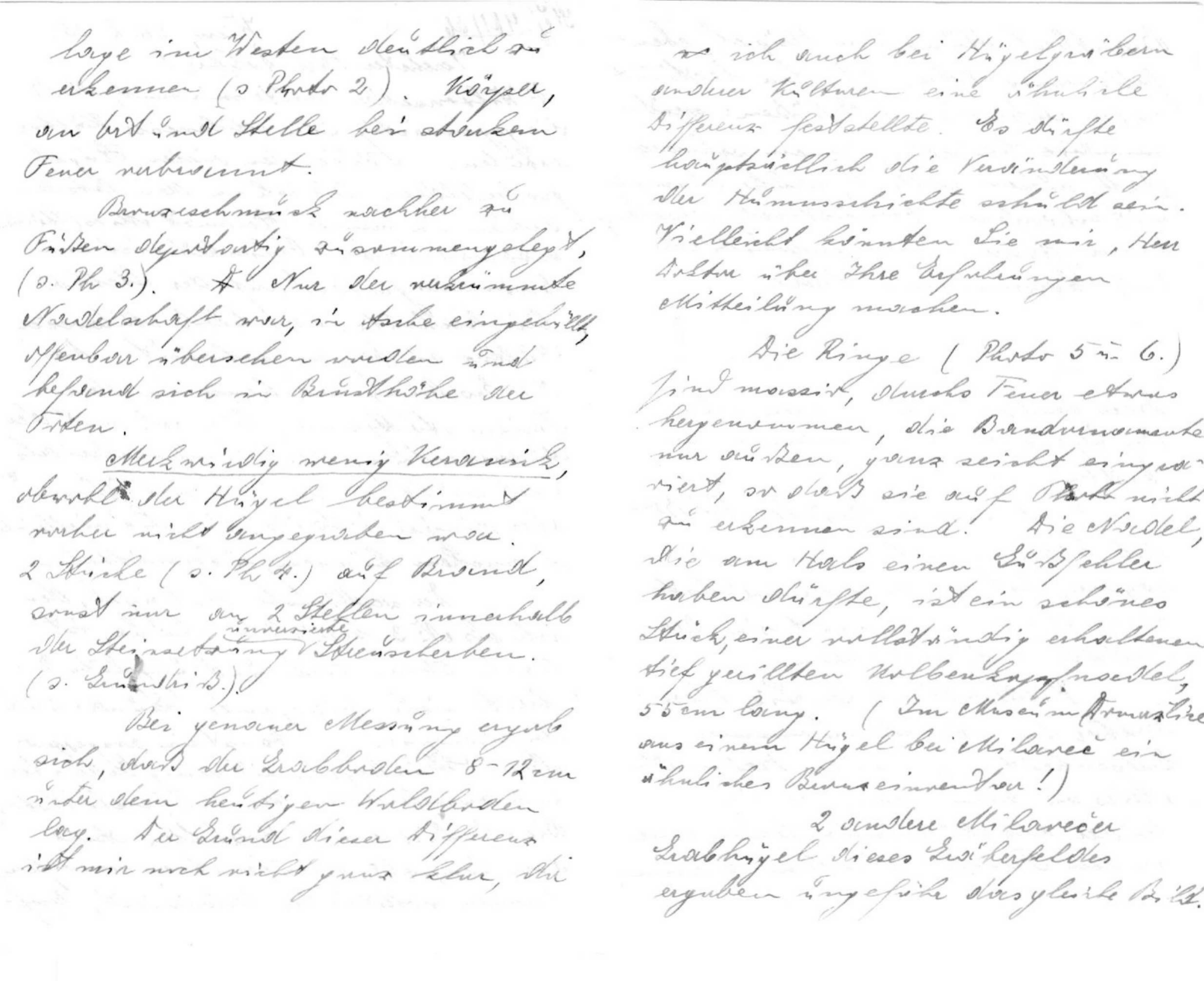}
        \caption{Dual page}
        \label{fig:e_text_hw_5}
    \end{subfigure}

    \vspace{0.5em}

    \begin{subfigure}{0.45\textwidth}
        \centering
        \includegraphics[width=\textwidth]{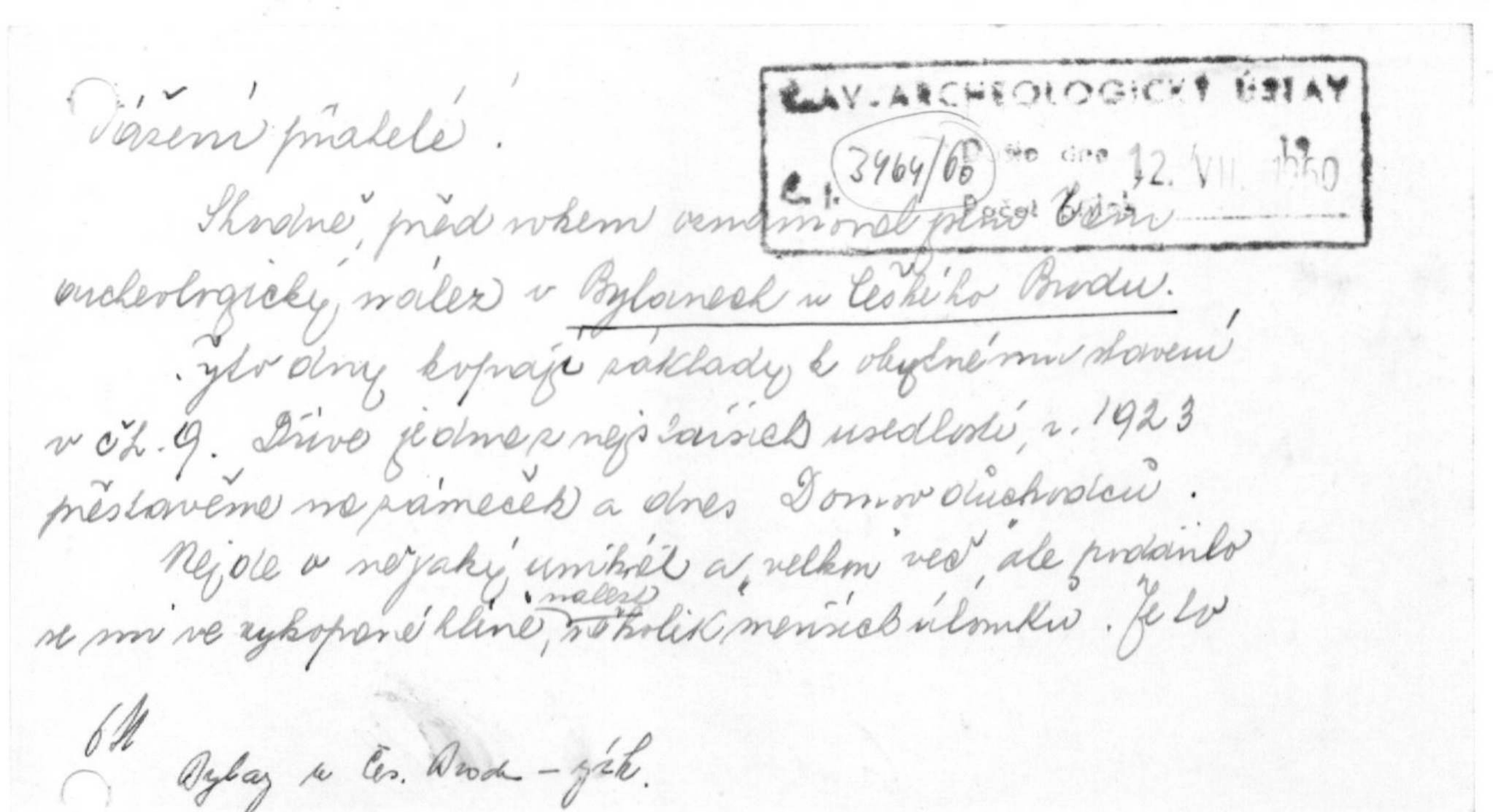}
        \caption{Simple stamp}
        \label{fig:e_text_hw_6}
    \end{subfigure}
    \begin{subfigure}{0.45\textwidth}
        \centering
        \includegraphics[width=\textwidth]{img/pages/example_TEXT_HW_water.pdf}
        \caption{Wet paper}
        \label{fig:e_text_hw_7}
    \end{subfigure}
\end{figure}

% TEXT_P
\begin{figure}[h]
    \centering
    \caption{Label TEXT\_P examples}
    \label{fig:example_text_p}
    \begin{subfigure}{0.3\textwidth}
        \centering
        \includegraphics[width=\textwidth]{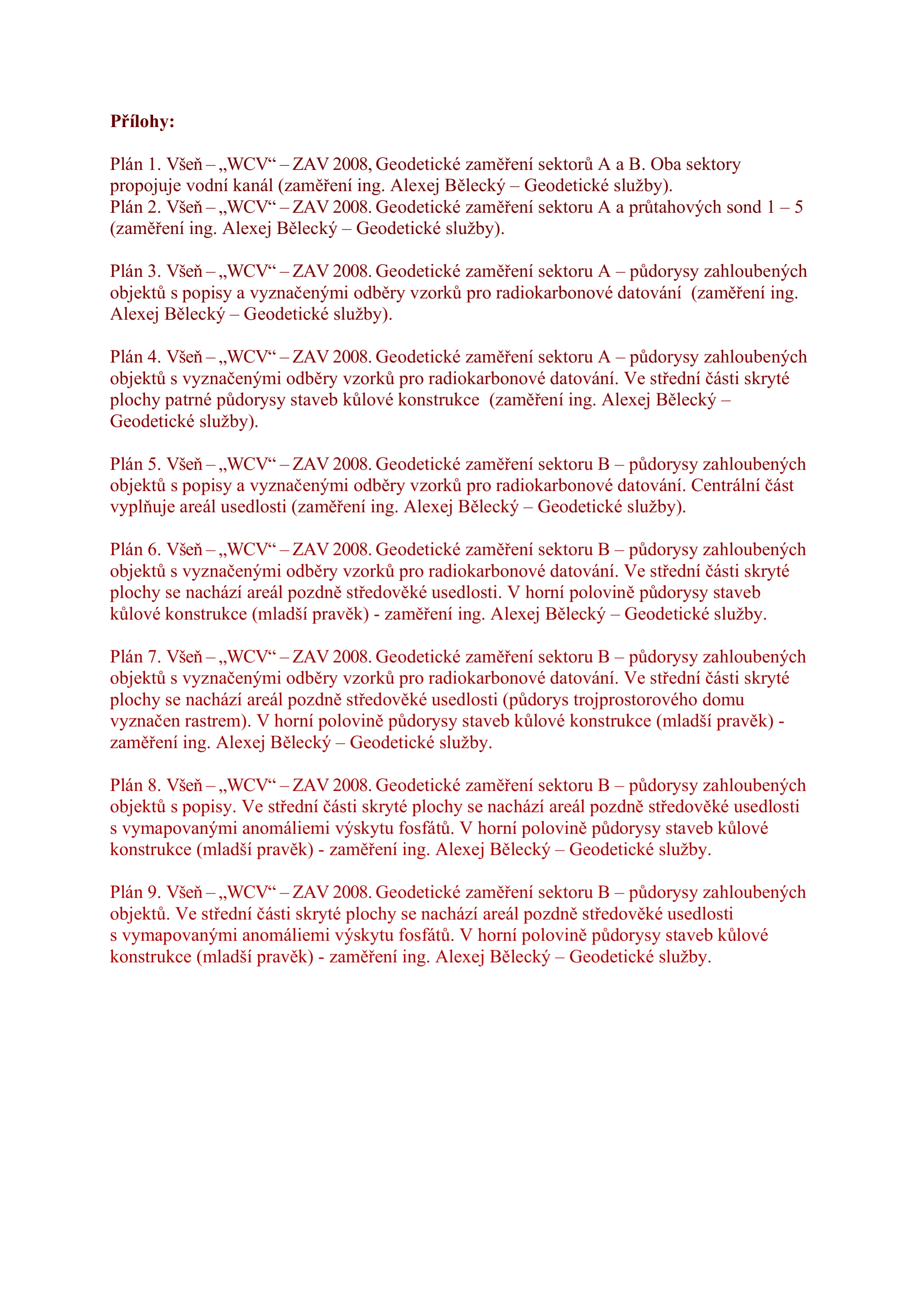}
        \caption{Color printed}
        \label{fig:e_text_p_1}
    \end{subfigure}
    \begin{subfigure}{0.3\textwidth}
        \centering
        \includegraphics[width=\textwidth]{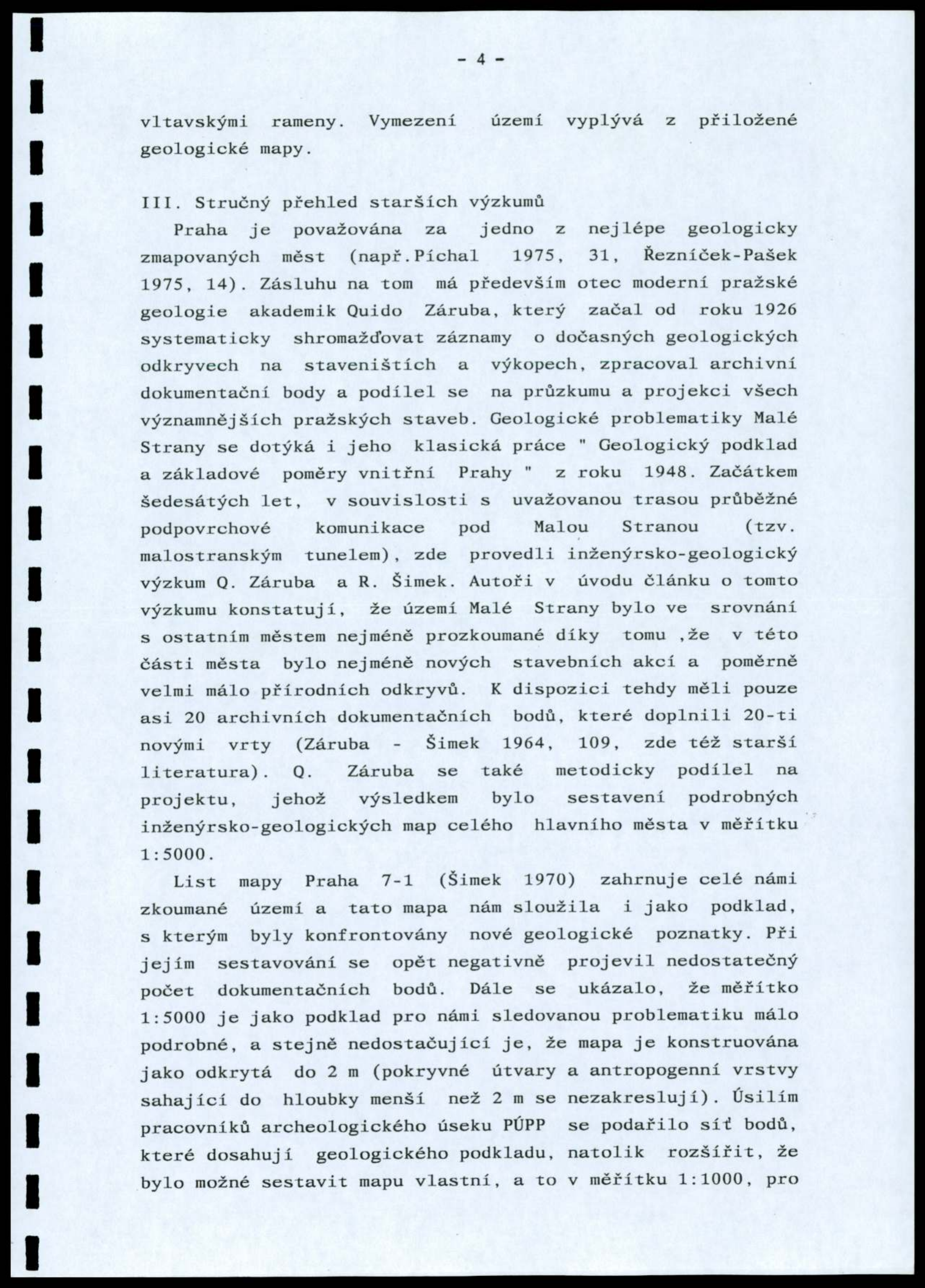}
        \caption{Journal scan}
        \label{fig:e_text_p_2}
    \end{subfigure}
    \begin{subfigure}{0.3\textwidth}
        \centering
        \includegraphics[width=\textwidth]{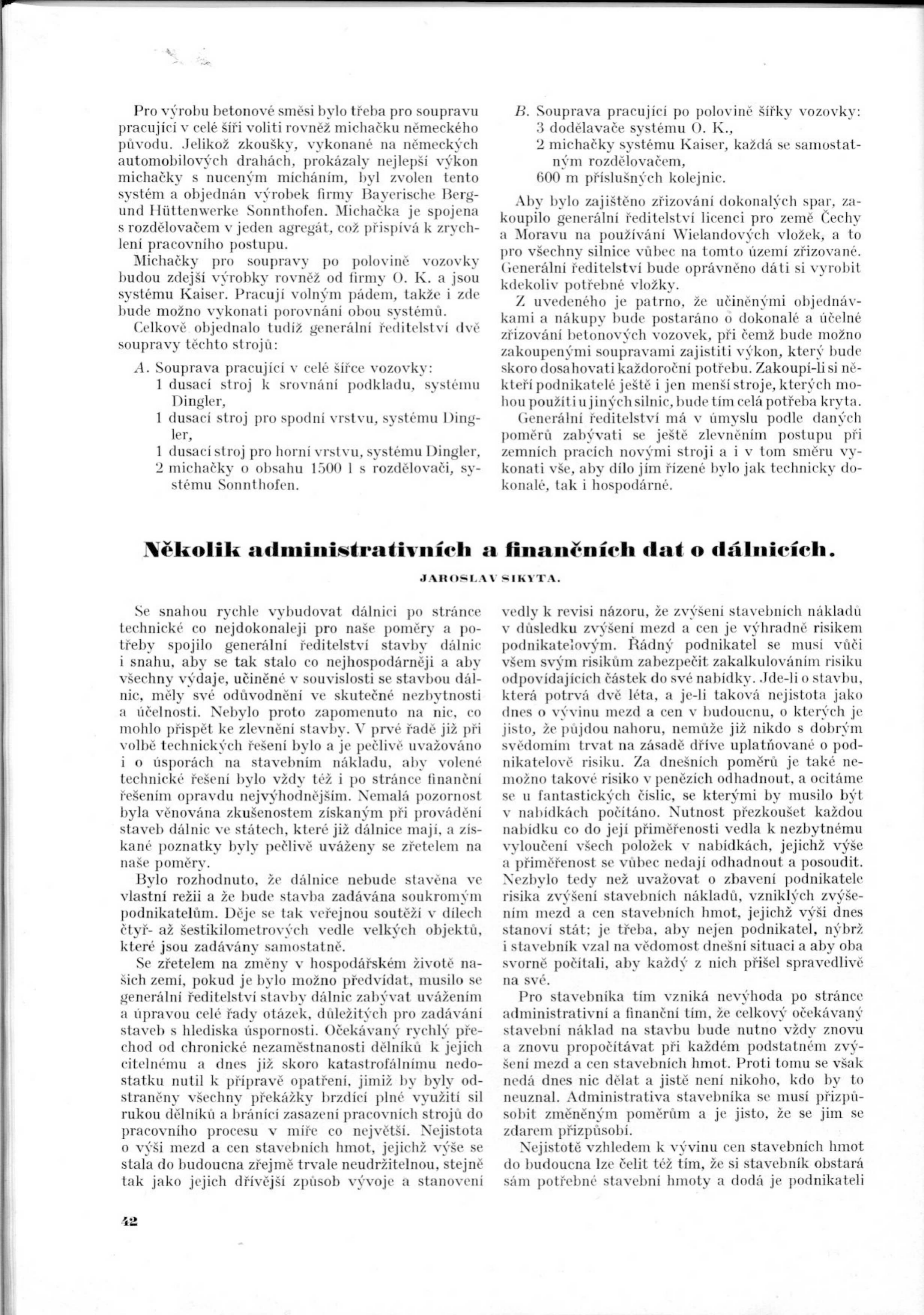}
        \caption{Book or article page}
        \label{fig:e_text_p_3}
    \end{subfigure}

    \vspace{0.5em}

    \begin{subfigure}{0.3\textwidth}
        \centering
        \includegraphics[width=\textwidth]{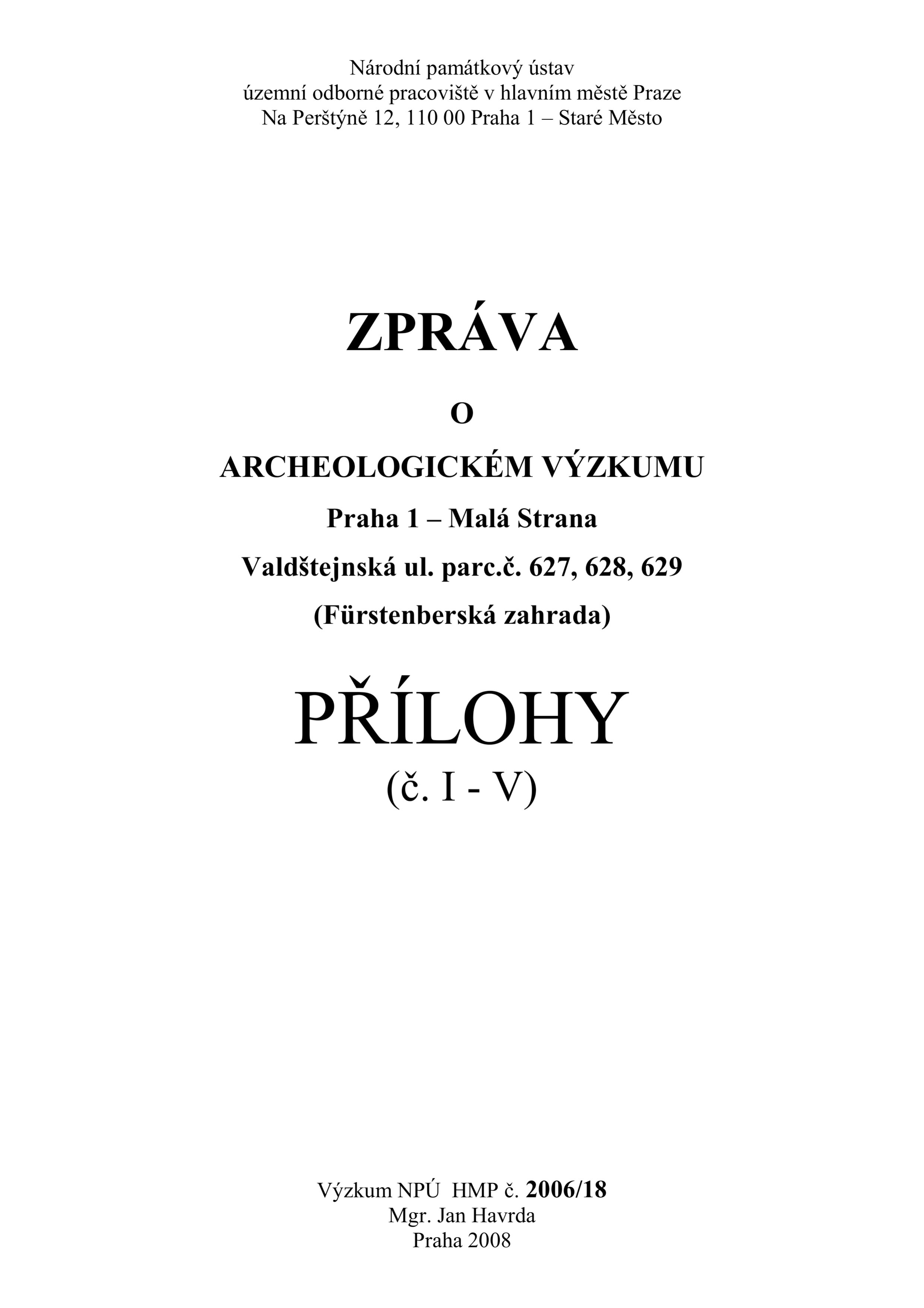}
        \caption{Title page}
        \label{fig:e_text_p_4}
    \end{subfigure}
    \begin{subfigure}{0.3\textwidth}
        \centering
        \includegraphics[width=\textwidth]{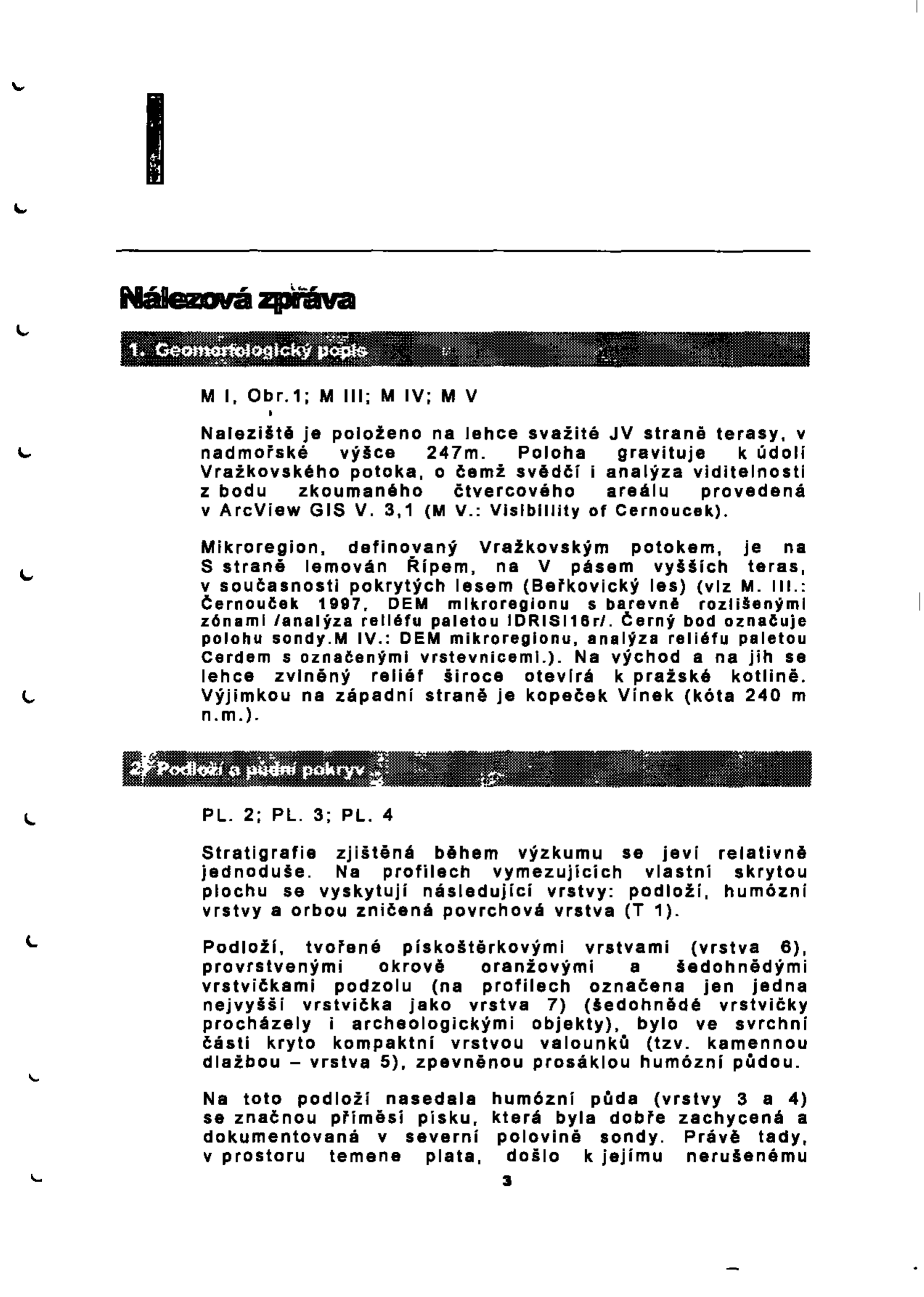}
        \caption{Decorative prints}
        \label{fig:e_text_p_5}
    \end{subfigure}
    \begin{subfigure}{0.3\textwidth}
        \centering
        \includegraphics[width=\textwidth]{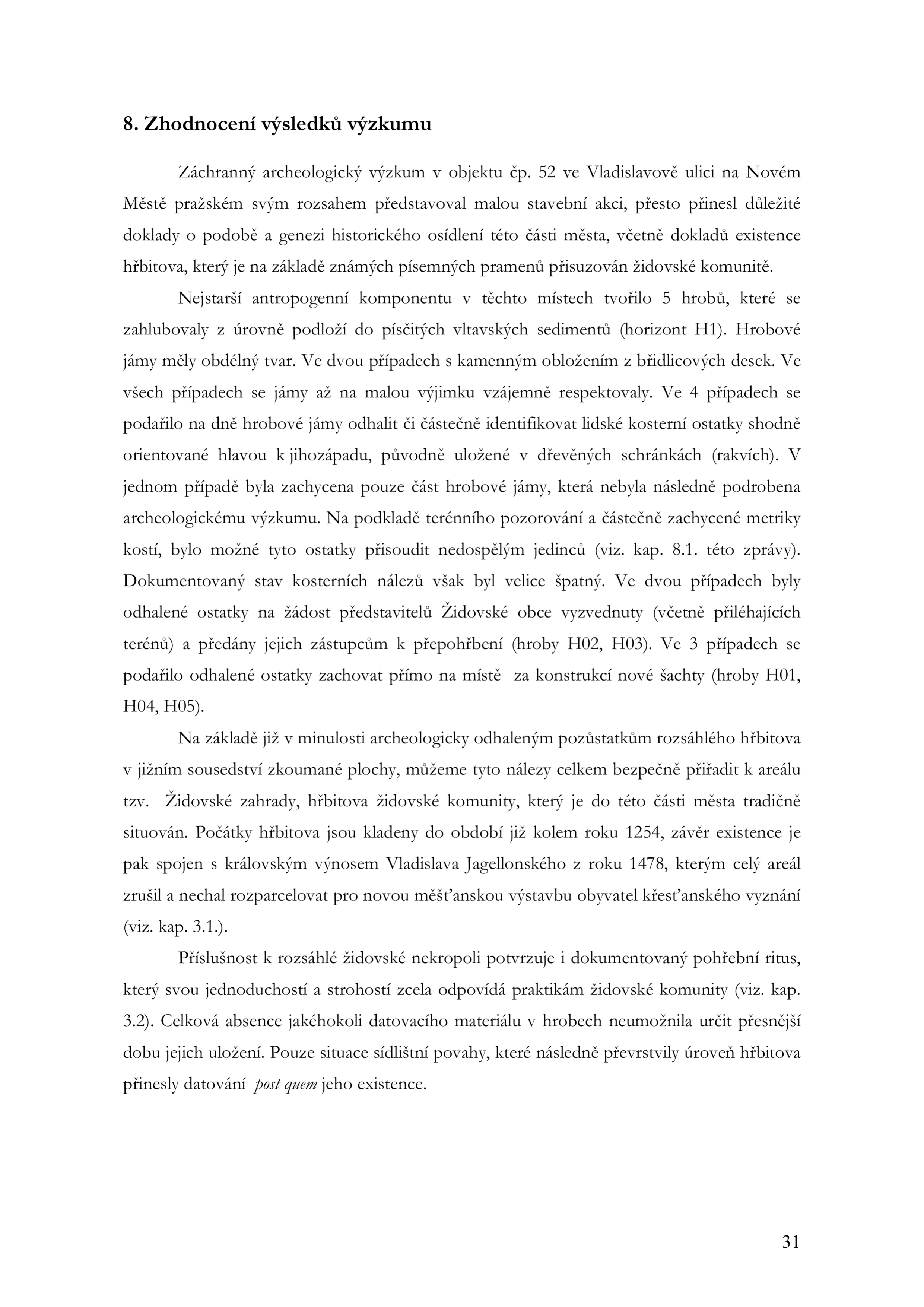}
        \caption{Simple text}
        \label{fig:e_text_p_6}
    \end{subfigure}

    \vspace{0.5em}

    \begin{subfigure}{0.3\textwidth}
        \centering
        \includegraphics[width=\textwidth]{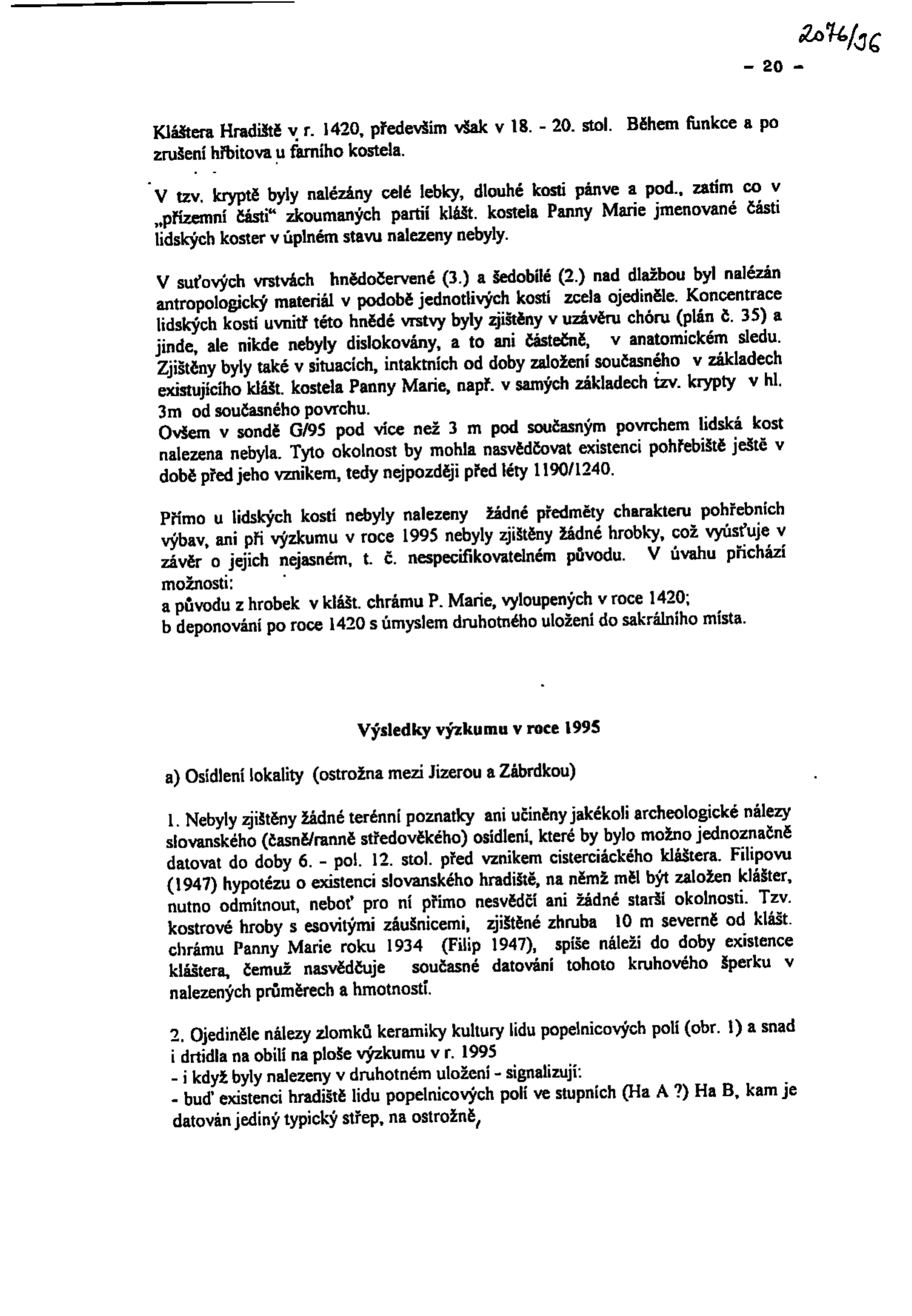}
        \caption{Skewed print}
        \label{fig:e_text_p_7}
    \end{subfigure}
    \begin{subfigure}{0.3\textwidth}
        \centering
        \includegraphics[width=\textwidth]{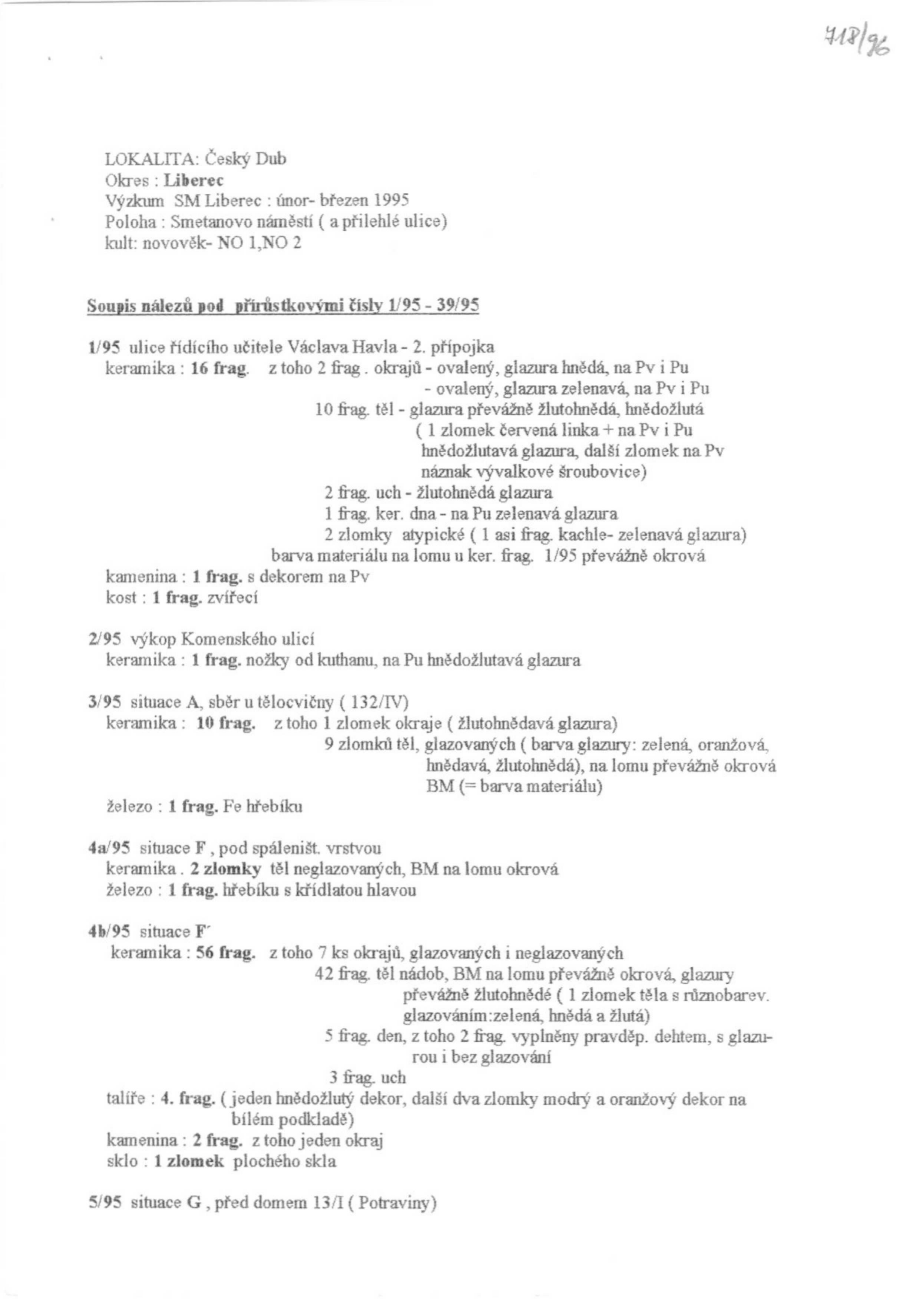}
        \caption{Shifted formatting}
        \label{fig:e_text_p_8}
    \end{subfigure}
    \begin{subfigure}{0.3\textwidth}
        \centering
        \includegraphics[width=\textwidth]{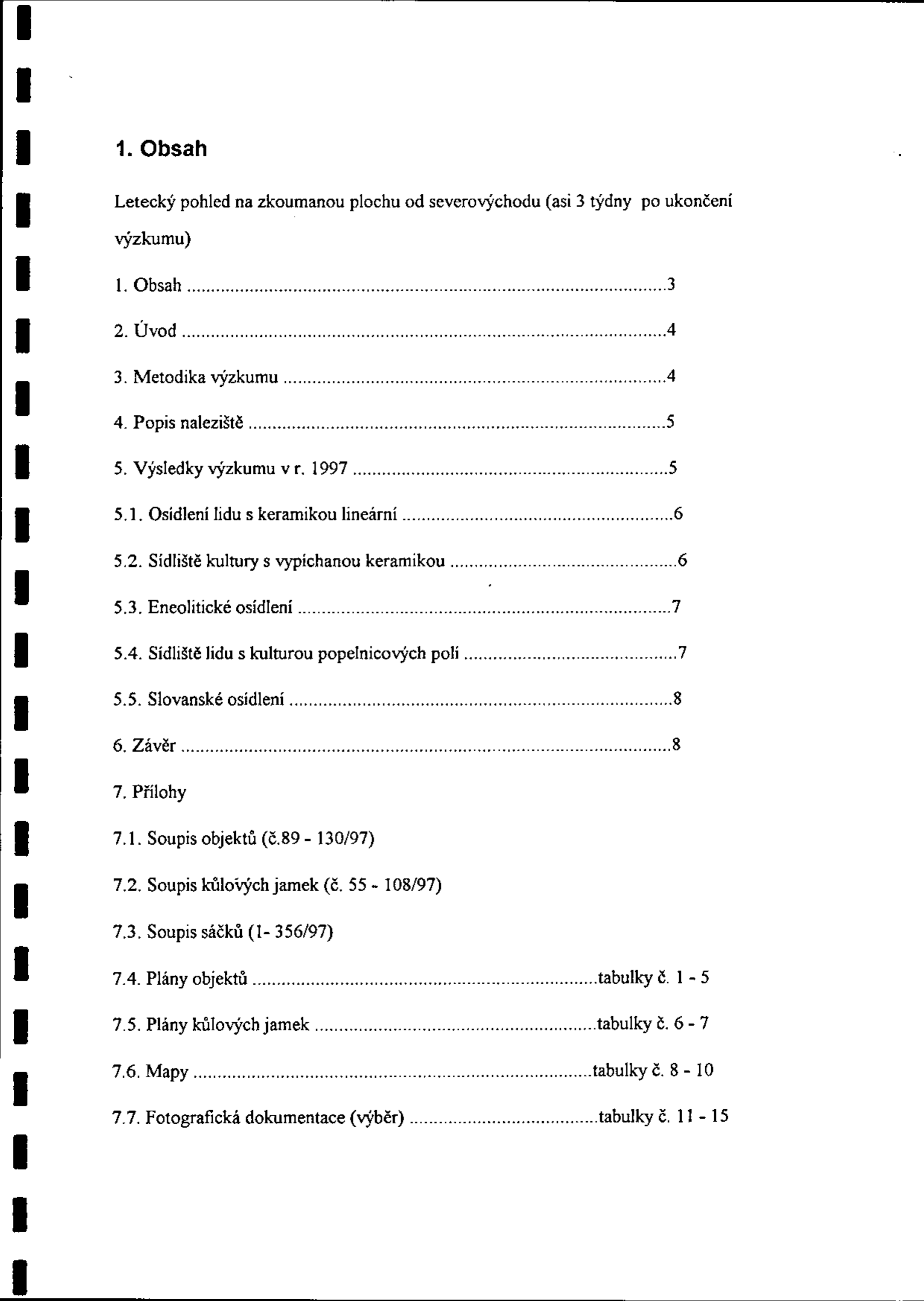}
        \caption{List-like column}
        \label{fig:e_text_p_9}
    \end{subfigure}
\end{figure}

% TEXT_T
\begin{figure}[h]
    \centering
    \caption{Label TEXT\_T examples}
    \label{fig:example_text_t}
    \begin{subfigure}{0.3\textwidth}
        \centering
        \includegraphics[width=\textwidth]{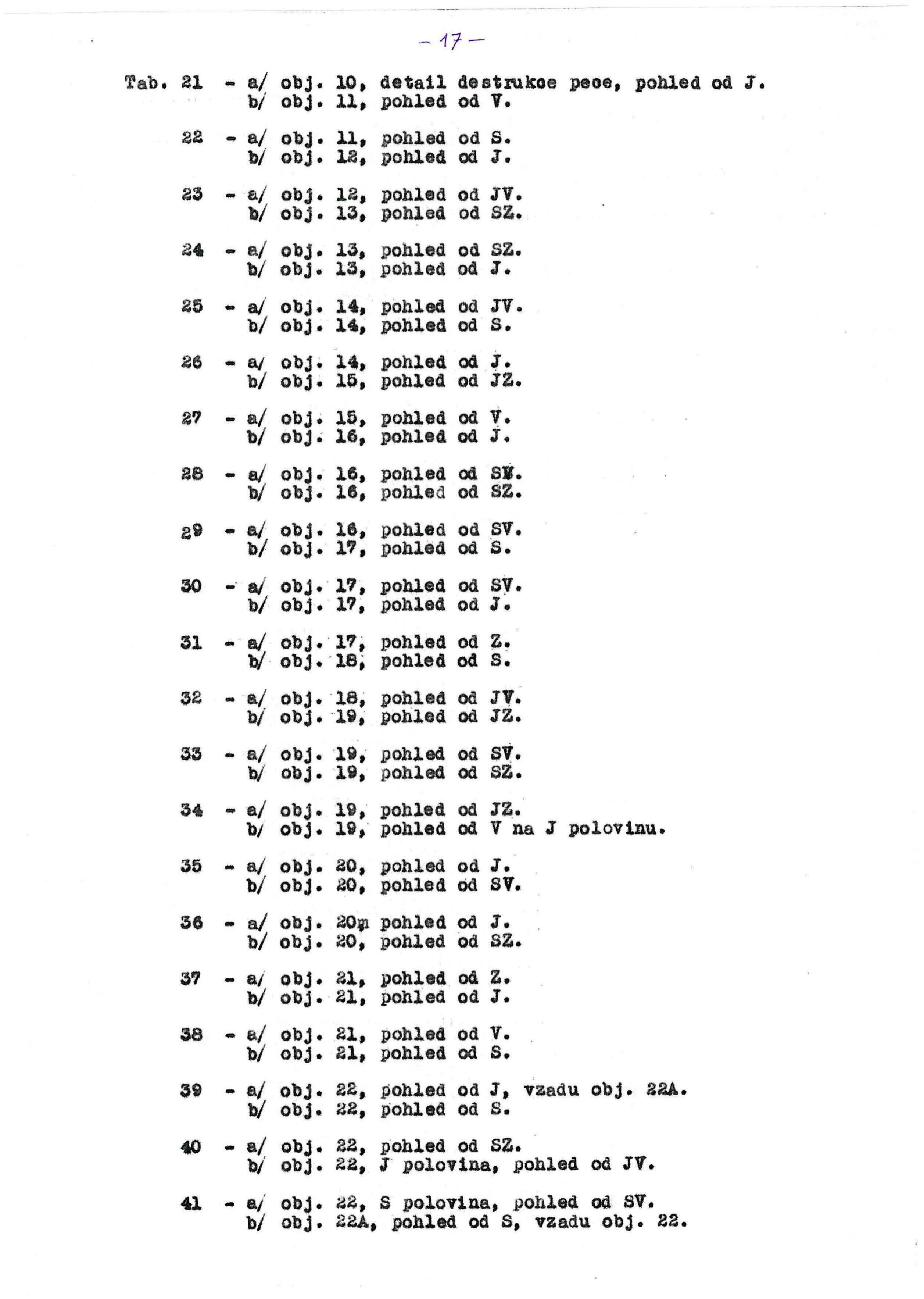}
        \caption{List-like text}
        \label{fig:e_text_t_1}
    \end{subfigure}
    \begin{subfigure}{0.3\textwidth}
        \centering
        \includegraphics[width=\textwidth]{img/pages/example_TEXT_T_yellow.pdf}
        \caption{Yellow journal page}
        \label{fig:e_text_t_2}
    \end{subfigure}
    \begin{subfigure}{0.3\textwidth}
        \centering
        \includegraphics[width=\textwidth]{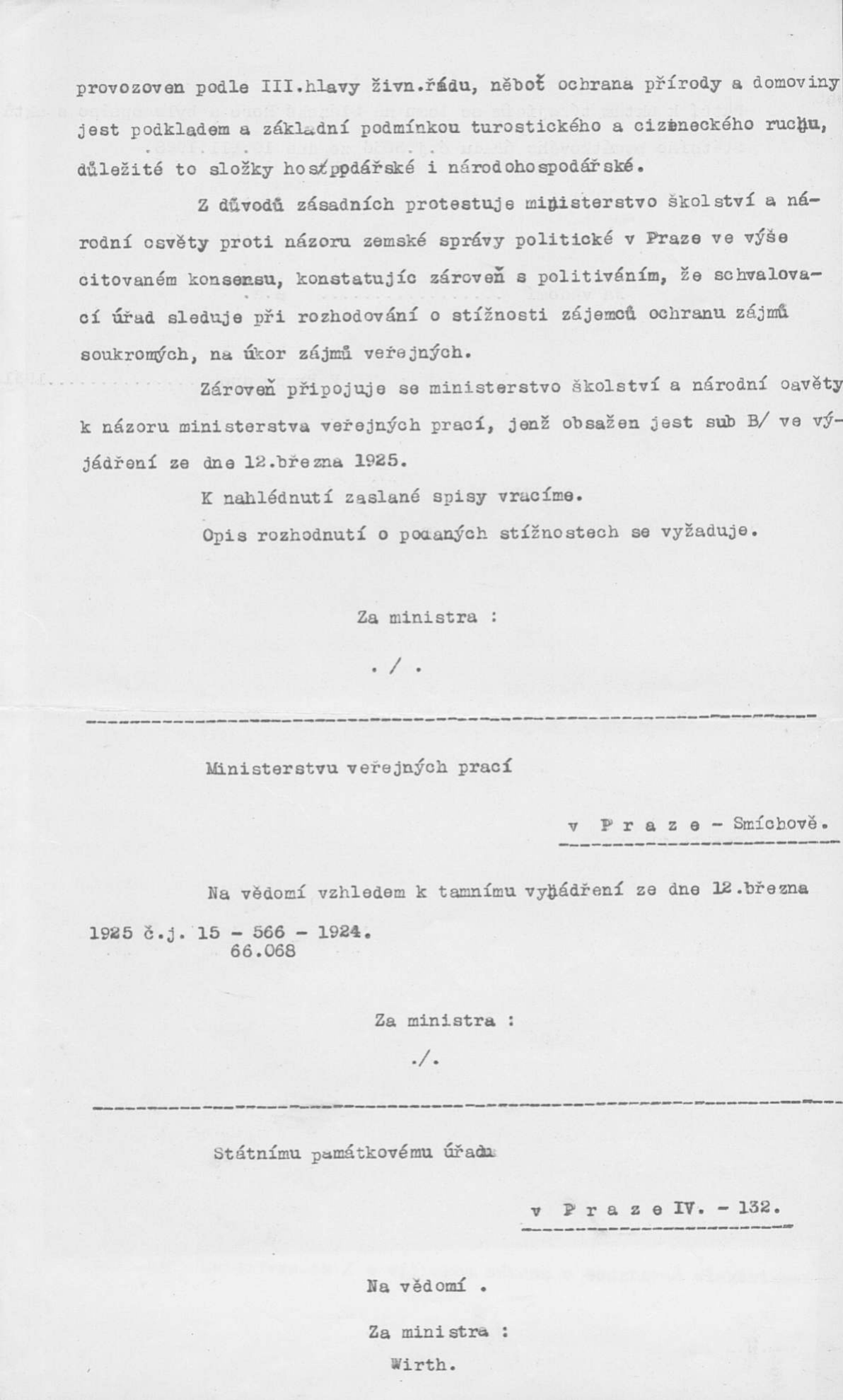}
        \caption{Decorative lines}
        \label{fig:e_text_t_3}
    \end{subfigure}

    \vspace{0.5em}

    \begin{subfigure}{0.3\textwidth}
        \centering
        \includegraphics[width=\textwidth]{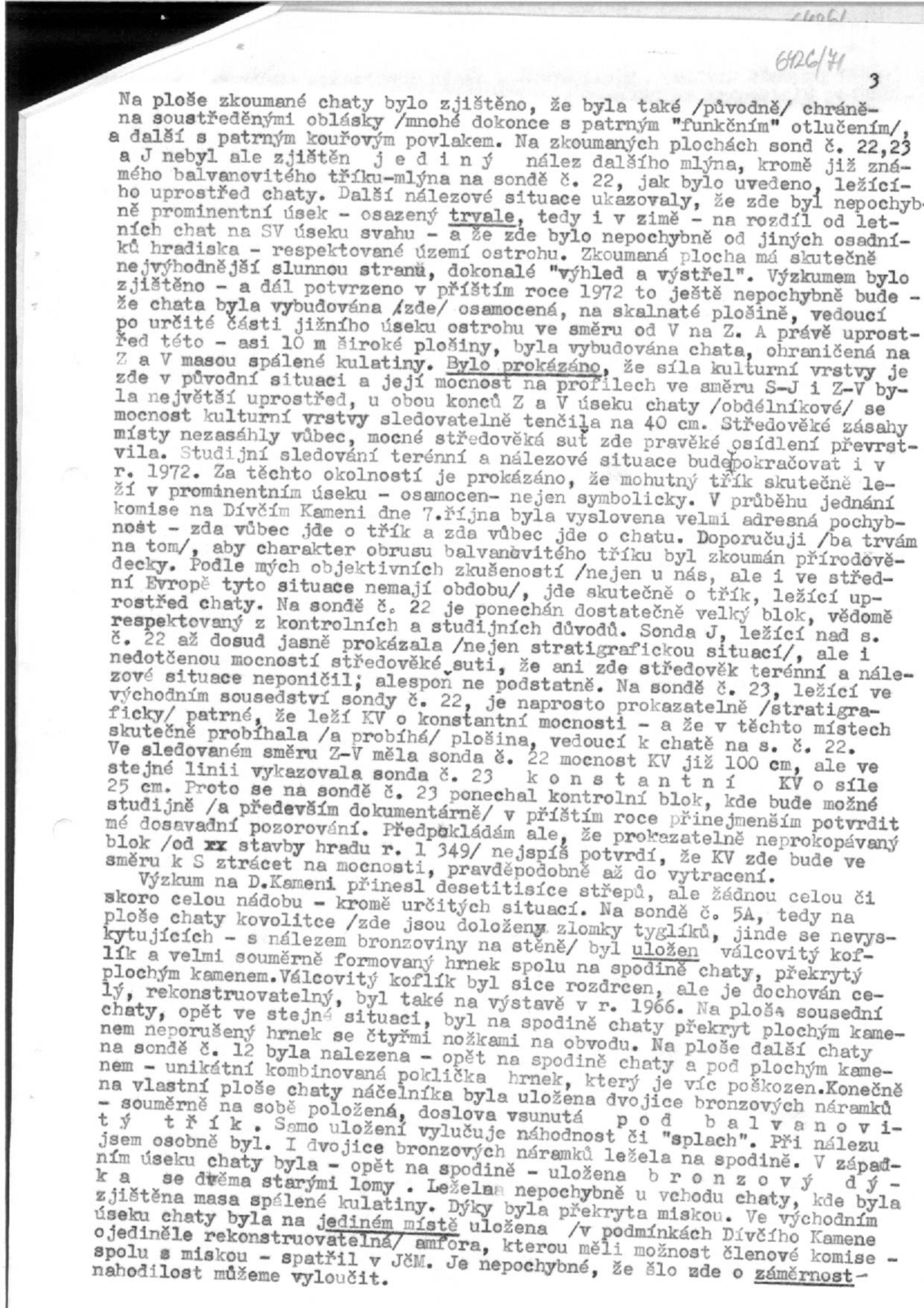}
        \caption{Corner is missing}
        \label{fig:e_text_t_4}
    \end{subfigure}
    \begin{subfigure}{0.3\textwidth}
        \centering
        \includegraphics[width=\textwidth]{img/pages/example_TEXT_T_light.pdf}
        \caption{Very light ink}
        \label{fig:e_text_t_5}
    \end{subfigure}
    \begin{subfigure}{0.3\textwidth}
        \centering
        \includegraphics[width=\textwidth]{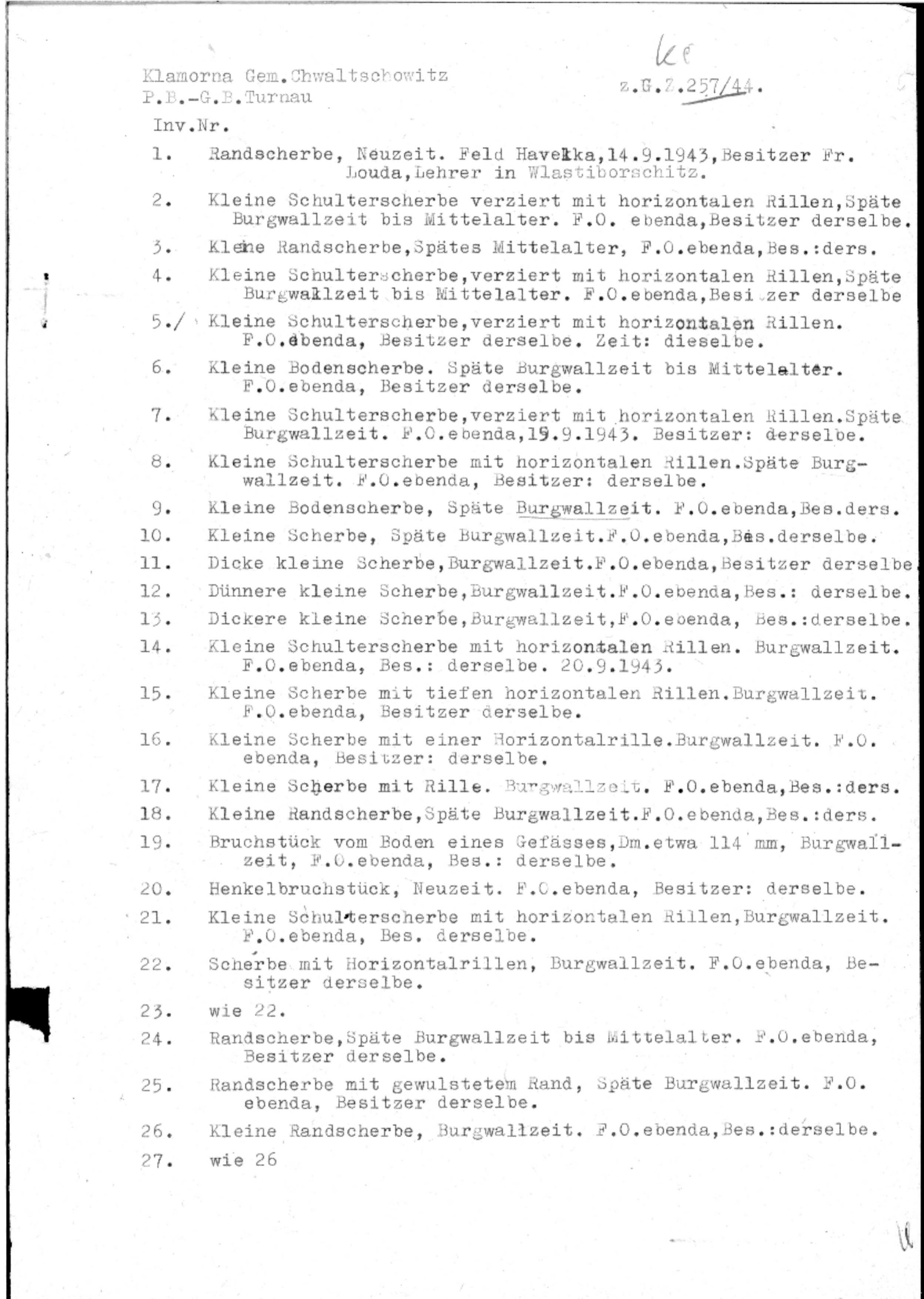}
        \caption{Page with a hole}
        \label{fig:e_text_t_6}
    \end{subfigure}

    \vspace{0.5em}

    \begin{subfigure}{0.3\textwidth}
        \centering
        \includegraphics[width=\textwidth]{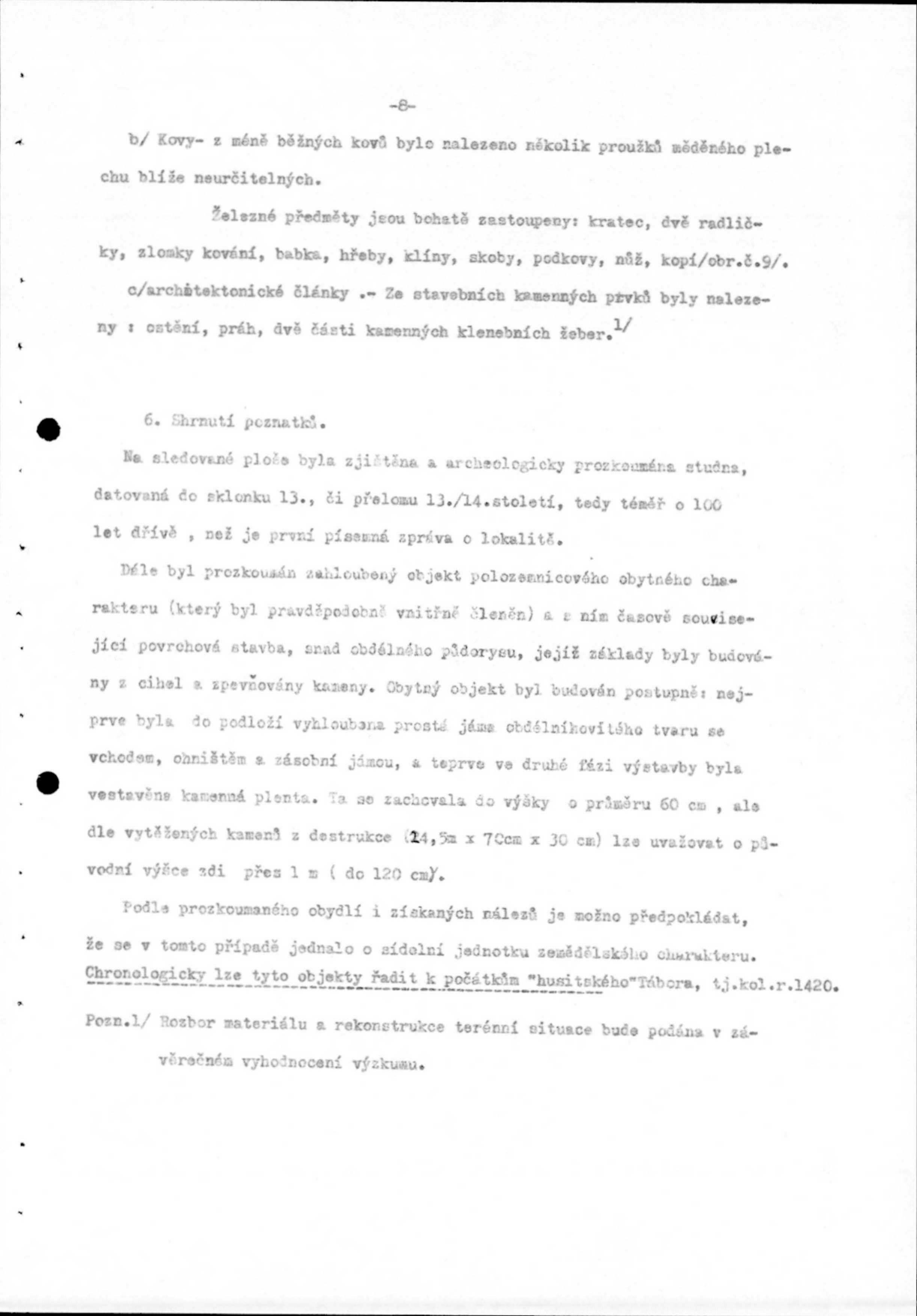}
        \caption{Skewed scan}
        \label{fig:e_text_t_7}
    \end{subfigure}
    \begin{subfigure}{0.3\textwidth}
        \centering
        \includegraphics[width=\textwidth]{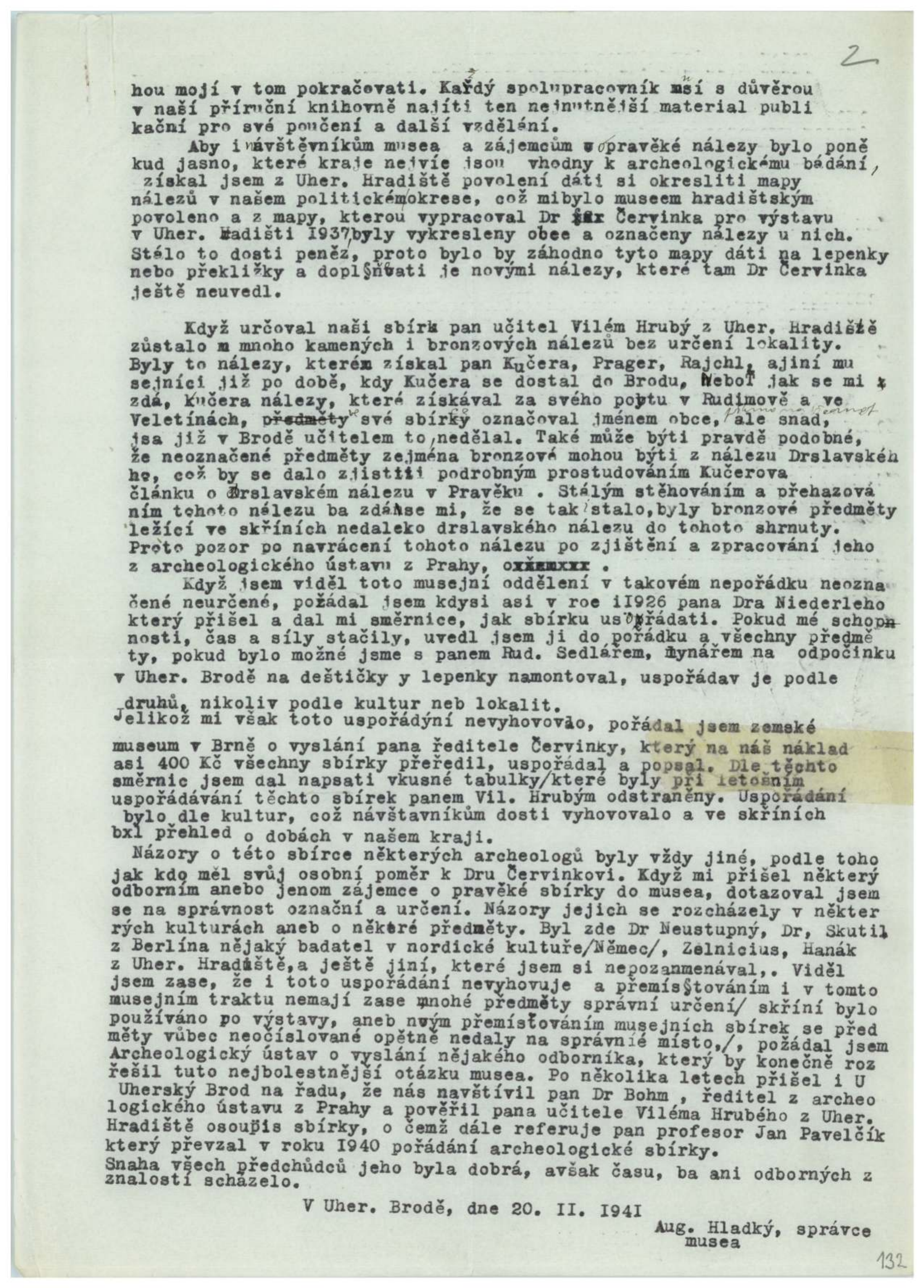}
        \caption{Stained paper}
        \label{fig:e_text_t_8}
    \end{subfigure}
    \begin{subfigure}{0.3\textwidth}
        \centering
        \includegraphics[width=\textwidth]{img/pages/example_TEXT_T_scan.pdf}
        \caption{Scan of the copy}
        \label{fig:e_text_t_9}
    \end{subfigure}
\end{figure}

\chapter{\acs{clip} category descriptions} \label{app-clip-cat}

\begin{table}[h!]
\centering
\caption{Initial Document Classification Labels}
\label{tab:init_classification}
\begin{tabular}{>{\raggedright\arraybackslash}p{2.2cm}>{\raggedright\arraybackslash}p{9.8cm}}
\toprule
\textbf{Label} & \textbf{Description} \\
\midrule
DRAW & drawings, maps, paintings, schematics, graphics with labels \\
DRAW\_L & drawings, maps, paintings, schematics, graphics with a table legend, inside a table or form \\
LINE\_HW & handwritten text lines inside a table or form \\
LINE\_P & printed text lines inside a table or form \\
LINE\_T & typed text lines inside a table or form \\
PHOTO & photos or cutouts from photos with labels \\
PHOTO\_L & photos with a table caption, inside a table or form \\
TEXT & mixed printed and handwritten texts \\
TEXT\_HW & handwritten text page \\
TEXT\_P & printed text page \\
TEXT\_T & typed document page \\
\bottomrule
\end{tabular}
\end{table}

\begin{table}[h!]
\centering
\caption{Short Document Classification Labels}
\label{tab:short_classification}
\begin{tabular}{>{\raggedright\arraybackslash}p{2.2cm}>{\raggedright\arraybackslash}p{9.8cm}}
\toprule
\textbf{Label} & \textbf{Description} \\
\midrule
DRAW & a drawings or a map or a diagram \\
DRAW\_L & a table and drawings or a map or a diagram \\
LINE\_HW & a table or a filled form of handwritten texts \\
LINE\_P & a table or a filled form of printed texts \\
LINE\_T & a table or a filled form of typed texts \\
PHOTO & photos or photo cutouts \\
PHOTO\_L & a table and photos or photo cutouts \\
TEXT & printed and handwritten text styles on a page \\
TEXT\_HW & a handwritten plain text page or handwritten text paragraphs \\
TEXT\_P & a printed plain text page or printed text paragraphs \\
TEXT\_T & a typed plain text document page or typed text paragraphs \\
\bottomrule
\end{tabular}
\end{table}

\begin{table}[h!]
\centering
\caption{Minimal Document Classification Labels}
\label{tab:minimal_classification}
\begin{tabular}{>{\raggedright\arraybackslash}p{2.2cm}>{\raggedright\arraybackslash}p{8cm}}
\toprule
\textbf{Label} & \textbf{Description} \\
\midrule
DRAW & a drawings or a map or a diagram \\
DRAW\_L & a table and a drawings or a map \\
LINE\_HW & a handwritten table or a filled form \\
LINE\_P & a printed table or a filled form \\
LINE\_T & a typed table or a filled form \\
PHOTO & photos on a page \\
PHOTO\_L & a table and photos \\
TEXT & printed text and handwritten comment \\
TEXT\_HW & a handwritten text \\
TEXT\_P & a printed text \\
TEXT\_T & a typed text \\
\bottomrule
\end{tabular}
\end{table}

\begin{table}[h!]
\centering
\caption{Page-based Document Classification Labels (Proposed by GPT Deep Research based on Chapter \ref{chap-data-explore} content)}
\label{tab:page_classification}
\begin{tabular}{>{\raggedright\arraybackslash}p{2.2cm}>{\raggedright\arraybackslash}p{9.8cm}}
\toprule
\textbf{Label} & \textbf{Description} \\
\midrule
DRAW & A page with a drawing, map, painting, or schematic \\
DRAW\_L & A page with a drawing or map presented in a table layout \\
LINE\_HW & A page with handwritten text in a table layout \\
LINE\_P & A page with printed text in a table layout \\
LINE\_T & A page with typed text in a table layout \\
PHOTO & A page with a photograph or photographic image \\
PHOTO\_L & A page with a photograph presented in a table layout \\
TEXT & A page with mixed handwritten, printed, or typed text \\
TEXT\_HW & A page with only handwritten text in paragraphs \\
TEXT\_P & A page with only printed text in paragraphs \\
TEXT\_T & A page with only typed text in paragraphs \\
\bottomrule
\end{tabular}
\end{table}

\begin{table}[h!]
\centering
\caption{Cropped out word ``page'' from Table \ref{tab:page_classification} — Document Classification Labels}
\label{tab:crop_classification}
\begin{tabular}{>{\raggedright\arraybackslash}p{2.2cm}>{\raggedright\arraybackslash}p{8cm}}
\toprule
\textbf{Label} & \textbf{Description} \\
\midrule
DRAW & a drawings or a map or a diagram \\
DRAW\_L & a drawings or a map or a diagram and a table \\
LINE\_HW & a table or a filled form of handwritten cells \\
LINE\_P & a table or a filled form of printed cells \\
LINE\_T & a table or a filled form of typed cells \\
PHOTO & photos or photo cutouts \\
PHOTO\_L & photos or photo cutouts and a table \\
TEXT & mixed text styles \\
TEXT\_HW & a handwritten text page \\
TEXT\_P & a printed text page \\
TEXT\_T & a typed document page \\
\bottomrule
\end{tabular}
\end{table}

\begin{table}[h!]
\centering
\caption{Long Document Classification Labels (Proposed by Gemini Deep Research based on Chapter \ref{chap-data-explore} content)}
\label{tab:long_classification}
\begin{tabular}{>{\raggedright\arraybackslash}p{2.2cm}>{\raggedright\arraybackslash}p{11cm}}
\toprule
\textbf{Label} & \textbf{Description} \\
\midrule
DRAW & A drawing, a map, a painting, a schematic, or a graphic with labels on a document page. \\
DRAW\_L & A drawing, a map, a painting, a schematic, or a graphic with an accompanying table or tabular legend on a document page. \\
LINE\_HW & A table or a filled form containing handwritten text lines on a document page. \\
LINE\_P & A table or a filled form containing printed text lines on a document page. \\
LINE\_T & A table or a filled form containing typed text lines on a document page. \\
PHOTO & A photo or a photo cutout attached to a document page. \\
PHOTO\_L & A photo or a photo cutout with an accompanying table or tabular legend on a document page. \\
TEXT & A document page with mixed printed and handwritten text styles, such as a newspaper or book cutout with handwritten comments. \\
TEXT\_HW & A handwritten plain text, handwritten text paragraphs, or handwritten list items on a document page. \\
TEXT\_P & A printed plain text, printed text paragraphs, or printed list items on a document page. \\
TEXT\_T & A typed plain text, typed text paragraphs, or typed list items on a document page. \\
\bottomrule
\end{tabular}
\end{table}

\begin{table}[h!]
\centering
\caption{Comprehensive Document Classification Labels with Examples}
\label{tab:comprehensive_classification} 
\begin{tabular}{>{\raggedright\arraybackslash}p{2.2cm}>{\raggedright\arraybackslash}p{11cm}}
\toprule
\textbf{Label} & \textbf{Description} \\
\midrule
DRAW & A drawing, a map, a painting, a sketch, a diagram, a schematic, or a graphic with labels on a document page. \\
DRAW\_L & A drawing, a map, a painting, a sketch, a diagram, a schematic, or a graphic with an accompanying table or tabular legend on a document page. \\
LINE\_HW & A table or a filled form with handwritten text lines on a document page. Such as a filled report \\
LINE\_P & A table or a filled form with printed text lines on a document page. \\
LINE\_T & A table or a filled form with typed text lines on a document page. Such as a filled report \\
PHOTO & A photo or a photo cutout with labels attached to a document page. \\
PHOTO\_L & A photo or a photo cutout labeled with an accompanying table or tabular legend on a document page. \\
TEXT & A document page with mixed printed and handwritten text styles, such as a newspaper or book cutout with handwritten comments. \\
TEXT\_HW & A handwritten plain text, handwritten text paragraphs, or handwritten list items on a document page. Such as a field diary scans and notebook notes. \\
TEXT\_P & A printed plain text, printed text paragraphs, or printed list items on a document page. Such as a clean book page scan or an article page scan. \\
TEXT\_T & A typed plain text, typed text paragraphs, or typed list items on a document page. Such as an old style field report scans \\
\bottomrule
\end{tabular}
\end{table}

\begin{table}[h!]
\centering
\caption{Enhanced Document Classification Labels with Detailed Annotations}
\label{tab:enhanced_classification}
\begin{tabular}{>{\raggedright\arraybackslash}p{2.2cm}>{\raggedright\arraybackslash}p{11cm}}
\toprule
\textbf{Label} & \textbf{Description} \\
\midrule

DRAW & A page with a drawing, map, painting, or schematic with text annotation \\
DRAW\_L & A page with a drawing or map presented in a table layout or with a legend table \\
LINE\_HW & A page with handwritten text in a table layout or a manually filled form \\
LINE\_P & A page with printed text in a table layout or a large printed tabled that takes a whole page \\
LINE\_T & A page with typed text in a table layout or a filled with typed text form \\
PHOTO & A page with a photograph or photographic image with text annotation \\
PHOTO\_L & A page with a photograph presented in a table layout or with a legend table \\
TEXT & A page with mixed handwritten, printed, or typed text such as a previously printed and then manually commented page from a book, a newspaper, or an article \\
TEXT\_HW & A page with only handwritten text in paragraphs such as a page from field journals \\
TEXT\_P & A page with only printed text in paragraphs, lists, and plain text layouts such as a page from an article or a thesis \\
TEXT\_T & A page with only typed text in paragraphs, lists, and plain text layouts such as a page from an article or a thesis \\
\bottomrule
\end{tabular}
\end{table}

\begin{table}[h!]
\centering
\caption{Extended Document Classification Labels}
\label{tab:extended_classification}
\begin{tabular}{>{\raggedright\arraybackslash}p{2.2cm}>{\raggedright\arraybackslash}p{11cm}}
\toprule
\textbf{Label} & \textbf{Description} \\
\midrule
DRAW & a drawings or a map or a diagram \\
DRAW\_L & a table or a tabular legend and drawings or a map or a diagram \\
LINE\_HW & a table or a filled form of handwritten texts on a page \\
LINE\_P & a table or a filled form of printed texts on a document page \\
LINE\_T & a table or a filled form of typed texts on a document page \\
PHOTO & photos or photo cutouts attached to the document page \\
PHOTO\_L & a table or a tabular legend and photos or photo cutouts on a document page \\
TEXT & printed and handwritten text styles on a page like a newspaper cutout or a book cutout commented in another way \\
TEXT\_HW & a handwritten plain text or handwritten text paragraphs on a page \\
TEXT\_P & a printed plain text or printed text paragraphs or printed list items on a document page \\
TEXT\_T & a typed plain text or typed text paragraphs or typed list items on a document page \\
\bottomrule
\end{tabular}
\end{table}

\pagebreak

% https://chatgpt.com/share/68fac5c0-56a0-8001-b7e3-5802d361817c
% https://gemini.google.com/share/7fb2a2d59428

\chapter{\acs{llm} prompts} \label{app-llm-prompt}

\noindent This appendix includes prompts for generation of category text embeddings used in \ac{clip} models, specifically Tables \ref{tab:page_classification} \& \ref{tab:long_classification}, referred to as \textbf{gpt} \& \textbf{gemini} category descriptions sets.

\paragraph{GPT}research prompt:

\begin{lstlisting}[style=promptstyle]
<table short>
or
<table init>
or
<table min>

based on the provided tables you must compose descriptions for the defined categories 

these descriptions will be used in few-shot finetuning of clip (ViT-L/14@336px and ViT-B/32) 

find best practices in defining such text embedding and return a single refined table

a table from the related article should help you as well:
<table 11-labels>

and also a list with data expert opinions
<final expert feedback criteria list>


To tailor the refined descriptions optimally for CLIP fine-tuning, could you please confirm:

1. Should the language be concise and technical, or is a more descriptive tone preferred?
2. Are the descriptions going to be shown to annotators, or are they strictly for model training?
3. Should we keep compatibility with the LaTeX formatting you showed (e.g., escaping underscores like \_)?
4. Do you have a maximum character length per description?
5. Once I have this, I can finalize the optimal descriptions for each label.


1. chose based on the found info 
2. only model training 
3. latex format is only for my own convenience, use raw text 
4. depends on the model input size, find it as well 
5. and importantly provide a refined table with label - description, not only found info
\end{lstlisting}

\noindent where \texttt{<table *>} lines refer to Tables \ref{tab:short_classification} \ref{tab:init_classification} \ref{tab:minimal_classification} and \ref{tab:11-label}.

\paragraph{Gemini}research prompt:

\begin{lstlisting}[style=promptstyle]
<table short>
or
<table init>
or
<table min>

based on the provided tables you must compose descriptions for the defined categories 

these descriptions will be used in few-shot finetuning of clip (ViT-L/14@336px and ViT-B/32) 

find best practices in defining such text embedding and return a single refined table


I've put together a research plan. If you need to update it in any way, let me know.
CLIP Text Embedding Refinement:

-Research Websites:

(1) Analyze the three provided sets of label descriptions to identify the core concept and common elements associated with each category (e.g., DRAW, LINE\_HW, TEXT\_P).
(2) Research best practices for crafting effective text descriptions or prompts for vision-language models like CLIP, specifically for few-shot fine-tuning in classification tasks.
(3) Evaluate how factors such as conciseness, specificity, the inclusion of key visual attributes, and the overall clarity of descriptions influence the quality of text embeddings for distinguishing document layout elements.
(4) Compare and contrast the strengths and weaknesses of the existing descriptions for each label, considering their potential impact on CLIP's ability to learn and generalize from limited examples.
(5) Synthesize a refined, concise, and highly discriminative description for each category, ensuring it accurately represents the visual content and is optimized for generating robust text embeddings for CLIP.
(6) Compile the refined descriptions into a single, clear table format, presenting the 'label' and its corresponding 'description'.

-Analyze Results
-Create Report


Start research
\end{lstlisting}

Notably, the same prompt as used for Gemini was tested with GPT as well. The generated table of category descriptions' text embeddings contained multiple complex for tokenization phrasings, thus an expanded version of the prompt including labels definition Table \ref{tab:11-label} and a list of expert-defined criteria from Section \ref{sec-dla-critical}.

\chapter{System architecture} \label{architecture-system}

\section{Finetuning to downstream task functionality} \label{sec-finetune}

\noindent The system implements finetuning of pretrained models on domain-specific historical document scans. This transfer learning approach adapts general-purpose vision and hybrid models to the specialized task of document page classification.

To address class imbalance during finetuning, the system uses a custom \textbf{BalancedBatchSampler} that extends PyTorch’s \textbf{BatchSampler}. The objective is to ensure each training batch contains a fixed number of classes and a fixed number of samples per class, thereby improving training stability. The sampler operates as follows:

\begin{itemize}
\item For a given tensor of labels (subset of the whole dataset), the sampler first builds an index list for each class and shuffles it.
\item On each iteration, it randomly selects a specified number of distinct classes (\texttt{n\_classes}).
\item It then takes a specified number of indices (\texttt{n\_samples}) for each selected class. When a class's index list is exhausted, it is reset and reshuffled.
\end{itemize}

This process generates class-balanced mini-batches of size $n\_classes \times n\_samples$ whose implementation logic is adapted from \href{https://github.com/b-hahn/CLIP/blob/3756210e1b4cec869c75eb3f2beb8016f8cfe7ec/balanced_batch_sampler.py}{the \ac{clip} finetuning repository}. Although the balanced-batch requirement is \ac{clip}-specific, the same algorithm was used in other models' finetuning.

Moreover, the cross-validation technique was used to finetune five folds (data subset selections) of different pretrained models, which were then averaged to get the generalized model weights. The averaging was performed for fold checkpoints matching a given pattern of the common base model. State dictionary of each model was loaded and added element-wise to other models' weights, which was then divided by the number of located folds. This procedure implements a simple, unweighted arithmetic average of all state-dict entries (including buffers such as batch-norm statistics), thus, this approach, which can be memory-intensive for large models, assumes that the parameterization is broadly compatible across folds.

\subsection{Transformation into model-friendly inputs} \label{sec-tokenize}
\noindent Before being processed by a model, input images undergo several transformations:

\begin{enumerate}
\item Image loading and validation for corrupted files (if safety load is enabled).
\item Resizing to the model's input resolution (\ac{eg} $224\times224$, $336\times336$ or $384\times384$ pixels)
\item Normalization of pixel values according to the model's preprocessor requirements, optionally, during finetuning, applying the image preprocessing transformations (detailed in Section \ref{sec-preproc}). 
\item Organization into batches for efficient processing, optionally, during finetuning, use a balanced batch sampler (mentioned in Section \ref{sec-finetune}).
\end{enumerate}

The inference batch size is configurable based on available hardware resources. For prediction tasks, only transformations provided by the model's preprocessor are applied to preserve the original image content, whereas finetuning always employs additional data augmentation techniques to improve model robustness.

\subsection{Hyper-parameters} \label{sec-hyperparam}
\noindent The finetuning process is controlled by several hyperparameters:

\begin{itemize}
    \item \textbf{Learning rate}: Default 1e-5 or 5e-5, configurable in the code
    \item \textbf{Epochs}: Default 3 or 7, adjustable in configuration
    \item \textbf{Batch size}: Variable based on memory constraints, set in configuration (\ac{eg} 12 or 16)
    \item \textbf{Warmup ratio}: 0.1 (10\% of training steps)
    \item \textbf{Evaluation strategy}: Per-epoch evaluation
    \item \textbf{Saving strategy}: Best model checkpoint saving
    \item \textbf{Best model selection}: Based on accuracy metric
\end{itemize}

These default parameters are optimized for datasets containing 10,000-50,000 page samples. The documentation recommends adjusting the number of epochs according to evaluation loss to prevent overfitting and modifying the learning rate if the training loss does not converge to values lower than 0.001 as expected.

\subsection{Preprocessing of images} \label{sec-preproc}
\noindent During finetuning, images undergo data augmentation to improve model robustness. The system applies the following transformations:

\begin{itemize}
    \item Brightness adjustment (probability factor: 0.5)
    \item Contrast modification (probability factor: 0.5)
    \item Saturation changes (probability factor: 0.5)
    \item Hue variation (probability factor: 0.5)
    \item Sharpness enhancement (random factor between 0.5-1.5)
    \item Gaussian blur (random radius between 0-2)
\end{itemize}

Notably, geometric transformations like rotation or flipping are avoided due to the orientation-sensitive nature of document pages and the importance of preserving structural layouts.

\subsection{Data split by category proportions} \label{sec-data-split}

\noindent To ensure robust evaluation, the system automatically splits the dataset into training (80\%) and evaluation (10\%), and test (10\%) sets while maintaining a proportional representation of all categories. This stratified approach ensures that evaluation results accurately reflect performance across all document types, even in the presence of class-imbalanced datasets.

The exact logic of the dataset split procedure was described in Section \ref{sec-split}, where the rejection of a standard randomized shuffling approach was justified. The specialized subset selection based on a randomized periodic step in the time-sorted dataset was developed to support the variability of each category representation in the evaluation and test subsets. The data distributions in Figure \ref{sec:annotation-distrib} illustrate the category-specific variability obtained in the chronologically ordered samples after applying the developed split procedure.

\section{Output} \label{sec-output}

\noindent The system generates outputs in multiple formats, designed for both direct user inspection and integration into downstream document processing pipelines. The output results are organized in standardized formats for further analysis.

\subsection{Directory level parsing} \label{sec-dir-lvl}
\noindent When processing directories of images, the system is optimized for efficiency and usability:
\begin{itemize}
    \item A list of image files to be processed can be divided into chunks of specified size if requested by the user (recommended for directories containing thousands of pages)
    \item Page files (of a chunk) are loaded in batches of scaled and preprocessed vectorized images to ensure memory efficiency; thus \texttt{batch\_size} controls how many images are loaded into memory when a single model prediction attempt is performed.
    \item Progress information (\ac{eg} \textbf{\{current\_iteration\}-th batch - Processed \{already\_predicted\} images in	\{minutes\_from\_start\} min} or directory evaluation \textbf{Testing: \{processed\}\% |\{dynamic\_progress\_bar\}| \{cur\_iter\}/\{total\} [\{time:spent\}<\{time:left\}, \{mean\_time\}s/it]}) is registered in the console output.
    \item Results are organized by file and page number, and consolidated into \ac{csv} tables (processed chunk's results are recorded, proceeding to the next chunk) with columns for predicted category labels and their scores.
\end{itemize}

% , or \texttt{Step <current\_iteration>/<total>, Loss: <loss>, lr: <learning\_rate>}

This approach to processing directories in chunked lists of images and loading them into memory in small batches balances memory usage with throughput, enabling the efficient handling of large document collections.

\subsection{Confusion matrix plot generation} \label{sec-conf-matrix}

\noindent The evaluation process automatically generates confusion matrices to visualize category-specific classification performance. These plots help users identify systematic classification errors and understand model limitations. Key features include:

\begin{itemize}
\item A clear representation of the relationship between predicted and true categories.
\item Diagonal elements representing correct classifications, and off-diagonal elements highlighting error patterns.
\item Separate matrices generated for different top-N configurations (\ac{eg} Top-1, Top-3).
\end{itemize}

In practice, confusion matrices across models reveal the most problematic category pairs. If such mistakes repeatedly occur in various models, then the annotated data should be examined for interchangeable samples in the most problematic cases (\ac{eg} category precision is constantly lower by 1-2\% compared to other categories). If the size of the category allows for reduction, samples exhibiting a higher amount of mixed features common to multiple categories must be removed entirely.

\begin{table}[hp!]
\centering
\caption{\ac{cli} options for \texttt{run.py}}
\label{tab:cli-options}
\begin{tabular}{p{1.2cm}p{2.4cm}p{1.1cm}p{8.3cm}}
\toprule
\textbf{Group} & \textbf{Flag} & \textbf{Alias} & \textbf{Description} \\
\midrule
\multicolumn{4}{l}{\textbf{General}} \\
& \texttt{--help} & \texttt{-h} & Displays available command options. \\
& \texttt{--hf} & & Retrieve or publish model to HuggingFace. \\
& \texttt{--revision} & \texttt{-rev} & Specifies the global model version. \\
& \texttt{--model}/\texttt{--base} & \texttt{-m}/\texttt{-b}  & Sets the base architecture (\ac{clip} / image-only). \\
& \texttt{--file} & \texttt{-f} & Provides a single image file to process. \\
& \texttt{--file\_format} & \texttt{-ff} & Files format of interest  \\

\midrule
\multicolumn{4}{l}{\textbf{Prediction}} \\
& \texttt{--file} & \texttt{-f} & Provides a single image file path to process. \\
& \texttt{--directory} & \texttt{-d} & Path to folder with pages to be processed. \\
& \texttt{--dir} & & Run a whole directory processing. \\
& \texttt{--inner} & & Include files from nested directories. \\
& \texttt{--chunk} & & Process input files and save results in chunks. \\
& \texttt{--topn} & \texttt{-tn} & Sets N in Top-N for all predictions. \\
& \texttt{--raw} & & Output raw scores for all categories. \\
& \texttt{--best} & & Output Top-1 predictions for every best model. \\
& \texttt{--safe} & & Safely load images, skipping corrupted ones. \\

\midrule
\multicolumn{4}{l}{\textbf{Training and Evaluation}} \\
& \texttt{--train} & & Run model finetuning (training). \\
& \texttt{--lr} & & Learning rate for the optimizer. \\
& \texttt{--epochs} & & Number of training epochs. \\
& \texttt{--max\_categ} & \texttt{-mc} & Maximum number of samples per category. \\
& \texttt{--folds} & & Number of folds for cross-validation. \\
& \texttt{--average} & & Run averaging of finetuned fold models. \\
& \multicolumn{2}{l}{\texttt{--average\_pattern} \texttt{-ap}} & Prefix of finetuned models' filenames. \\
& \texttt{--eval} & & Evaluate a saved model. \\

\midrule
\multicolumn{4}{l}{\textbf{\ac{clip}-specific}} \\
& \texttt{--avg} & & Average features of category description files. \\
& \texttt{--cat\_prefix} & & Prefix for category description TSV files. \\
& \texttt{--cat\_csv} & \texttt{-cc} & Prefix for category description TSV files. \\
& \texttt{--cat\_dir} & & Directory with category description files. \\
& \texttt{--model\_path} & & Path to model .pt checkpoint or folder. \\
& \texttt{--eval\_dir} & & Evaluate a directory of saved models. \\
& \texttt{--model\_dir} & & Path to the model checkpoints directory. \\
& \texttt{--zero\_shot} & & Run off-the-shelf \ac{clip} models. \\
& \texttt{--vis} & & Visualize model accuracy statistics. \\

\bottomrule
\end{tabular}
\end{table}

\chapter{Confusion matrices of all models} \label{confusion-matrices}

\begin{figure}[h]
    \centering
    \caption{EfficientNetV2 confusion matrices, three epochs}
    \label{fig:effnet}
    
    \begin{subfigure}[b]{0.32\textwidth}
        \centering
        \includegraphics[width=\textwidth]{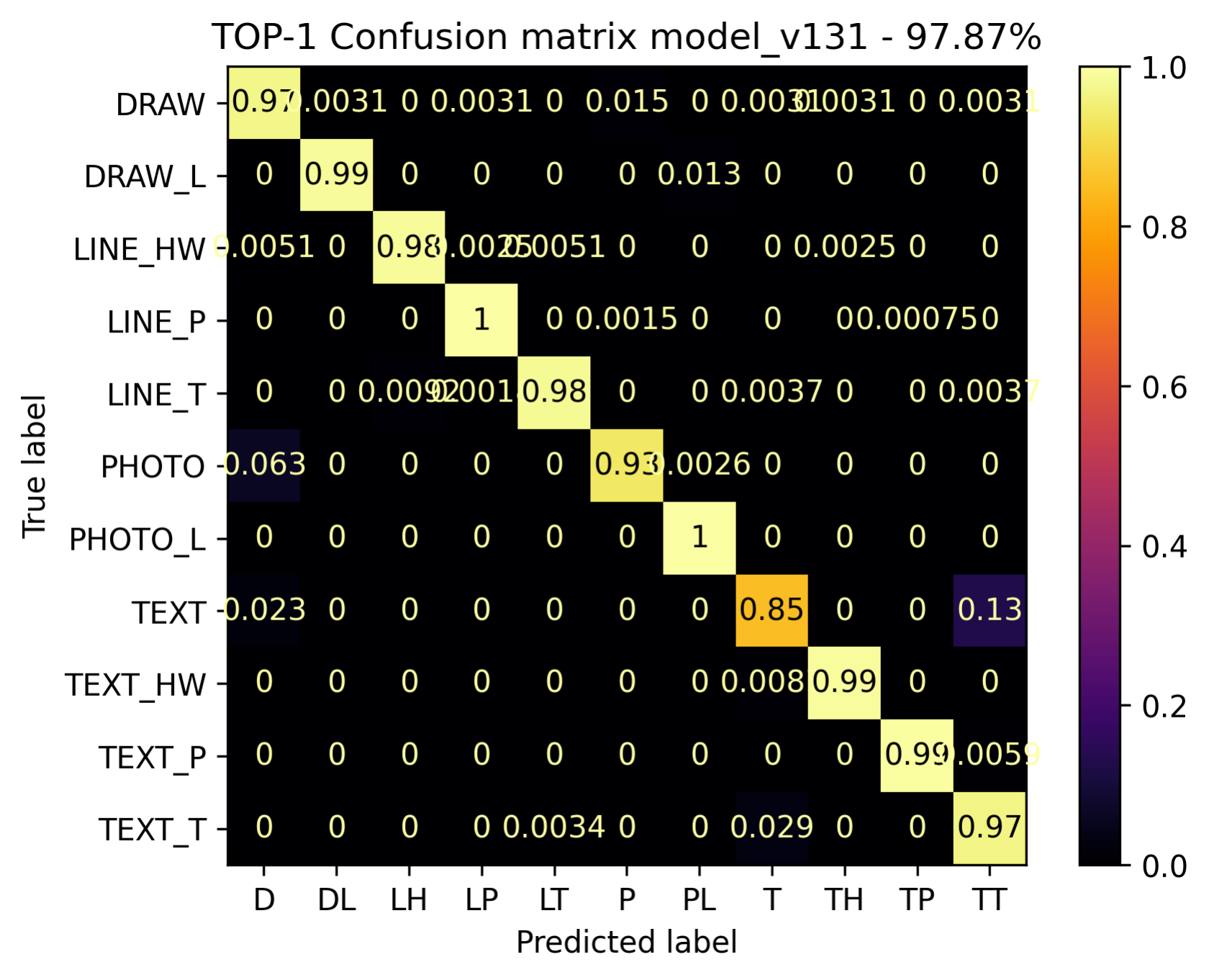}
        \caption{Size S — 97.87\%}
        \label{fig:eff-s}
    \end{subfigure}
    % \hfill
    \begin{subfigure}[b]{0.32\textwidth}
        \centering
        \includegraphics[width=\textwidth]{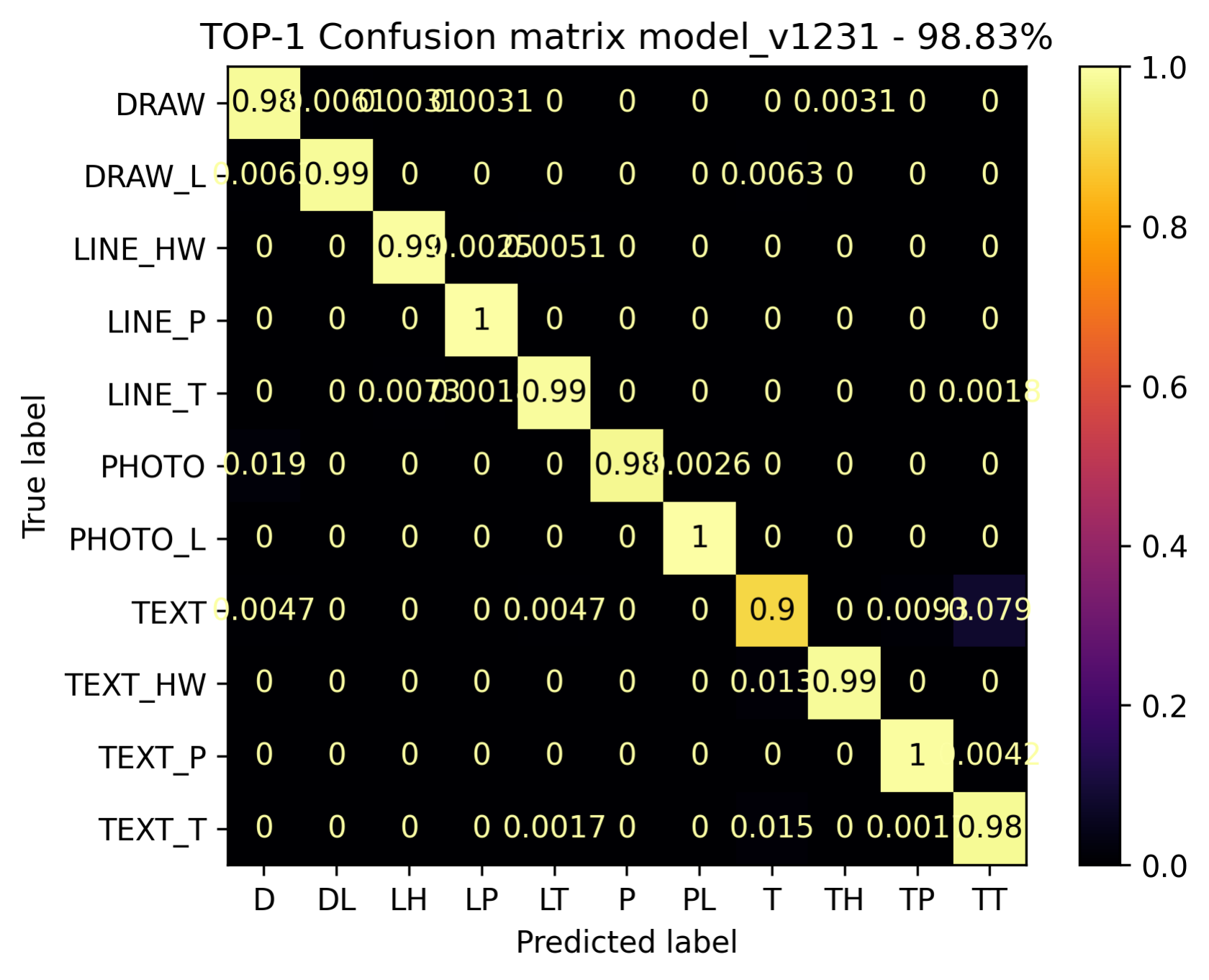}
        \caption{Size M — 98.83\%}
        \label{fig:eff-m}
    \end{subfigure}
    % \hfill
    \begin{subfigure}[b]{0.32\textwidth}
        \centering
        \includegraphics[width=\textwidth]{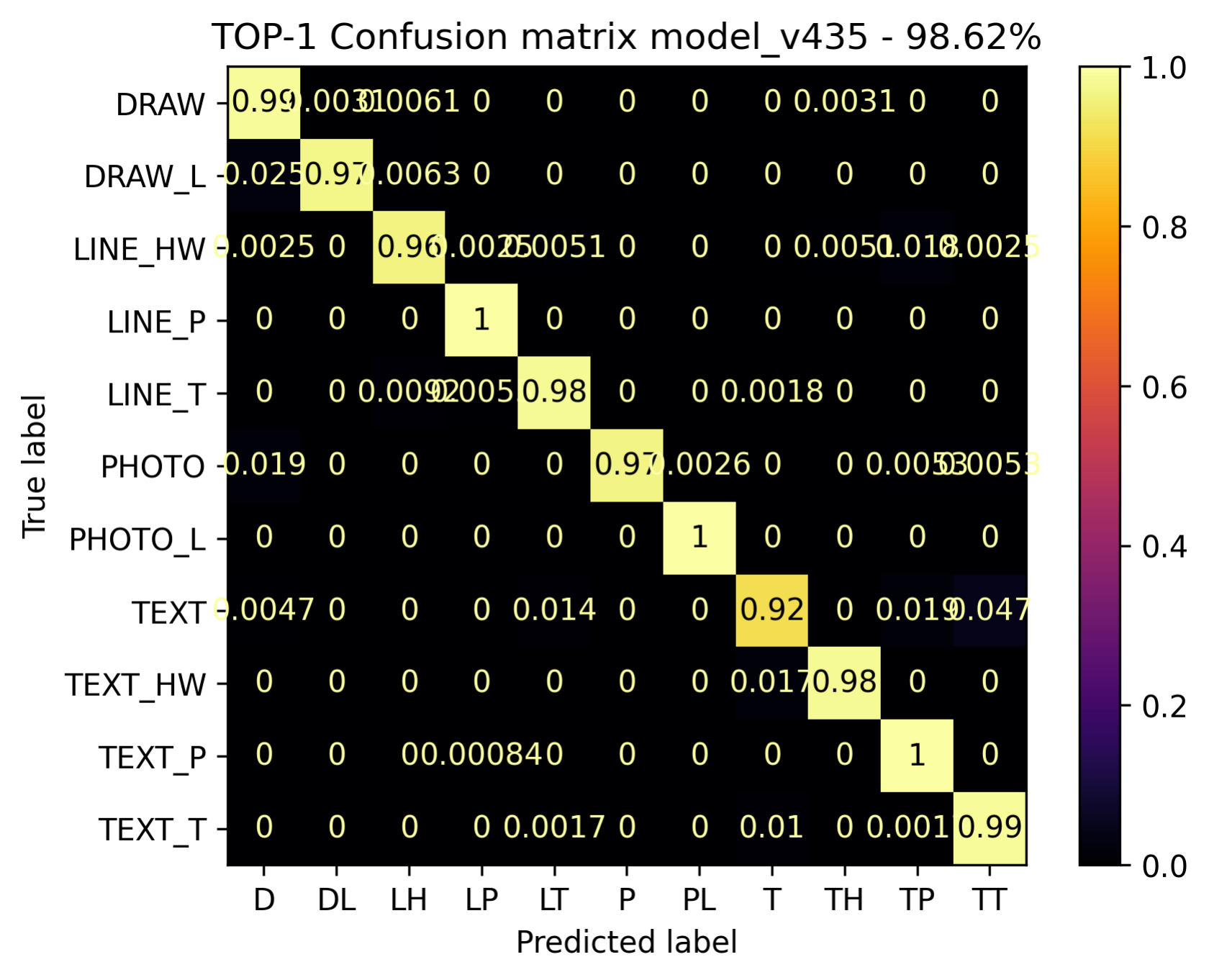}
        \caption{Size L — 98.62\%}
        \label{fig:eff-l}
    \end{subfigure}
    
\end{figure}

\begin{figure}[h]
    \centering
    \caption{RegNetY confusion matrices, three epochs}
    \label{fig:regnety}

    \begin{subfigure}[b]{0.32\textwidth}
        \centering
        \includegraphics[width=\textwidth]{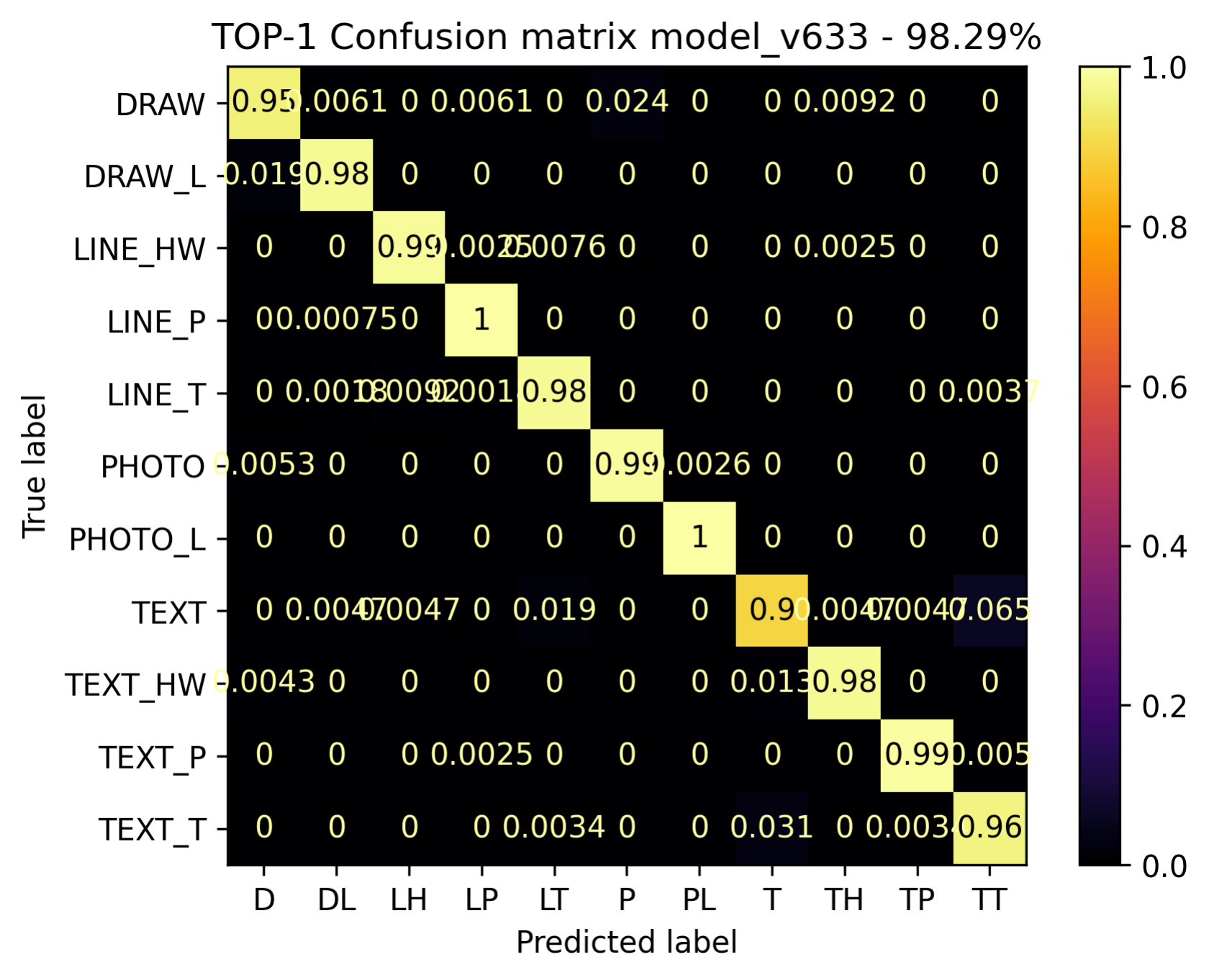}
        \caption{Size 12GF — 98.29\%}
        \label{fig:reg-12}
    \end{subfigure}
    % \hfill
    \begin{subfigure}[b]{0.32\textwidth}
        \centering
        \includegraphics[width=\textwidth]{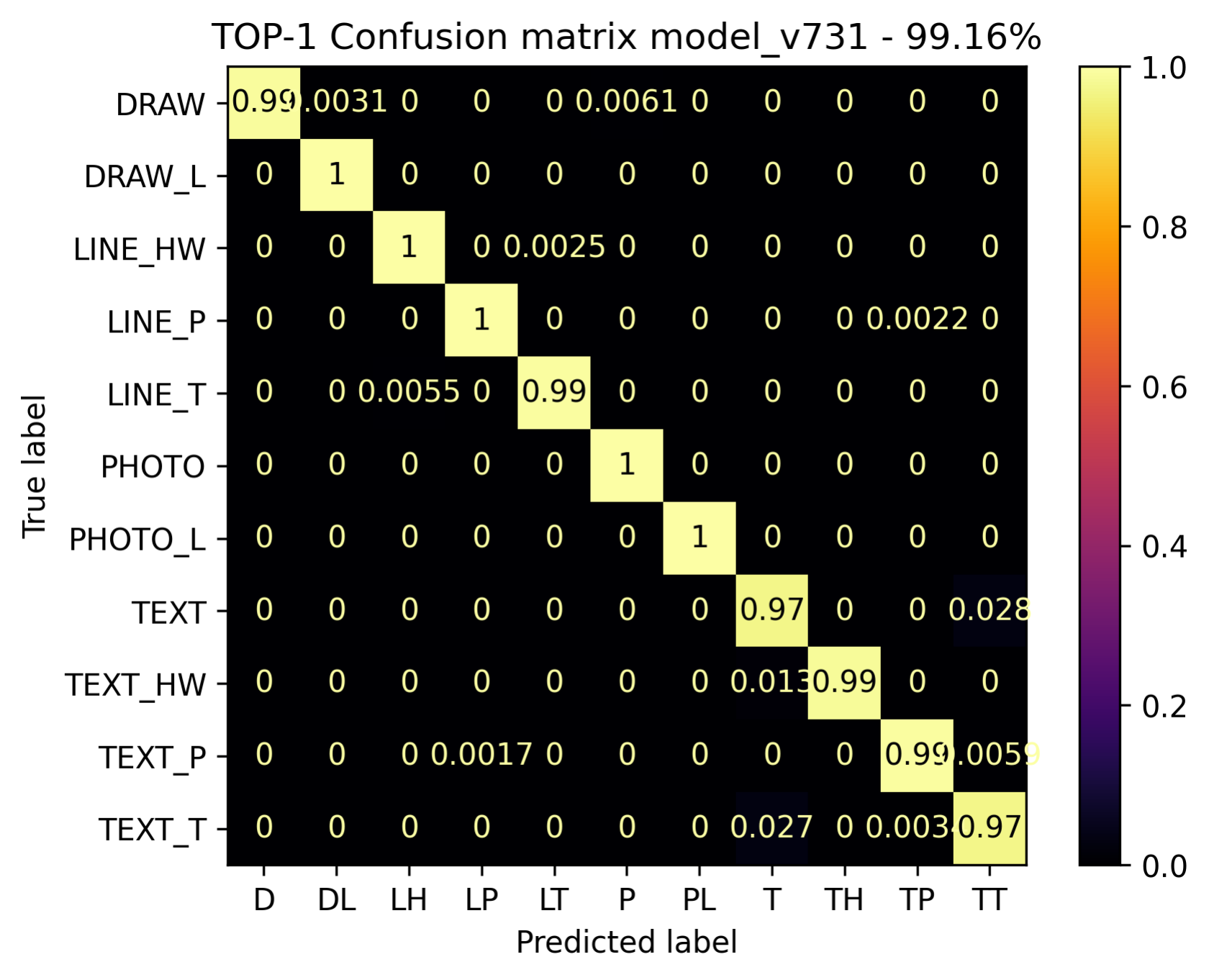}
        \caption{Size 16GF — 99.16\%}
        \label{fig:reg-16}
    \end{subfigure}
    % \hfill
    \begin{subfigure}[b]{0.32\textwidth}
        \centering
        \includegraphics[width=\textwidth]{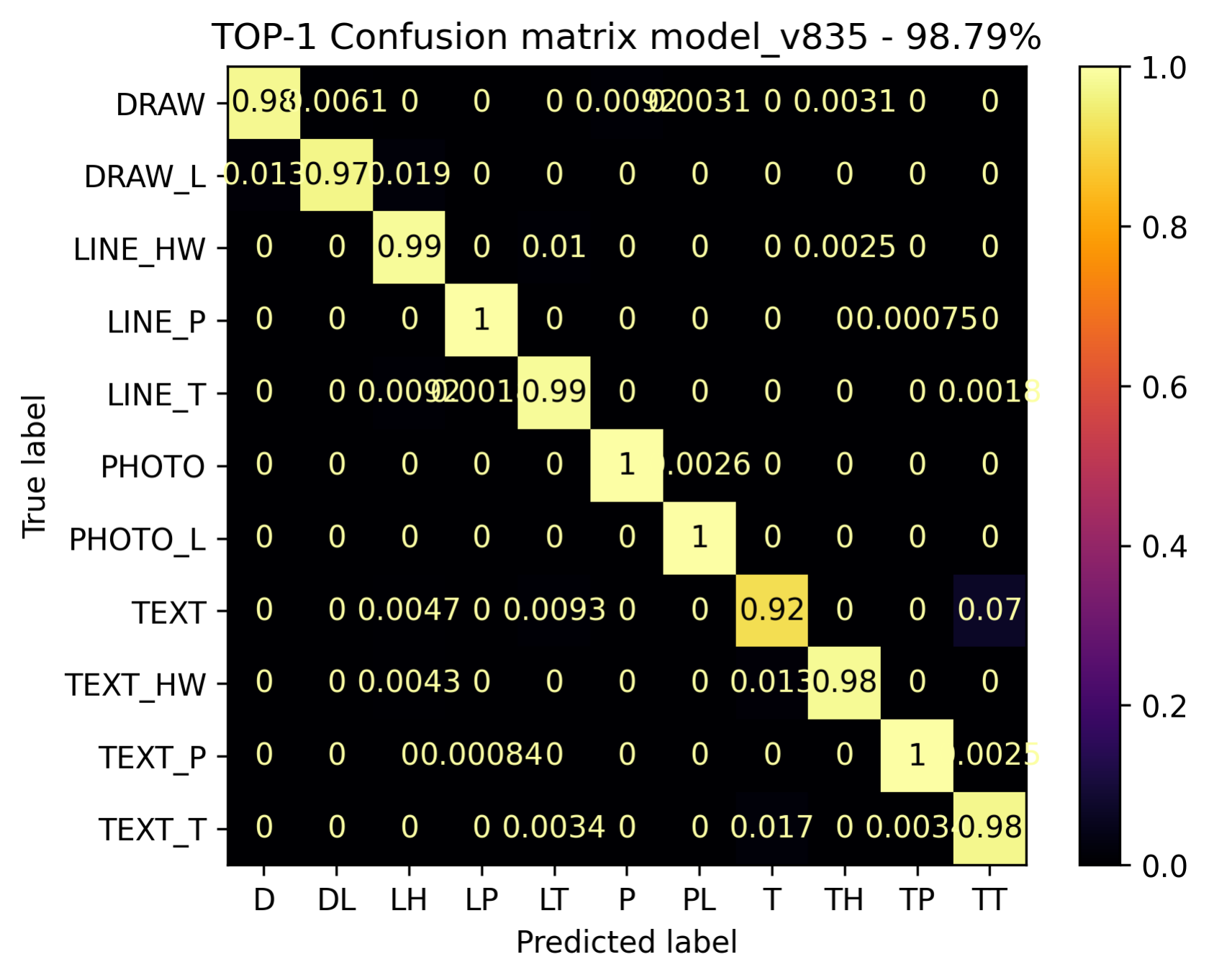}
        \caption{Size 64GF — 98.79\%}
        \label{fig:reg-64}
    \end{subfigure}
    
\end{figure}

\begin{figure}[h]
    \centering
    \caption{\ac{dit} confusion matrices, three epochs}
    \label{fig:dit}
    
    \begin{subfigure}[b]{0.32\textwidth}
        \centering
        \includegraphics[width=\textwidth]{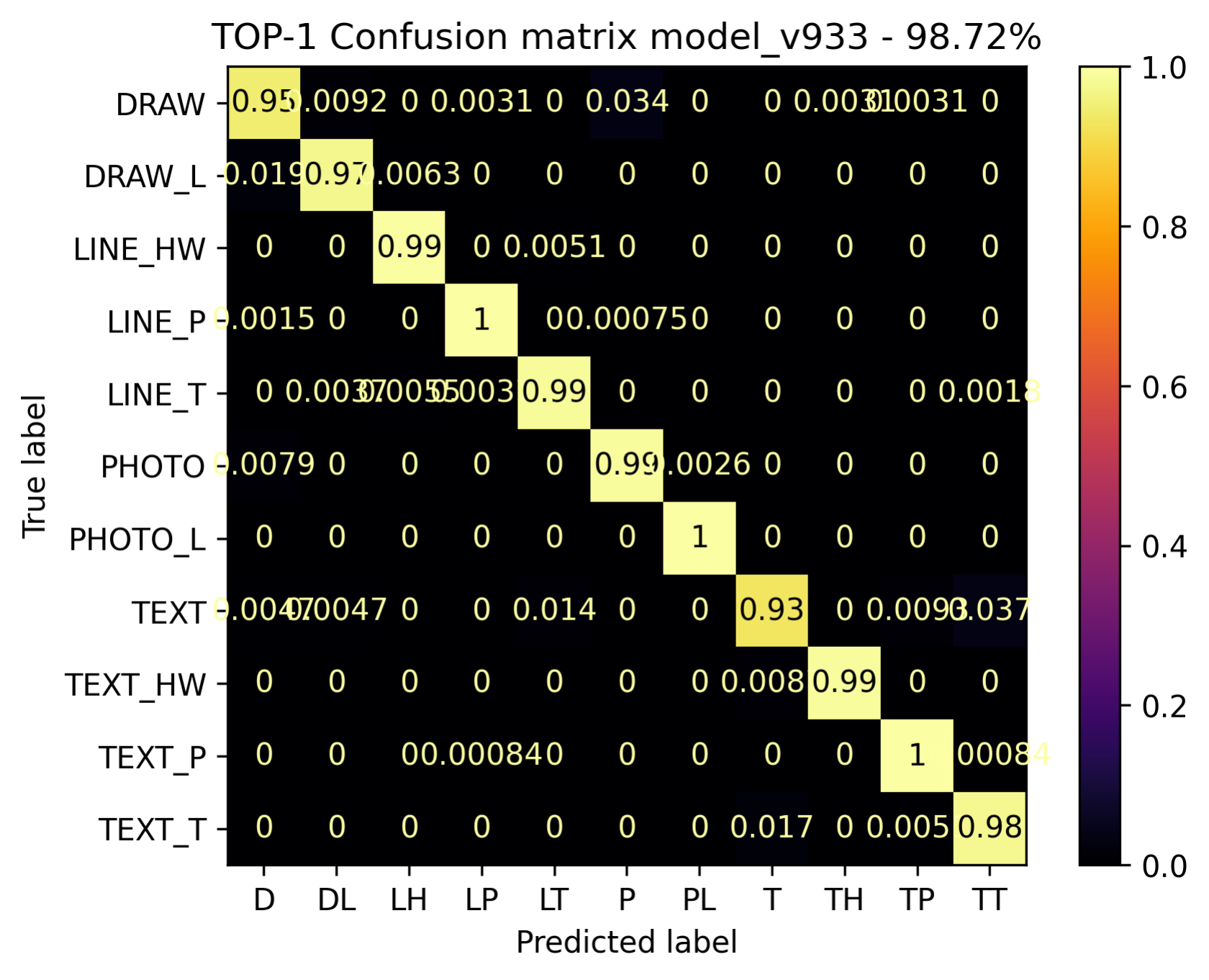}
        \caption{Base RVL — 98.72\%}
        \label{fig:dit-base-rlv}
    \end{subfigure}
    % \hfill
    \begin{subfigure}[b]{0.32\textwidth}
        \centering
        \includegraphics[width=\textwidth]{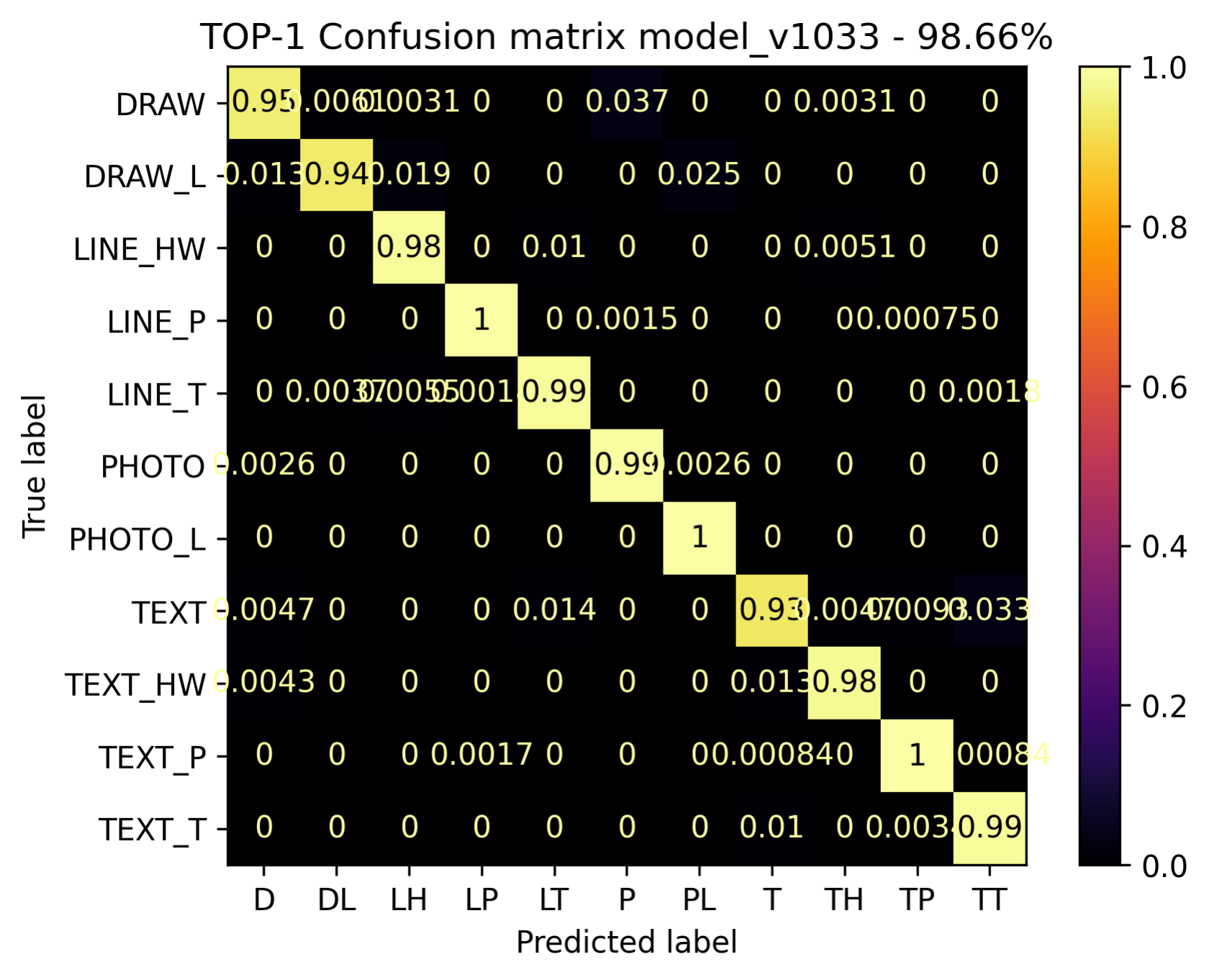}
        \caption{Large — 98.53\%}
        \label{fig:dit-large}
    \end{subfigure}
    % \hfill
    \begin{subfigure}[b]{0.32\textwidth}
        \centering
        \includegraphics[width=\textwidth]{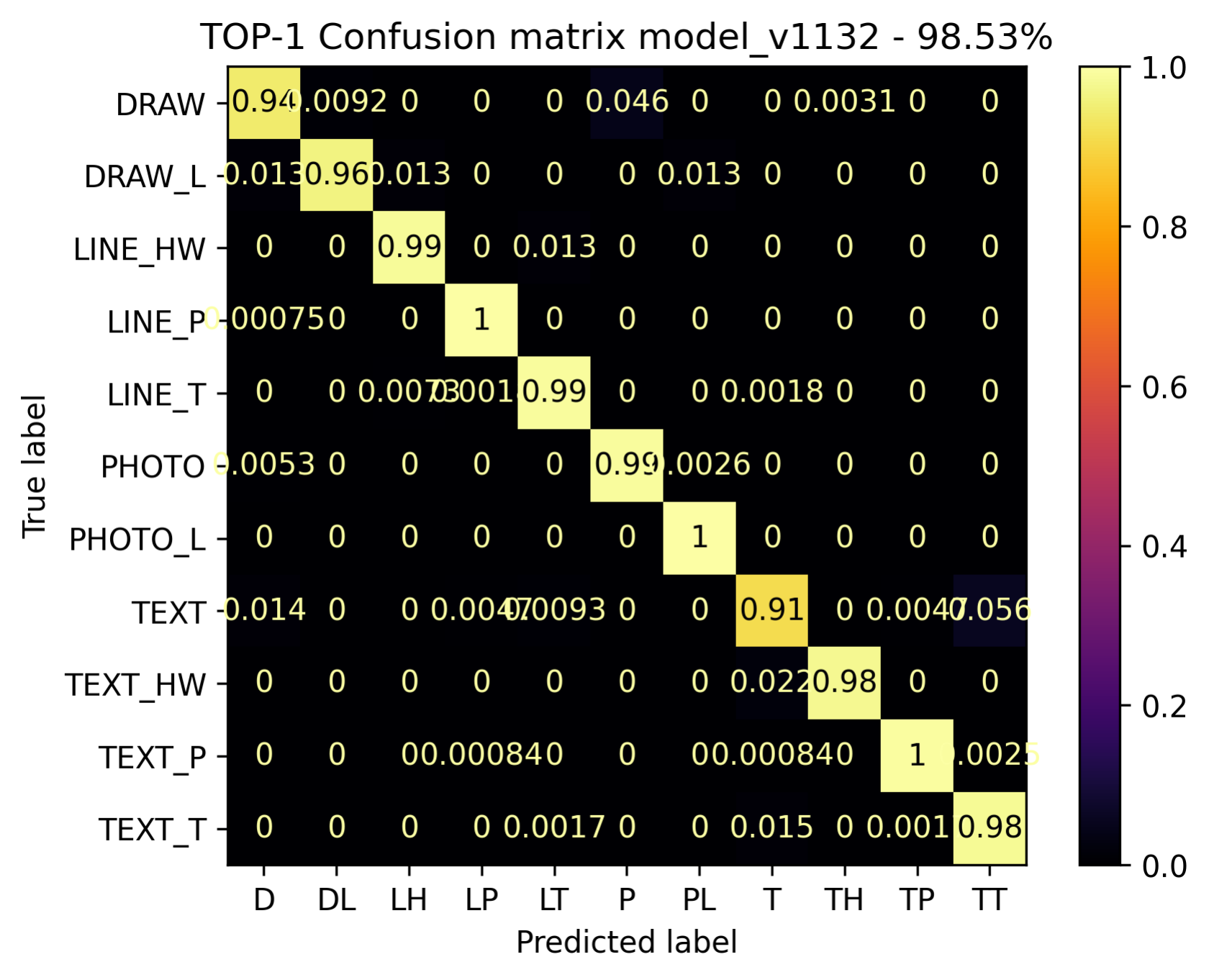}
        \caption{Large RVL — 98.66\%}
        \label{fig:dit-large-rlv}
    \end{subfigure}
    
\end{figure}

\begin{figure}[h]
    \centering
    \caption{\ac{vit} confusion matrices, three epochs}
    \label{fig:vit}
    
    \begin{subfigure}[b]{0.32\textwidth}
        \centering
        \includegraphics[width=\textwidth]{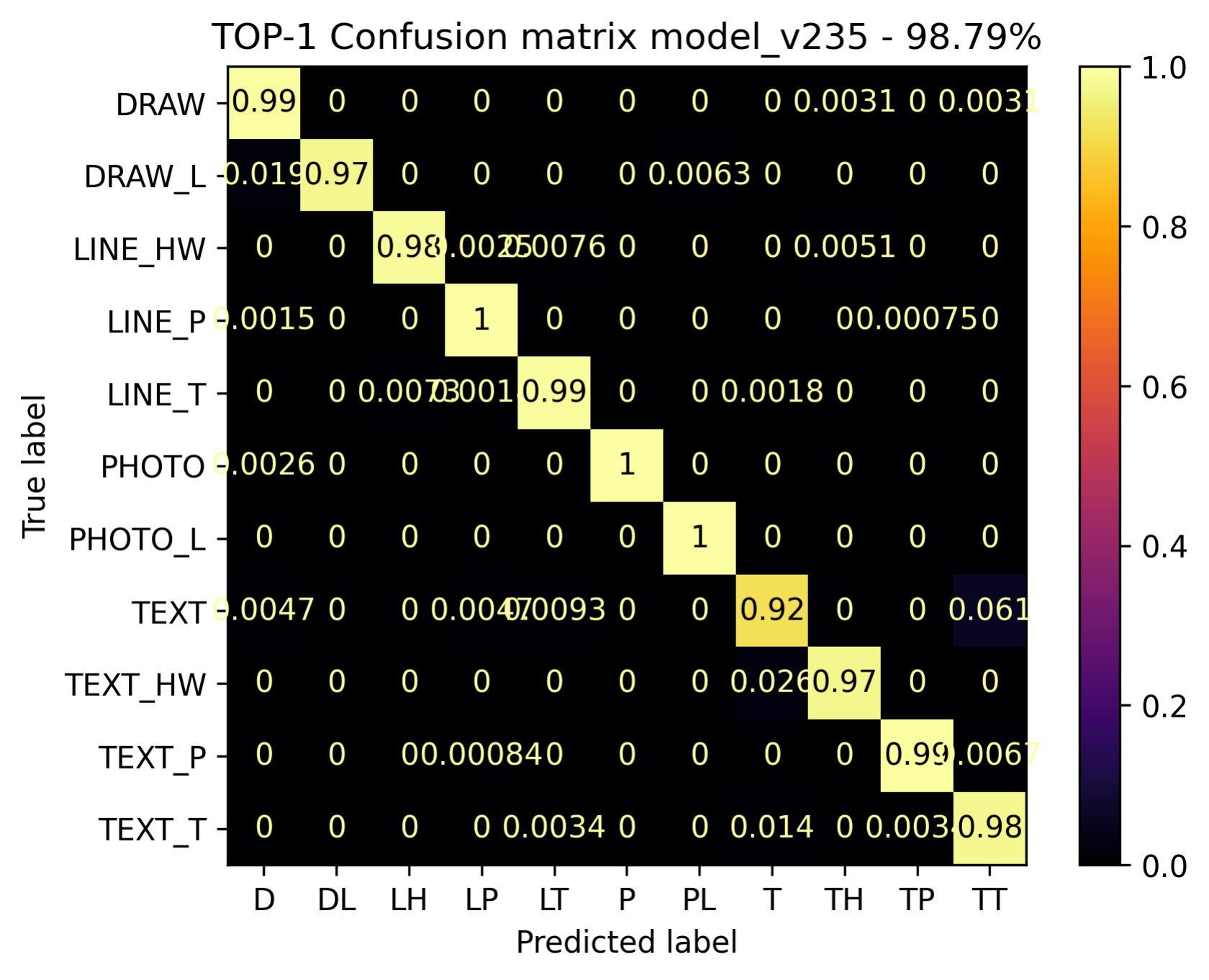}
        \caption{Base 224 — 98.79\%}
        \label{fig:vit-b224}
    \end{subfigure}
    % \hfill
    \begin{subfigure}[b]{0.32\textwidth}
        \centering
        \includegraphics[width=\textwidth]{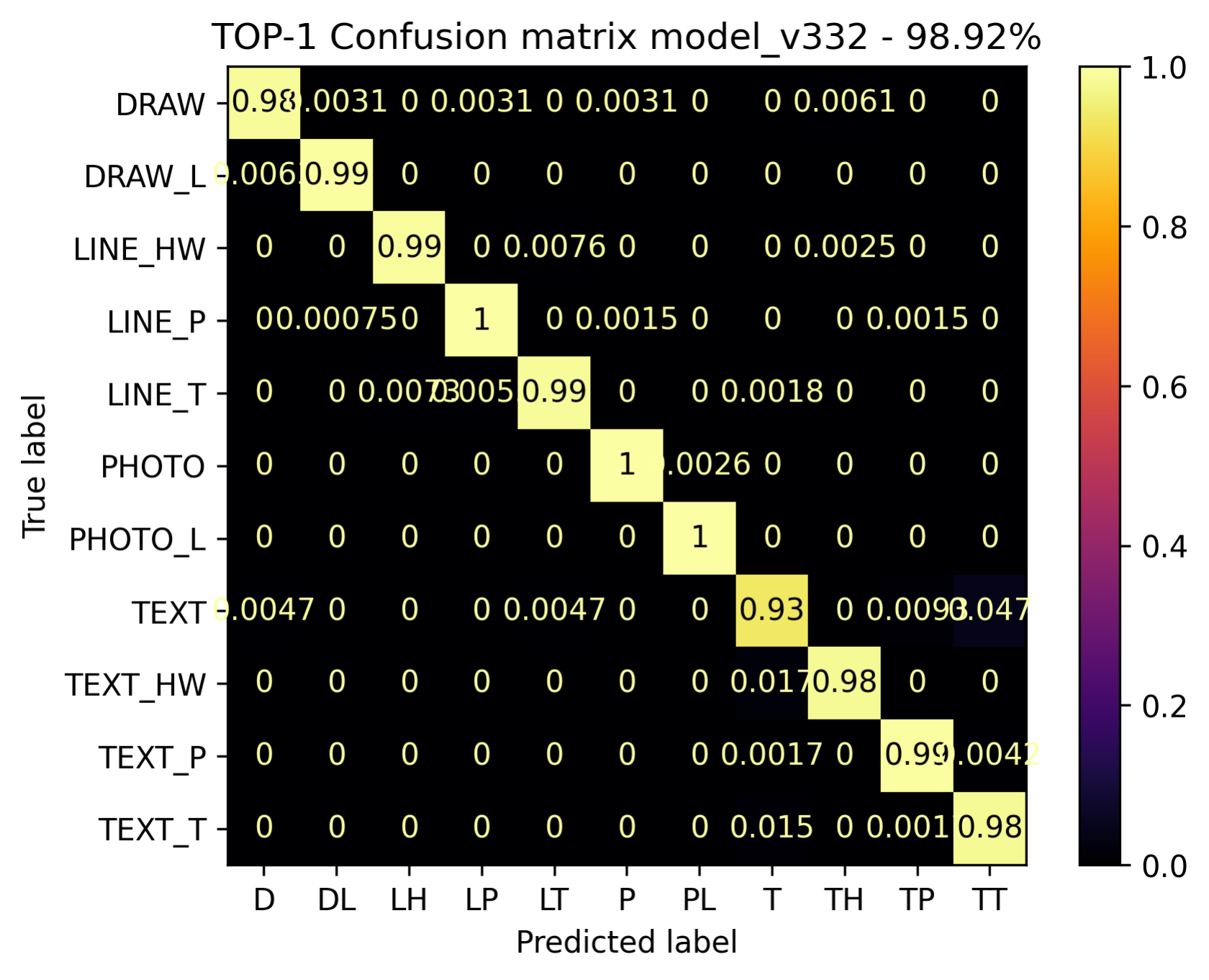}
        \caption{Base 384 — 98.92\%}
        \label{fig:vit-b384}
    \end{subfigure}
    % \hfill
    \begin{subfigure}[b]{0.32\textwidth}
        \centering
        \includegraphics[width=\textwidth]{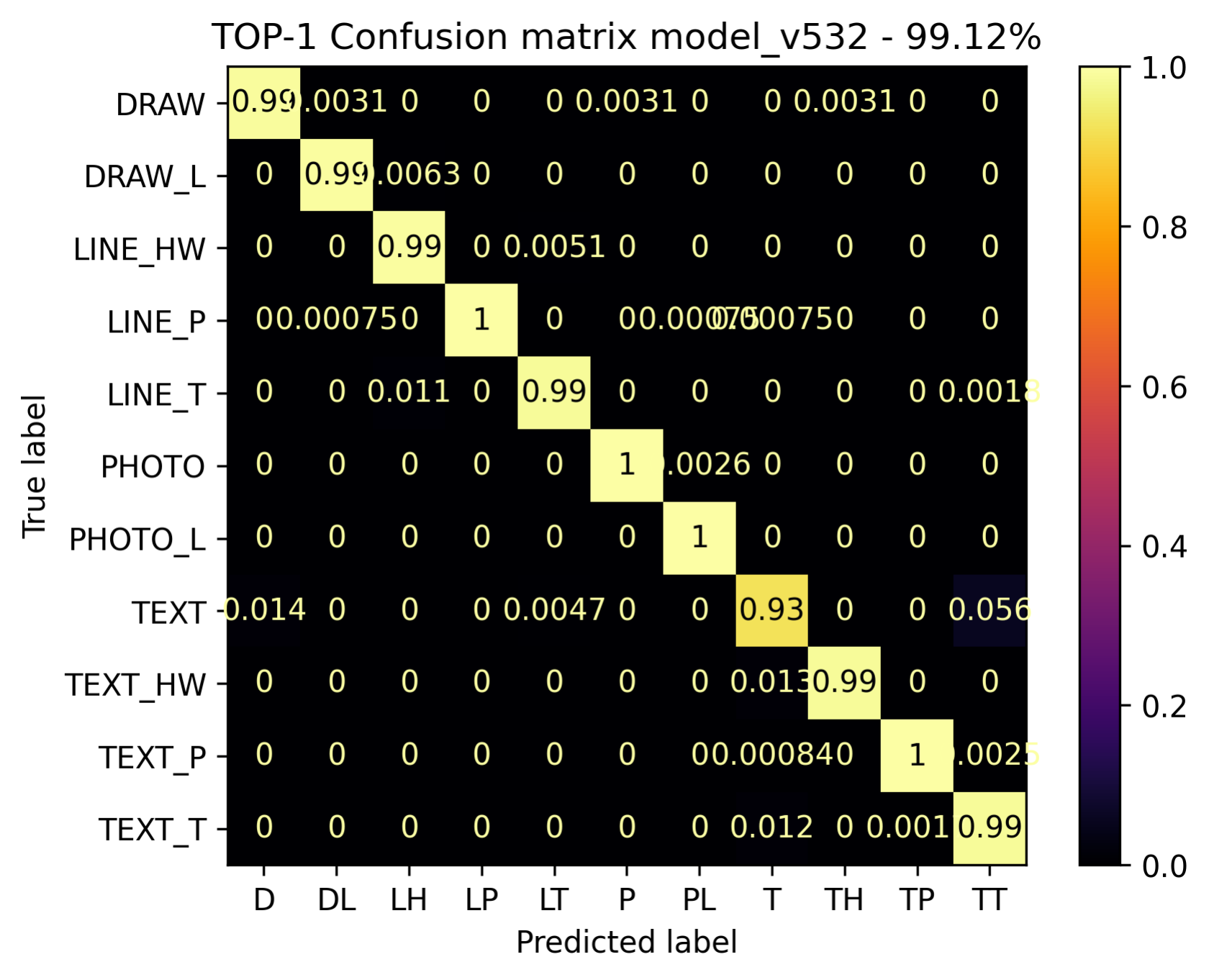}
        \caption{Large 384 — 99.12\%}
        \label{fig:vit-l384}
    \end{subfigure}
    
\end{figure}

\begin{figure}[h]
    \centering
    \caption{Finetuned \ac{clip} best models in each base model group confusion matrices, seven epochs}
    \label{fig:clip-base4}
    
    \begin{subfigure}[b]{0.45\textwidth}
        \centering
        \includegraphics[width=\textwidth]{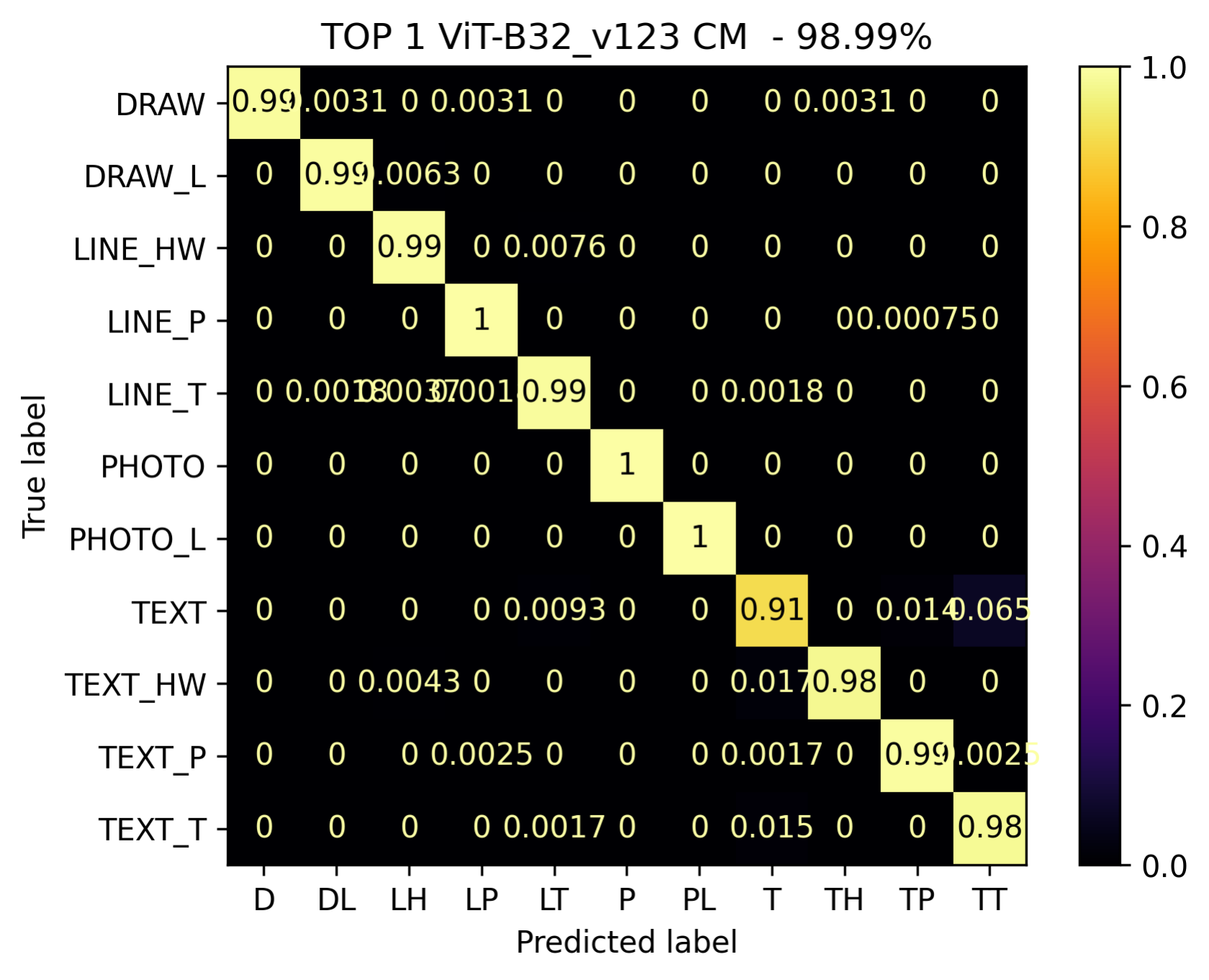}
        \caption{ViT-B/32 averaged — 98.99\%}
        \label{fig:avg-b32}
    \end{subfigure}
    % \hfill
    \begin{subfigure}[b]{0.45\textwidth}
        \centering
        \includegraphics[width=\textwidth]{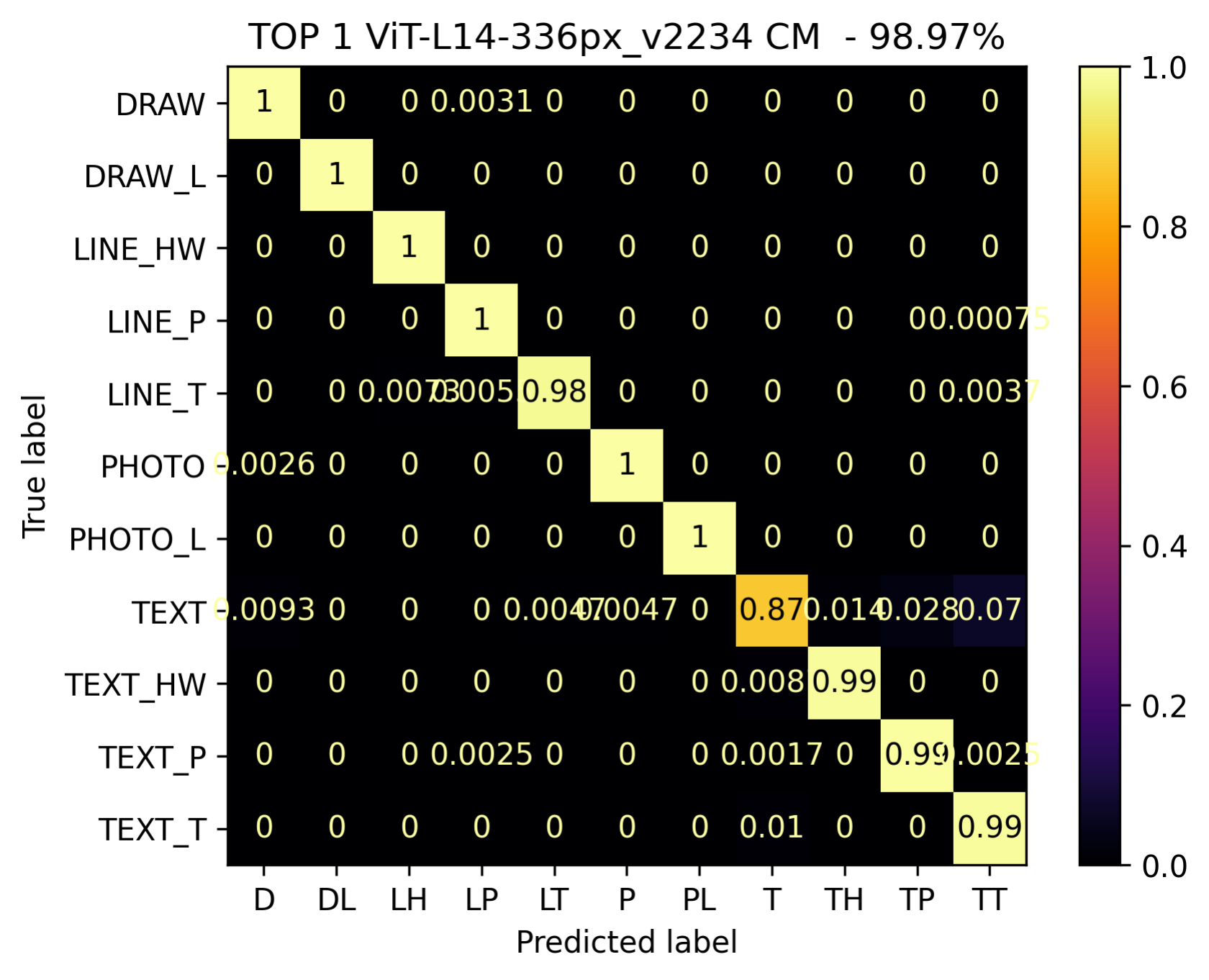}
        \caption{ViT-L/14-336 gemini \ref{tab:long_classification}  — 98.97\%}
        \label{fig:gemini-l336}
    \end{subfigure}

    \hspace{0.5em}

    \begin{subfigure}[b]{0.45\textwidth}
        \centering
        \includegraphics[width=\textwidth]{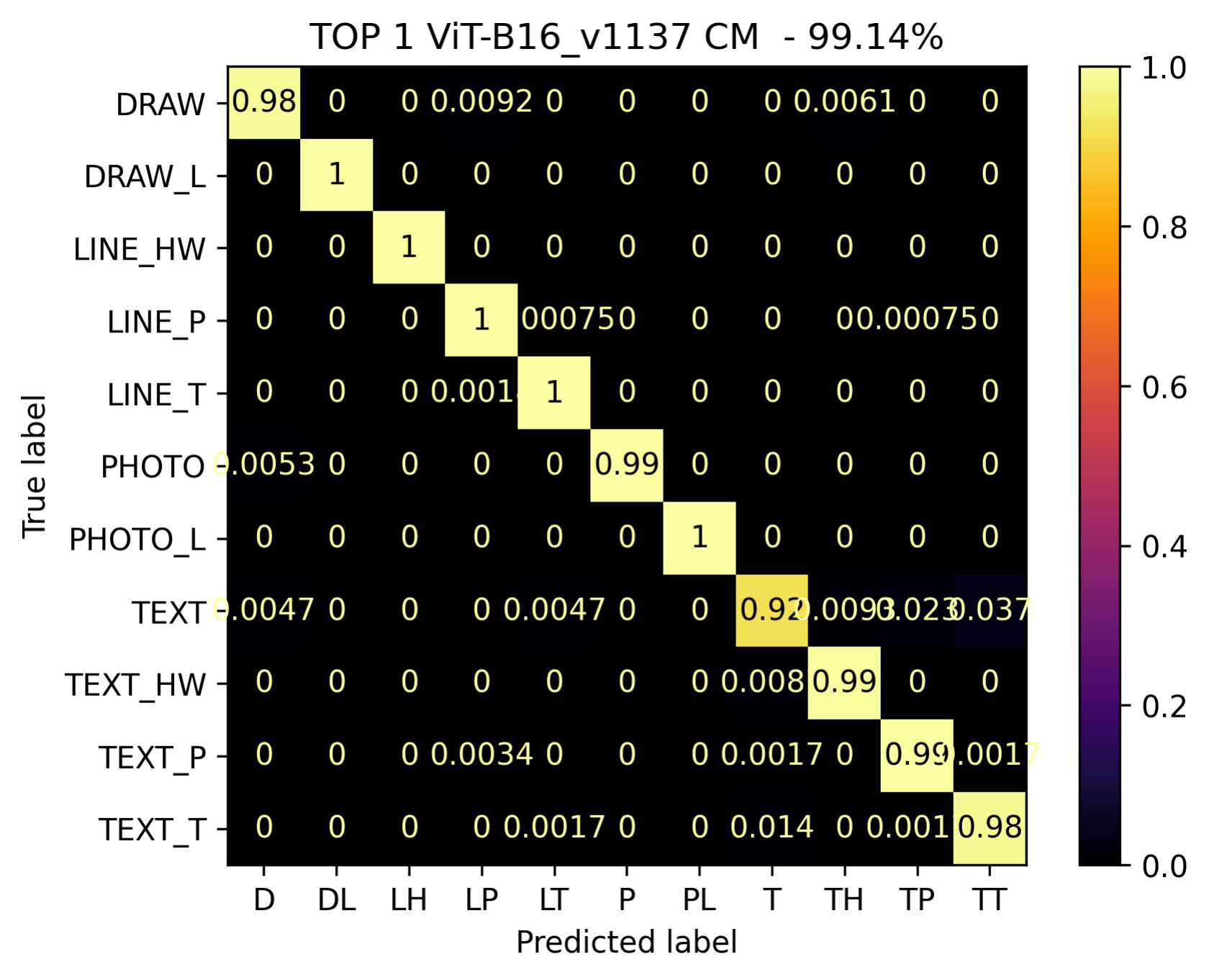}
        \caption{ViT-B/16 mid Table \ref{tab:short_classification} — 99.14\%}
        \label{fig:mid-b16}
    \end{subfigure}
    % \hfill
    \begin{subfigure}[b]{0.45\textwidth}
        \centering
        \includegraphics[width=\textwidth]{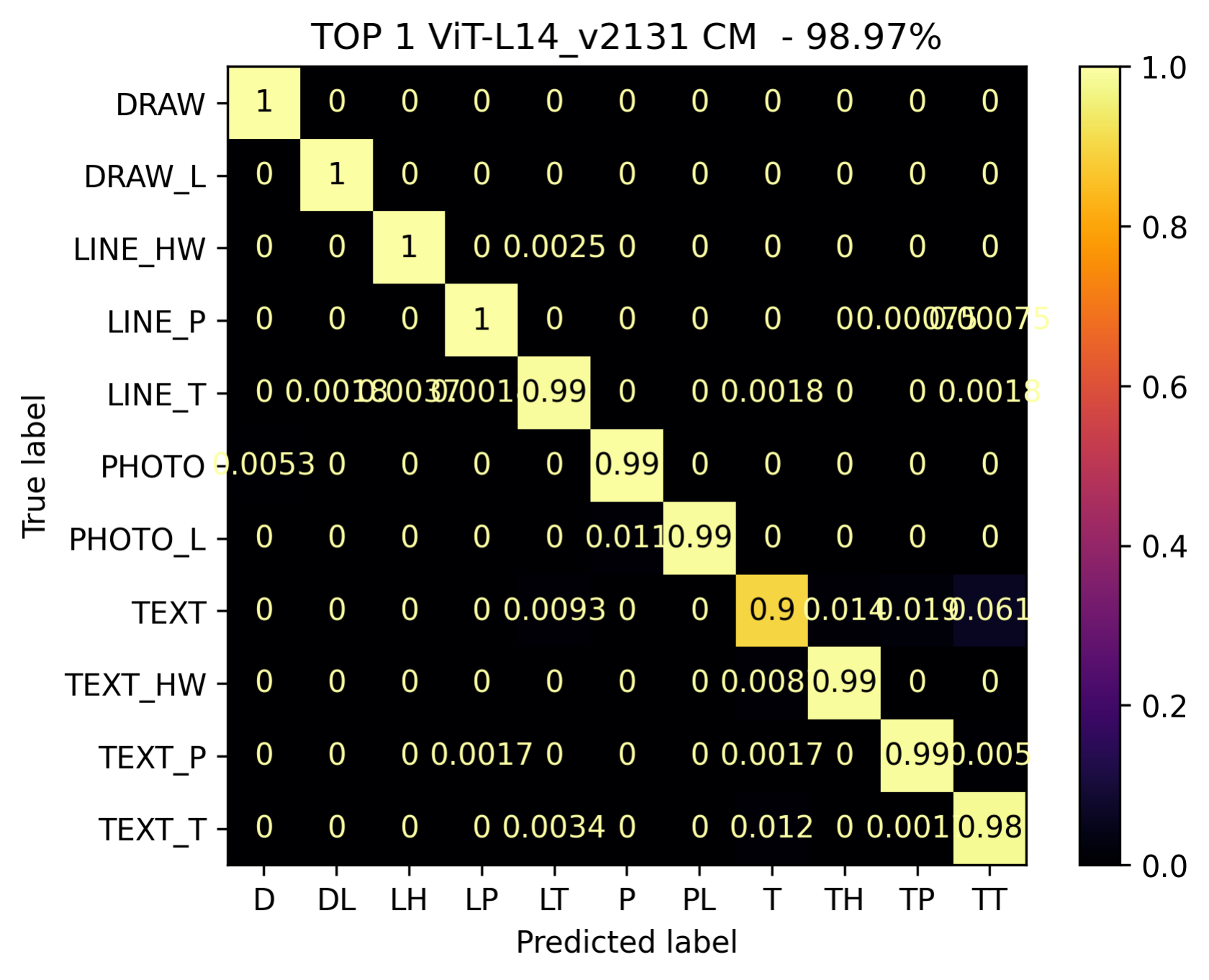}
        \caption{ViT-L/14 init Table \ref{tab:init_classification} — 98.97\%}
        \label{fig:init-l}
    \end{subfigure}
  
\end{figure}

\end{document}